\documentclass[a4paper,11pt]{article}
\title{Liouville Field Theory}
\author{Yu Nakayama}
\usepackage{amsmath}
\usepackage{amssymb}
\usepackage{graphicx}
\usepackage{hyperref}
\def\drawbox#1#2{\hrule height#2pt
        \hbox{\vrule width#2pt height#1pt \kern#1pt
              \vrule width#2pt}
              \hrule height#2pt}

\def\Fund#1#2{\vcenter{\vbox{\drawbox{#1}{#2}}}}
\def\Asym#1#2{\vcenter{\vbox{\drawbox{#1}{#2}
              \kern-#2pt       
              \drawbox{#1}{#2}}}}

\def\funda{\Fund{6.5}{0.4}}
\def\asymm{\Asym{6.5}{0.4}}

\def\symm{\funda\kern-0.4pt\funda}

\allowdisplaybreaks[0]

\newcommand{\sectiono}[1]{\section{#1}\setcounter{equation}{0}}

\def\Fus#1#2#3#4#5#6{
F_{#5#6}\left[
\begin{array}
[c]{cc}%
#3 & #2\\
#4 & #1%
\end{array}
\right]}
\begin{document}

\begin{titlepage}
\thispagestyle{empty}
\begin{flushright}
UT-04-02\\
hep-th/0402009\\
January, 2004 
\end{flushright}

\vskip 1.5 cm

\begin{center}
\noindent{\textbf{\LARGE{ Liouville Field Theory  \\\vspace{0.5cm}
 --- A decade after the revolution 
\vspace{0.5cm}\\
}}} 
\vskip 1.5cm
\noindent{\large{Yu Nakayama}\footnote{E-mail: nakayama@hep-th.phys.s.u-tokyo.ac.jp}}\\ 
\vspace{1cm}
\noindent{\small{\textit{Department of Physics, Faculty of Science, University of 
Tokyo}} \\ \vspace{2mm}
\small{\textit{Hongo 7-3-1, Bunkyo-ku, Tokyo 113-0033, Japan}}}
\end{center}
\vspace{1cm}
\begin{abstract}
We review recent developments (up to January 2004) of the Liouville field theory and its matrix model dual. This review consists of three parts. In part I, we review the bosonic Liouville theory. After briefly reviewing the necessary background, we discuss the bulk structure constants (the DOZZ formula) and the boundary states (the FZZT brane and the ZZ brane). Various applications are also presented. In part II, we review the supersymmetric extension of the Liouville theory. We first discuss the bulk structure constants and the branes as in the bosonic Liouville theory, and then we present the matrix dual descriptions with some applications. In part III, the Liouville theory on unoriented surfaces is reviewed. After introducing the crosscap state, we discuss the matrix model dual description and the tadpole cancellation condition. This review also includes some original material such as the derivation  of the conjectured dual action for the $\mathcal{N}=2$ Liouville theory from other known dualities and the comparison of the Liouville crosscap state with the $c=0$ unoriented matrix model. This is based on my master's thesis submitted to Department of Physics, Faculty of Science, University of Tokyo on January 2004.
\end{abstract}

\end{titlepage}

\tableofcontents
\newpage

\sectiono{Introduction}\label{sec:1}
Everything that has an end has a beginning. Joseph Liouville (1809-1882) studied his equation in order to understand the conformal property of the Riemann surface (especially the uniformization problem). In this review, we study the quantum Liouville theory, which he might or might not realize but naturally emerges in the quantization of the two dimensional gravity (or in the noncritical string theory). The Liouville field theory is defined as an irrational conformal field theory whose action is given by
\begin{equation}
S = \frac{1}{4\pi} \int d^2 z\sqrt{g} \left(g^{ab}\partial_a\phi\partial_b\phi + QR\phi + 4\pi\mu e^{2b\phi}\right),
\end{equation}
where $Q = b + b^{-1}$, and it becomes an essential ingredient in calculating the scattering amplitudes in the noncritical string theory. In the noncritical string theory, the quantum anomaly of the Weyl transformation gives rise to the dynamical Liouville mode. The noncritical string theory is defined perturbatively by the path integration over the Liouville field and the matter fields on various world sheet Riemann surfaces and the subsequent integration of the resultant correlators over the moduli space of Riemann surfaces.

However, this world sheet theory is difficult to solve. A naive perturbative calculation in $\mu$ yields only the limited class of correlators when the inserted momenta satisfy a special relation. This is because the Liouville momentum is conserved with the perturbative calculation in $\mu$, but actually we anticipate that the correlators do not vanish even when the Liouville momentum is not conserved. At the same time, the Liouville field theory is an irrational CFT, so the conventional CFT techniques used to solve the minimal models do not work.

Suppose, however, that we have solved the world sheet Liouville theory somehow. Even if the world sheet theory is solved, the integration over the moduli space of the Riemann surfaces seems hopeless unless a miracle happens. The miracle happens, for example, in the case of the topological gravity, where the integrand becomes a known cohomological object in the moduli space. As a consequence, we can integrate over the moduli space with less trouble. Surprisingly, we believe that this is the case in the Liouville field theory coupled to the minimal model. Indeed, the Liouville field theory coupled to the minimal model (or $c=1$ matter) is believed to be equivalent to the exactly solvable matrix model in the double scaling limit! The supporting argument for the equivalence was given by the discretization of the Riemann surface, which should yield the noncritical string theory.

This was one of the greatest achievements in the matrix revolution era (around 1990).\footnote{For the future reader's sake, we should mention that the ``Matrix Reloaded" and the ``Matrix Revolutions" are titles of the films which have caught on in the year 2003.} We had not only full genus amplitudes for the Liouville field theory coupled to the minimal model (and their generalizations), but also a nonperturbative description of the string theory. However, the matrix model description of the noncritical string theory is limited to the $d\le 2$ dimension only, and nobody at the time was certain whether the discovered nonperturbative effects are truly applicable to the higher dimensional critical string theories. At the same time, why the matrix model yields the nonperturbative description of the Liouville field theory has remained an open question (besides an intuitive discretization of the Riemann surface argument).

A decade has passed since then. 

There have been  steady developments in the Liouville field theory itself. For example, now we have a three-point function formula for general Liouville momenta  (what is called the DOZZ formula), and exact boundary (Cardy) states in the Liouville theory (what is called the FZZT brane and the ZZ brane) at hand. These discoveries have become the prototypical method to study other irrational CFTs and also shed new light on the noncritical string theories. One of the main themes of this thesis is to review these developments after the matrix revolution. 

The second theme of this thesis is to review the nonperturbative physics in the  noncritical string theories. Although some of these were found in the matrix revolution era, the physical meanings of them were not so clear at that time. Now after the second revolution of the string theory, we have a lot of examples of the nonperturbative effects in the critical string theories including
\begin{itemize}
	\item D-brane physics (\ref{6.1}, \ref{6.4}, \ref{6.6}, \ref{10.3})
	\item various dualities (T-duality, S-duality etc.) (\ref{9.1}, \ref{9.2}, \ref{11.3}, \ref{12.1}, \ref{12.3})
	\item black hole physics (\ref{6.5}, \ref{12.1})
	\item string theory under R-R field background (\ref{10.1}, \ref{10.4})
	\item nonperturbative moduli fixing --- R-R potential (\ref{10.4})
	\item connection with the topological string theory (\ref{6.3})
	\item gauge/gravity correspondence (\ref{6.1}, \ref{9.1}, \ref{9.2}, \ref{11.3}, \ref{12.2}, \ref{13.3}, \ref{14.3})
	\item holography principle (\ref{6.1}, \ref{9.1}, \ref{9.2}, \ref{11.3}, \ref{12.2}, \ref{13.3}, \ref{14.3})
	\item geometric transition (\ref{9.2})
	\item rolling tachyon --- Sen's conjecture (\ref{6.1}, \ref{6.2}, \ref{6.4})
	\item closed string theory as a vacuum string field theory (\ref{6.1}, \ref{6.4})
\end{itemize}
to name a few. Recent studies on the matrix-Liouville theory reveal that astonishingly \textit{all} of those listed above are realized in the matrix-Liouville theory.\footnote{Some of them are derived only from the matrix model and not yet from the Liouville theory. It is a challenging problem to obtain them from the Liouville perspective. Also the above list is as of January 2004, so it will  definitely extend in future.} In addition, they are exact and explicit descriptions, which is not always the case in the higher dimensional critical string theories. We have attached section numbers in the above list where we will discuss them in the matrix-Liouville context in this review. Of course, the range we will cover here is limited and we cannot discuss all of them in detail, but we hope that the above partial list is enough to convince that the matrix-Liouville theory is a good starting point to understand nonperturbative physics in the string theory.

Furthermore, by using various dualities, we can discuss some limiting properties of the higher dimensional critical string theory from the noncritical Liouville theory (and its matrix model dual). Some examples are
\begin{itemize}
	\item The universal nonperturbative physics of the supersymmetric gauge theory which can be obtained from the singular conifold. The Liouville partition function reproduces the Veneziano-Yankielowicz term and its graviphoton corrections.
	\item The (double scaling limit of the) little string theory, which is also related to the string theory on a singular Calabi-Yau space.
\end{itemize}
Therefore, the Liouville theory is not only relevant for the lower dimensional ($d\le 2$) string theories but also important in the higher dimensional (and more ``physical") string theories.

To conclude the introduction section, we sketch our organization of the thesis. This thesis consists of three distinct parts. In part I, we review the bosonic Liouville theory and its applications. In part II, we review the supersymmetric extension of the Liouville theory, where we also present the recent proposals for the matrix model duals and their consequent results and applications. In part III, we discuss the (both bosonic and super) Liouville theory on unoriented surfaces, part of which includes the review of the author's original paper.

Each part has several sections;

Part I:

In section \ref{sec:2} and \ref{sec:3}, we review the basic facts on the Liouville field theory and matrix model. A reader who is accustomed to these relatively older subjects (or the reader who has once read the ``Ginsparg and Moore") may wish to skip these sections. In section \ref{sec:4}, we obtain the basic structure constants of the Liouville theory on the sphere (the DOZZ formula), where we have provided their original derivation and the more elegant and useful derivation by Teschner which we will repeatedly use in the following sections. In section \ref{sec:5}, we discuss the boundary states in the Liouville theory. The FZZT brane and the ZZ brane mentioned above are introduced. In section \ref{sec:6}, we review some applications of the bosonic Liouville theory.

Part II:

In section \ref{sec:7}, we review the bulk physics of the $\mathcal{N}=1$ super Liouville theory. In section \ref{sec:8}, the boundary $\mathcal{N}=1$ super Liouville theory is discussed. In section \ref{sec:9}, we present the matrix dual of the $\mathcal{N}=1$ super Liouville theory by using the results from section \ref{sec:7} and \ref{sec:8}. In section \ref{sec:10}, we review the applications of the $\mathcal{N}=1$ super Liouville theory and its matrix model dual. In section \ref{sec:11}, the bulk and boundary physics of the $\mathcal{N}=2$ super Liouville theory is discussed. The application of the $\mathcal{N}=2$ super Liouville theory is reviewed in section \ref{sec:12}, where we include our original explanation of the conjectured duality of the $\mathcal{N}=2$ super Liouville theory.

Part III:

In section \ref{sec:13}, we discuss the bosonic unoriented Liouville theory and its matrix model dual. In section \ref{sec:14}, we review the unoriented $\mathcal{N}=1$ supersymmetric Liouville theory and its matrix model dual.

 In the concluding section \ref{sec:15}, we present our concluding remarks and the future outlook on the Liouville theory. It is complementary to this introduction in a sense and the reader may directly jump to the concluding section before he/she begins to read the main text. This thesis has two appendix sections. In section \ref{sec:A}, we collect our conventions and useful formulae. In section \ref{sec:B}, we collect miscellaneous topics which are helpful to understand the arguments in the main text or to follow some technical calculations.

Finally, this review is partially based on the author's note of the informal seminar held at Tokyo University. The papers used in the seminar were \cite{Seiberg:1990eb}, \cite{Fateev:2000ik}, \cite{Zamolodchikov:2001ah}. The author would like to thank the organizers, all the participants and lecturers in the seminar.

\subsection{Literature Guide}
At the end of each section, we provide a literature guide in order to show references of the subjects discussed in the section. The purpose of the literature guide section is twofold. We provide not only the direct sources of the arguments in the main text but also the related papers whose contents we cannot cover in the main text because of the lack of space or other reasons (especially in the application sections). In order to avoid the overlap, we only refer to the direct (not necessarily original) source of the subjects and the formulae whose derivations are omitted in the main text, though it does not mean at all that we have not borrowed other paper's results or explanations to make the argument clearer. 

There exist many great reviews on the Liouville theory and matrix model (or noncritical string theory). The older reviews (in the matrix revolution era) are \cite{Seiberg:1990eb}, \cite{Kutasov:1991pv}, \cite{Martinec:1991kn}, \cite{Klebanov:1991qa}, \cite{Ginsparg:is}, \cite{DiFrancesco:1993nw}, \cite{Jevicki:1993qn}, and \cite{Polchinski:1994mb} to name a few. The more recent (after the DOZZ formula) reviews are \cite{Teschner:2001rv}, \cite{Ponsot:2003ju}, \cite{Mukhi:2003sz}, \cite{Alexandrov:2003ut}. Also, the first half of the following papers include excellent reviews on the boundary (super) Liouville theory, \cite{Kostov:2003uh}, \cite{Douglas:2003up}.

For a general string theory background, we refer to GSW \cite{Green:1987sp,Green:1987mn}, Joe's big book \cite{Polchinski:1998rq,Polchinski:1998rr}, and the author's favorite Polyakov's book \cite{Polyakov:1987ez}.

\part{Bosonic Liouville Theory}
\sectiono{Basic Facts 1: World Sheet Theory}\label{sec:2}
We review the basic facts on the Liouville theory from the world sheet perspective in this section. The good references are \cite{Seiberg:1990eb}, \cite{Ginsparg:is}, \cite{DiFrancesco:1993nw}, \cite{Klebanov:1991qa}, \cite{Martinec:1991kn}. A reader who has once read these review articles may skip this section (and the next section) and jump into section \ref{sec:4}. 

The organization of this section is as follows.
In section \ref{2.1}, we derive the Liouville action from the quantization of the noncritical string (two dimensional quantum gravity), and discuss its basic properties as a CFT. In section \ref{2.2}, we reinterpret the Liouville theory from the critical string theory propagating in a nontrivial background with a linear dilaton and a tachyon condensation. In section \ref{2.3}, we quantize the Liouville theory canonically and study its properties. Also we perform the semiclassical path integration and discuss the semiclassical properties of the Liouville theory. In section \ref{2.4}, we briefly review the rolling tachyon system and discuss its connection with the Liouville theory. In section \ref{2.5}, we review the ground ring structure of the $c=1$ (which means that the target space is two dimensional) noncritical string theory. 
\subsection{2D Quantum Gravity}\label{2.1}

Following  David-Distler-Kawai (DDK) \cite{David:1988hj}, \cite{Distler:1989jt,Distler:1990jv}, we introduce the Liouville action from the Polyakov formalism \cite{Polyakov:1981rd} of the 2D (world sheet) gravity. We consider the two dimensional quantum gravity (or the quantization of the bosonic string). In the Polyakov formalism, the starting point is the partition function which is given by 
\begin{align}
Z = \int [\mathcal{D}g][\mathcal{D}X] e^{-S[X;g] - \mu_0 \int d^2 z \sqrt{g}},
\end{align}
where any matter field $X$ is allowed at this point, but we take $d$ free bosons for simplicity and definiteness. Then the matter action can be written as
\begin{equation}
S[X,g] = \frac{1}{4\pi}\int d^2 z \sqrt{g} g^{ab}\partial_aX^I\partial_b X^I. 
\end{equation}

While the path integral measures for the metric and bosons are invariant under the world sheet diffeomorphism, they are \textit{not} invariant under the Weyl transformation $ g_{ab}\to e^{\sigma}g_{ab}$. The anomaly for this transformation is given by
\begin{equation}
\mathcal{D}_{e^\sigma g} X = e^{\frac{d}{48\pi}S_L(\sigma)}\mathcal{D}_gX\label{eq:ano1}
\end{equation}
where $S_L$ is the famous (unrenormalized) Liouville action whose precise form is
\begin{equation}
S_L(\sigma) = \int d^2 z \sqrt{g}\left(\frac{1}{2}g^{ab}\partial_a\sigma\partial_b\sigma + R\sigma +\mu e^{\sigma}\right).
\end{equation}

Similarly the path integral measure for the metric is not invariant under the Weyl transformation. To carry out the path integral over the metric, we decompose the fluctuation of the metric $\delta g_{ab}$ into the diffeomorphism $v_a$, Weyl transformation $\sigma$ and the moduli $\Upsilon$. Since the measure and the action is invariant under the diffeomorphism by definition, we can regard it as a gauge symmetry. Dividing the path integral measure by the gauge (diffeomorphism) volume, we are left with the integration over the Weyl transformation freedom and the moduli. The Jacobian for this change of variables can be calculated via the Fadeev-Popov method, which is given by
\begin{equation}
\int \mathcal{D}b\bar{b}\mathcal{D}c\bar{c} e^{-\int d^2z \sqrt{g} ( b\bar{\nabla}c +\bar{b}\nabla \bar{c})}.
\end{equation}
The Weyl transformation of this measure becomes
\begin{equation}
\mathcal{D}_{e^\sigma g}(bc) = e^{-\frac{26}{48\pi} S_L(\sigma)}\mathcal{D}_g(bc).\label{eq:ano2}
\end{equation}

If $d=26$ then we are dealing with the critical string. In this particular case (after setting the cosmological constant to be zero), we can also regard the Weyl symmetry as a gauge symmetry and ignore it. However in the more general case, we cannot ignore its freedom because of the above anomaly (\ref{eq:ano1},\ref{eq:ano2}). Thus, in the conformal gauge $g_{ab} = e^\phi \hat{g}_{ab}$, the partition function of the 2D quantum gravity (with matters) can be written as
\begin{equation}
Z = \int d\Upsilon \mathcal{D}\phi_{e^\phi \hat{g}}\mathcal{D}(bc)_{e^\phi \hat{g}}\mathcal{D}X_{e^\phi \hat{g}} e^{-S[X,\hat{g}]-S[bc,\hat{g}]}.
\end{equation}
Naively speaking, as the Liouville action emerges from the path integral measure, only we have to do is to integrate over the Liouville mode. However, there is a subtlety here. The problem is the path integral measure for this Liouville field. Since it has been constructed diffeomorphism invariantly, the measure satisfies $||\delta\phi||^2_g = \int d^2z \sqrt{g}(\delta\phi)^2 = \int d^2z \sqrt{\hat{g}}e^\phi (\delta\phi)^2 $. It is very inconvenient to use this measure, for it is not the Gaussian measure nor invariant under the parallel translation in the functional space. Then we would like to transform it to the standard Gaussian measure:
\begin{equation}
|| \delta\phi||^2_{\hat{g}} = \int d^2 z \sqrt{\hat{g}}(\delta\phi)^2. 
\end{equation}
To do so, we need to obtain the Jacobian of this transformation and include it into the Liouville action. Since it is difficult to obtain this Jacobian from the first principle, we will guess, following DDK, the form of the ``renormalized Liouville action" by assuming its ``locality", ``diffeomorphism invariance" and ``conformal invariance".\footnote{The derivation of the Jacobian from the functional integral has been discussed in \cite{Mavromatos:1989nf}, \cite{D'Hoker:1990md,D'Hoker:1991ac}. The author would like to thank E.~D'Hoker for calling his attention to these papers.} 

This assumption leads to the following form of the action
\begin{equation}
S = \frac{1}{4\pi} \int d^2 z\sqrt{g} \left(g^{ab}\partial_a\phi\partial_b\phi + QR\phi + 4\pi\mu e^{2b\phi}\right).
\end{equation}
We would like to determine the unknown parameters $Q$ and $b$. First, by considering that the choice of the residual metric $\hat{g}_{ab}$ has been arbitrary, we find that the whole theory should be invariant under $\hat{g}_{ab}\to e^\sigma \hat{g}_{ab}$ and $\phi \to \phi - \sigma/2b $. For this to be a symmetry of the theory, the total central charge of the system should be zero from the previous argument, namely,
\begin{equation}
c_{tot} = c_\phi + c_X + c_{gh} = 0
\end{equation}
should hold. From this, we find $c_\phi = 26-d $. On the other hand, the central charge of $\phi$ can be calculated irrespective of $\mu$ by the Coulomb gas representation,\footnote{See appendix \ref{b-2} for the actual calculation.} which is given by
\begin{equation}
c_\phi = 1+ 6Q^2.
\end{equation}
Therefore we obtain
\begin{equation}
Q = \sqrt{\frac{25-d}{6}}. \label{eq:Q}
\end{equation}

Then we demand the conformal invariance. For this theory to be consistent as a conformal field theory, it is necessary that ``the interaction term" $e^{2b\phi}$ is a $(1,1)$ tensor. The calculation by the Coulomb gas representation shows
\begin{equation}
\Delta = b\left(Q-b\right) = 1,
\end{equation}
so we obtain the famous formula $Q = b+b^{-1}$. Furthermore, we notice that for the cosmological constant (the actual metric) to be real, the matter central charge must satisfy $ c_m \le 1$ ($c=1$ barrier).

We have some comments on the $c_m>1$ noncritical string theory. In this case, as we have seen, there should be a phase transition ($c=1$ barrier) from the DDK approach. Then the $c_m>1$ theory is believed to be a different continuum theory from the Liouville one. For example, Polyakov \cite{Polyakov:1998ju} conjectured that, in $1<c_m<25$, it becomes a string theory propagating in the warped space-time. On the other hand, when $c_m=25$, there is an interesting conjecture and it is important for the later application, so we introduce the conjecture here.\footnote{As far as the author knows, this was first discussed in \cite{Das:1988ds}.}

Formally, when we substitute $d=25$ into (\ref{eq:Q}), we obtain $Q=0$. This means $b=i$. Since the reality of the metric is lost for this value of $b$, we ``Wick rotate" the Liouville direction $\phi \to i \phi$. Then we observe that the kinetic term of $\phi$ becomes minus that of the ordinary boson, so it is natural to interpret $\phi$ as the ``time direction". Furthermore, setting the cosmological constant $\mu$ to be zero, we have an interesting interpretation ``the noncritical string propagating in the 25D Euclidean space is equivalent to the critical string propagating in the 26D Minkowski space-time". This mechanism seems an elegant scenario which naturally generates the time-like negative metric into the whole story, which is very suggestive and impressive. On the other hand, if we take the cosmological constant to be finite, we can interpret that the Liouville potential represents the world sheet description of the rolling tachyon. We will discuss in the later section whether and how this ``analytic continuation" actually works.

\subsection{2D Critical String Interpretation}\label{2.2}
There is another interpretation of the Liouville theory. In this section, we interpret the $c=1$ Liouville theory as a two dimensional \textit{critical} string theory. To begin with, we consider the sigma model description of the critical string in the general dimension with arbitrary backgrounds
\begin{equation}
 S = \frac{1}{4\pi} \int d^2z \sqrt{g}\left(g^{ab}G_{\mu\nu}(X)\partial_aX^\mu\partial_bX^\nu+2\Lambda^2T(X) +\frac{1}{2}R \Phi(X)\right),
\end{equation}
where $\Lambda$ is the cut-off scale of the world sheet, and we have assumed the Kalb-Ramond field $B$ is zero. For this action to satisfy the conformal invariance so that the background is consistent with the string equation of motion, the following beta functions \cite{Friedan:1980jf}, \cite{Callan:1985ia}, \cite{Fradkin:1985pq,Fradkin:1985ys} (in the first order approximation with respect to $\alpha'$) should vanish
\begin{align}
\beta_{\mu\nu}(g) &= R_{\mu\nu}+\nabla_\mu\nabla_\nu\Phi - \frac{1}{4}\partial_\mu T \partial_\nu T \cr
\beta(\Phi) &= -R + (\partial_\mu\Phi)^2 - \nabla^2\Phi + \frac{2(D-26)}{3} -T^2 +\frac{1}{6}T^3  \cr
\beta(T) &= \nabla^2 T -\partial_\mu\Phi\partial^\mu T+ 4T-T^2.\label{eq:beta}
\end{align}
These equations are equivalent to the equation of motions (in the string frame) which can be derived from the following effective action whose form is determined by the string tree level scattering amplitudes,
\begin{equation}
S = \int d^Dx \sqrt{G}e^{-\Phi} \left[R+(\partial_\mu\Phi)^2-\frac{2(D-26)}{3}-\frac{1}{4}(\partial_\mu T)^2 + T^2 -\frac{1}{6} T^3\right].
\end{equation}

Let us compare this sigma model with the $c=1$ Liouville action,
\begin{equation}
S_{L} = \frac{1}{4\pi} \int d^2z \sqrt{g}\left[g^{ab}\partial_a\phi\partial_b\phi + 2R\phi + 4\pi\mu e^{2\phi} +g^{ab}\partial_aX\partial_bX \right].
\end{equation}
The first thing to note is, when $\mu=0$, this can be indeed interpreted as the above sigma model where $d=2$, $G_{\mu\nu}=\eta_{\mu\nu}$, $\Phi = 4\phi$ and $T=0$. Next, we consider the $\mu \neq 0$ case. In this case, we can regard it as a sigma model with a further tachyon background $\Lambda^2 T = 2\pi \mu e^{2\phi}$. While this background satisfies the naive tachyon mass-shell equation of motion $\nabla^2 T -\partial_\mu\Phi\partial^\mu T+ 4T = 0 $, it does not satisfy the one-loop corrected beta function equation (\ref{eq:beta}). This is obvious since the Einstein equation does not hold under the flat space-time with a general scalar field expectation value. This means that under the naive perturbative treatment in $\mu$, the Liouville theory does not yield the conformal background at least at the one-loop level. However, as has been discussed in the last section, the Liouville theory is by definition conformally invariant, so this background should be conformally invariant (up to an arbitrariness of the renormalization scheme and the freedom of field redefinition), at all orders in $\alpha'$. The origin of this inconsistency is believed to lie in the failure of the perturbative treatment of the Liouville potential and the first order approximation of the beta function equations \cite{Martinec:1991kn}.\footnote{It is also argued in \cite{Tseytlin:1990mz} that in the lower dimension considered here, \eqref{eq:beta} should be modified to account for the kinematic restriction. The Liouville background solves the modified equation. The author would like to thank A.~Tseytlin for calling his attention to the paper.}

As is often said, compared with the effective action which is restricted to the massless sector, the effective action which includes the tachyon sector should include all the massive fields at the same time in order to be consistent. This is because there are cubic interactions which generate these massive states, so we should integrate them out by taking the extremum point of the effective action instead of simply setting them zero. Furthermore, in this case, the higher derivative terms cannot be neglected, for the space-time fluctuation scale of the tachyon field considered here is just the order of $\alpha'$. Considering these effects, we find that it is not easy to derive the effective equation of motion of the tachyon field as is studied in the string field theory (SFT). However, one thing we have learned from the Liouville theory is that this Liouville background is one of the consistent tachyon backgrounds at all orders of the world sheet $\alpha'$ and the string coupling $g_s$. This plays an important role in the relationship with the rolling tachyon which we will discuss later.

\subsection{Semiclassical Liouville Theory}\label{2.3}
In this section, we quantize the Liouville theory semiclassically \cite{Curtright:1982gt}, \cite{Seiberg:1990eb}, \cite{Zamolodchikov:1996aa}, \cite{Teschner:2001rv} (See \cite{D'Hoker:1982er,D'Hoker:1983is}, \cite{Curtright:1991qp} for other quantization approaches. One more quantization approach which has a long history is the quantum B\"acklund transformation though we will not review it here. Standard review articles are collected in the literature guide section). 

First, let us discuss the canonical quantization of the Liouville theory. By the conformal transformation $z= e^{-iw}$, we map the complex $z$ plane to the cylinder: $w=\sigma+i\tau = \sigma + t$ and perform the canonical quantization on the cylinder. The Lagrangian is given by
\begin{equation}
L = \frac{1}{4\pi} \partial_a \phi \partial^a \phi + \mu e^{2b\phi},
\end{equation}
so the conjugate momentum of $\phi$ becomes
\begin{equation}
\Pi = \frac{\partial_t \phi}{2\pi}.
\end{equation}
The equation of motion is given by
\begin{equation}
(\partial_t^2 - \partial_\sigma^2)\phi = -4\pi \mu b e^{2b\phi},
\end{equation}
and the equal time commutation relation is defined as
\begin{equation}
[\phi(\sigma),\Pi(\sigma')] = i\delta(\sigma-\sigma').
\end{equation}
When we Fourier transform the canonical field\footnote{Note $a_n$ and $b_n$ are time dependent Heisenberg representation operators.} as
\begin{eqnarray}
\phi &=& q+ i \sum_{n\neq 0} \frac{1}{n}[a_n e^{-in\sigma} + b_n e^{in\sigma}] \cr
\Pi &=& p + \sum_{n \neq 0} [a_n e^{-in\sigma} + b_n e^{in\sigma}],
\end{eqnarray}
we can rewrite the equal time commutation relation as
\begin{eqnarray}
[p,q] &=& -i \cr
[a_n,a_m] &=& \frac{n}{2} \delta_{n,-m} \cr
[b_n,b_m] &=& \frac{n}{2} \delta_{n,-m}.
\end{eqnarray}
Using this Fock representation, the Hamiltonian of the system is given by
\begin{equation}
H = \frac{1}{2}p^2 + 2\sum_{k >0}[a_{-k}a_{k} + b_{-k}b_k]_r + \mu \int_0^{2\pi} d\sigma [e^{2b\phi(\sigma)}]_r,
\end{equation}
where $[\cdots]_r$ means the need for quantum corrections or renormalizations.

The conventional step of the canonical quantization is to obtain eigenvalues and eigenstates of this Hamiltonian. For the time being, we ignore excitations of oscillator modes (or we can easily extend the result here to the case where we ignore only the interactions between oscillator modes) and just consider the zero-mode Hamiltonian. This approximation is often called ``minisuperspace approximation" after an analogy to the similar approximation method for the four dimensional canonical quantization of gravity. In this approximation the Hamiltonian can be written as (with the zero-point energy $N$)
\begin{equation}
H_0 = -\frac{1}{2} \partial_q^2 + 2\pi \mu e^{2bq} + N,
\end{equation}
where we have assumed that the Hilbert space of this theory is $L^2$,\footnote{Strictly speaking this is not true. We of course allow plane wave normalizable states.} and the conventional representation of the canonical commutation relation is given by $ p = -i\partial_q $ under the conventional inner product $\langle a| b \rangle =\int dq \Psi^*_a(q) \Psi_b(q)$.\footnote{Beyond the minisuperspace approximation, though the entire Hilbert space seems to be $L^2 \otimes F$ naively, this is not probably true \cite{Teschner:2001rv}.}

Eigenvalues of the Hamiltonian for the plane wave normalizable states are labeled by the continuous parameter $p>0$. As $q\to -\infty$, the interaction vanishes, so we can write the wavefunction as
\begin{equation}
\Psi_p(q) \sim e^{2i pq} + R(p) e^{-2i pq}.
\end{equation}
The reason that the Liouville ``momentum" is restricted to the positive value $p >0$ is that an incoming wave uniquely determines the amplitude of the reflected wave because the wave should damp in the potential barrier $q \to \infty$. The normalization is given by $\langle p | p' \rangle = \pi \delta(p-p')$.
\footnote{Usually, it is given by $2\pi \delta(p-p')$. However, the wave considered here is $e^{2ipq}$, so $\delta(2p-2p') = \delta(p-p')/2$.} The wave equation in this potential is solved analytically and the reflection amplitude in the $b\to0$ limit is given by
\begin{equation}
R(p) = - (\pi\mu b^{-2})^{-\frac{2ip}{b}} \frac{\Gamma(1+2ib^{-1}p)}{\Gamma(1-2ib^{-1}p)}. \label{eq:cref}
\end{equation}

\begin{figure}[htbp]
	\begin{center}
	\includegraphics[width=0.6\linewidth,keepaspectratio,clip]{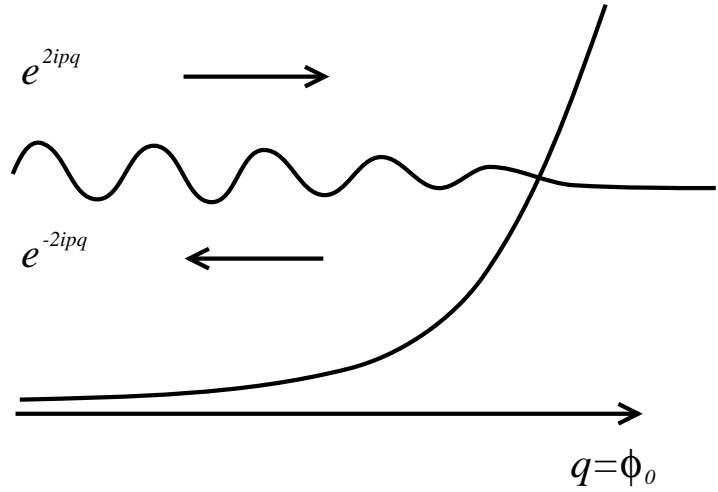}
	\end{center}
	\caption{An incoming wave is reflected from the Liouville wall and becomes an outgoing wave with a reflection coefficient $R(p)$.}
	\label{fig1}
\end{figure}

Let us consider here the state-operator mapping in order to see the relation of these states to the correlation functions on the $z$ plane. The claim is that the operator on the $z$ plane corresponding to the state which has a ``momentum" $p$ is $e^{2\alpha \phi}$, where $\alpha = \frac{Q}{2} + i p$. To see this, note that the weight of this operator on the $z$ plane is given by the CFT result, $L_0 = \alpha(Q-\alpha) $ and the relation to the Hamiltonian is given by $H_0 = L_0 + \bar{L}_0 -\frac{c}{12}$. Substituting $\alpha = \frac{Q}{2} + i p $ and $c= 1+6Q^2$ into this expression, we obtain $H_0 = 2p^2 -\frac{1}{12}$. This is just the energy which has an asymptotic momentum $p$. At the same time, the requirement of being a CFT has fixed the zero-point energy of the system to be $N = -\frac{1}{12}$. In addition, since the $p < 0 $ states are identified with the $p>0$ states up to an overall constant, the vertex operator $V_\alpha = e^{2\alpha \phi}$ is identified with $V_{Q-\alpha} $, too.

Taking this state-operator mapping into consideration, we are inclined to conclude that only states which can be written as $\alpha = \frac{Q}{2} + i p$, where $p$ is real, exist in the theory. However, what happened to the cosmological constant operator $e^{2b\phi}$, for instance? Is it not included in the theory? This is the very peculiar point in the state-operator mapping in the Liouville theory. Generally, the state corresponding to the operator $e^{2\alpha \phi}$, even if $\alpha$ is real, is believed to exist but considered as nonnormalizable one. Nonnormalizable as it is, there is a difficulty concerning the completeness of the states for example, but introducing the cut-off into $\phi$ removes some of the difficulties. In the minisuperspace approximation, this is equivalent to considering the wavefunction which formally has an imaginary momentum. However, we do not allow every imaginary momentum state. Taking into account that the asymptotic form of the wavefunction in the $q\to -\infty$ limit should be dominated by $e^{2ipq} $, the condition $ \mathrm{Re}(\alpha) \le \frac{Q}{2}$ is required. This is what is called the Seiberg bound \cite{Seiberg:1990eb}.

Seiberg \cite{Seiberg:1990eb} has named the normalizable states (operators) as nonlocal and the nonnormalizable states (operators) as local. The root of this naming is the behavior of the world sheet metric ($e^{2b\phi}$) under the inserted source in the WKB (semiclassical) approximation as we will see in the following. In fact, the source corresponding to the nonnormalizable states can be seen as the local curvature singularity from the world sheet point of view.

Leaving the canonical quantization, we next try to carry out the path integration by the WKB (semiclassical) approximation to calculate the Liouville correlation functions. The semiclassical limit corresponds to the $b \to 0$ limit. Before doing this, let us first consider the Ward-Takahashi (WT) identity which can be applied to any $n$-point function. The definition of the $n$-point function is
\begin{equation}
\langle e^{2\alpha_1 \phi} e^{2\alpha_2\phi} \cdots e^{2\alpha_N\phi} \rangle = \int \mathcal{D} \phi e^{2\alpha_1 \phi} e^{2\alpha_2\phi} \cdots e^{2\alpha_N\phi} e^{-S},
\end{equation}
where the action is given by
\begin{equation}
S = \frac{1}{4\pi} \int d^2 z\sqrt{g} \left(g^{ab}\partial_a\phi\partial_b\phi + QR\phi + 4\pi\mu e^{2b\phi}\right).
\end{equation}
Shifting $\phi$ as $\phi \to \phi - \frac{\log\mu}{2b} $, we can remove the $\mu$ dependence in front of the Liouville potential completely because we have made the path integral measure invariant under the translation as we have considered in the last section.
Then using the Gauss-Bonnet theorem which states $\frac{1}{4\pi} \int d^2z \sqrt{g}R = 2-2g$, we obtain the following exact $\mu$ dependence of the correlation function on the genus $g$ Riemann surface 
\begin{equation}
\langle e^{2\alpha_1 \phi} e^{2\alpha_2\phi} \cdots e^{2\alpha_N\phi} \rangle_g \propto \mu^{\frac{(1-g)Q-\sum_i \alpha_i}{b}} \label{eq:ewt},
\end{equation}
which is what is called the Knizhnik-Polyakov-Zamolodchikov (KPZ) scaling law \cite{Knizhnik:1988ak}.

Interestingly, if we consider the Liouville theory as the string theory, the string coupling constant $g_s$ is multiplied to the genus $g$ correlation function as $g_s^{2(g-1)}$. However, looking at the way the power of the string coupling constant enters into the amplitude, we observe that this is just the same way in which the power of $\mu^{Q/b}$ does (without vertices). That is to say, the partition function of the Liouville theory does not depend independently on $g_s$ and $\mu$, but it actually depends only on the particular combination $\mu_r^{-2} =g_s^2 \mu^{-Q/b}$. In addition, the power of the $\mu_r$ is determined as $\mu_r^{2-2g} $ in the usual manner. \footnote{Although for $g=0,1$, $\log$ correction is needed as the partition function diverges.}

Returning back to the semiclassical calculation on the sphere, we try to find the classical solution $\phi_{cl}$ of the equation of motion and substitute back into the integrand of the path integral so that we obtain the zeroth order approximation of the correlation functions. The classical equation of motion (or the saddle point equation) for $\langle e^{2\alpha_1 \phi} e^{2\alpha_2\phi} \cdots e^{2\alpha_N\phi} \rangle $ is given by
\begin{equation}
 \frac{2}{\pi} \partial\bar{\partial} \phi - \frac{1}{4\pi}RQ - 2\mu b e^{2b\phi} + \sum_i2\alpha_i \delta(z-z_i) = 0.
\end{equation}
To see the necessary condition of the existence of the real solution of this equation, we integrate this equation over the world sheet. Setting $\mu >0$ and applying the Gauss-Bonnet theorem, we obtain the following inequality,
\begin{equation}
 Q - \sum_i \alpha_i < 0.
\end{equation}
Under this condition, we will obtain the real semiclassical solution. If we can solve the Liouville equation with source, substituting the solution into the action gives the semiclassical correlation functions. However, the actual calculation has been done only for the three-point function. The calculation is complicated and not so illuminating, so we just quote the result \cite{Zamolodchikov:1996aa}. Setting $\alpha_i = \eta_i/b$ and $\langle e^{2\alpha_1 \phi} e^{2\alpha_2\phi}e^{2\alpha_3\phi} \rangle = \exp(-S_c(\eta_1,\eta_2,\eta_3)/b^2)$, we find
\begin{eqnarray}
S_c &=& -\left(\sum_i \eta_i-1\right)\log(\pi\mu b^2) - F(\eta_1+\eta_2+\eta_3-1) -F(\eta_1+\eta_2-\eta_3) - \cr
&-& F(\eta_2+\eta_3-\eta_1) - F(\eta_3+\eta_1-\eta_2) +F(0)+F(2\eta_1)+F(2\eta_2)+F(2\eta_3),
\end{eqnarray}
where $F(\eta) = \int_{1/2}^\eta \log\gamma(x)dx$. Since it is difficult to extract the physical intuition from this result, let us see the effect of the local vertex insertion instead. In the mean time, we will see the root of the local operator and the semiclassical meaning of the Seiberg bound. The solution of the classical Liouville equation around a vertex is easily found to be
\begin{equation}
e^{2b\phi} = \frac{1}{\pi \mu b^2} \frac{\nu^2|z|^{2\nu-2}}{(1-|z|^{2\nu})^2} 
\end{equation}
where $\mu >0 $ and $\alpha = \frac{1-\nu}{2b}$. If we regard this as the Weyl factor of the metric, we can interpret it as the conical curvature singularity at the insertion point, which is the reason why we call the operator with real $\alpha$ local. Since the deficit angle of the singularity is $\pi\nu$, it is necessary to have $\nu \ge 0$ to represent the semiclassical geometry. From this, the semiclassical bound $\frac{1}{2b} \ge \alpha $ is derived. Since $Q\sim \frac{1}{b}$ in the $b\to 0 $ limit where the WKB approximation is good, we find the semiclassical interpretation of the Seiberg bound  $\alpha \le \frac{Q}{2}$.

To conclude this section, we would like to make some comments on the perturbative treatment of the ``interaction" $\mu e^{2b\phi}$ and perform the one-loop calculation of the partition function. As we have seen above, the Liouville theory has an exact WT identity for the $\mu$ dependence of the correlation function, so the perturbation in $\mu$ does not make sense for general $\mu$. However, it is believed that when the power of $\mu$ becomes an integer, the perturbative treatment gives the correct answer. This is the assumption which appears throughout this review. As the simplest application of this assumption, let us calculate the one-loop partition function of the $c=1$ Liouville theory with a compactified target space whose radius is $R$.

The reason of the calculability of this partition function is that the power of $\mu$ just vanishes for the torus partition function. Thus, from the previous assumption, we can deal with the Liouville field as if they were free.\footnote{To see this more explicitly, we first integrate over the zero mode of $\phi$. Then the non-zero mode path integral becomes simply free. The only contribution from the zero-mode is given by the Liouville volume.} The partition function which we would like to obtain becomes
\begin{equation}
Z_1 = \int [d\Upsilon] \int \mathcal{D}X\mathcal{D}\phi \mathcal{D}b\mathcal{D}c e^{-S_0}.
\end{equation}
Taking the conventional torus moduli as the fundamental region $\mathcal{F}$, this partition function can be calculated as 
\begin{equation}
Z_1 = V_{\phi} \int_{\mathcal{F}} \frac{d^2\tau}{2\tau_2} |\eta(q)|^4 (2\pi \sqrt{\tau_2})^{-1}|\eta(q)|^{-2} Z(R,\tau),
\end{equation}
where $V_\phi = \int d\phi e^{-\mu e^{2b\phi}} = -\frac{1}{2b} \log\mu$ is the volume of the Liouville direction and $q=e^{2\pi i\tau}$. The measure of the moduli is from the Beltrami differential and the next $|\eta(q)|^4$ is from the ghost oscillator. $(2\pi \sqrt{\tau_2})^{-1}$ is from the integration of the Liouville momentum and the final $|\eta(q)|^{-2}$ comes from the oscillator of the Liouville mode. In addition, the partition function of the compactified boson $Z(R,\tau)$ is given by
\begin{equation}
Z(R,\tau) = 2\pi R \frac{1}{2\pi \sqrt{\tau_2}|\eta(q)|^2}\sum_{m,n=-\infty}^{\infty} \exp\left(-\frac{\pi R^2|n-m\tau|^2}{\tau_2}\right).\label{eq:1loop}
\end{equation}
Combining all these, we find the oscillator contribution cancels with that of the ghost:
\begin{equation}
Z_1 = V_\phi \frac{R}{4\pi} \int_{\mathcal{F}}\frac{d^2\tau}{\tau_2^2} \sum_{m,n}\exp\left(-\frac{\pi R^2|n-m\tau|^2}{\tau_2}\right).
\end{equation}

The integration over $\tau$ can be carried out.\footnote{The trick \cite{Polchinski:1986zf} is, instead of taking the summation over $m$, we can change the integration range of the moduli $\tau$ from the fundamental region to the $-1/2\le \tau_1 < 1/2$ region on the upper half plane $\tau_2>0$ (see figure \ref{fig2}). Then the calculation is easily done. See appendix \ref{b-8}.} Using the formula $\int_{\mathcal{F}}\frac{d^2\tau}{\tau_2^2} =\frac{\pi}{3}$, we obtain
\begin{equation}
Z_1 = V_\phi \frac{1}{12} \left(R+\frac{1}{R}\right).
\end{equation}
Note that this expression is invariant under the T-duality $R\to 1/R$ as expected. As we have seen from the zero-mode integration, it is appropriate to set $ V_\phi = -\frac{1}{2}\log \mu $ with an implicit cut-off. Therefore the final result becomes
\begin{equation}
Z_1 = - \frac{1}{24} \left(R+\frac{1}{R}\right)\log\mu,
\end{equation}
which reproduces the matrix model result (\ref{eq:matz}) as we will see later in the next section.

\begin{figure}[htbp]
	\begin{center}
	\includegraphics[width=0.6\linewidth,keepaspectratio,clip]{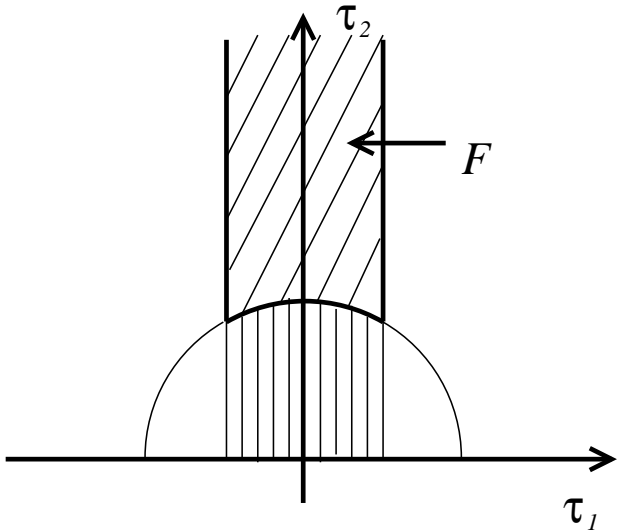}
	\end{center}
	\caption{The summation over $m$ can be effectively replaced with the integration over the horizontally hatched region.}
	\label{fig2}
\end{figure}

\subsection{Rolling Tachyon: Sen's Conjecture}\label{2.4}
In this section, we briefly review the connection between the world sheet description of the tachyon condensation and the Liouville theory.
In general non-BPS branes are unstable because of the tachyon living on them. Sen \cite{Sen:2002nu} has conjectured that, besides the unstable vacuum, another kind of vacuum solution exists in the full open string field theory which has the perturbative tachyon. In this ``true" vacuum, there exist only the closed string degrees of freedom without any open string excitation, and the energy difference from the unstable vacuum is simply the energy difference of the decaying D-brane.

For simplicity, let us discuss this conjecture by taking the bosonic D25-brane for example. We mainly concentrate on the world sheet description here, so we will not discuss the SFT (string field theory) approach much in detail. However, let us in advance point out the major problem of considering the time evolution of the tachyon by using the off-shell effective potential. For instance, suppose that we would like to construct the tachyon effective action from the on-shell scattering amplitudes (which are obtainable from the first quantized perturbative string theory). From the simple calculation, we find that the tachyon effective action at the three point level may be written as
\begin{equation}
S = \int d^{26} x \left[-\frac{1}{2}\partial_\mu\phi\partial^\mu \phi+ \frac{1}{2}\phi^2 +\frac{g_o}{3}\phi^3\right]
\end{equation}
from the scattering amplitude of three on-shell tachyons. Can we conclude from the above effective action, that the true vacuum exists at $\phi=-\frac{1}{g_o}$? We could imagine, however, another effective action which gives the same S matrix in this order such as
\begin{equation}
S = \int d^{26} x \left[-\frac{1}{2}\partial_\mu\phi\partial^\mu \phi+ \frac{1}{2}\phi^2 -\frac{g_o}{3}\phi^2\partial^2 \phi\right].
\end{equation}
If we take this effective action, the true vacuum is located at $\phi=0$. Of course the detailed study on the higher point functions and the unitarity fix this kind of ambiguity to some extent. Nevertheless, this kind of ambiguity persists in nature for the tachyon or massive modes unlike in the massless case.

To give a definite answer to this problem, we should assign an off-shell formulation of the string theory (which is not guaranteed to be unique). For example, Witten's open string field theory \cite{Witten:1986cc} fixes the effective potential for the tachyon to be \cite{Ohmori:2001am}, \cite{Taylor:2002uv}
\begin{equation}
V(\phi) = -\frac{1}{2}\phi^2 -g_o\bar{\kappa}\phi^3,
\end{equation}
where $\bar{\kappa} = 3^{7/2}/2^6$.\footnote{Of course, there are infinitely many momentum dependent three-point couplings in order to reproduce the correct on-shell S matrix.} To determine the true vacuum where $\phi \neq 0$, we should consider the three-point coupling of the tachyon to the other massive fields seriously. Only after integrating out all the massive fields, we will have the effective potential for the tachyon which we can extremize to obtain the true vacuum. Whether the energy difference of the potential agrees with the D-brane energy is a nontrivial question, but there is a numerical study which excellently confirms this (for a review, see \cite{Ohmori:2001am}, \cite{Taylor:2002uv}, \cite{DeSmet:2001af}, \cite{Arefeva:2001ps}).

\begin{figure}[htbp]
	\begin{center}
	\includegraphics[width=0.6\linewidth,keepaspectratio,clip]{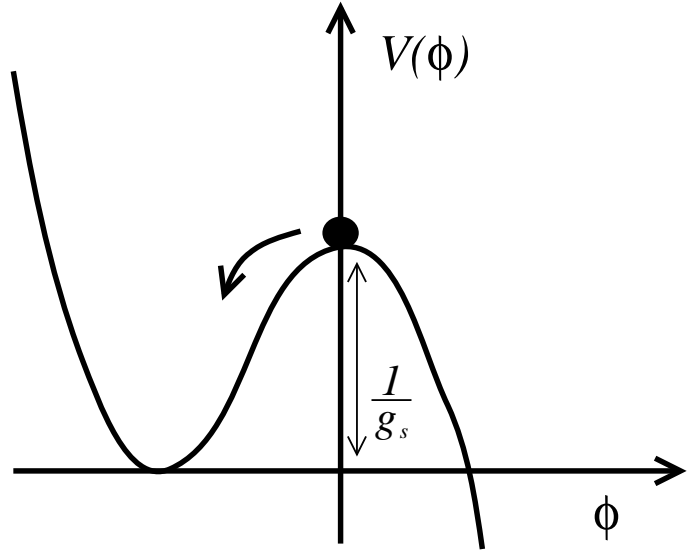}
	\end{center}
	\caption{Sen's conjecture: the closed string vacuum is the global minimum of the effective tachyon potential and the open string vacuum is the extremum of the potential. The potential difference corresponds to the D-brane tension.}
	\label{fig3}
\end{figure}

Let us next review the derivation of the effective action based on the boundary CFT (which is often called BSFT --- boundary string field theory). The basic purpose of this approach is to obtain the effective action whose equation of motion yields the condition of vanishing beta functions for the boundary perturbation on the world sheet. The claim \cite{Tseytlin:2000mt} is as follows. When we write the path integration on the disk as $\langle \ \rangle$, the partition function with the tachyon and vector background is given by 
\begin{equation}
Z[T,A,\epsilon] = \left\langle e^{-\int d\theta \left[ \frac{T(x)}{\epsilon} + iA_m(X) \dot{X}^m\right]} \right\rangle,
\end{equation}
where dot denotes a tangential differential and $\epsilon$ is introduced to imitate the Weyl scaling of the metric. This partition function is renormalizable by the power counting. Thus we should be able to define the renormalized partition function as
\begin{equation}
Z[T(\epsilon),A(\epsilon),\epsilon] = Z_R[T_R,A_R].
\end{equation}
Then the effective action \cite{Witten:1992qy,Witten:1993cr}, \cite{Shatashvili:1993kk,Shatashvili:1993ps} is given by
\begin{equation}
S = Z_R + \beta^T \frac{\delta}{\delta T} Z_R \label{eq:ea},
\end{equation}
where $\beta^T$ is the beta function for the tachyon.

Performing the path integration around the static background and renormalization, we obtain the final effective action
\begin{equation}
S = \frac{1}{g} \int d^{26}x e^{-T}\sqrt{\det(\eta_{\mu\nu} + 2\pi F_{\mu\nu})}\left[(1+T)[1+\frac{1}{2}\gamma^{\mu\nu}\partial_\mu T\partial_\nu T]\right],
\end{equation}
where $\gamma^{\mu\nu} = \left(\frac{1}{1+2\pi F}\right)^{\mu\nu}$.
We have some comments on this effective action. First, since we have expanded around $T= const$, we cannot trust this action for the on-shell S matrix while it is believed to be exact as the effective potential of $T$. It is just accidental that this effective action properly reproduces the three particle S matrix. Note that when $T\sim e^{ikx}, k^2 = 1$, all higher derivative terms should contribute to the scattering amplitude. Similarly, though $T = a + uX^2 $ is a solution of the equation of motion at this order, this is just an artefact of the approximation.\footnote{Because this solution makes $X$ massive, it is not a marginal perturbation. Nevertheless we can extract a useful information about the RG flow of the boundary field theory and obtain the effective action from this perturbation. This has been originally advocated in \cite{Witten:1992qy,Witten:1993cr}.}

It is important to note that the extremum of this potential is located at $T=\infty$ and the energy difference from the perturbatively unstable vacuum is remarkably given by just the D-brane energy. In addition, if we consider the world sheet interpretation of the $T=\infty$ vacuum, this shows that all the scattering amplitudes with boundaries on the world sheet becomes zero, which is naturally interpreted that after D-brane decays into the true vacuum, there are only closed strings and no open string excitations in the theory.



As a complementary method to these off-shell effective action computations, there is another idea that we follow the time evolution of the decaying D-brane as the on-shell dynamics of the tachyon. For instance, we consider the boundary action
\begin{equation}
 S_{half} = \int d\theta e^{X^0},
\end{equation}
or
\begin{equation}
S_{full} = \int d\theta \cosh(X^0).
\end{equation}
It is known (after the Wick rotation of $X^0$) that these are exactly marginal perturbations. \footnote{By the way, if these are really exactly marginal, they should be solutions of the BSFT equations of motion. However, even if we substitute this into the equation of motion, we should sum up all the higher derivative terms to confirm this. Even if we had known all the higher derivative terms, it would not be necessary for the beta functions to vanish, since the renormalization prescription could be different.} Since this interaction can be regarded as the time evolution of the rolling tachyon, it has been conjectured that the study of this theory leads to the understanding of the tachyon condensation. As we will see in section \ref{6.2}, this theory is very similar to the boundary Liouville theory (in a sense it is an ``analytic continuation"). Therefore the understanding of the Liouville theory is expected to result in the understanding of the time evolution of the  rolling tachyon.

The same notion can be applied not only to the open tachyon but also to the closed tachyon. However, the perturbation
\begin{equation}
S_{full?} = \int d^2 z \cosh(2X^0)
\end{equation}
is known to be not exactly marginal but marginally relevant \cite{Kutasov:2003er} after we Wick rotate $X^0$ and make it sine-Gordon theory. As a result this tachyon profile does not become a CFT (= on-shell). On the other hand,
\begin{equation}
S_{half} = \int d^2 z e^{2X^0} \label{eq:hafrt}
\end{equation}
is conjectured to be exactly marginal. The relation between this perturbation and the Liouville theory will be discussed in section \ref{6.2}, but let us review the idea quickly. As has been discussed in the previous section, the noncritical string in $d = 25$ has formally $b=i$ and there is no dilaton background because $Q=0$. In addition, the cosmological constant can be seen as the tachyon background $e^{2i\phi}$.
After Wick rotating $\phi$ as $i\phi = X^0$, we can interpret this system as the rolling tachyon system (\ref{eq:hafrt}). Therefore, very formally, the analytic continuation $b \to i$ of the Liouville theory describes the tachyon dynamics of the critical bosonic string if we believe the Liouville theory is analytic in $b$. The partial success and the problems are discussed in detail later in section \ref{6.2}.
\subsection{Ground Ring Structure}\label{2.5}
In this section, we discuss the BRST cohomology of the $c=1$ Liouville theory with the Euclidean $X$ boson. From the familiar discussion on the critical string theory (see also section \ref{3.2.2}), we have a massless tachyon vertex
\begin{equation}
T_q = c\bar{c}e^{iqX+(2-|q|)\phi}. \label{eq:tachop}
\end{equation}
At first sight, all other higher excitations are BRST exact with an experience in the critical string theory. However, the detailed study on the BRST cohomology reveals that this is not the case. For some special momenta, there exist BRST invariant physical discrete states (see e.g. \cite{Lian:1991gk,Lian:1991ju,Lian:1991ty,Lian:1992aj,Lian:1993mn}, \cite{Witten:1992zd,Witten:1992yj}, \cite{Bouwknegt:1992yg}, \cite{Ishikawa:1994ji}). For a ghost number 1 sector (usual (1,1) vertex operators), we can construct them as follows. We prepare the special primary fields of the form
\begin{equation}
V_{J,m} (\partial X, \partial^2 X \cdots) e^{2mi X(z)}
\end{equation}
with conformal dimension $J^2$. They form $SU(2)$ multiplets with total spin $J$ and $J_z = m$. Then we gravitationally dress them to obtain the dimension 1 operators as
\begin{equation}
V_{J,m}(z) = V_{J,m} e^{2m i X - 2(J-1)\phi} (z).
\end{equation}
These are remnants of the longitudinal modes of the higher dimensional string theory, and as we will see in the later section, they appear in the Euclidean scattering amplitudes as poles whenever the source momentum takes an integral value. This is because the OPE of two integer momentum tachyons is 
\begin{eqnarray}
e^{in X - (-2+|n|)\phi}(z) e^{-inX - (-2+|n|)\phi} (0) \sim \frac{1}{|z|^2} V_{|n|-1,0}\bar{V}_{|n|-1,0} + \cdots.
\end{eqnarray}

Actually, we have one more series of the BRST cohomology classes with ghost number 0 corresponding one to one to every $V_{J,m}$, which are discovered by Lian and Zuckerman. For example, at the low excitation level, we have
\begin{eqnarray}
O_{0,0} &=& 1 \cr
O_{1/2,1/2} &=& (cb+\partial\phi + i\partial X)(\bar{c}\bar{b}+\bar{\partial}\phi + i\bar{\partial} X)e^{iX-\phi}\cr
O_{1/2,-1/2} &=& (cb + \partial\phi - i\partial X)(\bar{c}\bar{b} + \bar{\partial}\phi - i\bar{\partial} X) e^{-iX-\phi}, 
\end{eqnarray}
which have dimension 0 and are BRST invariant. Furthermore we can show that $\partial O_{j,m}$ are BRST exact and hence the correlation function involving these states does not depend on their inserted points. Because of this property, we say these operators form a ground ring \cite{Witten:1992zd}. The OPE of these operators is given by the following form (modulo BRST exact terms)
\begin{equation}
O_{j_1,m_1} O_{j_2,m_2} = O_{j_1+j_2,m_1+m_2}
\end{equation}
if we have set the cosmological constat $\mu = 0$. When we turn on the cosmological constant, the structure itself remains the same, but the proportional factor changes somewhat. For example, we have \cite{Douglas:2003up}
\begin{equation}
O_{1/2,1/2} O_{1/2,-1/2} = \mu.
\end{equation}
To derive this equality (with a precise numerical factor 1), we have to utilize the exact structure constants of the Liouville OPE we will discuss in section \ref{sec:4}, so we will not derive this relation here (see \cite{Douglas:2003up}. See also \cite{Kazama:1993sj} for earlier study.). 

Another important feature of the ground ring structure is that the tachyon vertex operators form a module under the action of the ground ring. If we take $q>0$ in (\ref{eq:tachop}), we have
\begin{eqnarray}
O_{1/2,1/2} T_q &=& q^2 T_{q+1} \cr
O_{1/2,-1/2} T_{q} &=& \frac{\mu}{(q-1)^2} T_{q-1}.
\end{eqnarray}
On the other hand if we take $q<0$, we have
\begin{eqnarray}
O_{1/2,1/2} T_q &=& \frac{\mu}{(q+1)^2} T_{q+1} \cr
O_{1/2,-1/2} T_{q} &=& q^2T_{q-1}.
\end{eqnarray}

Though we will not discuss the ground ring structure any further, this ring structure is important both physically and mathematically. In the physical application, the ground ring shows a $W_{\infty}$ algebra \cite{Witten:1992yj}, \cite{Klebanov:1993ui}, \cite{Klebanov:1991hx} and reveals an underlying integrable structure of the theory (which should be closely related to the matrix model dual description). On the other hand, the geometrical picture of the cohomology is beautifully exposed in \cite{Witten:1992yj}, \cite{Lian:1993mn} particularly when we compactify $X$ at the selfdual radius. The recent applications of the ground ring to the Liouville theory and the matrix model dual can be found in \cite{Douglas:2003up}, \cite{Seiberg:2003nm}.

\subsection{Open-Closed Duality}\label{2.6}
In this section, we review the basic facts and ideas of the open/closed duality. Since this realm is so vast, we refer to other excellent reviews \cite{Aharony:1999ti}, \cite{Polyakov:2001af}, lecture notes (to name a few \cite{D'Hoker:2002aw}, \cite{Maldacena:2001dj}, \cite{Klebanov:2000me}) or original papers \cite{Maldacena:1998re}, \cite{Gubser:1998bc}, \cite{Witten:1998qj} on this subject for details.

The first idea of the open closed duality is very old. Consider a long cylinder diagram of the string theory. If we cut the diagram vertically, we obtain the cross-section which looks like a closed string. In fact, the cylinder diagram has a pole when the propagating momentum is on mass-shell of the \textit{closed} string spectrum. Therefore, the open string knows the existence of the closed string. However, this is not a duality. It simply states that the open string theory includes the closed string sector naturally.

\begin{figure}[htbp]
	\begin{center}
	\includegraphics[width=0.6\linewidth,keepaspectratio,clip]{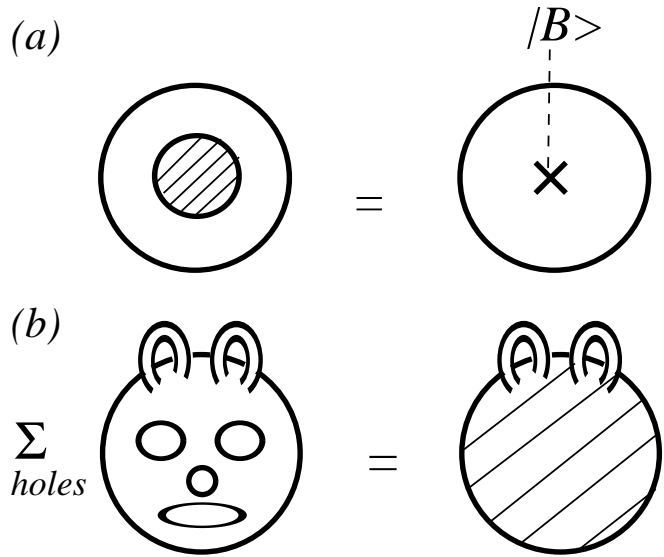}
	\end{center}
	\caption{(a) the boundary effect can be regarded as an insertion of the Ishibashi states in the closed string language. (b) in a certain limit, we expect that the existence of the boundaries can be replaced with a closed string theory on a nontrivial background. This picture is taken from \cite{Gopakumar:1999}.}
	\label{cat}
\end{figure}

However, this observation makes it possible to write boundary conditions of the open string theory in the closed string language, which is just the boundary states or Ishibashi states. Let us consider a world sheet with many holes. Formally, each hole can be represented by the closed string boundary states. 
\begin{equation}
V_H = \sum_n c_n(\tau) O_n,
\end{equation}
where $V_H$ is the ``hole operator" and $O_n$ is the suitable basis of the boundary operators and $c_n(\tau)$ is the moduli dependence (for large $\tau$, $c(\tau) \sim e^{-n\tau}$). We hope in some limiting case, the above replacement makes sense as a closed string theory after summing up all the hole contributions. The proper limit is necessary, since the boundary states usually does not possess a normalizable norm nor definite weight as the closed string states. The remaining world sheet theory is the closed string theory with modified backgrounds. Look at figure \ref{cat}, where in the left side, the cat's eyes, nose and mouth are holes where open strings have ends. After summing up those holes, we expect to obtain the right side figure. Cat remains but grin disappeared.

The first and the most fundamental realization of the idea above is the Maldacena conjecture \cite{Maldacena:1998re} of the AdS/CFT duality. We consider the $N$ D3-branes in the flat type IIB background, on which we have the $SU(N)$ $\mathcal{N} = 4$ super Yang-Mills theory as the low energy effective theory. On the other hand, in the closed string (gravity) perspective, the presence of the D3-branes curves the background and the metric becomes
\begin{equation}
ds^2 = h^{-1/2} (-dt^2 + dx_1^2 +dx_2^2 + dx_3^2) + h^{1/2} (dr^2 + r^2d\Omega_5^2)
\end{equation}
where $d\Omega_5^2$ is the metric of a unit 5-sphere and the warp factor is
\begin{equation}
h(r) = 1 + \frac{L^4}{r^4},
\end{equation}
where $L$ is given by
\begin{equation}
L^4 = 4\pi g_s N (\alpha')^2.
\end{equation}
Now we want to decouple the gravity from the gauge (open) side. To do this, we take the $r\to 0$ limit; then we obtain
\begin{equation}
ds^2 = \frac{L^2}{z^2} (-dt^2 + d\vec{x}^2 + dz^2) + L^2 d\Omega^2_5,
\end{equation}
where $z = \frac{L^2}{r}$. This describes the geometry of the direct product of $\mathrm{AdS}_5$ and $\mathrm{S}^5$.

The claim of the Maldacena conjecture is that the closed string theory on this $\mathrm{AdS}_5 \times \mathrm{S}^5$ background describes the gauge theory on the $N$ D3-branes. Since the parameter identifications are
\begin{eqnarray}
g_s &=& g_{YM^2} \cr
L^4 &=& 4\pi g_s N (\alpha')^2,
\end{eqnarray}
the large $N$ with fixed $\lambda = g^2_{YM} N$ limit ('t Hooft limit) corresponds to the classical type IIB string theory on $\mathrm{AdS}_5 \times \mathrm{S}^5$. Further taking the large $\lambda$ limit corresponds to the classical type IIB supergravity on $\mathrm{AdS}_5 \times \mathrm{S}^5$. Note that the symmetry of the both theories match, for the $\mathcal{N} = 4 $ super Yang-Mills theory actually has a superconformal symmetry which is isomorphic to the (supersymmetric extension of the) $\mathrm{AdS}_5 \times \mathrm{S}^5$ isometry group.

The Maldacena duality can be extended to other D-brane configurations -- D-branes on conifold or orbifold, wrapped branes around the cycles in the Calabi-Yau space etc. These gauge/gravity correspondence techniques have now become a strong tool to investigate the strongly coupled nonperturbative physics of the gauge theory.

Despite many efforts, however, the general proof of this duality along the line with the idea stated in the introductory part of this section is still missing. One of the major difficulties\footnote{Of course, the presence of the RR background is the most difficult part.} is that the perturbative expansion of the small 't Hooft coupling corresponds to the small radius limit where the world sheet description of the sigma model becomes ill-behaved. However, in the simpler setup, the open/closed duality has been proved.

\begin{figure}[htbp]
	\begin{center}
	\includegraphics[width=0.6\linewidth,keepaspectratio,clip]{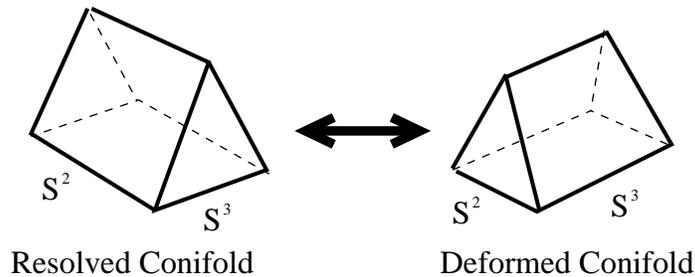}
	\end{center}
	\caption{The geometric transition: the topological open A-model on the deformed conifold with $N$ A-branes is dual to the closed topological A-model on the resolved conifold.}
	\label{geomt}
\end{figure}

The first setup \cite{Gopakumar:1998ki}, \cite{Ooguri:2002gx} is the duality between a topological open string theory and a closed string theory (and its embedding in the physical superstring theory \cite{Berkovits:2003pq}). This is sometimes called the ``geometric transition" or ``large $N$ duality" which we will very briefly review in section \ref{sec:6}. In the simplest case, the topological open A-model on the deformed conifold with $N$ A-branes is dual to the closed topological A-model on the resolved conifold (see figure \ref{geomt}). 
The key provability (besides the fact that the theory is exactly solvable) lies in that, unlike the $\mathrm{AdS}_5 \times \mathrm{S}^5$ case, we have the exact world sheet description (Landau-Ginzburg description) of the theory when the volume of the shrinking cycle $t$ tends to zero. Of course, more ingenuity is needed for the actual proof of the correspondence; see the original papers. 

The second setup is nothing but the Liouville theory. As is emphasized in \cite{Giveon:1999zm}, \cite{Polyakov:2001af}, \cite{McGreevy:2003kb}, the (unstable) D0-branes in the Liouville theory is the holographic dual of the Liouville theory itself. The world line theory on the D0-branes is given by the (gauged) matrix model and hence, we expect that the matrix model (``gauge" theory) describes the dual Liouville theory (``gravity" theory). As we will see later, this yields the physical explanation of the longstanding question why the matrix model describes the Liouville theory.\footnote{An alert reader may realize that this duality does not have a ``proof" because the  duality is based on our poor intuition that the double scaling matrix model reproduces the discretized Riemann surface. It is even more true in the type 0 matrix model we will discuss in section \ref{sec:9}, where we have no direct argument similar to the discretized Riemann surface. Very recently, the Kontsevich model, which describes originally the topological gravity but is equivalent to the physical two dimensional gravity, is reconsidered from the open string field theory perspective in \cite{Gaiotto:2003yb}. They have given a world sheet proof of the duality between the Kontsevich model, which emerges from the open string field theory and the closed topological (or physical) world sheet theory in line with the argument discussed in this section.}

To conclude this section, let us summarize what is important in the relation to the Liouville theory, some of which we would like to study further in the later section. The Liouville theory is the simplest example of the open/closed duality. The dual description of the Liouville theory (which is the matrix theory) yields the nonperturbative information of the theory. Furthermore, it is interesting to note that the topological string duality is related to the $c=1$ Liouville theory on the selfdual circle. We will study this connection in section \ref{6.3}. It is very surprising that, after embedding the topological string duality into the physical superstring theory \cite{Vafa:2000wi}, the nonperturbative physics of the $\mathcal{N}=1$ gauge theory is captured by the Liouville partition function.
\subsection{Literature Guide for Section 2}\label{2.7}
The complete reference list for the earlier studies on the Liouville theory is beyond the scope of this review. We can find them in the earlier reviews such as \cite{Seiberg:1990eb}, \cite{Ginsparg:is}, \cite{DiFrancesco:1993nw}, \cite{Klebanov:1991qa}, \cite{Martinec:1991kn}, \cite{Kutasov:1991pv}. The reviews on the string field theory and the tachyon potential can be found in \cite{Ohmori:2001am}, \cite{Taylor:2002uv}, \cite{DeSmet:2001af}, \cite{Arefeva:2001ps}. For the AdS/CFT correspondence, \cite{Aharony:1999ti} and \cite{Polyakov:2001af} are the standard review articles.

\sectiono{Basic Facts 2: Matrix Model}\label{sec:3}
In this section, we review the basic facts about the matrix model which is dual to the noncritical string theory (hence the Liouville theory). Good references are \cite{Ginsparg:is}, \cite{DiFrancesco:1993nw}, \cite{Klebanov:1991qa}, \cite{Martinec:1991kn}, \cite{Das:1992dm}. We mainly discuss the ``physical" double scaling limit matrix model, and give only a few words on the ``topological" matrix models which actually yield the same information of the noncritical string theory.

The organization of this section is as follows. In section \ref{3.1}, we review the basic Hermitian one matrix model in the double scaling limit which describes the $c<1$ noncritical string theory (the Liouville theory coupled to the minimal model). In subsection \ref{3.1.1}, we introduce the orthogonal polynomial method to solve the theory and in subsection \ref{3.1.2}, we check the results from the WKB approach. In subsection \ref{3.1.3}, we briefly discuss the integrable structure of the matrix models and try to interpret all the $c<1$ matrix models in a unified manner. 

In section \ref{3.2}, we turn to the $c=1$ matrix quantum mechanics. In subsection \ref{3.2.1} we derive the partition function of the theory, and in subsection \ref{3.2.2}, we discuss the tachyon scattering of the $c=1$ noncritical string theory from various points of view and compare their results. In subsection \ref{3.2.3}, we obtain the generating function of the tachyon S matrix which encodes all the scattering physics of the $c=1$ matrix model (the Liouville theory coupled to a single boson).

\subsection{$c < 1$ Matrix Model}\label{3.1}
In this section we solve the $c < 1$ string theory by the matrix model technique \cite{Brezin:1990rb}, \cite{Douglas:1990ve}, \cite{Gross:1990vs}.

We would like to define the noncritical string theory (2D gravity) as the continuum limit of the random lattice. For simplicity, we first concentrate on the pure 2D gravity without matter ($c=0$). The basic strategy is that we approximate the summation over all the metric by the random triangulations of the world sheet.

\begin{equation}
Z \sim \int \mathcal{D}g \to \lim_{cont} \sum_{random \ triangulations}
\end{equation}

It is known that the right-hand side summation can be done efficiently by the matrix model technique. Later in section \ref{sec:6}, we will review the physical meaning of this matrix. Let us consider the following matrix integral

\begin{equation}
e^Z = \int dM \exp\left[-N\left(\frac{1}{2}\mathrm{tr} M^2 + \frac{\kappa}{3!}\mathrm{tr} M^3\right)\right],
\end{equation}
where $M$ is an $N\times N$ Hermitian matrix. It is well-known that when one expands this integral in $1/N$ perturbatively, the expansion becomes the genus $g$ expansion of the topology of each Feynman diagram as follows
\begin{equation}
Z = \sum_g N^{2-2g}Z_g(\kappa).
\end{equation}

We would like to connect each of this expansion with the genus expansion of the partition function of the Liouville theory (2D gravity). Since the continuum limit seems to correspond to the $N\to\infty$ limit, we may conclude at first glance that the only genus $0$ amplitude survives in this limit. To avoid this difficulty, we take the $\kappa\to \kappa_c$ limit at the same time, which is called the double scaling limit, so that $Z_g(\kappa)$ diverges. It is known that the critical point $\kappa_c$ does not depend on the genus $g$. The way in which the partition function diverges is well-known and as we will see later, it is given by
\begin{equation}
Z_g(\kappa) \sim (\kappa_c-\kappa)^{(2-\Gamma)\chi/2},\label{eq:KPZs}
\end{equation}
where $\chi = 2-2g$ and $\Gamma$ is called the string susceptibility whose value will be explicitly shown soon. Thus, in the double scaling limit, the matrix model perturbative expansion changes from the double parameter expansion of ($1/N,\kappa$) to the single parameter expansion of $\mu_r \equiv N(\kappa-\kappa_c)^{(2-\Gamma)/2}$:
\begin{equation}
Z(\mu_r) = \sum_g \mu_r^{2-2g} f_g. 
\end{equation}
Considering that $1/N$ in the matrix model takes the role of the string coupling $g_s$ in the Liouville theory, this just corresponds to the fact that the actual expansion of the Liouville theory, contrary to the naive double expansion in $g_s$ and $\mu$, is a single expansion of the particular combination of $g_s$ and $\mu$. The purpose of the following calculation is to find $f_g$ or its non-perturbative completion $Z(\mu_r)$ by using the matrix model technique.

\begin{figure}[htbp]
	\begin{center}
	\includegraphics[width=0.6\linewidth,keepaspectratio,clip]{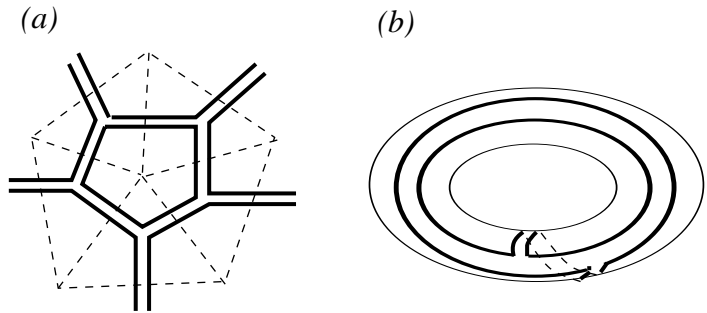}
	\end{center}
	\caption{(a) the random triangulations of the Riemann surface can be done effectively by the matrix Feynman diagrammatic summation. (b) this nonplanar diagram corresponds to the genus one Riemann surface.}
	\label{random}
\end{figure}
\subsubsection{Orthogonal polynomial method}\label{3.1.1}
In the double scaling limit, we can solve this matrix model completely. In this subsection, we review the solution which uses the orthogonal polynomials \cite{Bessis:1980ss}, \cite{DiFrancesco:1993nw}, \cite{Ginsparg:is}. First, we diagonalize the Hermitian matrix $M$: $M = U^\dagger \Lambda U$. Then we change the integration variable from $M$ to a unitary matrix $U$ and diagonal elements $\lambda_i,(i=1\cdots N)$. This can be done either by the Fadeev-Popov method or by the direct calculation. Here, we take the following step: set $dU = dT U$ and introduce $T$, and the metric can be written as
\begin{equation}
\mathrm{tr} dM^2 = \mathrm{tr} (U^\dagger (d\Lambda + [\Lambda,dT]) U)^2 = d\Lambda^2 + \sum_{i,j}(\lambda_i-\lambda_j)^2|dT_{ij}|^2.
\end{equation}
From this expression, we can see that the independent elements of $dM$ are $d\lambda_i$ and the imaginary and real part of the off-diagonal elements of $d T_{ij}$($i<j$). Then the Jacobian of this transformation becomes
\begin{equation}
\sqrt{G} = \prod_{i<j} (\lambda_i-\lambda_j)^2 = \Delta^2(\lambda) = (\det\lambda_i^{j-1})^2,
\end{equation}
which is the Vandermonde determinant. Substituting this transformation into the original integral, the dependence of $T$ drops out completely. Consequently, the integral over $T$ (or $U$ integration) simply yields the trivial factor. Dividing this factor (``gauge freedom"), we obtain \footnote{Since there is a residual gauge freedom which corresponds to the Weyl group of $U(N)$, further division by $N!$ is needed in fact. However this does not change the following argument at all.} 
\begin{equation}
e^Z = \int d\lambda_i \Delta^2(\lambda)e^{-V(\lambda_i)}. \label{eq:int}
\end{equation}
Note that the detail of the potential barely affects the final result in the double scaling limit as we are dealing with the critical behavior of the random lattice.\footnote{When we use the symmetric potential, the resulting partition function is just double as that when we use the generic potential. Though the physical interpretation of this fact becomes important in the $c=1$ matrix model, we do not discuss it for the time being here.}

Now let us introduce the orthogonal polynomials $P_m(\lambda)$ as follows
\begin{equation}
\int_{-\infty}^{\infty} d\lambda e^{-V(\lambda)}P_n(\lambda)P_m(\lambda) = h_n \delta_{nm},
\end{equation}
where the polynomials start like $P_n(\lambda) = \lambda^n + \cdots $. We can see from this 
\begin{equation}
\Delta(\lambda) = \det\lambda_i^{j-1} = \det P_{j-1}(\lambda_i).
\end{equation}

When we substitute this expression of the Vandermonde determinant into the integral (\ref{eq:int}), the integration is easily done because of the orthogonality, which results in 
\begin{equation}
e^Z = N! \prod_{i=0}^{N-1} h_i = N! h_0^N \prod_{k=1}^{N-1} f_k^{N-k},
\end{equation}
where $f_k \equiv h_k/h_{k-1}$.

In the double scaling limit, we take $N\to \infty$. In this limit $k/N$ can be regarded as a continuous parameter $0 \le\xi \le 1$. Also, $f_k/N$ can be regarded as a continuous function $f(\xi)$. As a result, we can write
\begin{equation}
\frac{1}{N^2}Z \sim \int_0^1 d\xi (1-\xi)\log f(\xi).
\end{equation}

In the following, in order to simplify the calculation, we take the symmetric potential $V(\lambda)$ and later we will divide the final result by $2$ if needed. We notice that the following recursion relation for the orthogonal polynomial holds
\begin{equation}
\lambda P_n = P_{n+1} + r_n P_{n-1} \label{eq:rec},
\end{equation}
where $r_n$ does not depend on $\lambda$. The right-hand side does not have $P_n$ term because the potential is even. The lower terms also vanish owing to the orthogonality. Thus, we obtain
\begin{equation}
\int d\lambda e^{-V} P_n \lambda P_{n-1} = r_n h_{n-1} = h_n,
\end{equation}
which leads to $f_n = r_n$. Similarly, by the partial integration, we can derive
\begin{equation}
nh_n = \int d\lambda e^{-V} P'_n \lambda P_n = \int d\lambda e^{-V} P'_n r_n P_{n-1} = r_n\int d\lambda e^{-V}V' P_n P_{n-1}. \label{eq:cr}
\end{equation}

For concreteness, let us consider the following potential
\begin{equation}
V(\lambda) = \frac{1}{2g}\left(\lambda^2 + \frac{\lambda^4}{N} +b\frac{\lambda^6}{N^2}\right),
\end{equation}
\begin{equation}
gV'(\lambda) = \lambda + 2\frac{\lambda^3}{N}+3b\frac{\lambda^5}{N^2}.
\end{equation}
We would like to obtain $r_n$ from this potential. First, we substitute (\ref{eq:rec}) into the right-hand side of (\ref{eq:cr}). Substituting the same relation several times, we finally obtain the integrations which are proportional to $h_{n-1}$, namely $\int e^{-V} P_{n-1}P_{n-1}$. Dividing this by $h_{n-1}$, we have
\begin{equation}
gn = r_n + \frac{2}{N} r_n(r_{n+1}+r_n+r_{n-1}) +\frac{3b}{N^2} (10\  rrr \ \mathrm{terms}).
\end{equation}
Then, taking the large $N$ limit, we can write the above expression as
\begin{eqnarray}
g\xi &=& W(r) + 2r(\xi) (r(\xi+\epsilon)+r(\xi-\epsilon) - 2r(\xi)) \cr
 &=& g_c + \frac{1}{2}W''|_{r=r_c}(r(\xi)-r_c)^2 + \epsilon^2 2r_c(1+15br_c)r''(\xi) + \cdots
\end{eqnarray}
where  $\epsilon = 1/N$, $W(r) \equiv r + 6r^2+30br^3 $ and $r_c$ is the critical point of $W(r)$. If we further tune the potential and obtain the higher multi critical point, we have a double scaling limit which describes the $(m,1)$ minimal model coupled to the two dimensional gravity.
  
It is important to note that, in the naive $N\to \infty$ limit, $(r-r_c) \sim (g_c-g\xi)^{1/2}$. Thus, comparison of the 0 loop result\footnote{$\frac{1}{N^2} Z= \int_0^1d\xi (1-\xi)(g_c-g\xi)^{1/2} = g_c^{3/2} + \int_0^1 d\xi (g_c-g\xi)^{3/2} \sim (g_c-g)^{5/2}$.} with the above general argument (\ref{eq:KPZs}) determines the string susceptibility $\Gamma=-1/2$. Now, we take the following double scaling limit,
\begin{eqnarray}
 g_c- g &=& g_c a^2 z \cr
 \epsilon &=& 1/N = a^{5/2}.
\end{eqnarray}
where $a \to 0$. In this limiting procedure, we fix $\mu_r = z^{5/4} = (g-g_c)^{5/4} N $. Note this limit is compatible with the general argument. For the actual calculation, it is convenient to change variables such that $g_c-g\xi = g_c a^2 x$ and $r(\xi) = r_c(1-au(x)) $. For the genus 0 amplitude, $u^2 \sim x$. With this change of variables (and rescaling of $x$), we obtain the following differential equation (Painlev\'e I equation)
\begin{equation}
 x = u^2 -\frac{1}{3}u''. \label{eq:PLI}
\end{equation}
On the other hand, the partition function can be written in this limit as
\begin{equation}
Z = \int_0^1 d\xi (1-\xi)\log r(\xi) \sim \int_{a^{-2}}^z dx(z-x) u(x). 
\end{equation}
Differentiating this expression twice, we obtain 
\begin{equation}
Z''(z) = - u(z).
\end{equation}

Therefore, if we know the solution of the Painlev\'e I equation, the 2D quantum gravity is solved. The perturbative solution in powers of $z^{-5/2}$ takes the form
\begin{equation}
u = \sqrt{z}(1-\sum_{k=1}^\infty u_{k}z^{-5k/2}),
\end{equation}
where $u_{k}$ are all positive. Note that this expansion is only asymptotic as we expect from the string perturbation theory and the full solution has a one parameter ambiguity. The nonperturbative part of this expression will be further studied in section \ref{6.6}.

\subsubsection{WKB method}\label{3.1.2}
In this subsection, we review the WKB (or steepest decent) method \cite{Brezin:1978sv} to solve the large $N$ matrix model. This may seem rather complicated and less efficient if we are only interested in the double scaling limit, but this method has a lot of applications including the recent Dijkgraaf-Vafa conjecture \cite{Dijkgraaf:2002fc,Dijkgraaf:2002vw,Dijkgraaf:2002dh}, \cite{Cachazo:2002ry,Cachazo:2002zk,Cachazo:2003yc}. For simplicity we consider here only the one-cut distribution of the eigenvalues which is sufficient for our application to the bosonic noncritical string theory. 

Originally, we have to calculate the partition function,
\begin{equation}
e^Z = \frac{1}{N!} \int \prod_{i=1}^N d\lambda_i e^{-N^2 S_{eff}(\lambda)},
\end{equation}
where the effective action is
\begin{equation}
S_{eff} = \frac{1}{N} \sum_{i=1}^N V(\lambda_i) - \frac{1}{N^2} \sum_{i\neq j}\log |\lambda_i - \lambda_j|.
\end{equation}
In the WKB (large $N$ steepest decent) approximation, we evaluate this by solving the equation of motion of the effective action and substituting back its solution into the partition function.

The equation of motion is
\begin{equation}
V'(\lambda_i) = \frac{2}{N} \sum_{j\neq i} \frac{1}{\lambda_i-\lambda_j}.
\end{equation}

In the large $N$ limit, we can define the density of states:
\begin{equation}
\rho(\lambda) = \frac{1}{N} \sum_{i=1}^{N} \delta(\lambda-\lambda_i) 
\end{equation}
as a continuum function. Then the equation of motion becomes
\begin{equation}
V'(\lambda) = 2\mathrm{P}\int \frac{\rho(\lambda')d\lambda'}{\lambda- \lambda'} \label{eq:meom}
\end{equation}
in the continuum language. As a result, the genus 0 partition function can be written as
\begin{equation}
Z = N^2\left(\int d\lambda \rho(\lambda) V(\lambda) - \int d\lambda d\lambda' \rho(\lambda)\rho(\lambda') \log |\lambda - \lambda'|\right).
\end{equation}

To solve (\ref{eq:meom}), let us introduce an auxiliary quantity, the resolvent:
\begin{equation}
 W(p) = \frac{1}{N} \langle \mathrm{Tr} \frac{1}{p-M}\rangle.
\end{equation}
In the WKB approximation, this can be expressed as
\begin{equation}
W_0(p) = \int_a^b d\lambda \frac{\rho(\lambda)}{p-\lambda}
\end{equation}
where we assumed that the support of $\rho(\lambda)$ is on the finite interval $[a,b]$ on the real line (one-cut distribution assumption). From the definition we can see
\begin{equation}
\rho(\lambda) = \frac{1}{2\pi i} (W_0(\lambda+i\epsilon) - W_0(\lambda-i\epsilon)).
\end{equation}
On the other hand, the equation of motion states
\begin{equation}
W_0(\lambda+i\epsilon) + W_0(\lambda-i\epsilon) = - V'(\lambda).
\end{equation}

This equation can be solved as 
\begin{equation}
W_0 (p) = \frac{1}{2}\oint_C \frac{dw}{2\pi i} \frac{V'(w)}{p-w} \left(\frac{(p-a)(p-b)}{(w-a)(w-b)}\right)^{\frac{1}{2}} \label{eq:lf},
\end{equation}
where the contour $C$ is circling around the interval $[a,b]$. To see this, we introduce $\hat{W}_0(p) \equiv W_0(p)/\sqrt{(p-a)(p-b)}$ and observe that this satisfies the discontinuity equation 
\begin{equation}
\hat{W}_0(\lambda+i\epsilon) - \hat{W}_0(\lambda-i\epsilon) = - \frac{V'(\lambda)}{\sqrt{(\lambda-a)(\lambda-b)}},
\end{equation}
then it is easy to understand (\ref{eq:lf}) holds. Furthermore, we can determine $a,b$ by demanding $W(p)$ should behave $\sim 1/p$ as $p$ tends to infinity, which means,
\begin{eqnarray}
\oint_C \frac{V'(w)}{\sqrt{(w-a)(w-b)}} \frac{dw}{2\pi i}&=& 0 \cr
\oint_C \frac{wV'(w)}{\sqrt{(w-a)(w-b)}} \frac{dw}{2\pi i}&=& 1.
\end{eqnarray}
In this way, we can evaluate $W_0$ by the change of variables: $w =1/z$. The resulting discontinuity of $W_0$ states
\begin{equation}
\rho(\lambda) = \frac{1}{4\pi} \sqrt{(\lambda-a)(b-\lambda)}\oint_C \frac{dz}{2\pi i} \frac{V'(1/z)}{1-\lambda z} \frac{1}{\sqrt{(1-az)(1-bz)}}. \label{eq:disc}
\end{equation}

\begin{figure}[htbp]
	\begin{center}
	\includegraphics[width=0.6\linewidth,keepaspectratio,clip]{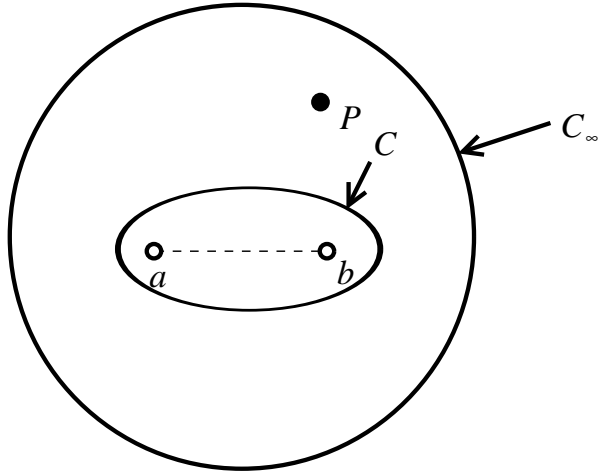}
	\end{center}
	\caption{The location of the cut and integration contours.}
	\label{cut}
\end{figure}

For example, take $V = \frac{1}{2 t} \mathrm{Tr}M^2 $. This leads to the famous Wigner distribution:
\begin{equation}
\rho(\lambda) = \frac{1}{2\pi t} \sqrt{ 4t- \lambda^2}.
\end{equation}

Next, we take the model $V = \frac{1}{2} \mathrm{Tr}M^2  + g\mathrm{Tr} M^4 $ which is relevant for the 2D gravity. In this case, the location of the cut becomes
\begin{equation}
b = -a = \sqrt{\frac{1}{6g}\left(-1+\sqrt{1+48g}\right)}.
\end{equation}
Calculating the density of states, we obtain
\begin{equation}
\rho(\lambda) = \frac{1}{\pi} \left(\frac{1}{2} + g a^2 + 2g\lambda^2\right)\sqrt{a^2-\lambda^2}.
\end{equation}
Observing this solution, we notice that, for a general value of $g$, $\rho(\lambda) \sim \sqrt{a^2-\lambda^2} $ around the edge of the distribution $a$, but if we take $g \to g_c = -1/48$ then $\rho(\lambda) \sim (a^2-\lambda^2)^{3/2}$ and this suggests the critical behavior of $g$ (double scaling limit will exist). To see this, we decompose the first term as $\frac{2g}{\pi}\left(\frac{1}{4g}+\frac{a^2}{2}+\lambda^2\right)$, and demand $\frac{1}{4g} = -\frac{3a^2}{2}$.

Finally, we can substitute this solution back into the effective action. Then the partition function becomes 
\begin{eqnarray}
Z_0 &=& N^2\left(\int d\lambda \rho(\lambda) V(\lambda) - \int d\lambda d\lambda' \rho(\lambda)\rho(\lambda') \log (\lambda - \lambda')\right) \cr
 &=& N^2 \int d\lambda \rho(\lambda) \left(\frac{V(\lambda)}{2}-\log|\lambda|\right) \cr
 &=& N^2 \left(\frac{1}{24}(a^2/4-1)(9-a^2/4) - \frac{1}{2} \log (a^2/4)\right) \cr
 &\sim& N^2(g-g_c)^{5/2}.\label{eq:WKBp}
\end{eqnarray}
Here, in the second line, we have used the following formula which can be obtained by integrating the equation of motion,
\begin{equation}
V(\lambda) = 2\int d\lambda' \rho(\lambda')\left(\log|\lambda'-\lambda|-\log|\lambda'|\right),
\end{equation}
where the integration constant is determined by demanding $V(0) =0$. In the last line of (\ref{eq:WKBp}), we actually mean 
\begin{equation}
Z_0 = c_0 + c_1(g-g_c) + c_2(g-g_c)^2 + c_{5/2} (g-g_c)^{5/2} + \cdots.
\end{equation}
The parts which have an integer power of $(g-g_c)$ do not show any singular behavior. Note that the absence of the term $c_{1/2}$ and $c_{3/2}$ is nontrivial.
This partition function shows the same critical behavior as what we have obtained in the last subsection by using the orthogonal polynomial technique. Thus we have reproduced the genus 0 partition function by the WKB method as well.

We can also derive the formula (\ref{eq:disc}) by using the loop equation. We concentrate on the one-cut model again. The loop equation is most easily derived from the WT identity of the matrix integral.
\begin{eqnarray}
0 &=& \int dM \mathrm{Tr}\left(\frac{\partial}{\partial M}\right) \frac{1}{p-M} \exp\left(-N\mathrm{Tr} V(M)\right) \cr
&=& \left\langle \left(\mathrm{Tr}\frac{1}{p-M}\right)^2 \right\rangle - N \left\langle \mathrm{Tr}\left(\frac{1}{p-M}V'(M)\right)\right\rangle .
\end{eqnarray}
In the large $N$ limit, the first term can be factorized as 
\begin{equation}
 \left\langle \left(\mathrm{Tr}\frac{1}{p-M}\right)^2 \right\rangle =  \left\langle \left(\mathrm{Tr}\frac{1}{p-M}\right)\right\rangle^2 = N^2W_0(p)^2.
\end{equation}
We decompose the second term as follows,
\begin{equation}
\left\langle \mathrm{Tr}\left(\frac{1}{p-M}V'(M)\right)\right\rangle = \left\langle \mathrm{Tr}\left(\frac{1}{p-M}(V'(M)-V'(p)+V'(p))\right)\right\rangle = NW_0(p)V'(p) + Nf(p),
\end{equation}
where $f(p)$ is an $(n-1)$ degree polynomial of $p$ which does not have a singularity.  

Then the loop equation becomes in the large $N$ limit
\begin{equation}
W_0(p)^2 = W_0(p) V'(p) + f(p),
\end{equation}
which can be solved as
\begin{equation}
W_0= \frac{1}{2}\left(V'- \sqrt{(V')^2 +4f}\right).
\end{equation}
The single cut ansatz states that the above square root factorizes as
\begin{equation}
\sqrt{(V'(p))^2 +4f(p)} = M(p)\sqrt{(p-a)(p-b)} 
\end{equation}
where $M(p)$ is a regular function which does not have a singularity or cut. We have assumed $a<b$ here. This ansatz means that the eigenvalue density behaves like 
\begin{equation}
\rho(\lambda) \sim M(\lambda) \sqrt{(a-\lambda)(\lambda-b)},
\end{equation}
which can be regarded as a generalization of the Wigner distribution.

Now we can completely determine $M(p)$ and hence $W_0(p)$ only with this information. Since $M(p)$ does not have a singularity except at $p=\infty$, the following relation holds,
\begin{equation}
M(p) = \oint_{C_\infty} \frac{dw}{2\pi i} \frac{M(w)}{p-w} = \oint_{C_\infty} \frac{dw}{2\pi i} \frac{V'(w)}{p-w}\frac{1}{\sqrt{(w-a)(w-b)}},
\end{equation}
where the integral involving $W(z)$ does not contribute because $W(z) \sim \frac{1}{z}$ as $z \to \infty$ from its definition. We can see this explicitly by the change of variable $z = 1/w$
\begin{equation}
M(p) = \oint_{C_0} dz \frac{V'(1/z) -2W_0(1/z)} {(1-pz)\sqrt{(1-az)(1-bz)}}
\end{equation}
Substituting this expression into the loop equation, we obtain
\begin{equation}
W_0(p) = \frac{1}{2} \left(V'(p) -\sqrt{(p-a)(p-b)}\oint_{C_0} dz \frac{V'(1/z)}{(1-pz)\sqrt{(1-az)(1-bz)}}\right)
\end{equation}
We may note that the first term $V'(p)$ is just the residue at $p=w$. This observation leads to the final expression,
\begin{equation}
W_0 (p) = \frac{1}{2}\oint_C \frac{dw}{2\pi i} \frac{V'(w)}{p-w} \left(\frac{(p-a)(p-b)}{(w-a)(w-b)}\right)^{\frac{1}{2}}
\end{equation}
which is the same formula as above (\ref{eq:lf}).
\subsubsection{Integrable hierarchy}\label{3.1.3}
General $c<1$ Liouville theory coupled to the minimal model reveals a very elegant structure when we see from the integrable hierarchy point of view. In this subsection we review only elementary facts following \cite{Ginsparg:is}, \cite{DiFrancesco:1993nw} (see also \cite{Mironov:1994wi}). For a complete description, we recommend the reader to consult these reviews and original papers cited therein. 

Let us begin with a one-matrix model and consider the possible generalization of the string equation heuristically. For this purpose, we rescale orthogonal polynomials as
\begin{equation}
\Pi_n (\lambda) = P_n(\lambda)/\sqrt{h_n}
\end{equation}
so that the orthogonality condition simplifies as
\begin{equation}
\int_{-\infty}^{\infty}d\lambda e^{-V} \Pi_n \Pi_m = \delta_{nm}.
\end{equation}
Then the recursion relation (\ref{eq:rec}) can be rewritten as
\begin{eqnarray}
\lambda \Pi_n &=& \sqrt{r_{n+1}} \Pi_{n+1} + \sqrt{r_n}\Pi_{n-1} \cr
	&=& Q_{nm}\Pi_m .
\end{eqnarray}
In the matrix notation, $Q$ is symmetric: $Q^T = Q$, and hence becomes an Hermitian operator in the continuum limit. Indeed, in the multicritical model, it can be written as
\begin{equation}
Q = 2r_c^{1/2} + \frac{a^{2/m}}{\sqrt{r_c}} (u +r_c \kappa^2 \partial^2_x),
\end{equation}
where the first constant part is a non-universal term. It is convenient to introduce another matrix $A$ by
\begin{equation}
\frac{\partial}{\partial \lambda} \Pi_n = A_{nm} \Pi_m,
\end{equation}
which satisfies
\begin{equation}
[A,Q] = 1, 
\end{equation}
by differentiating $\lambda \Pi = Q \Pi$ with respect to $\lambda$. We define antisymmetric matrix $P$ as
\begin{equation}
P = \frac{1}{2}(A-A^T),
\end{equation}
which also satisfies $[P,Q] =1$. This can be easily seen if we notice $A+A^T = V'(Q)$ which is derived from $ 0 = \int d\lambda \frac{\partial}{\partial \lambda} (\Pi_n\Pi_m e^{-V})$. In general if $V$ is of order $2l$, we have an order $2l-1$ differential operator $P$ in the continuum limit. The fundamental commutation relation $[P,Q] =1 $ is related to the Lax representation of the KdV equations.

For the simplest example, we set $l=2$ and $Q= d^2 -u$. Then the third order anti-Hermitian differential operator $P$ is given by $P = d^3 -\frac{3}{4} \{u,d\}$. Then the commutation relation yields
\begin{equation}
1 = [P,Q] = \left(\frac{3}{4} u^2 - \frac{1}{4}u''\right)',
\end{equation}
which is equivalent to the Painlev\'e I equation (\ref{eq:PLI}) with a trivial rescaling.

In general, it can be shown \cite{Douglas:1990dd} that the general differential operators satisfying $[P,Q] = 1$ yield the string equation which includes the Liouville theory coupled to $(p,q)$ minimal model. First, we take $Q$ as a $q$th order differential operator 
\begin{equation}
Q = d^q + \{v_{q-2}(x),d^{q-2}\} + \cdots +2v_0(x),
\end{equation}
where the constants $v_{n}$ correspond to various responses of the theory (e.g. $v_{q-2}$ is the specific heat). Then we define $p$th differential operator $P$ as $P = Q_+^{p/q}$, where $Q^{1/q}$ is a formally defined pseudodifferential operator and the subscript $+$ means that we take only the positive powers of $d$. In this way, we can in principle solve all the minimal models coupled to gravity. At the same time, the general theorem by \cite{Drinfeld:1984qv} states that in general, $P$ can be taken as 
\begin{equation}
P = \sum_{k \ge 1} -(1+k/q) t_{k+q} Q_{+}^{k/q},
\end{equation}
and physically the string equation following from $[P,Q] = 1$ corresponds to the minimal model coupled to gravity with the massive deformation. Since $t_n$ couples to the dual operators like $S_{int} = \int d^2z t_n O_n$, the derivative of the partition function with respect to $t_n$ yields the corresponding correlation function. Also it is important to note that the dependence of the susceptibilities $v_{n}$ on $t_k$ is governed by the generalized KdV flow (see \cite{DiFrancesco:1993nw}, \cite{Mironov:1994wi} and references therein for a review on this point). In a word, the partition function is given by the $\tau$ function of the generalized KdV hierarchy.

The integrable hierarchy of the double scaling limit of the (multi)matrix model is a good point to relate them to the topological matrix model. It is known that the topological matrix model and the physical double scaling matrix model yield the same integrable structure and hence the same physical observables. The topological gravity aspect of the generalized KdV hierarchy is reviewed in e.g. \cite{Dijkgraaf:1990qw}, \cite{Dijkgraaf:1990pb}, \cite{Dijkgraaf:1991qh}, \cite{Itzykson:1992ya}, \cite{Weis:1997kj}. See also \cite{Morozov:1994hh}, \cite{Alexandrov:2003pj} for reviews on the integrable structure of the matrix models.

\subsection{$c = 1$ Matrix Quantum Mechanics}\label{3.2}
In this section, we solve the $c=1$ noncritical string theory ($b=1$ Liouville theory) via the matrix quantum mechanics \cite{Klebanov:1991qa,Ginsparg:is}. 
\subsubsection{Partition function}\label{3.2.1}
As in the last section, we consider the random triangulations of the world sheet. To obtain the $c=1$ theory, we introduce free bosons on each lattice point. First, we consider the case where the boson is not compactified i.e. $R=\infty$. The partition function of the theory is given by
\begin{equation}
Z(g_0,\kappa) =\sum_g g_0^{2(g-1)} \sum_{\Lambda} \kappa^V\prod_{i=1}^{V} \int dX_i \prod_{\langle ij\rangle}G(X_i-X_j) \label{eq:mac}.
\end{equation}
We suppose this yields the $c=1$ noncritical string theory in the double scaling limit. While it is natural to take $G(X) = e^{-\frac{1}{2}X^2}$ as the Polyakov action, it is known \cite{Kazakov:1988ch} that the precise form of the propagator does not matter in the double scaling limit. Thus we here take $G(X) = e^{-|X|}$ for convenience. In fact, there exists a matrix model which reproduce this partition function perturbatively. Let us consider the following matrix quantum mechanics
\begin{equation}
e^{Z} = \int \mathcal{D}\Phi(t) \exp\left[-N\int_{-\infty}^{\infty} dt \mathrm{Tr} \left(\frac{1}{2}\left(\frac{d\Phi}{dt}\right)^2+\frac{1}{2}\Phi^2 -\frac{\kappa}{3!}\Phi^3\right)\right]
\end{equation}
where, $\Phi$ is an $N\times N$ Hermitian matrix and $\kappa = \sqrt{N/\beta}$.
Evaluating this partition function by the Feynman diagram expansion, we obtain
\begin{equation}
 Z = \sum_g N^{2-2g}\sum_{\Lambda}\kappa^V\prod_{i=1}^{V} \int dX_i \prod_{\langle ij\rangle}e^{-|X_i-X_j|}
\end{equation}
which is equivalent to (\ref{eq:mac}) up to the propagator difference.

Rewriting this path integral into the operator formalism of the quantum mechanics leads to the following calculation
\begin{equation}
 e^Z = \lim_{T\to \infty}\langle f| e^{-\beta HT} |i \rangle, 
\end{equation}
which means
\begin{equation}
 \lim_{T\to \infty} \frac{Z}{T} = -\beta E_0.
\end{equation}
Therefore, the evaluation of the partition function is equivalent to obtaining the lowest energy eigenvalue of this quantum system.

In order to know the energy eigenvalue, we canonically quantize this matrix quantum mechanics. As in the last section, we consider the diagonalization of the Hermitian matrix
\begin{equation}
\Phi(t) = \Omega^\dagger(t) \Lambda(t)\Omega(t).
\end{equation}
The different point from the matrix integration considered in the last section is the appearance of the $\Omega(t)$ dependent part from the kinetic term:
\begin{equation}
\mathrm{Tr}\dot{\Phi}^2 = \mathrm{Tr}\dot{\Lambda}^2 + \mathrm{Tr}[\Lambda,\dot{\Omega}\Omega^\dagger]^2.
\end{equation}

To go further, we decompose $\dot{\Omega}\Omega^\dagger$ as follows
\begin{equation}
\dot{\Omega}\Omega^\dagger = \frac{i}{\sqrt{2}}\sum_{i<j} T_{ij} \dot{\alpha}_{ij} + \sum_i^{N} \dot{\alpha}_i H_i,
\end{equation}
where $T_{ij}$ is a nondiagonal generator of $U(N)$ and $H_i$ is a generator of the Cartan subalgebra. Since the Cartan part decouples from the action, these degrees of freedom are simply the gauge freedom. In other words, when we decompose an Hermitian matrix which has $N^2$ degrees of freedom into a unitary matrix which has $N^2$ degrees of freedom and a diagonal matrix which has $N$ degrees of freedom, there apparently exists a redundant part. Of course this is just the Cartan generator of $U(N)$.

Then we substitute this decomposition into the Lagrangian, and we find
\begin{equation}
L = \sum_i \left(\frac{1}{2}\dot{\lambda}_i^2+U(\lambda_i) \right) + \frac{1}{2}\sum_{i<j} (\lambda_i-\lambda_j)^2 \dot{\alpha}_{ij}^2.
\end{equation}
To construct the Hamiltonian from this Lagrangian, we should be a little bit careful. Naive method would lead to a mistake just as it would lead to a mistake if we would try to obtain the Hamiltonian in the polar coordinate from the naive replacement $p_r \to -i\partial_r$ etc. The important point is that there is a Jacobian of the path integral measure which appears when we change the integration variables. We can obtain this Jacobian just as in the last section and it is given by $ \mathcal{D}\Phi = \mathcal{D}\Omega \prod_i d\lambda_i \Delta^2(\lambda) $.

The way to implement this Jacobian factor into the canonical quantization has been known since the Pauli's era. The gist is that we should write the Hamiltonian in terms of the generally covariant Laplacian as in the case of the polar coordinate. The metric in this case is given by $\sqrt{G} = \Delta^2(\lambda)$ as in the last section, so the kinetic term of the Hamiltonian is given by
\begin{equation}
-\frac{1}{2\beta^2} \frac{1}{\Delta^2(\lambda)} \frac{d}{d\lambda_i}\Delta^2(\lambda)\frac{d}{d\lambda_i}
\end{equation}
in this coordinate.

Taking the fact $\frac{d^2}{d\lambda_i^2} \Delta(\lambda) = 0$\footnote{This expression must be totally antisymmetric with respect to $\lambda_i$, but only the Vandermonde determinant has this property. However, the degree of this expression does not match with that of the Vandermonde determinant, so it must vanish identically.} into consideration, we can rewrite this expression as
\begin{equation}
-\frac{1}{2\beta^2\Delta(\lambda)} \frac{d^2}{d\lambda_i^2} \Delta(\lambda).
\end{equation}
Thus, the complete Hamiltonian finally becomes
\begin{equation}
H = -\frac{1}{2\beta^2\Delta(\lambda)} \frac{d^2}{d\lambda_i^2} \Delta(\lambda) + U(\lambda_i) + \frac{\Pi_{ij}^2}{(\lambda_i-\lambda_j)^2},
\end{equation}
where $\Pi_{ij}$ is the conjugate momentum of angular variable $\alpha_{ij}$.
Note that if we only need the lowest energy state of this system, the wavefunction should possess a trivial dependence on the angular parameter because of the last term in the Hamiltonian, which means we can focus only on the trivial representation of the $U(N)$. On the other hand, when we consider the finite temperature situation, we may feel the necessity of including the contribution from the non-singlet representations at first sight. In fact, however, we should \textit{discard} these contributions for the calculation of the \textit{ordinary} Liouville partition function as we will see.\footnote{We will see the natural interpretation of this fact from the D-brane point of view later in section \ref{sec:6}.} In any case, the singlet wavefunction is symmetric with respect to the argument $\lambda_i$. Then we define a new ``antisymmetric" wavefunction $\Psi(\lambda)$ as $\Psi(\lambda) \equiv \Delta(\lambda) \chi_{sym}(\lambda)$. For this new wavefunction, the Schr\"odinger equation is given in a simple form:
\begin{equation}
\left(-\frac{1}{2\beta^2}\frac{d^2}{d\lambda^2_i} + U(\lambda_i)\right)\Psi(\lambda) = E\Psi(\lambda).
\end{equation}

Since all the eigenvalues are diagonalized in this Hamiltonian, this is equivalent to the $N$ independent fermions system with the potential $U(\lambda)$. Note that the antisymmetry of the wavefunction makes the system fermionic. The lowest energy state is simply given by occupying the lowest $N$ states with fermions, so the lowest energy becomes
\begin{equation}
E_0 = \sum_{i=1}^N \epsilon_i,
\end{equation}
where $\epsilon_i$ is the energy of the $i$th excited state. 

For the actual calculation of this system, it is convenient to utilize the statistical mechanical method. To begin with, we introduce the density of states as
\begin{equation}
\rho(\epsilon) = \frac{1}{\beta} \sum_n\delta(\epsilon-\epsilon_n),
\end{equation}
Then we can write the total particle number and energy as
\begin{eqnarray}
\frac{N}{\beta} &=& \int_{-\infty}^{\mu_F} \rho(\epsilon) d\epsilon \cr
\beta E &=& \beta^2 \int_{-\infty}^{\mu_F} \rho(\epsilon) \epsilon d\epsilon,
\end{eqnarray}
where $\mu_F$ is the Fermi energy under which the states are all occupied. Differentiating the first equation and setting $\Delta = \pi(\kappa^2_c- \frac{N}{\beta})$, we obtain
\begin{equation}
\frac{\partial \Delta}{\partial \mu} = \pi \rho(\mu_F). \label{eq:mu}
\end{equation}
We can adjust the height of the potential freely to set the critical point to become $\mu_c=0$. Then we can write $\mu = -\mu_F$ as a function of $\Delta$. On the other hand, differentiating the second equation and leaving only the singular terms, we obtain
\begin{equation}
\frac{\partial E}{\partial \Delta} = \frac{1}{\pi}\beta \mu. \label{eq:en}
\end{equation}
In this way, the knowledge of $\rho(\epsilon)$ is enough to calculate the energy as a function of $\Delta$ and $\beta$ by solving these equations. Since the noncritical string theory can be obtained in the double scaling limit, we take $N\to \infty$ and $\beta \to \infty$ simultaneously. In this limit, the information of the detailed shape of the potential is almost lost and all we can see is the neighborhood around the critical point. Therefore there is no problem in approximating the potential with the inverted harmonic oscillator. Furthermore, for simplicity, we fill both sides of the potential with fermions (which corresponds to taking the even potential) and divide it by two at the end of the calculation.

Taking the inverted harmonic oscillator as the Hamiltonian: $h_0 = \frac{1}{2}p^2 - \frac{1}{2}x^2$, we can write the density of states as follows,
\begin{equation}
\rho(\mu) = \mathrm{Tr}\delta(h_0 + \beta\mu) = \frac{1}{\pi} \mathrm{Im}\ \mathrm{Tr} \left[\frac{1}{h_0+\beta\mu - i\epsilon}\right]. 
\end{equation}
The Green function for the harmonic oscillator was given by Feynman:
\begin{eqnarray}
\langle x|\frac{1}{\frac{1}{2}p^2 + \frac{1}{2}\omega^2 x^2 +\beta\mu-i\epsilon}|x\rangle &=& \int_0^\infty dTe^{-\beta\mu T} \int_{x(0)=x}^{x(T)=x}\mathcal{D}x(t) e^{-\int_0^Tdt \frac{1}{2}(\dot{x}^2 +\omega^2 x^2)} \cr
&=& \int_0^\infty dT e^{-\beta \mu T} \sqrt{\frac{\omega}{2\pi\sinh\omega T}} e^{-\omega\left[2x^2\cosh\omega T-2x^2\right]/2\sinh \omega T}. \cr
& &
\end{eqnarray}
In this case, $\omega = -i$, so it is convenient to ``analytically continue" the Euclid time to the Minkowski time i.e. $T \to iT $\footnote{$\omega \to -i$ is indeed an analytic continuation, but the transformation of $T$ is actually not.  Because of the $i\epsilon$ term, this is not a analytic continuation but just a contour deformation, so the result does not change in this procedure.}.

Substituting this into the expression for the density of states, we obtain
\begin{equation}
\frac{\partial \rho}{\partial (\beta \mu)} = \frac{1}{\pi} \mathrm{Im} \int_0^\infty dT e^{-i\beta\mu T} \frac{T/2}{\sinh(T/2)},
\end{equation}
after the $x$ integration. Though this is a correct expression perturbatively in the $\mu^{-1}$ expansion, we should interpret this result as one of the nonperturbative completions for the order $e^{-\beta\mu}$ correction. Actually, it is known that the nonperturbative terms do depend on the precise form of the potential far from the critical point. For example, the third order potential, which is supposed to correspond to the bosonic noncritical string theory, is nonperturbatively unstable. 

Expanding perturbatively in $\mu^{-1}$ and carrying out the integration over $T$, we obtain\footnote{Note this is the expansion in terms of $\mu^{-1}$ not in $\mu$. Since this is a nontrivial asymptotic expansion, we review the details of the calculation in appendix \ref{b-3}.}
\begin{equation}
\rho(\mu) = \frac{1}{\pi}\left[-\log\mu + \sum_{m=1}^\infty (2^{2m-1}-1)\frac{|B_{2m}|}{m}(2\beta\mu)^{-2m}\right].
\end{equation}

The task left is to integrate (\ref{eq:mu}) and express $\mu$ as the function of $\Delta$. However, in hindsight, the final expression does not have a suitable $\Delta$ expansion in the double scaling limit, rather $\mu_0$ is the actual expansion parameter, which is defined as
\begin{equation}
\Delta \equiv -\mu_0 \log\mu_0.
\end{equation}
 This special renormalization is related to the fact that $c=1$ is the phase transition point of the noncritical string theory \footnote{In fact, there is a simple heuristic method for this calculation. $\frac{\partial^2 Z(\mu_r)}{\partial \mu_r^2} = \rho(\mu_r)$ holds. Note this is just the Legendle transformation.}. Accepting this ansatz, we find $\mu$ is given by
\begin{equation}
\mu = \mu_0 \left[1-\frac{1}{\log\mu_0}\sum_{m=1}^\infty (2^{2m-1}-1)\frac{|B_{2m}|}{m(2m-1)}(2\beta\mu_0)^{-2m} +\mathcal{O} \left(\frac{1}{\log^2\mu_0}\right)\right],
\end{equation}
which can be confirmed by differentiating it directly. Integrating (\ref{eq:en}) next, we find
\begin{equation}
-\beta E = \frac{1}{8\pi} \left[(2\beta\mu_0)^2\log\mu_0 - \frac{1}{3}\log\mu_0 + \sum_{m=1}^{\infty} \frac{(2^{2m+1}-1)|B_{2m+2}|}{m(m+1)(2m+1)}(2\beta \mu_0)^{-2m}\right].
\end{equation}
In the double scaling limit, we take the $\beta \to \infty$ and $\mu \to 0$ limit while fixing $\beta \mu_0 = \mu_r$. In this limit, the sphere and torus contribution (logarithmically) diverges, which is also expected from the continuum Liouville analysis. Up to the zero-mode factor for $X$ (and $1/2$ for the bosonic string), this reproduces the correct partition functions for the sphere and the torus which are calculated from the Liouville approach. We cannot calculate the comparable higher genus Liouville partition function in the continuum approach yet.

Now we consider the case where the target space $X$ is compact (with radius $R$). In this case, the matrix quantum mechanical expression becomes
\begin{equation}
e^{Z_R} = \int \mathcal{D}\Phi(t) \exp\left[-\beta\int_{0}^{2\pi R} dt \mathrm{Tr} \left(\frac{1}{2}\left(\frac{d \Phi}{d t}\right)^2+U(\Phi)\right)\right].
\end{equation}
This can be rewritten in the operator formalism as 
\begin{equation}
e^{Z_R} = \mathrm{Tr} e^{-2\pi R\beta H},
\end{equation}
which describes the finite temperature system with the same Hamiltonian at $R=\infty$. Generally speaking, the calculation for the finite temperature system becomes more complicated than the zero-temperature case. However, if we limit ourselves in the singlet sector\footnote{As we will see in section \ref{sec:6}, there is a physical ground which justifies this restriction. Practically, if we include the non-singlet sector correction, undesirable properties such as the breakdown of the T-duality emerge \cite{Gross:1991md}.}, the calculation is not so difficult as it may seem.

First, we introduce the chemical potential as a convenient prescription to deal with the finite temperature system in the $N \to \infty$ limit:
\begin{equation}
\frac{N}{\beta} = \int_{-\infty}^\infty \rho(\epsilon) d\epsilon \frac{1}{1+e^{2\pi R\beta(\epsilon - \mu_F)}}.
\end{equation}
Then, in the thermodynamical limit, the free energy (the string partition function) can be written as
\begin{equation}
\frac{\partial Z_R}{\partial N} = \mu_F.
\end{equation}
Setting the critical point of the potential to be zero: $\mu_c=0$, we obtain by differentiating the above equation,
\begin{eqnarray}
\frac{\partial \Delta}{\partial \mu} &=& \pi \tilde{\rho}(\mu) = \pi \int_{-\infty}^{\infty} de \rho (e) \frac{\partial}{\partial \mu} \frac{1}{1+e^{2\pi R\beta(\mu-e)}} \cr
\frac{\partial Z_R}{\partial \Delta} &=& \frac{1}{\pi} \beta \mu,\label{eq:renritu}
\end{eqnarray}
where $\mu=-\mu_c$ and $e=-\epsilon$. The point is that only the density of states effectively receives some corrections compared with the zero-temperature case.

Substituting the zero-temperature density of states $\rho(e)$ into (\ref{eq:renritu}) and carrying out the integration, we obtain
\begin{equation}
\frac{\partial \Delta}{\partial \mu} = \mathrm{Re} \int_\mu^\infty \frac{dt}{t}e^{-it} \frac{t/2\beta\mu}{\sinh{t/2\beta\mu}}\frac{t/2\beta\mu R}{\sinh(t/2\beta\mu R)}.
\end{equation}

Just as in the $R=\infty$ case, we introduce $\mu_0$ and perform the integration perturbatively. After some algebra, the partition function is given by
\begin{equation}
Z = \frac{1}{4}\left[(2\beta\mu_0\sqrt{R})^2\log\mu_0 - 2f_1(R)\log\mu_0 + \sum_{m=1}^\infty \frac{f_{m+1}(R)}{m(2m+1)}(2\beta\mu_0\sqrt{R})^{-2m}\right], \label{eq:matz}
\end{equation}
where $f_n$ is a function of $R+R^{-1}$ and manifestly invariant under the exchange of $R$ and $R^{-1}$. The actual form is given by $f_m(R) = (2m-1)!\sum_{k=0}^m|2^{2k}-2||2^{2(m-k)}-2|\frac{|B_{2k}||B_{2(m-k)}|}{(2k)![(2(m-k)]!}R^{m-2k}$.  This is the celebrated Gross-Klebanov formula \cite{Gross:1990ub}. After we introduce $\mu_r= \beta\mu_0$ and take the double scaling limit (and divide it by two), we finally obtain the all-genus partition function of the compactified two dimensional noncritical string theory which reproduces the Liouville computation for the genus zero and one (see (\ref{eq:1loop})).

Note that this expression is manifestly invariant under the following T-duality (up to diverging $\log$ terms):
\begin{eqnarray}
 R &\to& \frac{1}{R} \cr
 \mu_r &\to& R \mu_r.
\end{eqnarray}
The transformation of the string coupling constant is familiar, for the T-duality transforms the dilaton expectation value because of the one-loop correction.
Finally, we explicitly calculate the $R=1$ partition function where the radius is selfdual under the T-duality. This case is important for the application to the topological string as we will review in section \ref{sec:6}. 

Since $\tilde{\rho}(\mu_r)$ is given, we calculate the second derivative of the partition function via the convenient method based on the Legendle transformation
\begin{equation}
\frac{\partial^2 Z}{\partial \mu_r^2} = \mathrm{Re} \int_0^\infty \frac{dt}{t}e^{-i\mu_r t} \left(\frac{t/2}{\sinh t/2}\right)^2.
\end{equation}
Carrying out the integration, we obtain
\begin{equation}
Z(\mu_r) = \frac{1}{2}\mu_r^2\log\mu_r -\frac{1}{12}\log\mu_r +\sum_{g=2}^\infty (-1)^g\frac{B_{2g}}{2g(2g-2)}\mu_r^{2-2g}. \label{eq:cr1p}
\end{equation}
\subsubsection{Tachyon scattering}\label{3.2.2}
Next we review the tachyon scattering in the two dimensional string theory. Since there are many good review articles \cite{Klebanov:1991qa}, \cite{Jevicki:1993qn}, \cite{Ginsparg:is}, \cite{Polchinski:1994mb} on the subject, the description here will be brief. In the following, we derive the tachyon scattering amplitudes in three different but (should-be) equivalent ways: the direct Liouville calculation, the direct matrix calculation and the collective field theory of the Fermi surface. 

Let us consider the on-shell vertex of the two dimensional (Liouville $+$ time-like boson) string theory. After Wick rotating the $X$ direction as $X \to it$, the on-shell tachyon vertex is given by
\begin{equation}
V_{\omega}^{\pm} = e^{\pm i\omega t + i\omega \phi +2\phi},
\end{equation}
which has the weight $(1,1)$ as it should be. The minus sign corresponds to the incoming rightmover and the plus sign corresponds to the outgoing leftmover. In the actual calculation, it is convenient to do in the Euclidean signature. This is done by 
\begin{equation}
-i\omega \to |q|
\end{equation}
so that the vertices become
\begin{eqnarray}
V_\omega^- &\to& e^{-i|q|X +(2-|q|)\phi} \cr
V_\omega^+ &\to& e^{+i|q|X +(2-|q|)\phi}.
\end{eqnarray}
These can be combined as $V_q = e^{iqX+(2-|q|)\phi}$, which represents the incoming wave when $q <0$ and outgoing wave when $q>0$ respectively.\footnote{Note that the normalization of the momentum here is somewhat different from what is used in the most of the other sections: $q_{here} = 2p_{there}$.}

The direct Liouville calculation is the most challenging program. We here consider the particular amplitude where the violation of the Liouville charge is an integer; then we analytically continue the result \cite{Goulian:1991qr}, \cite{DiFrancesco:1992ud}. Of course, the calculation here is limited to the tree level amplitude.

The setup is the following. We consider the situation where one incoming wave is scattered into $N$ outgoing waves whose (Euclidean) momenta are $(q_1, \cdots, q_N)$ respectively. The simplest scattering occurs if we set the momentum of the incoming wave $q_{N+1}$ to be $1-N$ so that the Liouville momentum is conserved without any insertion of the Liouville potential. Here we have also used the ``energy" conservation: $\sum_{i=1}^{N+1} q_i= 0$.

Using the free field correlator, we can evaluate the $N$-point function explicitly. In addition, the integration over the sphere can be carried out \cite{DiFrancesco:1991ss,DiFrancesco:1992ud}. The result is given by
\begin{eqnarray}
\langle V_{q_1} V_{q_2} \cdots V_{q_{N+1}} \rangle &=& \lim_{s\to 0} \frac{\mu^s}{2}\Gamma(s) \frac{\pi^{N-2}}{(N-2)!} \prod^{N}_{i=1}\gamma(1-|q_i|) \cr
&=& \frac{\pi^{N-2}}{2} \prod^{N}_{i=1}\gamma(1-|q_i|)\frac{\Gamma(1-(N-1))}{\Gamma(N-1)} (N-2)! \cr
&=& \frac{(N-2)!}{2} \prod^{N+1}_{i=1}\gamma(1-|q_i|) \cr
&=& \lim_{s\to 0}\frac{1}{2} \prod^{N+1}_{i=1}\gamma(1-|q_i|) \left(\frac{\partial}{\partial \mu}\right)^{N-2}\mu^{s+N-2}.
\end{eqnarray}
Taking the appropriate decoupling limit, we can obtain the similar expression for non-zero (but integer) $s = 1-N +|q_{N+1}|$ with  $q_i >0, i=1\cdots N$.
\begin{equation}
\langle V_{q_1} V_{q_2} \cdots V_{q_{N+1}} \rangle = \frac{1}{2} \prod^{N+1}_{i=1}\gamma(1-|q_i|) \left(\frac{\partial}{\partial \mu}\right)^{N-2}\mu^{s+N-2}
\end{equation}
This form has an obvious analytic continuation for non-integer $s$ which is given by,
\begin{equation}
\langle V_{q_1} V_{q_2} \cdots V_{q_{N+1}} \rangle = \frac{1}{2} \prod^{N+1}_{i=1}\gamma(1-|q_i|) \left(\frac{\partial}{\partial \mu}\right)^{N-2}\mu^{s+N-2}.
\end{equation}
In the above formula, the pole from the integer $s$ should be renormalized as $\lim_{s\to 0}\mu^s\Gamma(s) = \log\mu$. In this case, we call the process as the ``bulk scattering" because the scattering amplitude is proportional to the Liouville volume. In this terminology, only the bulk scattering has been calculated in the direct Liouville approach. The other amplitudes are conjectured by the analytic continuation.

We next consider the direct matrix model calculation. The natural object which corresponds to the tachyon vertex is given by the following puncture operator
\begin{equation}
P(q) \sim \lim_{l \to 0} l^{-|q|}O(q,l),
\end{equation}
where the loop operator $O(q,l)$ is defined as
\begin{equation}
O(q,l) = \int dx e^{iqx} \mathrm{Tr} e^{-l\Phi(x)}.
\end{equation}
We can easily see that the insertion of this operator into the matrix model path integral corresponds to eliminating the loop of length $l$. We naturally expect the following correspondence
\begin{equation}
V_q \iff P(q).
\end{equation}
The direct matrix model evaluation of the correlator is performed in the second quantized formalism, where we take the second quantized action as 
\begin{equation}
S = \int d\lambda dx \psi^\dagger \left(-\frac{\partial}{\partial x} + \frac{\partial^2}{\partial \lambda^2} + \frac{\lambda^2}{4} + \mu\right)\psi.
\end{equation}
We rewrite the loop operator in this formalism as 
\begin{equation}
O(q,l) = \int dx e^{iqx} \int d\lambda e^{-l\lambda} \psi^\dagger\psi(x,\lambda).
\end{equation}
The calculation left is the application of the Wick theorem (at least in principle). The actual evaluation is highly technical, so we simply quote some of the lowest order results \cite{Ginsparg:is}\footnote{Actually, there exists a very elegant graphical rule to calculate these matrix elements. We will briefly review the method in the next subsection.}
\begin{equation}
\langle P(q) P(k) \rangle = - \delta(q+k) \Gamma(1-|q|)^2 \mu^{|q|}\left(\frac{1}{|q|} +\cdots \right)
\end{equation}
\begin{equation}
\langle P(q_1) P(q_2) P(q_3)\rangle = \delta(q_1 +q_2 +q_3) \prod_{i=1}^3 \left(\Gamma(1-|q_i|) \mu^{|q_i|/2} \right) \left(\frac{1}{\mu} + \cdots\right),
\end{equation}
where the omitted terms are the higher loop terms (which can be obtained explicitly in this formalism). Comparing these amplitudes with the previous direct Liouville calculation, we observe that the extra wavefunction renormalization factor $\frac{1}{\Gamma(|q|)}$ is needed to reproduce the amplitude. Hence, the exact correspondence is given by
\begin{equation}
T(q) = \frac{P(q)}{\Gamma(|q|)},
\end{equation}
which is only confirmed at the tree level. However, the correspondence is believed to remain correct at the higher genus. 

Finally, let us review the collective field method which is sometimes called the Das-Jevicki field theory \cite{Das:1990ka}. This can be obtained in various ways. Gross and Klebanov \cite{Gross:1991st} have derived it from the bosonization of the non-relativistic fermion. Recall that the matrix model is equivalent to the non-relativistic fermionic system whose second quantized Hamiltonian is given by
\begin{equation}
H = \int d\lambda \left[\frac{1}{2}\partial_\lambda \psi^\dagger \partial_\lambda \psi -\frac{\lambda^2}{2} \psi^\dagger\psi + \mu \psi^\dagger \psi\right].\label{eq:second}
\end{equation}
First, we introduce new fermionic variables $\Psi_L$ and $\Psi_R$ as
\begin{equation}
\psi(\lambda,t) = \frac{e^{i\mu t}}{\sqrt{v(\lambda)}}\left[ e^{-i\int^\lambda d\lambda' v(\lambda') +i\pi/4} \Psi_L(\lambda,t)+ e^{i\int^\lambda d\lambda' v(\lambda') -i\pi/4}\Psi_R(\lambda,t)\right],
\end{equation}
where $v(\lambda) = \frac{d\lambda}{d \tau} = \sqrt{\lambda^2 -2\mu}$, which also determines a new variable $\tau$ in terms of $\lambda$. This makes the Hamiltonian almost relativistic. Then we bosonize the chiral fermions as 
\begin{eqnarray}
\Psi_L(\tau,t) = \frac{1}{\sqrt{2\pi}} :\exp\left[ i\sqrt{\pi}\int ^\tau (\Pi_S-S')d\tau'\right]: \cr
\Psi_R(\tau,t) = \frac{1}{\sqrt{2\pi}} :\exp\left[ i\sqrt{\pi}\int ^\tau (\Pi_S+S')d\tau'\right]:,\label{eq:bos}
\end{eqnarray}
where $S$ is a massless bosonic field and $\Pi_S$ is its canonical conjugate momentum. With these variables, the Hamiltonian is now
\begin{equation}
H = \frac{1}{2}\int_0^\infty d\tau \left( \Pi_S^2 + (S')^2 - \frac{\sqrt{\pi}}{2\mu \sinh^2\tau} \left[ \Pi_S S' \Pi_S + \frac{1}{3}(S')^3 \right] + tadpole \right), \label{eq:bosonized}
\end{equation}
where the tadpole term can be found in the literature \cite{Gross:1991st}, \cite{Klebanov:1991qa}. We can calculate the S matrix by using this Hamiltonian. In this sense, $S$ can be regarded as the space-time tachyon field (up to the wavefunction renormalization we will see later).

What is the physical meaning of the Hamiltonian? This can be clearly seen from the original derivation of this collective Hamiltonian by Das and Jevicki \cite{Das:1990ka}, \cite{Jevicki:1993qn}. Going back to the original matrix Lagrangian:
\begin{equation}
L = \frac{1}{2}\mathrm{Tr}\left(\dot{\Phi}^2 + \Phi^2\right),
\end{equation}
we introduce the collective density field as 
\begin{equation}
\phi(\lambda ,t) = \sum_{i=1}^N \delta(\lambda-\lambda_i(t)).
\end{equation}
Then the action can be rewritten as follows
\begin{equation}
S = \int dt d\lambda \left[ \frac{1}{2} \frac{\partial^{-1}_\lambda\dot{\phi}\partial^{-1}_\lambda\dot{\phi}}{\phi} -\frac{\pi^2}{6}\phi^3 + (-\mu + \frac{\lambda^2}{2})\phi\right].\label{eq:Das}
\end{equation}
In order to treat the tadpole term correctly, we consider the fluctuation around the classical solution 
\begin{equation}
\phi = \frac{1}{\pi} (v - \frac{\sqrt{\pi}}{v} \partial_\tau S),
\end{equation}
where $\tau$ and $v$ have already been introduced: $v(\lambda) = \frac{d\lambda}{d \tau} = \sqrt{\lambda^2 -2\mu}$.
Substituting this back into the action (\ref{eq:Das}), we rederive the bosonized action (\ref{eq:bosonized}) up to the tadpole term which is related to the normal ordering and renormalization needed for the collective field theory to be finite. The physical interpretation of the collective field theory is now clear. The bosonic field $S$ describes the small fluctuation of the Fermi surface (eigenvalue density).

We can use this Hamiltonian to calculate the scattering amplitude to any order of the perturbation. Instead of doing that, we would like to obtain the tree level (classical) scattering amplitude following Polchinski \cite{Polchinski:1994mb} (see also \cite{Gutperle:2003ij}) by using the hidden symmetry of the theory ($W_\infty$ symmetry). Given the classical equation of motion
\begin{equation}
\dot{\lambda} = p, \ \ \ \ \dot{p} = \lambda,
\end{equation}
we have conserved quantities
\begin{equation}
v = (-\lambda - p) e^{-t}, \ \ \ \ w = (-\lambda+p) e^t.
\end{equation}
Furthermore, arbitrary powers of these are conserved\footnote{If we define $C_{m,n} \equiv v^m w^n$, they constitute the generators of $W_\infty$ algebra. We can easily see that $\{C_{m,n},C_{m',n'}\} = 2(m'n-n'm) C_{n+n'-1,m+m'-1}$ hold under the classical Poisson bracket.}. Let us consider the following conserved quantity
\begin{equation}
v_{mn} = \int_{F-F_0} \frac{dp dx}{2\pi} v^m w^n,
\end{equation}
where $F_0$ is the static Fermi surface. When $t$ tends to $-\infty$, the rightmoving fluctuation of the Fermi surface is given by $\epsilon_+(\tau,t) = \epsilon(t-\tau) = \sqrt{\pi}( \Pi_S-\partial_\tau S).$\footnote{It is important to note that $\tau = -\log(-\lambda)$ and $\epsilon_{\pm} = \pm(p\pm \lambda)$ in this limit. We will use this fact later in section \ref{sec:6}.} Evaluating the conserved quantity in this limit, we obtain
\begin{equation}
 v \to e^{\tau-t} \epsilon_+(\tau-t),\ \ \ \ w \to 2 e^{-\tau +t}.
\end{equation}
Similarly, in the $t \to \infty$ limit, we have
\begin{equation}
 v \to 2e^{-\tau-t},\ \ \ \ w \to e^{\tau +t} \epsilon_-(\tau + t),
\end{equation}
where $\epsilon_-(\tau,t) = \epsilon(t+\tau) = \sqrt{\pi}( -\Pi_S-\partial_\tau S)$. In these variables, the conserved quantity can be written as 
\begin{eqnarray}
v_{mn} &=& \frac{2^n}{2\pi(m+1)} \int_{-\infty}^\infty dt e^{(n-m)(t-\tau)}\left((\epsilon_+(t-\tau))^{m+1} -\mu^{m+1}\right) \cr
&=& \frac{2^m}{2\pi(n+1)} \int_{-\infty}^\infty dt e^{(n-m)(t+\tau)}\left((\epsilon_-(t+\tau))^{n+1} -\mu^{n+1}\right) \label{eq:vmn}
\end{eqnarray}
This expression enables us to calculate the tree level S matrix. To do this, we consider the mode expansion of the incoming (outgoing) wave as 
\begin{eqnarray}
S(\tau,t) = \int_{-\infty}^\infty \frac{dk}{\sqrt{8\pi^2k^2}}\left[a_k e^{-i|k|t+ik\tau} +a^\dagger_k e^{i|k|t-ik\tau}\right] \label{eq:modet}
\end{eqnarray}
where incoming $k>0$ operators and outgoing $k<0$ operators are not independent because of the Liouville wall. Substituting this into (\ref{eq:vmn}) with $m=0$ and $n=i\omega$, we have the nonlinear expression which reveals the relation between incoming modes and outgoing modes.
\begin{eqnarray}
a^\dagger_k = (\frac{1}{2}\mu)^{-ik} \sum_{n=1}^\infty \frac{1}{n!} \left(\frac{i}{\sqrt{2\pi}\mu}\right)^{n-1} \frac{\Gamma(1-ik)}{\Gamma(2-n-ik)} \cr
\int_{-\infty}^0 dk_1\cdots dk_n (a_{k_1}^\dagger -a_{k_1})\cdots (a_{k_n}^\dagger -a_{k_n}) \delta(\pm |k_1| \cdots \pm |k_n|-k), \label{eq:smat}
\end{eqnarray}
where $\pm$ sign in the delta function is related to whether we have chosen a creation operator or an annihilation operator in the braces before.
Mode operators satisfy the canonical commutation relation
\begin{equation}
[a_k,a^\dagger_{k'}] = 2\pi |k|\delta(k-k').\label{eq:canoc}
\end{equation}
Defining the in-states as
\begin{equation}
|k_1\cdots k_n:in\rangle = a^\dagger_{k_1} \cdots a^\dagger_{k_n} |0\rangle, \ \ \ \ k_i >0,
\end{equation}
and similarly for the out-states, we can calculate the tree level S matrix using the canonical commutation relation (\ref{eq:canoc}). For example, the now familiar $1\to n$ amplitude is given by
\begin{equation}
\langle k_1 \cdots k_n;out| k;in\rangle = \left(\frac{i}{\sqrt{2\pi}\mu}\right)^{n-1} \frac{\Gamma(1+ik)}{\Gamma(2-n+ik)} 2\pi \delta (k_1+\cdots +k_n -k)\prod_{i=1}^n 2\pi k_i.
\end{equation}
which agrees with the previous calculation if we further introduce the leg factor $ -i(\mu)^{-i\frac{k}{2}} \frac{\Gamma(ik)}{\Gamma(-ik)}$ for each external leg (for the Euclidean S matrix, this becomes $-\mu^{|q|/2}\frac{\Gamma(-|q|}{\Gamma(|q|)}$). Note that this leg pole factor is simply a pure phase in the Minkowski signature. Therefore this factor does not affect the cross-section unless we prepare the superposition states of the different momenta. 

To see the origin of the leg factor, let us reconsider what we should expect as the external line attached to the rest of the Feynman diagram in the collective field theory. Let us recall that the tachyon vertex corresponds to the puncture operator in the matrix model. Then, the corresponding operator in the collective field theory becomes
\begin{eqnarray}
O(l,q) &=& \int dx e^{iqx} \mathrm{Tr} e^{-l\Phi} \cr
       &=& \int dx e^{iqx} \int d\lambda e^{-l\lambda} \partial_\tau S \frac{\partial \tau}{\partial \lambda} \cr
       &=& \int dx e^{iqx} \int d\tau e^{-l\lambda(\tau)} \partial_\tau S,
\end{eqnarray}
where we have used the correspondence between the matrix model and the collective field theory: $\mathrm{Tr} O(\Phi) \sim \int d\lambda O(\lambda) S'(\lambda) $. We perform the Fourier transformation as $O(l,q) = i\int dk F(k,l) k S(q,k)$, where
\begin{align}
F(k,l) &= \int_{-\infty}^\infty d\tau e^{-l\lambda(\tau)} \cos(k\tau) \cr
S(x,\tau) &= \int dq e^{-iqx} \int dk \sin(k\tau) S(q,k).
\end{align}
Evaluating this with the classical motion $\lambda(\tau) = \sqrt{2\mu} \cosh \tau$, we find
\begin{equation}
F(k,l) = \frac{\pi}{2\sin(ik\pi)} \left(I_{-ik}(l\sqrt{2\mu}) - I_{ik}(l\sqrt{2\mu})\right)
\end{equation}
We need the small $l$ limit of this function where we can replace it with 
\begin{equation}
I_\nu (l\sqrt{2\mu}) \to (l\sqrt{2\mu}/2)^\nu \frac{1}{\Gamma(\nu +1)}.
\end{equation}
In calculating the Euclidean Feynman diagram, every external line should include this factor with the propagator $\frac{1}{q^2+k^2}$ and the momentum integration over $k$ and $q$. We deform the $k$ integration in a special way \cite{Klebanov:1991qa}: for $I_{-ik}$ we pick up the pole at $k =i|q|$, while for $I_{ik}$ we pick up the pole at $k=-i|q|$. This seems strange but it is actually necessary in order to ensure the convergence of the integral. Consequently we obtain the desired external leg factor
\begin{equation}
-(l\sqrt{\mu/2})^{|q|} \Gamma (-|q|).
\end{equation}
Finally, we must not forget the additional leg factor needed to connect the matrix model puncture operator with the Liouville vertex operator, $T(q) = \frac{P(q)}{\Gamma(|q|)}$. This completes the explanation of the leg factor in the collective field theory.

Let us summarize what we have learned in this (rather lengthy) subsection. There are three different (but supposed to be equivalent) methods to calculate the tachyon scattering in the 2D noncritial string theory. 
\begin{enumerate}
	\item The direct Liouville calculation of the correlation functions of the vertex operators: this is conceptually straightforward but has only been performed for the tree level (special) amplitudes.
	\item The direct matrix model calculation of the correlation functions of the puncture operators with the leg factor $\frac{1}{\Gamma(|q|)}$: this provides the finite full genus answer, but the actual calculation is involved.
	\item The collective field calculation of the S matrix for the massless $S$ field with the leg factor $-\mu^{|q|/2}\frac{\Gamma(-|q|)}{\Gamma(|q|)}$: $S$ describes the fluctuation of the Fermi surface. However this method somewhat hides the nonperturbative nature of the matrix model.
\end{enumerate}

\subsubsection{From MPR formula to DMP formula}\label{3.2.3}
In the last subsection, we studied how to calculate the tachyon S matrix from the matrix quantum mechanical point of view. Actually, there is a very elegant graphical method to evaluate the tachyon S matrix even nonperturbatively. This has been done by Moore, Plesser and Ramgoolam (MPR) \cite{Moore:1992zv}. We briefly review their philosophy and the elegant and intuitive results without delving into the detailed derivation.

The main claim is the following intuitive formula:
\begin{equation}
S_{CF} = \iota_{f\to b} \circ S_{FF} \circ \iota_{b\to f}, \label{eq:MPR}
\end{equation}
where $\iota_{f\to b}$ is the bosonization map and $\iota_{b\to f}$ is its inverse, and $S_{FF}$ is the S matrix for the free fermion (under the inverted harmonic oscillator potential). The tachyon S matrix is related to the collective field S matrix $S_{CF}$ via multiplying the leg pole factors (see the last subsection). The intuitive meaning of this formula is as follows. We first fermionize the incoming bosonic collective field by the asymptotic fermionization formula\footnote{For the nonrelativistic fermion, bosonization-fermionization is only possible in the asymptotic region where by suitable transformation we can treat them as if they were relativistic (see the last subsection), but for the S matrix, only the asymptotic region is necessary.}
\begin{equation}
a(\omega) \to \int_{-\infty}^{\infty} d\xi b(\mu +\xi)b^\dagger(\mu -(\omega-\xi)),
\end{equation}
where the incoming boson is normalized as $[a(\omega),a(\omega')] = \omega \delta(\omega + \omega')$ and $a^\dagger(\omega) = a(-\omega)$. Then we calculate the S matrix for the fermion. This is very easy since the fermion is free in this representation. Therefore the S matrix (just the reflection amplitude) is diagonal
\begin{equation}
b_{out} (\omega) = R(\omega) b_{in} (\omega) = S b_{in}(\omega) S^{-1}, 
\end{equation}
where 
\begin{equation}
S  = \exp \left[-\int_{-\infty}^{\infty} d\omega \log R(\omega) b^\dagger_{in}(\omega) b_{in}(\omega)\right]. \label{Smat}
\end{equation}
Then we rebosonize again the scattered fermion, which yields the final state.

Thus, the only nontrivial information needed to calculate the S matrix is the reflection amplitude $R(\omega)$. This can be calculated by solving the quantum mechanical scattering problem. For example, the inverted harmonic oscillator which is restricted to the left side only\footnote{This is the theory I in the language of \cite{Moore:1992zv}. In the modern sense, this theory is rather ad hook. We will see in part II of this review, how these stable (and unitary) theories emerge from the supersymmetric extension of the Liouville theory.} has the following reflection amplitude
\begin{equation}
R(\omega) = \mu^{i\omega} i\sqrt{\frac{1+ie^{-\pi(\mu +\omega)}}{1-ie^{-\pi(\mu +\omega)}}} \sqrt{\frac{\Gamma(\frac{1}{2}-i(\mu+\omega))}{\Gamma(\frac{1}{2}+i(\mu+\omega))}}, 
\end{equation}
where we have omitted the irrelevant $\omega$ independent phase factor.

The important corollary of the MPR formula (\ref{eq:MPR}) is that the collective S matrix is unitary if and only if $S_{FF}$ is unitary and the bosonization map yields the complete set of states. For example, if we allow the tunneling amplitude which goes from the left-hand side of the potential into the right-hand side of the potential and only consider the left-hand side as the physical bosonic excitation (theory II in the language of \cite{Moore:1992zv}), we have a non-unitary theory, which leads to the conclusion that the naive bosonic Liouville theory is not unitary nonperturbatively.

\begin{figure}[htbp]
	\begin{center}
	\includegraphics[width=0.6\linewidth,keepaspectratio,clip]{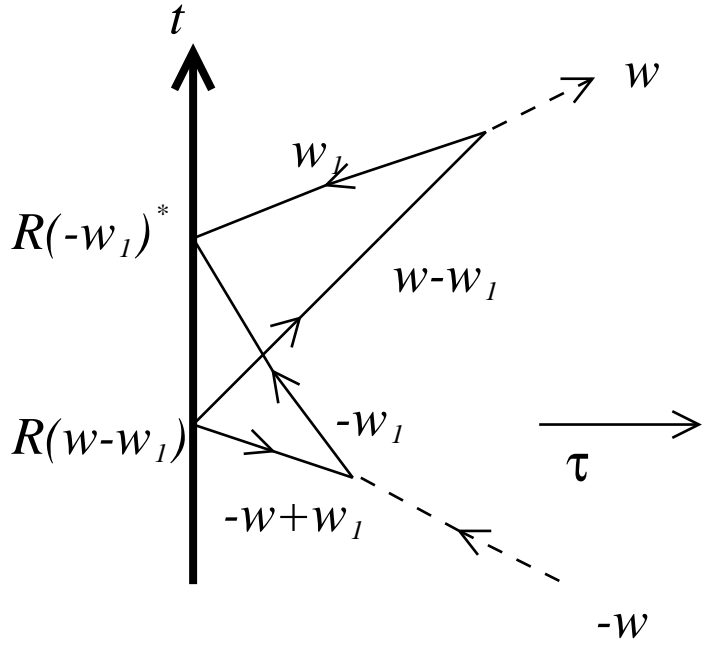}
	\end{center}
	\caption{The diagram for the MPR formula.}
	\label{MPR}
\end{figure}

Let us go into the actual evaluation of the amplitude. There is a very intuitive diagrammatic rule to calculate the S matrix. We just list the rule (in the Minkowski signature for definiteness) here. To each incoming and outgoing boson, associate a vertex in the $(t,\tau)$ space. Connect points via line segments to form a one-loop graph. Each line segment carries a momentum (energy) and an arrow. Once the line hits the ``reflection line" the propagator is reflected as in figure \ref{MPR} with the reflection factor $R$.
These lines are joined according to the following rule:
\begin{enumerate}
	\item Lines with positive (negative) momenta slope upwards to the right (left),
	\item At any vertex arrows are conserved and momentum is conserved as time flows upwards. In particular momentum $\omega_i$ is inserted at the vertex as in figure \ref{MPR}.
	\item Outgoing vertices at $(t_{out}, \tau_{out})$ all have later times than incoming vertices $(t_{in}, \tau_{in})$: $t_{out} > t_{in}$.
\end{enumerate}

To each graph we associate an amplitude, with reflection factors $R$ and $\pm$ for upwards (downwards) sloping direct propagators. Finally, we sum over graphs and integrate over kinematically allowed momenta, schematically
\begin{equation}
S = i^n \sum_{graph}\pm \int d\omega \prod_{reflection} R(\omega)R^*(-\omega)
\end{equation}

\begin{figure}[htbp]
	\begin{center}
	\includegraphics[width=0.8\linewidth,keepaspectratio,clip]{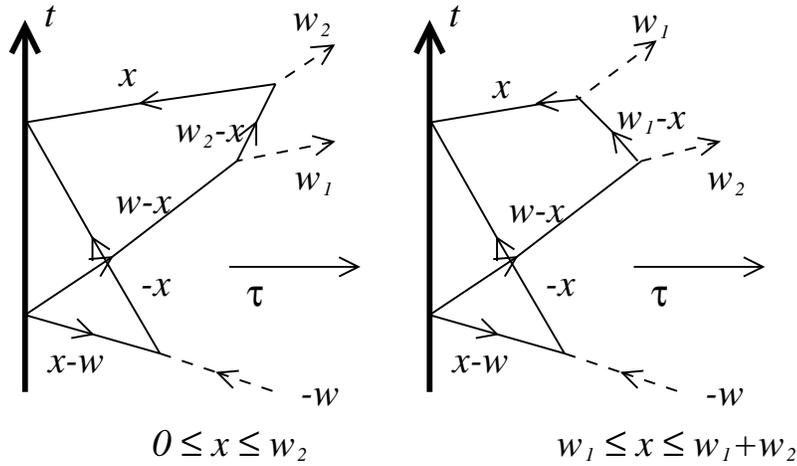}
	\end{center}
	\caption{The three body scattering calculation from the MPR formula.}
	\label{MPR2}
\end{figure}

For example, $1\to 2$ scattering is represented by two diagrams (figure \ref{MPR2}); then we have
\begin{equation}
S(\omega \to \omega_1 + \omega_2) = \int_0^{\omega_2} dx R(\omega-x)R^*(-x) - \int_{\omega_1}^{\omega} dx R(\omega -x) R^*(-x).
\end{equation}
In the actual evaluation, we can expand the reflection amplitude in the series of $1/\mu$ as 
\begin{equation}
R(\omega) = 1 + \sum_k\frac{Q_{k}(\omega)}{\mu^k}.
\end{equation}
Then we have the perturbative expansion of the S matrix to any desired order.

Now that we have learned how to calculate any S matrix, let us consider the generating functional of the tachyon amplitudes in the Liouville theory. First, we define the renormalized vertex operator as 
\begin{equation}
\tilde{V}^{\pm}_\omega = \frac{\Gamma(-i\omega)}{\Gamma(i\omega)} \mu^{1+i\omega/2} V_{\omega}^{\pm}, \label{eq:veren}
\end{equation}
and define the following generating functional of the vertex correlators:
\begin{equation}
\mu^2 F[t(\omega),\bar{t}(\omega)] \equiv \left\langle\left\langle e^{\int_0^\infty d\omega t(\omega) \tilde{V}_{\omega}^+} e^{\int_0^\infty d\omega \bar{t}(\omega) \tilde{V}_{\omega}^-}\right\rangle \right\rangle,
\end{equation}
where $\left\langle\left\langle \cdots \right\rangle \right\rangle$ means a sum over genus and integration over the moduli space. On the other hand, the matrix model duality and the MPR formula suggest that the same generating functional can be written as
\begin{equation}
\mu^2 F[t(\omega),\bar{t}(\omega)] = \langle 0| e^{\mu \int_0^\infty d\omega t(\omega) \alpha(-\omega)} S e^{\mu\int_0^\infty \bar{t} (\omega) \alpha(\omega)} |0\rangle,
\end{equation}
where $\alpha(\omega)$ is a bosonic creation operator and $S$ is the fermionic S matrix \eqref{Smat}. This concise formula was first derived in \cite{Dijkgraaf:1993hk} by Dijkgraaf, Moore and Plesser (DMP). Furthermore it has a simple compactified Euclidean analog. With a compactified radius $\beta$, we have
\begin{eqnarray}
\mu^2 F &\equiv& \left\langle\left\langle e^{\sum_{n\ge 1}t_n\tilde{V}_{n/\beta} + \sum_{n \ge 1} \bar{t}_n \tilde{V}_{-n/\beta} }\right\rangle \right\rangle  \cr
&=& -\frac{1}{\beta} \langle 0| e^{i\mu \sum_{n \ge 1}t_n \alpha_n} S e^{i\mu \sum_{n\ge 1} \bar{t}_n \alpha_{-n}}|0\rangle,
\end{eqnarray}
where 
\begin{equation}
S = :\exp \left(\sum_{m} \log R_{p_m} \psi^{out}_{-(m+1/2)} \bar{\psi}^{out}_{m+1/2}\right):
\end{equation}
and $\alpha_n$ and $\psi_n$ are related by the usual bosonization formula: $\partial \phi = \psi \bar{\psi}$.

Now we take the special radius, namely the selfdual radius $\beta = 1$. Using the explicit expression for $R$, we can show (see \cite{Dijkgraaf:1993hk} for the derivation) that $Z(t,\bar{t}) = e^{F(t,\bar{t})}$ is a $\tau$-function of the Toda hierarchy, satisfying an infinite set of constraints that form a $W_\infty$ algebra. Surprisingly these constraints can be integrated to another matrix model! We are not going into the derivation here but just quote the final result \cite{Imbimbo:1995yv}.

We make use of the large $N\times N$ matrix $A$ instead of $t_k$ whose relation is given by the Miwa transformation
\begin{equation}
t_k = \frac{1}{\nu k} \mathrm{Tr} A^{-k},
\end{equation}
where $\nu = -i\mu$.
Then the generating function is given by
\begin{equation}
Z(t,\bar{t}) = e^{F(t,\bar{t})} = \int dM \exp\left[\mathrm{Tr}\left(-\nu M + (\nu -N) \log M - \nu \sum_{k=1}^\infty \bar{t}_k (MA^{-1})^k\right)\right].
\end{equation}
This is what is called the $W_\infty$ matrix model. However we should note that this generating function does not include the winding mode of the theory. It is not yet known how to implement them into the matrix form.
\subsection{Literature Guide for Section 3}\label{3.3}
In this section, we have provided only the minimal knowledge of the matrix model which is necessary to follow the arguments in the later sections, and the complete reference list for the matrix model is beyond the scope of this review. We can find the reference list of the matrix model in the reviews such as \cite{Ginsparg:is}, \cite{DiFrancesco:1993nw}, \cite{Klebanov:1991qa}, \cite{Martinec:1991kn}, \cite{Das:1992dm}.

While in the main text, we have mainly discussed the \textit{physical} matrix models, which are derived from the discretization of the Riemann surface, there is another \textit{topological} approach (the Kontsevich model and the Penner model) to the noncritical string theory (topological matrix model). Surprisingly they give exactly the same result as the physical matrix models. The recent review on the topological matrix model and relation to the physical matrix model (emphasizing the $c=1$ $R=1$ model) is \cite{Mukhi:2003sz}. The topological matrix model yields a discretization of the moduli space of the Riemann surface  in a sense \cite{Kontsevich:1992ti}, \cite{Penner}, \cite{Distler:1991mt} as opposed to the Riemann surface itself, and it is a finite matrix model as opposed to the double scaling limit of the \textit{physical} matrix model.

The connection between these two different perspective\footnote{Yet another formulation of $c=1$ noncritical string theory has been given in \cite{Alexandrov:2003qk}, where they have used the ``normal matrix" as a building block.} is curious and has caught a lot of attention recently (e.g. \cite{Mukhi:2003sz}, \cite{Aganagic:2003qj}). The former discusses the D-instanton physics which may connect $c=1$, $R=1$ matrix model and the Penner model (or the $W_\infty$ matrix model). The latter proposes that the topological $B$ model on $\mathrm{CY}_3$ is the right language to unify all these formulations of the noncritical string theory (and others).

\sectiono{DOZZ Formula}\label{sec:4}
In this section we derive the DOZZ formula \cite{Dorn:1994xn}, \cite{Zamolodchikov:1996aa} and study its properties. The DOZZ formula is a proposal for the Liouville three-point function on the sphere which is the fundamental building block of the Liouville theory as a CFT. 

The organization of this section is as follows. In section \ref{4.1}, we review its original derivation by \cite{Zamolodchikov:1996aa}, where the analytic continuation of the perturbatively calculable amplitudes are utilized. In section \ref{4.2}, we study the reflection property of the DOZZ formula and its physical implication. In section \ref{4.3}, we rederive the DOZZ formula by using what is called Teschner's trick \cite{Teschner:1995yf}, where the properties of degenerate operators are fully utilized. This trick will be repeatedly used in this review to derive many other Liouville amplitudes. In section \ref{4.4}, we review the higher equations of motion as an application of the DOZZ formula, which may lead to the understanding of the integrability of the minimal model coupled to the two dimensional gravity from the continuum Liouville perspective.

\subsection{Original Derivation}\label{4.1}
Though the three-point functions (or the structure constants) in the Liouville theory have not been calculated from the first principle, there is a proposal by Dorn-Otto and Zamolodchikov-Zamolodchikov. We review their derivation and its properties in this section. 

To begin with, we should note that for any CFT (on the sphere), the three-point function takes the following general form
\begin{equation}
\langle V_{\alpha_1}(z_1) V_{\alpha_2}(z_2) V_{\alpha_3}(z_3)\rangle = |z_{12}|^{2\Delta_{12}} |z_{23}|^{2\Delta_{23}} |z_{31}|^{2\Delta_{31}} C(\alpha_1,\alpha_2,\alpha_3),
\end{equation}
where $z_{ij} = z_i-z_j$, $\Delta_{12} = \Delta_3-\Delta_2-\Delta_1$ and $\Delta_i$ is the weight of $V_i$.

The claim of the DOZZ formula is
\begin{eqnarray}
& &C(\alpha_1,\alpha_2,\alpha_3) = \left[\pi\mu\gamma(b^2)b^{2-2b^2}\right]^{(Q-\sum_{i=1}^3 \alpha_i)/b} \cr &\times& \frac{\Upsilon'(0)\Upsilon(2\alpha_1)\Upsilon(2\alpha_2)\Upsilon(2\alpha_3)}{\Upsilon(\alpha_1+\alpha_2+\alpha_3-Q)\Upsilon(\alpha_1+\alpha_2-\alpha_3)\Upsilon(\alpha_1-\alpha_2+\alpha_3)\Upsilon(-\alpha_1+\alpha_2+\alpha_3)}, \label{eq:DOZZ}
\end{eqnarray}
where $V_\alpha = e^{2\alpha \phi} $. We refer to appendix \ref{a-3} for the special functions such as $\Upsilon$ function. We use the standard notation: $\gamma(x) \equiv \Gamma(x)/\Gamma(1-x)$. It may be just a trivial check, but note that the power of $\mu$ satisfies the exact WT identity or the KPZ scaling law (\ref{eq:ewt}) .

The basic strategy to derive this three-point function is as follows. As we have seen in the last section, the power of $\mu$ is determined from the exact WT identity for the general correlation functions in the Liouville theory. Thus, in general, the perturbation of $\mu$ is not reliable at all. 
However, if we adjust $\alpha$ or $b$ properly, the power of $\mu$ becomes an integer, so there is a possibility that the perturbative calculation is valid. Here, we ``assume" that the three-point function can be calculated perturbatively when the power of $\mu$ becomes an integer.\footnote{This reminds us of the instanton calculation for the supersymmetric gauge theory (for reviews, see \cite{Shifman:1999mv}, \cite{Dorey:2002ik}). In general, the naive instanton (not fractional) calculation does not yield the answer which satisfies the condition suggested  by the symmetry argument, but in the particular situation where the symmetry argument and the nonvanishing instanton amplitude matches, the instanton calculation yields the exact answer. For other cases (in which we cannot use the naive instanton calculation), we can ``analytically continue" the answer for the exact case because of the holomorphy of the theory.}
For the other part of the parameter region, we will define it by the analytic continuation of $s \equiv (Q-\sum_i \alpha_i)/b $. In the actual calculation, we immediately observe the following basic fact. Since we calculate the perturbative Liouville theory by the Coulomb gas integral, the charge conservation yields the zero-mode divergence of the Liouville direction whenever the power of $\mu$ becomes an integer. We regard this divergence as the pole of the analytically continued three point structure constant in terms of $s$. This yields a very strict constraint on the analytic property of $C(\alpha_1,\alpha_2,\alpha_3)$.

In order to study the actual structure of the poles, we try to obtain the three-point function by applying the ``perturbation theory",
\begin{eqnarray}
\langle V_{\alpha_1}(z_1) V_{\alpha_2}(z_2) V_{\alpha_3}(z_3)\rangle &=& \int\mathcal{D} \phi \prod_{i=1}^3e^{2\alpha_i \phi(z_i)} e^{-S_L}  \cr
&\sim & \int[\mathcal{D} \phi]_{free} \prod_{i=1}^3e^{2\alpha_i \phi(z_i)} \sum_{n=0}^{\infty} \frac{(-\mu)^n}{n!}\left(\int d^2z e^{2b\phi}\right)^ n e^{-S_{CG}}.
\end{eqnarray}
We separate the zero mode of the path integration over $\phi(z)$ from the non-zero mode: $\phi(z) = \phi_0 + \bar{\phi}(z)$, and integrate over the zero mode first,
\begin{equation}
\langle V_{\alpha_1}(z_1) V_{\alpha_2}(z_2) V_{\alpha_3}(z_3)\rangle \sim -\sum_{n=0}^{\infty} \frac{(-\mu)^n}{n!} \frac{1}{2b(s-n)} \int[\mathcal{D} \bar{\phi}]_{free} \prod_{i=1}^3e^{2\alpha_i \bar{\phi}(z_i)}\left(\int d^2z e^{2b\bar{\phi}}\right)^ n e^{-S_{CG}}. \label{eq:pert}
\end{equation}

We assume this yields the pole structure of the Liouville three-point functions. We can calculate the residue at $s=n$ by using the formula for the correlation functions of the Coulomb gas representation:
\begin{equation}
\langle e^{2\alpha_1 \phi_1} \cdots e^{2\alpha_N \phi_N} \rangle_{C.G;Q} = \delta\left(Q- \sum_i \alpha_i\right) \prod_{i>j} |z_{ij}|^{-4\alpha_i\alpha_j}  
\end{equation}
after the elimination of the delta function (which becomes the pole instead). After a long calculation \cite{Dotsenko:1984nm,Dotsenko:1985ad} involving the nontrivial integration which we will not reproduce here (the final result is summarized in \eqref{eq:DFF}), we obtain the residues of the correlator (\ref{eq:pert}), calling it $G^n_{\alpha_1,\alpha_2,\alpha_3}(z_1,z_2,z_3)$,
\begin{equation}
G^n_{\alpha_1,\alpha_2,\alpha_3}(z_1,z_2,z_3) = |z_{12}|^{2\Delta_{12}}|z_{23}|^{2\Delta_{23}}|z_{31}|^{2\Delta_{31}} I_n(\alpha_1,\alpha_2,\alpha_3)
\end{equation}
\begin{equation}
I_n(\alpha_1,\alpha_2,\alpha_3) = \left(\frac{-\pi\mu}{\gamma(-b^2)}\right)^n \frac{\prod_{j=1}^n \gamma(-jb^2)}{\prod_{k=0}^{n-1} \gamma(2\alpha_1b + kb^2)\gamma(2\alpha_2b+kb^2)\gamma(2\alpha_3b + kb^2)},\label{eq:preDOZZ2}
\end{equation}
where $\sum_{i=1}^3\alpha_i = Q- nb$.

We would like to analytically continue this expression in $s$ and determine $C(\alpha_1,\alpha_2,\alpha_3)$. As we will review later, this analytic continuation is unique if we demand the $b \to b^{-1}$ duality \cite{Teschner:1995yf}. Thus we only check that the DOZZ formula actually satisfies this relation.

First, note that the residues satisfy the following functional relations,
\begin{equation}
\frac{I_{n-1}(\alpha_1+b,\alpha_2,\alpha_3)}{I_{n}(\alpha_1,\alpha_2,\alpha_3)} = -\frac{\gamma(-b^2)}{\pi\mu}\frac{\gamma(b(2\alpha_1+b))\gamma(2b\alpha_1)\gamma(b(\alpha_2+\alpha_3-\alpha_1-b))}{\gamma(b(\alpha_1+\alpha_2+\alpha_3-Q))\gamma(b(\alpha_1+\alpha_2-\alpha_3))\gamma(b(\alpha_1+\alpha_3-\alpha_2))}. \label{eq:preDOZZ}
\end{equation}

To see this, we substitute the definition
$$ (LHS)=-\frac{\gamma(-b^2)}{\pi\mu} \frac{\gamma(2\alpha_1b+ nb^2)\gamma(2\alpha_2b+(n-1)b^2)\gamma(2\alpha_3+(n-1)b^2)\gamma(2\alpha_1b)\gamma(2\alpha_1b + b^2)}{\gamma(-nb^2)},$$
and use the relation $\gamma(x)=1/\gamma(1-x)$.

It is natural to assume that this functional relation holds for general $s$. The claim is
\begin{equation}
\frac{C(\alpha_1+b,\alpha_2,\alpha_3)}{C(\alpha_1,\alpha_2,\alpha_3)} = -\frac{\gamma(-b^2)}{\pi\mu}\frac{\gamma(b(2\alpha_1+b))\gamma(2b\alpha_1)\gamma(b(\alpha_2+\alpha_3-\alpha_1-b))}{\gamma(b(\alpha_1+\alpha_2+\alpha_3-Q))\gamma(b(\alpha_1+\alpha_2-\alpha_3))\gamma(b(\alpha_1+\alpha_3-\alpha_2))}. \label{eq:funrel}
\end{equation}
It is easy to see (\ref{eq:DOZZ}) satisfies this functional relation by using the difference formula of $\Upsilon$ function \eqref{eq:updif} which we have derived in appendix \ref{a-3}. Furthermore, the fact that the DOZZ formula has a pole at $s=n$ follows immediately from the structure of the zeros of the $\Upsilon$ functions. Therefore, the DOZZ formula indeed satisfies the desired properties at the pole $s=n$ which have been proposed by the perturbative calculation.

We have several comments on the DOZZ formula. As is remarked above on the uniqueness of the analytic continuation with respect to $s$, this formula has the following (quantum) symmetry,
\begin{eqnarray}
b &\to& \tilde{b} =b^{-1} \cr
\mu &\to& \tilde{\mu} = \frac{(\pi\mu\gamma(b^2))^{b^{-2}}}{\pi\gamma(b^{-2})}. \end{eqnarray}
This means that the DOZZ formula also has poles at
\begin{equation}
 sb = Q -\alpha_1-\alpha_2-\alpha_3 = nb+ mb^{-1},
\end{equation}
where $\Upsilon$ function indeed has a zero. This corresponds to the fact that the Liouville field theory has degenerate states besides $V_{-nb/2}$, namely the dual series $V_{-mb^{-1}/2}$ as we will see in the alternative derivation of the DOZZ formula (what is called Teschner's trick). In terms of the path integral, it can be interpreted as follows. When we split the path integral measure $\mathcal{D}\phi(z)$ into the zero mode and non-zero mode, there should be a renormalization of the action which involves the Jacobian of the transformation (because the path integral is not free, this is nontrivial). We expect this generates the ``dual cosmological constant term\footnote{Since this does not actually contribute in the semi classical limit $b\to 0$, it is appropriate to think of it as a quantum correction. Also, including this term in the action makes the above duality manifest. It is important to note that the dual cosmological constant $\tilde{\mu}$ does not spoil the KPZ scaling law only if it scales as $\mu^{b^{-2}}$.}"
\begin{equation}
\Delta S = \int d^2 z \tilde{\mu} e^{2\tilde{b}\phi(z)}
\end{equation}
However, this term violates the Seiberg bound as is discussed in section \ref{sec:2}. Therefore some subtleties exist when we include this term in the action.

On the other hand, this duality, with some further technical assumptions, enables us to prove the uniqueness (up to a constant multiplicity) of the solution of (\ref{eq:funrel}) \cite{Teschner:1995yf}.\footnote{Without this duality, there are actually infinitely many solutions of the functional relation. For instance, we can take any function $C(\alpha)$ which is defined in the interval $(0,b)$ and satisfies $C(0)=C(b)=0$. Then we \textit{define} the entire $C(\alpha)$ by the functional relation itself. These functions obviously satisfy the functional relation.} We sketch the proof here.

The assumptions we need are the continuity of $C$ with respect to $\alpha$ and the irrationality of $b^2$. Suppose we have another solution of the functional relation which we call $D$. If we define $R = D/C$, $R$ has a period $b$ and $b^{-1}$ at the same time. However, when $b^2$ is an irrational number, the function $R$ must be a constant (which was first proven by Liouville himself! \cite{Liouville:1990}). This theorem can be proven by the Fourier analysis. Thus the structure constant $C$ is unique up to a constant multiplication.

When $b^2$ is a rational number, though this case has important applications such as a coupling to the minimal model, the above proof fails. We just assume the continuity of $C$ with respect to $b$ and claim it to be unique for the time being. Besides, when $c>1$, $b$ being a complex number, there exists a double periodic function in this case, so the derivation is completely invalid. As a special case, when $c=25$, $b$ becomes pure imaginary, and the above derivation becomes a little subtle and we will reconsider it with special care later in section \ref{sec:6}. 

\subsection{Reflection Amplitude}\label{4.2}

The DOZZ formula has the following remarkable reflection property:
\begin{equation}
C(Q-\alpha_1,\alpha_2,\alpha_3) = C(\alpha_1,\alpha_2,\alpha_3) S(i\alpha_1-iQ/2), \label{eq:bref}
\end{equation}
where $S(P)$ is what is called the ``reflection amplitude"
\begin{equation}
S(P) = - (\pi\mu\gamma(b^2))^{-2iP/b} \frac{\Gamma(1+2iP/b)\Gamma(1+2iPb)}{\Gamma(1-2iP/b)\Gamma(1-2iPb)}.
\end{equation}
In the $b \to 0$ limit, this reproduces the result from the minisuperspace approximation (\ref{eq:cref}).

Let us discuss some physical implications of this result. The first thing to note is that we can identify $V_{Q-\alpha}$ with $S(i\alpha_1-iQ/2) V_{\alpha}$ even quantum mechanically, which explains the quantum Seiberg bound. The second thing to note is that it is unitary when $P$ is real in the sense that $S^\dagger S = \langle \alpha|\alpha \rangle \langle \alpha-Q|\alpha-Q\rangle =1$. Since $S(P)$ represents the scattering matrix of a particle from the Liouville wall, this should be so. Note the condition that $P$ is real is needed to have a normalizable wave to test the scattering. 

The last thing is the connection between the reflection amplitude and the spectrum (density of states) of the theory. At first sight, there is no connection between them, so let us explain this in the simpler quantum mechanics setup (see e.g. \cite{Maldacena:2000kv}, \cite{Ponsot:2001gt}). Consider the one-dimensional scattering problem from a potential $U(x)$, where $U(x) \to \infty$ as $x\to \infty$ and $U(x) \to 0$  as $x\to -\infty$. The energy eigenstate of this system can be written as follows ($x\to -\infty$)
\begin{equation}
\psi(x) \sim e^{2ipx} + e^{-2ipx + i\delta(p)},
\end{equation}
where $\delta(p)$ is the phase shift of the outgoing wave with respect to the incoming wave. To relate the phase shift to the density of states, we introduce a perfectly reflecting wall at $x=-V$. The Dirichlet boundary condition at $x=-V$ quantizes the spectrum as
\begin{equation}
4pV + \delta(p) = 2\pi\left(n+\frac{1}{2}\right).
\end{equation}
In general situation, $p$ and $n$ are one-to-one, so we define the density of states as $ \rho = \frac{dn}{dp}$. In this setup, it becomes
\begin{equation}
\rho(p) = \frac{1}{2\pi}\left(4V + \frac{d\delta(p)}{dp}\right)
\end{equation}
This formula shows the connection between the phase shift (or the logarithm of the reflection amplitude) and the density of states of the spectrum.

Now let us apply this formula to the Liouville reflection amplitude,
\begin{equation}
\rho(P) = \frac{1}{2\pi} \left(4V -\frac{2}{b}\log(\pi \mu \gamma(b^2)) -i\log\left(\frac{\Gamma(1+2iP/b)\Gamma(1+2iPb)}{\Gamma(1-2iP/b)\Gamma(1-2iPb)}\right)\right).
\end{equation}
At first sight, in the Liouville theory, it may seem that the volume is $V\sim 1/2b \log\mu$ and the first two terms cancel out. However, as we will see, this is not the case. Rather, the volume factor makes $\log\mu \to \log(\mu/\Lambda)$ providing the cut-off.

Note that the density of states possess a nontrivial dependence on the Liouville momentum $P$. How is this fact consistent with the free path integral one-loop calculation done in section \ref{sec:2}, where we have treated the density of states as if it did not depend on $P$? The answer is simple. When we only consider the $\log\mu$ dependent factor in the one-loop partition function, this does not make any difference. The integration over the $P$ dependent density of states simply yields a finite number which can be absorbed into the cut-off $\log\Lambda$.\footnote{At the same time, recall we have neglected the term $\log(\int d^2z e^{2b\hat{\phi}})$ which can be obtained after the integration over the zero-mode of $\phi$. This may provide a certain number whose direct calculation seems impossible, which corresponds to the cut-off $\Lambda$} However, when we treat the tadpole cancellation etc, we must be careful about those finite parts. We will return to this issue later.

\begin{figure}[htbp]
	\begin{center}
	\includegraphics[width=0.6\linewidth,keepaspectratio,clip]{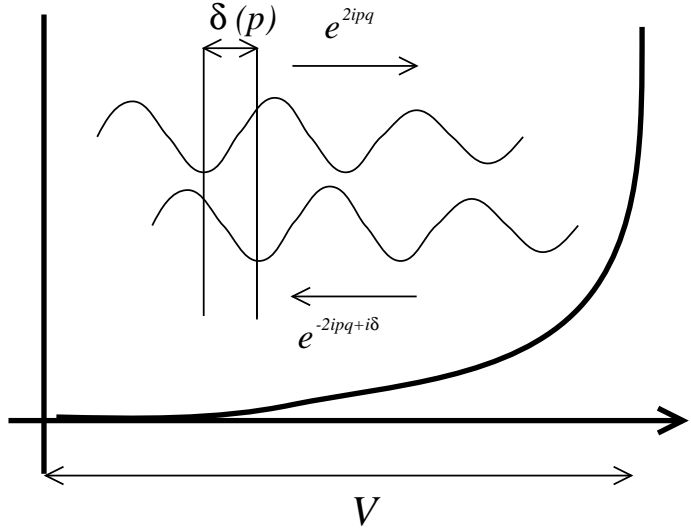}
	\end{center}
	\caption{The reflection amplitude is related to the density of states.}
	\label{reflec}
\end{figure}

\subsection{Teschner's Trick}\label{4.3}
In the following, we would like to rederive the DOZZ formula by what is called Teschner's trick \cite{Teschner:1995yf} which makes use of the degenerate representation of the Virasoro algebra in the Liouville theory. For this purpose, we can study the bulk degenerate states by the Kac formula \cite{Kac:1978ge}, \cite{Feigin:1988se}.\footnote{Since the Liouville theory is not compact nor the degenerate operators are normalizable, some of the assumptions needed to derive the Kac formula are not satisfied. However the equation obtained in this way is consistent and correct a posteriori. This is close to the spirit of the Coulomb gas representation.} The Kac formula states that the degenerate representation exists if and only if the weight of the highest operator satisfies
\begin{equation}
\Delta_{rs} = \frac{c-1}{24} +\frac{1}{4}(r\alpha_++s\alpha_-)^2,
\end{equation}
where $\alpha_{\pm} = \frac{1}{\sqrt{24}}(\sqrt{1-c}\pm\sqrt{25-c})$. For the Liouville theory, we have $c= 1 +6Q^2$ and $Q= b+b^{-1}$. For instance, setting $(r,s) =(1,2)$, we can immediately see $V_{-b/2}$ is the degenerate representation. Similarly setting $(r,s) = (1,3)$, we obtain $V_{-b}$. In general, we have a degenerate representation if and only if $\alpha = -\frac{nb}{2}-\frac{m}{2b}$, where $(n,m)$ are positive integers. We can also derive the differential equation which the degenerate representation satisfies. For $(r,s) =(1,2)$, this is written as
\begin{equation}
0 = \left(L_{-2} - \frac{3}{2(2\Delta+1)}L_{-1}^2\right) \phi_{21}.
\end{equation}
In the Liouville theory, it becomes
\begin{equation}
\left(\frac{1}{b^2} \partial^2 + T(z)\right) V_{-b/2} = 0. \label{eq:a1}
\end{equation}
For $(r,s) =(1,3)$, the differential equation is
\begin{equation}
0 = \left[ L_{-3}-\frac{2}{\Delta+2}L_{-1}L_{-2}+\frac{1}{(\Delta+1)(\Delta+2)}L^3_{-1}\right] \phi_{31},
\end{equation}
and applying this to the Liouville theory we obtain
\begin{equation}
\left(\frac{1}{2b^2} + 2T(z)\partial +(1+2b^2)\partial T(z) \right)V_{-b} = 0. \label{eq:a2}
\end{equation}
Note that although the equation for $(1,2)$ state might be obtained by using the semiclassical equation of motion, the equation for $(1,3)$ state has an order $b^2$ ambiguity of the operator ordering if we attempt to derive it from the equation of motion. This method which uses the degenerate conformal states removes this kind of ambiguity.

In order to derive the general three-point functions, we consider an auxiliary four-point function first.  Because of the conformal invariance, the four-point function essentially depends only on the cross-ratio:
\begin{equation}
z = \frac{z_{21}z_{43}}{z_{31}z_{42}}.
\end{equation}
Then the four-point function becomes
\begin{align}
& \langle V_{\alpha_4}(z_4) V_{\alpha_3}(z_3) V_{\alpha_2}(z_2) V_{\alpha_1}(z_1)\rangle \cr
&= |z_{42}|^{-4\Delta_2}|z_{41}|^{2(\Delta_3+\Delta_2-\Delta_1-\Delta_4)}|z_{43}|^{2\Delta_1+\Delta_2-\Delta_3-\Delta_4)}|z_{31}|^{2(\Delta_4-\Delta_1-\Delta_2-\Delta_3)} G_{\alpha_4\alpha_3\alpha_2\alpha_1}(z,\bar{z})
\end{align}
This four-point function can be calculated either by first taking the OPE of $(4,3)$ and $(1,2)$ or by taking the OPE of $(1,3)$ and $(2,4)$, which is the crossing symmetry. In terms of the three-point function, this states
\begin{eqnarray}
G_{\alpha_4\alpha_3\alpha_2\alpha_1}(z,\bar{z}) &=& \sum_{\alpha} C(\alpha_4,\alpha_3,\alpha) C(Q-\alpha,\alpha_2,\alpha_1)\left|\Fus{\alpha_1}{\alpha_2}{\alpha_3}{\alpha_4}{s}{}\right|^2 \cr
&=& |z|^{-4\Delta_2}G_{\alpha_1\alpha_3\alpha_2\alpha_4}(1/z,1/\bar{z}),\label{eq:boot}
\end{eqnarray}
where we implicitly set the Zamolodchikov metric of the vertex operator to be $\langle Q-\alpha|\alpha\rangle =1$.

In general we cannot solve this bootstrap equation because the number of the intermediate states is infinite. The trick here is to set $\alpha_2 = -\frac{b}{2}$ so that it becomes a degenerate operator. Thanks to the differential equation derived above, the four-point function satisfies 
\begin{equation}
\left(-\frac{1}{b^2} \frac{d^2}{dz^2} +\left(\frac{1}{z-1}+\frac{1}{z}\right)\frac{d}{dz} - \frac{\Delta_3}{(z-1)^2}- \frac{\Delta_1}{z^2} + \frac{\Delta_3+\Delta_2+\Delta_1-\Delta_4}{z(z-1)}\right)G(z,\bar{z}) = 0, \label{eq:4hg}
\end{equation}
which means there are only two intermediate states in (\ref{eq:boot}), $\alpha = \alpha_1 + sb/2, s= \pm1$. Furthermore the differential equation determines $F_s = F_{\alpha_1+sb/2}$ in terms of the hypergeometric functions as
\begin{equation}
F_s(z) = z^{a_s} (1-z)^\beta F(\alpha_s,\beta_s,\gamma_s,z), 
\end{equation}
where $a_s = \Delta_{\alpha_1+sb/2}-\Delta_2 -\Delta_1$, $\beta = \Delta_{\alpha_3-b/2}-\Delta_3-\Delta_2$ and 
\begin{eqnarray}
\alpha_s &=& - sb(\alpha_1-Q/2) +b(\alpha_3+\alpha_4-b)-1/2 \cr
\beta_s &=& - sb(\alpha_1-Q/2) +b(\alpha_3-\alpha_4)+1/2 \cr
\gamma_s &=& 1- sb(2\alpha_1-Q).
\end{eqnarray}

\begin{figure}[htbp]
	\begin{center}
	\includegraphics[width=0.8\linewidth,keepaspectratio,clip]{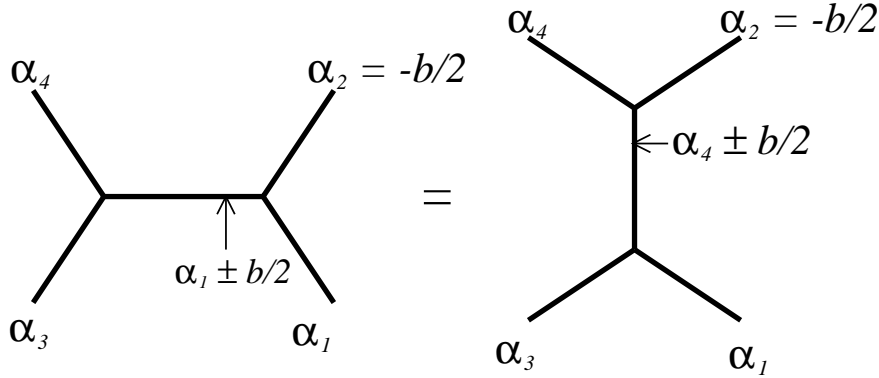}
	\end{center}
	\caption{The conformal bootstrap enables us to obtain the functional relation for the three-point functions.}
	\label{teschner}
\end{figure}

To obtain the functional relation from the conformal bootstrap equation, we utilize the inversion formula of the hypergeometric function (\ref{eq:inv}), which yields the following relation
\begin{equation}
\Fus{\alpha_1}{\alpha_2}{\alpha_3}{\alpha_4}{s}{}(z)= z^{-2\Delta_2} \sum_{t=\pm} B_{st}\Fus{\alpha_4}{\alpha_2}{\alpha_3}{\alpha_1}{t}{}(1/z).
\end{equation}
Substituting this relation into the crossing symmetry equation (\ref{eq:boot}) and comparing the cross terms of the $F_+$ and $F_-^*$, which are absent in the right-hand side, we find  the structure constant must satisfy
\begin{equation}
\frac{C(\alpha_4,\alpha_3,\alpha_1+b/2)}{C(\alpha_4,\alpha_3,\alpha_1-b/2)} = -\frac{C(\alpha_1,-b/2,Q-(\alpha_1-b/2))}{C(\alpha_1,-b/2,Q-(\alpha_1+b/2))}\frac{B_{-+}\bar{B}_{--}}{B_{++}\bar{B}_{+-}}.\label{eq:Doz}
\end{equation}

The right-hand side of this expression is actually calculable, as $B$'s are known from the inversion formula
\begin{eqnarray}
B_{+-}= \frac{\Gamma(\gamma_+)\Gamma(\beta_+-\alpha_+)}{\Gamma(\beta_+)\Gamma(\gamma_+-\alpha_+)} \cr
B_{--}= \frac{\Gamma(\gamma_-)\Gamma(\beta_--\alpha_-)}{\Gamma(\beta_-)\Gamma(\gamma_--\alpha_-)} \cr
B_{++}= \frac{\Gamma(\gamma_+)\Gamma(\alpha_+-\beta_+)}{\Gamma(\alpha_+)\Gamma(\gamma_+-\beta_+)} \cr
B_{-+}= \frac{\Gamma(\gamma_-)\Gamma(\alpha_--\beta_-)}{\Gamma(\alpha_-)\Gamma(\gamma_--\beta_-)} 
\end{eqnarray}
 and $C$ in the right-hand side can be calculated by the perturbative Coulomb gas integral under the perturbative saturation assumption. First, $C(\alpha_1,-b/2,Q-(\alpha_1-b/2)) =1$ because this satisfies the conservation of the Liouville charge and we do not need any insertion of the Liouville potential. To calculate $C(\alpha_1,-b/2,Q-(\alpha_1+b/2))$, we just need one insertion of the Liouville potential:
\begin{eqnarray}
C(\alpha_1,-b/2,Q-(\alpha_1+b/2)) &=& -\mu\left\langle e^{2\alpha_1\phi}(1) e^{-b\phi}(0) e^{2(Q-\alpha_1-b/2)\phi}(\infty) \int d^2z e^{2b\phi}(z)\right\rangle_{Q,free} \cr
&=& -\mu \int d^2z |1-z|^{-4\alpha_1 b}|z|^{2b^2} \cr
&=& -\pi\mu \frac{1}{\gamma(-b^2)\gamma(2\alpha_1b)\gamma(2+b^2-2b\alpha_1).}
\end{eqnarray}

Substituting this and the actual form of $B$'s into the equation (\ref{eq:Doz}), we obtain the desired functional relation for the structure constant after a little algebra,
\begin{equation}
\frac{C(\alpha_1+b,\alpha_2,\alpha_3)}{C(\alpha_1,\alpha_2,\alpha_3)} = -\frac{\gamma(-b^2)}{\pi\mu}\frac{\gamma(b(2\alpha_1+b))\gamma(2b\alpha_1)\gamma(b(\alpha_2+\alpha_3-\alpha_1-b))}{\gamma(b(\alpha_1+\alpha_2+\alpha_3-Q))\gamma(b(\alpha_1+\alpha_2-\alpha_3))\gamma(b(\alpha_1+\alpha_3-\alpha_2))},
\end{equation}
where we have shifted $\alpha_1$. This is the same functional relation we have already derived in (\ref{eq:funrel}). The rest of the calculations to derive the DOZZ formula are the same and we will not repeat them here. 

By the way, note that we can redo the same reasoning above by replacing $\alpha_2 = -b/2$ with $\alpha_2 = -1/(2b)$. In this way, we seem to obtain the second functional relation which determines the structure constant uniquely. However this is not so automatic as it may seem. The point is if we replace $b \to b^{-1}$, we cannot calculate $C(\alpha_1,-1/(2b),Q-(\alpha_1+1/(2b)))$ without a further assumption. Of course, the natural guess is that the quantum correction provides the dual cosmological constant term $\Delta S = \int d^2 z \tilde{\mu} e^{2\tilde{b}\phi(z)}$ in the functional integration as we have discussed above\footnote{It is interesting to note that $\mu e^{2b\phi} = \mu S(b)e^{2b^{-1}\phi} = \frac{\tilde{\mu}}{b^2} e^{2b^{-1}\phi}$.}, but this is rather an assumption or a conjecture. Therefore Teschner's trick suggests the $b \to b^{-1}$ duality of the Liouville theory, but does not prove it. Alternatively saying, to fix the ambiguity of the quantization, we should demand this duality to define the quantum Liouville theory. However, the general Liouville bootstrap suggests that the DOZZ formula is compatible with the CFT requirement, so perhaps the duality symmetry is the fundamental requirement for the Liouville theory to be consistent as a CFT. The nature of the duality in the Liouville theory has been further studied in \cite{O'Raifeartaigh:1998da}, \cite{O'Raifeartaigh:1999zf} and \cite{O'Raifeartaigh:2000nd}.

\subsection{Higher Equations of Motion}\label{4.4}
As an application of the DOZZ formula, we review the higher equations of motion in the Liouville theory, which is proposed in \cite{Zamolodchikov:2003yb}. The rough idea can be seen from the simplest Liouville equation of motion. The Liouville equation of motion is given by
\begin{equation}
\partial \bar{\partial}\phi = \pi b \mu e^{2b\phi}.
\end{equation}
As a result, the derivative of any correlation function with respect to the cosmological constant becomes
\begin{equation}
\frac{\partial}{\partial\mu} \langle U_1 \cdots U_n\rangle = \frac{-1}{\pi \mu b} \int d^2z \langle U_1 \cdots U_n \partial \bar{\partial}\phi(z) \rangle.
\end{equation}
The integration over $d^2z$ turns to the boundary integration which results in the KPZ scaling relation:
\begin{equation}
\frac{\partial}{\partial\mu} \langle U_1 \cdots U_n\rangle = -\frac{\sum_i{\alpha_i} - (1-g)Q}{\mu b}\langle U_1 \cdots U_n \rangle.
\end{equation}

The purpose of this section is to obtain the higher derivative counterparts of the above argument. To begin with, let us introduce the logarithmic degenerate field as 
\begin{equation}
V'_{\alpha} = \frac{1}{2}\frac{\partial}{\partial \alpha} V_{\alpha} = \phi e^{2\alpha \phi},
\end{equation}
and we take $\alpha$ as the degenerate representations: $V'_{mn} = V'_{\alpha} |_{\alpha = \alpha_{mn}}$. The first proposition \cite{Zamolodchikov:2003yb} is that $D_{m,n}\bar{D}_{m,n} V'_{m,n}$ is a primary field. Here $D_{m,n}$ denotes the differential operator which annihilates the degenerate primary operator $V_{m,n}$ (like in \eqref{eq:a1} and \eqref{eq:a2}). 

The proof of this proposition is as follows. Consider $\bar{D}_{m,n} V_{\alpha}$ in the vicinity of $\alpha = \alpha_{m,n}$. From the analyticity in $\alpha$, we have $\bar{D}_{m,n} V_{\alpha} = (\alpha - \alpha_{m,n}) A_{m,n} + \mathcal{O}((\alpha-\alpha_{m,n})^2)$, where $A_{m,n}$ is an operator of dimension $(\Delta_{m,n}, \Delta_{m,n} + mn)$ which is no more right primary but still a left primary with the \textit{same} dimension of the degenerate operator.\footnote{We have a different left and right notation from \cite{Zamolodchikov:2003yb}. The left primary which has the same dimension as the degenerate operator may seem to vanish, as in the rational conformal theory, when we act $D_{m,n}$, but this is not the case here. It remains to be a nonvanishing left primary field. The same thing happens in the boundary operator such as $B_{-\frac{b}{2}}$ as we will see in the next section.} Because of the degenerate dimension, $D_{m,n} A_{m,n} = 2 D_{m,n} \bar{D}_{m,n} V'_{m,n}$ is also a left primary. Since we could inverse the roles of $D_{m,n}$ and $\bar{D}_{m,n}$, $D_{m,n} \bar{D}_{m,n} V'_{m,n}$ is a primary field of dimension $(\Delta_{m,n}+mn,\Delta_{m,n}+mn)$. That is to say
\begin{equation}
{D}_{m,n}\bar{D}_{m,n} V'_{m,n} = B_{m,n} V_{\tilde{\alpha}_{m,n}}, \label{eq:heq}
\end{equation}
where $\tilde{\alpha}_{m,n} = -(m-1)b^{-1}/2 + (n+1)b/2$. $B_{m,n}$ can be calculated from the DOZZ formula \eqref{eq:DOZZ} by applying $\eqref{eq:heq}$ to the three-point function. The direct calculation shows that it is given by
\begin{align}
B_{m,n} &= \left(\pi\mu\gamma(b^2)b^{2-2b^2}\right)^n \frac{\Upsilon'(2\alpha_{m,n})}{\Upsilon(2\tilde{\alpha}_{m,n})} \cr
&= \left(\pi\mu\gamma(b^2)\right)^n b^{1+2n-2m} \gamma(m-nb^2) \prod_{k=1-n,l=1-m, (k,l)\neq (0,0)}^{n-1,m-1} (lb^{-1}+kb)
\end{align}

As a possible application of this formula, \cite{Zamolodchikov:2003yb} has considered the two dimensional minimal gravity theory. The $(p,q)$ minimal model has a central charge
\begin{equation}
c_M = 1- 6(b^{-1} - b)^2,
\end{equation}
where $b = \sqrt{p/q}$ and the same $b$ is used for the Liouville part. The matter primary field $\Phi_{m,n}$ has a conformal dimension 
\begin{equation}
\Delta_{m,n}^M =  1 -\Delta_{m,n} - mn.
\end{equation}
Thus the gravitationally dressed $(1,1)$ primary fields are given by 
\begin{equation}
U_{m,n} = \Phi_{m,n}^M \tilde{V}_{m,n},
\end{equation}
where $\tilde{V}_{m,n} = V_{\tilde{\alpha}_{m,n}}$ is the operator which appears in the higher equations of motion \eqref{eq:heq}. Furthermore, since $\Phi_{m,n}^M$ is the degenerate field, $D_{m,n}^M \Phi_{m,n}^M = \bar{D}_{m,n}^M \Phi_{m,n}^M  = 0$. If we define the joint operators $\mathcal{D}_{mn}$ as
\begin{equation}
\mathcal{D}_{m,n} = D_{m,n} - D^{M}_{m,n},
\end{equation}
we finally have the following identity for the correlation function in the minimal gravity,
\begin{equation}
\int d^2 z \langle U_{m,n}(z) \cdots \rangle_{MG} = \frac{1}{B_{m,n}} \int d^2z \langle  \mathcal{D}_{m,n}\bar{\mathcal{D}}_{m,n} \Theta'_{m,n}(z) \cdots \rangle_{MG},
\end{equation}
where $\Theta'_{m,n} = \Phi^M_{m,n} V'_{m,n}$ and the subscript $MG$ stands for the path integration over the minimal model, the Liouville field and the conformal ghost. 

The further conjecture by \cite{Zamolodchikov:2003yb} is that $\mathcal{D}_{m,n}\bar{\mathcal{D}}_{m,n}\Theta'_{m,n}(z)$ is actually BRST exact up to the total derivative:
\begin{equation}
\mathcal{D}_{m,n}\bar{\mathcal{D}}_{m,n} \Theta'_{m,n}(z) =\partial \bar{\partial} (\bar{H}_{m,n} H_{m,n} \Theta'_{m,n}) + \mathrm{BRST} \ \mathrm{exact}, \label{eq:BRSTex}
\end{equation}
where $H_{m,n}$ are operators of level $mn-1$ with ghost number $0$ constructed from $L_n, L_n^{M}$ and ghosts (see e.g. \cite{Bouwknegt:1992yg} for explicit construction).

If this conjecture is true, the evaluation of any BRST closed correlation function and its integration over the moduli space is reduced to the boundary terms. However, we expect that this is the case from the matrix model integrability of the minimal gravity. Indeed, the equivalence of the physical matrix model and the topological gravity states that the integration of BRST closed correlation functions over the moduli space \textit{should} be reduced to the cohomological integration. We expect that this observation and higher equations of motion play significant roles in proving the integrability of the minimal gravity from the continuum Liouville approach. This has not yet been done, but it is an important problem and we hope it will be solved in the near future.

\subsection{Literature Guide for Section 4}\label{4.5}
For a general review of the subjects discussed in this section, we recommend an excellent review \cite{Teschner:2001rv}. In this section and the followings, we assume the basic knowledge about the conformal field theory. For the standard references of the CFT, we refer to \cite{Belavin:1984vu}, \cite{Ginsparg:1988ui} and \cite{DiFrancesco:1997nk}.

The original proposal of the DOZZ formula has been given independently in \cite{Dorn:1994xn}, \cite{Zamolodchikov:1996aa}. In the main text, we have not done consistency checks, but of course, we have many supporting checks of the proposal in the literature. Besides the original checks done in \cite{Zamolodchikov:1996aa}, where the semiclassical limit (see also \cite{Hadasz:2003he}) and the numerical conformal bootstrap are discussed, consistency checks with the general fusion and braiding property of the chiral part of the Liouville operators proposed by Gervais and his collaborators \cite{Gervais:1996rv,Cremmer:1994kc,Cremmer:1994qf,Gervais:1993fh,Gervais:1994is,Gervais:1994ec} have been done (those authors have used the similar technique to Teschner's trick). The perturbative check of the DOZZ formula can be found in \cite{Thorn:2002am} by applying the method in \cite{Braaten:1984np}. For the canonical quantization approach, see \cite{Jorjadze:2002nj,Jorjadze:2003cg}.

With these developments, the consistency of the DOZZ formula as a CFT is finally given (with a mild assumption) by proving the general conformal bootstrap in \cite{Ponsot:1999uf}, which is reviewed in \cite{Teschner:2001rv}, \cite{Teschner:2003en} and \cite{Ponsot:2003ju}. This is important because at least we have a nontrivial irrational CFT defined by the DOZZ formula irrespective of whether or not the CFT is actually describing the quantization of the Liouville action. Unfortunately, we have not been able to discuss this issue in the main text, so an interested reader should consult these papers.

 The higher equations of motion and possible application to the Liouville theory coupled to the minimal model is proposed in \cite{Zamolodchikov:2003yb}. The logarithmic behavior is also observed in \cite{Yamaguchi:2002rt}. We expect that this observation is the first important step to prove the integrability of the minimal gravity from the continuum Liouville perspective. It may be possible in the near future that the integration over the higher genus Liouville correlator dressing the minimal matter is explicitly reduced to the topological gravity without a help of the matrix model.

\sectiono{Boundary Liouville Theory}\label{sec:5}
In this section, we discuss the boundary Liouville theory. The Liouville theory admits two kinds of boundary states, the FZZT (Fateev-Zamolodchikov-Zamolodchikov \cite{Fateev:2000ik}, Teschner \cite{Teschner:2000md}) brane and the ZZ (Zamolodchikov-Zamolodchikov \cite{Zamolodchikov:2001ah}) brane. They correspond to the brane extending in the Liouville direction and the one localized in the Liouville direction respectively. We review their basic properties here. The organization of this section is as follows.

In section \ref{5.1}, we discuss the FZZT brane. In subsection \ref{5.1.1}, we derive the one-point function by the boundary bootstrap method. In subsection \ref{5.1.2}, we derive the boundary two-point function. In subsection \ref{5.1.3}, and subsection \ref{5.1.4} we show the bulk-boundary structure constant and the boundary three-point function.

In section \ref{5.2}, we discuss the ZZ brane. In subsection \ref{5.2.1}, we derive the bulk one-point function by the boundary bootstrap method as has been done in the FZZT brane. In subsection \ref{5.2.2}, we present a unified way to view various boundary states from the modular bootstrap method, which also reveals the nature of the open strings propagating on these branes. In subsection \ref{5.2.3} we show the bulk-boundary structure constant.

\subsection{FZZT Brane: D1-Brane}\label{5.1}

The FZZT brane is the brane which exists in the open sector of the Liouville theory \cite{Fateev:2000ik}, \cite{Teschner:2000md}. This brane is extending in the Liouville direction, so we can think of it as a D1-brane on which the open strings have the Neumann boundary condition (if we take $c=1$ and regard $X$ as time).

The starting point is the Liouville action for the open sector, namely on the world sheet with boundaries:
\begin{equation}
S = \int_{\Gamma} d^2z \sqrt{g} \left(\frac{1}{4\pi}g^{ab}\partial_a \phi\partial_b\phi+ \frac{1}{4\pi}QR\phi+\mu e^{2b\phi}\right) + \int_{\partial \Gamma} dx g^{1/4}\left(\frac{QK\phi}{2\pi} +\mu_Be^{b\phi}\right),
\end{equation}
where $K$ is the extrinsic curvature on the boundary. Here, we discuss only the case of the disk topology. The boundary interaction $e^{b\phi}$ is determined by the geometrical requirement from the power of $g^{1/4}$, and the coupling to $K$ is uniquely fixed by the relationship between the perturbative expansion of the string interaction and the Euler number of the world sheet which can be counted by the Gauss-Bonnet theorem with boundaries:
\begin{equation}
\frac{1}{4\pi} \int_\Sigma \sqrt{g} R + \frac{1}{2\pi}\int_{\partial \Sigma} g^{1/4} K = \chi = 2-2g-h.
\end{equation}
In the actual calculation, it is convenient to take the upper half side of the complex plane as the world sheet coordinate. 

The primary fields which we would like to consider now are the bulk operator $V_\alpha = e^{2\alpha \phi} $ and the boundary operator $B_{\beta} = e^{\beta\phi}$. The central discussion below is to obtain the correlation functions involving these. The conformal weight of the boundary primary operator $B_\beta$ is given by 
\begin{equation}
\Delta_\beta = \beta(Q-\beta).
\end{equation}
We can see this from the fact that the ``charge conjugated operator" (which means it has a unit Zamolodchikov metric with respect to the original one) $B_{Q-\beta}$ should have the same dimension as $B_\beta$ and that its dimension should become $\Delta_\beta =\beta^2$ when $Q=0$.

Moreover, since the path integral measure is invariant under the change of variable: $\phi(z) \to \phi(z) -\frac{1}{2} \log\mu$, the exact WT identity for the correlation functions holds,
\begin{equation}
\langle V_{\alpha_1}\cdots V_{\alpha_n} B_{\beta_1}\cdots B_{\beta_m} \rangle = \mu^{(Q-2\sum_i \alpha_i - \sum_j \beta_j)/2b}F\left(\frac{\mu_B^2}{\mu}\right).
\end{equation}
Note the power of $\mu$, which was $Q-\sum_i\alpha_i$ for the sphere, but now becomes $Q-2\sum_i\alpha_i-\sum_j\beta_j$ for the disk. This is because the Gauss-Bonnet theorem requires the ``charge" of $Q$ to be proportional to the world sheet Euler number.

Solving this boundary CFT amounts to determining the following correlation functions in principle.

1. Bulk one-point function (FZZ) \cite{Fateev:2000ik}.
\begin{equation}
 \langle V_\alpha(z) \rangle = \frac{U(\alpha|\mu_B)}{|z-\bar{z}|^{2\Delta_\alpha}}
\end{equation}

2. Boundary two-point functions (FZZ) \cite{Fateev:2000ik}.
\begin{equation}
\langle B_\beta^{\mu_1\mu_2}(x)B_\beta^{\mu_2,\mu_1}(0)\rangle = \frac{d(\beta|\mu_1,\mu_2)}{|x|^{2\Delta_\beta}},
\end{equation}
where we have introduced the Chan-Paton Hilbert space so that we can think of the possibility that the vertex insertion changes the boundary condition to that of attaching to the other D-branes. $\mu_1$ and $\mu_2$ here denote the different cosmological constants (the expectation value of the tachyon field) on each brane.

3. Bulk-boundary structure constant \cite{Hosomichi:2001xc}.
\begin{equation}
\langle V_\alpha(z) B_\beta(x)\rangle = \frac{R(\alpha,\beta|\mu)}{|z-\bar{z}|^{2\Delta_\alpha - \Delta_\beta}|z-x|^{2\Delta_\beta}}
\end{equation}

4. Boundary three-point function \cite{Ponsot:2001ng}.
\begin{equation}
\langle B_{\beta_1}^{\mu_2\mu_3}(x_1)B_{\beta_2}^{\mu_3\mu_1}(x_2)B_{\beta_3}^{\mu_1\mu_2}(x_3)\rangle = \frac{c(\beta_1,\beta_2,\beta_3|\mu_1,\mu_2,\mu_3)}{|x_{12}|^{\Delta_1+\Delta_2-\Delta_3}|x_{23}|^{\Delta_2+\Delta_3-\Delta_1}|x_{31}|^{\Delta_3+\Delta_1-\Delta_2}}
\end{equation}

We will derive these constants in the following. We can obtain them in principle by the analytic continuation of the perturbative treatment which is valid when the Liouville charge is ``conserved" up to the perturbation of the Liouville potential. However this is cumbersome in practice. Instead, we use Teschner's trick for this purpose. The trick has been already used in the last section to derive the DOZZ formula. The crucial point of Teschner's trick is to insert the degenerate operator into the correlator we would like to calculate, which yields the functional relations constraining  the structure constants. With the dual relation which can be obtained from $b\to b^{-1}$, we find the solution is unique.

\begin{figure}[htbp]
	\begin{center}
	\includegraphics[width=0.8\linewidth,keepaspectratio,clip]{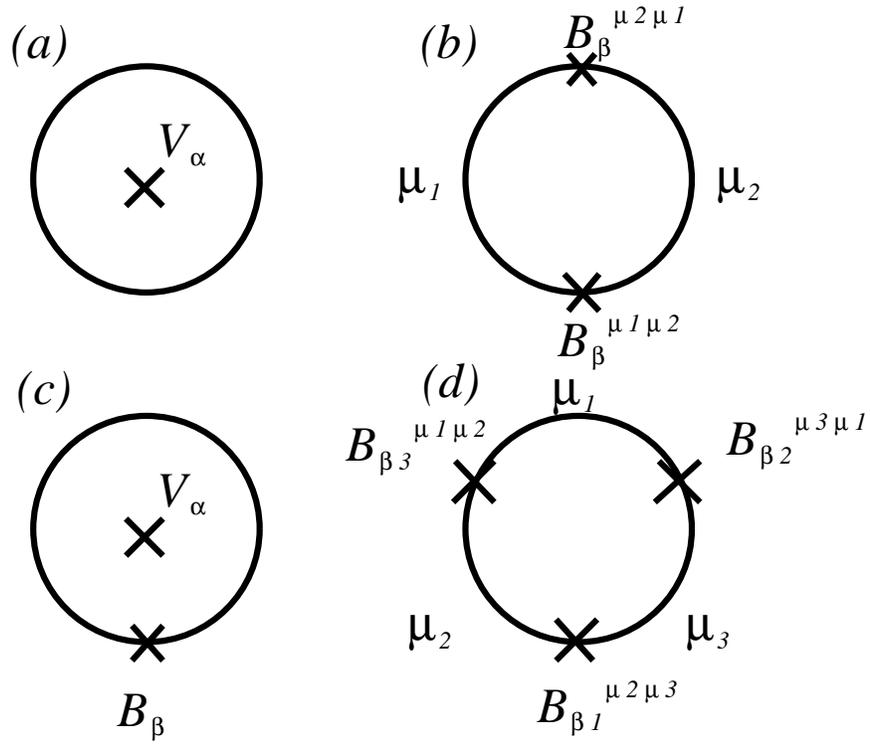}
	\end{center}
	\caption{The basic structure constants for the boundary Liouville theory. (a) the bulk one-point function. (b) the boundary two-point function. (c) the bulk-boundary two-point function. (d) the boundary three-point function.}
	\label{boundary}
\end{figure}

\subsubsection{Bulk one-point function}\label{5.1.1}
In order to obtain the bulk one-point function, we consider the following auxiliary two-point function,
\begin{equation}
G_{-b/2,\alpha} (z,x) = \langle V_{-b/2}(z)V_{\alpha}(x) \rangle .
\end{equation}
To calculate this two-point function, we use the $z \to x $ OPE. Since $V_{-b/2}$ belongs to the degenerate representation, its OPE with any operator begins only with the two primary fields such as
\begin{equation}
V_{-b/2} V_{\alpha} = C_+[V_{\alpha-b/2}] + C_-[V_{\alpha + b/2}].
\end{equation}
To prove this, note the more general correlator satisfies
\begin{equation}
0 = \langle :\left(\frac{1}{b^2} \partial^2 + T(z)\right) V_{-b/2}(z):V_{\alpha}(z_1) \cdots \rangle. 
\end{equation}
Taking the $z\to z_1$ limit and substituting the general OPE relation
\begin{equation}
V_{-b/2}(z)V_{\alpha}(z_1) \sim (z-z_1)^{\Delta_{\alpha'}-\Delta_{\alpha}-\Delta_{-b/2}} V_{\alpha'} + \cdots,
\end{equation}
we obtain the second order equation for $\Delta_{\alpha'}$. The solution of the equation yields $\alpha' = \alpha \pm b/2 $.

Next, we would like to know the fusion coefficient $C_{\pm}$. Fortunately (or deliberately), since the mismatch of the Liouville charge conservation is just the integral multiple of what can be provided by the perturbative treatment of the Liouville potential, we can calculate this by assuming the perturbative calculation saturates the amplitude. Considering the Zamolodchikov innerproduct between the OPE and the ``charge conjugated" out state,
\begin{eqnarray}
C_+ &=& \langle V_{Q-\alpha+b/2}(\infty) V_{-b/2}(1) V_{\alpha}(0) \rangle \cr
&=& \langle V_{Q-\alpha+b/2}(\infty) V_{-b/2}(1) V_{\alpha}(0) \rangle_{Q,free} \cr
&=& \delta(0) \infty^{2\Delta_{\alpha-b/2}} \equiv 1, \label{eq:OPE1}
\end{eqnarray}
where in the last line, we have normalized our result properly. Under the same normalization, we obtain
\begin{eqnarray}
C_{-} &=& \langle  V_{Q-\alpha-b/2}(\infty) V_{-b/2}(1) V_{\alpha}(0) \rangle \cr 
&=&  -\mu\int d^2 z \langle V_{Q-\alpha-b/2}(\infty)e^{2b\phi(z)} V_{-b/2}(1) V_{\alpha}(0) \rangle \cr
&=& -\mu \frac{\pi\gamma(2b\alpha-1-b^2)}{\gamma(-b^2)\gamma(2b\alpha)}.\label{eq:OPE2}
\end{eqnarray}

Collecting these pieces of information, we can determine $G_{-b/2,\alpha}$ as follows
\begin{equation}
G_{-b/2,\alpha} = \frac{|x-\bar{x}|^{2\Delta_\alpha-2\Delta_{-b/2}}}{|z-\bar{x}|^{4\Delta_\alpha}} [C_+ U(\alpha-b/2) \mathcal{F}_+(\eta)+C_-U(\alpha+b/2)\mathcal{F}_-(\eta)]\label{eq:G1},
\end{equation}
where we have introduced the $SL(2,\mathbf{R})$ invariant cross-ratio:
\begin{equation}
\eta = \frac{(z-x)(\bar{z}-\bar{x})}{(z-\bar{x})(\bar{z}-x)}.
\end{equation}
Since the whole two-point function obeys the second order differential equation, we can calculate $\mathcal{F}_\pm(\eta)$. 
 This is essentially the hypergeometric function whose precise form was calculated by BPZ \cite{Belavin:1984vu}, and it is written as
\begin{eqnarray}
\mathcal{F}_+(\eta) &=& \eta^{\alpha b}(1-\eta)^{-b^2/2} F(2\alpha b-2b^2-1,-b^2,2\alpha b-b^2,\eta) \cr
\mathcal{F}_-(\eta) &=& \eta^{1+b^2-\alpha b}(1-\eta)^{-b^2/2} F(-b^2,1-2\alpha b,2+b^2-2\alpha b,\eta) \label{eq:bhg}
\end{eqnarray}

On the other hand, to calculate the same correlation function, we can also do it by expanding the bulk operator by the boundary OPE. Since the intermediate states have only two primary fields, the degenerate operator $V_{-b/2}$ has only two OPE expansion with respect to the boundary operator, namely, $B_0$ and $B_{-b}$. Particularly, if we concentrate on the coupling to $B_0$, the fusion coefficient with respect to $V_{\alpha}$ is of course the one-point function $U(\alpha)$ and the fusion coefficient with respect to $V_{-b/2}$ is given by $R(-b/2,Q)$ which can be interpreted as the innerproduct between $V_{-b/2}$ and the ``charge conjugated" boundary operator $B_{Q}$. This bulk-boundary structure constant $R(-b/2,Q)$ can be calculated perturbatively because the charge conservation is violated by just the integral multiple of $b$. Assuming the perturbative saturation, we obtain 
\begin{eqnarray}
R(-b/2,Q) &=& -2^{2\Delta_{12}}\mu_B \int dx \langle V_{-b/2}(i) B_{b}(x) B_{Q} (\infty) \rangle_{Q,free}  \cr 
&=& -\frac{2\pi\mu_B \Gamma(-1-2b^2)}{\Gamma(-b^2)^2}.
\end{eqnarray}
The coupling to the descendant is also determined as above. Solving the differential equation, we obtain
\begin{equation}
G_{-b/2,\alpha} = \frac{|x-\bar{x}|^{2\Delta_\alpha - 2\Delta_{-b/2}}}{|z-\bar{x}|^{4\Delta_\alpha}} [B^{+} \mathcal{G}_{+}(\eta) +B^{-} \mathcal{G}_{-}(\eta) ]\label{eq:G2},
\end{equation}
where
\begin{eqnarray}
\mathcal{G}_+(\eta) &=& \eta^{\alpha b}(1-\eta)^{-b^2/2} F(-b^2,2\alpha b -2b^2-1,-2b^2,1-\eta) \cr
\mathcal{G}_-(\eta) &=& \eta^{\alpha b}(1-\eta)^{1+3b^2/2} F(1+b^2,2\alpha b,2+2b^2,1-\eta),
\end{eqnarray}
and the information available so far states $B^{-} = U(\alpha)R(-b/2,Q)$.

\begin{figure}[htbp]
	\begin{center}
	\includegraphics[width=0.8\linewidth,keepaspectratio,clip]{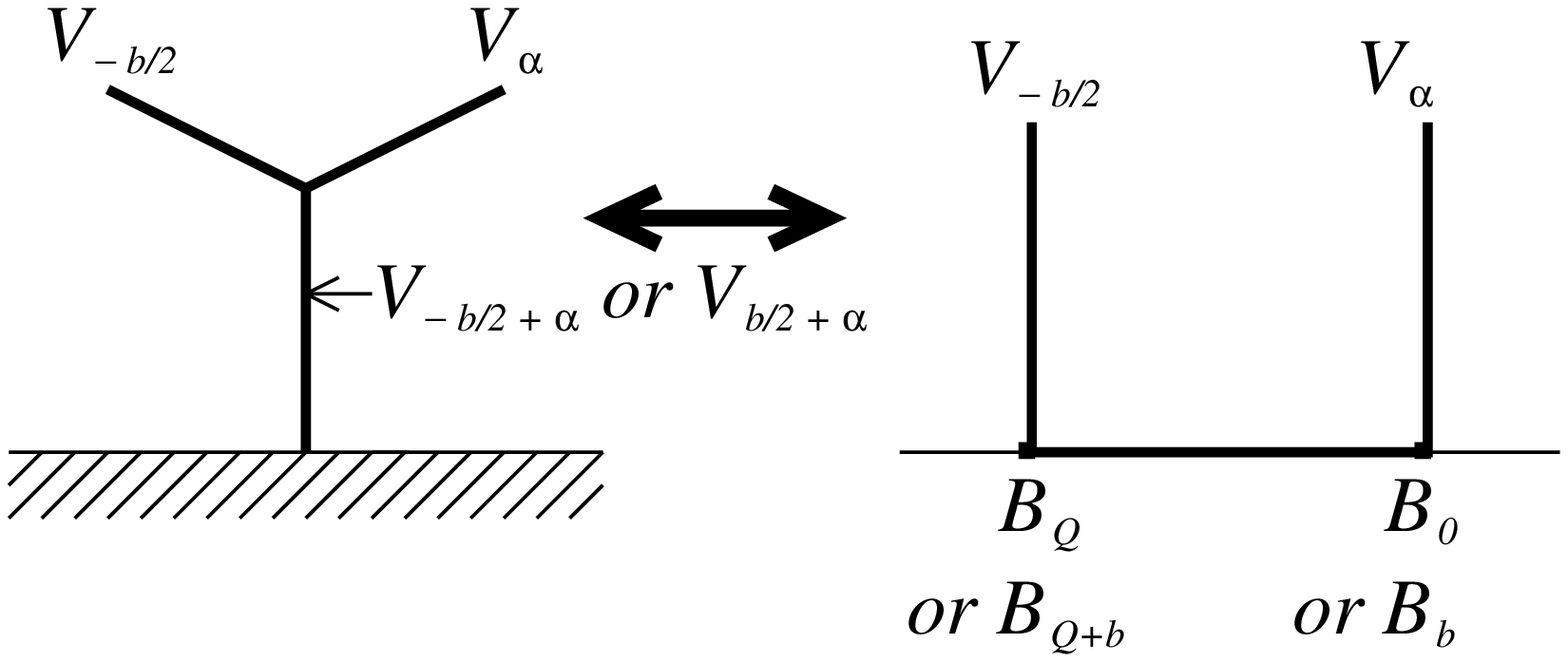}
	\end{center}
	\caption{The auxiliary bulk two-point function can be calculated in two different ways.}
	\label{bulk2p}
\end{figure}

Comparing the two different calculation for the same two-point function: (\ref{eq:G1}) and (\ref{eq:G2}), the functional relation for $U(\alpha)$ can be derived. In order to do that, we rewrite $\mathcal{F}_\pm$ in terms of $\mathcal{G}_\pm$. Using the inversion formula for the hypergeometric function (\ref{eq:hyperg}), we can rewrite $\mathcal{F}_\pm$ as
\begin{eqnarray}
\mathcal{F}_+(\eta) &=& \frac{\Gamma(2\alpha b-b^2) \Gamma(1+2b^2)}{\Gamma(1+b^2)\Gamma(2\alpha b)} \mathcal{G}_+(\eta) + \frac{\Gamma(2\alpha b-b^2)\Gamma(-1-2b^2)}{\Gamma(2\alpha b - 2b^2 -1)\Gamma(-b^2)}\mathcal{G}_{-}(\eta) \cr
\mathcal{F}_-(\eta) &=& \frac{\Gamma(2+b^2-2\alpha b) \Gamma(1+2b^2)}{\Gamma(1+b^2)\Gamma(2+2b^2-2\alpha b)} \mathcal{G}_+(\eta) + \frac{\Gamma(2+b^2-2\alpha b)\Gamma(-1-2b^2)}{\Gamma(1-2\alpha b)\Gamma(-b^2)}\mathcal{G}_{-}(\eta).
\end{eqnarray}
Substituting this and comparing the term proportional to $\mathcal{G}_-(\eta)$,
we obtain the functional relation for $U(\alpha)$:
\begin{equation}
-\frac{2\pi\mu_B}{\Gamma(-b^2)} U(\alpha) = \frac{\Gamma(-b^2+2b\alpha)}{\Gamma(-1-2b^2+2b\alpha)} U(\alpha -b/2) - \frac{\pi\mu\Gamma(-1-b^2+2b\alpha)}{\gamma(-b^2)\Gamma(2b\alpha)}U(\alpha +b/2), \label{eq:frelf}
\end{equation}
after substituting $C_-$ and working with some algebra.

The solution of this functional relation is given by
\begin{equation}
U(\alpha) = \frac{2}{b}(\pi\mu\gamma(b^2))^{(Q-2\alpha)/2b}\Gamma(2b\alpha - b^2)\Gamma\left(\frac{2\alpha}{b}-\frac{1}{b^2}-1\right)\cosh(2\alpha-Q)\pi s,\label{FZZT1p}
\end{equation}
where, instead of $\mu_B$, we have introduced a new parameter $s$:
\begin{equation}
\cosh ^2 \pi b s = \frac{\mu^2_B}{\mu}\sin \pi b^2.
\end{equation}
We sometimes use the related quantity $\sigma$:
\begin{equation}
\cos^2\left(2\pi b(\sigma-Q/2)\right) =  \frac{\mu^2_B}{\mu}\sin \pi b^2 \label{eq:ssig}
\end{equation}
The calculation for showing that this quantity satisfies the functional relation \eqref{eq:frelf} is straightforward. For instance, after dividing the both sides by $U(\alpha)$, use the formula $\cosh(a-b)+\cosh(a+b) = 2\cosh a \cosh b$, and we have $\cosh\pi b s$. Then, writing this as $\sqrt{\sin\pi b^2}$, we cancel this term with the remaining factor $\Gamma(1-b^2)\Gamma(b^2) = \sin(\pi b^2) /\pi $. The rest of the calculation is simply the application of the $\Gamma$ function formula.

Let us comment on the properties of this solution. The solution is invariant under the dual transformation $b\to b^{-1}$, $\mu \to \tilde{\mu}$, $s \to s$, $\mu_B \to \tilde{\mu}_B$ (the last transformation for $\mu_B$ is defined such that $s$ is invariant under this transformation). With the same reasoning as in the DOZZ formula, this solution is unique up to an overall normalization factor. The overall normalization is determined by the requirement that the residue at $2\alpha = Q- nb$ reproduces the perturbative calculation.

\subsubsection{Boundary two-point function}\label{5.1.2}
Next, we proceed to the determination of the boundary two-point function by Teschner's trick. Although the basic strategy is the same --- we derive the functional equation from the OPE of the boundary degenerate operator, there is a difference concerning the boundary degenerate operator.

For example, it may seem that we have a null vector if we take the boundary operator $B_{-nb/2}$ for a positive integer $n$ because the central charge and the conformal dimension suggest so from the Kac formula. However, when we take $n=1$, we find from the semiclassical ($b\to 0$) analysis, 
\begin{equation}
T(x) = -\frac{1}{4}\phi_x^2 +\frac{1}{2b}\phi_{xx} + \pi (\pi \mu_B^2 b^2-\mu)e^{2b\phi},
\end{equation}
and consequently, the left-hand side does not vanish as follows,
\begin{equation}
\left(\frac{d^2}{dx^2}+ b^2T \right) e^{-b\phi/2} = \pi b^2 (\pi \mu_B^2b^2-\mu)e^{3b\phi/2}.
\end{equation}
Since we are dealing with nonnormalizable states, this kind of peculiarity could happen. Therefore, we cannot use the $n=1$ degenerate state. However, the $n=2$ degenerate state $B_{-b}^{ss}$ can be used because this satisfies the following third order homogeneous differential equation (we have labeled its cosmological constant by $s$ instead of $\mu_B$), 
\begin{equation}
\left(\frac{1}{2b^2}\partial^3 + 2T(z)\partial +(1+2b^2)\partial T(z) \right)B_{-b}^{ss} = 0.
\end{equation}
The intuitive reason why the third order differential equation is valid is that $B_{-b}^{ss}$ can be regarded as the limit of the bulk degenerate operator $V_{-b/2}$ to the boundary. Since the bulk degenerate operator satisfies the corresponding differential equation we expect the boundary limit obeys the similar equation unlike the $n=1$ case.\footnote{This can be directly seen semiclassically by  using the equation of motion.} Using this degenerate operator, we can derive the fusion formula:
\begin{equation}
B_{-b}^{ss} B_{\beta}^{ss'} = c_+[B_{\beta-b}] + c_0[B_\beta] + c_-[B_{\beta+b}],
\end{equation}
which is based on the same reasoning as in the one-point function.

Let us consider the auxiliary three-point function $\langle B_{-b}^{ss}(x_1) B_{\beta}^{ss'}(x_2)B_{\beta+b}^{s's}(x_3)\rangle $. First, comparing the result from the $x_1\to x_2$ OPE with that from the $x_1 \to x_3$ OPE, we obtain
\begin{equation}
d(\beta+b|s,s')c_{-}(\beta) = c_{+}(\beta+b)d(\beta|s,s').
\end{equation}
We set $c_+$ =1 because this fusion coefficient can be calculated from the zeroth order perturbation theory. For $c_-$, since the charge non-conservation is $2$, we can calculate this, under the perturbative saturation assumption, as the sum of the the bulk first order perturbation contribution and the boundary second order perturbation contribution. The former contribution is given by
\begin{eqnarray}
c_-^{v} &=& -\mu \int_{\mathrm{Im}z>0} d^2z \langle e^{2b\phi(z)} B_\beta^{ss'}(0)B^{s's'}_{-b}(1) B^{s's}_{Q-\beta-b} (\infty)\rangle_{Q,free} \cr
&=& \mu\int_{\mathrm{Im}z>0} d^2z \frac{|1-z|^{4b^2}}{|z|^{4b\beta}|z-\bar{z}|^{2b^2}},
\end{eqnarray}
and the latter contribution is given by
\begin{eqnarray}
c_{-}^b &=& \sum_{ij}\frac{\mu_i\mu_j}{2}\int_{C_i}\int_{C_j} dx_1dx_2 \langle e^{b\phi(x_1)}e^{b\phi(x_2)} B_\beta^{ss'}(0)B^{s's'}_{-b}(1) B^{s's}_{Q-\beta-b} (\infty)\rangle_{Q,free} \cr
&=& \sum_{ij}\frac{\mu_i\mu_j}{2}\int_{C_i}\int_{C_j} dx_1dx_2\frac{|(1-x_1)(1-x_2)|^{2b^2}}{|x_1-x_2|^{2b^2}|x_1x_2|^{2b\beta}}.
\end{eqnarray}
Performing the integration, we obtain

\begin{eqnarray}
c_-(\beta) &=& 4\mu \sin\left(\pi b \frac{2\beta+i(s+s')}{2}\right)\sin\left(\pi b \frac{2\beta-i(s+s')}{2}\right) \cr
&\times& \sin\left(\pi b \frac{2\beta+i(s-s')}{2}\right) \sin\left(\pi b \frac{2\beta-i(s-s')}{2}\right) I_0(\beta) 
\end{eqnarray}
where 
\begin{equation}
I_0(\beta) = -\frac{\gamma(1+b^2)}{\pi}\Gamma(1-2b\beta)\Gamma(2b\beta-1)\Gamma(1-b^2-2b\beta)\Gamma(2b\beta-b^2-1).
\end{equation}
Substituting this into the functional relation, we have to solve
\begin{equation}
\frac{d(\beta+b|s,s')}{d(\beta|s,s')} = c_-^{-1}(\beta).
\end{equation}
As usual, enforcing the dual relation $b\to b^{-1}$ makes the solution unique up to an overall normalization. Using the special function $\Gamma_b(x)$ and $S_b(x)$ which have been introduced in appendix \ref{a-3}, we can write the solution as 
\begin{equation}
d(\beta|s,s') = \frac{(\pi\mu\gamma(b^2)b^{2-2b^2})^{(Q-2\beta)/2b}\Gamma_b(2\beta-Q)\Gamma_b^{-1}(Q-2\beta)}{S_b(\beta+i(s+s')/2)S_b(\beta-i(s+s')/2)S_b(\beta+i(s-s')/2)S_b(\beta-i(s-s')/2)} \label{eq:blk2p}
\end{equation}
To check this, we use the difference relation of $\Gamma_b(x)$ and $S_b(x)$ (four $\sin$s are from the shift of $S_b$. Since each $\Gamma_b(x)$ is shifted twice, we obtain four $\Gamma$ functions, which reproduces $I_0(\beta)$). The normalization of the two-point function is determined by the ``unitarity" \footnote{The meaning of the unitarity is as follows. We regard this relation as $\langle \beta| \beta \rangle\langle Q-\beta|Q-\beta\rangle $. If we realize $|Q-\beta\rangle = S|\beta \rangle$, the above relation means $SS^\dagger = 1$. $S$ can be interpreted as the reflection amplitude of the open string state. Note that the $S$ matrix is diagonal in this representation.}:
\begin{equation}
d(\beta|s,s')d(Q-\beta|s,s') = 1.
\end{equation}

\subsubsection{Bulk-boundary correlator}\label{5.1.3}
We consider the bulk boundary correlation function. We present the final result \cite{Hosomichi:2001xc} first. We introduce the Fourier transformation of $R(\alpha,\beta,s)$, calling it $\tilde{R}(\alpha,\beta,p)$, as 
\begin{equation}
\tilde{R}(\alpha,\beta,p) = \frac{1}{2}\int_{-\infty}^{\infty} dse^{4\pi sp} R(\alpha,\beta,s),
\end{equation}
then the result is given by
\begin{eqnarray}
\tilde{R}(\alpha,\beta,p) = 2\pi(\mu\pi\gamma(b^2)b^{2-2b^2})^{(Q-2\alpha-\beta)/2b}\frac{\Gamma_b(Q-\beta)^3 \Gamma_b(2Q-2\alpha-\beta)\Gamma_b(2\alpha-\beta)\Upsilon(2\alpha)}{\Gamma_b(Q)\Gamma_b(\beta)\Gamma_b(Q-2\beta)} \cr
\times S_b(p+\frac{\beta+2\alpha-Q}{2})S_b(p+\frac{\beta-2\alpha+Q}{2})S_b(-p+\frac{\beta+2\alpha-Q}{2})S_b(-p+\frac{\beta-2\alpha+Q}{2}).\label{eq:bb2}
\end{eqnarray}

Let us sketch the proof. As we discussed before, $B_{-b/2}$ does not satisfy the fusion rule of the degenerate operator. However FZZ \cite{Fateev:2000ik} conjectured that $B_{-b/2}^{s,s\pm ib}$ does satisfy the second order differential equation. Actually the two-point functions which can be obtained independently from this conjecture are
consistent with the result of the last subsection. Here, we just assume this conjecture and set
\begin{equation}
B_{-b/2}^{s,s\pm ib} B_{\beta}^{s\pm ib,s'} = [B_{\beta-b/2}^{s,s'}] + c_{\pm}[B_{\beta+b/2}^{s,s'}],
\end{equation}
where we have set the fusion coefficient to be $1$ as usual in the case where the Liouville charge is conserved. $c_{\pm}$ is calculated perturbatively and given by
\begin{eqnarray}
c_{\pm} &=& 2\left(-\frac{\mu}{\pi\gamma(-b^2)}\right)^{1/2} \Gamma(1-2b\beta)\Gamma(2b\beta-b^2-1) \cr
&\times& \sin\pi b(\beta \mp ib(s_1+s_2)/2)\sin\pi b^2(\beta \mp ib(s_1-s_2)/2).
\end{eqnarray}

Now, consider the auxiliary three-point function $\langle V_\alpha(z) B_\beta(x)^{ss'} B_{-b/2}(y)^{s's} \rangle$, where $s\pm s' = \pm ib$. Using the conformal invariance, we can map the upper half plane to the disk of a unit radius. Then the three-point function can be written as
\begin{eqnarray}
 & &\langle V_\alpha(z) B_\beta(x)^{ss'} B_{-b/2}(y)^{s's} \rangle^{u.h.p}  \cr
 &=& (x-z)^{-\Delta_\beta-\Delta_{-b/2}}(x-\bar{z})^{-\Delta_\beta + \Delta_{-b/2}}(z-\bar{z})^{\Delta_\beta-2\Delta_\alpha+\Delta_{-b/2}}(y-\bar{z})^{-2\Delta_{-b/2}} \cr
 &\times& \langle V_\alpha(0)B_\beta(1)^{ss'}B_{-b/2}^{s's}(\eta)\rangle^{disk},
\end{eqnarray}
where $\eta = \frac{(y-z)(x-\bar{z})}{(y-\bar{z})(x-z)}$. We would like to derive the functional relation taking the $\eta \to 1$ OPE. At first sight, we cannot solve $R$ from this equation because we cannot evaluate the left-hand side directly. The trick is that when $\eta$ approaches from the upper and lower, the answer is different so that we can obtain the closed equations (in other words, there is a monodromy and $B$ is not local with each other). 

The equations are
\begin{eqnarray}
& &\langle V_\alpha(0)B_\beta(1)^{ss'}B_{-b/2}^{s's}(\eta)\rangle^{disk} \cr
&=& R(\alpha,\beta-b/2;s')G_{\alpha,\beta}(\eta)e^{i\pi b\beta/2} + c_-(\beta,s,s') R(\alpha,\beta+b/2;s')G_{\alpha,Q-\beta}(\eta) e^{i\pi b(Q-\beta)/2} \cr
&=& R(\alpha,\beta-b/2;s)G_{\alpha,\beta}(e^{-2\pi i}\eta)e^{-i\pi b\beta/2} + c_-(\beta,s',s)R(\alpha,\beta+b/2;s) G_{\alpha,Q-\beta}(e^{-2\pi i}\eta) e^{-i\pi b(Q-\beta)/2}, \cr
\ & &
\end{eqnarray}
where $G$ is given by the hypergeometric function as
\begin{equation}
G_{\alpha,\beta}(\eta) = \eta^{b\alpha}(1-\eta)^{b\beta} F(b(2\alpha+\beta-Q-b/2),b(\beta-b/2),b(2\beta-b),1-\eta).
\end{equation}
Solving this equation, we finally obtain $R$. The answer has been given at the beginning of this subsection.

\subsubsection{Boundary three-point function}\label{5.1.4}
We present the boundary three-point function here. The result is \cite{Ponsot:2001ng}
\begin{eqnarray}
& &C^{\sigma_3\sigma_2\sigma_1}_{(Q-\beta_3)\beta_2\beta_1} \cr
&=&\frac{\Gamma_b(2Q-\beta_1-\beta_2-\beta_3)\Gamma_{b}(\beta_2+\beta_3-\beta_1)\Gamma_b(Q+\beta_2-\beta_1-\beta_3)\Gamma_b(Q+\beta_3-\beta_1-\beta_2)}{\Gamma_b(2\beta_3-Q)\Gamma_b(Q-2\beta_2)\Gamma_b(Q-2\beta_1)\Gamma_b(Q)} \cr
&\times& (\pi\mu\gamma(b^2)b^{2-2b^2})^{(\beta_3-\beta_2-\beta_1)/2b}\frac{S_b(\beta_3+\sigma_1-\sigma_3)S_b(Q+\beta_3-\sigma_3-\sigma_1)}{S_b(\beta_2+\sigma_2-\sigma_3)S_b(Q+\beta_2-\sigma_3-\sigma_2)} \cr
&\times& \frac{1}{i}\int_{-i\infty}^{i\infty} dt \frac{S_b(U_1+t)S_b(U_2+t)S_b(U_3+t)S_b(U_4+t)}{S_b(V_1+t)S_b(V_2+t)S_b(V_3+t)S_b(Q+t)},
\end{eqnarray}
where we have used the notation:
\begin{eqnarray}
U_1 &=& \sigma_1+\sigma_2-\beta_1, \ \ \ \ \ \ \ \ V_1 \ = \ Q+\sigma_2-\sigma_3-\beta_1+\beta_3 \cr 
U_2 &=& Q-\sigma_1+\sigma_2-\beta_1, \ \ V_2 \ = \ 2Q+\sigma_2-\sigma_3-\beta_1-\beta_3 \cr 
U_3 &=& \beta_2+\sigma_2-\sigma_3, \ \ \ \ \ \ \ \ V_3 \ = \ 2\sigma_2 \cr 
U_4 &=& Q-\beta_2+\sigma_2-\sigma_3. 
\end{eqnarray}
See (\ref{eq:ssig}) for the relation among $\sigma$, $s$ and $\mu_B$.

We will not delve into the detailed calculation here, but the basic strategy is the same. We use the conformal bootstrap method to find the functional relation which can be derived by inserting the degenerate operator $B_{-b}$.

\subsection{ZZ Brane: D0-Brane}\label{5.2}
Zamolodchikov brothers showed \cite{Zamolodchikov:2001ah} that in the Liouville theory there exists a D0-brane which is localized in the Liouville direction. We call it the ZZ brane. Semiclassically, this boundary condition corresponds to the solution of the Liouville equation on the disk which describes the Poincar\'e disk metric (or the Euclidean AdS$_2$ metric on the upper half plane).
The solution of the classical Liouville equation: $\Delta \varphi = 2R^{-2} e^{\varphi}$ is given by
\begin{equation}
e^{\varphi(z)} = \frac{R^2}{(\mathrm{Im} z)^2},
\end{equation}
which means $\varphi \to \infty$ at the boundary.\footnote{Note that the Liouville equation originally describes the constant negative curvature metric. In the semiclassical limit $b \to 0$, we identify $\varphi = 2b\phi$ and $R^{-2} = 4\pi\mu b^2$.} Another explanation why the D0-brane is localized in $\phi \to \infty$ is that there is a linear dilaton background so that the brane tension which is inversely proportional to the string coupling constant becomes smaller as the brane goes into the deeper $\phi$ region. The purpose of this section is to understand the quantum description of this boundary condition.

\subsubsection{Bulk one-point function}\label{5.2.1}
As in the FZZT brane case discussed in the last section, we consider the one-point function which is given by
\begin{equation}
 \langle V_\alpha(z) \rangle = \frac{U(\alpha|\mu_B)}{|z-\bar{z}|^{2\Delta_\alpha}}.
\end{equation}
The basic strategy is the same as before, so we consider the auxiliary two-point function $\langle V_{-b/2}(z)V_\alpha(x)\rangle$. We expand this in two different ways and compare them to obtain the functional relations which constrain $U(\alpha)$.
First, using the bulk OPE, we obtain
\begin{equation}
G_{-b/2,\alpha} = \frac{|x-\bar{x}|^{2\Delta_\alpha-2\Delta_{-b/2}}}{|z-\bar{x}|^{4\Delta_\alpha}} [C_+ U(\alpha-b/2) \mathcal{F}_+(\eta)+C_-U(\alpha+b/2)\mathcal{F}_-(\eta)]
\end{equation}
Second, using the boundary OPE, we obtain
\begin{equation}
G_{-b/2,\alpha} = \frac{|x-\bar{x}|^{2\Delta_\alpha - 2\Delta_{-b/2}}}{|z-\bar{x}|^{4\Delta_\alpha}} [B^{+} \mathcal{G}_{+}(\eta) +B^{-} \mathcal{G}_{-}(\eta) ]\end{equation}
These equations are almost the same as those obtained in the last section. The only difference is $B^{-}(\alpha)$. If we assume that our results reproduce the Poincar\'e disk metric semiclassically, the physical distance between the two point will diverge in the limit $\mathrm{Im}z\to 0$.  Then we expect, from the cluster decomposition theorem, that the two-point function factorizes to the product of the one-point functions of each operator. Since the surviving intermediate state on the boundary here is the identity only, we expect 
\begin{equation}
B^{-} (\alpha) = U(\alpha) U(-b/2)
\end{equation}
to hold.\footnote{More precisely, it is $U(\alpha) U(-b/2) D(0)$ from the correspondence with the FZZT brane discussed above. While we take the innerproduct between the boundary operators to be $\langle Q| 0\rangle = 1$ in the last section, in this case we have set $D(0) = \langle 0|0\rangle =1$. This is because we cannot define the natural charge conjugation of the boundary operator in this boundary condition unlike the FZZT boundary condition.} Substituting the inversion formula for the hypergeometric function, we obtain the functional relation for $U(\alpha)$ which states
\begin{equation}
\frac{\Gamma(-b^2)U(\alpha)U(-b/2)}{\Gamma(-1-2b^2)\Gamma(2\alpha b -b^2)} = \frac{U(\alpha -b/2)}{\Gamma(2\alpha - 2b^2 -1)} - \frac{\pi\mu\Gamma(1+b^2)U(\alpha+b/2)}{(2\alpha b -b^2 -1)\Gamma(-b^2) \Gamma(2\alpha b)}.
\end{equation}

The solutions (after demanding the $b\to1/b$ duality as usual) of this equation  are labeled by two positive integers $m$ and $n$ and can be written as
\begin{equation}
U_{m,n}(\alpha) = \frac{\sin(\pi b^{-1}Q)\sin(\pi mb^{-1}(2\alpha -Q))\sin(\pi b Q)\sin(\pi n b (2\alpha -Q))}{\sin(\pi mb^{-1}Q)\sin(\pi b^{-1}(2\alpha -Q))\sin(\pi nb Q)\sin(\pi b (2\alpha -Q))} U_{1,1}(\alpha), \label{eq:umn}
\end{equation}
where the fundamental solution $U_{1,1}(\alpha)$ is given by
\begin{equation}
U_{1,1}(\alpha) = \frac{[\pi\mu\gamma(b^2)]^{-\alpha/b}\Gamma(2+b^2)\Gamma(1+1/b^2)}{\Gamma(2+b^2-2b\alpha)\Gamma(1+b^{-2}-2\alpha/b)}. \label{eq:zz11}
\end{equation}
Since the functional relation here is \textit{non}linear unlike the ones treated so far, it is not obvious (at least to the author) that there are no other solutions. In any case, we can check these solutions satisfy the functional relation. For example, we transform all the gamma functions into the trigonometric functions by the formula $\Gamma(1-x)\Gamma(x) = \frac{\pi}{\sin \pi x}$; then we can see the both sides of the equations are equivalent.

Although many solutions labeled by $(m,n)$ exist, only $m=1$ survives in the $b\to 0$ semiclassical limit. This fact suggests that the $m \neq 1$ solutions are difficult to interpret as the quantization of the Poincar\'e disk. Among other solutions, it is believed  that the $(1,1)$ solution has a special meaning. Its remarkable property is vanishing of $B^{+}_{1,1}(\alpha) $, which means that this boundary couples to the bulk only through the identity operator (and its descendants). This is very interesting. If we regard these boundary conditions as the D0-brane localized in $\phi\to\infty$, it is natural to consider the $(1,1)$ state as the ground state of the D0-brane and other states as the discrete excited states.

\subsubsection{Boundary states and modular bootstrap}\label{5.2.2}
The one-point function obtained above is normalized as $U_{m,n}(0) = 1$. Though the superficial reason is that the functional relation is nonlinear and determines the normalization uniquely, the more fundamental reason is that we have used the cluster decomposition theorem to derive the functional relation. Clearly, for the relation $\langle 1 \rangle \langle 1 \rangle = \langle 1 \rangle $ to hold, we should have 
$\langle 1 \rangle =1$. In some cases, however, we would like to know the unnormalized one-point function which is not divided by $\langle 1\rangle_{unnormalized} $. In the following, we consider this case.

For this purpose, we use the open/closed correspondence of the cylinder diagram. First, we define the open Virasoro character as
\begin{equation}
\chi_P(\tau) = \frac{q^{P^2}}{\eta(\tau)}.
\end{equation}
On the other hand the character for the $(m,n)$ degenerate representation is given by
\begin{equation}
\chi_{m,n}(\tau) = \frac{q^{-(m/b+nb)^2/4}- q^{-(m/b-nb)^2/4}}{\eta(q)}.
\end{equation}
The modular transformation of this character to the exchange channel is given by
\begin{equation}
\chi_{m,n}(\tau') = 2\sqrt{2} \int_{-\infty}^\infty \chi_P(\tau) \sinh(2\pi m P/b) \sinh(2\pi nbP) dP,
\end{equation}
where $\tau'= -1/\tau$. Especially the modular transformation of $\chi_{1,1}$ which corresponds to $U_{1,1}$ can be written as
\begin{equation}
\chi_{1,1}(\tau') = 2\sqrt{2} \int_{-\infty}^\infty \chi_P(q) \sinh(2\pi b P) \sinh(2\pi P/b) dP.
\end{equation}
Then we can see this cylinder amplitude from the closed channel exchange point of view as
\begin{equation}
\chi_{1,1}(\tau') = \int_{-\infty}^\infty \Psi_{1,1}(P) \Psi_{1,1}(-P) \chi_P(q) dP = \langle 1,1|q^\frac{H}{2}| 1,1\rangle, 
\end{equation}
where the ``wavefunction" for the $(1,1)$ state is given by
\begin{equation}
\Psi_{1,1}(P) = \frac{2^{3/4}2i\pi P}{\Gamma(1-2iPb)\Gamma(1-2iP/b)} (\pi\mu\gamma(b^2))^{-iP/b}. \label{eq:11wf} 
\end{equation}
Comparing this with the normalized one-point function $U_{1,1}(\alpha)$ \eqref{eq:zz11}, we find that the normalization is different for the $\alpha$ independent constant. From the derivation above, we should regard \eqref{eq:11wf} as the unnormalized disk one-point function which is not normalized by $\langle 1\rangle_{unnormalized}$. To fix the ($P$ dependent) phase of the wavefunction, we have required the wavefunction to satisfy the bulk reflection property \eqref{eq:bref}, which completely determines the unnormalized disk one-point function up to truly meaningless $P$ independent total phase. This wavefunction can also be seen as the expansion coefficient of the boundary state by the Ishibashi states
\begin{equation}
\langle 1,1 | = \int_{-\infty}^\infty \Psi_{1,1}(P) \langle P| dP,
\end{equation}
where the Ishibashi states \cite{Ishibashi:1989kg} are defined as
\begin{equation}
\langle P|q^\frac{H}{2} | P'\rangle = \delta(P-P') \chi_{P}. 
\end{equation}

From this $(1,1)$ wavefunction, we can obtain the $(m,n)$ wavefunction. The $P$ dependence of the one-point function derived above suggests
\begin{equation}
\Psi_{m,n} = \Psi_{1,1}(P) \frac{\sinh(2\pi mP/b)\sinh(2\pi n bP)}{\sinh(2\pi P/b)\sinh(2\pi bP)} \label{eq:zzmn}
\end{equation}
up to an over all normalization factor. To confirm the normalization to be correct, we compare \eqref{eq:zzmn} with the open loop diagram which can be obtained by the modular transformation
\begin{equation}
\chi_{m,n}(\tau') = \int_{-\infty}^\infty \Psi_{m,n}(P) \Psi_{1,1}(-P) \chi_P(q) dP = \langle m,n|q^\frac{H}{2}|1,1\rangle,
\end{equation}
which shows that the open/closed duality works with this normalization. Note, as a bonus, we have learned that the open string spectrum which stretches between $(1,1)$ brane and $(m,n)$ brane contains only the $(m,n)$ degenerate state. This is the interpretation of the $(m,n)$ brane from the CFT point of view.

Similarly we can obtain the wavefunction for the FZZT brane, which is given by
\begin{equation}
\Psi_s(P) = \frac{2^{-1/4} \Gamma(1+2ibP) \Gamma(1+2iP/b)\cos(2\pi sP)}{-2i\pi P} (\pi\mu\gamma(b^2))^{-iP/b}.
\end{equation}
This expression is of course proportional to the previous FZZT one-point function (\ref{FZZT1p}). The exact normalization can be determined as follows, consider the general non-degenerate character with $P =s/2$,
\begin{equation}
\chi_{s/2}(q') = \sqrt{2}\int_{-\infty}^\infty \chi_{P}(q) \cos(2\pi sP) dP.
\end{equation}
This can be interpreted as follows
\begin{equation}
\chi_{s/2}(q') = \int_{-\infty}^\infty \Psi_{1,1}(P) \Psi_s(-P)\chi_{P}(q) dP,
\end{equation}
which means that the open string stretching from the FZZT brane with the boundary parameter $s$ to the $(1,1)$ ZZ brane is the nondegenerate state with the momentum $P=s/2$ whose weight is then $\Delta = \frac{Q^2}{4} + \frac{s^2}{4}$.

Now, using the above boundary states, we can obtain the spectrum stretching between various branes considered so far. For example, the partition function between $(m,n)$ and $(m',n')$ brane becomes
\begin{eqnarray}
Z_{(m,n)(m',n')} &=& \int_{-\infty}^\infty \Psi_{m,n}(P) \Psi_{m',n'}(-P) \chi_P(q) dP \cr
                &=& \sum_{k=0}^{\mathrm{min}(m,m')-1}\sum_{l=0}^{\mathrm{min}(n,n')-1} \chi_{m+m'-2k-1,n+n'-2l-1}(q'), \label{eq:mnmn}
\end{eqnarray}
which is just the fusion algebra of the degenerate representations. In this way, we can obtain the operator contents of the $(m,n)$ ZZ brane. They consist of the finite degenerate operators. Furthermore we can obtain the density of states of the FZZT brane by calculating the cylinder diagram 
\begin{eqnarray}
Z_{s,s'} &=& \int_{-\infty}^\infty \Psi_s(P) \Psi_{s'}(-P) \chi_P (q) dP \cr
 & =& \int_{-\infty}^{\infty} \rho(P') \chi_{P'} (q') dP', \label{eq:densitys}
\end{eqnarray}
where the density of states is
\begin{eqnarray}
\rho(P') &=& 2\sqrt{2} \int_{-\infty}^\infty \Psi_s(P) \Psi_{s'}(-P) e^{-4\pi i PP'} dP \cr
&=& \int_{-\infty}^{\infty} \frac{2 \cos(st)\cos(s' t)}{\sinh(bt)\sinh(t/b)} e^{-2\pi iP' t} \frac{dt}{2\pi}.
\end{eqnarray}
This expression has a divergence as $t \to 0$ which signals that the FZZT brane is extending along the $\phi$ direction (on the other hand we can see that the wavefunction for the ZZ brane vanishes as $P \to 0$ which means that the ZZ brane localizes in the $\phi$ direction). If we properly regularize this expression we should have $ \rho \sim \frac{1}{2b}\log(\mu) + \rho_r(p)$, where $\rho_r(p)$ is finite. Instead of doing that, we consider the relative density of states here which is defined as
\begin{equation}
\rho_{rel}(P,s,s') = \rho(P,s,s') - \rho(P,s_{ref}),
\end{equation}
where the reference boundary parameter $s_{ref}$ has been introduced. From the discussion in section \ref{4.2}, we have a nontrivial consistency check about the relationship between the (relative) density of states and the boundary two-point function:
\begin{equation}
\rho_{rel}(P,s,s') = -\frac{i}{2\pi} \frac{d}{dP} \log \left(\frac{d(P|s,s')}{d(P|s_{ref})}\right).
\end{equation}
Substituting (\ref{eq:blk2p}) into the left-hand side and using the integral formula for the logarithm of the double sine function (\ref{eq:intdsin}), we can see that the relation actually holds.

However, the change of integration order in (\ref{eq:densitys}) is not legitimate if $s$ is pure imaginary \cite{Teschner:2000md}. In this case, we deform the contour of the modular transformation formula as
\begin{equation}
\chi_{\alpha}(q') = \sqrt{2}\int_{r + i \mathbf{R}} d\beta e^{4\pi i(\alpha-\frac{Q}{2})(\beta-\frac{Q}{2})} \chi_\beta(q),
\end{equation}
where $r$ is in the range $r> \mathrm{max}(|\sigma_+|,|\sigma_-|,\frac{Q}{2})$ with $\sigma_\pm = i(s_2\pm s_1) \in \mathbf{R}$. Now we can change the integration order, but if we deform the integration in the density of states to the original range $\frac{Q}{2}+i\mathbf{R}$, we have other contributions from poles in $S_b$, which yield the discrete spectrum. The discrete spectrum is unitary only if $|\sigma_\pm| < Q$, in which case, the discrete spectrum is just the $(1,1)$ degenerate state. On the other hand, if $|\sigma_\pm| > Q$, the nonunitary $(m,n)$ degenerate states appear in the spectrum. There is no discrete spectrum when $|\sigma_\pm| < Q/2$.

To conclude this subsection, let us make some comments on the boundary states and spectrum of the open modes.

First, as has been noticed by Martinec \cite{Martinec:2003ka}, the $(m,n)$ boundary states are formally related to the FZZT boundary states as
\begin{equation}
|m,n\rangle = |\mathrm{FZZT}; s(m,n)\rangle - |\mathrm{FZZT}; s(m,-n)\rangle, \label{eq:marti}
\end{equation}
where the boundary parameter $s(m,n)$ is defined as
\begin{equation}
s(m,n) = i \left(\frac{m}{b} + nb\right).
\end{equation}
It is interesting to note that the boundary cosmological constants for $s(m,\pm n)$ have actually the same value
\begin{equation}
\mu_B(m,n) = (-1)^m \sqrt{\mu} \frac{\cos(\pi n b^2)}{\sqrt{\sin(\pi b^2)}}.
\end{equation}
This is possible when $s$ takes the pure imaginary value. In a sense, we can make different FZZT branes of the same boundary cosmological constant by adding $(m,n)$ ZZ branes, which is one of the fundamental reasons why we have the additional discrete degenerate spectrum in the open strings stretching between the FZZT branes with an imaginary boundary parameter $s$. This viewpoint and its application to the boundary perturbation theory has been further developed in \cite{Teschner:2003qk} (see also \cite{Klebanov:2003wg} and \cite{Seiberg:2003nm} for the discussion on the nature of the monodromy).

Second, for the future application, we consider the spectrum of the open string which stretches between the ZZ branes when $c=1$ with a time-like $X$ boson which has the Neumann boundary condition. The mass formula for the $(m,n)$ degenerate state is given by
\begin{equation}
L_0^{total} = -\frac{(n+m)^2}{4},
\end{equation}
which means that the ground state becomes more tachyonic with increasing $m$ or $n$. While the dynamics of $(1,1)$ tachyon will be further studied in the later section, the physical meaning of the remaining ``heavy" tachyon is rather mysterious.\footnote{However see section \ref{6.6} for the nonperturbative contribution from these states.}

Let us further study the spectrum on the $(1,1)$ brane. In the two dimension, the oscillator partition function becomes
\begin{equation}
Z = q^{H_0}\eta(q)^2 (1-q) \frac{1}{\eta(q)^2} = q^{H_0}(1-q) 
\end{equation}
Because of the nature of the degenerate state, the partition function looks non-unitary. \footnote{At first glance, there is a contradiction with the generalized no ghost theorem \cite{Asano:2000fp,Asano:2003jn}. However, the situation here just corresponds to the exceptional cases considered there.} The low energy on-shell spectrum consists of a tachyon whose mass is $(-1)$, a time-like vector, the ghost and anti-ghost whose energy is zero (recall no ghost decouples). Since the dynamics is one-dimensional, it is not obvious whether this has a deep meaning or not. It is important to note, however, that there is no degree of freedom which moves the ZZ brane in the $\phi$ direction, which guarantees that the ZZ brane is fixed in the $\phi \to \infty$ region. 

Finally let us consider the spectrum of the open string stretching between the FZZT brane with boundary parameter $s$ and the $(m,n)$ ZZ brane. The character is given by
\begin{equation}
\int_{-\infty}^\infty \Psi_{m,n}(P) \Psi_s(-P) \chi_p(q)dP = \sum_{k=1-m,2}^{m-1} \sum_{l=1-n,2}^{n-1} \chi_{(s+i(k/b+lb))/2} (q'),
\end{equation}
where $\sum_{k=1-n,2}^{n-1}$ denotes the summation over the set $k =\{ -n+1,-n+3\cdots,n-1\}$. However, the right-hand side shows that the mass (weight) of the open string becomes imaginary unless $s$ is pure imaginary. Even in that case, the mass is tachyonic. This further shows the difficulty of the physical interpretation of $(m,n) \neq (1,1)$ branes. The only meaningful possibility may be the $(1,1)$ brane. When we take $c=1$, the open string state between the FZZT brane and the $(1,1)$ brane can be either tachyonic or massive, which is given by
$m = \frac{s^2}{4}$. 

\subsubsection{Bulk-boundary structure constant}\label{5.2.3}
The bulk-boundary two-point function on the ZZ brane is derived by Ponsot \cite{Ponsot:2003ss}. We review his derivation and result here. The bulk boundary two-point function on $(m,n)$ brane is given by
\begin{equation}
\langle V_\alpha(z) B_\beta(x)\rangle_{m,n} = \frac{R_{m,n}(\alpha,\beta)}{|z-\bar{z}|^{2\Delta_\alpha - \Delta_\beta}|z-x|^{2\Delta_\beta}},
\end{equation}
where the notation here is $\beta = -ub - vb^{-1}$ with positive integers $(u,v)$ which corresponds to the $(2v+1,2u+1)$ degenerate operator. Note that the notation is just for the convenience and $B_{\beta}$ has nothing to do with any exponential operator of the Liouville field. Also, recall that the $(m,n)$ boundary state has only $u<n$ and $v<m$ operators in the spectrum. For simplicity we concentrate on the $(1,2u+1)$ operator here (for more general cases, see \cite{Ponsot:2003ss}). The direct derivation  of the two-point function is as follows. 

Consider an auxiliary bulk two-point function $\langle V_{-ub/2} V_{\alpha} \rangle$. This auxiliary two-point function can be calculated in two ways: by taking the OPE of the two bulk operators first or by taking the boundary OPE first. Equating the $B_{-ub}$ contribution of the both calculations, we obtain
\begin{align}
\sum_{k=0}^u C(\alpha, -ub/2,Q-\alpha+ub/2-kb) U_{m,n}(\alpha-ub/2 + kb) \Fus{\alpha}{-ub/2}{-ub/2}{\alpha}{\alpha-ub/2+kb}{,-ub/2} \cr
 = R_{m,n} (\alpha,-ub) R_{m,n} (-ub/2,-ub) D_{m,n} (-ub). \label{eq:bbrel}
\end{align}
Almost all the structure constants in \eqref{eq:bbrel} have been calculated. Thus $R_{m,n} (\alpha,-ub)$ is calculable from the expression above. First, the three-point function with $\alpha_1+\alpha_2+\alpha_3 = Q-kb$ is given by 
\begin{equation}
C_k(\alpha_1,\alpha_2,\alpha_3) = \left(\frac{-\pi\mu}{\gamma(-b^2)}\right)^k \frac{\prod_{j=1}^k \gamma(-jb^2)}{\prod_{l=0}^{k-1} \gamma(2\alpha_1b + lb^2)\gamma(2\alpha_2b+lb^2)\gamma(2\alpha_3b + lb^2)},
\end{equation}
from the (pre) DOZZ formula (\ref{eq:preDOZZ2}). Second, the one-point function $U_{m,n}(\alpha)$ is given by (\ref{eq:umn}). Third, $D_{m,n}(-ub)$ and $R_{m,n}(-ub/2,-ub)$ can be set to one as long as the corresponding operator exists in the boundary spectrum (otherwise zero). \footnote{The former is simply the normalization convention of our boundary operators. Note that the nontrivial reflection amplitude does not exist in this case. The latter is considered to be the `no screening charge case' which means the case where the spin just adds up with the coefficient $1$. The author would like to thank B.~Ponsot for clarifying this point to the author.}

Finally, the fusion matrix $\Fus{\alpha}{-ub/2}{-ub/2}{\alpha}{\alpha-ub/2+kb}{,-ub/2}$ is the transformation matrix between $t$ channel conformal block and $s$ channel conformal block. For the simplest case, we have derived it from the inversion formula of the hypergeometric function. In principle, we can redo the same procedure to obtain the degenerate fusion matrix. In practice, it is easier to obtain it from reading off the residue of poles of the known general results. The general fusion matrix has been found in \cite{Ponsot:1999uf}, which we are not going to derive here. It is given by
\begin{align}
\Fus{\alpha}{-ub/2}{-ub/2}{\alpha}{\alpha-ub/2+kb}{,-ub/2} = \prod_{l=1}^u \Gamma(bQ+ub^2+(l-1)b^2) \prod_{l=k}^{u-1} \frac{\Gamma(2b\alpha-ub^2+(l+k)b^2)}{\Gamma(2b\alpha + lb^2)\Gamma(bQ+lb^2)} \cr
\times \prod_{j=1}^k \frac{\Gamma(bQ+(j-1)b^2)\Gamma(2bQ-2b\alpha+ub^2-2kb^2+(j-1)b^2)}{\Gamma(bQ+ub^2-jb^2)^2\Gamma(2bQ-2b\alpha+ub^2-jb^2)} .
\end{align}

Combining these factors, we can calculate $R_{m,n} (\alpha,-ub)$. The result can be also expressed in somewhat different form as (perhaps up to factor $\pi^2$)
\begin{eqnarray}
\lefteqn{R_{m,n} (\alpha,-ub) =} \cr
 &&-\frac{1}{b}(\pi\mu\gamma(b^2))^{\frac{1}{2b}(-2\alpha+ub)}\frac{\sin(\pi b^{-1} Q) \sin(\pi b Q)}{\sin(\pi m b^{-1} Q)\sin(\pi n b Q)} \Gamma(bQ)\Gamma(Q/b) Q b^{-ub(b-b^{-1})} \cr
&& \times \Gamma(2b\alpha-b^2)\Gamma(2b^{-1}\alpha-1-b^{-2}) \prod_{j=0}^{u-1} \frac{\Gamma(2b\alpha-bQ-ub^2+jb^2)\Gamma(bQ+ub^2+jb^2)}{\Gamma(2b\alpha +ub^2 - (j+1)b^2)\Gamma(bQ+jb^2)} b^{1-b(-ub+2j)} \cr
&& \times \sin \pi m b^{-1}(2\alpha - u b - Q) \cr
&& \times \left(\sum_{k=0}^u \sin \pi n b (2\alpha-ub - Q + 2kb) \prod_{j=0}^{k-1} \frac{\sin\pi b(2\alpha-ub-Q+jb)\sin\pi b (-ub+jb)}{\sin\pi b (2\alpha + jb) \sin \pi b(Q+jb)}\right).
\end{eqnarray}
We can check that the expression above becomes zero whenever $u>n$. Since there is no such boundary operator in this case, this is obvious physically, but the actual cancellation occurs after summing over $k$ and it is nontrivial. This yields a further consistency check on the spectrum of the $(m,n)$ boundary operators.

Finally we note that the above expression is shown to satisfy Hosomichi's proposal \cite{Hosomichi:2001xc}. Roughly speaking, if we Fourier transform the two-point function of the FZZT brane with respect to the boundary parameter $s$ as
\begin{equation}
\tilde{R}(\alpha,\beta,p) = \int_{-\infty}^{\infty} e^{4\pi s p} R_s(\alpha,\beta) ds
\end{equation}
and perform the further transformation
\begin{equation}
\int_{-i\infty}^{+i\infty} \sin(2\pi n p b) \sin (2\pi m pb^{-1}) \tilde{R} (\alpha,\beta,p) dp,
\end{equation}
then we can obtain the two-point function on the $(m,n)$ ZZ brane up to the factor
\begin{equation}
-(\pi\mu\gamma(b^2))^{-Q/2b} \frac{\sin(\pi b^{-1} Q)\sin(\pi b Q)}{\sin(\pi m b^{-1} Q)\sin(\pi n b Q)}\Gamma(bQ) \Gamma(Q/b) Q.
\end{equation}
Using the explicit formula of the $R_s(\alpha,\beta)$ (\ref{eq:bb2}), we can reproduce the direct result in this way.

To conclude this section, let us list the remaining structure constants which have not been derived yet.
\begin{itemize}
	\item Boundary two-point functions. However, on the same brane, they are trivially set to one (unless zero because of the weight mismatch). On the different branes, they can be nontrivial (though we think it is natural to have one also in this case).
	\item Boundary three-point functions. Knowing these constants will certainly improve our understanding of the $(m,n) \neq (1,1)$ ZZ branes.

\end{itemize}

\subsection{Literature Guide for Section 5}\label{5.3}
The general boundary conformal theory was pioneered by Cardy \cite{Cardy:1984bb,Cardy:1986gw,Cardy:1989ir}. These ideas were applied to the boundary Liouville theory in \cite{Fateev:2000ik}, \cite{Teschner:2000md}. Besides the various consistency checks done in the original papers, the comparison of those amplitudes reviewed in the main text with the loop gas approach \cite{Kostov:1989eg,Kostov:1991xt,Kostov:1992cg} to the two dimensional gravity has been done in \cite{Kostov:2002uq,Kostov:2003uh}. 

Before the work of ZZ \cite{Zamolodchikov:2001ah}, the pseudosphere geometry was studied in \cite{D'Hoker:1982er,D'Hoker:1983ef}. While the ZZ brane is the most natural brane existing in the irrational CFT with a continuum spectrum from the open string perspective\footnote{The author feels that one cannot imagine any other simple possibility of the open spectrum satisfying the Cardy condition with only the conformal symmetry.}, the geometrical interpretation of the pseudosphere is surprising. Though we have not treated in the main text, the direct way to see its connection is to calculate correlation functions on the pseudosphere by the background field path-integral, which means that we expand the Liouville field around the classical pseudosphere background as a perturbation in $b$. In their original paper \cite{Zamolodchikov:2001ah}, they have verified that the calculation reproduces the $(1,1)$ brane solution up to the two loop order. The three loop calculation has been done in \cite{Menotti:2003km}.

\sectiono{Applications}\label{sec:6}
In this section, we review (only a few) applications of the bosonic Liouville theory. Some of them are taken from the recent developments and some of them are older ones. Since the realm of the subject is so vast, we omit the detailed derivation of the many statements in this section. The reader can find them in the original papers. Even without details, we cannot review all the related works in the main text. We will list other works in the literature guide section. The organization of this section is as follows.

In section \ref{6.1}, we revisit the matrix model as a holographic (gauge/gravity) dual of the open string theory on the unstable ZZ branes, which provides us a new interpretation and insight of the old matrix models. In section \ref{6.2}, we discuss the time-like Liouville theory which is obtained from the analytic continuation of the Liouville theory to $c=25$. In section \ref{6.3}, we discuss a close relation between the topological string theory and the Liouville theory. After reviewing the basic properties of the topological string theory, we will see that the Liouville theory coupled to the $c=1$, $R=1$ compactified boson can be described by the topological string theory via various dualities. 

In section \ref{6.4}, we return to the matrix model description and discuss the decaying D-brane in the two dimensional string theory from the matrix (and Liouville) point of view. In section \ref{6.5}, we discuss the two dimensional black hole which is closely related to the Liouville theory. The dual matrix model and the brane in this model are also discussed there. In section \ref{6.6}, we discuss the nature of the nonperturbative effect encoded in the matrix model partition function from the continuum Liouville perspective, where the ZZ type branes play an important role. In section \ref{6.7}, we discuss the connection between the branes in the minimal model coupled to the two dimensional gravity (Liouville theory) and the degenerate Riemann surface.

\subsection{Matrix Reloaded}\label{6.1}

In section \ref{sec:3}, we have shown that the closed Liouville theory can be described by the double scaling limit of the matrix quantum mechanics. There, we have given the interpretation that the $c=1$ Liouville theory emerges in the continuum limit of the free-boson on the discretized world sheet. However, the new physically deeper interpretation of the matrix quantum mechanics has been proposed recently by McGreevy and Verlinde \cite{McGreevy:2003kb} (see also \cite{Giveon:1999zm}, \cite{Polyakov:2001af} for an earlier argument). We would like to review the idea quickly.

The idea is that we regard the matrix quantum mechanics as the quantum mechanics of the tachyon field which lives on the $N$ unstable  D0-branes (ZZ branes).
As is discussed in section \ref{sec:5}, the ZZ branes have a tachyon which transforms as an adjoint representation of the Chan-Paton gauge group $U(N)$. Its effective action is given by
\begin{equation}
S_{eff} = \int dt \frac{1}{2}\mathrm{Tr}(\mathcal{D}_t T)^2 - \mathrm{Tr}[V(T)] + \cdots,
\end{equation}
where the kinetic term has been gauged since the ZZ branes also have a non-dynamical gauge field. This action yields clearly the (gauged) matrix quantum mechanics! From the open/closed duality, we expect that the open string dynamics of the D0-branes describes the pure closed string physics (here, the Liouville theory) in an appropriate limit. The appropriate limit is nothing but the double scaling limit. Recall that the double scaling limit means the simultaneous limit of $g_s \to 0$ and $N \to \infty$ with the fixed Fermi level $\mu$.\footnote{By the way, as long as we fix $\mu$, the physics does not change if we change $N$ or $g_s$. This is called the ``new duality" in \cite{McGreevy:2003kb}.} In this limit, the physics of D0-branes completely decouples from the closed string, namely ``gravity". Thus, this observation yields the foundation that the effective action on the D0-brane is the decoupled tachyon system above (the Maldacena limit!). In fact, since the Liouville theory is given by the double scaling limit of the tachyon matrix quantum mechanics, we can say that the open/closed correspondence of this system is proved in some sense.

Before considering the physical implication of this new interpretation, we would like to comment on the tachyon potential. While the field theory on the D0-branes is generally unknown, the D0-branes considered here have a free Neumann boson $X^0$ whose spectrum is well-known. The open spectrum is given by the tachyon $1$ and the (non-dynamical) gauge field $\dot{X^0}$. \footnote{The subtlety concerning the ghost has been discussed in the previous section.} Since the gauge field does not have a kinetic term in the space of dimension one, the actual effect is simply truncating the states of the matrix quantum mechanics onto the gauge invariant states. As we will see later, this has an important meaning which gives the strong evidence of the open/closed duality. However, the problem is the tachyon potential. Around $T=0$, it is given by the tachyonic ($-1$ in our unit) mass term from the tree level spectrum, but the higher form of the potential is difficult to obtain. In principle, we can calculate the perturbative S matrix from the boundary correlation functions, but the evaluation of the effective potential is rather difficult. It is natural to assume the boundary CFT type effective potential which has been reviewed in section \ref{sec:2} (Actually, we expect $T=e^{X^0}$ becomes the time-like CFT). In any case, in the double scaling limit, the detailed shape of the potential fortunately does not matter.

Let us list evidence and comments about this duality, some of which we will discuss further later in this section.

\begin{itemize}
	\item The gauging of the tachyon by coupling to the world line gauge field plays a significant role when we consider the partition function of the compactified $X^0$ theory. As we have seen in section \ref{sec:3}, the matrix quantum mechanics does not describe the continuum Liouville theory unless we restrict the spectrum to the gauge invariant sector. Otherwise the T-duality becomes ruined for example. We will discuss the physical interpretation of including the twisted sector later in this section in relation to the two dimensional black hole. As for the Liouville theory, the gauging of the matrix quantum mechanics naturally solves this problem.
	\item From the discretized surface point of view, it is ambiguous whether we should take the parity even potential or the more general potential which does not preserve the parity. Empirically, the matrix-Liouville correspondence chooses the general potential (or we should divide the final answer by two in the parity even case). However, the proposal here naturally yields the left-right asymmetric potential, so this is indeed consistent with the Liouville theory choice.
	\item As we will see in part II of this review, there is stronger evidence of this kind of duality. The same reasoning also applies to the supersymmetric extension of the Liouville theory. Particularly the matrix model dual for the $\mathcal{N}= 1$ super Liouville theory coupled to $\hat{c}=1$ matter has been proposed in Takayanagi-Toumbas \cite{Takayanagi:2003sm} and Douglas et.al. \cite{Douglas:2003up}. The idea is that we regard the matrix quantum mechanics of the tachyon on the unstable D0-branes which exist in the (type-0A or type-0B) super Liouville theory as the dual description of the parent super Liouville theory. In fact, it can be shown that the one-loop super Liouville partition function and the proper scattering amplitudes are exactly reproduced from the matrix quantum mechanics which is different for 0A and 0B.
	\item While closed tachyon fluctuations become the collective excitations of the Fermi surface, the extra addition of an eigenvalue corresponds to the addition of an unstable D0-brane into the theory in this new interpretation. Actually, the matrix model description of the decay and the Liouville theory direct calculation match \cite{Klebanov:2003km}, \cite{McGreevy:2003ep}, \cite{Gutperle:2003ij} as we will see later in this section. In particular, the emission rate of the closed string and the final state distribution beyond the lowest order are given by the semiclassical matrix quantum mechanics.

	\item Let us rephrase this in the string field theory language \cite{Sen:2003iv}, \cite{Mandal:2003tj}. The D0-brane quantum mechanics is in a sense an open string field theory because we treat the theory fully quantum mechanically (not only on mass shell). The new interpretation of the duality states that the open string field theory does know the closed string theory as the collective field theory of the D0-brane Fermi surface. In this sense, Sen's conjecture is marvelously realized in the Liouville theory though the open string field theory is the ``holographic dual" of the closed string theory. It is important to note that the rolling tachyon can be regarded as the rolling eigenvalue of the fermion. In the later time, this can be interpreted as the disturbance of the Fermi surface, i.e. the closed string background. The tachyon matter \cite{Sen:2002in} is just the closed tachyon coherent states and conserves the energy. We will see this issue more closely in the following sections. 
	\item The somewhat mysterious leg factor now has a new interpretation. It appears just in the decaying amplitude of the D0-brane.

	\item From the correspondence discussed above, the open-closed string theory with the space-filling (FZZT) D1-branes should be described by the suitable matrix vector model \cite{Martinec:2003ka}, \cite{Klebanov:2003km}. The candidate action is
\begin{equation}
S = \int_{-\infty}^\infty dt \left[\mathrm{Tr}\left(\frac{1}{2}(D_t \Phi)^2 + U(\Phi)\right)+ \frac{1}{2} (D_t V)^\dagger D_t V + \frac{1}{2} m^2 V^\dagger V + gV^\dagger \Phi V\right],\label{eq:mmv}
\end{equation}
where $V$ is a complex vector (fundamental representation of $U(N)$). From the  spectrum of the D0-D1 string in the last section, we observe that $m^2 \propto s^2$ and can be either tachyonic or massive. \footnote{When the mass becomes too tachyonic, the D1-D1 string has extra tachyonic states which break the unitarity (see the arguments given in section \ref{5.2.2}). It would be interesting to see what happens in this dual matrix vector model.} In order to obtain the double scaling limit, $g$ should be tuned appropriately. This model has been studied in \cite{Yang:1991sc}, \cite{Minahan:1993bz}, \cite{Kazakov:1992pt}, \cite{Avan:1996sp}, \cite{Avan:1997vi} (some of which consider the fundamentals to be fermionic though), however, the qualitative comparison with the exact FZZT result is still lacking.
	\item While the double scaling matrix model has been reinterpreted as a holographic dual theory of the bulk Liouville theory as we have seen so far in this section, there is yet another matrix model which describes the noncritical string theory equivalently well --- the Kontsevich matrix model. Very recently, \cite{Gaiotto:2003yb} has given a dual explanation of this matrix model as an OSFT on the FZZT branes in the $c=-2$ string theory. Their explanation resembles the world sheet proof of the gauge/gravity correspondence in \cite{Ooguri:2002gx}, \cite{Berkovits:2003pq}. Also in \cite{Aganagic:2003qj}, the reason why the double scaling matrix model and the  Kontsevich model describe the same physics is studied in the topological B-model point of view. Essentially, the same CY space is dual to the compact B-brane physics which yields the double scaling matrix model and, at the same time, dual to the noncompact B-brane physics which yields the Kontsevich model. We cannot review these very exciting results any further, so we refer to their original papers.
\end{itemize}

\begin{figure}[htbp]
	\begin{center}
	\includegraphics[width=0.6\linewidth,keepaspectratio,clip]{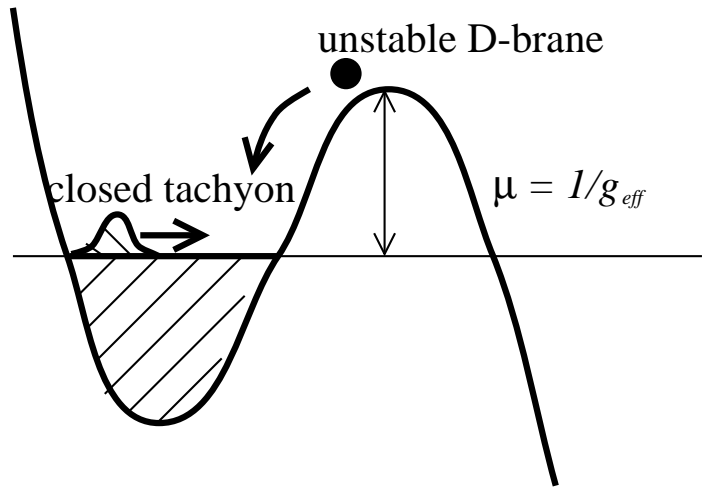}
	\end{center}
	\caption{The schematic picture of the decaying D-brane and the closed tachyon excitation from the open string field ($=$ fermion) perspective.}
	\label{matrixrelo}
\end{figure}

\subsection{Time-like Liouville Theory}\label{6.2}
\subsubsection{Open case}\label{6.2.1}
As we have stated in section \ref{sec:2}, if we perform the analytic continuation of the Liouville theory to $c=25$, we obtain the time-like CFT. Particularly, since this system seems to represent the half S-brane solution of the rolling tachyon, this is an interesting application of the Liouville theory to the time evolution of the theory including the tachyon. In this subsection, we would like to concentrate on the boundary Liouville theory and discuss its analytic continuation \cite{Gutperle:2003xf}, \cite{Constable:2003rc}.\footnote{The closed Liouville theory is a well defined theory (at least for the Liouville sector) at all orders of the string coupling. On the other hand, the open boundary Liouville theory actually has a divergence in the moduli integral at the higher genus. Therefore, we need to change the background via the Fischler-Susskind mechanism \cite{Fischler:1986ci,Fischler:1986tb}.}

The time-like boundary Liouville theory we would like to analyze is given by
\begin{equation}
S_{TBL} = -\frac{1}{\pi} \int d^2z \partial X\bar{\partial}X + \mu_B\int dx e^X. \label{eq:TBL}
\end{equation}
On the other hand, $c \le 1$ space-like boundary Liouville theory (with $\mu =0$) is given by
\begin{equation}
S_{SBL} = \frac{1}{\pi} \int d^2z \left(\partial \phi\bar{\partial}\phi + \frac{QR\phi}{4}\right)+\int dx\left( \frac{QK\phi}{2\pi} + \mu_Be^{b\phi}\right). 
\end{equation}
Thus the formal analytic continuation $b\to i$ and Wick rotation $X = i\phi$ lead to (\ref{eq:TBL}). The reason why we call it ``formal" is that it is not trivial at all whether we can analytically continue the $c \le 1$ Liouville theory to $b \to i$. Actually this ``analytic continuation" has an ambiguity which seems to correspond to the different choice of the vacuum states in the time-dependent quantum theory.

We here consider the bulk one-point function and the boundary two-point function each of which has a specific physical significance. The former can be interpreted as the interaction between the closed string states and the boundary (brane) states. In the space-like Liouville theory ($c \le 1$), the latter can be interpreted as the reflection amplitudes. This goes as follows. In the minisuperspace approximation, the zero-mode wavefunction is given by
\begin{equation}
\Psi_p(\phi) \to e^{+ip\phi} + R_b (p) e^{-ip\phi}, \ \ \ \ \  \ \phi\to - \infty,
\end{equation}
where $\phi$ is regarded as space and $R_b$ is what is called the reflection amplitude which can be read off from the boundary two-point function as
\begin{equation}
R_b(p) = d_b(Q/2 - ip).
\end{equation}
Intuitively, we can imagine the process where an incoming wave $e^{ip\phi}$ reflected from the Liouville potential becomes an outgoing wave with the phase shift (or S matrix) $ R_b(p)$. Note that we have required that the wave should damp as $\phi \to \infty $ in order to obtain this wavefunction uniquely. Now we try to ``analytically continue" this to $b \to i$. Wick rotating the coordinate as $\phi \to - iX$ and the energy as $p \to -i\omega $, we obtain
\begin{equation}
\Psi_\omega(\phi) \to e^{-i\omega X} + d_i (\omega) e^{i\omega X}, \ \ \ \ \  \ X\to - \infty,
\end{equation}
where $d_i(\omega)$  is given by 
\begin{equation}
d_i(\omega) = \langle e^{-i\omega X} e^{-i\omega X} \rangle_{TBL},
\end{equation}
in this analytic continuation procedure. To see the meaning of this amplitude, we consider the Wick-rotated zero-mode Schr\"odinger (Wheeler-DeWitt) equation which becomes
\begin{equation}
\left( \frac{\partial^2}{\partial X^2} + 2\mu_B e^X + N-1 +\vec{p}^2 \right) \psi(X) = 0,
\end{equation}
where $N$ is the level of the excitation and $\vec{p}$ is the spatial momentum.
This has, in general, two asymptotic solutions in the $X\to \infty $ region, namely the positive frequency solution $\propto e^{-x/4-2i\sqrt{2\mu} e^{x/2}} $ and the negative frequency solution $\propto e^{-x/4+2i\sqrt{2\mu} e^{x/2}}$. However, the analytically continued solution only has the positive frequency solution in the $X \to \infty$ region because the original solution had only this type. This corresponds to choosing the ``out vacuum" condition in the time dependent string theory. On the other hand, in the $X\to - \infty$ region, the out vacuum wavefunction can be written as the superposition of the positive and the negative frequency solutions. The coefficient is nothing but the Bogoliubov coefficient. Thus the physical meaning of the disk two-point function $d_i(\omega)$ is
\begin{equation}
d_i(\omega) = \frac{\beta_\omega}{\alpha_\omega},
\end{equation}
where we have introduced the standard Bogoliubov coefficient notation \cite{Gutperle:2003xf}: $\psi_p^{out} = \alpha_p\psi^{in}_p + \beta_p \psi^{in*}_{-p}$, $\psi_p^{in} = \alpha^*_p \psi^{out}_p - \beta_p \psi^{out*}_{-p}$. This enables us to understand the pair production of the string mode in the rolling tachyon background.

\begin{figure}[htbp]
	\begin{center}
	\includegraphics[width=0.8\linewidth,keepaspectratio,clip]{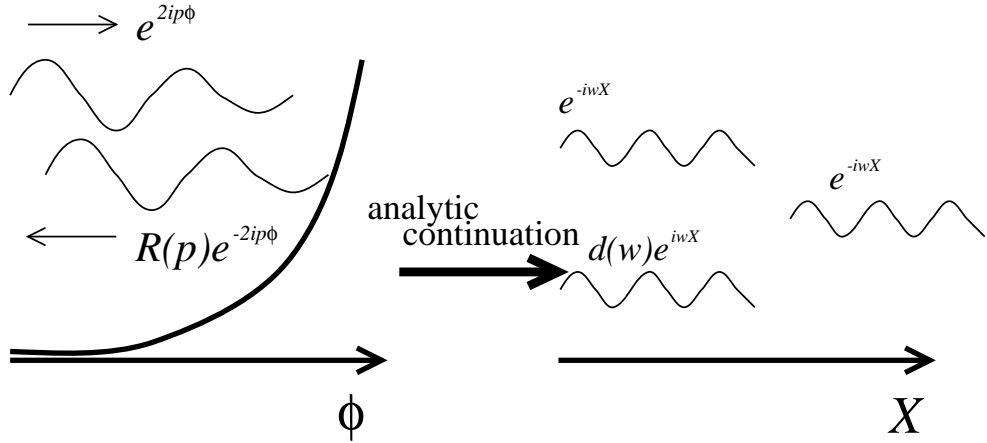}
	\end{center}
	\caption{The analytic continuation of the Liouville scattering yields the Bogoliubov coefficients.}
	\label{wick}
\end{figure}

Let us start the actual analytic continuation with the bulk one-point function $U(\alpha)$, where we encounter no difficulty or subtlety. From the result obtained in section \ref{sec:5}, the one-point function is given by
\begin{equation}
U(\alpha) = \frac{2}{b}(\pi\mu\gamma(b^2))^{(Q-2\alpha)/2b}\Gamma(2b\alpha - b^2)\Gamma\left(\frac{2\alpha}{b}-\frac{1}{b^2}-1\right)\cosh(2\alpha-Q)\pi s.
\end{equation}
Considering the limit $\mu \to 0$ and $b\to i$ which corresponds to $s\to \infty$, we have
\begin{equation}
\lim_{b\to i}\lim_{s\to \infty} U(\alpha) = \pi(2\pi \mu_B)^{2i\alpha}\frac{1}{\sinh(2\pi \alpha)}. \label{eq:tll1p}
\end{equation}
Let us check the correctness of this procedure by comparing this result with what can be obtained from the boundary perturbation of the time-like boundary Liouville theory which we have reviewed in appendix \ref{b-5}. Of course the perturbation theory can be used only to evaluate $\langle e^{-nX} \rangle $ for a positive integer $n$. This can be calculated as 
\begin{eqnarray}
\langle e^{-nX}(z) \rangle_{TBL} &=& \mu_B^n \langle e^{-nX}(z)\int dx_1\cdots dx_n \prod_ie^X(x_i) \rangle_{free} \cr
&=& |z-\bar{z}|^{-n^2/2} \mu_B^n (2\pi)^n \Gamma(n+1) \Gamma(-n),
\end{eqnarray}
where we have used $\int dx e^{-nx - e^{x}} = \Gamma(-n) $ as the zero-mode integration. Analytically continuing this result for general momentum, we expect to obtain
\begin{equation}
\langle e^{-i\omega X} (z) \rangle_{TBL} = |z-\bar{z}|^{-2\Delta_\omega} (2\pi\mu_B)^{i\omega} \frac{\pi}{i\sinh\pi\omega},
\end{equation}
which is the same amplitude we have obtained as the analytic continuation of the boundary Liouville theory.

Then we consider the analytic continuation of the boundary two-point function, which is a little bit subtler. From the $b \le 1$ result in section \ref{sec:5}, we have
\begin{equation}
d(\beta|s,s') = \frac{(\pi\mu\gamma(b^2)b^{2-2b^2})^{(Q-2\beta)/2b}\Gamma_b(2\beta-Q)\Gamma_b^{-1}(Q-2\beta)}{S_b(\beta+i(s+s')/2)S_b(\beta-i(s+s')/2)S_b(\beta+i(s-s')/2)S_b(\beta-i(s-s')/2)}.
\end{equation}
Taking the $s=s' \to \infty$ limit, we obtain\footnote{For simplicity, we consider one kind of brane or a diagonal element of the several kinds of branes. However, the inclusion of the off-diagonal elements does not affect most of the following argument. Note that $s-s'$ is \textit{finite }in the $\mu \to 0$ limit.}
\begin{equation}
\lim_{s\to\infty} d(\omega) = \left(\frac{\pi 2\mu b^{1-b^2}}{\Gamma(1-b^2)}\right)^{(Q-2\omega)/b} \frac{\Gamma_b(-Q+2\omega)}{\Gamma_b(-2\omega+Q)} \frac{1}{S_b(\omega)^2}. \label{eq:sd}
\end{equation}

The problem is that $\Gamma_b(x)$ is not defined in the $b\to i$ limit. In fact, $\Gamma_b(x)$ has a simple pole at $ x = -mb-n/b$ for positive integers $n$ and $m$. In the $b\to i$ limit, it is easy to see that infinitely many poles accumulate to the integer multiple of $i$. Therefore (\ref{eq:sd}) appears to be meaningless at first sight.
 
Nevertheless, we can formally transform (\ref{eq:sd}) so that we obtain an integral expression which we can interpret physically by choosing the proper contour.
The existence of the contour dependence is because we are dealing with the time-dependent theory. From the formula found in appendix \eqref{a-3}, we have
\begin{eqnarray}
& &\frac{\Gamma_b(2\omega-Q)}{\Gamma_b(Q-2\omega)}\frac{1}{S_b(\omega)^2} \cr
&=& Y_b(\omega)b^{2\omega/b-2b\omega-1/b^2+b^2} \frac{\Gamma(2\omega/b-1/b^2)\Gamma(2\omega b-1-b^2)\Gamma(-2\omega/b+1)\Gamma(-2\omega b)}{\Gamma^2(-\omega/b+1)\Gamma^2(-b\omega)},
\end{eqnarray}
where 
\begin{equation}
Y_b(\omega) \equiv \frac{\Gamma_b(2\omega) \Gamma_b(-\omega)^2}{\Gamma_b(-2\omega) \Gamma_b(\omega)^2} = \prod_{m,n \ge 0} \left(\frac{2\omega+\Omega}{-2\omega+\Omega}\right)\left(\frac{-\omega+\Omega}{\omega+\Omega}\right)^2.
\end{equation} 
We have used the infinite product formula \eqref{eq:infpg} for the $\Gamma_b$ function found in appendix \ref{a-3} with $\Omega = mb+n/b$. 

To give a precise meaning to the $Y_b(\omega)$ in the $b\to i$ limit, we first take $b = i\beta $ with real $\beta$ and then we take $\beta \to 1$ limit. In this limiting prescription, $Y_b(\omega)$ keeps to be a pure phase \cite{Gutperle:2003xf}. The actual analytic continuation keeping this property can be found in the original paper. Omitting the detailed calculation, we present the final result 
\begin{equation}
Y_i(\omega) = -e^{i\theta(\omega)},
\end{equation}
where the phase is given by
\begin{equation}
\theta(\omega) = \frac{1}{2} \int_0^\infty \frac{d\tau}{\tau} \frac{\sin(4\omega\tau)-2\sin(2\omega\tau)}{\sinh^2(\tau)}.
\end{equation}
Therefore, the whole two-point function is now analytically continued to
\begin{equation}
d_i(\omega) = \frac{(2\pi\mu)^{2i\omega} e^{i\theta}}{4\cosh^2(\pi\omega)}.
\end{equation}
The remarkable property of this expression is that this reproduces the result of the minisuperspace approximation in the large $\omega$ limit \cite{Gutperle:2003xf} (though it is not clear at least to the author why the minisuperspace approximation becomes better in the large $\omega$ limit). Note that we would have had a different result which does not reproduce the minisuperspace approximation if we had taken the limit of $Y(\omega)$ from the other side. 

Thus the analytic continuation of the space-like boundary Liouville theory to $c=25$ time-like boundary Liouville theory has a subtlety concerning the analytic continuation of the special functions. In addition, there is no physically satisfying principle how to remove the ambiguity of the analytic continuation procedure for three- or higher multi-point functions. 

\subsubsection{Closed case}\label{6.2.2}
Let us now consider the analytic continuation of the bulk Liouville correlation functions to $c=25$ \cite{Strominger:2003fn}, \cite{Schomerus:2003vv}.
The time-like Liouville action which we would like to study is given by
\begin{equation}
S_{TL} = \frac{1}{4\pi} \int d^2z \left(-4\partial X \bar{\partial} X + 4\pi\mu e^{2X}\right).
\end{equation}
We would like to understand the theory from the analytic continuation $b\to i$ of the space-like Liouville theory
\begin{equation}
S = \frac{1}{4\pi} \int d^2 z\sqrt{g} \left(g^{ab}\partial_a\phi\partial_b\phi + QR\phi + 4\pi\mu e^{2b\phi}\right)
\end{equation}
as in the boundary Liouville theory considered in the last subsection. Here, we specifically focus on the two-point function and the three-point function. As we will see, the two-point function can be analytically continued without much trouble, but the three-point function has many subtle points. 

Let us first study the two-point function. For $c \le 1$, the two-point function is given by
\begin{equation}
D(\alpha) = (\pi \mu \gamma(b^2))^{(Q-2\alpha)/b} \frac{\gamma(2b\alpha -b^2)}{b^2\gamma(2-2\alpha/b + 1/b^2)}.
\end{equation}
Setting $b= i\beta$, we take the $\beta\to 1 -0$ limit. Then, we have
\begin{equation}
d(\omega) = \lim_{\beta \to 1-0} -(\pi \mu \gamma(-\beta^2))^{i\omega/\beta} \frac{\Gamma(1-i\omega/\beta)\Gamma(1+i\omega\beta)}{\Gamma(1+i\omega/\beta)\Gamma(1-i\omega\beta)}.
\end{equation}
Since the first term diverges in this limit, we introduce the renormalized cosmological constant as $\mu_R \equiv |\mu \gamma(-\beta^2)|$. Fixing the value of $\mu_R$, we can now take the $\mu \to 0$, $\beta \to 1-0$ limit:
\begin{equation}
d(\omega) \to -(\pi\mu_R)^{i\omega} e^{-\pi \omega},
\end{equation}
which reproduces the minisuperspace amplitude \cite{Strominger:2003fn}.\footnote{While the renormalization of the Liouville potential is also needed for $b=1$, the physical interpretation in $b=i$ seems not so clear to the author. Particularly, in the open case, the boundary cosmological constant never needs to be taken infinitesimally. Note that the value of the boundary cosmological constant does have a physical meaning for the full S-brane ($\mu_B=1$ corresponds to the no tachyon vacuum). This may or may not  result from the fundamental difference between the open string tachyon and the closed string tachyon.}

Let us go on to the three-point function. The $c\le 1$ result is given by the DOZZ formula
\begin{eqnarray}
& &C(\alpha_1,\alpha_2,\alpha_3) = \left[\pi\mu\gamma(b^2)b^{2-2b^2}\right]^{(Q-\sum_{i=1}^3 \alpha_i)/b} \cr &\times& \frac{\Upsilon'(0)\Upsilon(2\alpha_1)\Upsilon(2\alpha_2)\Upsilon(2\alpha_3)}{\Upsilon(\alpha_1+\alpha_2+\alpha_3-Q)\Upsilon(\alpha_1+\alpha_2-\alpha_3)\Upsilon(\alpha_1-\alpha_2+\alpha_3)\Upsilon(-\alpha_1+\alpha_2+\alpha_3)}.
\end{eqnarray}
 Along the same line of reasoning and calculation, Strominger and Takayanagi \cite{Strominger:2003fn} proposed the following expression for the $b=i$ three-point functions
\begin{equation}
\langle e^{2i\omega_1 X}e^{2i\omega_2 X}e^{2i\omega_3 X}\rangle = C(\omega_1,\omega_2,\omega_3)
\end{equation}
where the obvious conformal factors are suppressed and the structure constant is given by
\begin{eqnarray}
C_{ST}(\omega_1,\omega_2,\omega_3) &=& (\pi\mu_R)^{i\sum_j \omega_j} e^{2\pi \tilde{\omega}} \times \cr
&\times& \exp\left[\int_0^\infty \frac{d\tau}{\tau} \frac{1}{\sinh^2\tau/2}\left(\sin^2\tilde{\omega}\tau + \sum_j (\sin^2\tilde{\omega}_j\tau -\sin^2\omega_j\tau)\right)\right],
\end{eqnarray}
where $\tilde{\omega}_j$s are defined as
\begin{eqnarray}
\tilde{\omega} = \frac{1}{2} (\omega_1 +\omega_2 + \omega_3), \ \ \ \ \ \tilde{\omega}_1 = \frac{1}{2} (-\omega_1 +\omega_2 + \omega_3) \cr
\tilde{\omega}_2 = \frac{1}{2} (\omega_1 -\omega_2 + \omega_3), \ \ \ \ \ \tilde{\omega}_3 = \frac{1}{2} (\omega_1 +\omega_2 - \omega_3). 
\end{eqnarray}
The first factor comes from the prefactor of the DOZZ formula and the exponential factor comes from the careful evaluation of the $b\to i$ limit of the Upsilon functions. The puzzle of this formula appears when we take the naive limit $\omega_1\to 0$. We expect this reproduce the two-point function derived above. However this does not hold. The fact that the correlation functions with $\omega_2\neq\omega_3$ survives in this limit is especially puzzling, for this breaks the conformal invariance.

In \cite{Schomerus:2003vv}, Schomerus has made an attempt to solve this puzzle. His strategy is basically as follows. When we take the $b \to i$ limit, the obvious obstacle is the nonexistence of the $\Gamma_b$. However, the functional recursion relation itself does not suffer from this subtlety. Then we solve the recursion relation for the \textit{pure imaginary} $b$, which is unique if we demand the dual relation at the same time. This corresponds to solving the ``Liouville theory" whose central charge is less than $1$. Now we can take the limit $b\to i$ as we take the limit $b\to 1$ in the usual DOZZ formula. The limit exists, but it is different according to whether we consider the Euclidean continuation or the Minkowski continuation. 

In the Euclidean limit, we have the Runkel-Watts theory \cite{Runkel:2002yb} as a limit of the unitary minimal models. In the Minkowski limit, the essential difference from the Strominger-Takayanagi proposal is the additional factor 
\begin{equation}
P(\omega_1,\omega_2,\omega_3) = \left(1+\Theta(\tilde{\omega})\prod_{i=1}^3 \Theta(\tilde{\omega}_j)\right),
\end{equation}
where we have introduced the step function
$$
\Theta(x) = 
\left\{
\begin{array}{l}
-1 \ \ \ \mathrm{for} \ \ \ x <0  \\
+1 \ \ \ \mathrm{for} \ \ \ x >0.
\end{array}
\right.
$$
This factor does not make difference if all the $\omega_j$ are positive. Remarkably the factor solves the puzzling issue stated above as it gives zero unless $\omega_2 = \omega_3$ in the $\omega_1 \to 0$ limit. Can we obtain the correct reflection amplitude in this limit? This seems difficult, for the identity operator is not the simple $\alpha \to 0$ limit of $\Phi_\alpha = e^{i\omega X}$ as in the Runkel-Watts theory. However, we can read off the reflection amplitude from the behavior under $\omega_1 \to -\omega_1$, which yields $d(\omega) = (\pi\mu_R)^{2i\omega} e^{-2\pi\omega}$ as desired.

\subsection{Topological String Theory}\label{6.3}
\subsubsection{Topological sigma model, topological Landau-Ginzburg model}\label{6.3.1}
The easiest construction of the topological theory is twisting the supersymmetric theory. Since the action of the supersymmetric theory is invariant under the fermionic transformation, the change of its spin makes it a scalar BRST invariant theory. In addition, the energy momentum tensor can be represented as the anticommutator of the SUSY generator, which follows from the SUSY algebra. Therefore, the energy momentum tensor is automatically BRST exact, which guarantees the topological nature. In the following, we twist two dimensional $(1,1)$ supersymmetric theories to make topological world sheet theories. If we couple these twisted theories to the (topological) gravity in some sense, they become topological string theories.

First of all, we consider twisting the supersymmetric sigma model (see e.g. \cite{Witten:1988xj,Witten:1993yc}). In this case, there are two different ways to twist the theory. What we call the A-model is given by the following action
\begin{equation}
S = 2t \int d^2 z \left(\frac{1}{2}g_{IJ} \partial_z \phi^I \partial_{\bar{z}}\phi^J + i\psi^{\bar{i}}_zD_{\bar{z}}\chi^i g_{\bar{i}i}+ i\psi^{i}_{\bar{z}} D_z\chi^{\bar{i}}g_{\bar{i}i}- R_{i\bar{i}j\bar{j}}\psi^i_{\bar{z}}\psi^{\bar{i}}_z\chi^j\chi^{\bar{j}}\right),
\end{equation}
where $\psi^{\bar{i}}_z $ is $(1,0)$ form and $\chi^i$ is scalar as fields on the world sheet. We take $\mathrm{CY}_3$ as the target space. The supersymmetric (BRST) transformation is given by
\begin{eqnarray}
\delta\phi^i &=& i\alpha \chi^i \cr
\delta\phi^{\bar{i}} &=& i\bar{\alpha}\chi^{\bar{i}} \cr
\delta\chi^i &=& \delta \chi^{\bar{i}} = 0 \cr
\delta \psi_z^{\bar{i}} &=& -\alpha\partial_z \phi^{\bar{i}}-i\bar{\alpha}\chi^{\bar{j}}\Gamma^{\bar{i}}_{\bar{j}\bar{m}}\psi_z^{\bar{m}} \cr
\delta \psi_{\bar{z}}^{i} &=& -\bar{\alpha}\partial_{\bar{z}} \phi^{i}-i\alpha\chi^{j}\Gamma^{i}_{jm}\psi_{\bar{z}}^{m}. 
\end{eqnarray}
The action is BRST invariant up to the surface term from the general argument of the supersymmetric theory. This can be rewritten in a manifestly BRST invariant way as
\begin{equation}
S= it \int_{\sigma} d^2z\{Q,V\} + t\int\Phi^*(K), 
\end{equation}
where $V = g_{i\bar{j}}(\psi^{\bar{i}}_z\partial_{\bar{z}}\phi^j+\partial_z\phi^{\bar{i}}\psi^j_{\bar{z}})$. The total derivative term is the pull back of the K\"ahler form of the target space which is given by
\begin{equation}
\int\Phi^*(K) = \int d^2z (g_{i\bar{j}}\partial_z\phi^i\partial_{\bar{z}}\phi^{\bar{j}}-g_{i\bar{j}}\partial_{\bar{z}}\phi^i\partial_z\phi^{\bar{j}}).
\end{equation}
This depends only on the K\"ahler class of the original target space and the degrees of map from the world sheet. Therefore, the contribution is simply from the world sheet instanton. The remarkable feature of this model is that the physical amplitudes do not depend on the complex structure of the targetspace at all (Indeed the variation comes only from the BRST exact term $\{Q,V\}$). On the other hand, the dependence on the K\"ahler moduli is determined from the world sheet instanton contribution.

Similarly, if we twist the left and right fermions on the world sheet differently, the twisted theory becomes what we call the B-model. The action is given by
\begin{equation}
S = 2t \int d^2 z \left(\frac{1}{2}g_{IJ} \partial_z \phi^I \partial_{\bar{z}}\phi^J + i\eta^{\bar{i}}(D_z\rho^i_{\bar{z}} + D_{\bar{z}}\rho^i_z) g_{\bar{i}i}+ i\theta_i(D_{\bar{z}}\rho^{i}_z-D_z\rho^i_{\bar{z}})+ R_{i\bar{i}j\bar{j}}\rho^i_{z}\rho^{j}_{\bar{z}}\eta^{\bar{i}}\theta_k g^{k\bar{j}}\right).
\end{equation}
Its BRST transformation is given by
\begin{eqnarray}
\delta\phi^i &=& 0 \cr
\delta\phi^{\bar{i}} &=& i\alpha \eta^{\bar{i}} \cr
\delta\eta^{\bar{i}} &=& \delta\theta_i \ = \ 0 \cr
\delta\rho^i &=& -\alpha d\phi^i.
\end{eqnarray}
The action can be rewritten in a manifestly BRST invariant form. Defining $V = g_{i\bar{j}}(\rho^i_z\partial_{\bar{z}}\phi^{j}+\rho^i_{\bar{z}}\partial_z\phi^{\bar{j}})$, we obtain
\begin{equation}
S = it \int d^2 z\{Q,V\} + t W,
\end{equation}
where $W$ is given by
\begin{equation}
W = \int d^2z \left( -\theta_iD\rho^i- \frac{i}{2}R_{i\bar{i}j\bar{j}}\rho^i\wedge\rho^j\eta^{\bar{i}}\theta_kg^{k\bar{j}}\right).
\end{equation}
We can see that the B-model does not depend on the K\"ahler moduli but only depends on the complex moduli of the target space.

The other model we would like to discuss is the Landau-Ginzburg model \cite{Vafa:1991mu}. Using superfields $X^i$, we have the following action
\begin{eqnarray}
S &=& \int d^2z d^4\theta X^i\bar{X}^{\bar{i}} + \int d^2z d^2\theta W(X) + c.c. \cr
&=& \int d^2z (|\partial x^i|^2 +|\partial_i W|^2) + (\rho^i\bar{\partial}\psi^{\bar{i}} + \bar{\rho}^i\partial\bar{\psi}^{\bar{i}} +\rho^i\partial_i\partial_j W\bar{\rho}^{j} + \psi^{\bar{i}}\partial_{\bar{i}}\partial_{\bar{j}} \bar{W} \psi^{\bar{j}}),
\end{eqnarray}
where we have considered the B-model so that $\psi$ is a world sheet scalar and $\rho$ is a world sheet one-form. The BRST transformation is given by
\begin{eqnarray}
\delta x^{\bar{i}} &=& \psi^{\bar{i}} + \bar{\psi}^{\bar{i}} \cr
\delta \bar{\psi}^{\bar{i}} &=& -\partial_i W g^{i\bar{i}} \cr
\delta\psi^{\bar{i}} &=& \partial_i W g^{i\bar{i}} \cr
\delta\rho^i &=& -\partial x^i \cr
\delta\bar{\rho}^i &=& - \bar{\partial}x^i. \cr
\delta x^i &=& 0 
\end{eqnarray}
This model is not a CFT even at the classical level, but the energy momentum tensor is BRST exact. Therefore, the correlation function does not depend on the two dimensional metric. Then, rescaling the metric as $g\to\lambda^2 g$, we can see the path integral localizes at the points where $dW = 0$. As a result, any correlation function can be evaluated at the extrema of $W$. The nontrivial contribution we should evaluate in the path integral is simply the one-loop determinant. However, this cancels out almost completely except the zero-mode because of the supersymmetry. From the evaluation of the Gaussian integral $\int d^2 x^i \exp(-|\lambda \partial_i W|^2) $, the contribution from the bosonic zero-mode $x_i^0$ is given by 
\begin{equation}
Z_b = \lambda^{-2n} (H\bar{H})^{-1}, 
\end{equation}
where $H$ is the Hessian of $W$. On the genus $g$ world sheet, the number of the fermionic zero-modes are $g$ for each $\rho$ and  one for each $\psi$ from the index theorem, so their contribution is given by
\begin{equation}
Z_f = \lambda^{2n} H^g \bar{H}. 
\end{equation}
Combining them, we finally have the total contribution from the zero-mode: $Z = H^{g-1}$. Applying this calculation, we can easily obtain the correlation functions on the world sheet with any genus. Assuming one superfield for simplicity, we can write them explicitly
\begin{equation}
\langle F_1 \cdots F_N \rangle_g = \oint\frac{F_1 \cdots F_N (W'')^{g}}{W'}, \label{eq:LGC}
\end{equation}
where the integration contour is a large circle which encircles all the poles. Intuitively speaking, $W''$ can be regarded as the operator which inserts a handle into the world sheet.
\subsubsection{Landau-Ginzburg model for $c=1$ Liouville theory}\label{6.3.2}
Let us consider the $c=1$ Liouville theory compactified on the circle whose radius is $R=1$. The first claim here \cite{Ghoshal:1994qt,Ghoshal:1995rs} is that this theory is equivalent to the Landau-Ginzburg topological string theory with the superpotential $W= -\mu X^{-1}$. This superpotential is singular, but we can also regard this theory as a twisted Kazama-Suzuki $SU(2)/U(1)$ coset theory with level $k$ formally analytically continued to $k=-3$. The primary fields of this Landau-Ginzburg theory can be written as $X^{k-1}$. On the other hand, the (physical) primary field of the Liouville theory coupled to the $c=1$, $R=1$ compactified boson can be classified by the integer momentum $k$ because the $X$ space is compactified and the momentum becomes quantized. Then the proposed correspondence is $T_k = X^{k-1}$ up to leg factors.\footnote{However, the leg factor always diverges when $R=1$, so it is natural to define the renormalized operator whose leg factor is omitted as $T_k = X^{k-1}$. We have done the same renormalization in the DMP formula. See (\ref{eq:veren}).} The claim is that the correlation functions (scattering amplitude) calculated from the Landau-Ginzburg theory are the same as those obtained by the matrix model.

By using the basic facts about the Landau-Ginzburg theory that we have introduced in the last subsection, let us see the simplest example. Consider the tachyon three-point function on the sphere $\langle T_{k_1}T_{k_2}T_{k_3}\rangle_0 $. The previous result (\ref{eq:LGC}) yields
\begin{equation}
\langle T_{k_1}T_{k_2}T_{k_3}\rangle_0 = \mathrm{Res} \left(\frac{x^{k_1-1}x^{k_2-1}x^{k_3-1}}{x^{-2}}\right) = \delta_{k_1+k_2+k_3,0}.
\end{equation}
Though this example seems almost trivial, it is generally known that these calculations reproduce the matrix model results \cite{Ghoshal:1994qt,Ghoshal:1995rs}. In addition, the loop calculations can be basically done by using the handle operator which we have introduced in the last subsection. However, to perform the integration over the moduli space properly, we should introduce the contact terms. Furthermore, we should change the picture of the vertex operator with the negative momenta $k$ in order to pick out the correct gravitational descendant which is necessary to integrate over the moduli space.\footnote{On the very intuitive level, this can be explained as follows. Just as we practically carry out the supermoduli integration by the picture change in the superstring calculation, we replace the integration over the moduli space with the calculation of correlation functions in Witten's topological gravity \cite{Witten:1988xi,Labastida:1988zb} by the picture change.} Considering all these, we can show that all correlation functions including the partition function are the same as those from the matrix model calculation at any loop. However, the derivation of the contact term from the first principle is still lacking, so we will not delve into the detail any further.

Assuming this correspondence, we would like to show \cite{Ghoshal:1995wm} that the $c=1$ Liouville theory can be understood as the topological B-model on a certain $\mathrm{CY}_3$ via the correspondence between the Landau-Ginzburg theory and the topological string on CY \cite{Witten:1993yc}. As we will see later, the duality chain enables us to calculate the partition function of the topological string on this $\mathrm{CY}_3$ exactly, which matches the all-order matrix model calculation. This means that we have proved the Liouville Landau-Ginzburg correspondence indirectly. Since the critical dimension of the topological string is $3$, we consider the following superpotential
\begin{equation}
G(X) = - X_0^{-1} + X_1^2 + X_2^2 + X_3^2 + X_4^2.
\end{equation}
The addition of massive fields does not change the behavior in the IR from the Landau-Ginzburg point of view. Therefore nothing has changed as long as we consider the behavior of $X_0$. If we attribute the $U(1)$ charge as $q_0 = -2$, $q_i=1$, $(i\neq 0)$, then the superpotential $G(X)$ has a charge $2$. Since we would like to gauge this $U(1)$ charge, we further introduce a new field $P$ whose charge is $-2$ in order to cancel $U(1)$ anomaly. This gauged supersymmetric theory has the following potential which can be obtained from the direct dimensional reduction of the $d=4$ superQED:
\begin{equation}
U = \frac{1}{2e^2}D^2 + |G|^2 + |p|^2|\partial_i G|^2 + (A_2^2+A_3^2) \left( 4|p|^2 + 4|x_0|^2 + \sum_{i=1}^4 |x_i|^2\right),
\end{equation}
with the  $D$ term condition
\begin{equation}
D = -e^2 \left( -2(|p|^2 + |x_0|^2 )+ \sum_{i=1}^4|x_i|^2 -r \right).
\end{equation}
where $r$ is the F.I. parameter and $A_2$, $A_3$ are from the second and third component of the $4d$ vector. 

Let us first consider the $r\gg 0$ case. In this case we have the vacuum condition $p=0$, $A_2=A_3=0$ and $\sum_{i=1}^4 |x_i|^2-2|x_0|^2 = r $, so $G(X) = 0$. Since the fluctuation becomes massive unless this condition is satisfied, the IR theory is on $G(X) = 0$. Further division by the $U(1)$ gauge symmetry of $X$ suggests that they are naturally regarded as the inhomogeneous coordinates on $\mathrm{WCP}^4_{-2,1,1,1,1}$. That is to say, this theory is just the sigma model on the toric $\mathrm{CY}_3$. In addition, from the condition $\sum_{i=1}^4 |x_i|^2-2|x_0|^2 = r$ and the fact that the metric of the sigma model is given by the pull back of the original flat metric (though the renormalization makes it flow into the Ricci flat one), we can see that the size of the CY manifold is given by $r$. The actual form of the $\mathrm{CY}_3$ can be seen as follows. Fixing $X_0 =1$\footnote{Since $X_0$ is the singular point of the potential, it is infinitely far away in the field space. Therefore, there is no obstruction to suppose $X_0 \neq 0$.}, we obtain
\begin{equation}
1 = X_1^2 +X_2^2 + X_3^2 + X_4^2.
\end{equation}
This is the deformed conifold whose original singularity is $A_1$.

On the other hand, by studying the behavior when $r\ll 0$, we can show this theory is equivalent to the Landau-Ginzburg theory with the superpotential $G(X)$.
To see this, we first note that the vacuum is given by $x_{a} = A_2 =A_3 = 0$  with $a=1\cdots 4$ and  $x_0 \to \infty$ with finite $|p|$ in this limit. Considering the fluctuation around the vacuum solution, we obtain the effective superpotential
\begin{equation}
W(X) = \langle p \rangle G(X)
\end{equation}
if we ignore the massive excitations.

The important point here is the fact that the B-model does not depend on the K\"ahler moduli at all. However, the F.I. parameter $r$ is the size of the manifold in the sigma model perspective, so this is just the K\"ahler moduli. We can also see explicitly that the F.I. term is indeed a BRST exact term in the gauged non-linear sigma model action. Therefore, although this gauged theory looks different in the different $r$ limit, these are equivalent as the B-model. This proves the equivalence of the B-model on the deformed conifold and the topological Landau-Ginzburg theory with the superpotential $X_0^{-1}$. Combining this fact with the Liouville Landau-Ginzburg correspondence, we can see the following Ghoshal-Vafa conjecture \cite{Ghoshal:1995wm} ``The $c=1$, $R=1$ Liouville theory is equivalent to the topological B-model on the deformed conifold." to be plausible.

\subsubsection{Gopakumar-Vafa duality}\label{6.3.3}
Let us pause here for a moment and review the Gopakumar-Vafa duality \cite{Gopakumar:1998vy,Gopakumar:1998ii,Gopakumar:1998jq} briefly. To begin with, let us see the partition function of the A-model topological string more closely.
The partition function with $g>1$ is defined as 
\begin{equation}
F(t) = \int_{M_g} d^{3g-3} \Upsilon d^{3g-3} \bar{\Upsilon} \langle \prod_{k=1}^{3g-3} \int \mu_k G^{-} \int \bar{\mu}_k \bar{G}^{-} \rangle,
\end{equation}
where $\mu$ is the Beltrami differential and $G^{-}$ is the supersymmetric current. From the BRST (SUSY) algebra, we regard the supersymmetric current as the antighost $b$ of the bosonic string. The index theorem states that the complex dimension of the target space should be three in order to absorb all the fermionic zero-modes only by these insertions. This shows that the critical dimension of the topological string is three.

Using the duality between M-theory \cite{Gopakumar:1998ii,Gopakumar:1998jq}, Gopakumar-Vafa\footnote{Recently Ooguri-Vafa \cite{Ooguri:2002gx} gave the purely world sheet derivation of this calculation.} calculated the partition function of the topological A-model whose target space is the resolved conifold. For $g>0$, we have
\begin{equation}
F_g = g_s^{2g-2} \left[(-1)^{g-1} \chi_g \frac{2\zeta(2g-2)}{(2\pi)^{2g-2}}-\frac{\chi_g}{(2g-3)!}\sum_{n=1}^\infty n^{2g-3} e^{-2\pi n s} \right],
\end{equation}
where $\chi_g$ is the (virtual) Euler characteristic of the moduli space of the genus $g$ Riemann surface whose precise value is given by $\chi_g = \frac{B_{2g}}{2g(2g-2)}$. $s$ is the area of the $\mathrm{S}^2$ which is the deformation parameter (K\"ahler moduli) from the conifold. Similarly, the lower genus calculation shows
\begin{equation}
F_0 = \frac{1}{g_s^2} \left[-\zeta(3)+\frac{\pi^2}{6}s+\frac{i}{4}\pi s^2 -\frac{1}{12}s^3 +\sum_{n=1}^{\infty} \frac{e^{-2\pi n s}}{n^3}\right]
\end{equation}
and 
\begin{equation}
F_1 = \frac{1}{24}s + \frac{1}{12}\log(1-e^{-2\pi s}).
\end{equation}

The Gopakumar-Vafa duality claims that this closed A-model partition function (and other physical quantities) transforms into the open A-model on the deformed conifold. This is also called the ``large $N$ duality" 
 because of the natural connection with the AdS/CFT (or gauge/gravity) correspondence. From the AdS/CFT point of view, the rough idea is as follows. We consider wrapping topological A-branes (which becomes D6-branes in the superstring embedding) on the $\mathrm{S}^3$ in the deformed conifold. Taking into account the fact that the gauge theory living on the branes should be topological, it is given by the $U(N)$ Chern-Simons theory on the $\mathrm{S}^3$. In addition, Witten \cite{Witten:1995fb} conjectured that this topological open A-model decouples from the closed sector completely. This fact is equivalent to the automatic Maldacena limit from the AdS/CFT perspective. Furthermore, since the size of $\mathrm{S}^3$ does not affect the topological A-model at all, we can take the conifold limit freely. Now, from the AdS/CFT analog, the gravity dual should correspond to the closed topological string theory in the deformed background without branes. The claim of Gopakumar-Vafa is that it is the closed A-model on the resolved conifold. The parameter correspondence is given by
\begin{eqnarray}
s &=& i\lambda \cr
g_s &=& i\frac{\lambda}{N},
\end{eqnarray}
where $\lambda$ is the 't Hooft coupling of the gauge theory.

To check the duality, we compare the partition function of the Chern-Simons theory. Since the Chern-Simons theory has been solved, this comparison is possible.
The partition function of the Chern-Simons theory is given by
\begin{eqnarray}
e^{-Z} = \frac{1}{(N+k)^{N/2}} \prod_{j=1}^{N-1} \left[ 2\sin\left(\frac{j\pi}{N+k}\right)\right]^{N-j},
\end{eqnarray}
where the renormalized 't Hooft parameter is given by $\lambda = \frac{2\pi N}{k+N}$. We can see the actual equivalence of the partition function by expanding this expression and performing the resummaion.

The particular point to which we would like to pay attention is the nonperturbative contribution of the Chern-Simons theory to its partition function. This is the part where we cannot write the contribution as the 't Hooft (or Feynman) diagram, and it corresponds to the lowest order in $s$ for each genus. For $g>1$, the nonperturbative part is given by
\begin{equation}
F_{g,n.p.} = \frac{B_{2g}}{2g(2g-2)s^{2g-2}}.
\end{equation}
For $g=0,1$, it is given by
\begin{equation}
F_{0,n.p.} = \frac{s^2}{2}\log(s) 
\end{equation}
and 
\begin{equation}
F_{1,n.p.} = -\frac{1}{12} \log(s),
\end{equation}
respectively. Actually this is from the measure factor of the path integral in the Chern-Simons theory. When we expand the path integral around $A=0$, the residual \textit{global} gauge freedom is $U(N)$. Therefore, the measure factor contribution to the partition function is essentially $\log (\mathrm{Vol}(U(N)))$. The volume of the group was calculated by a mathematician \cite{Macdonald:1980}, which is given by
\begin{equation}
\mathrm{Vol}(U(N)) = \frac{(2\pi)^{N^2/2 + N/2} }{G_2(N+1)},
\end{equation}
where $G_2(x)$ is a kind of double gamma function whose defining property is $G_2(x+1) = \Gamma(x) G_2(x)$, $G_2(1)=1$. In our notation, this is essentially the inverse of $\Gamma_b(x)$ with $b=1$ (times $(2\pi)^{x/2}$).\footnote{Note that $b=1$ means $c=1$ in the Liouville theory, which seems to indicate the relation to the $c=1$ Liouville theory.} The logarithm of this expression can be expanded asymptotically as
\begin{equation}
\log G_2(N+1) = \frac{N^2}{2}\log N -\frac{1}{12}\log N -\frac{3}{4}N^2 +\frac{1}{2}N\log 2\pi + \zeta'(-1) + \sum_{n=2}^\infty \frac{B_{2g}}{2g(2g-2)N^{2g-2}},
\end{equation}
which clearly explains the nonperturbative part of the partition function of the Chern-Simons theory.

\subsubsection{Relation to the Liouville theory}\label{6.3.4}
Contemplating on the previous result, we notice that the partition function of the $c=1$ Liouville theory on the $R=1$ circle \eqref{eq:cr1p} has a striking resemblance to the logarithm of the volume of the $U(N)$ group which represents the non-perturbative contribution to the gauge theory. We would like to consider the duality which connects these.

The duality between $R=1$ Liouville theory and the topological string theory has already been discussed. It is given by the B-model on the deformed conifold. Now, we apply the large $N$ conifold transition here. Then the closed B-model on the deformed conifold should be equivalent to the open B-model on the resolved conifold (holomorphic Chern-Simons theory on $\mathrm{CP}_1$). In fact, this is just the mirror symmetry of the A-model large $N$ transition. However, note that this duality is stronger in the A side because the mirror of the deformed conifold on B is the $s \to 0$ limit of the A-model on the resolved conifold.\footnote{The exact mirror of finite $s$ is also known \cite{Marino:2002fk}, \cite{Aganagic:2002wv}. In this case, the B-model theory becomes the matrix model with the unconventional measure factor, which becomes $\sin(\lambda_{ij})$ after the diagonalization.} Hence, by the mirror symmetry, the $s\to 0$ limit, i.e. the nonperturbative part of the A-model, reproduces the partition function of the closed topological B-model on the deformed conifold which yields the Liouville partition function.

Though the above statement may sound very complicated, the gist is that the Liouville partition function is given by the $\log$ of the $U(N)$ group volume from the chain of dualities. By the way, according to the Dijkgraaf-Vafa proposal, we can argue, without resorting to the mirror symmetry, that the partition function of the closed string on the deformed conifold is given by the Gaussian matrix model (which is \textit{not} the $c=1$ matrix quantum mechanics). Since the usual matrix model is defined with the division by the $U(N)$ volume, its $\log$ naturally gives the Liouville partition function. In addition, the Dijkgraaf-Vafa correspondence yields the more general matrix duals of the conifold deformations. It is interesting to see whether these deformations can be understood from the $c=1$ Liouville language. See e.g \cite{Dijkgraaf:2002vw,Aganagic:2003qj} for a recent discussion on the subject.

\subsection{Decaying D-brane}\label{6.4}
Generally, D-branes in the bosonic string theory are unstable. The on-shell description of the unstable D-brane has been reviewed in section \ref{sec:2}. We would like to apply this technique to the branes existing in the Liouville theory. Though the method can be applied to both the ZZ brane and the FZZT brane, we consider the decay of the (ZZ) D0-brane here anticipating the relation between D0-brane decay and the rolling fermion from the top of the potential \cite{Klebanov:2003km}, \cite{Gutperle:2003ij}, \cite{deBoer:2003hd}.\footnote{However, note that the classical probe eigenvalue has been discussed in \cite{McGreevy:2003kb} to be described by the decaying FZZT brane with a special boundary parameter. See also \cite{McGreevy:2003ep}, \cite{Boyarsky:2003jb} for a related discussion.}

The idea is very simple. Let us assume that the decaying D0-brane boundary state is given by the direct product of the ZZ brane boundary state and the rolling tachyon boundary state \cite{Sen:2002nu}, \cite{Lambert:2003zr,Karczmarek:2003xm} :
\begin{equation}
|B\rangle = |B(t)\rangle \otimes |ZZ\rangle.
\end{equation}
Then the emission rate of the closed string in the one-loop approximation can be calculated via the optical theorem, which is given by
\begin{equation}
 N = 2 \mathrm{Im} Z_{Cylinder} = \int_0^\infty \frac{dE}{2E} |A(E)|^2, 
\end{equation}
where $A(E)$ is the on-shell decaying amplitude given by the innerproduct between  $\langle E=2P| \otimes \langle P|$ and the D-brane boundary state.

Now, we start by constructing $|B(t)\rangle $. For definiteness we take the $X^0$ boundary tachyon profile as that of the half S-brane \cite{Strominger:2002pc}, \cite{Larsen:2002wc}, i.e. $\lambda e^{X^0}$. \footnote{Alternatively, we could take the S-brane profile $\lambda \cosh{X^0}$ with the Hartle-Hawking contour, which provides the same result \cite{Lambert:2003zr}. See also \cite{deBoer:2003hd} for a discussion on the various contours and their physical differences.} In order to obtain the corresponding boundary state, we calculate the partition function of the deformed theory with $\delta S = \int dt \lambda e^{X^0}$ except the zero-mode contribution. Splitting $X^0 = x^0 + \hat{X}^0$, we have
\begin{equation}
\langle 1 \rangle_{half S} = \int dx^0\left \langle \exp\left(-\lambda e^{x^0}\int dt e^{\hat{X}^0}\right) \right\rangle_{free \hat{X}^0},
\end{equation}
where we have assumed that this theory can be defined perturbatively. The perturbative expansion yields
\begin{equation}
\langle 1 \rangle_{half S} = \int dx^0\sum_{n=0}^\infty \frac{(-2\pi\lambda e^{x^0})^n}{n!} \int_0^{2\pi} \frac{dt_1}{2\pi}\cdots\frac{dt_n}{2\pi} \left\langle e^{\hat{X}(t_1)}\cdots e^{\hat{X}(t_n)}\right\rangle_{free,\hat{X}}.
\end{equation}
The non-zero mode path integral is just that of the free field, so it can be directly performed as 
\begin{equation}
 \left\langle e^{\hat{X}(t_1)}\cdots e^{\hat{X}(t_n)}\right\rangle_{free,\hat{X}}= \prod_{i<j} |e^{it_i}-e^{it_j}|^2 = 4^{\frac{n(n-1)}{2}}\prod_{i<j} \sin^2\left(\frac{t_i-t_j}{2}\right).
\end{equation}
The integral over $t$'s can be carried out \cite{Larsen:2002wc}\footnote{Making use of the Haar measure $dU$, we can write $1=\frac{1}{\mathrm{Vol}U(N))}\int dU = \frac{1}{n!} \int \prod_i \frac{dt_i}{2\pi} \Delta^2(t)$, where $\Delta(t) = \prod_{i<j} 2\sin\left(\frac{t_i-t_j}{2}\right)$.}
\begin{equation}
\int_0^{2\pi} \frac{dt_1}{2\pi} \cdots \frac{dt_n}{2\pi} 4^{\frac{n(n-1)}{2}}\prod_{i<j} \sin^2\left(\frac{t_i-t_j}{2}\right) = n!.
\end{equation}
Therefore, we finally obtain
\begin{equation}
\langle 1 \rangle_{half S} = \int dx^0 \frac{1}{1+2\pi\lambda e^{x^0}}.
\end{equation}

In order to obtain $\langle e^{iEX^0} \rangle_{half S} = \langle E| B(t)\rangle$, we perform the Fourier transformation, which yields 
\begin{equation}
i \int dt \frac{1}{1+2\pi\lambda e^{t}} e^{iEt} = e^{-iE \log(2\pi\lambda)} \frac{\pi}{\sinh\pi E}.
\end{equation}
See also the time-like Liouville calculation \eqref{eq:tll1p}. On the other hand, the amplitude from the Liouville part is given by
\begin{equation}
A_L = \frac{2}{\sqrt{\pi}}i\sinh(2\pi P) \mu_r^{-iP} \frac{\Gamma(2iP)}{\Gamma(-2iP)},
\end{equation}
from the calculation in section \ref{sec:5}, where we have defined the renormalized cosmological constant as $\mu_r = \pi \mu_0\gamma(b^2), \ b^2\to1$.
Combining these, we obtain
\begin{equation}
A = A_t(E=2P) A_L(P) = 2\sqrt{\pi} i e^{-iE\log(2\pi\lambda)}e^{i\delta(P)}.
\end{equation}
Note that we have reproduced the proper leg factor $e^{i\delta(P)}$. Since it is a pure phase, the emission rate is not affected. Now, we can calculate the emission rate and energy,
\begin{equation}
N = \int_0^\infty\frac{dE}{2E} |const|^2, 
\end{equation}
which diverges logarithmically in the infrared and ultraviolet. In addition, the emitted energy diverges linearly (because the emitted energy is given by $E=\int dE |const|^2$). Note that the final state produced by the decay is a coherent state of the form
\begin{equation}
|\psi \rangle \sim e^{\int_0^\infty \frac{dp}{\sqrt{2E} }a^\dagger_p A_p}|0\rangle .\label{eq:cohe}
\end{equation}

If Sen's conjecture is correct, the decaying D-brane energy changes into some different form. The above lowest emission rate calculation indicates that it becomes closed string tachyons. However, the result has an apparent divergence. We will see in the following the nature of this divergence from the matrix quantum mechanical point of view. Anticipating the result, we will see that the quantum treatment of the D-brane cures the divergence. The divergence here is just from localizing the eigenvalue at a point, whose energy should be divergent in the quantum mechanics.

\begin{figure}[htbp]
	\begin{center}
	\includegraphics[width=\linewidth,keepaspectratio,clip]{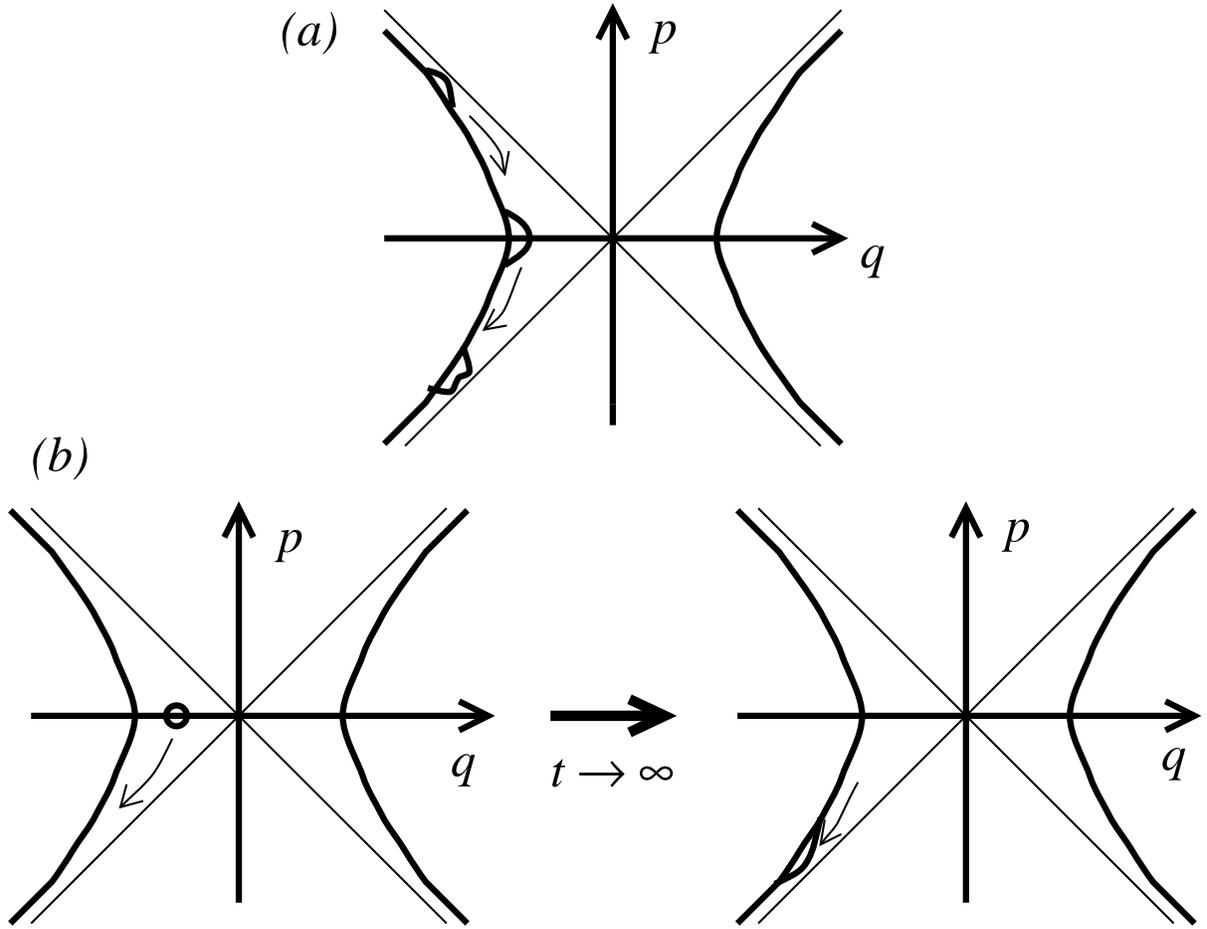}
	\end{center}
	\caption{(a) the scattering of closed string tachyons can be interpreted from the collective field perspective as the transformation of the collective in-field into the out-field. (b) the decaying D-brane can be regarded as the excitation of the fermion (matrix eigenvalue). In the later time ($t \to \infty$), the excitation of the fermion can be regarded as the collective excitation of the Fermi sea and hence the closed tachyon final state.}
	\label{decay}
\end{figure}

Let us move on to the matrix model calculation. The useful picture to keep in mind is figure \ref{decay}. The tachyon scattering is described by the collective excitation of the Fermi surface and the D0-brane is described by the excitation of the eigenvalue. In the later time, the decaying D0-brane approaches the Fermi surface and can be regarded as the fluctuation of the Fermi surface. This is the intuitive picture of the rolling D0-brane into the closed tachyon background from the matrix quantum mechanical point of view. Note that we have used here the wave packet picture deliberately to ensure the semiclassical picture.\footnote{The semiclassical description of the fermionic eigenvalue from the \textit{bosonic} collective field has a small subtlety. See \cite{Das:2004rx} for a recent discussion on this point.}

Suppose we would like to add a new rolling eigenvalue (D0-brane) to the matrix quantum mechanical system. From section \ref{sec:3}, this is described by the relativistic (right) chiral fermion field operator (similar consideration can be done for the left part)
\begin{equation}
\Psi_R(\tau,t) = \frac{1}{\sqrt{2\pi}} :\exp\left[i\sqrt{\pi}\int^\tau \Pi_S -S'd\tau'\right]:.
\end{equation}
Single fermion states are obtained by acting with the operator on the filled Fermi sea. However, if we consider the wave packets which have very small overlap with the Fermi sea, we can effectively treat them as acting on the Fock vacuum. The state $\Psi_R(\tau,t) |0\rangle$ can be expanded by the bosonization formula (\ref{eq:bos}) as the superposition of the incoming tachyon states or the outgoing tachyon states. We are interested in the outgoing expansion. Using the mode expansion (\ref{eq:modet}) and the semiclassical S matrix elements (\ref{eq:smat}), we obtain
\begin{eqnarray}
\sqrt{2\pi}\Psi_R(t-\tau+\log\frac{\mu}{2}) |0\rangle = \exp\left[2i\sqrt{\pi}\sum_{n=1}^\infty \frac{1}{n!} \int_0^\infty \frac{dk}{\sqrt{8\pi^2 k^2}} \int_{-\infty}^0 dk_1 \cdots dk_n e^{i\theta(k)} \right.\cr
\left.\left(\frac{i}{\sqrt{2\pi}\mu}\right)^{n-1} \frac{\Gamma(1-ik)}{\Gamma(2-n-ik)} :(a_{k_1}^\dagger-a_{k_1})\cdots(a_{k_n}^\dagger-a_{k_n}):\delta(\pm|k_1|\cdots \pm|k_n|-k)\right] |0\rangle, 
\end{eqnarray}
where $e^{i\theta(k)} = -2^{ik}e^{-i\log\lambda}$ and $t-\tau= -\log\frac{\lambda}{2}$. The lowest term is given by
\begin{equation}
\exp\left[2i\sqrt{\pi} \int_{-\infty}^0 \frac{dk_1}{\sqrt{8\pi^2k_1^2}}e^{i\theta(|k|)a_{k_1}^\dagger}\right]|0\rangle.
\end{equation}
Recalling that the collective field $S$ and the tachyon $T$ differs by the leg pole factor, this is the same final state calculated from the continuum Liouville field approach (\ref{eq:cohe}).

What about the rest of the terms in the expansion? If we only keep terms with all creation operators, the amplitude corresponds to the disk diagrams with any number of tachyon vertex operators. That is to say, it describes the multi-particle decaying process. Note this is a prediction from the matrix model, and whether this amplitude can be obtained from the Liouville disk amplitude is an interesting open question. Furthermore, terms with both creation and annihilation operators have more $\mu^{-1}$ than the inserted vertices. These terms are naturally regarded as the contribution of the world sheet with many boundaries. Since the closed S matrix has been treated semiclassically, the contribution of the world sheet with handles has not been considered above. For a more detailed argument, see the original paper \cite{Gutperle:2003ij}.

Though the matrix model calculates the higher order effects, the situation does not seem improved at first sight because the emitted energy remains infinite. As we stated above, however, this is a wrong setup quantum mechanically. This issue has been studied in \cite{Klebanov:2003km}, \cite{Gutperle:2003ij}, \cite{Ambjorn:2003ne}. Actually, the appropriate wavefunction should be multiplied to the chiral fermion creation operator. For example, if we consider the definite energy state, we obtain
\begin{equation}
\int d\tau e^{iE(t-\tau)} \Psi_R(t-\tau)|0\rangle = \int d\tau e^{iE(t-\tau)} e^{i2\sqrt{\pi} \int \frac{dp}{2\pi\sqrt{2p}} e^{-ip(t-\tau)} a_p^\dagger A_p}|0\rangle,
\end{equation}
which simply projects the final state onto a definite energy state. Therefore, the energy conservation is manifest. The lesson from here is that it is necessary to treat D-brane quantum mechanically in order to obtain the proper rolling tachyon energy which should be finite.  

Let us finally consider the effect of (unstable but) static D0-brane on the scattering amplitudes. In the matrix model language, this corresponds to the localized eigenvalue at $\lambda=0$. By the collective field method, the effective change in the Hamiltonian is shown to be
\begin{equation}
H = \frac{1}{2}p^2 - \frac{1}{2}\lambda^2 + \frac{1}{\lambda^2} + \mu.
\end{equation}
See \cite{Gutperle:2003ij} for the derivation of this Hamiltonian. Intuitively speaking, the extra eigenvalue (fermion) produces the repulsive effect which is the Calogero-Moser type. The equation of motion under the potential is
\begin{equation}
\frac{d^2}{dt^2} \lambda - \lambda- \frac{2}{\lambda^3} = 0
\end{equation}
which can be solved exactly
\begin{equation}
\lambda = a\sqrt{\cosh^2(t-\sigma) + b}, \ \ \ \ b= -\frac{1}{2}\left(1-\sqrt{1-\frac{8}{a^4}}\right),
\end{equation}
where $a$ and $\sigma$ are integration constants. From this explicit solution, we can calculate the time delay of the scattering. Concentrating on the large (negative) $\lambda$ region where we can write $\tau = -\log(-\lambda)$, the time delay between the given $x$ and back can be calculated from the above explicit solution as
\begin{equation}
t' - t  = 2\tau + \log \left(\frac{a^4}{4}\right) \label{eq:timedelay}
\end{equation}
Using the relation $\epsilon_{\pm} = \pm(p\pm \lambda)\lambda$ in the large $\lambda$ region, we have
\begin{equation}
\epsilon_{-}(t+\tau) = \frac{a^2}{2}\sqrt{1-\frac{8}{a^4}}.\label{eq:cmeom}
\end{equation}
In order to obtain the S matrix elements\footnote{Alternatively we can just use the MPR formalism here. We will use this formalism to obtain the loop scattering amplitudes of the same model in the different context in section \ref{10.1}.}, we use the relation $\epsilon_-(t+\tau) = \epsilon_+(t'-\tau)$ which follows from the definition of $t'$ (i.e. at the time $t'$, the particle is located at $\lambda$ but with the minus momentum as opposed to the incoming particle). Then, substituting (\ref{eq:timedelay}) and (\ref{eq:cmeom}), we obtain the non-linear relation between incoming and outgoing waves:
\begin{equation}
\epsilon_{-}(t+\tau) = \epsilon_+\left(t+\tau + \log(\frac{1}{2}\sqrt{\epsilon_-(t+\tau)^2 + 2})\right)
\end{equation}
We can obtain the S matrix by expanding $\epsilon$ as has been done in section \ref{sec:3}. The surprising consequence is that the S matrix is modified only at the order $g^2$, which means that the disk amplitudes with any on-shell closed vertex operators are actually zero. It is an interesting open question whether this property can be derived from the continuum Liouville field approach. The vanishing of the one-point function can be explained as follows. Since the Neumann $X$ direction forces the energy and momentum of the on-shell vertex operator to be zero, the vanishing property of the one-point function on the ZZ brane in the $p \to 0$ limit makes the whole amplitude vanish. For the T-dualized version of this conjecture, see section \ref{6.6} (particularly around the equation \eqref{eq:mpmtr}). 

\subsection{2D Black Hole}\label{6.5}
In this section, we review the physics of the 2D black hole and its relation to the Liouville theory.
\subsubsection{$SL(2,\mathbf{R})/U(1)$ coset model}\label{6.5.1}
Witten \cite{Witten:1991yr} has discovered that the $SL(2,\mathbf{R})/U(1)$ WZW model admits the semiclassical properties of the 2D black hole and he proposed that a string theory propagating in this exactly solvable CFT describes the string theory in the 2D black hole. Here, we briefly review this string theory and its connection with the Liouville theory (or, strictly speaking, a certain kind of 2D noncritical string theory).

To begin with, we introduce the gauged WZW model. The action for the ungauged $SL(2,\mathbf{R})$  WZW model at level $k$ is given by
\begin{equation}
S_{WZW} = \frac{k}{8\pi} \int d^2 z \sqrt{h}h^{ab}\mathrm{tr}(g^{-1}\partial_a g g^{-1}\partial_b g) +\frac{ik}{12\pi}\int_B d^3y \epsilon^{abc}\mathrm{tr}(g^{-1}\partial_a g g^{-1}\partial_b g g^{-1}\partial_c g),
\end{equation}
where $g$ is an element of the $SL(2,\mathbf{R})$ and $B$ is a three dimensional manifold whose boundary is given by the world sheet Riemann surface. We would like to gauge the following $U(1)$ subgroup of the global symmetry of this theory $g\to a g b^{-1}$:
\begin{equation}
a = b^{-1} = h = 1 + i \epsilon \sigma_2.
\end{equation}
To do this, we introduce a connection field $A_i$ which transforms under the gauge transformation as $\delta A_i = -\partial_i \epsilon$. Then we replace the ordinary derivative with the covariant derivative so that the space dependent transformation $\epsilon(z)$ becomes a symmetry of the theory. As a consequence we obtain the gauged action
\begin{equation}
S_{gauged} = S_{WZW} + \frac{k}{\pi} \int d^2z (\bar{A}\mathrm{tr}(i\sigma_2 g^{-1} \partial g) + A \mathrm{tr}(i\sigma_2 \bar{\partial} g g^{-1}) + A\bar{A} (-2+ \mathrm{tr}(-\sigma_2 g \sigma_2 g^{-1}))). \label{eq:WZW}
\end{equation}
We fix the gauge by imposing the Lorentz (Landau) gauge condition, so we demand $\partial^a A_a = 0$. This can be solved in the two dimension as
\begin{eqnarray}
A &=& \partial X \cr
\bar{A} &=& -\bar{\partial} X.
\end{eqnarray}
Substituting this into the action and picking up the ghost action from the Jacobian of the change of variables from $A$ to $X$, we obtain the following gauge fixed action
\begin{equation}
S_{gf} = S_{WZW} + \frac{k}{\pi} \int d^2 z \partial X \bar{\partial} X + \frac{1}{\pi}\int d^2z (B\bar{\partial} C + \bar{B}\partial \bar{C}),
\end{equation}
where the ghost system is a spin  $(1,0)$ fermion system. In addition, since the gauged $U(1)$ is compact, $X$ is compactified on the circle of radius $\sqrt{k}$ (after rescaling the kinetic term as a conventionally normalized boson). To obtain this action, we have used the global transformation of $g$ \cite{Dijkgraaf:1992ba}.

So far, the treatment has been exact in the quantum mechanical sense. However, to see the connection between this action and the 2D black hole, it is better to do the different gauge fixing. Since the following gauge fixing is semiclassical, we should understand the procedure is exact only in the limit $k\to \infty$. The gauge fixing is done by requiring the following form of the group action
\begin{equation}
g = \cosh{r} + \sinh{r}\left(
\begin{array}{cc}
\cos\theta & \sin\theta \\
\sin\theta & -\cos\theta
\end{array}
\right).
\end{equation}
After substituting this into the above action (\ref{eq:WZW}) and performing the Gaussian integral, we obtain the gauge fixed action
\begin{equation}
S = \frac{k}{\pi} \int d^2z (\partial r\bar{\partial}r + \tanh^2 r \partial \theta\bar{\partial} \theta),
\end{equation}
where the determinant of the Gaussian integral has been neglected. Therefore we can see that the semiclassical $SL(2,\mathbf{R})/U(1)$ WZW model is equivalent to the nonlinear sigma model whose metric is given by
\begin{equation}
ds^2 = k(dr^2 + \tanh^2 r d\theta^2). \label{eq:2DB}
\end{equation}
However, the consistent condition for the string background requires the one-loop Ricci flatness of the target space, and this metric \textit{does not} satisfy this condition. To overcome this difficulty, we should recall the neglected determinant of the Gaussian integral over $A$, which probably provides the dilaton background. We determine the induced dilaton background by requiring that the one-loop Einstein equation $R_{\mu\nu} = -2\nabla_\mu\nabla_\nu \Phi$ should hold. Then the dilaton background is given by $\Phi = 2\log\cosh r$. Of course, the procedure stated above is valid only semiclassically and higher corrections are needed when $k$ is small which may drive us from the 2D black hole description.

By the way, to obtain the Minkowski version of the above metric, we rotate $\theta$ as $\theta =it$ and take the periodicity for $t$ to be infinite. Thus the Minkowski metric becomes $ds^2 = dr^2 -\tanh^2 r dt^2$ which indeed has a (Minkowski) black hole property after extending the coordinate system. From the gauged WZW model point of view, this corresponds to the gauging of the noncompact $U(1)$ subgroup. Dijkgraaf-Verlinde-Verlinde \cite{Dijkgraaf:1992ba} has studied this semiclassical equivalence quantum mechanically by using the minisuperspace approximation. Though it is interesting, here we will not delve into the detail any further.

Returning to the $SL(2,\mathbf{R})/U(1)$ WZW model, we try to obtain the correlation functions of the theory. Since the direct use of the action is inconvenient, we introduce the free field (Wakimoto) representation of the algebra \cite{Bershadsky:1991in}. We first introduce a (1,0) $\beta\gamma$ superghost system and a Coulomb gas free boson as follows
\begin{equation}
S = \int d^2z \left(\frac{1}{\pi}\partial\phi\bar{\partial}\phi + \frac{QR}{4\pi}\phi + \frac{1}{\pi}(\beta\bar{\partial}\gamma + \bar{\beta}\partial\bar{\gamma})\right),
\end{equation}
where the background charge is given by $Q^2=\frac{1}{k-2}$. Using this action, we define the following current
\begin{eqnarray}
J_+ &=& i\beta \cr
J_3 &=& \beta\gamma + \frac{1}{Q}\partial \phi \cr
J_- &=& -i\beta\gamma^2 - i\frac{2}{Q}\gamma\partial\phi + ik\partial \gamma.
\end{eqnarray}
Via the free field OPE (for our convention, see appendix \ref{a-1}), we can check the following current algebra
\begin{eqnarray}
J_+(z)J_-(0) &=& \frac{k}{z^2} - \frac{2J_3(0)}{z} + \cdots \cr
J_3(z)J_\pm(0) &=& \pm \frac{J_\pm(0)}{z} + \cdots \cr
J_3(z)J_3(0) &=& -\frac{\frac{k}{2}}{z^2} + \cdots 
\end{eqnarray}
which indeed yield the level $k$ $SL(2,\mathbf{R})$ algebra.\footnote{Our convention is  $t^1=\sigma_1/2,t^2=\sigma_2/2,t^3=i\sigma_3/2$ and $J_\pm = J_1 \pm iJ_2$.} The energy momentum tensor of this theory can be calculated by the Sugawara construction (or we can directly calculate it because they are free),
\begin{equation}
T = -\beta\partial\gamma -(\partial\phi)^2 + Q\partial^2 \phi.
\end{equation}
The central charge is given by $ c = \frac{3k}{k-2}$.

Now we would like to define the vertex operator in order to calculate the correlation function. Since $SL(2,\mathbf{R})$ does not have a finite dimensional unitary representation, there might be some difficulties. Requiring that the Casimir $J_3$ and $J^2 = -\frac{1}{2}(J_+J_- + J_-J_+) + J_3J_3$ have a real eigenvalue, we find that the corresponding vertex operator is given by
\begin{equation}
T_{jm}(z) = \gamma^{j-m}e^{-2Qj\phi},
\end{equation}
where $j$ takes the same value as in the Liouville momentum, namely $j$ is real or $j = -\frac{1}{2} + ip$ where $p$ is real. Furthermore $m$ takes a real value but the available values are determined by a given $j$. For example, for $j = -\frac{1}{2} + ip$ with real $p$, we can take any real $m$, but for real $j$, we have two different representations. One is the highest or lowest representation with $j+m$ or $j-m$ integer respectively, and the other is the supplementary series with neither $j+m$ nor $j-m$ integer. The conformal dimension of this operator is given by
\begin{equation}
\Delta_{jm} = -\frac{j(1+j)}{k-2}.
\end{equation}

However, this is not the end of the story. While the current algebra and the vertex operators represented by the free field indeed provide the correct OPE of the current algebra, most of the correlation functions turn out to be zero if we calculate the actual correlation function. This is because of the ``momentum" and ghost number conservation from the zero-mode path integral of the boson and the ghost. To remedy this difficulty, we take the screening charge trick. We add the following term to the action
\begin{equation}
\delta S = Q = \int d^2 z J(z) = \int d^2z\beta \bar{\beta} e^{2Q\phi},
\end{equation}
whose weight is one and this does not spoil the differential equation which the current algebra should satisfy (Knizhnik-Zamolodchikov equation). This can be shown by the following OPE
\begin{eqnarray}
J_3(z) J(w) &\sim& reg \cr
J_+(z) J(w) &\sim& reg \cr
J_-(z) J(w) &\sim& \frac{\partial}{\partial w} \left(\frac{e^{2Q\phi}}{z-w}\right).
\end{eqnarray}
Inserting this term into the correlation function, (at least as long as the charge nonconservation is given by the integer unit) we can render them non-zero.

Now we are ready to write down the action of the free field (Wakimoto) representation for the $SL(2,\mathbf{R})/U(1)$ model. After gauge fixing by the Landau gauge, it is given by
\begin{eqnarray}
S = \int d^2z\left(\frac{1}{4\pi}(\partial_a\phi)^2 + \frac{QR}{4\pi}\phi + \frac{1}{\pi}(\beta\bar{\partial}\gamma + \bar{\beta}\partial\bar{\gamma}) + M\beta \bar{\beta} e^{2Q\phi} + \right. \cr
+ \left.\frac{1}{4\pi}(\partial_a X)^2 + \frac{1}{\pi}(B\bar{\partial}C + \bar{B}\partial\bar{C}) + \frac{1}{\pi}(b\bar{\partial}c + \bar{b}\partial\bar{c} ) \right),
\end{eqnarray}
where $X$ is compactified on the circle of radius $\sqrt{k}$. For this whole action to be conformally invariant, the total central charge should vanish, which yields the condition $k = \frac{9}{4}$ and $Q=2$.

The BRST symmetry of this theory comes from the diffeomorphism of the world sheet and the gauged $U(1)$, whose BRST charge are given by

\begin{eqnarray}
Q^{Diff} &=& \oint c(z)\left(T_{matter} + \frac{1}{2}T_{bc} \right) \cr
Q^{U(1)} &=& \oint C(z)\left(J_3 + i\sqrt{k} \partial X\right).
\end{eqnarray}
The BRST invariance under $U(1)$ determines the physical vertex operator as
\begin{equation}
V_{jm} = \gamma^{j-m} e^{-2Qj\phi} e^{\frac{2im}{\sqrt{k}} X}. \label{bkv}
\end{equation}
Moreover, the usual diffeomorphism BRST invariance and the compactness of $X$ requires
\begin{eqnarray}
2j+1 &=& \pm\frac{2}{3}m \cr
m &=& \frac{1}{2}(n_1+n_2 k) \cr
\bar{m} &=& -\frac{1}{2}(n_1 - n_2k),
\end{eqnarray}
where $n_1$ and $n_2$ are integers.

At this point, we would like to see the (semiclassical) relation between this free field representation and the sigma model whose target space is the 2D black hole. To do so we bosonize the $\beta \gamma$ ghost. \footnote{See appendix \ref{a-1} for our conventions} This can be written as $\beta = i\partial v e^{iv-u}$, $\gamma=e^{u-iv}$. Then the vertex operator for the screening charge becomes $V = i\partial v e^{iv-u+2Q\phi}$ (for the left moving part). Now, using the fact that
\begin{equation}
[Q^{U(1)},B] = i\partial v + \frac{1}{Q}\partial\phi + i\sqrt{k} \partial X
\end{equation}
holds, we can rewrite the vertex operator for the screening charge as 
\begin{equation}
V \simeq e^{-u+iv} e^{-\bar{u}+i\bar{v}} \left(\frac{1}{Q}\partial\phi+i\sqrt{k}\partial X\right)\left(\frac{1}{Q} \bar{\partial}\phi + i\sqrt{k}\bar{\partial} X\right) e^{2Q\phi} \label{eq:sqf}
\end{equation}
up to BRST exact terms. However, since the bosonized ghost $u$ and $v$ always appear as the combination $-u+iv$, the both contribution cancel in every contraction of these operators. Therefore we can ignore the factor in practice. Then the action becomes
\begin{equation}
S = \int d^2z \left[\frac{1}{4\pi}(\partial_a\phi)^2 + \frac{QR\phi}{4\pi} + \frac{1}{4\pi}(\partial_a X)^2 + M \left(\frac{1}{Q} \partial\phi + i\sqrt{k}\partial X\right)\left(\frac{1}{Q} \bar{\partial}\phi + i\sqrt{k}\bar{\partial} X\right) e^{2Q\phi}\right]. \label{eq:dual1}
\end{equation}
This action can be seen from the noncritical string point of view (recall $Q=2$.) as the 2D string perturbed not by the cosmological constant (tachyon) but by the discrete state operator.

On the other hand, we can regard this as a sigma model whose metric is given by
\begin{eqnarray}
G_{\phi\phi} &=& 1 + \frac{M\pi}{Q^2}e^{2Q\phi} \cr
G_{\phi X} &=& iM\frac{\pi\sqrt{k}}{Q}e^{2Q\phi} \cr
G_{XX} &=& 1- M\pi ke^{2Q\phi}.
\end{eqnarray}
In addition if we introduce $\theta = X + if(\phi)$, $f'(\phi) = \frac{1}{e^{2\phi} -1}$ and $\phi = \log\cosh r$ we reproduce the semiclassical black hole metric (\ref{eq:2DB}) which has been introduced by Witten. We have two comments here. First, the last transformation needs a quantum correction. Second, the transformation to the sigma model is given by the ``complex" transformation which seems a little bit strange. However, if we stick to the Minkowski signature, this is not a problem as we can see from the way $i$ enters the game \cite{Eguchi:1993tp}.

Let us move on to another interpretation of the 2D black hole or $SL(2,\mathbf{R})/U(1)$ model. FZZ \cite{FZZ} and Kazakov-Kostov-Kutasov (KKK) \cite{Kazakov:2000pm} conjectured that the model is also equivalent to the sine-Liouville theory. We first introduce the sine-Liouville theory and show several pieces of evidence of this equivalence. \footnote{The direct proof can be given for the supersymmetric extension of this duality, which we will review in the later section \ref{sec:12}. It is also interesting to note that the claim (\ref{eq:dual1}) and (\ref{eq:dual2}) describe the same physics is the bosonic analog of the conjectured duality in the $\mathcal{N} =2$ super Liouville theory. See section \ref{11.2} and \ref{12.3}.} The action of the sine-Liouville model is given by
\begin{equation}
S = \frac{1}{4\pi} \int d^2 z\left[(\partial_a\phi)^2 + (\partial_a X)^2 + QR\phi + 4\pi \lambda e^{b\phi}\cos \sqrt{k}(X_L-X_R)\right], \label{eq:dual2}
\end{equation}
where $Q^2 = \frac{1}{k-2}$ and $b=\frac{1}{Q}$ as in the above case. Similarly $X$ is compactified on the circle of the radius $\sqrt{k}$. As in the Liouville theory $Q=2$ is the critical dimension. The corresponding vertex operator is given by
\begin{equation}
V_{j;m\bar{m}} \iff e^{ip_L X_L + ip_R X_R -2Qj\phi},
\end{equation}
where, with the given integers $n_1$, $n_2$, $m$ and $\bar{m}$, the momenta are
\begin{eqnarray}
p_L &=& \frac{n_1}{\sqrt{k}} + n_2\sqrt{k} \cr
p_R &=& \frac{n_1}{\sqrt{k}} -n_2 \sqrt{k}.
\end{eqnarray}

We would like to see the equivalence of the two-point functions $\langle V_{j_1m_1}V_{j_2m_2}\rangle$ on the sphere. If we take $m=\bar{m}$, the conservation law enforces $m_1 = -m_2$ and $j_1=j_2$. We first calculate this two-point function in the free field representation. The ghost conservation, with consideration of the anomaly on the sphere, or the conservation of the zero-mode of $\phi$ which gives the same condition, leads to 
\begin{eqnarray}
\langle V_{j,m}(1)V_{j,-m}(0)\rangle = \left\langle \gamma^{j-m}\bar{\gamma}^{j-m}e^{-2Qj\phi} e^{2im/\sqrt{k} X}(1) \gamma^{j+m}\bar{\gamma}^{j+m}e^{-2Qj\phi}e^{-2im/\sqrt{k} X}(0)\right. \cr
\times \left. M^s\beta\bar{\beta}e^{2Q\phi}(\infty) \left[\int d^2z \beta\bar{\beta} e^{2Q\phi}\right]^{s-1}\right\rangle,
\end{eqnarray}
where $s = 2j+1$ and we have fixed three points by using the $SL(2,\mathbf{C})$ invariance of the world sheet. Since the contribution of the $bc$ ghost is trivial, we have not apparently included. Although, this expression is only meaningful when $s$ is an integer, the pole at the integer $s$ is believed to be given by this expression as in the Liouville theory.\footnote{We can explicitly see this by integrating over the zero-mode of $\phi$.} Because this expression is given in terms of the free field, the calculation is straightforward. The result is given by
\begin{equation}
\langle V_{j,m}(1)V_{j,-m}(0)\rangle = (-\pi M\gamma(Q^2))^s\gamma(1+j-m)\gamma(1+j+m)s\gamma(1-s)\gamma(-Q^2s)
\end{equation}
Now we concentrate on the $\gamma(-Q^2 s)$ term. Since this term is proportional to $\Gamma(1-\frac{2j+1}{k-2})$, it has a pole at
\begin{equation}
1- \frac{2j+1}{k-2} = -n, \label{eq:pole}
\end{equation}
where $n$ is a positive integer. Around the pole, the amplitude can be written as
\begin{equation}
\langle V_{j,m}(1)V_{j,-m}(0)\rangle \sim \frac{(-\pi M \gamma(\frac{1}{k-2}))^s \Gamma(-2j-1)\Gamma(j-m+1)\Gamma(1+j+m)}{n!(n+1-\frac{2j+1}{k-2})\Gamma(\frac{2j+1}{k-2})\Gamma(2j+2)\Gamma(-j-m)\Gamma(m-j)}.
\end{equation}

On the other hand, we consider the sine-Liouville calculation. Performing the path integral over the zero-mode of $\phi$, we obtain
\begin{eqnarray}
\langle V_{j,m}(1)V_{j,-m}(0)\rangle = \ \ \ \ \ \ \ \ \ \ \ \ \ \ \ \ \ \ \ \ \ \ \ \ \ \ \ \ \ \ \ \ \ \ \ \ \ \ \ \ \ \ \ \ \ \ \ \ \ \ \ \ \ \ \ \ \ \ \ \ \ \ \ \ \ \ \ \ \ \ \ \ \ \ \ \ \ \ \ \ \ \ \ \ \ \ \cr
= \lambda^{s'}\Gamma(-s')\left\langle e^{ip_LX_L+ip_RX_R-2jQ\phi}e^{ip_LX_L+ip_RX_R-2jQ\phi}\left[\int d^2ze^{b\phi}\cos\sqrt{k}(X_L-X_R)\right]^{s'}\right\rangle_{free},
\end{eqnarray}
where $s'=\frac{2(2j+1)}{k-2}$. If we take $s'$ to be a positive integer, there exists a pole from $\Gamma(-s')$. This is the same condition (\ref{eq:pole}) calculated from the free field $SL(2,\mathbf{R})/U(1)$ model. The rest of the path integral is free and we can explicitly confirm that the residue at the pole is also same.

Similarly, the match of the three-point function has been discussed in the literature \cite{FZZ}, \cite{Kazakov:2000pm}, \cite{Fukuda:2001jd}. The explicit calculation shows that they match exactly. In addition, the equivalence of the spectrum of the theory has been calculated there.

So far, we have only considered the case where the integral screening charge saturates the violation of the charge conservation. This restriction has enabled us to calculate the correlation functions by using the free path integral. In more general situations, the ``analytic continuation" which corresponds to the DOZZ formula and Teschner's trick is needed (see e.g. \cite{Fukuda:2001jd}). 
\subsubsection{Matrix model dual}\label{6.5.2}
Let us consider the matrix model dual of the 2D black hole which is quantum mechanically given by the $SL(2,\mathbf{R})/U(1)$ coset model \cite{Kazakov:2000pm}. The matrix model we will consider here is the direct dual of the sine-Liouville theory which we have discussed above to be equivalent to the 2D black hole. Taking $Q=2$, the action of the sine-Liouville model is given by
\begin{equation}
S = \frac{1}{4\pi}\int d^2z\left[(\partial_a\phi)^2 + (\partial_a X)^2 + 2R\phi + \lambda e^{b\phi}\cos \sqrt{k}(X_L-X_R)\right].\label{eq:sLe}
\end{equation}
Since we know the matrix model dual of the Liouville theory in detail, we regard the sine-Liouville model as the perturbation of the usual 2D Liouville theory. The perturbed action is 
\begin{equation}
S = \frac{1}{4\pi}\int d^2z\left[(\partial_a\phi)^2 + (\partial_a X)^2 + 2R\phi + \mu e^{2\phi} + \lambda e^{(2-r)\phi}\cos r(X_L-X_R)\right],
\end{equation}
where $r$ is the radius of the compactification circle. If we set $r=3/2$ and $\mu =0$, we reproduce the sine-Liouville action. As usual, we can derive the exact WT identity governing the power of the Liouville coupling $\mu$ and $\lambda$. Shifting $\phi \to \phi -\frac{1}{2}\log\mu$, we observe the genus $g$ partition function can be written as
\begin{equation}
F_g = \mu^{2-2g} F_g\left(\lambda \mu^{\frac{r-2}{2}}\right).
\end{equation}

First, we consider the unperturbed action. Then the partition function is given by the Gross-Klebanov formula\footnote{The first term has an extra minus here because of the Legendle transformation. The string partition function can be obtained by reversing the sign of the first term.} whose derivation was reviewed in section \ref{sec:3},
\begin{equation}
F(\mu,\lambda=0) = -\frac{r}{2}\mu^2\log\mu - \frac{1}{24}\left(r+\frac{1}{r}\right)\log\mu + r\sum_{g=2}^\infty \mu^{2-2g}C_g(r) + \mathcal{O}(e^{-2\pi\mu}) +  \mathcal{O}(e^{-2r\pi\mu}), 
\end{equation}
where the precise form of $C_g(r)$ can be found in section \ref{sec:3} (see equation \eqref{eq:matz}). The omitted term is the asymptotic completion of the series. 

For the perturbed partition function, the above argument for the exact WT identity suggests the following expansion 
\begin{align}
F_0(\lambda,\mu) &= -\frac{r}{2}\mu^2\log\mu + \mu^2 A_0(z) \cr
F_1(\lambda,\mu) &= -\frac{1}{24}\left(r+\frac{1}{r}\right)\log\mu + A_1(z) \cr
F_g(\lambda,\mu) &= r\mu^{2-2g} A_{g}(z),
\end{align}
where we have defined the actual expansion parameter $z \equiv \sqrt{r-1}\lambda \mu^{\frac{r-2}{2}}$. Clearly $A_g(0) = C_g(r)$. Note that the 2D black hole limit corresponds to $z\to \infty$, so this perturbative expansion is not useful (or the existence of the limit is not a priori apparent). 

The matrix model dual of this perturbed theory has been conjectured to be the twisted finite temperature matrix quantum mechanics. The partition function is given by
\begin{equation}
Z_N = e^{F} = \int \mathcal{D}\Omega \exp(1+\lambda \mathrm{Tr}(\Omega+\Omega^{-1})) \int_{M(2\pi r) = \Omega^\dagger M(0)\Omega} \mathcal{D}M \exp\left[-\int_0^{2\pi r}dt \left(\frac{1}{2}\dot{M}^2 -\frac{1}{2} M^2\right)\right],
\end{equation}
where we have implicitly taken the double scaling limit. $\Omega$ is a Unitary matrix which represents the twist or the vortex in the time evolution. This matrix quantum mechanics looks complicated. For the actual evaluation of this matrix integral, we refer to the original paper. The net result of \cite{Kazakov:2000pm} is summarized in the following differential equation (often called the Toda differential equation) which the free energy of the matrix quantum mechanics should satisfy
\begin{equation}
\frac{1}{4}\lambda^{-1}\partial_\lambda \lambda \partial_\lambda F(\mu,\lambda) + \exp\left[-4\sin^2\left(\frac{1}{2}\partial_\mu\right) F(\mu,\lambda)\right] = 1.
\end{equation}
The initial condition of this differential equation is given by the Gross-Klebanov formula. By solving the Toda differential equation, we can in principle obtain the perturbed partition function. Then we hope that the black hole limit will exist. Actually the black hole limit $z\to \infty$ exist when $1<r<2$, so we can find the partition function of the 2D critical black hole in this way.

Let us see some of the limiting solutions of the Toda differential equation. First, the perturbative solution of the genus $0$ partition function is given by
\begin{equation}
F_0(\lambda) = -\frac{r\mu^2}{2}\log\mu + r\mu^2 \sum_{n=1}^\infty\frac{1}{n!}((r-1)\mu^{r-2}\lambda^2)^n \frac{\Gamma(n(2-r)-2)}{\Gamma(n(1-r)+1)}
\end{equation}

On the other hand, the partition function in the black hole limit is given by\begin{equation}
F(\lambda,\mu=0) = -\frac{1}{4}(2-r)^2(r-1)^{\frac{r}{2-r}}\lambda^{\frac{4}{2-r}} - \frac{r+r^{-1}}{48} \log(\lambda^{\frac{4}{2-r}}) + \cdots.
\end{equation}
Relating the sine-Liouville parameter $\lambda$ with the black hole mass $M$ by the formula $M = \lambda^8 (r-1)^4$ which can be obtained from the comparison of the free field calculations, we have finally reached the matrix model conjecture for the partition function of the 2D black hole (with $r = \frac{3}{2}$)
\begin{equation}
F = -\frac{1}{8}M -\frac{13}{288}\log M + \cdots.
\end{equation}

The validity of this formula has been discussed in the literature. Especially, the torus partition function agrees with the continuum calculation \cite{Kazakov:2000pm}.

\subsubsection{Branes in 2D black hole}\label{6.5.3}
\begin{figure}[htbp]
	\begin{center}
	\includegraphics[width=0.6\linewidth,keepaspectratio,clip]{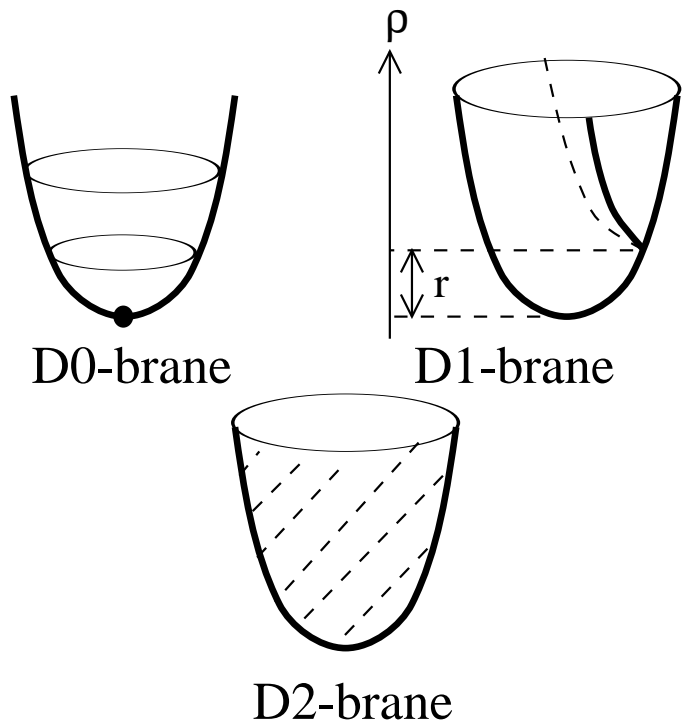}
	\end{center}
	\caption{The branes in the two dimensional Euclidean black hole.}
	\label{blkbrane}
\end{figure}

Recently in \cite{Ribault:2003ss}, exact boundary states of the D-branes living in the 2D black hole are proposed. We will review their proposal without a derivation. We have three kinds of D-branes in the 2D black hole (or $SL(2,\mathbf{R})/U(1)$ coset model), namely the D0-, D1- and D2-brane. Since we are in the Euclidean signature, the D0-brane means the point-like brane in the space and so on. See figure \ref{blkbrane}. 

The one-point function ($\Phi^{j}_{n_1n_2}$ is the vertex oprator introduced in \eqref{bkv} with the right-hand part) on the D0-brane is given by
\begin{eqnarray}
\langle \Phi^{j}_{n_1n_2}(z,\bar{z})\rangle_m^{D0} = \delta_{n_1,0} \left(\frac{b^2}{2}\right)^{\frac{1}{4}} \frac{\sin\pi b^2 m}{\sqrt{2\pi \sin \pi b^2}} (-1)^{m n_2} \left(\frac{k}{2}\right)^{\frac{1}{4}} \frac{\Gamma(-j+\frac{k n_2}{2})\Gamma(-j - \frac{k n_2}{2})}{\Gamma(-2j-1)} \cr
\times \frac{\sin\pi b^2}{\sin\pi b^2 m} \frac{\sin \pi b^2 m (2j+1)}{\sin\pi b^2(2j+1)} \frac{\Gamma(1+b^2)\nu_b^{j+1}}{\Gamma(1-b^2(2j+1))} \frac{1}{|z-\bar{z}|^{h^j_{n_1n_2} + \bar{h}^j_{n_1n_2}}},
\end{eqnarray}
where $b^2 = Q^2 = \frac{1}{k-2}$ and $\nu_b = \frac{\Gamma(1-b^2)}{\Gamma(1+b^2)}$ have been introduced, $m$ is a positive integer parameter which characterizes the nature of the D0-brane. Note the resemblance between this D0-brane and the ZZ brane in the Liouville theory. It would be interesting to see whether this brane describes the matrix model dual of the 2D black hole which has been reviewed in the last section, just as is the case in the Liouville theory.

The second brane is the D1-brane which descends from the $\mathrm{AdS}_2$ brane in $\mathrm{H}_3$, where $\mathrm{H}_3$ is the Wick rotated version of $SL(2,\mathbf{R})$. The one-point function is given by
\begin{eqnarray}
\langle \Phi^{j}_{n_1n_2}(z,\bar{z})\rangle_{r\theta}^{D1} = \delta_{n_2,0} (8b^2)^{-\frac{1}{4}} e^{in_1\theta} (2k)^{-\frac{1}{4}} \frac{\Gamma(2j+1)}{\Gamma(1+j+\frac{n_1}{2})\Gamma(1+j-\frac{n_1}{2})} \cr
\times (e^{-r(2j+1)} + (-1)^{n_1} e^{r(2j+1)}) \Gamma(1+b^2(2j+1))\nu_b^{j+\frac{1}{2}} \frac{1}{|z-\bar{z}|^{h^j_{n_1n_2} + \bar{h}^j_{n_1n_2}}},
\end{eqnarray}
where $r$ and $\theta$ parameterize the D1-brane.

The final brane is the D2-brane which expands along the whole space. The one-point function is given by
\begin{eqnarray}
\langle \Phi^{j}_{n_1n_2}(z,\bar{z})\rangle_{\sigma}^{D2} = \delta_{n_1,0} (8b^2)^{-\frac{1}{4}} \left(\frac{\Gamma(-j+\frac{kn_2}{2})}{\Gamma(j+1+\frac{kn_2}{2})}e^{i\sigma(2j+1)} +\frac{\Gamma(-j-\frac{kn_2}{2})}{\Gamma(j+1-\frac{kn_2}{2})}e^{-i\sigma(2j+1)}\right) \cr
\times (k/2)^{1/4} \Gamma(2j+1)\Gamma(1+b^2(2j+1))\nu_b^{j+\frac{1}{2}} \frac{1}{|z-\bar{z}|^{h^j_{n_1n_2} + \bar{h}^j_{n_1n_2}}}, 
\end{eqnarray}
where the parameter $\sigma$ takes values in the interval $[0,\frac{\pi}{2}(1+b^2)]$. In order to introduce two different kinds of D2-branes consistently, the parameter $\sigma$ must satisfy the following condition:
\begin{equation}
\sigma - \sigma' = 2\pi \frac{m}{k-2}
\end{equation}
with an integer $m$. The physical meaning of this restriction and the nature of the parameter $\sigma$ have been studied in \cite{Ribault:2003ss}. By studying the open spectrum stretching between these two kinds of branes, we can interpret $m$ as the extra D0-brane charge attached to the D2-brane. This reminds us of the FZZT brane with an imaginary boundary parameter which has an interpretation as the superposition of the FZZT brane and the ZZ brane. However, we should note that the D2-brane boundary state here is not so simple as the superposition of these branes.\footnote{In the dual sine-Liouville model, some results on the possible brane solutions have been obtained in \cite{Lukyanov:2003nj}, where W algebra plays an important role.} 
 
\subsection{Nonperturbative Effects}\label{6.6}
The matrix model calculation gives not only the perturbative string coupling $g_s$ expansion of the Liouville theory, but also (one of) its nonperturbative completion(s). This is because the matrix model calculation yields a finite answer while the expansion in $g_s$ or $\mu_r^{-1}$ is only the asymptotic series and not convergent. In this section, we would like to show that the nonperturbative contribution of the matrix model comes from the ZZ brane in the continuum approach \cite{Alexandrov:2003nn}, \cite{Alexandrov:2003un}. This strongly supports the general observation that the D-brane plays an important role in the nonperturbative effects of the string theory. It also suggests that the nonperturbative completion by the matrix model \cite{David:1991sk,David:1993za}, \cite{Eynard:1993sg}, \cite{Shenker:1990uf} is very plausible.

To see the nature of the nonperturbative effects, we take $c=0$ model for simplicity. The matrix model prediction is summarized in the Painlev\'e I equation
\begin{equation}
u^2(\mu) -\frac{1}{6}u''(\mu) = \mu,
\end{equation}
where we are using the convention which corresponds to filling the only half side of the Fermi surface. The string partition function is given by $Z''(\mu) = -u(\mu)$, where $\mu$ is related to the coefficient of the Liouville potential up to a multiplicative normalization. This equation has an asymptotic solution $u = \sqrt{\mu} + \cdots$, but it is not unique. Let us take another solution $\tilde{u}$ which is different from $u$ but has the same asymptotic behavior. Then the difference between the solutions $\epsilon = \tilde{u}-u$ satisfies the differential equation $\epsilon'' = 12 u\epsilon$. Substituting the asymptotic form of $u$, we obtain
\begin{equation}
\frac{\epsilon'}{\epsilon} = r\sqrt{u} + \cdots, 
\end{equation}
where $r$ is a parameter which expresses the nonperturbative effects. In this case it is given by $r=-2\sqrt{3}$. Thus the nonperturbative effects can be written as
\begin{equation}
\epsilon \sim e^{-\frac{8\sqrt{3}}{5}\mu^{\frac{5}{4}}}.
\end{equation}
On the other hand, considering the instanton effects of the D-brane, we find the nonperturbative correction can be written as  $\epsilon \sim e^{Z_{disk}}$ from Polchinski's combinatoric argument \cite{Polchinski:1994fq}. Comparing these, we find the prediction for $r$, which is given by
\begin{equation}
r = \frac{\frac{\partial}{\partial \mu}\log\epsilon}{\sqrt{u}} = \frac{\frac{\partial}{\partial\mu}Z_{disk}}{\sqrt{-Z''(\mu)}}.
\end{equation}
The important point from the calculational perspective is that this combination does not depend on the multiplicative normalization of $\mu$.

However, since it depends on the overall normalization of the partition function (because the denominator is square root and the degree is different from the numerator), it is more desirable to compare this factor with the theory which has a sequence of parameters. Fortunately, the prediction of $r$ from the multi-matrix model which corresponds to the 2D gravity coupled to the unitary minimal model is known. For the $p$th minimal model whose central charge is $c= 1-\frac{6}{p(p+1)}$, the parameters which show the nonperturbative ambiguity of the partition function are known \cite{Eynard:1993sg} to be labeled as 
\begin{equation}
r_{m,n} = - 4\sin\frac{\pi m}{p}\sin\frac{\pi n}{p+1}, 
\end{equation}
where $m$ and $n$ are positive integers. Let us compare this result with the Liouville theory.

Since the source of the nonperturbative correction is supposed to be D-branes, we first consider the contribution from the D-brane in the minimal model. The D-branes in the minimal model are well-known \cite{Cardy:1989ir} and correspond to the boundary states labeled by two positive integers $(m,n)$, whose disk partition function is given by
\begin{equation}
Z_{m,n} = \left(\frac{8}{p(p+1)}\right)^{\frac{1}{4}}\frac{\sin\frac{\pi m}{p} \sin\frac{\pi n}{p+1}}{\left(\sin \frac{\pi}{p}\sin\frac{\pi}{p+1}\right)^{\frac{1}{2}}},
\end{equation}
respectively.

Next, we consider the disk contribution from the Liouville direction. The candidate brane is the ZZ brane. The ZZ brane also has two positive integer parameters $(m,n)$. However, the brane which contributes to the leading nonperturbative effects is only the basic $(1,1)$ brane in hindsight. Whether the other branes contribute to the higher correction or not is an interesting question which requires a further argument.\footnote{As we will see later, at least the $(1,n)$ branes contribute to the $c=1$ Liouville theory.} In any case, the contribution from $(1,1)$ brane is 
\begin{equation}
\frac{\partial}{\partial \mu} Z_{disk} = Z_{m,n}\times \langle V_b\rangle_{(1,1)}.
\end{equation}
To calculate this quantity, we need the unnormalized one-point function which is not divided by $\langle 1\rangle$. From the result in section \ref{sec:5}, we find
\begin{equation}
\langle V_b \rangle_{(1,1)} = C \frac{1}{\sqrt{2\pi}} \Psi_{(1,1)}\left(P=\frac{b-b^{-1}}{2i}\right),
\end{equation}
where $C$ is a constant which we will determine by comparing it with the matrix model calculation. Substituting the actual form, we obtain 
\begin{equation}
\langle V_b \rangle_{(1,1)} = - C \frac{2^{1/4}\sqrt{\pi}[\pi\mu\gamma(b^2)]^{\frac{1}{2}(b^{-2}-1)}}{b\Gamma(1-b^2)\Gamma(b^{-2})}.
\end{equation}

Now, we move on to the calculation of the numerator. This is essentially the two-point function on the sphere $\langle V_b V_b \rangle_{sphere}$. To avoid the subtlety concerning the $SL(2,\mathbf{C})$ gauge fixing, we first calculate the three-point function and then we integrate the result by $-\mu$. The three-point function is given by the DOZZ formula,
\begin{equation}
\langle V_bV_bV_b\rangle_{sphere} = b^{-1}[\pi\mu]^{b^{-2}-2}[\gamma(b^2)]^{b^{-2}}\gamma(2-b^{-2}).
\end{equation}
Integrating this, we obtain
\begin{equation}
\langle V_b V_b \rangle_{sphere} = \frac{b^{-2}-1}{\pi b} [\pi \mu \gamma(b^2)]^{b^{-2}-1}\gamma(b^2)\gamma(1-b^{-2}).
\end{equation}
Combining these results, we can calculate $r$ from the continuum theory, which is given by
\begin{equation}
r_{m,n} = - 2C \sin\frac{\pi m}{p} \sin\frac{\pi n}{p+1}.
\end{equation}
If we have taken $C=2$, this reproduces the matrix model prediction. Note that it is nontrivial that $C$ does not depend on $p$, $n$, and $m$.

We next consider the nonperturbative effect for the $c=1$ Liouville theory. Following \cite{Alexandrov:2003nn}, \cite{Alexandrov:2003un} we consider the sine-Liouville perturbation of the Liouville theory which is given by the action
\begin{equation}
S = \frac{1}{4\pi}\int d^2z\left[(\partial_a\phi)^2 + (\partial_a X)^2 + 2\hat{R}\phi + \mu e^{2\phi} + \lambda e^{(2-R)\phi}\cos R(X_L-X_R)\right],
\end{equation}
where $\hat{R}$ is the world sheet curvature and $R$ is the compactification radius. As has been studied in the last section, the partition function is given by the solution of the Toda differential equation
\begin{equation}
\frac{1}{4}\lambda^{-1}\partial_\lambda \lambda \partial_\lambda F(\mu,\lambda) + \exp\left[-4\sin^2\left(\frac{1}{2}\partial_\mu\right) F(\mu,\lambda)\right] = 1 \label{eq:Todad}
\end{equation}
whose initial condition is given by the Gross-Klebanov formula (\ref{eq:matz}). Note that the Gross-Klebanov formula admits nonperturbative effects of order $\mathcal{O}(e^{-2\pi\mu})$ and $\mathcal{O}(e^{-2\pi R\mu})$. Let us consider the effect of $\lambda$ perturbation for these nonperturbative effects. Interestingly, the $\lambda$ correction to the first kind of nonperturbative effects is actually none. This is because the nonperturbative correction 
\begin{equation}
\Delta F = \sum_{n=1}^\infty C_n e^{-2\pi n\mu}
\end{equation}
provides an exact solution of the full equation (\ref{eq:Todad}). Thus this kind of correction is not modified by the $\lambda$ perturbation. On the other hand, the second type nonperturbative effect has a $\lambda$ correction. To see this, consider the non-perturbative correction $\epsilon(\mu,y) \sim e^{-\mu f(y)}$, where $y$ is the actual expansion parameter $y = \mu(\lambda\sqrt{R-1})^{\frac{2}{R-2}} = z^{\frac{2}{R-2}}$. Substituting the asymptotic expansion and the nonperturbative correction $\epsilon(\mu,y)$ into the Toda differential equation (\ref{eq:Todad}), we obtain the form of the nonperturbative correction $f(y)$. Setting the initial condition $\lim_{\lambda \to 0} f_k(y) = 2\pi R k$, we find
\begin{equation}
f_k(y) = 2\pi k R + 4\sin(\pi Rk) \mu^{-\frac{2-R}{2}} \lambda+ \cdots.
\end{equation}

Can we interpret this nonperturbative effect from the continuum Liouville point of view? The answer is yes. Actually, the $k$th correction comes from the $(1,k)$ ZZ brane. The matrix model prediction for the quantity $r$ is given by
\begin{equation}
r_k = -\frac{2k\pi \sqrt{R}}{\sqrt{-\log\mu}}.
\end{equation}
On the other hand, from the Liouville analysis, the disk partition function for the $X$ boson is given by
\begin{equation}
Z_{Neumann} = 2^{-1/4} \sqrt{R}
\end{equation}
and the disk one-point function of the Liouville theory is given by
\begin{equation}
\langle {V_{1}(0)}\rangle_{1,k} = - \lim_{b\to 1}k \frac{2^{5/4}\sqrt{\pi}}{\Gamma(1-b^2)}.
\end{equation}
The two-point function in the denominator can be calculated as 
\begin{equation}
\partial^2_\mu Z_0 = -\lim_{b\to 1} \frac{\log\mu}{\pi\Gamma^2(1-b^2)}.
\end{equation}
Combining all these factors, we reproduce $r_k$ correctly. We can also compare the $\mathcal{O}(\lambda)$ term. To do so, we first define the dimensionless quantity $\rho$ as 
\begin{equation}
\rho_k = \lim_{\lambda \to 0} \frac{\frac{\partial}{\partial \lambda} \log\epsilon_k}{\sqrt{-\partial_\lambda^2 Z_0}}.
\end{equation}
Then the matrix model calculation predicts $\rho = -2\sqrt{2}\sin(k\pi R)$. On the other hand, the Liouville counterpart is given by
\begin{equation}
\rho = \frac{\langle \cos(R(X_L-X_R))\rangle_X \langle V_{1-\frac{R}{2}}\rangle_{(1,k)}}{\sqrt{-\langle \cos(R(X_L-X_R)) V_{1-\frac{R}{2}}\cos(R(X_L-X_R)) V_{1-\frac{R}{2}}\rangle_{sphere}}}.
\end{equation}
Straightforward calculation reveals that this reproduces the matrix model prediction correctly \cite{Alexandrov:2003nn}, \cite{Alexandrov:2003un}.

We have two comments in order. First, the same nonperturbative effects could have been obtained by the $(k,1)$ brane instead of $(1,k)$ brane, for there is no difference between them when $b=1$. Whether $(k,1)$ brane makes sense quantum mechanically (of course not semiclassically) is an intriguing question. Second, the nonexistence of the $\lambda$ perturbation in the $(1,k)\times$ Dirichlet boundary nonperturbative effect (D-instanton correction) predicts a remarkable property of the disk amplitude, i.e.
\begin{equation}
\left\langle \left(\int d^2z e^{(2-R)\phi}\cos(R(X_L-X_R))\right)^n\right\rangle_{(1,1)\times Dirichlet} = 0. \label{eq:mpmtr}
\end{equation}
Though this is trivial when $n$ is odd, it is a nontrivial statement even for $n=2$. The T-dual version of this statement is that the closed tachyon scattering amplitude on the $(1,k)$ disk is zero, which we have addressed in the previous section \ref{6.4} as a conjecture.

\subsection{Branes and Riemann Surfaces}\label{6.7}
In this section we closely follow the discussion of \cite{Seiberg:2003nm}, and we review their interpretation of the FZZT branes and the ZZ branes in the Liouville theory coupled to the minimal model (``minimal gravity") as certain integrals over the degenerate Riemann surfaces. 

To begin with, let us review the ground ring structure of the $(p,q)$ minimal gravity (see also \cite{Kutasov:1992qx}. Very recently \cite{Kostov:2003cy} has discussed the boundary ground ring structure). The tachyon is defined as an on-shell vertex operator:
\begin{equation}
T_{r,s} = c\bar{c} \Phi_{r,s} V_{\beta_{r,s}},
\end{equation}
where $\Phi_{r,s}$ is the minimal model $(r,s)$ primary operator and $V_{\beta_{r,s}} = e^{2\beta_{r,s} \phi}$ is the corresponding Liouville dressing whose Liouville momentum is given by\begin{equation}
2\beta_{r,s} = \frac{p+q-rq+sp}{\sqrt{pq}}, \ \ \ rq-sp \ge 0.
\end{equation}
The ground ring structure is given by
\begin{equation}
O_{r,s} = H_{r,s}\bar{H}_{r,s} \Theta_{r,s},
\end{equation}
where $\Theta_{r,s} = \Phi_{r,s} V_{\alpha_{r,s}}$ and $H_{r,s}$ is defined in \eqref{eq:BRSTex}. It is easy to see that the Liouville momentum of $O_{r,s}$ is given by
\begin{equation}
2\alpha_{r,s} = \frac{p+q-rq-sp}{\sqrt{pq}}.
\end{equation}
The ring structure can be summarized as follows (see \cite{Seiberg:2003nm} for a derivation, but in principle we can calculate them from the Liouville and minimal model OPE). Set the fundamental objects $\hat{x}$ and $\hat{y}$ as
\begin{align}
\hat{y} &= \frac{1}{2\sqrt{\tilde{\mu}}}O_{2,1} \cr
\hat{x} &= \frac{1}{2\sqrt{\mu}}O_{1,2},
\end{align}
where we have normalized the dual cosmological constant as $\tilde{\mu} = \mu^{1/b^2} = \mu^{q/p}$, which is slightly different from the other part of this review. Then the ring elements are expressed as 
\begin{equation}
O_{r,s} = \mu^{\frac{q(r-1)+p(s-1)}{2p}} U_{r-1}(\hat{y}) U_{s-1}(\hat{x}),
\end{equation}
and the relations 
\begin{align}
U_{q-1}(\hat{x}) &= 0 \cr
U_{p-1}(\hat{y}) &= 0, \label{eq:ring1}
\end{align}
where $U_{r-1}(\cos\theta) = \frac{\sin(r\theta)}{\sin \theta} $ are the Chebyshev polynomials of the second kind. It is important to note that the tachyons form a (nonfaithful) module under the ring structure (see also section \ref{2.5}), which means
\begin{equation}
T_{r,s} = \mu^{1-s} O_{r,s} T_{1,1} = \mu^{\frac{q(r-1) - p(s-1)}{2p}} U_{r-1}(\hat{y})U_{s-1}(\hat{x}) T_{1,1}.
\end{equation}
Since there are only $(p-1)(q-1)/2$ tachyons in contrast to $(p-1)(q-1)$ ring elements, the following additional condition on the module exists,
\begin{equation}
(U_{q-2}(\hat{x}) - U_{p-2}(\hat{y})) T_{r,s} = 0. \label{eq:ring2}
\end{equation}
It is easy to see that the condition \eqref{eq:ring2} with the condition \eqref{eq:ring1} is summarized as 
\begin{equation}
(T_{q}(\hat{x}) - T_p(\hat{y})) T_{r,s} = 0,
\end{equation}
with the Chebyshev polynomials of the first kind $T_{p}(\cos\theta) = \cos(p\theta)$.

The next ingredients are branes in the minimal gravity. The FZZT brane in the minimal gravity is written as 
\begin{align}
|\sigma;k,l\rangle &= \sum_{k',l'} \int_{-\infty}^{\infty} dP \cos(2\pi P\sigma)\frac{\Psi(-P) S(k,l;k',l')}{\sqrt{S(1,1;k',l')}} |P\rangle |k',l'\rangle.
\end{align}
Here, the wavefunction $\Psi(P)$ and $S(k,l;k',l')$ are given by
\begin{align}
\Psi(P) &= \mu^{-ipb^{-1}} \frac{\Gamma(1+2iPb^{-1})\Gamma(1+2iPb)}{2i\pi P} \cr
S(k,l;k',l') &= (-1)^{kl'+k'l} \sin(\frac{\pi p ll'}{q})\sin(\frac{\pi q k k'}{p}),
\end{align}
and $|P\rangle$ are the Liouville Ishibashi states and $|k,l\rangle$ are the minimal model Ishibashi states. These branes are actually not independent. We can show \cite{Seiberg:2003nm} that
\begin{equation}
|\sigma;k,l\rangle = \sum_{m',n'}|\sigma + \frac{i(m'q+n'p)}{\sqrt{pq}};1,1\rangle 
\end{equation}
holds up to BRST exact states, where $(m',n')$ runs over
\begin{eqnarray}
m'&=& -(k-1),-(k-1)+2, \cdots, k-1 \cr
n'&=& -(l-1),-(l-1)+2, \cdots, l-1.
\end{eqnarray}

The ZZ brane is subtler because of the rationality of $b^2$. Particularly, for the higher degenerate ZZ branes, we should be more careful about the subtraction of the degenerate module (see e.g. \cite{Lian:1992aj}), for the subtraction done in section \ref{sec:5} is now overcounting. The careful analysis shows, however, independent branes can be labeled by the degenerate states label $1\le m \le p-1$, $1\le n \le q-1$ with $mq-np >0$ as in the usual ZZ branes. They call these $(p-1)(q-1)/2$ branes as principal ZZ branes. The explicit boundary states can be written as
\begin{align}
|m,n\rangle &= \sum_{k',l'} \int_{-\infty}^{\infty} dP \sinh(\frac{2\pi mP}{b}) \sinh(2\pi n Pb) \Psi^*(P)\sqrt{S(1,1;k',l')}|P\rangle |k',l'\rangle.
\end{align}
It is also interesting to note that the principal ZZ brane has the following one-point functions for the ground ring operator:
\begin{equation}
\langle O_{r,s} | m,n\rangle = U_{r-1}\left((-1)^n\cos\frac{\pi q m}{p}\right) U_{s-1} \left((-1)^m \cos\frac{\pi p n}{q}\right) \langle 1 | m,n\rangle,
\end{equation}
which is consistent with the ring multiplication rule and shows that the principal ZZ branes are eigenstates of the ring generator
\begin{align}
\hat{x} | m,n\rangle &= x_{m,n} | m,n\rangle = (-1)^m \cos \frac{\pi np}{q} | m,n\rangle \cr
\hat{y} | m,n\rangle &= y_{m,n} | m,n\rangle = (-1)^n \cos \frac{\pi mq}{p} | m,n\rangle. \label{eq:FZZTp}
\end{align}

With these preparations, we can state the following geometrical interpretation of the FZZT branes and the ZZ branes in the minimal gravity proposed in \cite{Seiberg:2003nm} (see also \cite{Klebanov:2003wg}). With some rescaling of the cosmological constant, we can easily show from the result in section \ref{sec:5} that the partition function on the disk with the FZZT boundary condition satisfies
\begin{align}
\partial_{\mu_B}Z &= \sqrt{\tilde{\mu}} \cosh \left(\pi \sigma \sqrt{\frac{q}{p}}\right) \cr
\mu_B &= \sqrt{\mu}\cosh \left(\pi \sigma \sqrt{\frac{p}{q}}\right).
\end{align}
This equation can be rewritten as 
\begin{equation}
F(x,y) = T_{q}(x) - T_{p} (y) = 0, \label{eq:rieman}
\end{equation}
with $x = \frac{\mu_B}{\sqrt{\mu}}$ and $y = \frac{\partial_{\mu_B}Z}{\sqrt{\tilde{\mu}}}$. The equation \eqref{eq:rieman} defines a degenerate Riemann surface whose singularities are located at $(p-1)(q-1)/2$ distinct points. Surprisingly, the singularities are located just at the positions of the principal ZZ branes \eqref{eq:FZZTp}. 

The disk partition function of the FZZT brane and the ZZ brane now have a geometrical interpretations on the degenerate Riemann surface \eqref{eq:rieman}. The FZZT disk partition function is given by
\begin{equation}
Z(\mu_B) = \mu^{\frac{p+q}{2p}}\int^{x(\mu_B)} y dx,
\end{equation}
which follows directly from the definition of $x$ and $y$.
On the other hand, the principal ZZ brane partition function is written as 
\begin{equation}
Z_{m,n} = \mu^{\frac{p+q}{2p}} \oint_{B_{m,n}} y dx,
\end{equation}
where $B_{m,n}$ is a cycle through the singularity $(x_{m,n}, y_{m,n})$. This is the direct consequence of the relation \eqref{eq:marti} and the above FZZT brane's result.

\subsection{Literature Guide for Section 6}\label{6.8}
The observation that the matrix model dual of the Liouville theory is the holographic dual of the ZZ brane is first made in \cite{Polyakov:2001af}, and it is further developed in \cite{McGreevy:2003kb}, \cite{McGreevy:2003ep}, \cite{Klebanov:2003km}. The connection with the string field theory is emphasized in \cite{Sen:2003iv}, \cite{Mandal:2003tj}. On the other hand, the Kontsevich matrix model is proposed to be dual to the open string field theory of the FZZT brane in the $c=-2$ theory in \cite{Gaiotto:2003yb}. In \cite{Aganagic:2003qj}, the equivalence of the description from the double scaling matrix model and the Kontsevich matrix model is studied from the topological string perspective.

The time-like Liouville theory is studied in \cite{Gutperle:2003xf}, \cite{Constable:2003rc}, \cite{Schomerus:2003vv}, \cite{Strominger:2003fn}. The relation between the topological string theory on the conifold and $c=1$ Liouville theory has been studied in \cite{Imbimbo:1995yv}, \cite{Ghoshal:1994qt}, \cite{Ghoshal:1995rs}. The decaying D-brane in the two dimensional string has been studied in 
 \cite{Klebanov:2003km}, \cite{Gutperle:2003ij}, \cite{deBoer:2003hd}, \cite{Ambjorn:2003ne}, \cite{Boyarsky:2003jb}.

The string theory on the two dimensional black hole has attracted  much attention not only because the theory is tractable but also it is a good toy model to understand the physical properties of the quantum black hole. After the seminal work by Witten \cite{Witten:1991yr}, much work has been done on the subject including \cite{Dijkgraaf:1992ba}, \cite{Tseytlin:1991ht}, \cite{Bershadsky:1991in}, \cite{Seiberg:1992bj}, \cite{Jack:1993mk}, \cite{Bershadsky:1991in}, \cite{Eguchi:1993tp}, \cite{Becker:1994at}. For a review, see e.g. \cite{Witten:1992fp}, \cite{Becker:1994vd}, \cite{Kazama:1994se}, \cite{Grumiller:2002nm}. On the matrix model dual side, the contribution of the non-singlet sector has been studied in \cite{Gross:1990ub}, \cite{Gross:1991md}, \cite{Boulatov:1993xz}, and the connection with the gravitational sine-Gordon model has been studied in \cite{Moore:1992ga}, \cite{Hsu:1993cm}.
While the conventional Liouville theory is not a two dimensional black hole as we have seen in the main text, \cite{Mukhi:1993zb} has discovered that the two dimensional \textit{topological} black hole  ($\mathcal{N}=2$ twisted $SL(2,\mathbf{R})/U(1)$ model) is equivalent to the conventional $c=1$ Liouville theory, which also reveals the topological nature of the $c=1$ Liouville theory.

The parent theory of the two dimensional black hole is the $SL(2,\mathbf{R})$ WZW model, which is closely related to the $\mathrm{AdS}_3$. Since the string theory on $\mathrm{AdS}_3$ is related to some $\mathrm{CFT}_2$ from the AdS/CFT correspondence, there have been many studies on the subject (note that unlike in the $\mathrm{AdS}_5$ case, we do not necessarily have the R-R flux background, so the string theory is more manageable). Some references are \cite{Bars:1999ik}, \cite{Berenstein:1999gj}, \cite{Giveon:1999jg}, \cite{Sugawara:1999fq}, \cite{Petropoulos:1999nc}, \cite{Teschner:1999ug}, \cite{Maldacena:2000hw}, \cite{Kato:2000tb}, \cite{Ishibashi:2000fn}, \cite{Hosomichi:2000bm}, \cite{Maldacena:2000kv}, \cite{Larsen:2000yw}, \cite{Maldacena:2001km}, \cite{Satoh:2001bi}, \cite{Hosomichi:2001fm}, \cite{Giribet:2001ft}, \cite{Giveon:2001up}. The brane in the AdS$_3$ has been studied in \cite{Giveon:2001uq}, \cite{Parnachev:2001gw}, \cite{Lee:2001gh}, \cite{Ponsot:2001gt}.

The study of the D-brane decay from the boundary CFT was initiated by Sen \cite{Sen:2002nu,Sen:2002in,Sen:2002an}. More detailed discussion on the D-brane decay can be found in \cite{Gutperle:2002ai}, \cite{Sen:2002vv}, \cite{Mukhopadhyay:2002en}, \cite{Strominger:2002pc}, \cite{Sen:2002qa}, \cite{Larsen:2002wc}, \cite{Gutperle:2003xf}, \cite{Maloney:2003ck}, \cite{Chen:2002fp}, \cite{Rey:2003xs}, \cite{Moeller:2002vx}, \cite{Sugimoto:2002fp}, \cite{Minahan:2002if}, \cite{Kluson:2002te}, \cite{Kluson:2002av}, \cite{Okuda:2002yd}, \cite{Aref'eva:2003qu}, \cite{Ishida:2003cj}, \cite{Lambert:2003zr}, \cite{Gaiotto:2003rm}, \cite{Demasure:2003av}, \cite{Kluson:2003rd}, \cite{Kluson:2003sh}, \cite{Nagami:2003yz}, \cite{Karczmarek:2003xm}, \cite{Kluson:2003sr}, \cite{Dasgupta:2003kk}, \cite{Dasgupta:2003xn}, \cite{Nagami:2003mr}.

Other topics which we have not covered in the main text include
 \begin{itemize}
 	\item The time dependent two dimensional background from the matrix model point of view, \cite{Alexandrov:2002fh}, \cite{Kostov:2002tk}, \cite{Alexandrov:2003uh}, \cite{Karczmarek:2003pv}.
 	\item The application of the recent development of the Liouville theory to the quantum Riemann surfaces, \cite{Verlinde:1990ua}, \cite{Teschner:2002vx}, \cite{Teschner:2003em}, \cite{Teschner:2003at} (see also \cite{Matone:1995tj}, \cite{Takhtajan:1993vt}, \cite{Matone:1994nf}).
 	\item The application to the \textit{three} dimensional quantum gravity, \cite{Krasnov:2001ui}, \cite{Klemm:2002ir}. 
	\item A nonperturbative formulation related to Gromov-Witten and
instanton theories, \cite{Matone:1997pz}, \cite{Bonelli:1996nu}.
\end{itemize}

\part{Supersymmetric Liouville Theory}
The Part II of this review deals with the supersymmetric extension of the Liouville theory. In this part $\alpha' =2$ unless otherwise stated.

\sectiono{$\mathcal{N}=1$ Super Liouville Theory}\label{sec:7}
In this section, we discuss the $\mathcal{N}=1$ supersymmetric extension of the  Liouville theory focusing on the bulk physics. The organization of this section is as follows.

In section \ref{7.1}, we review its basic setup and the properties of the theory as an SCFT. In section \ref{7.2}, we obtain various bulk structure constants of the theory including two-point functions (reflection amplitudes) and three-point functions. In section \ref{7.3}, we derive the torus partition function of the $\hat{c} =1$ (two dimensional type 0 noncritical) string theory.

\subsection{Setup}\label{7.1}
The $\mathcal{N}=1$ super Liouville theory can be introduced by the quantization of the two dimensional supergravity \cite{Polyakov:1981re}, \cite{Distler:1990nt} as we have reviewed in appendix \ref{b-1}. 

Its action is given by
\begin{eqnarray}
S &=& \frac{1}{2\pi} \int d^2z d\theta d\bar{\theta} D \Phi \bar{D}\Phi + 2i\mu \int d^2z d\theta d\bar{\theta} e^{b\Phi} \cr
  &=&  \frac{1}{2\pi} \int d^2z \left[\partial\phi\bar{\partial}{\phi} + \frac{QR\phi}{4} + \psi\bar{\partial}\psi+\bar{\psi}\partial\bar{\psi}-F^2 \right]+ 2i\mu \int d^2z \left[ibF e^{b\phi} + b^2 \psi\bar{\psi} e^{b\phi}\right]
\end{eqnarray}
See appendix \ref{a-2} for our superfield notation. After eliminating the auxiliary field $F$ by the equation of motion, we have
\begin{equation}
S_{int} = \int d^2z [2\pi \mu^2 b^2 :e^{b\phi}:^2 + 2i\mu b^2 \psi \bar{\psi} e^{b\phi}].
\end{equation}
Note that eliminating $F$ introduces contact terms in the correlator. In practice, however, we can simply neglect contact terms such as $:e^{b\phi}:^2$ in general correlators \cite{DiFrancesco:1991ss}, \cite{Green:1988qu}, \cite{Dine:1988vf}. Therefore as long as we stick to the superconformal properties at every step, we do not need to take care of the contact term problem for the time being (in any case, we do not have any chance to use the $:e^{b\phi}:^2$ term below).

 Although the relation $Q=b+1/b$ has been confirmed in the superfield formalism in appendix \ref{b-1}, we would like to check it again in the component formalism here. To see this, let us calculate the conformal dimension of the interaction term and confirm that it is a marginal perturbation. For the weight of the $\psi \bar{\psi} e^{b\phi}$ term to be $(1,1)$, we demand
\begin{equation}
1 = \frac{1}{2} + \frac{b}{2} \left(\frac{Q}{2} - \frac{b}{2}\right) 2,
\end{equation}
which leads to $Q=b+1/b$. Recall that we use $\alpha'=2$ notation here in contrast to the bosonic case in part I of this review. On the other hand, the dimension of $:e^{b\phi}:^2$  becomes
\begin{equation}
\Delta(:e^{b\phi}:^2) = 2\frac{b}{2}\left(\frac{Q}{2} - \frac{b}{2}\right) 2,
\end{equation}
which is also $(1,1)$ if we take $Q=b+1/b$. It is important to note that the normal ordering is necessary to obtain this result correctly.

The stress energy tensor and the superconformal current of this theory is given by
\begin{eqnarray}
T &=& -\frac{1}{2} (\partial \phi \partial \phi - Q\partial^2 \phi + \psi \partial \psi) \cr
G &=& i(\psi\partial \phi - Q\partial \psi),
\end{eqnarray}
and its central charge is $c = \frac{3}{2} \hat{c} = \frac{3}{2} + 3 Q^2$.

The superconformal algebra is written as 
\begin{eqnarray}
[L_m,L_n] &=& (m-n) L_{m+n} + \frac{c}{12}(m^3-m)\delta_{m,-n}\cr
\{G_r,G_s\}&=& 2L_{r+s} +\frac{c}{12}(4r^2-1)\delta_{r,-s} \cr
[L_m,G_r] &=& \frac{m-2r}{2} G_{m+r},
\end{eqnarray}
where $r,s$ takes an integer for the Ramond algebra and a half integer for the Neveu-Schwarz algebra.

Let us move on to the operator contents of the super Liouville theory.
For the primary states of this theory, we introduce the following notation. For NS-NS primary operators, we have just $V_\alpha \equiv e^{\alpha \phi}$ whose weight is $\Delta_\alpha = \frac{\alpha}{2}(Q-\alpha)$. To construct R-R operators, we first introduce the left and right spin operators (we follow the notation of \cite{Fukuda:2002bv}) as
\begin{eqnarray}
\psi(z)\sigma^{\pm}(0) &\sim& \frac{\sigma^{\mp}(0)}{\sqrt{2}z^{1/2}} \cr
\bar{\sigma}^{\pm}(0) \bar{\psi}(\bar{z}) &\sim& \frac{i\bar{\sigma}^{\mp}(0)}{\sqrt{2}\bar{z}^{1/2}}.
\end{eqnarray}
Then we define the left and right part of the R operator as follows
\begin{eqnarray}
\Theta^{\pm}_\alpha &=& \sigma^{\pm} e^{\alpha \phi}(z) \cr
\bar{\Theta}^{\pm}_\alpha &=& \bar{\sigma}^{\pm} e^{\alpha \phi}(\bar{z}).
\end{eqnarray}
The R-R operator itself is defined as
\begin{equation}
\Theta_\alpha^{\epsilon,\bar{\epsilon}} (z,\bar{z}) \equiv \sigma^{\epsilon} \bar{\sigma}^{\bar{\epsilon}} e^{\alpha \phi} (z,\bar{z}),
\end{equation}
where $\epsilon$ and $\bar{\epsilon}$ take either $+$ or $-$. Let us assume here that $\sigma^+$ and $\bar{\sigma}^+$ commute with fermions and  $\sigma^-$ and $\bar{\sigma}^-$ anticommute with fermions. Then $\Theta_\alpha^{\pm\pm}$ commutes with fermions and $\Theta_\alpha^{\pm\mp}$ anticommutes with fermions. 

In the free case ($\mu = 0$), the normalization of the spin field is given by
\begin{eqnarray}
\langle \sigma^{\pm\pm}(z)\sigma^{\pm\pm} \rangle_{free} &=& |z|^{-1/4} \cr
\langle \sigma^{\pm\mp}(z)\sigma^{\pm\mp} \rangle_{free} &=& i|z|^{-1/4}, 
\end{eqnarray}
while other correlators vanish because of their grassmannian nature. Then the dimension of the R-R operator $\Theta_\alpha^{\epsilon,\bar{\epsilon}} (z,\bar{z})$ is $\frac{1}{16} + \frac{\alpha}{2}(Q-\alpha)$.

As in the bosonic Liouville theory, the vertex operator which has the same dimension has to be identified. Anticipating the result of the next section, we have
\begin{eqnarray}
V_\alpha &=& D(\alpha) V_{Q-\alpha} \cr
\epsilon\bar{\epsilon} \Theta_{\alpha}^{\epsilon\bar{\epsilon}} &=& \tilde{D}(\alpha) \Theta_{Q-\alpha}^{-\epsilon-\bar{\epsilon}}, \label{eq:sref}
\end{eqnarray}
where the reflection amplitudes $D(\alpha)$ and $\tilde{D}(\alpha)$ can be obtained from the two-point functions
\begin{eqnarray}
\langle V_{\alpha_1}(z_1) V_{\alpha_2}(z_2) &=& |z_{12}|^{-4\Delta_{\alpha_1}} \left[\delta(p_1+p_2 ) + D(\alpha_1) \delta(p_1-p_2)\right] \cr
-i\langle \Theta_{\alpha_1}^{\pm \mp}(z_1)\Theta_{\alpha_2}^{\pm \mp}(z_2)\rangle =\langle \Theta_{\alpha_1}^{\pm \pm}(z_1)\Theta_{\alpha_2}^{\pm \pm}(z_2)\rangle &=& |z_{12}|^{-4\Delta_{\alpha_1} -1/4} \delta(p_1+p_2) \cr
i\langle \Theta_{\alpha_1}^{\pm \mp}(z_1)\Theta_{\alpha_2}^{\mp \pm}(z_2)\rangle =\langle \Theta_{\alpha_1}^{\pm \pm}(z_1)\Theta_{\alpha_2}^{\mp \mp}(z_2)\rangle &=& |z_{12}|^{-4\Delta_{\alpha_1} -1/4} \tilde{D}(\alpha_1)\delta(p_1-p_2), 
\end{eqnarray}
with the standard Liouville momentum notation $\alpha = \frac{Q}{2} + ip$.

In order to make use of the (supersymmetric version of) Teschner's trick, we need to know the degenerate operators in this theory. The supersymmetric version of the Kac formula \cite{Bershadsky:1985dq}, \cite{Friedan:1985rv}, \cite{Nam:1986qe} states that the degenerate operators appear if and only if the condition
\begin{equation}
\alpha_{m,n} = \frac{1}{2}(Q-mb- nb^{-1})
\end{equation}
satisfies. In addition, for R operators, $m+n$ must be odd and for NS operators $m+n$ must be even. The simplest and trivial example is $(1,1)$ degenerate operator, namely NS identity. The first nontrivial and most useful degenerate operator is the $(2,1)$ degenerate operator, which is given by the R operator $\Theta_{-\frac{b}{2}}^{\pm}$. It satisfies the following equation
\begin{equation}
\left(L_{-1} - \frac{2b^2}{2b^2+1} G_{-1}G_0 \right) \Theta_{-\frac{b}{2}}^{\pm} = 0.
\end{equation}
In the later sections, we heavily utilize this operator to derive the various structure constants as we have done in the bosonic Liouville theory.

\subsection{Bulk Structure Constants}\label{7.2}
In this section, we review the derivation of the supersymmetric counterpart of the DOZZ formula. That is to say, we derive the various three-point functions in the super Liouville theory. The formula has been proposed in \cite{Rashkov:1996jx} and \cite{Poghosian:1997dw} by using different methods. The former uses the analytic continuation of the pole structure as in the original derivation of the DOZZ formula which we have reviewed in section \ref{4.1} and the latter uses Teschner's trick which we have reviewed in section \ref{4.3}. Here, we follow the argument given in \cite{Fukuda:2002bv} in which Teschner's trick is used.

First, we derive the two-point functions (or reflection amplitudes). In this simple setup, Teschner's trick works quite well and we can learn the basic strategy which we will repeatedly use. For this purpose, we assume that the OPE between  the $(2,1)$ degenerate operator $\Theta^{++}_{-b/2}$ and a general primary operator yields just two primary operators
\begin{equation}
\Theta^{++}_{-b/2} V_{\alpha} \sim C_{+} \Theta_{\alpha-b/2}^{++} + C_{-} \Theta_{\alpha+b/2}^{--}.
\end{equation}
As in the bosonic Liouville theory, we can determine $C_{+}$ and $C_{-}$ via the free field method. We have $C_{+} = 1$ because there is no need for the perturbation. On the other hand, we need one super Liouville insertion to calculate $C_{-}$ as
\begin{eqnarray}
C_{-} &=& \langle \Theta^{--}_{Q-\alpha-b/2}(\infty) \Theta^{++}_{-b/2} (0) V_{\alpha}(1) \int d^2z [-2i\mu b^2 \psi \bar{\psi} e^{b\phi} (z)] \rangle_{free} \cr
     &=& \mu \pi b^2 \gamma(\frac{bQ}{2})\gamma(1-b\alpha) \gamma(b\alpha-\frac{bQ}{2}),
\end{eqnarray}
where we have used the formula (\ref{eq:kbnl}) and the free field correlator for the spin field\footnote{Note that if we had taken other spin components in the second term in the OPE, $C_{-}$ would be zero.}
\begin{equation}
\langle \sigma^{\pm\pm}(z_1) \sigma^{\mp\mp}(z_2) \psi\bar{\psi}(z_3)\rangle_{free} = i\langle \sigma^{\pm\mp}(z_1) \sigma^{\mp\pm}(z_2) \psi\bar{\psi}(z_3)\rangle_{free} =\frac{i}{2}|z_{12}|^{3/4}|z_{13}z_{23}|^{-1}.
\end{equation}
Similarly with the more general spin, we have
\begin{equation}
\Theta^{\epsilon\bar{\epsilon}}_{-b/2} V_{\alpha} \sim \Theta_{\alpha-b/2}^{\epsilon\bar{\epsilon}} + \epsilon\bar{\epsilon}C_{-} \Theta_{\alpha+b/2}^{-\epsilon-\bar{\epsilon}}.
\end{equation}
Substituting the reflection property (\ref{eq:sref}), we obtain the following functional relations
\begin{equation}
D(\alpha) = C_{-}(\alpha) \tilde{D}(\alpha + b/2), \ \ \ \ \ \tilde{D}(\alpha-b/2) = C_{-}(Q-\alpha) D(\alpha).
\end{equation}
With further assumptions, i.e. the unitarity and the $b\to b^{-1}$ duality, the solution of the above functional relations is uniquely fixed as 
\begin{eqnarray}
D(\alpha) &=& -(\mu\pi\gamma(\frac{bQ}{2}))^{\frac{Q-2\alpha}{b}} \frac{\Gamma(b(\alpha-\frac{Q}{2})) \Gamma(b^{-1}(\alpha - \frac{Q}{2}))}{\Gamma(-b(\alpha-\frac{Q}{2}))\Gamma(-b^{-1}(\alpha-\frac{Q}{2}))} \cr
\tilde{D}(\alpha) &=& (\mu\pi\gamma(\frac{bQ}{2}))^{\frac{Q-2\alpha}{b}} \frac{\Gamma(\frac{1}{2}+b(\alpha-\frac{Q}{2})) \Gamma(\frac{1}{2}+b^{-1}(\alpha - \frac{Q}{2}))}{\Gamma(\frac{1}{2}-b(\alpha-\frac{Q}{2}))\Gamma(\frac{1}{2}-b^{-1}(\alpha-\frac{Q}{2}))} .
\end{eqnarray}
We can obtain the same functional relations, hence the consistency check if we consider the degenerate OPE with general R-R operators. For example, we have
\begin{eqnarray}
-2 \Theta^{\pm\mp}_{-b/2} \Theta^{\mp\pm}_\alpha & = & 2i \Theta^{\pm\pm}_{-b/2} \Theta^{\mp\mp}_\alpha \cr
&\sim& \Tilde{C}_+(\alpha)\psi\bar{\psi} V_{\alpha-b/2} + \tilde{C}_- V_{\alpha+b/2}, \label{eq:ttope}
\end{eqnarray}
where the structure constant is given by
\begin{eqnarray}
\tilde{C}_+ &=& 1 \cr
\tilde{C}_- &=& 2i\mu\pi b^2\gamma(\frac{bQ}{2}) \gamma(\frac{1}{2}-b\alpha)\gamma(-\frac{b^2}{2}+b\alpha), \label{eq:ttoa}
\end{eqnarray}
which can be obtained from the perturbative calculation.

Now that we understand the basic strategy for using the supersymmetric extension of Teschner's trick, our next task is to obtain various three-point functions. The relevant three-point functions we would like to determine are
\begin{eqnarray}
\langle V_{\alpha_1} V_{\alpha_2} V_{\alpha_3} \rangle &=& C_{1}(\alpha_1,\alpha_2,\alpha_3), \ \ \ \ \ \ \langle V_{\alpha_1} \Theta_{\alpha_2}^{\pm\pm}\Theta_{\alpha_3}^{\mp\mp}\rangle \ = \ \Tilde{C_{1}}(\alpha_1;\alpha_2,\alpha_3) \cr
-\alpha^2_1\langle \psi\bar{\psi}V_{\alpha_1} V_{\alpha_2} V_{\alpha_3} \rangle &=& C_{2}(\alpha_1,\alpha_2,\alpha_3), \ \ \ \ \ \ \langle V_{\alpha_1} \Theta_{\alpha_2}^{\pm\pm}\Theta_{\alpha_3}^{\pm\pm}\rangle \ = \ \Tilde{C_{2}}(\alpha_1;\alpha_2,\alpha_3), 
\end{eqnarray}
where the obvious conformal factors are omitted.

To utilize Teschner's trick, we introduce an auxiliary four-point function $\langle V_{\alpha_3} V_{\alpha_2}\Theta^{++}_{-\frac{b}{2}} \Theta^{--}_{\alpha_1}\rangle$. Evaluating it in various channels and comparing the results, we obtain the functional relation between structure constants.

By using the $\Theta \Theta$ OPE as in (\ref{eq:ttope}), the four-point function is given by
\begin{eqnarray}
\langle V_{\alpha_3} V_{\alpha_2}\Theta^{++}_{-\frac{b}{2}} \Theta^{--}_{\alpha_1}\rangle &\sim& -\frac{1}{(\alpha_1-b/2)^2}\tilde{C}_+(\alpha_1) C_2(\alpha_1-b/2,\alpha_2,\alpha_3)|G_1(\alpha_1,\alpha_2,\alpha_3;\eta)|^2\cr
&+& \tilde{C}_-(\alpha_1) C_1(\alpha_1+b/2,\alpha_2,\alpha_3) |G_2(Q-\alpha_1,\alpha_2,\alpha_3;\eta)|^2,
\end{eqnarray}
where $\eta = \frac{z_{01}z_{23}}{z_{03}z_{21}}$ is the cross-ratio and the conformal factors are omitted for brevity. $G_i$ are determined from the superconformal property and the differential equation for the degenerate operator whose explicit form is given in the original literature \cite{Rashkov:1996jx}, \cite{Fukuda:2002bv} (essentially the hypergeometric function). $\tilde{C}_\pm(\alpha)$ are calculable from the free field method whose actual value is given in (\ref{eq:ttoa}). To obtain the functional relation, we use the crossing symmetry condition which fixes the relative factor in front of the $G_i$ independently of the OPE argument. Requiring these\footnote{We consider the $\eta\to 1/\eta$ transformation and require that the cross term vanish as in the bosonic case. See section \ref{4.3}.}, we obtain
\begin{equation}
P = -\frac{1}{(\alpha_1-b/2)^2} \frac{\tilde{C}_+(\alpha_1) C_2(\alpha_1-b/2,\alpha_2,\alpha_3)}{\tilde{C}_-(\alpha_1)C_1(\alpha_1+b/2,\alpha_2,\alpha_3)}. \label{eq:fr11}
\end{equation}
where the crossing symmetry condition states
\begin{eqnarray}
P &=& -(\frac{1}{2}+ibp_1)^{-2}\gamma(\frac{1}{2}-ibp_1)^2 \cr
& &\gamma(\frac{3}{4}+\frac{ib}{2}(p_1+p_2+p_3))\gamma(\frac{3}{4}+\frac{ib}{2}(p_1+p_2-p_3)) \cr
&\times&\gamma(\frac{3}{4}+\frac{ib}{2}(p_1-p_2+p_3))\gamma(\frac{3}{4}+\frac{ib}{2}(+p_1-p_2-p_3))\label{eq:fr12}
\end{eqnarray}
In addition, taking another auxiliary four-point function $\langle V_{\alpha_3} V_{\alpha_2}\Theta^{++}_{-\frac{b}{2}} \Theta^{++}_{\alpha_1}\rangle$ into consideration, we have another functional relation which states
\begin{equation}
P(p_1 \to -p_1) = \frac{1}{(2Q-2\alpha_1-b)^2} \frac{\tilde{C}_-(\alpha_1) C_2(\alpha_1+b/2,\alpha_2,\alpha_3)}{\tilde{C}_+(\alpha_1)C_1(\alpha_1-b/2,\alpha_2,\alpha_3)}.
\end{equation}
Now we can solve these functional relations by further requiring the $b \to b^{-1}$ duality. The results are
\begin{align}
C_1 &= \left(\mu\pi\gamma(\frac{bQ}{2})b^{1-b^2}\right)^{\frac{Q-\sum_i \alpha_i}b} \frac{\Upsilon'_{NS}(0) \Upsilon_{NS}(2\alpha_1) \Upsilon_{NS}(2\alpha_2) \Upsilon_{NS}(2\alpha_3)}{ \Upsilon_{NS}(\alpha_{1+2+3}-Q) \Upsilon_{NS}(\alpha_{1+2-3}) \Upsilon_{NS}(\alpha_{2+3-1}) \Upsilon_{NS}(\alpha_{3+1-2})} \cr
C_2 &= i\left(\mu\pi\gamma(\frac{bQ}{2})b^{1-b^2}\right)^{\frac{Q-\sum_i \alpha_i}b} \frac{2\Upsilon'_{NS}(0) \Upsilon_{NS}(2\alpha_1) \Upsilon_{NS}(2\alpha_2) \Upsilon_{NS}(2\alpha_3)}{ \Upsilon_{R}(\alpha_{1+2+3}-Q) \Upsilon_{R}(\alpha_{1+2-3}) \Upsilon_{R}(\alpha_{2+3-1}) \Upsilon_{R}(\alpha_{3+1-2})}, \label{eq:SDOZZ1}
\end{align}
where the notations $\Upsilon_{NS} (x) = \Upsilon(\frac{x}{2})\Upsilon(\frac{x+Q}{2}) $,  $\Upsilon_{R} (x) = \Upsilon(\frac{x+b}{2})\Upsilon(\frac{x+b^{-1}}{2}) $ and $\alpha_{1+2-3} \equiv \alpha_1+\alpha_2-\alpha_3$ have been introduced as in \cite{Fukuda:2002bv}.

Similarly we can take the $ V \Theta$ OPE to calculate the rest of the structure constants.
\begin{align}
\langle V_{\alpha_3} V_{\alpha_2}\Theta^{++}_{-\frac{b}{2}} \Theta^{--}_{\alpha_1}\rangle &\sim C_+(\alpha_2) \tilde{C_1}(\alpha_3;\alpha_2-b/2,\alpha_1)|H_1 (\alpha_1,\alpha_2,\alpha_3;1-\eta)|^2 \cr
&+ C_-(\alpha_2) \tilde{C}_2(\alpha_3;\alpha_2+b/2,\alpha_1) |H_2(\alpha_1,Q-\alpha_2,\alpha_3;1-\eta)|^2, 
\end{align}
The ratio of $\tilde{C_1}$ and $\tilde{C}_2$ is determined from the crossing symmetry, so we have a functional relation which is similar to (\ref{eq:fr11})(\ref{eq:fr12}). Solving these functional relations, we obtain 
\begin{align}
\tilde{C}_1(\alpha_1;\alpha_2,\alpha_3) &= \left(\mu\pi\gamma(\frac{bQ}{2})b^{1-b^2}\right)^{\frac{Q-\sum_i \alpha_i}b} \frac{\Upsilon'_{NS}(0) \Upsilon_{NS}(2\alpha_1) \Upsilon_{R}(2\alpha_2) \Upsilon_{R}(2\alpha_3)}{ \Upsilon_{R}(\alpha_{1+2+3}-Q) \Upsilon_{NS}(\alpha_{1+2-3}) \Upsilon_{R}(\alpha_{2+3-1}) \Upsilon_{NS}(\alpha_{3+1-2})} \cr
\tilde{C}_2(\alpha_1;\alpha_2,\alpha_3) &= \left(\mu\pi\gamma(\frac{bQ}{2})b^{1-b^2}\right)^{\frac{Q-\sum_i \alpha_i}b} \frac{\Upsilon'_{NS}(0) \Upsilon_{NS}(2\alpha_1) \Upsilon_{R}(2\alpha_2) \Upsilon_{R}(2\alpha_3)}{ \Upsilon_{NS}(\alpha_{1+2+3}-Q) \Upsilon_{R}(\alpha_{1+2-3}) \Upsilon_{NS}(\alpha_{2+3-1}) \Upsilon_{R}(\alpha_{3+1-2})}. \label{eq:SDOZZ2}
\end{align}

(\ref{eq:SDOZZ1}) and  (\ref{eq:SDOZZ2}) are the final form of the supersymmetric analog of the DOZZ formula.

\subsection{$\hat{c}=1$ Torus Partition Function}\label{7.3}
In this section, we calculate the torus partition function of $\hat{c} = 1$ super Liouville theory \cite{Douglas:2003up}. In the later section \ref{sec:9}, we calculate the same partition function from the matrix quantum mechanics and compare the results. 

Generally speaking, the fermionic string theory needs a proper GSO projection to become a modular invariant string theory. In the $\hat{c} = 1 $ string, the basic GSO projection is the type 0A and the type 0B theory, whose spectrum is given by
\begin{eqnarray}
\mathrm{type \ 0A :} \ \ \ \ \ \ (NS-,NS-) \oplus (NS+,NS+) \oplus (R+,R-) \oplus (R-,R+) \cr
\mathrm{type \ 0B :} \ \ \ \ \ \ (NS-,NS-) \oplus (NS+,NS+) \oplus (R+,R+) \oplus (R-,R-), 
\end{eqnarray}
where $\pm$ denotes allowed (left and right) fermion number $(-1)^F$. We could also imagine two dimensional type IIA and type IIB theory. However, they project out tachyon, so the super Liouville potential is impossible in these theories and we will not consider them here.

The space-time physics of these type 0 theories is as follows. In the type 0A theory, we have an NS-NS massless tachyon and two (nondynamical) R-R vectors.\footnote{The constant flux background is possible. However, as is shown \cite{Douglas:2003up}, only one kind of R-R vectors can yield the flux background. Moreover, the charge is quantized in this case, as we will see in section \ref{sec:9}. In this section, we only consider the case where no flux is introduced.} In the type 0B theory, we have an NS-NS massless tachyon and an R-R scalar (actually, it is the combination of selfdual and anti-selfdual scalar potential).

The world sheet description of this system is $\hat{c} = 1, b =1 $ super Liouville theory and the free $X$,$\chi$ superconformal theory which provides the (Euclidean) time direction.
Let us calculate the torus partition function with $X$ compactified on the circle of radius $R$. As has been done in the bosonic theory, we ignore the nontrivial density of states in the super Liouville theory and concentrate on the singular part which is proportional to $V_{\phi} = -\frac{1}{\sqrt{2}}\log\mu $.\footnote{The Liouville volume factor for the supersymmetric Liouville theory is $\sqrt{2}$ times that of the bosonic Liouville theory \cite{Douglas:2003up}. This $\sqrt{2}$ is necessary to obtain the correct matrix model result, but the physical nature of this factor is not clear to the author.} For the three even spin structures $(r,s) = (0,1),(1,0),(1,1)$, where $(r,s)$ represents the spin structure $(e^{i\pi r}, e^{i\pi s})$, the contribution from the free field $X$,$\chi$ is given by
\begin{equation}
Z(R,\tau \bar{\tau}) = \frac{R}{\sqrt{\alpha'}} \frac{1}{\sqrt{\tau_2}}|D_{r,s}|^2 \sum_{m,n} e^{-S_{m,n}}.
\end{equation}
where $D_{r,s}$ is the corresponding fermionic determinants divided by the square root of the scalar determinant and $\sum_{m,n} e^{-S_{m,n}}$ is given by the momentum and winding summation of the zero-mode (see (\ref{eq:1loop})). The contribution of the super Liouville theory is $\frac{V_\phi}{2\pi \sqrt{\tau_2}}|D_{r,s}|^2$, while the contribution from the superghost (with the Beltrami differential) is $\frac{1}{2\tau_2}|D_{r,s}|^{-4}$. Combining all these, and supplementing further $1/2$ because of the projection, we have
\begin{equation}
Z_{even} (R) = \frac{1}{2} \frac{R V_\phi}{2\pi \sqrt{\alpha'}} \int_{\mathcal{F}} \frac{d^2 \tau}{2\tau_2^2} \left( 3\sum_{m,n} e^{-S_{m,n}} \right) = \frac{3}{2} \frac{-\log\mu}{12\sqrt{2}} \left(\frac{R}{\sqrt{\alpha'}} + \frac{\sqrt{\alpha'}}{R} \right),\label{eq:even}
\end{equation}
where we have performed the summation over $(m,n)$ and integration over $\tau$ as has been done in the bosonic Liouville theory. Note that all oscillator summations are canceled in the two dimension.

 However, this is not the end of the story. We have to deal with the odd spin structure sector. In the higher dimensional theory, we have simply set its contribution zero because of the supertrace over the Ramond sector. In the two dimension, in contrast, we have $\frac{0}{0}$ owing to the superghost zero-mode. Anticipating the results, its contribution is
\begin{equation}
Z_{odd}(R) = \pm \frac{-\log\mu}{24\sqrt{2}}\left(\frac{R}{\sqrt{\alpha'}} - \frac{\sqrt{\alpha'}}{R}\right),\label{eq:odd}
\end{equation}
where $\pm$ corresponds to 0A and 0B respectively. Before giving its proof, we consider the physical implication first. The part of the partition function which is proportional to $1/R$ corresponds to the $T^2$ coefficient of the free energy of the theory.\footnote{The free energy and the string partition function is related via the relation: $F= -TZ$, $T=\frac{1}{2\pi R}$.} This part of the free energy (per unit volume) just counts the propagating degrees of freedom in the free theory, so the coefficient in the type 0B should be double that of the type 0A or bosonic theory as we can verify by adding (\ref{eq:odd}) to (\ref{eq:even}). In addition, we can check that the T-duality holds with this odd spin structure contribution. The whole partition function is
\begin{align}
Z_{0B} &= -\frac{\log\mu}{12\sqrt{2}}\left(\frac{R}{\sqrt{\alpha'}} + 2\frac{\sqrt{\alpha'}}{R}\right) \cr
Z_{0A} &= -\frac{\log\mu}{12\sqrt{2}}\left(2\frac{R}{\sqrt{\alpha'}} + \frac{\sqrt{\alpha'}}{R}\right). \label{eq:cparti}
\end{align}
We can easily see that $R \to \alpha'/R$ exchanges these theories with each other.

Now, we provide the proof of the (\ref{eq:odd}) following the method 1 in \cite{Douglas:2003up}. An alternative derivation, which we will not discuss here, is also given there. We have four bosonic zero-modes $\gamma, \bar{\gamma}, \beta,\bar{\beta}$ and four fermionic zero-modes $\chi,\bar{\chi}, \psi, \bar{\psi}$. The geometric nature of $\gamma,\bar{\gamma}$ is the covariantly constant spinor and that of $\psi,\bar{\psi}$ is the conformal Killing spinor on the torus. Therefore we expect that these zero-modes precisely cancel with each other. On the other hand, the $\beta$ and $\bar{\beta}$ zero-mode is related to the gravitino freedom which is not fixed by the superconformal gauge in this odd spin structure sector. As in the $b$ field case, we absorb this factor by inserting the left and right supercharges 
\begin{equation}
G(z) \bar{G}(\bar{w}) = (\chi \partial X(z) + \psi \partial \phi(z)-2\partial \psi(z))(\bar{\chi} \bar{\partial} X(\bar{w}) + \bar{\psi} \bar{\partial} \phi(\bar{w})- 2\bar{\partial} \bar{\psi}(\bar{w})).
\end{equation}
Note that the insertion of this operator does not depend on the position $z$ or $\bar{w}$ after integrating over the moduli.
Now that all fermion zero-modes are absorbed, we can calculate the odd spin structure contribution to the partition function. The remaining nontrivial contribution is proportional to
\begin{equation}
\langle \partial X (z) \bar{\partial}X(\bar{w})\rangle,
\end{equation}
where the path integral is $X$ field only. Setting $w=z$ and integrating over $z$, we have
\begin{equation}
Z_{odd} \propto \langle \int d^2z \partial X(z) \bar{\partial} X(\bar{z}) \rangle_X = R\frac{\partial}{\partial R}\langle 1\rangle_X \sim R\frac{\partial}{\partial R} \log\mu \left(\frac{R}{\sqrt{\alpha'}} + \frac{\sqrt{\alpha'}}{R}\right)=  \log\mu \left(\frac{R}{\sqrt{\alpha'}} - \frac{\sqrt{\alpha'}}{R}\right)
\end{equation}
The proportional factor can be fixed by the physical consideration as above.

\subsection{Literature Guide for Section 7}\label{7.4}
The super Liouville theory has been studied for decades and the earlier references on the subject include \cite{Arvis:1983tq}, \cite{Arvis:1983kd}, \cite{D'Hoker:1983zy}, \cite{Curtright:1984sb}, \cite{Babelon:1984nz}, \cite{Babelon:1985gw}, \cite{Zamolodchikov:1988nm}, \cite{Abdalla:1992hp}, \cite{Aoki:1992xr}, \cite{Dalmazi:1993sc}. As we have discussed in the main text, the basic structure constants which are analogous to the DOZZ formula have been proposed in \cite{Rashkov:1996jx} and \cite{Poghosian:1997dw}. The partition function for the super Liouville theory has been studied originally in \cite{Bershadsky:1991zs}, which is reviewed in \cite{Douglas:2003up}.

\sectiono{Branes in $\mathcal{N}=1$ Super Liouville Theory}\label{sec:8}
In this section, we review the properties of the branes in the $\mathcal{N}=1$ super Liouville theory. The organization of this section is as follows.

In section \ref{8.1}, we review the basic setup and discuss the classical boundary condition and the Ishibashi states which become the building blocks of the boundary states. In section \ref{8.2}, we derive the super ZZ brane states and discuss their properties. In section \ref{8.3}, we derive the super FZZT brane states and discuss their properties. In both cases, we mainly focus on the modular bootstrap method. We also use the conformal bootstrap method with Teschner's trick to check the results, but the detailed calculations are left to the original papers.

\subsection{Setup and Preparation}\label{8.1}
We consider the super Liouville theory on the world sheet with boundaries. For simplicity, we concentrate on the super Liouville theory on the disk. The bulk action is given by
\begin{align}
S &= \frac{1}{2\pi} \int_\Sigma d^2z \left(\partial \phi \bar{\partial}\phi + \frac{QR\phi}{4}+ \psi\bar{\partial}\psi+ \bar{\psi}\partial\bar{\psi} - F^2 \right)\cr
 &+\left( 2i\mu \int d^2z ibF e^{b\phi} + b^2 \psi \bar{\psi} e^{b\phi}\right).
\end{align}
This action is supersymmetric up to the boundary term: 
\begin{equation}
\delta S = -i\int_{-\infty}^\infty dx (\zeta L|_{\bar{\theta}} + \bar{\zeta}L|_\theta) |_{y=0},
\end{equation}
where we have taken the upper half plane as the world sheet with a boundary.
To preserve some supersymmetry, we should add the boundary action:
\begin{equation}
S_B = i\eta \int_{-\infty}^\infty dx L|_{\theta =\bar{\theta} =0},
\end{equation}
with $\eta = \pm1$. Then one of the combination of the supercharge $Q+ \eta\bar{Q}$ is conserved. If the boundary condition preserves the conformal invariance, the supercurrent satisfies $G+\eta\bar{G} = 0$.

For the total variation $\delta S + \delta S_B$ to vanish, two kinds of boundary condition can to be imposed.
\begin{enumerate}
	\item Dirichlet boundary condition (super ZZ brane). We set $\phi(y=0) = \infty, \psi (y=0) + \eta\bar{\psi}(y=0) = 0$. As in the bosonic Liouville theory, we should place the brane in the deep $\phi$ region so as to preserve the conformal invariance. In the brane language, the linear dilaton background enforces the brane to localize at the strong coupling region. 
	\item Neumann boundary condition (super FZZT brane). We set $\psi(y=0) - \eta \bar{\psi}(y=0) =0$. On the other hand, for the bosonic part, we could add the following boundary term \cite{Douglas:2003up} without breaking the superconformal invariance
	\begin{equation}
	S'_B = -\frac{1}{2} \int^{\infty}_{-\infty} dx \left[\gamma \partial_x \gamma -\mu' b\gamma (\psi + \eta \bar{\psi}) e^{\frac{b}{2}\phi} + (\mu')^2 :e^{\frac{b}{2}\phi}:^2\right].
	\end{equation}
	Thus we have a final expression for the boundary interaction of the supersymmetric extension of the FZZT brane as
	\begin{equation}
	S_B = \int_{-\infty}^\infty dx \left[\frac{QK\phi}{4\pi} + \gamma\mu_B b\psi e^{b\phi/2} \right],\label{eq:cFZZT}
	\end{equation}
	where we have reintroduced the boundary curvature and we have used the bulk fermion boundary condition. We treat the boundary fermion $\gamma$ as nondynamical and normalize it as $\gamma^2 =1$ here. In addition, contact terms such as $ :e^{\frac{b}{2}\phi}:^2$ are neglected for the same reason as in the bulk structure constants calculation.
\end{enumerate}

So far, we have dealt with the super Liouville theory with a boundary only classically. To make the precise quantum description of the branes, we use the boundary state formalism. Let us first review the supersymmetric extension of the Ishibashi states which are building blocks for constructing appropriate boundary states.

The conformal boundary condition states
\begin{equation}
(L_{n} - \bar{L}_{-n}) |B,\eta \rangle = (G_{r} + i\eta\bar{G}_{-r}) |B,\eta \rangle,
\end{equation}
where $r$ takes a value in integers for R states and half integers for NS states. The formal solution of this equation can be written from closed primary states as 
\begin{eqnarray}
|h, NS,\eta \rangle &=& |h,NS\rangle_L |h,NS \rangle_R + \mathrm{descendants} \cr
|h, R,\eta \rangle &=& |h,R^+\rangle_L |h,R^+ \rangle_R +i\eta |h,R^-\rangle_L |h,R^- \rangle_R+ \mathrm{descendants},
\end{eqnarray}
which satisfies the following defining property
\begin{eqnarray}
\langle h,NS,\eta |e^{i\pi \tau_c (L_0 + \bar{L}_0 -\frac{c}{12})}|k,NS,\eta'\rangle &=& \delta_{h,k} \mathrm{Tr}_{NS}[e^{2i\pi \tau_c (L_0-\frac{c}{24})} (\eta\eta')^F] \cr
\langle h,R,\eta |e^{i\pi \tau_c (L_0 + \bar{L}_0 -\frac{c}{12})}|k,R,\eta'\rangle &=& \delta_{h,k} \mathrm{Tr}_{R}[e^{2i\pi \tau_c (L_0-\frac{c}{24})} (\eta\eta')^F] .
\end{eqnarray}
Our conventions and the properties of these modular functions are reviewed in appendix \ref{a-4}. In the super Liouville theory, we take the following normalization
\begin{eqnarray}
\langle P,NS,\eta |e^{i\pi \tau_c (L_0 + \bar{L}_0 -\frac{c}{12})}|P',NS,\eta'\rangle &=& \delta(P-P')\chi_{P(NS)}^{\eta\eta'} (\tau_c)\cr
\langle P,R,\eta |e^{i\pi \tau_c (L_0 + \bar{L}_0 -\frac{c}{12})}|P',R,\eta'\rangle &=& \frac{1}{\sqrt{2}}\delta(P-P')\chi_{P(R)}^{\eta\eta'} (\tau_c).
\end{eqnarray}

Using these Ishibashi states, the Cardy states should be expressed as
\begin{eqnarray}
|h,NS,\eta\rangle &=& \int_{-\infty}^\infty dP\Psi_{NS}(-P,h,\eta)|P,NS,\eta\rangle \cr
|h,R,\eta\rangle &=& \int_{-\infty}^\infty dP \Psi_{R}(-P,h,\eta)|P,R,\eta\rangle,
\end{eqnarray}
where we have assumed $\Psi_{R,NS}(P,h)^\dagger = \Psi_{R,NS}(-P,h)$. These wavefunctions can be regarded as the unnormalized one-point functions on the disk as in the bosonic Liouville theory. In the following sections, we would like to derive the wavefunctions $\Psi(P)$ which correspond to the super FZZT and super ZZ branes \cite{Fukuda:2002bv}, \cite{Ahn:2002ev}. 
\subsection{Super ZZ Brane}\label{8.2}
We focus on the super ZZ brane first. This is because the $(1,1)$ boundary state has only identity operator in the spectrum and it is useful to construct the FZZT boundary state by utilizing the property of the $(1,1)$ boundary state.

For degenerate representations, the characters are given by the following subtraction,
\begin{eqnarray}
\chi_{m,n(NS)}^+ &=& \chi^+_{\frac{i}{2}(mb+nb^{-1}) (NS)} -  \chi^+_{\frac{i}{2}(mb-nb^{-1}) (NS)} \cr
\chi_{m,n(NS)}^- &=& \chi^-_{\frac{i}{2}(mb+nb^{-1}) (NS)} -  (-1)^{mn}\chi^-_{\frac{i}{2}(mb-nb^{-1}) (NS)} \cr
\chi_{m,n(R)}^+ &=& \chi^+_{\frac{i}{2}(mb+nb^{-1}) (R)} -  \chi^+_{\frac{i}{2}(mb-nb^{-1})(R)}. \label{eq:degem}
\end{eqnarray}
Note that the alternating sign in the second line is due to the $(-1)^F$ insertion.
With an analogy to the bosonic Liouville theory, we expect that the ZZ branes correspond to the degenerate representations. Thus our guess is that the open spectrum stretching between $(1,1)$ ZZ brane and $(m,n)$ ZZ brane consists of the $(m,n)$ degenerate state. In the language of the wavefunction, it becomes
\begin{eqnarray}
\Psi_{NS}(P,(m,n),\eta) \Psi_{NS} (-P,(1,1),\eta)  &=& \sinh(\pi P m b)\sinh(\pi Pn/b) \cr
\Psi_{R}(P,(m,n),\eta) \Psi_{R} (-P,(1,1),\eta)  &=& \sinh(\pi P m b+ \frac{i\pi mn}{2})\sinh(\pi Pn/b-\frac{i\pi mn}{2}) 
\end{eqnarray}
where we have modular transformed (\ref{eq:degem}) via the modular transformation formula (\ref{eq:modult}). To fix the normalization, we use the following reflection property
\begin{eqnarray}
\Psi_{NS} (P,h,\eta) &=& D(\frac{Q}{2} + iP) \Psi_{NS} (-P,h,\eta) \cr 
\Psi_{R} (P,h,\eta) &=& \eta\tilde{D}(\frac{Q}{2} + iP) \Psi_{R} (-P,h,\eta).
\end{eqnarray}
Note the $\eta$ dependent reflection property. This is because our reflection also reverses the spin and the R Ishibashi states are defined $\eta$ dependently as\footnote{We will not consider the mixed spin components for the time being. But we hope the extension in this case is not so different.}
\begin{equation}
|P,R,\eta\rangle = |\Theta^{++}_{Q/2-iP}\rangle + \eta|\Theta^{--}_{Q/2-iP}\rangle + \mathrm{descendants}.
\end{equation}
These equations determine the wavefunctions as
\begin{eqnarray}
\Psi_{NS}(P,(1,1),\eta) &=& \pi (\mu \pi\gamma(\frac{bQ}{2}))^{-iP/b} \left(iP\Gamma(-iPb) \Gamma(-iP/b)\right)^{-1} \cr
\Psi_{NS}(P,(m,n),\eta) &=& -\frac{1}{\pi} (\mu \pi\gamma(\frac{bQ}{2}))^{-iP/b}iP\Gamma(iPb) \Gamma(iP/b) \sinh(\pi P m b) \sinh(\pi P n/b) \cr
\Psi_{R}(P,(1,1),+) &=& \pi (\mu \pi\gamma(\frac{bQ}{2}))^{-iP/b} \left(\Gamma(\frac{1}{2}-iPb) \Gamma(\frac{1}{2}-iP/b)\right)^{-1} \cr
\Psi_{R}(P,(m,n),+) &=& \frac{1}{\pi} (\mu \pi\gamma(\frac{bQ}{2}))^{-iP/b}\Gamma(\frac{1}{2}+iPb) \Gamma(\frac{1}{2}+iP/b) \cr
& & \times \sinh(\pi P m b+\frac{i\pi mn}{2}) \sinh(\pi P n/b-\frac{i\pi mn}{2}).\label{eq:szzw}
\end{eqnarray}
It is impossible to obtain $\eta = -1$ R wavefunction from the above argument. As explained in \cite{Fukuda:2002bv}, \cite{Douglas:2003up}, this is because there is no $(1,1)$ state with $\eta = -1$.\footnote{Precisely speaking, what we mean here is $\eta = -\mathrm{sign}(\mu)$. That is because, if we change $\psi \to -\psi$ with $\bar{\psi}$ fixed, which changes $\eta \to -\eta$, we effectively cancel this redefinition by changing the sign of $\mu$ in the action \cite{Douglas:2003up}. In the following discussion, we fix the convention of the sign of the fermion by demanding $\mu >0$ implicitly.} Actually, we should have $(-1)^{m+n} = \eta$ to obtain the degenerate Cardy states. By solving the functional relation for the bulk one-point function as we will briefly review in the following, we can obtain the R wavefunction for $\eta = -1$
\begin{eqnarray}
\Psi_{R}(P,(m,n),-) &=& -i^{m+n}\frac{1}{\pi} (\mu \pi\gamma(\frac{bQ}{2}))^{-iP/b}\Gamma(\frac{1}{2}+iPb) \Gamma(\frac{1}{2}+iP/b) \cr
& & \times \sin(\pi m(\frac{1}{2}+iPb)) \sin(\pi n (\frac{1}{2} + iP/b)).
\end{eqnarray}
All above formulae are valid when $(-1)^{m+n} = \eta$.

\begin{figure}[htbp]
	\begin{center}
	\includegraphics[width=0.8\linewidth,keepaspectratio,clip]{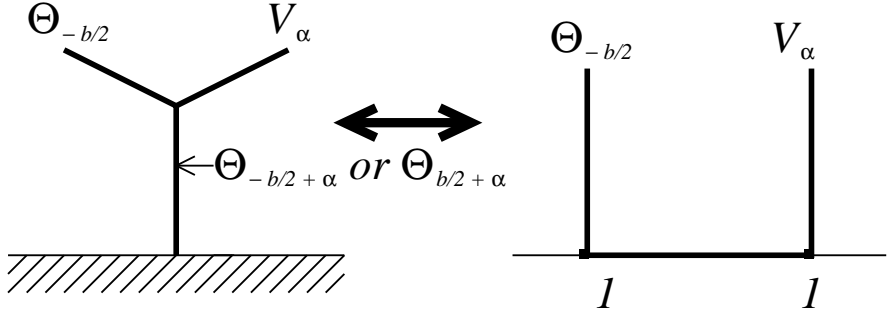}
	\end{center}
	\caption{The auxiliary two-point function can be calculated in two different ways.}
	\label{std}
\end{figure}

Let us now check the above one-point function, and more importantly, derive the  R wavefunction for $\eta= -1$ by the conformal bootstrap method. The basic strategy is the same as in the bosonic Liouville theory. We consider auxiliary bulk two-point functions $\langle V_{\alpha} \Theta^{\epsilon\epsilon}_{-b/2} \rangle $ and $\langle \Theta^{-\epsilon-\epsilon}_{-b/2} \Theta^{\epsilon\epsilon}_{\alpha} \rangle $. Using the notation:
\begin{eqnarray}
\langle V_\alpha (z)\rangle_\eta &=& U_{NS} (\alpha) |z-\bar{z}|^{-2\Delta_\alpha} \cr
\langle \Theta_\alpha^{\epsilon\epsilon}\rangle_\eta &=&  U_{R} (\alpha,\epsilon,\eta) |z-\bar{z}|^{-2\Delta_\alpha-\frac{1}{8}},
\end{eqnarray}
we can derive the following functional relations by evaluating them in the s-channel and t-channel and comparing their results (see figure \ref{std}, we also use the cluster decomposition argument and concentrate on the identity intermediate state contribution as in the bosonic ZZ brane case).
\begin{eqnarray}
U_{R}(-\frac{b}{2},\epsilon) U_{NS}(\alpha) &=& C_+(\alpha) U_{R}(\alpha -\frac{b}{2},\epsilon) \frac{\Gamma(b\alpha + \frac{1-b^2}{2})\Gamma(-b^2)}{\Gamma(b\alpha - b^2) \Gamma(\frac{1-b^2}{2})} \cr
& & + C_{-}(\alpha) U_{R}(\alpha +\frac{b}{2},-\epsilon) \frac{\Gamma(\frac{3+b^2}{2}-b\alpha)\Gamma(-b^2)}{\Gamma(1-b\alpha) \Gamma(\frac{1-b^2}{2})} \cr
U_{R}(-\frac{b}{2},-\epsilon) U_{R}(\alpha,\epsilon) &=& \eta \tilde{C}_+(\alpha) U_{NS}(\alpha-\frac{b}{2}) \frac{\Gamma(b\alpha -\frac{b^2}{2})\Gamma(-b^2)}{\Gamma(b\alpha - b^2-\frac{1}{2}) \Gamma(\frac{1-b^2}{2})} \cr
& & + \tilde{C}_{-}(\alpha) U_{NS}(\alpha +\frac{b}{2}) \frac{\Gamma(1+\frac{b^2}{2}-b\alpha)\Gamma(-b^2)}{2i\Gamma(\frac{1}{2}-b\alpha) \Gamma(\frac{1-b^2}{2})}, 
\end{eqnarray}
where the combinations of gamma functions come from the transformation from the t-channel conformal block to the s-channel conformal block.
With the $b\to 1/b$ duality condition, we can solve these functional relations if and only if $(-1)^{m+n} = \eta$ holds. The solutions are given by
\begin{eqnarray}
U_{NS}(\alpha,(m,n)) &=& \frac{\Psi_{NS}(-i(\alpha-\frac{Q}{2}),(m,n),\eta)}{\Psi_{NS}(\frac{iQ}{2},(m,n),\eta)} \cr
U_{R}(\alpha,(m,n),\eta) &=& \frac{\Psi_{R}(-i(\alpha-\frac{Q}{2}),(m,n),\eta)}{\Psi_{NS}(\frac{iQ}{2},(m,n),\eta)}
\end{eqnarray}
with the proper cluster decomposition normalization.

Let us comment on the peculiar feature that only one type of branes is available as the supersymmetric ZZ brane. In the type 0 setup, we naively expect two kinds of branes --- electric one and magnetic one (in addition to the usual brane and its anti-brane). The nature of this doubling comes from the fermionic boundary conditions $\psi_L = \psi_R$ or $\psi_L = -\psi_R$. Actually, the two types of D-branes are related by the transformation $\psi_L \to -\psi_L$. However, this transformation is no longer a symmetry once the Liouville superpotential is turned on. Therefore it is possible that one kind of brane becomes infinitely heavy and decouples from the theory. We will see the space-time interpretation of this result in the later section \ref{sec:9}.  

As in the bosonic theory, only $(1,1)$ brane has a semiclassical description  as the Poincar\'e disk metric. In this case, the spectrum of $\hat{c} = 1$ theory can be easily derived by suitability tensoring the $X,\chi$ SCFT boundary states and the superghost boundary states. The spectrum depends on the GSO projection (or the combination of the boundary states), but essentially the low energy spectrum consists of the tachyon with $m^2 = -\frac{1}{2\alpha'}$ and massless (nondynamical) gauge boson (provided they are not projected out). As in the bosonic theory, the movement of the ZZ-brane in the Liouville direction does not exist regardless of the GSO projection. We will see in section \ref{sec:9}, these branes play a significant role in the holographic dual description of the super Liouville theory.

\subsection{Super FZZT Brane}\label{8.3}
Next, we consider the quantum description of the super FZZT brane whose classical boundary condition is given by (\ref{eq:cFZZT}). Our natural guess is that the non-degenerate state in the open channel is related to the spectrum of the string stretching between the super FZZT brane and the super $(1,1)$ ZZ brane as in the bosonic Liouville theory. By comparing its wavefunction and disk one-point function, we will obtain the precise relation between the weight of the non-degenerate state and the boundary cosmological constant $\mu_B$.

With this assumption we can write the following equations for the FZZT wavefunction $\Psi(P,s,\eta)$ as
\begin{eqnarray}
\Psi_{NS}(P,s,\eta) \Psi_{NS}(-P,(1,1)),\eta) &=& \frac{1}{2}\cos(2\pi Ps) \cr
\Psi_{R}(P,s,\eta) \Psi_{R}(-P,(1,1),\eta) &=& \frac{1}{2}\cos(2\pi Ps) .
\end{eqnarray}
We can solve these equations from the knowledge of the $(1,1)$ wavefunctions. The solutions are
\begin{eqnarray}
\Psi_{NS}(P,s,\eta) &=& -\frac{1}{2\pi} (\mu\pi\gamma(\frac{bQ}{2}))^{-iP/b} i P \Gamma(iPb) \Gamma(iP/b) \cos(2\pi Ps) \cr
\Psi_{R}(P,s,+) &=&  \frac{1}{2\pi} (\mu\pi\gamma(\frac{bQ}{2}))^{-iP/b}\Gamma(\frac{1}{2}+iPb) \Gamma(\frac{1}{2}+iP/b) \cos(2\pi Ps) \label{eq:onep1}
\end{eqnarray}
As in the super ZZ brane case, we cannot obtain the R wavefunction for $\eta = -1$ by this method because there is no $(1,1)$ state. However, using the functional relation for the one-point functions on the disk, we will obtain the $\eta = -1$ R wavefunction as 
\begin{equation}
\Psi_{R}(P,s,-) =  \frac{1}{2\pi} (\mu\pi\gamma(\frac{bQ}{2}))^{-iP/b}\Gamma(\frac{1}{2}+iPb) \Gamma(\frac{1}{2}+iP/b) \sin(2\pi Ps). \label{eq:onep2}
\end{equation}
It is important to note that the FZZT branes have indeed a second kind of brane unlike the ZZ branes. The magnetic and electric branes have different R wavefunctions, hence the different RR charge.

Let us now check the above one-point function and derive the R wavefunction for $\eta= -1$ by the conformal bootstrap method. The basic strategy is the same as in the ZZ case. We calculate auxiliary two-point functions $\langle V_{\alpha} \Theta^{\epsilon\epsilon}_{-b/2} \rangle $ and $\langle \Theta^{-\epsilon-\epsilon}_{-b/2} \Theta^{\epsilon\epsilon}_{\alpha} \rangle $ in two ways and compare the results. The only difference from the super ZZ case is that we calculate the bulk-boundary structure constant by the perturbation method
\begin{eqnarray}
\Theta^{\epsilon\epsilon}_{-b/2} &\to& r_{+}\psi B_{-b} + r_- B_0 \cr
r_- &=& -\gamma b \mu_B \int dx\langle \Theta^{\epsilon\epsilon}_{-b/2}(i) \psi B_b(x) B_Q(\infty)\rangle_{free} = \pm \sqrt{2}\pi b \mu_B\Gamma(-b^2) \Gamma(\frac{1-b^2}{2})^{-2},
\end{eqnarray}
where the last sign is somewhat subtle and we determine it from the consistency conditions. 
Then the functional relation becomes
\begin{eqnarray}
r_- U_{NS}(\alpha,s) &=& C_+(\alpha) U_{R} (\alpha-\frac{b}{2},\epsilon,s)\frac{\Gamma(b\alpha+\frac{1-b^2}{2})\Gamma(-b^2)}{\Gamma(b\alpha-b^2)\Gamma(\frac{1-b^2}{2})} \cr
& &  + C_-(\alpha) U_{R}(\alpha+\frac{b}{2},-\epsilon,s)\frac{\Gamma(\frac{3+b^2}{2}-b\alpha)\Gamma(-b^2)}{\Gamma(1-b\alpha)\Gamma(\frac{1-b^2}{2})} \cr
\sqrt{2}\lambda^{-2} i r_- U_{R} (\alpha,\epsilon,s) &=& 2i\eta \tilde{C}_+(\alpha) U_{NS} (\alpha -\frac{b}{2},s) \frac{\Gamma(b\alpha-\frac{b^2}{2})\Gamma(-b^2)}{\sqrt{2}\Gamma(b\alpha-b^2-\frac{1}{2})\Gamma(\frac{1-b^2}{2})} \cr
& & + \tilde{C}_-(\alpha) U_{NS} (\alpha + \frac{b}{2},s)\frac{\Gamma(1+\frac{b^2}{2}-b\alpha)\Gamma(-b^2)}{\sqrt{2}\Gamma(\frac{1}{2}-b\alpha)\Gamma(\frac{1-b^2}{2})},
\end{eqnarray}
where $\lambda^{-2}$ is an ambiguity parameter from the four spin operator mixing we will not discuss here \cite{Fukuda:2002bv}. The consistency condition states it should be unity. Solving these equations with the usual duality constraint $b\to1/b$, we can obtain the one-point functions (\ref{eq:onep1}) and (\ref{eq:onep2}). Moreover, we have the following relation between the boundary parameter $s$ and the boundary cosmological constant $\mu_B$:
\begin{equation}
\mu_B = \left(\frac{2\mu}{\cos(\pi b^2/2)}\right)^{1/2} \cosh(\pi s b) 
\end{equation}
for $\eta = 1$, and 
\begin{equation}
\mu_B = \left(\frac{2\mu}{\cos(\pi b^2/2)}\right)^{1/2} \sinh(\pi s b) 
\end{equation}
for $\eta = -1$. It is interesting to note that the relation is different for the different choice of $\eta$.

In the literature \cite{Fukuda:2002bv}, another structure constant, namely the boundary two-point function is derived, too. As in the bosonic Liouville theory, we expect that the logarithmic derivative of two-point functions is related to the density of states on the strip, which we can directly calculate from the boundary states derived above. Indeed it has been shown this is satisfied up to a divergent term which needs a regularization (or we can consider the relative density of states as we have done in the bosonic case in section \ref{5.1.2}).

The derivation of the boundary two-point function is somewhat complicated but conceptually straightforward. As is the case with the bosonic Liouville theory, we make use of Teschner's trick to calculate OPE of the degenerate boundary operator and a general boundary operator; then we substitute the reflection ansatz into the OPE (recall that the reflection amplitude is nothing but the boundary two-point function), and we obtain functional relations between these two-point functions. Solving these functional relations, we obtain the desired two-point functions. An interested reader will consult the original paper for the further explanation and the actual calculation.

For the later purpose, we calculate here just one kind of density of states for illustration. We consider the partition function bounded by the following Cardy states
\begin{equation}
|\mathrm{FZZT},s,+\rangle = \int_{-\infty}^{\infty} dP \left[\Psi_{NS}(-P,s,+)|P,NS,+\rangle + \Psi_{R}(-P,s,+)|P,R,+\rangle\right].
\end{equation}
The partition function becomes
\begin{eqnarray}
Z &=& \int_{-\infty}^{\infty} dP \Psi_{NS}(P,s,+)\Psi_{NS}(-P,s,+) \chi^{+}_{P(NS)}(\tau_c) + \frac{1}{\sqrt{2}}\Psi_{R}(P,s,+)\Psi_{R}(-P,s,+) \chi^{+}_{P(R)}(\tau_c) \cr
&=& \int_{-\infty}^{\infty}dP\int_{-\infty}^{\infty}dP' e^{2\pi i PP'}(\Psi_{NS}(P,s,+)\Psi_{NS}(-P,s,+) \chi^{+}_{P'(NS)}(\tau_o) \cr
& &  + \frac{1}{\sqrt{2}}\Psi_{R}(P,s,+)\Psi_{R}(-P,s,+) \chi^{-}_{P'(NS)}(\tau_o) ) \cr
&=& \frac{1}{2} \int_{-\infty}^{\infty}dP' \left[\rho(P',s) (\chi_{P'(NS)}^+ + \chi_{P'(NS)}^-)(\tau_o) + \rho'(P',s) (\chi_{P'(NS)}^+ - \chi_{P'(NS)}^-)(\tau_o)\right],
\end{eqnarray}
where the open modular parameter $\tau_o$ is defined as $\tau_o = -1/\tau_c$, and the density of states here is given by
\begin{eqnarray}
\rho (P',s) &=& \int_{-\infty}^{\infty} dP \frac{e^{2\pi i PP'} \cosh(\pi Q P) \cos(2\pi s P)\cos(2\pi s P)}{\sinh(2\pi bP) \sinh(2\pi P/b)} \cr
\rho' (P',s) &=& \int_{-\infty}^{\infty} dP \frac{e^{2\pi i PP'} \cosh(\pi (b-b^{-1}) P) \cos(2\pi s P)\cos(2\pi s P)}{\sinh(2\pi bP) \sinh(2\pi P/b)}.
\end{eqnarray}
These density of states are divergent, so we must impose an infrared regularization. It is interesting to note that the R contribution to the density of states is very small because the $P \to 0$ divergence mostly comes from the NS-NS exchange. As a consequence, the open GSO projection is not imposed in this spectrum as opposed to the naive free field expectation.\footnote{In the free field case, both NS wavefunction and R wavefunction behave as $\delta(P)$, so $\rho'(P)$ actually vanishes and the open GSO projection is imposed for this Cardy state. Of course, in the free field, if we consider an ``unstable" brane whose wavefunction is only NS, the GSO projection is not imposed.} Note, in addition, even the infrared bulk singular behavior $P \to 0$ which corresponds to the $\log\mu$ term is not reproduced with this wavefunction.
\subsection{Literature Guide for Section 8}\label{8.4}
On the structure constants of the boundary super Liouville theory, the bulk one-point function has been studied in \cite{Fukuda:2002bv}, \cite{Ahn:2002ev}, and the boundary two-point function has been studied in \cite{Fukuda:2002bv}. The remaining structure constants which have not been obtained so far are the bulk boundary two-point function and the boundary three-point function. These cannot be obtained from the modular bootstrap and require a thorough investigation of the boundary bootstrap, which seems to be a hard work.

\sectiono{Matrix Model Dual}\label{sec:9}
In this section, we consider the matrix model dual of the supersymmetric Liouville theory whose continuum properties have been discussed in section \ref{sec:7}. The fundamental observation behind this duality lies in the holographic perspective we have discussed in section \ref{6.1}. We will construct the matrix model duals from the tachyon dynamics on the unstable ZZ branes. The organization of this section is as follows.

In section \ref{9.1}, we review the $\hat{c}=1$ matrix quantum mechanics. In subsection \ref{9.1.1}, we recall the possible branes in the two dimensional type 0 noncritical string theory. In subsection \ref{9.1.2}, we discuss the dual matrix model for the type 0B theory. In subsection \ref{9.1.3}, we discuss the dual matrix model for the type 0A theory. We will see that the matrix model reproduces the continuum Liouville results in some examples. At the same time, we obtain the matrix model description which cannot be handled from the world sheet NS-R formulation, e.g. the theory with the flux background.

In section \ref{9.2}, we review the $\hat{c}=0$ matrix models. In subsection \ref{9.2.1}, we discuss the unitary matrix model which is the dual matrix model for the type 0B theory. In subsection \ref{9.2.2}, we discuss the complex matrix model which is the dual matrix model for the type 0A theory. The comparison with the continuum super Liouville theory will be done in subsection \ref{9.2.3}.

\subsection{$\hat{c}=1$ Matrix Quantum Mechanics}\label{9.1}
In this section, we focus on the $\hat{c}=1$ super Liouville theory whose dual matrix theory should be a certain matrix quantum mechanics. First, we review the brane contents in the $\hat{c} =1$ super Liouville theory and then we present the dual matrix model proposal made in \cite{Takayanagi:2003sm}, \cite{Douglas:2003up}.
\subsubsection{Branes in $\hat{c} =1$ theory}\label{9.1.1}
We have briefly reviewed the space-time spectrum of the type 0A and type 0B theory in section \ref{7.3}, so let us now list the D-branes in this setup from the space-time point of view.

For the type 0B theory, we have two kinds of massless scalar fields (NS-NS and R-R) in the spectrum. The closed GSO projection determines the allowed boundary states in the theory. In the type 0B theory, any D-brane allows NS boundary states, but only the odd space dimension branes allow R boundary states. Therefore, for the ``stable brane" we have only the D1-brane in the Minkowski signature.\footnote{In the Minkowski signature, we only consider the time-like brane, i.e. we take the Neumann boundary condition for $X$.} It is charged under the R-R two-form which does not propagate. Therefore, we should consider only D1-$\bar{\mathrm{D}}$1 pair for the R-R tadpole cancellation (NS-NS tadpole can be canceled by the Fischler-Susskind mechanism if one wants). Here, by antibrane $\bar{\mathrm{D}}$, we mean the brane whose R wavefunction is reversed:
\begin{eqnarray}
|\textrm{D-brane}, \pm\rangle  &=& |NS,\pm\rangle + |R,  \pm\rangle\cr
|\mathrm{\bar{D}}\textrm{-brane}, \pm \rangle &=& |NS, \pm \rangle - |R, \pm \rangle . 
\end{eqnarray}
Therefore there are four kinds of brane. We distinguish $\pm$ by calling electric and magnetic respectively. Note that in the super Liouville theory they are actually different branes.

However, there is a small subtlety here. The natural candidate for this brane is the FZZT $\otimes$ $X$ Neumann brane in the exact super Liouville description. We recall that there are two kinds of FZZT R wavefunction corresponding to $\eta = \pm$, and whose cylinder divergence from the R-R exchange is much weaker than the usual NS-NS exchange divergence. Particularly, for $\eta = -$, the R-R tadpole cylinder diagram actually converges. The author is not quite sure whether the R-R tadpole (non)divergence should be canceled in this case or not. In addition, by saying ``stable brane", we usually mean that the (massless) tachyon on them are projected out by the open GSO projection. As we have seen in the last section, however, the GSO projection on them actually does not work because of the wavefunction difference for NS sector and R sector even if we take $\mu_B =0$. The true meaning of this observation is unclear to the author. Also there is a subtlety concerning the allowed D1-brane. See appendix \ref{b-6} for a further discussion on the discrepancy between the free field observation and the super Liouville treatment.

Now let us go on to the unstable brane whose dynamics we expect to be the dual description of the bulk system. We have D0-brane in the type 0B theory whose boundary state is given by the NS (1,1) ZZ brane (no R wavefunction). With $N$ such branes, we have $U(N)$ adjoint tachyons whose mass is $m^2 = -\frac{1}{2\alpha'}$. The conjecture made in \cite{Takayanagi:2003sm}, \cite{Douglas:2003up} is that the matrix quantum mechanics realized by these tachyons in the double scaling limit provides the holographic dual description of the type 0B super Liouville theory.

For the type 0A, we have one NS-NS tachyon and two R-R nondynamical vectors as the closed string spectrum. The closed GSO projection determines allowed branes in the type 0A theory. For stable branes, we have a D0-brane which is the source of the R-R one-form. Because of the GSO projection (recall that for the ZZ brane, the open GSO projection works properly), we have to consider D0-$\bar{\mathrm{D}}0$ pair to have nontrivial tachyon dynamics. The proposal made in \cite{Douglas:2003up} is that the quantum mechanics on $N$ D0 branes and $(N+q)$ $\bar{\mathrm{D}}0$ antibranes provides the holographic dual of the super Liouville theory with $q$ background R-R charge installed. Note that the ZZ brane in this case actually has one kind (no magnetic one) because the $(1,1)$ ZZ brane necessarily has $\eta = +$.

In the type 0A theory, we also have the ``unstable" D1-brane whose property is similar to the D1-$\bar{\mathrm{D}}$1 pair in the 0B case. The natural extension of the bosonic Liouville theory (\ref{eq:mmv}) leads to the matrix vector theory for this case. The study on this matrix-vector model, however, is only limited up to date.
\subsubsection{Type 0B matrix quantum mechanics}\label{9.1.2}
The type 0B matrix quantum mechanics is proposed as the dynamics of the tachyon matrix on the $N$ D0-branes. The mass of the tachyon is given by $m^2 = -1/2\alpha'$ and more importantly, the potential for the tachyon is symmetric under $M \to -M$. The symmetry under $M \to -M$ comes from the world sheet symmetry $(-1)^{F_L}$, where $F_L$ is the space-time fermion number, and the both sides of the Fermi sea should be filled with. Therefore, the matrix quantum mechanics for the type 0B theory is expected to be stable as opposed to the bosonic Liouville theory which is believed to be nonperturbatively unstable. In this way, we are naturally led to the two-cut Hermitian matrix model. 
\begin{equation}
S = \int_{-\infty}^{\infty} dt \mathrm{Tr} \left(\frac{1}{2} (D_t M)^2 + V(M)\right)
\end{equation}

In the double scaling limit, which corresponds to the Maldacena limit in the AdS/CFT language, we consider the following inverted harmonic oscillator potential
\begin{equation}
V(\lambda) = -\frac{1}{4\alpha'} \lambda^2,
\end{equation}
where we have diagonalized the Hermitian matrix. In the following we will check the proposal by comparing two quantities, which can be also derived in the continuum super Liouville formalism, the finite temperature free energy and the scattering amplitudes.

The finite temperature free energy (or the string partition function with compactified Euclidean $X$) can be obtained very easily \cite{Douglas:2003up} once we know the result for the matrix dual of the bosonic Liouville theory. The difference from the bosonic Liouville theory is simply the tachyon mass $-\frac{1}{\alpha'} \to -\frac{1}{2\alpha'}$ and the final division by two. Thus we can immediately borrow the result for the bosonic theory which is summarized in the Gross-Klebanov formula (\ref{eq:matz}) and replace $\alpha' \to 2\alpha'$ and not divide by two\footnote{As in the bosonic Liouville theory, since the adjoint tachyon is gauged, only the singlet representation contribute to the thermal path integral.}:
\begin{equation}
\frac{\partial\tilde{\rho}}{\partial\mu} = \frac{\sqrt{2\alpha'}}{\pi \mu} \mathrm{Im}\int_0^\infty dt e^{-it} \frac{t/(2\beta\mu\sqrt{2\alpha'})}{\sinh[t/2\beta\mu\sqrt{2\alpha'}]} \frac{t/(2\beta\mu R)}{\sinh[t/2\beta\mu R]},
\end{equation}
where the partition function itself is given by 
\begin{equation}
\frac{ \partial^2 Z}{\partial \mu^2} = \tilde{\rho}.
\end{equation}
For example, the torus partition function can be calculated as 
\begin{equation}
Z = -\frac{\log\mu}{12} \left(\frac{R}{\sqrt{2\alpha'}} + \frac{\sqrt{2\alpha'}}{R}\right),
\end{equation}
which reproduces the continuum result (\ref{eq:cparti}). 

Now let us go on to the comparison of the S matrix. The important point we should consider first is the leg pole factor which is the momentum dependent, hence non-local, wavefunction renormalization factor connecting the collective field operators with the spacetime tachyon operators. As in the bosonic Liouville theory, this factor can be calculated from the decaying amplitude of the unstable (ZZ) D0-brane. 

For the time part, we use the following boundary action \cite{Kutasov:2000aq}, \cite{Takayanagi:2003sm}, \cite{Douglas:2003up}:
\begin{equation}
\delta S = \lambda \int_{\partial \Sigma} \eta \psi e^{\frac{1}{2}X^0},
\end{equation}
where $\eta$ is the anticommuting boundary fermion with $\eta^2 = 1$. The NS-NS tachyon amplitude for the $X^0$ part is given by
\begin{equation}
A_t = \langle e^{iEX^0}\rangle = -\frac{1}{\sqrt{2\pi}}(\sqrt{2}\lambda \pi)^{-2iE}\frac{\pi}{\sinh(\pi E)}.
\end{equation}
On the other hand, the NS-NS one-point function for the $(1,1)$ ZZ brane is given by (see (\ref{eq:szzw})),
\begin{equation}
A_\phi = \langle e^{(iP+1)\phi}\rangle = - i(\mu_r)^{-iP} \frac{\Gamma(iP)\sinh(\pi P)}{\Gamma(-iP)}
\end{equation}
where we have introduced a renormalized cosmological constant $\mu_r = \mu_0 \gamma(1) \pi$. With the mass-shell condition $E=P$, we have
\begin{equation}
A_{NS} = A_t A_\phi = \mu_r^{-iP} \frac{\Gamma(iP)}{\Gamma(-iP)}e^{-iP\log(2\pi^2\lambda^2)},\label{eq:legns}
\end{equation}
which yields the NS-NS leg pole factor $e^{i\delta_{NS}}$ times the time delay factor $e^{-iP\log(2\pi^2\lambda^2)}$. We can check that the tachyon scattering in the continuum super Liouville formalism factorizes with this factor attached to every external leg as in the bosonic Liouville theory \cite{DiFrancesco:1992ud}. For the R-R sector, the only difference in the $X^0$ part is the factor $\cosh(\pi E)$ instead of $\sinh(\pi E)$. With the $(1,1)$ ZZ amplitude for the R-R wavefunction, we have \begin{equation}
A_R =  \mu_r^{-iP} \frac{\Gamma(\frac{1}{2}+iP)}{\Gamma(\frac{1}{2}-iP)}e^{-iP\log(2\pi^2\lambda^2)},\label{eq:legr}
\end{equation}
which is known to provide the correct leg pole factor for the R-R field.\footnote{This can be obtained by analyzing the R-R correlator when the Liouville momentum is conserved up to perturbative insertions as in the bosonic Liouville theory in section \ref{3.2.2}. The reason why zero momentum R-R field does not decouple is that the R-R zero-mode minisuperspace wave equation (Dirac equation) has a normalizable solution for the zero energy as opposed to the NS-NS sector. At the same time, we should recall that the $(-1/2,-1/2)$ vertex considered here is associated with the R-R field strength rather than the potential itself.}

The matrix model for the type 0B theory fills both sides of the Fermi sea. Therefore, the small disturbances of the Fermi sea on the left side $T_L(k)$ and the right side $T_R(k)$ are perturbatively independent. Actually the same thing holds in the super Liouville calculation on the sphere. This naturally leads to the following identification:
\begin{eqnarray}
T(k) &=& e^{i\delta_{NS}} \left(T_R(k) + T_L(k)\right) \cr
V(k) &=& e^{i\delta_R}  \left(T_R(k) - T_L(k)\right).
\end{eqnarray}
We can also see that the S matrix for the odd number $V(k)$ insertion is zero. In the original matrix model language, the corresponding vertex operator is proposed to be \cite{Takayanagi:2003sm}\footnote{Following \cite{Takayanagi:2003sm}, we use the convention $\alpha' = 1/2$ here in order to borrow the bosonic theory formulae in section \ref{sec:3}. In the end of the calculation we set $p, E \to 2p, 2E$ to obtain the $\alpha' = 2$ results.} 
\begin{eqnarray}
T(q) \sim P_{NS}(q) &=& \lim_{l\to 0} l^{-|q|/2} O_{NS}(q,l) \cr
V(q) \sim P_{R}(q) &=&  \lim_{l\to 0} l^{(-|q|+1)/2} O_{R}(q,l), 
\end{eqnarray}
where the NS and R puncture operators are 
\begin{eqnarray}
O_{NS} (q,l) &=& \int dx e^{iqx} \mathrm{Tr}[ e^{-l M(x)^2}] \cr
O_{R} (q,l) &=& \int dx e^{iqx} \mathrm{Tr} [M(x)e^{-l M(x)^2}].
\end{eqnarray}
Now, the central problem is whether this vertex shows the correct leg pole factor after fermionization and rebosonization as in the bosonic Liouville theory which has been treated in section \ref{3.2}. 

\begin{figure}[htbp]
	\begin{center}
	\includegraphics[width=0.8\linewidth,keepaspectratio,clip]{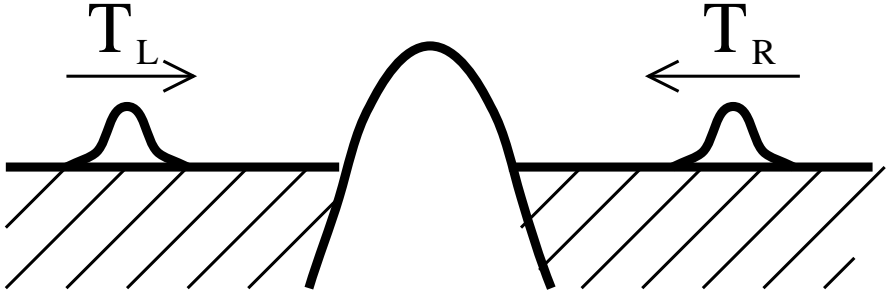}
	\end{center}
	\caption{The matrix model dual of the type-0B is given by the double scaling matrix quantum mechanics with both sides of the potential filled.}
	\label{0bm}
\end{figure}

Let us consider the NS operator first. The bosonization procedure is just the same as in the bosonic theory\footnote{Strictly speaking, the bosonization deals just the half part of the Fermi surface, so we should have two kinds of bosonization at the same time. For the NS part, the diagonal matrix is implicitly suppressed in the following formula.}, so we write the puncture operators as 
\begin{equation}
O_{NS} (q,l) = i\int dk F(k,l) k S(q,k),
\end{equation}
where 
\begin{equation}
F(k,l) = \int d\tau e^{-ly^2(\tau)} \cos (k\tau).
\end{equation}
Evaluating $F(k,l)$ with the classical trajectory $y(\tau) = \sqrt{2\mu} \cosh(\tau)$, we obtain
\begin{equation}
F(k,l) = \frac{e^{-l\mu}}{2} K_{ik/2} (l\mu).
\end{equation}
In the small $l$ limit, we find
\begin{equation}
F(k,l) \sim \frac{\pi}{4\sin(\frac{ik\pi}{2})}\left((l\mu/2)^{-ik/2}\frac{1}{\Gamma(-ik/2+1)} - (l\mu/2)^{ik/2}\frac{1}{\Gamma(ik/2+1)}\right). 
\end{equation}
As in the bosonic Liouville theory, we should add this factor to every external line of the Euclidean Feynman diagram. Picking up the pole for the integration over $k$, we obtain the leg factor
\begin{equation}
(l\mu/2)^{|q|/2} \Gamma(-|q|/2).
\end{equation}
Further multiplying by $1/\Gamma(|q|/2)$, which yields the precise operator mapping from the super Liouville theory to the matrix puncture operator, we have reproduced the full leg pole factor (\ref{eq:legns}) for the $S$ bosonic theory
\begin{equation}
\mu^{-iP} \frac{\Gamma(iP)}{\Gamma(-iP)},
\end{equation}
where we have translated the final result into the $\alpha'=2$ notation.

Now let us consider the RR sector. The corresponding wavefunction for the bosonization is given by
\begin{equation}
F(k,l) = \int d\tau y e^{-l y^2} \cos (k\tau).
\end{equation}
For $y= \sqrt{2\mu} \cosh \tau$, we find
\begin{equation}
F(k,l) = \frac{1}{2}\sqrt{\frac{\mu}{2}}e^{-l\mu} \left(K_{(1+ik)/2}(l\mu)+K_{(-1+ik)/2}(l\mu)\right).
\end{equation}
In the small $l$ limit, we have
\begin{equation}
F(k,l) = \sqrt{\frac{\mu}{2}}(l\mu/2)^{-1/2} \left[\frac{\pi}{4\sin((ik+1)\pi/2)}(l\mu/2)^{-ik/2} \frac{1}{\Gamma((-ik-1)/2+1)} + cc\right],
\end{equation}
which yields the following external leg factor for every external line of $S$ boson
\begin{equation}
\sqrt{\frac{\mu}{2}} (l\mu/2)^{(|q|-1)/2} \Gamma((1-|q|)/2).
\end{equation}
Returning to the Lorentzian signature and multiplying $1/\Gamma(1/2-iP)$, we have the correct leg pole factor (\ref{eq:legr})
\begin{equation}
\mu^{-iP} \frac{\Gamma(1/2+iP)}{\Gamma(1/2-iP)}.
\end{equation}
They are associated with the $S$ boson scattering amplitude which decouples in the zero momentum limit. To relate this amplitude with the continuum calculation of the $(-1/2,-1/2)$ picture R-R vertex insertion, further division by $2iP$ is needed because the string calculation deals with the field strength.

\subsubsection{Type 0A matrix quantum mechanics}\label{9.1.3}
The matrix dual for the type 0A theory can be obtained similarly from the double scaling limit of the matrix quantum mechanics on the unstable $N$ D0 branes and $(N+q)$ $\bar{\mathrm{D}}0$ antibranes system \cite{Douglas:2003up}. The open string theory on these branes is from the tachyon which transforms as the $U(N)\times U(N+q)$ bifundamental. The Lagrangian (in the double scaling limit) is given by 
\begin{equation}
L = \mathrm{Tr} \left[(D_t T)^\dagger D_t T + \frac{1}{2\alpha'} T^\dagger T\right],
\end{equation}
where $T$ is an $N\times (N+q)$ complex matrix. 

To solve this system, it is convenient to diagonalize $T$. However, the special care is needed because it is not an Hermitian matrix as has been exclusively studied so far. For the time being, let us concentrate on the $q=0$ case. Recall that any $N\times N$ complex matrix $T$ can be decomposed as 
\begin{equation}
T = V x W^\dagger, \label{eq:decom}
\end{equation}
where $V$ and $W$ are unitary matrices and $x$ is a real and nonnegative diagonal matrix. The change of variables to $V,x,W$ needs a Jacobian \cite{DiFrancesco:2002ru}, \cite{Morris:1991cq}, \cite{Dalley:1992kd}. To obtain this Jacobian $J$, we use the following trick (see e.g. \cite{Morris:1991cq}). First, we write the infinitesimal variation of (\ref{eq:decom}) as 
\begin{equation}
V^\dagger\delta T W = \delta S x - x \delta M + \delta x,
\end{equation}
where $\delta S = V^\dagger \delta V$ and $\delta M = W^\dagger \delta W$. On the other hand, with the flat measure for $\delta T, \delta M$ and $\delta S$, we have
\begin{eqnarray}
Const &=& \int d(\delta T) \exp\left[-\mathrm{Tr} \delta T^\dagger \delta T\right]\cr
&=& \int d(\delta S)d(\delta M) d^N(\delta x) J \exp \left[-\mathrm{Tr} (x\delta S - \delta M x - \delta x)(\delta S x -x\delta M + \delta x)\right]. \label{eq:sgaus}
\end{eqnarray}
Making the linear substitution 
\begin{equation}
\delta S_{\alpha\beta} \to \delta S_{\alpha\beta} + \frac{2\delta M_{\alpha\beta} x_{\alpha} x_{\beta}}{x_{\alpha}^2 + x_{\beta}^2},
\end{equation}
we end up with 
\begin{equation}
\mathrm{Tr} \delta T^\dagger \delta T = \sum_{\alpha\beta} \left( (x_\alpha^2 + x_\beta^2)\delta S_{\alpha\beta}\delta S_{\alpha\beta}^* + \delta M_{\alpha\beta} \delta M_{\alpha\beta}^* \frac{(x_\alpha^2 -x_\beta^2)^2}{(x^2_\alpha + x^2_\beta)}\right) + \sum_{\alpha}\delta x_{\alpha}^2.
\end{equation}
Carrying out the Gaussian integration (\ref{eq:sgaus}), we have
\begin{equation}
J = const \prod_{\alpha} x_\alpha \prod_{\alpha < \beta} (x_\alpha^2 -x_\beta^2)^2
\end{equation}
as the Jacobian. Note that the symmetric part of $\delta M$ is the redundant variable. 

Then we can write down the Schr\"odinger equation for the diagonal elements via Pauli's method. Setting $\chi = \prod_{i<j} (x^2_i - x_j^2) \psi(x)$, we obtain
\begin{equation}
\left(-\frac{1}{2x_i}\frac{\partial}{\partial x_i} x_i \frac{\partial}{\partial x_i} - \frac{1}{4\alpha'} x_i^2 \right)\chi = E\chi.
\end{equation}
The simple generalization for the $q \neq 0$ yields
\begin{equation}
\left(-\frac{1}{2x^{1+2q}_i}\frac{\partial}{\partial x_i} x_i^{1+2q} \frac{\partial}{\partial x_i} - \frac{1}{4\alpha'} x_i^2 \right)x^{-q}\chi = Ex^{-q}\chi. \label{eq:Phami}
\end{equation}
Or equivalently,
\begin{equation}
\left(-\frac{1}{2x_i}\frac{\partial}{\partial x_i}x_i\frac{\partial}{\partial x_i} + \frac{q^2}{2x_i^2}- \frac{1}{4\alpha'}x^2_i\right) \chi = E \chi.
\end{equation}

\begin{figure}[htbp]
	\begin{center}
	\includegraphics[width=0.6\linewidth,keepaspectratio,clip]{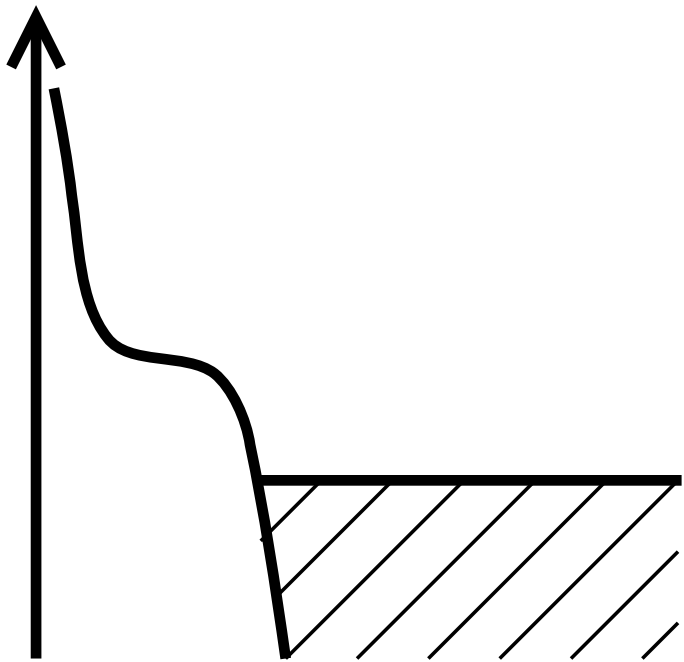}
	\end{center}
	\caption{The matrix model dual of the type-0A is given by the complex rectangular matrix quantum mechanics. In the eigenvalue basis, the potential is slightly deformed.}
	\label{0am}
\end{figure}

This quantum mechanical problem is exactly solvable and the finite temperature free energy is calculable.\footnote{To see this explicitly, it is convenient to redefine $\psi = x^{1/2} \chi$.} In \cite{Douglas:2003up}, it is given by
\begin{equation}
\frac{\partial \tilde{\rho}}{\partial \mu} = \frac{\sqrt{\alpha'/2}}{\pi\mu} \mathrm{Im} \int_0^\infty dt e^{-it} \frac{t/(2\beta\mu\sqrt{\alpha'/2})}{\sinh[t/(2\beta\mu\sqrt{\alpha'/2})]} \frac{t/2\beta\mu R}{\sinh[t/2\beta\mu R]} e^{-qt/(2\beta|\mu| \sqrt{\alpha' /2})} . \label{eq:densi}
\end{equation}
The one-loop contribution becomes
\begin{equation}
Z = \log\mu \left[-\frac{1}{24}\left(4\frac{R}{\sqrt{2\alpha'}} + \frac{\sqrt{2\alpha'}}{R}\right) + \frac{1}{2} q^2 \frac{R}{\sqrt{2\alpha'}}\right]
\end{equation}
The first term reproduces the continuum results (\ref{eq:cparti}) and the second term which is proportional to $q^2$ yields the background charge contribution to the partition function as was studied in \cite{Douglas:2003up} from the space-time point of view.
To derive the expression (\ref{eq:densi}), we need to know the zero energy density of states as in the bosonic Liouville theory (see section \ref{3.2}). The zero energy density of states is easily obtained by calculating the phase shift of the scattering event
\begin{equation}
\rho(e) = \frac{1}{2\pi} \frac{\partial \delta(e)}{\partial e}.
\end{equation}
See section \ref{4.2} for the explanation of this formula. By the suitable change of variable, the phase shift can be exactly calculable for the Hamiltonian (\ref{eq:Phami}). An interested reader should consult appendix B of \cite{Douglas:2003up}.

\subsection{$\hat{c}=0$ Matrix Model}\label{9.2}
In \cite{Klebanov:2003wg}, the matrix model dual for the $\hat{c} = 0$ (or more generally, some classes of the superconformal $\hat{c} <1$ matter coupled to the world sheet supergravity) string theory is proposed. The analogy to the last section's argument suggests that the type 0A theory is dual to the double scaling limit of the Hermitian matrix models with two-cut eigenvalue distribution (or the unitary matrix model) and the type 0B theory is dual to the rectangular complex matrix model. In this section, we exclusively deal with the simplest $\hat{c} = 0$ model, namely the pure 2D supergravity theory. For the multi critical matrix model which is related to the $\hat{c} <1$ theory coupled to minimal matters, we refer to \cite{Klebanov:2003wg}, \cite{Johnson:2003hy}.
\subsubsection{Unitary matrix model}\label{9.2.1}
The double scaling limit of the unitary matrix model is first solved in \cite{Periwal:1990gf,Periwal:1990qb}. First, we review its solution for the simplest case. The string partition function is given by
\begin{equation}
e^{Z} = \int dU \exp\left(-\frac{N}{\gamma} \mathrm{Tr} V(U + U^\dagger)\right).\label{eq:umat}
\end{equation}
where $U$ is unitary and the simplest potential which we consider here is given by $V(x) = x$. We diagonalize the unitary matrix as $ V^{^\dagger}U V= \mathrm{diag} (e^{\alpha_1},\cdots e^{\alpha_N})$. With these variables, the measure becomes
\begin{equation}
dU = \prod_{i} d\alpha_i\Delta(\alpha)\bar{\Delta}(\alpha), 
\end{equation}
where the Jacobian is given by
\begin{equation}
\Delta = \prod_{i<j} \left(\exp(i\alpha_i) - \exp(i\alpha_j)\right).
\end{equation}
Using the notation $z_i = e^{i\alpha_i}$, we can write the matrix integral (\ref{eq:umat}) as
\begin{equation}
e^Z = \int d\alpha \Delta(\alpha)\bar{\Delta}(\alpha) \exp\left[-\frac{N}{\gamma} \sum_i V(z_i + z_i^{-1})\right].
\end{equation}
As we have done in the Hermitian matrix model in section \ref{3.1}, we make use of the orthogonal polynomial technique. The orthogonal polynomials in this case are normalized as
\begin{equation}
\int_0^{2\pi} d\mu p_n(z) p_m(z^{-1}) = h_n \delta_{n,m},
\end{equation}
where $d\mu \equiv \frac{dz}{2\pi iz} \exp(-V(z+z^{-1})) $ and $p_m(z) = z^m + \cdots + R_{m-1}$. From the definition, we can derive the following recursion relation
\begin{equation}
p_{n+1}(z) = zp_n(z) + R_n z^n p_n(z^{-1}).
\end{equation}
By multiplying $p_n$ or $p_{n+1}$ and integrating over $d\mu$, we can easily obtain the relation $h_{n+1}/h_{n} = 1- R_n^2$. Then as in the Hermitian matrix model (see section \ref{3.1.1}), we can rewrite the partition function as 
\begin{equation}
e^Z \propto N! \prod_i (1-R_{i-1})^{N-i}.
\end{equation}
Therefore the remaining task is to obtain the expression for $R$ in the double scaling limit. In this limit the partition function is given by $-Z''=f^2$ where $f$ is proportional to $R$. Just as we have obtained (\ref{eq:cr}), we can use the partial integration to obtain
\begin{equation}
(n+1) (h_{n+1} - h_n) = -\int d\mu V'(z+z^{-1}) (1-1/z^2) p_{n+1}(z) p_n(z^{-1}).\end{equation}
With the simplest potential $V' = \frac{N}{\gamma}$, we obtain the discrete ``string equation":
\begin{equation}
\frac{\gamma}{N} (n+1) R_n^2 = R_n (R_{n+1} + R_{n-1}) (1- R_n^2).\label{eq:stri}
\end{equation}
From this expression, we can see that the critical point is given by $\gamma = 2$, $\frac{n}{N} = 1$ and $R_n = 0$. In order to take the double scaling limit, we introduce the following scaling ansatz:
\begin{eqnarray}
 \xi &=& \frac{n}{N} \cr
 y &=& (2-\gamma) N^{2/3} \cr
 z &=& (2-\gamma \xi) N^{2/3} \cr
 f &=& N^{1/3} R .
\end{eqnarray}
Fixing $y$ and taking the $N \to \infty$, $\gamma \to 2$ limit of (\ref{eq:stri}), we can obtain\footnote{The calculation is the same as has been done in section \ref{3.1.1}.} the following continuum string equation
\begin{equation}
- 2 z f + 2 f^3  = \partial^2_zf,
\end{equation}
which is what is called the Painlev\'e II equation with a rescaling of $y$ by a numerical factor.\footnote{This equation can be also obtained from the double cut Hermitian matrix model as it should be. The application to the gauge theory in the context of the Dijkgraaf-Vafa proposal has been recently studied in \cite{Fuji:2003wf}. However we should think seriously the existence of poles in the solution of the Painlev\'e equation in this case. This is because the Dijkgraaf-Vafa matrix model is a holomorphic matrix model, i.e. the coefficient of the potential is complex in nature. The physical meaning of this kind of singularity in the double scaling limit of the prepotential (which should correspond to the nonperturbative completion of the nonperturbative graviphoton correction) is not clear so far. Perhaps, there may be other branches of solution which is regular.} The solution of the equation yields the partition function after we integrate it twice.
 
The slight extension of the model has been considered in \cite{Klebanov:2003wg} (see also \cite{Brower:1993mn}). Without a derivation (see \cite{Klebanov:2003wg}), we just assume the following extended string equation:
\begin{equation}
f'' -\frac{1}{2}f^3 + \frac{1}{2}z f + \frac{q^2}{f^3} = 0.\label{eq:defst}
\end{equation}
We would like to see its consequence and the physical meaning of the $q^2$ term in the following.

The perturbative solution of (\ref{eq:defst}) \footnote{Since $z$ is related to the cosmological constant, hence the string coupling, this is the weak coupling expansion.} for large positive $z$ is given by
\begin{eqnarray}
u(z) = f^2(z)/4 = \frac{z}{4} + \left(q^2-\frac{1}{4}\right) \left[\frac{1}{2z^2} + \left (q^2-\frac{9}{4}\right) \left[-\frac{2}{z^5} + \cdots\right]\right],
\end{eqnarray}
whereas, for the large negative $z$, the solution is given by
\begin{eqnarray}
u(z) = f^2(z)/4 = \frac{|q|\sqrt{2}}{4|z|^{1/2}} - \frac{q^2}{4|z|^2} + \frac{5|q|(1+4q^2)\sqrt{2}}{64|z|^{7/2}} - \frac{q^2(7+8q^2)}{8|z|^5} \cdots,
\end{eqnarray}
The important point is that $u(z)$ is not an analytic function of $q$. This leads the authors of \cite{Klebanov:2003wg} to the conclusion that for $z <0$, the physical R-R vertex which turns on the $q$ perturbation does not exist. We will see in section \ref{9.2.3} its explanation from the super Liouville theory. 

The physical interpretation of this (second derivative of the) partition function is as follows. First, we note that the world sheet genus expansion is given by the term proportional to $|z|^{1-3(g+b/2)}$ with $g$ handles and $b$ boundaries. From this (and the analytic property of $q$ dependence which we have seen above), we can guess the physical interpretation of the $q$ deformation. When $ z \sim \mu > 0$, we have the closed string theory, where the R-R flux operator is inserted which produces the $q$ contribution. On the other hand, when $z \sim \mu <0$, we have the open string theory where $|q|$ yields the number of branes. With this interpretation, it is natural to consider that $q$ is quantized.

As is emphasized in \cite{Klebanov:2003wg}, we have actually the smooth description when we change $\mu$ from the large negative value to the large positive value. This is because the original matrix model (or the extended action) is smooth in the change of $z$ as long as $q>0$. However, the perturbative description is very different --- on one hand we have the open string theory, and on the other hand we have the closed string theory with the R-R flux. This suggests the geometric transition type of duality \cite{Gopakumar:1998ki}, \cite{Klebanov:2000hb}, \cite{Maldacena:2000yy}, \cite{Vafa:2000wi}. We have encountered the exact description of the sophisticated duality of the string theory in the Liouville - matrix model setup again!

While the last statement holds perturbatively in $q$, the large $q$ 't Hooft limit also has been studied in \cite{Klebanov:2003wg}. Here by the 't Hooft limit, we mean $q \to \infty$ and $z \to \pm \infty$ with $t = q^{-2/3} z$ fixed. In this limit, the second derivative of the partition function behaves as 
\begin{eqnarray}
v(t) &=& \frac{\sqrt{2}}{|t|^{1/2}} - \frac{1}{|t|^2} + \frac{5\sqrt{2}}{4|t|^{7/5}}+ \cdots,  \ \ \ \ \ \ t<0 \cr
v(t) &=& t + \frac{2}{t^2} - \frac{8}{t^5} + \frac{56}{t^8} - \cdots \ \ \ \ \ t>0.
\end{eqnarray}
This expression has the obvious geometric transition type interpretation. For negative $t$, we have D-branes and the 't Hooft limit is given by the planer graph. For positive $t$, we have no D-branes, but instead we have R-R flux background. It is important to note that the latter expansion is based on the R-R flux $t^{-3}$ expansion and not the world sheet (loop) expansion.

\subsubsection{Complex matrix model}\label{9.2.2}
The natural candidate for the matrix model dual of the type 0A $\hat{c} = 0$ string theory, which is equivalent to the 2D pure supergravity, is the double scaling limit of the rectangular complex matrix model. In this subsection, we review its solution and the physical meaning of the result \cite{Klebanov:2003wg}.

The string partition function is given by the following matrix integral \cite{Morris:1990bw}, \cite{Dalley:1992br,Dalley:1992vr}, \cite{Lafrance:1994wy}, \cite{DiFrancesco:2002ru}, \cite{Klebanov:2003wg}:
\begin{eqnarray}
e^Z = \int dM dM^\dagger \exp\left[-\frac{N}{\gamma} V(MM^\dagger)\right],
\end{eqnarray}
where $M$ is given by the complex $N\times (N+q)$ complex matrix.
The diagonalization of this matrix and the associated Jacobian have been reviewed in section \ref{9.1.3}. Setting $y = x^2$, we have
\begin{equation}
e^Z = \prod_{i=1}^N \int_0^\infty dy_i y_i^q e^{-\frac{N}{\gamma} V(y_i)} \Delta(y)^2,
\end{equation}
where $\Delta(y) = \prod_{i<j}(y_i-y_j)$ is the usual Vandermonde determinant. Then we define the orthogonal polynomials $P_n$ as 
\begin{equation}
\int d\mu P_n P_m = \int_0^\infty dy y^q e^{-\frac{N}{\gamma}V} P_n P_m = \delta_{nm},
\end{equation}
which begins like $P_n = y^n + \cdots$. For simplicity, we show the calculation for the $q = 0$ case in the following. First, from the definition of the orthogonal polynomials, we have the recursion relation
\begin{equation}
yP_n = P_{n+1} + s_n P_n + r_n P_{n-1} \label{eq:recre}
\end{equation}
with $r_n = h_n / h_{n-1}$. Next, by using the partial integration we can derive the following relation:
\begin{eqnarray}
h_n^{-1} \int d\mu P_n V' P_n &=& \Omega_n \cr
h_{n-1}^{-1} \int d\mu P_{n-1} V' P_n &=& x_n +\tilde{\Omega}_n,
\end{eqnarray}
where we have introduced $x_n$, $\Omega_n$ and $\tilde{\Omega}_n$ as
\begin{eqnarray}
x_n &\equiv & \frac{\gamma n}{N} \cr
\Omega_n &\equiv & \frac{\gamma}{N} \frac{P_n(0)^2}{h_n} e^{-\frac{N}{\gamma} V(0)} \cr
\tilde{\Omega}_n &\equiv & \frac{\gamma}{N} \frac{P_{n-1}(0)P_{n}(0)}{h_{n-1}} e^{-\frac{N}{\gamma} V(0)}.
\end{eqnarray}
From its definition, we immediately obtain
\begin{equation}
r_n \Omega_n \tilde{\Omega}_{n-1} = \tilde{\Omega}_n^2.
\end{equation}
In addition, by multiplying $V'P_n$ to (\ref{eq:recre}) and integrating over $d\mu$, we have
\begin{equation}
\tilde{\Omega}_n + \tilde{\Omega}_{n+1} = - s_n\Omega_n.
\end{equation}
With the simplest potential $V(y) = -y + \frac{y^2}{2}$, which corresponds to the pure supergravity theory, we have the following string equations:
\begin{eqnarray}
r_n - x_n &=& \tilde{\Omega}_n \cr
\tilde{\Omega}_n + \tilde{\Omega}_{n+1} &=& - (\Omega_n +1 )\Omega_n \cr
r_n \Omega_n \Omega_{n-1} &=& \tilde{\Omega_n}^2.
\end{eqnarray}
As has been shown in the literature \cite{Klebanov:2003wg}, the double scaling ansatz is given by
\begin{equation}
r = \frac{1}{4} - \epsilon (\tilde{u}+\tilde{z}/2), \ \ \ x = \frac{1}{4} - \epsilon \tilde{z}, \ \ \ \Omega = \epsilon \hat{\Omega}, \ \ \ \tilde{\Omega} = - \epsilon \tilde{R} , \ \ \ N = \gamma\epsilon^{-3/2}.
\end{equation}
With these variables, the continuum limit of the string equations and the partition function become
\begin{eqnarray}
0  &=& 8\tilde{u} \tilde{R}^2 - \frac{1}{2}\tilde{R}\tilde{R}'' + \frac{1}{4}(\tilde{R}')^2\cr
\tilde{R} &=& \tilde{u} - \tilde{z}/2 \cr
Z'' &=& - 4 \tilde{u}.
\end{eqnarray}

The extension for nonzero $q$ is straightforward, and the result is
\begin{eqnarray}
\frac{q^2}{4}  &=& 8\tilde{u} \tilde{R}^2 - \frac{1}{2}\tilde{R}\tilde{R}'' + \frac{1}{4}(\tilde{R}')^2\cr
\tilde{R} &=& \tilde{u} - \tilde{z}/2 \cr
Z'' &=& - 4 \tilde{u}.
\end{eqnarray}
It is convenient to rescale the parameter as $ u = 2^{5/3} \tilde{u}$, $z = 2^{2/3} \tilde{z}$ and $R = 2^{5/3} \tilde{R}$. Substituting $u(z) = f(z)^2 + z$, we have
\begin{equation}
\partial_z^2 f - f^3 - zf + \frac{q^2}{f^3} = 0.
\end{equation}
Surprisingly, up to the trivial rescaling and important sign in front of $z$, this is just the Painlev\'e II equation derived in the last subsection for the matrix model dual for the type 0B theory. This leads the authors of \cite{Klebanov:2003wg} to the conjecture that the $\hat{c} = 0$ super Liouville theory has the duality: the type 0A with $\mu > 0$ is dual to the type 0B with $\mu <0$ and the type 0A with $\mu <0$ is dual to the type 0B with $\mu >0$. Typically, the unitary matrix model and the complex matrix model have a corresponding model with each other. Therefore we expect the similar duality holds for the general $\hat{c} <1 $ model.

At the same time, this duality strongly suggests the D-brane like contribution of the quantity $q$ in the partition function because we regard the double scaling limit of the complex matrix theory as the super Liouville theory with $q$ extra D-branes background.
 
\subsubsection{Comparison with super Liouville theory}\label{9.2.3}
So far, we have solved the unitary and complex matrix model without any reference to the underling super Liouville theory. Here, we list some of the comparisons made in \cite{Klebanov:2003wg} which confirms the matrix model proposal for the super Liouville theory.

\begin{itemize}
	\item The first point is the physical state for the one dimensional noncritical fermionic string theory. To describe the theory in the super Liouville language, we take $ Q = \frac{3}{\sqrt{2}}$ and $b = \frac{1}{\sqrt{2}}$. The cosmological constant scales as $\mu^{3(2-2g)/2}$ in the partition function. The physical vertex operators are as follows. For the NS-NS sector, we have $e^{b\phi}$ in the $(-1,-1)$ picture. For the R-R sector, the naive free field guess is that we have $V_{\pm} = \sigma^{\pm} e^{Q\phi/2}$ in the $(-1/2,-1/2)$ picture.\footnote{We have used the two dimensional diagonal representation here. See appendix \ref{b-6} for the zero-mode algebra. The dimension of the operator is obtained in the following manner: the ghost and superghost contribute $-\frac{5}{8}$ and the spin field contribute $\frac{1}{16}$, finally the Liouville part is given by $\frac{1}{2}\frac{Q}{2} (Q-\frac{Q}{2}) = \frac{9}{16}$, hence the total dimension is zero.} In the nonzero energy sector, the both $\pm$ vertices appear, but in the zero energy sector considered here, the one of the zero-mode wavefunction does not damp in the $\phi \to \infty$ region in the minisuperspace approximation. The behavior is correlated with the sign of $\mu$. Thus for $\mu <0$, only $V_{+}$ is acceptable while for $\mu >0$ only $V_{-}$ is acceptable, which means that in the type 0A theory there is one R-R operator for $\mu<0$ and no operator for $\mu >0$. In the type 0B, the situation is reversed. This fact is compatible with the claim that $\mu \to -\mu$ exchanges 0A and 0B theory.
	
	\item In \cite{Klebanov:2003wg}, the connection between the FZZT brane and the matrix resolvent has been studied. The dependence of the boundary and bulk cosmological constant of the FZZT brane reveals that the $\eta= -1$ brane in the type 0B theory is related to the resolvent of the two cut matrix model. On the other hand, the FZZT brane with $\eta= 1$ is related to the resolvent of the complex matrix model. See also \cite{Seiberg:2003nm} for the interpretation of the ZZ and FZZT branes in the super Liouville theory coupled to the super minimal models from the Riemann surface viewpoint.
	\item The torus partition functions have been calculated and compared with the matrix model prediction \cite{Klebanov:2003wg}. Although the odd spin structure partition function is difficult to evaluate, the other calculable partition functions match with each other.

\end{itemize}
\subsection{Literature Guide for Section 9}\label{9.3}
The matrix quantum mechanics dual of the $\hat{c} =1$ super Liouville theory (two dimensional fermionic string theory) has been proposed in \cite{Takayanagi:2003sm}, \cite{Douglas:2003up}. In addition to the check we have reviewed in the main text, the ground ring structure has been studied in \cite{Douglas:2003up}. The earlier references on the BRST cohomology and the ground ring for the two dimensional fermionic string theory include \cite{Itoh:1992ix2}, \cite{Bouwknegt:1992am}, \cite{Aldazabal:1994ae}, \cite{Itoh:1992ix}.

The earlier studies on the $\hat{c} <1$ unitary matrix model (or the double cut Hermitian matrix model) can be found in \cite{Gross:1980he}, \cite{Periwal:1990gf}, \cite{Dalley:1992br}, \cite{Periwal:1990qb}, \cite{Nappi:1990bi}, \cite{Crnkovic:1991ms}, \cite{Crnkovic:1992wd}, \cite{Hollowood:1992xq}, \cite{Brower:1993mn}, some of which (e.g. \cite{Crnkovic:1992wd}) connect the multiple critical behavior with the Zakharov-Shabat hierarchy. Since we have focused on the simplest example, we have not emphasized the point in the main text, but this viewpoint is useful when we extend the string equation to the multicritical theory which is related to the super Liouville theory coupled to the super minimal model. 

On the other hand, the earlier studies on the complex rectangular matrix model can be found in \cite{Dalley:1992vr,Dalley:1992yi}, \cite{Morris:1990bw}, \cite{Lafrance:1994wy}. More recently, extending the type 0A string interpretation of this model proposed in \cite{Klebanov:2003wg}, \cite{Johnson:2003hy} has argued the more general gravitational coupling to the super minimal models and has studied particularly the tricritical Ising model coupled to the world sheet supergravity  (super Liouville theory) in detail.
 
\sectiono{Applications}\label{sec:10}
In this section we review various applications of the $\mathcal{N}=1$ super Liouville theory and its matrix model dual. The organization of this section is as follows.

In section \ref{10.1}, we discuss the scattering amplitudes in the R-R background, which cannot be easily obtained from the world sheet theory. In section \ref{10.2}, the unitarity of the type 0B S matrix will be discussed, where we will find that the $\theta$ vacua like degeneration emerges. In section \ref{10.3}, the hole interpretation of the matrix model from the boundary states is reviewed. The hole states are interpreted as the boundary states with an extra minus sign which has a negative energy if the plus sign is assigned. In section \ref{10.4}, we review the nonperturbative generation of the R-R potential, which stabilizes the moduli.

\subsection{Scattering in R-R Background}\label{10.1}
Matrix model dual prediction enables us to calculate the scattering amplitudes in an R-R background for the type 0A theory \cite{Kapustin:2003hi}. The type 0A matrix model has one integer parameter $q$ which represents the number of extra stable D0-branes (hence, the remaining R-R flux in the double scaling limit \textit{closed} theory). After the diagonalization of the complex rectangular matrix, the effective potential for the eigenvalue becomes
\begin{equation}
V(\lambda) = -\frac{\lambda^2}{2\alpha^2} + \frac{q^2-\frac{1}{4}}{2\lambda^2}, \label{eq:popo}
\end{equation}
where we have introduced $\alpha = \sqrt{2\alpha'}$ to relate the potential to the bosonic convention. It is important to note that the eigenvalue is restricted to $\lambda >0$. 

By the way, this potential reminds us of the bosonic Liouville theory with an unstable D-brane on top of the potential (see section \ref{6.4}), where we have the same potential. There we have observed that the scattering amplitudes are not modified at the tree level or the disk level. Since the potential is the same, we expect the same result.

The only information needed to obtain the S matrix in the MPR formalism (see section \ref{3.2.3}) is just the reflection amplitude of the potential (\ref{eq:popo}). Since the theory is solvable quantum mechanically (see section \ref{9.1.3} or \cite{Demeterfi:1993cm}), we can find that the amplitude is given by
\begin{equation}
R(p) =\left|\frac{{q}^2-\frac{1}{4}+\alpha^2\mu^2}{4}\right|^{-\alpha |p|/2} \frac{\Gamma(\frac{1}{2}+\frac{|q|}{2}-\frac{i}{2}\alpha \mu + \frac{1}{2}\alpha |p|)}{\Gamma(\frac{1}{2}+\frac{|q|}{2}+\frac{i}{2}\alpha \mu -\frac{1}{2}\alpha |p|)}. \label{eq:refla}
\end{equation}
(We have considered the Euclidean amplitudes here. To obtain the Minkowski amplitudes we simply substitute $|p| \to i\omega$). In the following we set $\alpha = 1$.

Following the MPR rule, we can calculate any S matrix via the reflection amplitude (\ref{eq:refla}). For example, the three-point amplitude is given by
\begin{equation}
A_3(p_1,p_2;-p) = -i\left(\int_{p_1}^{p} R(p-x)R^*(x) dx + \int_{p_2}^{p} R(p-x)R^*(x) dx -\int_{0}^{p} R(p-x)R^*(x) dx\right),
\end{equation}
where we have suppressed the momentum conservation factor $\delta (p_1 + p_2 = p)$. The one-loop result is given by
\begin{align}
A_3(p_1,p_2;-p) &= \frac{-p_1p_2p}{\mu}\left(1-\frac{1}{24}(p^4-2p^3(p_1+2) + 2p^2(p_1^2 + 3p_1 + 2) \right.\cr
&- \left. p(6p_1^2 + 4p_1 -1) + 4(p_1^2-2) + 24q^2)\mu^{-2} + \cdots\right).
\end{align}

Usually, the scattering amplitudes in the R-R background is difficult to calculate because the sensible R-NS formulation in the R-R background perturbation theory is limited. The world sheet derivation of these amplitudes is a challenging problem.\footnote{However, note that this amplitude is nothing but a normal bosonic Liouville theory amplitude with boundaries in the approximation taken in section \ref{6.8}.}

With the exact MPR formula, we can calculate the scattering amplitudes in various limits. For example, in \cite{Kapustin:2003hi}, the following two limits have been considered. The strong field limit is the limit where we fix $q = f \mu$ and expand in $1/\mu$. The three-body S matrix becomes
\begin{equation}
A_3(p_1,p_2,-p) = \frac{-p_1p_2p}{(1+f^2)\mu} + \cdots,
\end{equation}
It is interesting to observe that the renormalization 
\begin{equation}
\mu \to \tilde{\mu} = (1+f^2) \mu,
\end{equation}
yields the same tree level amplitudes for any correlation function (but not for the loop correction).

The dual string limit is defined as follows. We set $g'_s = 1/q$ and $q' = \mu$ with fixed $f' = \frac{1}{f}$. Then the three-point function becomes
\begin{equation}
A_3(p_1,p_2;-p) = -f'g'_s p_1p_2p +\cdots
\end{equation}
and other amplitudes also admit a $g'_s$ expansion under the dual R-R flux $f'$. The nature of this dual theory is not well-known.\footnote{In \cite{Kapustin:2003hi} they have argued that the field content of the theory should be a massless R-R scalar only.}

Actually, the more radical limit is possible as long as the partition function is concerned. In \cite{Strominger:2003tm}, the type 0A potential without a harmonic term, namely $V(\lambda) = -\frac{q^2}{\lambda^2}$ is proposed to be dual to the $\mathrm{AdS}_2$ geometry. This is plausible since this quantum mechanical system is known to be $\mathrm{CFT}_1$ and hence we can regard it as $\mathrm{AdS}_2$/$\mathrm{CFT}_1$ correspondence (see e.g. \cite{deAlfaro:1976je}, \cite{Akulov:1983uh}, \cite{Fubini:1984hf}). See also \cite{Danielsson:2003yi}, \cite{Thomson:2003fz}, \cite{Gukov:2003yp} for a related discussion, where the classical solution is studied in \cite{Thomson:2003fz} and the $\mu \to 0$ limit is studied in \cite{Gukov:2003yp}. The $\mu \to 0$ singularity is removed once we turn on $q$.

\subsection{Unitarity of Type 0B S Matrix}\label{10.2}
We consider the unitarity of the type 0B S matrix from the matrix model point of view \cite{DeWolfe:2003qf}. The central formula for the S matrix of the matrix model is given by the MPR formula (\ref{eq:MPR}): 
\begin{equation}
S_{CF} = \iota_{f\to b} \circ S_{FF} \circ \iota_{b\to f}.
\end{equation}
The major difference in the type 0B theory is that we have two fluctuations of the Fermi surface $T_{L,R} = \frac{1}{\sqrt{2}}(T \pm V)$ which correspond to the left and right Fermi surface fluctuation respectively. Therefore, we consider two kinds of fermion field $b_+$ and $b_-$ with
\begin{equation}
\psi (t,\lambda) = \int_{-\infty}^{\infty} d\omega e^{i\omega t} [ b_+ (\omega) \psi^+(\omega, \lambda) + b_-(\omega) \psi^{-}(\omega, \lambda)],
\end{equation}
where $\psi^{\pm}$ is the even(odd) eigenstates with respect to $\lambda \to -\lambda$ for the inverted harmonic oscillator Hamiltonian:
\begin{equation}
H = -\frac{1}{2}\frac{d^2}{d\lambda^2} - \frac{1}{2\alpha^2} \lambda^2.
\end{equation}

The fermionic mode operator $b$ satisfies
\begin{equation}
\{b_\epsilon(\omega), b^\dagger_{\epsilon'}(\omega')\} = \delta_{\epsilon,\epsilon'} \delta(\omega -\omega').
\end{equation}
We also use their linear combination:
\begin{equation}
b_{L,R} = \frac{1}{\sqrt{2}}(b_+ \pm b_-).
\end{equation}

\begin{figure}[htbp]
	\begin{center}
	\includegraphics[width=0.8\linewidth,keepaspectratio,clip]{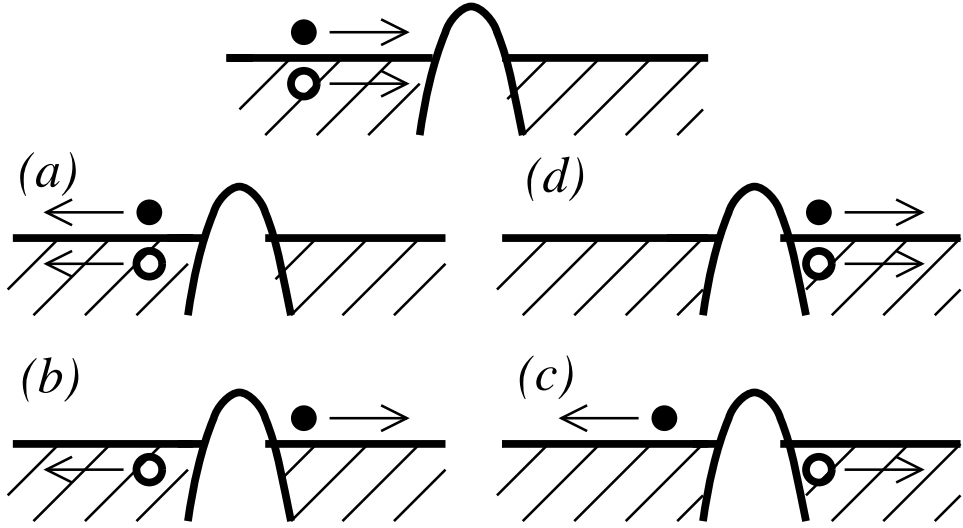}
	\end{center}
	\caption{The schematic picture of the tachyon scattering in the type 0B theory. (a)-(d) is the possible final states.}
	\label{0bsca}
\end{figure}

Now we state the central problem of the 0B S matrix raised in \cite{DeWolfe:2003qf} and solved there. Actually, there is a small (suppressed as $e^{-\mathcal{O}(1/g)}$) probability that the fermion/hole tunnels into the other side of the Fermi sea. See figure \ref{0bsca} for example. The process (a) and (d) are easy to interpret: a single quantum of $T_L$ or $T_R$ emerges from the Liouville wall (reflection or tunneling) and goes out to infinity. However what happens in the process (b) and (c)? From the bosonic fluctuation point of view, it seems difficult to interpret them because one fermion or hole does not related to a single outgoing massless boson. At the same time, we should note that just throwing away these possibilities ruins the unitarity as we can easily see from the MPR formula.

The solution of this problem is given in \cite{DeWolfe:2003qf}, where they have introduced the concept of $k$ vacuum in which the fermion number is nonzero in the fermionic language. In the physical perspective, they are related to the existence of the D-instanton much like the $n$ vacuum in the four dimensional Yang Mills theory. In the following, we would like to briefly review their idea and results.

For simplicity, we consider the compactified (Euclidean) version of the bosonization. For a chiral part we have 
\begin{equation}
\partial X(z) = -i\sum_m \frac{a_m}{z^{m+1}}
\end{equation}
and 
\begin{equation}
\psi(z) = \sum_{r=\mathbf{Z} +1/2} z^{-r-1/2}b_r,
\end{equation}
where the bosonization formula is 
\begin{equation}
a_n = \sum_{r=\mathbf{Z}+1/2} :b_r b^\dagger_{r-n}:.
\end{equation}
In this case, the Hilbert space of the bosonic sector is labeled by the momentum $k$ as $\mathcal{H} = \oplus_k \mathcal{H}_k$, where $a_0 |k\rangle = k|k\rangle$. When we perform the fermionization, we have  distinct vacua whose fermion number is counted by $k$.
These $k$ vacua is produced by the solitonic excitation as $|k\rangle = e^{ikX}(0) |0\rangle$.

\begin{figure}[htbp]
	\begin{center}
	\includegraphics[width=0.8\linewidth,keepaspectratio,clip]{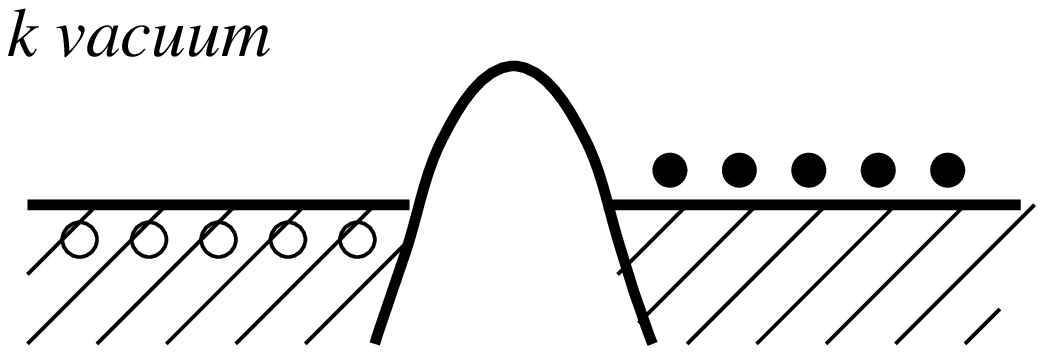}
	\end{center}
	\caption{The $k$ vacuum is degenerate in energy but has a different fermion number.}
	\label{kvacuum}
\end{figure}

The important point is that under the nonperturbative tunneling effect, the separate left or right fermion number is not conserved any more. While we have a strict conservation law and nothing prevents us from truncating the spectrum onto the $k=0$ sector in the type 0A theory, we should include all these sectors in the type 0B theory in order to preserve unitarity. We illustrate the intuitive picture of the $k$ vacuum in figure \ref{kvacuum}. Note that in the space-time point of view, these $k$ vacua are degenerate in energy in the continuum limit.

The space-time interpretation of $k$ vacua is as follows. First we note that the R-R scalar $C$ has the following shift symmetry to all orders in the string perturbation
\begin{equation}
C \to C + a.
\end{equation}
However, since the $k$ vacua are related to the $0$ vacuum as 
\begin{equation}
|k \rangle = e^{i\sqrt{2} k C} (0) |0\rangle,
\end{equation}
where we should recall that the leg pole factor is unity if we only consider the zero-mode. Therefore the shift symmetry is reduced to the discrete one:
\begin{equation}
C \sim C + \sqrt{2} \pi.
\end{equation}
This is due to the fact that the $k$ vacua have nonzero RR charge. The transition between these states are induced by nothing but the D-instanton. All these facts have a close analogy to familiar $n$ vacua in the four dimensional pure Yang Mills theory.

With the analogy to the Yang Mills theory, we may expect $\theta$ vacua like states which diagonalize the Hamiltonian. In \cite{DeWolfe:2003qf} they have named such states $c$ vacua:
\begin{equation}
|c \rangle \equiv \sum_k e^{-i\sqrt{2}kc} |k\rangle.
\end{equation}
For these $c$ vacua to make sense, the $k$ vacua are indistinguishable from each other. This is certainly the case when we speak of the Yang Mills $n$ vacua. Here we have seen that the energy of the $k$ vacua is degenerate in the continuum limit, but this is not enough. In \cite{DeWolfe:2003qf}, they have calculated several simple scattering amplitudes and found that the transition amplitudes between different soliton sector ($k$ vacuum) are plagued by the infrared divergences. With a particular regularization scheme used there, they found that $k$ vacua are not indistinguishable, so the diagonalization of the Hamiltonian is difficult. It would be interesting to study further to see whether there is a proper regularization scheme which treats $k$ vacua on equal footing.

\subsection{Hole Interpretation from Boundary States.}\label{10.3}
We will briefly review the interpretation of hole states in the matrix model from the boundary states perspective, which has been studied in \cite{Douglas:2003up}, \cite{Gaiotto:2003yf}. Most of the statements here are applicable to both the bosonic Liouville theory and the super Liouville theory. We have so far considered the excitation of an eigenvalue in the matrix model as an extra unstable D0-brane in the Liouville theory. However, we have another kind of excitation in the matrix model, namely the hole excitation. What is the interpretation of the hole from the boundary state point of view?

First, we recall the description of the rolling (positive) eigenvalue in the boundary states. This can be described by the ZZ brane tensored with the Sen's rolling boundary state\footnote{In this section, we consider the S-brane type $\lambda \cosh X$ interaction with the Hartle-Hawking contour.} with $0\le \lambda \le \frac{1}{2}$ for the bosonic theory. For the fermionic string, we have $-\frac{1}{2}\le\lambda\le \frac{1}{2}$. The energy in the matrix model eigenvalue and the parameter $\lambda$ is related via 
\begin{equation}
E = \mu \cos^2 \pi \lambda. \label{eq:eener}
\end{equation}
Therefore, the above parameter region only deals with the positive energy states above the Fermi sea. Nevertheless it can be shown that the formal analytic continuation of $\lambda = \frac{1}{2} + i\alpha$, where $\alpha$ is real\footnote{In the fermionic case, we have $\lambda = \pm\frac{1}{2} + i\alpha$. }, actually works fine. At first sight, this does not make sense because the boundary interaction becomes complex. However, the actual boundary state is a function of $(\lambda-1/2)$ and hence remains real. In this parameter region, the energy given by (\ref{eq:eener}) becomes negative, so we expect that this boundary state describes the negative energy state below the Fermi sea.

However, this is not the end of the story. The hole state is not a negative energy state but a positive energy state. In the second quantized language, the negative energy wavefunction should be treated as an annihilation operator of holes. In order to include this property, it has been proposed in  \cite{Douglas:2003up}, \cite{Gaiotto:2003yf} that the boundary state associated with a hole is obtained by tensoring the ZZ brane with the minus of Sen's boundary state.

\begin{figure}[htbp]
	\begin{center}
	\includegraphics[width=0.8\linewidth,keepaspectratio,clip]{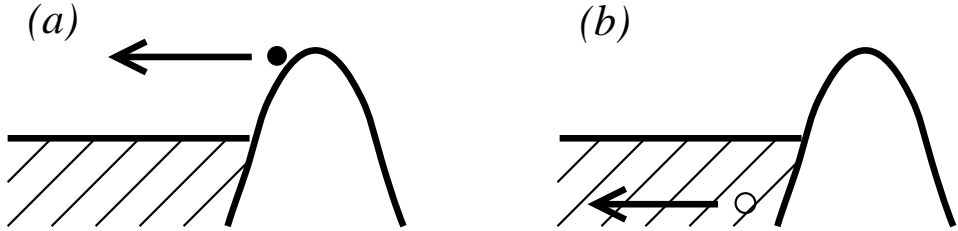}
	\end{center}
	\caption{(a) the particle excitation is interpreted as the decaying D0-brane. (b) what corresponds to the hole?}
	\label{hole-particle}
\end{figure}

This is expected and can be checked from the relation between fermionization formula of collective field theory and the rolling-tachyon decaying profile. The final state for the (positive) rolling eigenvalue is schematically given by
\begin{equation}
|out\rangle \sim \exp\left(\int_0^\infty\frac{dp}{\sqrt{E}} a^\dagger A(p)\right) |0\rangle,
\end{equation}
where $A$ is determined from the boundary states described above (see section \ref{6.4}). From the collective field theory point of view, this is just the fermionization formula $\psi^\dagger = :\exp(i\phi):$. To describe a fermionic annihilation operator, we should have $\psi = :\exp(-i\phi):$. In the boundary state calculation, this is equivalent to using the boundary state with an extra minus sign.

Finally let us comment on the matrix theory on the particle and hole excitation. Since the open strings propagating between a particle boundary state and a hole boundary state have a space-time minus sign in the loop channel, they become fermions. Alternatively saying, the gauge theory with $N$ particle states and $M$ hole states is given by the supergroup $U(N|M)$. It should be remarked however,  that the description here presumes the existence of the Fermi sea and condensation of negative energy particles. We do not claim that the double scaling limit of the supergroup theory becomes a continuum Liouville(-like) theory.

\subsection{Nonperturbative Generation of R-R Potential}\label{10.4}
In section \ref{10.2}, we have discussed the nature of the degenerate R-R vacua, where we have seen that in the double scaling limit, these vacua are energetically degenerate. However, it has been pointed out in \cite{Gross:2003zz} (see also \cite{Gukov:2003yp} for a noncritical type 0 string theory with a flux background), that with a finite amount of the R-R flux background, the degeneration over the zero-mode $C$ can be removed. The rough idea is to put the Fermi surface left right asymmetrically; then the $k$ vacuum is necessarily lifted (though in the following we do not use the concept of $k$ vacua but instead $C$ vacua). This is essentially nonperturbative effect and much like the fate of the $\theta$ vacua in QCD. In this section, we will briefly sketch the idea of \cite{Gross:2003zz} and discuss their main results.

\begin{figure}[htbp]
	\begin{center}
	\includegraphics[width=0.8\linewidth,keepaspectratio,clip]{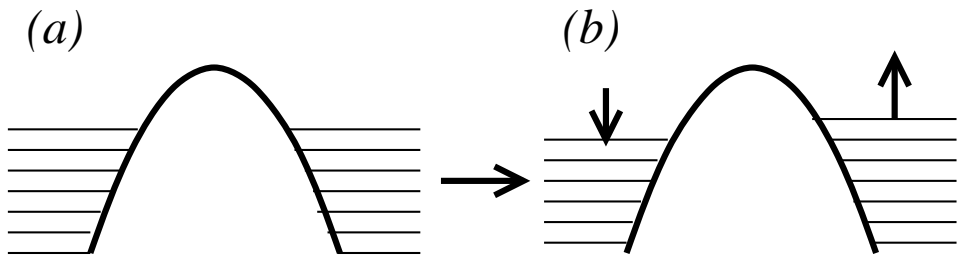}
	\end{center}
	\caption{Turning on the zero-mode of $C$ corresponds to raising and lowering the left and right eigenvalues respectively.}
	\label{con}
\end{figure}

Let us first review the $Q=0$ case which we have discussed in section \ref{10.2} from the WKB point of view again --- in this case following \cite{Gross:2003zz}. The Bohr-Sommerfeld quantization condition is given by
\begin{equation}
\pi n \hbar = \int^{\lambda_*} \sqrt{2(\epsilon_n-V(\lambda))} d\lambda,
\end{equation}
where the potential is given by $V \sim -\frac{1}{2}\lambda^2$ in the double scaling limit (with $\alpha' = 1/2$ convention) and $\lambda_*$ is given by the turning point of the potential. The Planck constant here is related to the bare (open) string coupling constant as $\hbar \sim g_s$. Then the density of states is given by
\begin{equation}
\rho = \left|\frac{dn}{d\epsilon}\right| = \frac{\log(2\pi \hbar n)}{2\pi \hbar}
\end{equation}
in this limit. If we consider the zero-mode translation $C\to C + 2\alpha$, the fermion obtains an extra phase $\alpha$, which raises (lowers) the right (left) hand side of the energy level (see figure \ref{con}). We normalize that the $2\pi$ shift of $C$ does nothing to the system, so we have
\begin{equation}
\frac{dn}{dC} = \frac{1}{2\pi},
\end{equation}
which shifts the energy by
\begin{equation}
\frac{d\epsilon_{L/R}}{dC} = \pm\frac{1}{2\pi \rho},
\end{equation}
in the double scaling limit where $\rho$ diverges as $\log\hbar$. From this, when the original Fermi level in the left and right potential is the same, the shift of $C$ does not change the total energy, which is compatible with what we have learned in section \ref{10.2}.

\begin{figure}[htbp]
	\begin{center}
	\includegraphics[width=0.8\linewidth,keepaspectratio,clip]{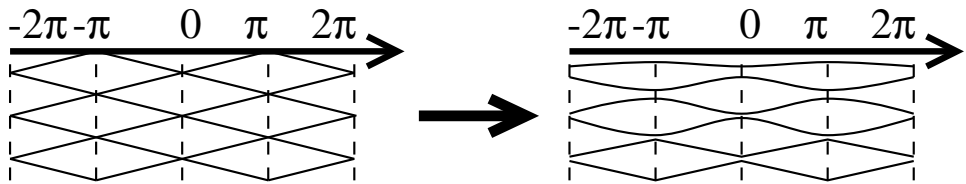}
	\end{center}
	\caption{When perturbed eigenvalues cross, the repulsion of eigenvalues occurs.}
	\label{repel}
\end{figure}

Let us now fill the eigenvalues asymmetrically. Set $\mu_L = \mu +Q$ and $\mu_R=\mu -Q$ with a positive $Q<\mu$. Later, we will identify the parameter $Q$ as the background flux of the theory.\footnote{Note that the introduction of $Q$ requires an infinite energy, and we should regard it as a different superselection sector as opposed to the value of the infinitesimal deformation by the zero-mode $C$. This is much like the background flux $q$ in the type 0A theory. Indeed they are perturbatively related by the T-duality.} In this background, the deformation by the shift of $C$ changes the energy of the system, which is now given by
\begin{equation}
V(C) = \int_{\mu_R}^{\mu_L} d\epsilon \rho (\epsilon_R(C)-\epsilon_R(0)) = -\frac{CQ}{\pi},
\end{equation}
in the double scaling limit. This is a perturbative result and requires a nonperturbative completion, for we should have a periodic potential because of the original identification $C \sim C + 2\pi$. In other words, when $C$ approaches $\pi$, the perturbed eigenvalues cross and the repulsion occurs as is always the case with the perturbation theory in the one dimensional quantum mechanics (see figure \ref{repel}).

The identification of the Fermi level difference $Q$ as the background flux can be understood from the coupling of the flux to the world line tachyon, which is described by
\begin{equation}
S = \int \mathcal{F}(T) \wedge \tilde{F},
\end{equation}
where $\mathcal{F}(T)$ is in general an odd function of $T$ and $\tilde{F}$ is the dual field strength made from $C$ \cite{Sen:1999mg}. The contribution of this effect to the second quantized Hamiltonian for the fermion becomes
\begin{equation}
\delta H = \int d\lambda Q \Psi^\dagger(\lambda) \Psi (-\lambda),
\end{equation}
which directly implies that the (anti)symmetric part of $\Psi$ has a Fermi level $\mu_{\pm} = \mu \pm Q$. The symmetric/antisymmetric classification used here is more suitable than the left/right classification in the nonperturbative definition because of the tunneling effect. Also, it shows the stability of the nonzero $Q$ background though energetically unfavored. It can be also seen from the fact that the T-dual of this configuration is the type 0A theory which has an imaginary R-R flux, where a different (real) R-R flux means a different nonperturbatively stable background.

In the following, we derive the full shape of the potential from the matrix model consideration. The Schr\"odinger equation for the eigenvalue in the double scaling limit can be written as
\begin{equation}
\left(\frac{d^2}{dx^2} +\frac{x^2}{4} - a\right)\psi (a,x) = 0,
\end{equation}
where $\lambda = \sqrt{\frac{\hbar}{2}} x$ and $\epsilon = \frac{1}{2}-\hbar a$. The Fermi level mismatch can be emulated by setting the system in a box whose range is between $x_{L/R} = \mp\sqrt{\frac{2}{\hbar}} + \sqrt{\frac{\hbar}{2}} C $. Using the explicit solutions of the Schr\"odinger equation, we can find that the energy $a_\pm$ which corresponds to the even/odd mode respectively is given by
\begin{equation}
\phi_0(a_{\pm}) = \mp \frac{1}{2} \arccos\left[\frac{\cos C}{\sqrt{1+e^{-2\pi a_{\pm}}}}\right] + n\pi,
\end{equation}
for each integer $n$, where $\phi_0(a) = \frac{1}{2\hbar} -2a\log\sqrt{2/\hbar} +\frac{\pi}{2} + \arg\Gamma(\frac{1}{2} + ia)$. From this result, we can obtain the energy difference which depends only on $C$ in the double scaling limit (other parts are divergent as $\log\hbar$ in this limit). 
\begin{equation}
V(C) = \frac{1}{2\pi} \int_{\mu-Q}^{\mu+Q} da \arccos\left[\frac{\cos C}{\sqrt{1+e^{-2\pi a}}}\right],
\end{equation}
which shows a correct periodicity and weak coupling limit $\mu \to \infty$.

Some comments are in order.
\begin{itemize}
	\item At first sight, $C$ is very massive with $V'' \sim e^{\pi\mu}$. But as it is emphasized in \cite{Gross:2003zz}, this does not mean that the space-time field $C$ is massive. This is because the potential here is obtained after integrating over the Liouville volume $V_{\phi} \sim \log\mu$, and hence the actual ``mass" in the space-time is zero.
	\item We can do the same calculation with the compactified Euclidean time (finite temperature matrix mode). The perturbative calculation shows the T-duality to the type 0A matrix model with an imaginary flux. Imaginary here should be related to the Wick rotation.\footnote{We have encountered the imaginary flux also in the $\hat{c} = 0$ model. See \cite{Brower:1993mn}, \cite{Klebanov:2003wg}.} However, it has been pointed out in \cite{Gross:2003zz} that the nonperturbative completion does not show the naive T-duality. On the other hand, the ``S-duality" which involves $\mu \to -\mu$ and $C \to \tilde{C}$ and a particle/hole duality which exchanges NS and RR background are strict symmetries. See also \cite{Gukov:2003yp}, \cite{Yin:2003iv} for discussions on the T-duality. In the latter paper, the T-duality is studied from the duality between the double scaling matrix quantum mechanics and the instanton matrix model.
	\item All those considered in this section are derived from the matrix model. It would be interesting to see its connection with the super Liouville theory. The D-instanton effects (with ZZ brane) must be one of the major sources. 
\end{itemize}

\subsection{Literature Guide for Section 10}\label{10.5}
While the study on the matrix model corresponding to the type 0 noncritical string theory has a long history over a decade, its application to the type 0 theory has begun only recently \cite{Klebanov:2003wg}, \cite{Kapustin:2003hi}, \cite{DeWolfe:2003qf}, \cite{Douglas:2003up}, \cite{Gaiotto:2003yf}, \cite{Johnson:2003hy}, \cite{Gross:2003zz}, \cite{Seiberg:2003nm}, \cite{Danielsson:2003yi}, \cite{Strominger:2003tm}, \cite{Gukov:2003yp}, \cite{Thomson:2003fz}, \cite{Yin:2003iv}. However, we have already discovered many concrete (and exact) realizations of the nonperturbative string physics including Sen's conjecture, the geometric transition, the nonperturbative generation of the R-R potential, various dualities which we have not met so far, and so on. We hope we will find more examples of these phenomena by investigating matrix models or the Liouville theory. 

\sectiono{$\mathcal{N}=2$ Super Liouville Theory}\label{sec:11}
In this section, we discuss the $\mathcal{N}=2$ supersymmetric Liouville theory. The $\mathcal{N}=2$ Liouville theory is in a sense very different from the $\mathcal{N} = 0,1$ Liouville theory discussed so far. One of the major disinctions is the nonrenormalization of the cosmological constant operator and the consequent disappearances of the $c (\hat{c}) = 1$ barrier. Because of this lack of the barrier, we can use the $\mathcal{N}=2$ super Liouville theory as an ``internal SCFT" in the general $d>2$ superstring theory. In the next section, we use the $\mathcal{N}=2$ super Liouville theory as a building block of the superstring theory propagating in the singular CY space. The organization of this section is as follows.

In section \ref{11.1}, we discuss the bulk theory. In subsection \ref{11.1.1}, we review the basic setup of the $\mathcal{N}=2$ super Liouville theory and its CFT properties. In subsection \ref{11.1.2}, we attempt to obtain the bulk structure constants but face a difficulty which we have not met in the $\mathcal{N} = 0,1$ theory. The dual action proposal will be discussed to overcome this difficulty.

In section \ref{11.2}, branes in the $\mathcal{N}=2$ super Liouville theory are discussed. After reviewing the basic modular transformation properties of the $\mathcal{N}=2$ characters in subsection \ref{11.2.2}, and \ref{11.2.3}, we construct the boundary states by using the modular bootstrap method. 

In section \ref{11.3}, we review the matrix model proposal of the $\mathcal{N}=2$ super Liouville theory. In subsection \ref{11.3.1} and \ref{11.3.2} we discuss the basic properties of the Marinari-Parisi model which is a supersymmetric matrix model and proposed to be the dual of the $\mathcal{N}=2$ super Liouville theory. In subsection \ref{11.3.3}, we discuss its connection with the world sheet $\mathcal{N}=2$ super Liouville theory.

\subsection{Bulk Theory}\label{11.1}
\subsubsection{Setup}\label{11.1.1}
The $\mathcal{N}=2$ supersymmetric Liouville theory \cite{Ivanov:1983wp} can be seen in two ways. The first one is that we regard it as the quantization of the $\mathcal{N}=2$ supergravity on the worldsheet (see e.g. \cite{Distler:1990nt}, \cite{Antoniadis:1990mx}, \cite{Ketov:1996es}). The second one is that we simply regard it as a particular (irrational) CFT with an accidental $\mathcal{N}=2$ supersymmetry \cite{Kutasov:1990ua}, \cite{Kutasov:1991pv}, \cite{Murthy:2003es}. Though the first perspective seems natural from the experience with the $\mathcal{N}=0,1$ super Liouville theory we have discussed so far, we will take the second point of view. That is because if we take the first standpoint, we have to gauge the $\mathcal{N}=2$ algebra, which leads to the so called $(2,2)$ (noncritical) string. While it is mathematically interesting, the physical application of the $(2,2)$ string is less clear.\footnote{However, see \cite{Aharony:2003vk} for a relation to the \textit{topological} LST which involves the $\mathcal{N}=2$ Liouville sector.} Therefore, to utilize our familiar physical intuition about the usual fermionic string, we take the second view point, where we regard the $\mathcal{N}=2$ algebra as an accidental symmetry of the world sheet theory. It is also important to note that we should have at least $(2,0)$ world sheet supersymmetry in order to have a space-time supersymmetry.

The action of the $\mathcal{N}=2$ super Liouville theory is given by
\begin{equation}
S_0 = \frac{1}{2\pi} \int d^2z d^4\theta S\bar{S},
\end{equation}
for the free kinetic part, and the interaction part is given by
\begin{eqnarray}
S_+ &=& \mu\int d^2 z d^2 \theta e^{bS} \cr
S_- &=& \bar{\mu} \int d^2 z d^2 \bar{\theta} e^{b\bar{S}}, \label{eq:chid}
\end{eqnarray}
where $S$ is a chiral superfield (for the bosonic part $S= \phi + iY$) and there is a suppressed background charge $1/b$ for $\phi$ --- the real part of the bosonic field. Without ruining the $\mathcal{N}=2$ superconformal symmetry, another deformation is possible,
\begin{equation}
S_{nc} = \int d^2z d^4\theta e^{\frac{1}{2b} (S+\bar{S})}.\label{eq:nchid}
\end{equation}
In the next section, we will see that the chiral perturbation (\ref{eq:chid}) corresponds to the complex moduli deformation of the singularity of the CY space and the nonchiral perturbation (\ref{eq:nchid}) corresponds to the K\"ahler moduli deformation (resolution) of the singularity. However, we should note that the nonchiral perturbation is beyond the Seiberg bound.

One of the most significant features of the $\mathcal{N}=2$ theory is the nonrenormalization theorem. In this case, the background charge $Q=\frac{1}{b}$ is not renormalized quantum mechanically \cite{Distler:1990nt}. We can easily see this in the component setup. After eliminating the auxiliary fields, the kinetic term is given by
\begin{eqnarray}
S_0 &=& \int d^2z \frac{1}{4\pi} \left(\partial S \bar{\partial}S^* + \partial S^* \bar{\partial} S + \psi^+ \bar{\partial}\bar{\psi}^+ + \bar{\psi}^+ \bar{\partial}\psi^+ \right.\cr
&& \ \ \ \ \ \ \ \ \ \ \ + \left. \psi^{-}\partial \bar{\psi}^- + \bar{\psi}^-\partial\psi^-\right). 
\end{eqnarray}
See appendix \ref{a-2} for our notation; the bar on the fermion here is the complex conjugation which does not change the chirality (thus $\psi^+$ is constructed by the two independent Majorana Weyl fermions $\psi^+ = \psi_1 + i\psi_2$ and $\bar{\psi}^+ = \psi_1 - i\psi_2$), and the chirality is governed by the superscript $\pm$.
The chiral interaction part is given by
\begin{equation}
S_+ + S_- = \int d^2z \left(\mu b^2 \psi^+ \psi^- e^{bS} + \bar{\mu} b^2 \bar{\psi}^+\bar{\psi}^- e^{b S^*} + \pi \mu\bar{\mu} b^2 :e^{bS}::e^{bS^*}:\right),
\end{equation}
and the nonchiral interaction is given by
\begin{equation}
S_{nc} = \tilde{\mu}\int d^2 z (\partial \phi -i\partial Y -\frac{1}{2b}\psi^+\bar{\psi^+})(\bar{\partial}\phi + i\bar{\partial}Y -\frac{1}{2b}\psi^-\bar{\psi}^-) e^{\frac{1}{b}\phi}.
\end{equation}
With the background charge $1/b$ for $\phi$, we can verify that the interaction is indeed $(1,1)$ tensor. The dimension of the chiral part is given by
\begin{equation}
\Delta(\psi^+ e^{b (\phi + iY)}) = \frac{1}{2} + \frac{b(\frac{1}{b}-b)}{2} + \frac{b^2}{2} = 1,
\end{equation}
and that of the nonchiral interaction, for example for the first term, is given by 
\begin{equation}
\Delta(\partial \phi e^{\frac{1}{b}\phi}) = 1 + \frac{\frac{1}{b}(\frac{1}{b}-\frac{1}{b})}{2} = 1.
\end{equation}
Therefore, the background charge is not renormalized unlike in the bosonic or $\mathcal{N} =1$ Liouville theory. Because of this property, there is no analog of the $c =1$ barrier in the $\mathcal{N} =2$ super Liouville theory. Note that the central charge of this theory is given by $c= 3\tilde{c} = 3 + \frac{3}{b^2}$ and $b$ can be chosen as an arbitrary real number. Consequently, the central charge can become small enough to introduce other transverse dimensions as opposed to the bosonic or $\mathcal{N}=1$ Liouville theory. Owing to this property, we can embed the $\mathcal{N}=2$ super Liouville theory into the superstring theory propagating in the nontrivial background whose dimension is larger than two as we will see later in section \ref{12.2}.

Let us show the explicit realization of the $\mathcal{N}=2$ world sheet superalgebra of this theory:
\begin{eqnarray}
T &=& -\frac{1}{2}(\partial Y)^2 - \frac{1}{2}(\partial \phi)^2 + \frac{1}{2b} \partial^2 \phi -\frac{1}{4}(\psi^+\partial \bar{\psi}^+ - \partial \psi^+ \bar{\psi}^+) \cr
G^+ &=& -\frac{1}{2} \psi^+ (i\partial Y + \partial \phi) + \frac{1}{2b}\partial \psi^+ \cr
G^- &=& -\frac{1}{2}\bar{\psi}^+ (i\partial Y -\partial \phi) - \frac{1}{2b}\partial \bar{\psi}^+ \cr
J &=& \frac{1}{2}\psi^+ \bar{\psi}^+ + \frac{1}{b}i\partial Y.
\end{eqnarray}
The $\mathcal{N}=2$ mode algebra is given by
\begin{eqnarray}
[L_m,G^{\pm}_r] &=& \left(\frac{m}{2}-r\right) G^{\pm}_{m+r} \cr
[L_m,J_n] &=& -nJ_{m+n} \cr
\{G_r^+, G_s^-\} &=& 2L_{r+s} + (r-s) J_{r+s} + \frac{c}{3}\left(r^2 -\frac{1}{4}\right)\delta_{r,-s} \cr
\{G_r^{\pm}, G_s^{\pm}\} &=& 0 \cr
[J_n,G^{\pm}_r] &=& \pm G_{r+n}^{\pm} \cr
[J_m,J_n] &=& \frac{c}{3} m \delta_{m,-n},   
\end{eqnarray}
where the central charge is given by $c= 3\tilde{c} = 3 + \frac{3}{b^2}$. For the later purpose, it is convenient to define the spectral flow generators as
\begin{eqnarray}
U^{-1}_\eta L_m U_\eta &=& L_m + \eta J_m + \frac{\tilde{c}}{2}\eta^2 \delta_{m,0} \cr
U^{-1}_\eta J_m U_\eta &=& J_m + \tilde{c}\eta \delta_{m,0} \cr
U^{-1}_\eta G^{\pm}_r U_\eta &=& G^{\pm}_{r\pm \eta}. \label{eq:spcfl}
\end{eqnarray}

Anticipating the application to the superstring theory, we have a remark on the compactification of the imaginary part of the superfield, $Y$. For the chiral interaction part of the action to be single-valued, $Y$ should be compactified on the radius $R = \frac{n}{b}$ with an integer $n$. Since the $U(1)$ current measures the momentum of $Y$, we have an integral $U(1)$ charge for the NS sector if we have $n=1$, the smallest radius. On the other hand, following the standard procedure, we can construct a \textit{spacetime} supercharge operator from the $\mathcal{N}=2$ worldsheet superalgebra (see e.g. \cite{Kutasov:1990ua}, \cite{Kutasov:1991pv}).
\begin{equation}
Q_\alpha = \oint dz e^{-\frac{\sigma}{2}}e^{-\frac{i}{2}(H-QY)}S_\alpha,
\end{equation}
where $H$ is the bosonized Liouville fermion and $S_\alpha$ is the flat space spinor operator.
 However, this supercharge operator becomes local only if the $U(1)$ current has an integral spectrum in the unit $1/Q$. Therefore, for the application to the spacetime superstring theory, the integral charge quantization of the $NS$ sector is necessary. Of course, if we give up the spacetime SUSY (for example, the type 0 string), this condition is not necessary.

Let us now go on to the primary states of the $\mathcal{N} =2 $ Liouville theory \cite{Mussardo:1989av}, \cite{Ahn:2002sx}. The fundamental building block is given by the NS vertex operator
\begin{equation}
N_{\alpha \bar{\alpha}} = e^{\alpha S + \bar{\alpha} S^*} = e^{(\alpha+\bar{\alpha}) \phi + i(\alpha-\bar{\alpha})Y},
\end{equation}
and the R vertex operator 
\begin{equation}
R^{\pm}_{\alpha\bar{\alpha}} = \sigma^{\pm} e^{\alpha S + \bar{\alpha} S^*} = \sigma^{\pm}e^{(\alpha+\bar{\alpha}) \phi + i(\alpha-\bar{\alpha})Y},
\end{equation}
where $\sigma^{\pm} $ is the (left) spin operator and made up with the two dimensional left fermionic zero-mode algebra.\footnote{For a complete vertex operator, we should tensor it with the right part properly according to the GSO projection.} The conformal dimensions of these operators are given by
\begin{eqnarray}
\Delta_N &=& -2\alpha \bar{\alpha} + \frac{1}{2b} (\alpha + \bar{\alpha}) \cr
\Delta_R &=& -2\alpha \bar{\alpha} + \frac{1}{2b} (\alpha + \bar{\alpha}) + \frac{1}{8} 
\end{eqnarray}
and the $U(1)$ charges are given by
\begin{eqnarray}
w_N &=& \frac{1}{b} (\alpha -\bar{\alpha}) \cr
w^\pm_R &=& w \pm \frac{1}{2}. \label{eq:u1charge}
\end{eqnarray}
It is important to note that if $\bar{\alpha} = 0$, we have $2\Delta_{N} = w_N$, hence the chiral primary operator. Introducing the momentum $P$ as 
\begin{equation}
\alpha + \bar{\alpha} = \frac{1}{2b} + iP, \label{eq:2momen}
\end{equation}
we can rewrite the conformal dimension as
\begin{equation} 
\Delta_N = \frac{P^2}{2} + \frac{1}{8b^2}  + \frac{b^2w^2}{2}. \label{eq:masdim}
\end{equation}

As we can see from (\ref{eq:u1charge}) (\ref{eq:masdim}), the $U(1)$ charge and the conformal dimension does not change if we reverse the momentum $p \to -p$. As in the bosonic and $\mathcal{N} =1$ Liouville theory, we should identify these operators up to a multiplicative constant, namely the reflection amplitude \cite{Ahn:2002sx}. In the original exponential form, the identification map is given by
\begin{eqnarray}
\alpha &\to& \frac{1}{2b} - \bar{\alpha} \cr
\bar{\alpha} &\to& \frac{1}{2b} -\alpha.
\end{eqnarray}

\subsubsection{Bulk structure constants and dual action}\label{11.1.2}
The fundamental goal in this subsection must be, of course, to derive various structure constants which characterize this SCFT. However, because of the technical difficulty which we will explain below, the derivation of the structure constant seems difficult without a further assumption. The most difficult point is that we do not have a duality such as $b \to b^{-1}$. This is because the background charge is not renormalized and $b\to b^{-1}$ \textit{changes} the central charge as a result. This duality has played a significant role in specifying the structure constant uniquely from the functional relations. Even in the $\mathcal{N}=2$ theory, we are actually able to derive a functional relation by using Teschner's trick, but we have only one relation, which is not enough to determine the structure constant uniquely. 

To overcome the situation, a new kind of duality is proposed in \cite{Ahn:2002sx} (see also \cite{Fateev:1996ea}, \cite{Baseilhac:1998fd}, \cite{Baseilhac:1998eq}). The claim is that the $\mathcal{N}=2$ super Liouville theory perturbed by the chiral interaction $S_+ + S_-$ with the cosmological constant $\mu$ is equivalent to the $\mathcal{N}=2$ super Liouville theory perturbed by the nonchiral interaction $S_{nc}$ with the dual cosmological constant $\tilde{\mu}$. Of course, we use the same background charge $1/b$ to ensure that the central charge remains the same. In section \ref{12.3}, we make an attempt to give its proof after reviewing the necessary background, but we have some remarks here.
\begin{itemize}
	\item In the neutral sector where the $U(1)$ current $J = 0$, this conjecture seems to yield the correct (or at least consistent) bulk two-point function as we will see \cite{Fateev:1996ea}. 
	\item From the viewpoint of the superstring propagating in the singular CY space \cite{Ooguri:1996wj}, \cite{Aharony:1998ub}, \cite{Giveon:1999px}, \cite{Giveon:1999tq}, \cite{Eguchi:2000cj}, \cite{Eguchi:2000tc}, the chiral deformation corresponds to the complex moduli deformation and the nonchiral deformation corresponds to the K\"ahler moduli deformation. Therefore, if these deformations yield the same results, we should expect a mirror type duality. Perhaps, this conjectured duality, if any, should involve the flipping of the GSO projection, or may be true only in the limited sector. Furthermore, if we restrict ourselves to the strings propagating in the singular ALE space, the deformation of the singularity by the complex deformation and the K\"ahler deformation is actually the same thing (see appendix \ref{b-7} for a discussion). Therefore we expect the deformation by the real cosmological constant $\mu$ is related to the nonchiral deformation of the K\"ahler potential.\footnote{However, it is not clear whether we can deform the theory by the complex ($\theta$ term like) K\"ahler parameter without violating the superconformal properties.}

\end{itemize}

In the following, we assume this duality and obtain the reflection amplitude by using Teschner's trick. Concentrating on the neutral sector $\alpha = \bar{\alpha}$, we have basic degenerate operators
\begin{eqnarray}
N_{-\frac{b}{2}} &=& e^{-b\phi} \cr
R_{-\frac{1}{4b}}^\pm &=& \sigma^{\pm}e^{-\frac{1}{2b}\phi}.
\end{eqnarray}
The OPE of any NS field with the degenerate operator $N_{-\frac{b}{2}} $ is given by
\begin{equation}
N_{\alpha} N_{-b/2} = N_{\alpha-b/2} + C^N_-(\alpha) N_{\alpha+b/2},
\end{equation}
where the fusion coefficient $C^N_-$ can be calculated by the usual perturbation saturation assumption as 
\begin{eqnarray}
C^N_- &=& \mu \bar{\mu} \langle e^{(\frac{1}{b} -2 \alpha -b)\phi}(\infty) e^{2\alpha\phi}(1) e^{-b\phi}(0) \cr
&& \int d^2z_2d^2z_1 \psi^+(z_1)\psi^-(z_1) e^{b(\phi + iY)}(z_1) \bar{\psi}^+(z_2)\bar{\psi}^-(z_2) e^{b(\phi -iY)} (z_2)\rangle_{free}.
\end{eqnarray}

Performing the free field path integral and integration over $z_1,z_2$ (the Dotsenko-Fateev formula \eqref{eq:DFF} is useful to evaluate the integration), we have
\begin{equation}
C^N_-(\alpha) = \kappa_1 \gamma (1-2\alpha b) \gamma(1/2 - 2\alpha b -b^2) \gamma(-1/2 + 2\alpha b)\gamma(2\alpha b + b^2),
\end{equation}
where $\kappa_1$ is an $\alpha$ independent constant which is proportional to $\mu\bar{\mu}$. The two-point functions we would like to determine can be written as 
\begin{eqnarray}
\langle N_{\alpha}(z) N_{\alpha}(0) \rangle &=& \frac{D^N(\alpha)}{|z|^{4\Delta^N_\alpha}} \cr
\langle R_{\alpha}^+(z) R_{\alpha}^-(0) \rangle &=& \frac{D^R(\alpha)}{|z|^{4\Delta^R_\alpha}}. 
\end{eqnarray}

 If we consider an auxiliary three-point function $\langle N_{\alpha+b/2} N_{\alpha} N_{-b/2} \rangle$, we have the following functional relation:
\begin{equation}
C_-^N(\alpha) D^N(\alpha+b/2) = D^N(\alpha).
\end{equation}
Of course its solution is not unique. To obtain the dual functional relation, we need a dual action assumption. Consider the following OPE
\begin{eqnarray}
N_{\alpha} R^+_{-1/4b} &=& R^+_{\alpha-1/4b} + \tilde{C}_-^N(\alpha) R^+_{\alpha+1/4b} \cr
R^-_{\alpha} R^+_{-1/4b} &=& N_{\alpha-1/4b} + \tilde{C}_-^R(\alpha) N_{\alpha+1/4b}. 
\end{eqnarray}
The first order insertion of the dual action determines the structure constants as 
\begin{eqnarray}
\tilde{C}^N_- (\alpha) &=& \kappa_2 \frac{\gamma(2\alpha/b-1/2b^2)}{\gamma(2\alpha/b)} \cr
\tilde{C}^R_- (\alpha) &=& \kappa_2 \frac{\gamma(2\alpha/b-1/2b^2+1)}{\gamma(2\alpha/b+1)}, 
\end{eqnarray}
where $\kappa_2$ is proportional to the dual cosmological constant $\tilde{\mu}$. If we consider auxiliary three-point functions, $\langle R^-_{\alpha+1/4b} N_\alpha R^+_{-1/4b} \rangle $ and $\langle N_{\alpha+1/4b} R^-_\alpha R^+_{-1/4b} \rangle $, we have 
\begin{eqnarray}
\tilde{C}^N_- (\alpha) D^R(\alpha+1/4b) &=& D^N(\alpha) \cr
\tilde{C}^R_- (\alpha) D^N(\alpha +1/4b) &=& D^R(\alpha).
\end{eqnarray}
These are the second functional relations. Solving these, we obtain
\begin{eqnarray}
D^N(P) &=& \hat{\mu}^{-2i P/b} \frac{\Gamma(1+\frac{iP}{b}) \Gamma(1+iPb) \Gamma(\frac{1}{2}-iPb)}{\Gamma(1-\frac{iP}{b})\Gamma(1-iPb)\Gamma(\frac{1}{2}+iPb)} \cr
D^R(P) &=& \hat{\mu}^{-2i P/b} \frac{\Gamma(1+\frac{iP}{b}) \Gamma(1-iPb) \Gamma(\frac{1}{2}+iPb)}{\Gamma(1-\frac{iP}{b})\Gamma(1+iPb)\Gamma(\frac{1}{2}-iPb)}, 
\end{eqnarray}
where $\hat{\mu}$ is a renormalized cosmological constant which is proportional to $\mu$, whose precise form can be found in the original paper \cite{Ahn:2002sx}, and we have introduced the momentum $P$ as in (\ref{eq:2momen}). For the later purpose, it is important to note that this can be rewritten as
\begin{equation}
D^N(P) =  \hat{\mu}^{-2i P/b} \frac{\Gamma(\frac{iP}{b}) \Gamma(1+2iPb) \Gamma(\frac{1}{2}-iPb)^2}{\Gamma(-\frac{iP}{b})\Gamma(1-2iPb)\Gamma(\frac{1}{2}+iPb)^2} \label{eq:refn2}
\end{equation}
by using the Legendle duplication formula (\ref{eq:ledgam}) up to an irrelevant sign.

Some consistency checks of the results are done in the original paper \cite{Ahn:2002sx}. We will see in the next section that this reflection property is consistent with the one-point function on the disk.

\subsection{Branes in $\mathcal{N}=2$ Super Liouville Theory}\label{11.2}
In this section, we derive the Cardy states for the branes in the $\mathcal{N}=2$ super Liouville theory. The essential difficulty here is the lack of the $b \to b^{-1}$ duality, so Teschner's trick does not determine the one-point function uniquely. Nevertheless, we recall that in the bosonic and $\mathcal{N}=1$ Liouville theory, we have another approach which yields the same result without resorting to the duality argument. This has been done by what is called the modular bootstrap method \cite{Eguchi:2003ik}, \cite{Ahn:2003tt}.

The basic strategy of the modular bootstrap (see section \ref{5.2.2}, \ref{8.2} and \ref{8.3} for example) is that we first assume the open spectrums and then modular transform them into the closed exchange channel to obtain the Cardy boundary states (the boundary wavefunctions, or equivalently the disk one-point functions). The identity representation was modular transformed into the $(1,1)$ ZZ brane and the general nondegenerate representation was transformed into the FZZT brane in our previous case. Of course, this method does not guarantee that these candidate Cardy states really satisfy the Cardy condition. We have to check this in the end, but from the experience with the bosonic and $\mathcal{N}=1$ Liouville theory, we expect this is the case.

\subsubsection{Modular transformation of $\mathcal{N}=2$ character}\label{11.2.1}
To utilize the modular bootstrap method to obtain the Cardy states, we need to know the modular transformation properties of the basic $\mathcal{N}=2$ representations. In this section we review their properties \cite{Boucher:1986bh}, \cite{Dobrev:1987hq}, \cite{Eguchi:2003ik}. For simplicity, we concentrate on the NS sector because the characters of other spin structures can be obtained by the spectral flows. We denote the conformal dimension and $U(1)$ charge as $h,Q$ and set $ q=e^{2\pi i\tau}$, $y = e^{2\pi iz}$.

The unitary representation\footnote{To construct a physically meaningful theory, we here deal with the unitary representation exclusively. Note that this requirement excludes the branes such as general $(m,n) \neq(1,1)$ ZZ branes.} of the $\mathcal{N} =2 $ algebra is classified into the following three types.

\begin{enumerate}
	\item massive representations :
	\begin{equation}
	\mathrm{ch}^{NS}(h,Q;\tau,z) = q^{h-(\tilde{c}-1)/8} y^Q \frac{\theta_3(\tau,z)}{\eta(\tau)^3}
	\end{equation}
which is spanned by operating $L_{-n}$,$J_{-n}$,$G^{\pm}_{-r}$ to the highest primary states which is not annihilated by any of these. The unitarity demands $h>|Q|/2$, $0\le |Q| < \tilde{c}-1$. In the closed sector, the general vertex operator $e^{\alpha S + \bar{\alpha} S^*}$ belongs to this representation.\footnote{The second unitarity condition seems to restrict the allowable $U(1)$ charge, but actually it is not the case. By using the integral spectral flow (\ref{eq:spcfl}), we obtain $Q \to Q +(\tilde{c}-1)n$, and this generates other $U(1)$ charge sectors. Also in the chiral sector, we have only finite operators in the physical theory because of this bound. However, once we twist the theory, this bound is removed, so for instance the analysis in \cite{Mukhi:1993zb} is consistent with the unitarity.}
	\item massless matter (chiral or anti-chiral) representations :
	\begin{equation}
	\mathrm{ch}_M^{NS}(Q;\tau,z) = q^{\frac{|Q|}{2}-(\tilde{c}-1)/8} y^Q \frac{\theta_3(\tau,z)}{(1+y^{\mathrm{sgn}(Q)}q^{1/2})\eta(\tau)^3}
	\end{equation}
which is defined by $G^{+}_{-1/2} |Q\rangle = 0$ or $G^{-}_{-1/2} |Q\rangle = 0$ for chiral and anti-chiral respectively. The unitarity demands $h=|Q|/2$, $0\le |Q| < \tilde{c}$. In the closed sector, the general vertex operator $e^{\alpha S}$ belongs to this representation.
\item graviton (vacuum) representations :
	\begin{equation}
	\mathrm{ch}_G^{NS}(\tau,z) = q^{-(\tilde{c}-1)/8} \frac{(1-q)\theta_3(\tau,z)}{(1+yq^{1/2})(1+y^{-1}q^{1/2})\eta(\tau)^3}
	\end{equation}
which is defined by $G^{\pm}_{-1/2} |1\rangle = 0$ and $L_{-1} |1\rangle = 0$. This corresponds to the unique identity operator in the theory.
\end{enumerate}

The more general representations are generated by the spectral flows (\ref{eq:spcfl}). Under these flows, the characters transform as 
\begin{equation}
\mathrm{ch}^{NS}(n;\tau,z) = q^{\frac{\tilde{c}}{2}n^2} y^{\tilde{c}n} \mathrm{ch}^{NS} (\tau,z+n\tau),
\end{equation}
for an integer $n$.

The modular transformation of these characters are given in \cite{Miki:1990ri}, \cite{Eguchi:2003ik}. The easiest one is the massive representation, where we can simply use the modular transformation formula in appendix \ref{a-4}. We label the massive representation as $ h = \frac{p^2}{2} + \frac{w^2b^2}{2} + \frac{1}{8b^2}$ as in (\ref{eq:masdim}). Then the modular transformation is given by
\begin{eqnarray}
\mathrm{ch}^{NS} \left(\frac{p^2}{2} + \frac{w^2b^2}{2} + \frac{1}{8b^2},w;-\frac{1}{\tau},\frac{z}{\tau}\right) = e^{i\pi\frac{\tilde{c}z^2}{\tau}} b \int_{-\infty}^{\infty} dw' \int_{-\infty}^\infty dp'\cr
e^{-2\pi i ww'b^2} \cos (2\pi pp') \mathrm{ch}^{NS} \left(\frac{{p'}^2}{2} + \frac{{w'}^2b^2}{2} + \frac{1}{8b^2},w';\tau,z\right).
\end{eqnarray}
On the other hand, the modular transformations of the massless matter and graviton characters are quite nontrivial\footnote{At first sight, one might try to expand the denominator as $\frac{1}{1+yq^{1/2}} = \sum_{n=0}^\infty (-1)^n(yq^{1/2})^n $ and modular transform term by term (this can be done by the usual formula) and resum again. However, this does not yield the correct results because the expanded characters contain infinitely heavy tachyon (imaginary momentum) contributions and the modular transformation is ill-defined and divergent (see the similar argument in section \ref{5.2.2}). Indeed, the careful evaluation of the modular transformation requires an additional massless matter sector contribution. One way to see its necessity is to make a spectral flow and consider the $\tilde{R}$ sector. The massless and graviton sector can possess a nonzero Witten index, whereas on the other hand, if it were not for the contribution from the massless sector, the right-hand side would have a zero Witten index from the massive sector only.} and given by (see \cite{Miki:1990ri}, \cite{Eguchi:2003ik} for the details)
\begin{align}
\mathrm{ch}_M^{NS} \left(\lambda,n;-\frac{1}{\tau},\frac{z}{\tau}\right) = e^{i\pi\frac{\tilde{c}z^2}{\tau}} \left[\frac{b}{2} \int_{-\infty}^{\infty} dw' \int_{-\infty}^\infty dp' e^{-2\pi i(b^2\lambda + n)w'} \right.\cr
\frac{\cosh(2\pi p'b(1+\frac{1}{2b^2}-\lambda)) +e^{2\pi i b^2 w'} \cosh(2\pi p'b(\lambda-\frac{1}{2b^2}))}{2|\cosh\pi(p'b+iw'b^2)|^2} \mathrm{ch}^{NS} \left(\frac{{p'}^2}{2} + \frac{{w'}^2b^2}{2} + \frac{1}{8b^2},w';\tau,z\right) \cr
\left.+ i \sum_{n' \in \mathbf{Z}} \int_{\frac{1}{2b^2}}^{1+\frac{1}{2b^2}} d\lambda' e^{-2\pi ib^2\{(\lambda+n/b^2)(\lambda' + n'/b^2)-(\lambda-1/2b^2)(\lambda' -1/2b^2)\}}\mathrm{ch}_M^{NS} \left(\lambda',n';\tau,z\right)\right],
\end{align}
where $n \in \mathbf{Z}$ specifies the integer spectral flow and $\lambda$ is the $U(1)$ charge.

Similarly for the graviton representations, we have
\begin{eqnarray}
\mathrm{ch}_G^{NS}\left(n;-\frac{1}{\tau},\frac{z}{\tau}\right) = e^{i\pi\frac{\tilde{c}z^2}{\tau}}\left[\frac{b}{2}\int_{-\infty}^{\infty} dw' \int_{-\infty}^{\infty} dp' e^{-2\pi i nw'} \right. \cr
\frac{\sinh(\pi p'/b)\sinh(2\pi bp')}{|\cosh\pi(p' b + iw'b^2)|^2}  \mathrm{ch}^{NS} \left(\frac{{p'}^2}{2} + \frac{{w'}^2b^2}{2} + \frac{1}{8b^2},w';\tau,z\right) \cr 
\left. +2\sum_{n'\in \mathbf{Z}} \int_{\frac{1}{2b^2}}^{1+\frac{1}{2b^2}} d\lambda' \sin(\pi(\lambda'-1/2b^2)) e^{-2\pi i n(\lambda'+n'/b^2)}\mathrm{ch}_M^{NS} \left(\lambda',n';\tau,z\right)\right].
\end{eqnarray}

However, this is not the end of the story. As has been stressed in \cite{Eguchi:2003ik}, the right-hand side of the modular transformation above has a continuous $U(1)$ charge spectrum and is not compatible with the local spacetime supercharge. Let us explain this point further. Suppose we compactify $Y$ and hence we have a discrete $U(1)$ charge in the closed sector (as we have stated in the last subsection, this is necessary to obtain the local spacetime supercharge operator). In this case, the Ishibashi states made up with the closed states must have a discrete $U(1)$ charge, but if we have such an open character as we have seen above, the $U(1)$ charge in the \textit{closed} exchange channel necessarily contains a continuous $U(1)$ charge, which is not what we want. It seems hopeless to obtain a discrete $U(1)$ charge in the exchange channel by combining the open character by trial and error. Fortunately, for the rational central charge $\tilde{c} = 1+\frac{2K}{N}$ with integers $K$ and $N$, the ``extended character" which preserves the discrete $U(1)$ charge under the modular transformation has been introduced in \cite{Eguchi:2003ik} (see also \cite{Eguchi:1988wf}, \cite{Eguchi:1988af}). The key idea is that we sum over the spectral flow.

The extended characters are defined as
\begin{eqnarray}
\chi^{NS}(h_0,r,j_0;\tau,z) &\equiv& \sum_{n\in r +N \mathbf{Z}} q^{\frac{\tilde{c}}{2}n^2} y^{\tilde{c}n} \mathrm{ch}^{NS}\left(h_0,Q=\frac{j_0}{N};\tau,z+n\tau\right) \cr
\chi_M^{NS}(r,s;\tau,z) &\equiv& \sum_{n\in r +N \mathbf{Z}} q^{\frac{\tilde{c}}{2}n^2} y^{\tilde{c}n} \mathrm{ch}_M^{NS}\left(Q=\frac{s}{N};\tau,z+n\tau\right) \cr
\chi_G^{NS}(r;\tau,z) &\equiv& \sum_{n\in r +N \mathbf{Z}} q^{\frac{\tilde{c}}{2}n^2} y^{\tilde{c}n} \mathrm{ch}_G^{NS}\left(\tau,z+n\tau\right),
\end{eqnarray}
where $r,j_0, s$ are the reference spectral flow parameters and $U(1)$ charge and their ranges are 
\begin{equation}
r \in \mathbf{Z}_N, \ \ 0 \le j_0 \le 2K-1, \ \ 1\le s \le N+2K-1.
\end{equation}
Furthermore, we take $j_0$ and $s$ as integers in order to ensure the locality of the spectral flow generators. For the massive representations, we use the following convenient parameterization:
\begin{eqnarray}
h &\equiv& h_0 + \frac{rj_0}{N} + \frac{Kr^2}{N} = \frac{p^2}{2} +\frac{j^2+K^2}{4NK}\cr
j &\equiv& j_0 + 2Kr,
\end{eqnarray}

The modular transformations of the extended characters are given by
\begin{equation}
\chi^{NS}\left(p,j;-\frac{1}{\tau},\frac{z}{\tau} \right)= e^{i\pi\frac{\tilde{c}z^2}{\tau}} \frac{1}{\sqrt{2NK}} \sum_{j' \in \mathbf{Z}_{2NK}} e^{-2\pi i \frac{jj'}{2NK}} \int_{-\infty}^{\infty} dp' \cos(2\pi pp') \chi^{NS}(p',j';\tau,z),
\end{equation}
for the extended massive representations, and
\begin{eqnarray}
\chi_M^{NS}\left(r,s;-\frac{1}{\tau},\frac{z}{\tau}\right) =  e^{i\pi\frac{\tilde{c}z^2}{\tau}} \left[\frac{1}{2\sqrt{2NK}} \sum_{j' \in \mathbf{Z}_{2NK}} e^{-2\pi i\frac{(s+2Kr)j'}{2NK}} \int_{-\infty}^{\infty} dp' \right. \cr
\frac{\cosh(2\pi bp'\frac{N+K-s}{N}) + e^{i\frac{\pi j'}{K}} \cosh (2\pi bp'\frac{s-K}{N})}{2|\cosh\pi(p'b + i\frac{j'}{2K})|^2} \chi^{NS}(p',j';\tau,z) \cr
+\frac{i}{N} \sum_{r'\in \mathbf{Z}_N} \sum_{s'=K+1}^{N+K-1} e^{-2\pi i\frac{(s+2Kr)(s'+2Kr')-(s-K)(s'-K)}{2NK}} \chi_M^{NS}(r',s';\tau,z) \cr
+\left.\frac{i}{2N} \sum_{r'\in \mathbf{Z}_N} e^{-2\pi i\frac{(s+2Kr)(2r'+1)}{2N}} \{\chi_M^{NS}(r',K;\tau,z) - \chi_M^{NS}(r',N+K;\tau,z) \}\right]
\end{eqnarray}
for the massless matter representations, and finally
\begin{align}
\chi^{NS}_G\left(r;-\frac{1}{\tau},\frac{z}{\tau}\right) =  e^{i\pi\frac{\tilde{c}z^2}{\tau}} \left[\frac{1}{2\sqrt{2NK}} \sum_{j' \in \mathbf{Z}_{2NK}} e^{-2\pi i\frac{rj'}{N}} \int_{-\infty}^{\infty} dp' \frac{\sinh(\pi p'/b) \sinh(2\pi p'b)}{|\cosh\pi (p' b+i\frac{j'}{2K})|^2} \chi^{NS}(p',j';\tau ,z) \right.\cr
\left.+\frac{2}{N} \sum_{r' \in \mathbf{Z}_N} \sum_{s'=K+1}^{N+K-1} \sin\left(\frac{\pi(s'-K)}{N}\right) e^{-2\pi i\frac{r(s'+2Kr')}{N}} \chi^{NS}_M(r',s';\tau,z) \right]
\end{align}
for the graviton representations. Using these modular transformation properties, we will derive the disk one-point functions in the following subsections.
\subsubsection{Boundary conditions and Ishibashi states}\label{11.2.2}
The $\mathcal{N} = 2 $ superconformal symmetry allows the following two types of boundary conditions \cite{Ooguri:1996ck}; for the A-type, we have
\begin{eqnarray}
(J_n - \bar{J}_{-n}) |B\rangle &=& 0\cr
(G^{\pm}_r - i\eta\bar{G}_{-r}^{\mp}) |B\rangle &=& 0,
\end{eqnarray}
with $\eta$ is either $+1$ or $-1$. For the B-type, we have
\begin{eqnarray}
(J_n + \bar{J}_{-n}) |B\rangle &=& 0\cr
(G^{\pm}_r - i\eta\bar{G}_{-r}^{\pm}) |B\rangle &=& 0.
\end{eqnarray}
Both of them are compatible with the $\mathcal{N} =1 $ superconformal symmetry which we would like to gauge
\begin{eqnarray}
(L_n - \bar{L}_{-n}) |B\rangle &=& 0\cr
(G_r - i\eta\bar{G}_{-r})|B\rangle &=& 0,
\end{eqnarray}
where $G = G^+ + G^{-}$. Roughly speaking, the $U(1)$ condition states that the A type imposes the ``Dirichlet boundary condition" on the $Y$ coordinate, and the $B$ type imposes the ``Neumann boundary condition" on the $Y$ coordinate. The $\eta$ dependence simply yields the spin structures and they are almost trivial, so we ignore these factors in the following.\footnote{Note that if the (type II) space-time supersymmetry is possible, we should have both $\eta = \pm$ sector with the \textit{same} wavefunction. In a sense, it is guaranteed by the spectral flow unlike in the $\mathcal{N}=1$ theory.}

The Ishibashi states of the extended character are defined as (for example, we take the A-type brane, for the B type we should replace $J_0+\bar{J}_0$ with $J_0-\bar{J}_0$.)
\begin{eqnarray}
\langle p,j|e^{-\pi t_c H_c } e^{i\pi z(J_0+\bar{J}_0)} |p',j'\rangle &=& \delta(p-p') \delta^{(2NK)}_{j,j'} \chi(p,j;it_c,z) \cr
\langle r,s;M|e^{-\pi t_c H_c } e^{i\pi z(J_0+\bar{J}_0)} |r',s';M\rangle &=& \delta_{r,r'}^{(N)}\delta_{s,s'}\chi_M(r,s;it_c,z) \cr
\langle p,j|e^{-\pi t_c H_c } e^{i\pi z(J_0+\bar{J}_0)} |r,s;M\rangle &=& 0,
\end{eqnarray}
which are constructed from the primary states appearing in the right-hand side of the modular transformation formulae (particularly, we have $K+1 \le s \le K+N-1$). Then the Cardy boundary states should be expanded as 
\begin{equation}
\langle B,\xi | = \int_{-\infty}^{\infty}dp \sum_{j\in \mathbf{Z}_{2NK}} \Psi_\xi (p,j)\langle p,j| + \sum_{r\in \mathbf{Z}_N} \sum_{s=K+1}^{N+K-1} C_\xi (r,s) \langle r,s;M|.
\end{equation}
In the next subsection, we determine the wavefunctions $\Psi_\xi(p,j)$ and $C_\xi(r,s)$ via the modular bootstrap method. To relate the results to the Liouville boundary parameter, we need the suitable boundary action for the $\mathcal{N} =2 $ Liouville theory. This is given in \cite{Ahn:2003wy} and we just quote the result:
\begin{eqnarray}
S_B &=& \int_{-\infty}^{\infty} dx \left[ -\frac{i}{8\pi}(\bar{\psi}^+\psi^- + \psi^+\bar{\psi}^-) + \frac{1}{4} a\partial_x\bar{a} \right. \cr
& &-\frac{1}{4}e^{bS^*/2} \left(\mu_Ba +\frac{\mu b^2}{2\mu_B} \bar{a}\right) (\psi^+ + \psi^-) - \frac{1}{4}e^{bS/2} \left(\mu_B\bar{a} +\frac{\mu b^2}{2\mu_B} a\right) (\bar{\psi}^+ + \bar{\psi}^-) \cr
& &\left.-\frac{1}{b^2}\left(\mu_B^2 + \frac{\mu^2 b^4}{4\mu_B^2}\right) e^{b(S+S^*)/2}\right],
\end{eqnarray}
where $a$ and $\bar{a}$ are (complex) boundary fermion operators and we have assumed the B-type boundary condition here. We should also note that this boundary action is not enough to yield the FZZT type disk one-point function unlike in the bosonic or $\mathcal{N}=1$ Liouville theory due to the lack of the duality. However we could use this boundary action to relate the boundary parameter $\mu_B$ to the Cardy states continuous parameter as we will see in the next subsection.
\subsubsection{Cardy boundary states}\label{11.2.3}
Now it is time to obtain the Cardy boundary states for the $\mathcal{N}=2$ super Liouville theory. The modular bootstrap assumption\footnote{Whether these branes obtained via the modular bootstrap satisfy the Cardy condition or not is a nontrivial problem. In \cite{Eguchi:2003ik}, various cylinder amplitudes are calculated and the positivity of the spectral density is confirmed (at least for the particular parameter region).} is given by
\begin{eqnarray}
e^{\pi\frac{\tilde{c}z^2}{\tau_c}} \langle B;O|e^{-\pi \tau_c H_c} e^{i\pi z(J_0+\bar{J}_0)} |B;\xi\rangle &=& \chi_{\xi}(it_o,z') \cr
e^{\pi\frac{\tilde{c}z^2}{\tau_c}} \langle B;O|e^{-\pi \tau_c H_c} e^{i\pi z(J_0+\bar{J}_0)} |B;O\rangle &=& \chi_G(r=0;it_o,z'), \label{eq:mobta}
\end{eqnarray}
where $t_o = \frac{1}{\tau_c}$ and $z' = -it_oz$. By using the modular transformation properties in section \ref{11.2.1}, we have \cite{Eguchi:2003ik}, \cite{Ahn:2003tt}
\begin{eqnarray}
\Psi_O(p,j) &=& \frac{b}{2} \left(\frac{2}{NK}\right)^{1/4} \frac{\Gamma(\frac{1}{2}+\frac{j}{2K}-ipb)\Gamma(\frac{1}{2}-\frac{j}{2K}-ipb)}{\Gamma(-ip/b)\Gamma(1-2ipb)} \cr
C_O(r,s) &=& \sqrt{\frac{2}{N}} \sqrt{\sin(\frac{\pi(s-K)}{N})},
\end{eqnarray}
which is determined up to a (possibly $p$ dependent) phase factor and a complex conjugation. This can be fixed by demanding the reflection property which states
\begin{equation}
\Psi_O(p,j) = R(p,j)\Psi_O(-p,j),
\end{equation}
where the reflection amplitude $R(p,0)$ is given in (\ref{eq:refn2}). For nonzero $j$, it is given by 
\begin{equation}
R(p,j) =  \hat{\mu}^{-2i p/b} \frac{\Gamma(\frac{ip}{b}) \Gamma(1+2ipb) \Gamma(\frac{1}{2}-ipb+\frac{j}{2K})\Gamma(\frac{1}{2}-ipb-\frac{j}{2K})}{\Gamma(-\frac{ip}{b})\Gamma(1-2ipb)\Gamma(\frac{1}{2}+ipb+\frac{j}{2K})\Gamma(\frac{1}{2}+ipb-\frac{j}{2K})},
\end{equation}
which was first derived in \cite{Baseilhac:1998eq}.
 It also determines the overall phase to be $\tilde{\mu}^{-ipb^{-1}}$, where $\tilde{\mu}$ is the renormalized cosmological constant which is proportional to $\mu$. This boundary state corresponds to the $(1,1)$ ZZ brane in the bosonic Liouville theory and plays a fundamental role to derive other boundary states. In \cite{Eguchi:2003ik}, three classes of boundary states are proposed: the class 1 corresponds to the graviton representations, the class 2 corresponds to the massive matter representations and the class 3 corresponds to the massless matter representations.

We can easily derive the other boundary wavefunctions from the modular bootstrap (\ref{eq:mobta}) and the actual form of the modular transformation found in subsection \ref{11.2.1}. For example, the trivial one is given by the class 1 general graviton representations $|B;r\rangle$ which is given by
\begin{equation}
\Psi_r(p,j) = e^{-2\pi i rj}\Psi_O(p,j).
\end{equation}
The class 2 boundary wavefunction (for the continuum part) is given by
\begin{equation}
\Psi(p',j';p,j) = \tilde{\mu}^{-ipb^{-1}} \frac{1}{2b}\left(\frac{2}{NK}\right)^{1/4} \frac{\Gamma(ipb^{-1})\Gamma(1+2ipb)}{\Gamma(\frac{1}{2}+\frac{j}{2K}+ipb)\Gamma(\frac{1}{2}-\frac{j}{2K}+ipb)} e^{-2\pi i\frac{jj'}{2NK}} \cos(2\pi p p'),
\end{equation}
which satisfies the reflection property. The relation between the boundary cosmological constant $\mu_B$ and the boundary parameter $p'$ is given in \cite{Ahn:2003tt} by comparing the pole structure with the perturbative calculation; for $j'=0$, we have
\begin{equation}
\left(\mu_B^2 + \frac{\mu^2b^4}{4\mu_B^2}\right) =\frac{\mu b}{32\pi}\cosh(2\pi p' b).
\end{equation}

We will discuss the application of these boundary states to the supersymmetric cycle of the singular Calabi-Yau space in the next section, but for the moment in this subsection, we discuss the $\tilde{c} = 5$ theory (two dimensional noncritical string theory) either in type 0 or type II \cite{Eguchi:2003ik}.

For the type 0 string, we can choose the world sheet fermion GSO projection:
\begin{equation}
J_0 + \bar{J}_0 \in 2 \mathbf{Z}
\end{equation}
for the type 0A theory or
\begin{equation}
J_0 - \bar{J}_0 \in 2 \mathbf{Z}
\end{equation}
for the type 0B theory. For the type 0B theory, we have R-R charged A-branes and non R-R charged B-branes. The A-branes have Dirichlet boundary condition on the $Y$ direction\footnote{We can (and should in the supersymmetric case) compactify the $Y$ direction, and regard it as the Euclidean time \cite{Sugawara:2003xt}. The Wick rotation in this case is very subtle especially in the supersymmetric case.}, so the class 1 A-brane is the D0-instanton and class 2,3 A-brane is the D1-instanton. Schematically, we have 
\begin{equation}
|B \pm\rangle_A = |B\rangle^{NS} \pm |B\rangle^R,
\end{equation}
where $+$ corresponds to the BPS(-like\footnote{Since space-time SUSY is missing in the type 0 theory, the ``BPS" here does not have any deeper meaning than branes without an open tachyon.}) brane and $-$ corresponds to the antibrane.
These are stable in the sense that the tachyon is projected out by the open GSO projection.\footnote{Unlike in the $\mathcal{N}=1$ theory, the open GSO projection for the class 2,3 branes can be imposed properly. The essential reason is the existence of the spectral flow. At first sight, since the wavefunction for the R sector is different from the NS sector (for example, compare eq (4.12) and (4.15) in \cite{Ahn:2003tt}), the proper GSO projection does not seem to be imposed. However, if we actually calculate the density of states, the integration (summation) over the $U(1)$ charge makes them equal (up to a sign which is related to the more involved GSO projection involving the $U(1)$ charge). The author would like to thank Y.~Sugawara for teaching the author this point.} Of course, if we consider the D-$\bar{\mathrm{D}}$ system, we have open ``tachyon" (for the class 2, this is actually a massless tachyon). We should note that we have both even and odd dimensional ``stable" branes in this case as we have seen in the $\hat{c}=1$, $\mathcal{N}=1$ super Liouville theory.

The B-branes have Neumann boundary condition on the $Y$ direction, and in the type 0B theory, we cannot have R-R Ishibashi states for this boundary condition. Therefore, the class 1 states which correspond to the D0-branes have a genuine tachyon in the spectrum and this should yield the matrix dual of the type 0B $\mathcal{N} = 2$ super Liouville theory in the double scaling limit. As we will see in the next section, for the minimum radius $R=2$ $(M=1)$, the dual theory is proposed to be the symmetric version of the KKK matrix model \cite{Giveon:2003wn} though we cannot see any reason why the vortex sector should be included from this holographic perspective.
For the type 0A theory, the ``stable" and ``unstable" branes are just reversed. We have stable (with R-R sector) branes for B-branes (D0 and D1) and unstable (without R-R sector) branes for A-branes (D0 and D1 instanton).

Finally we consider the type IIB case. If we impose the spacetime SUSY, we should have the minimum radius $R=2$. The boundary states for the A-branes are BPS and their properties are similar to those of the type 0B theory (with fermion sector). However note that the BPS branes exist both for odd dimension and even dimension as opposed to the naive free field guess. The more interesting boundary states are the B-branes which are non-BPS. For the class 1 brane, which is the D0-brane and supposed to play a matrix holographic dual role in the double scaling limit, we find an open tachyon and an open massless fermion in the spectrum \cite{McGreevy:2003dn}, \cite{Eguchi:2003ik}. As we will discuss in the next section, this is just the spectrum of the Marinari-Parisi super matrix model expanded around the nonsupersymmetric vacuum \cite{McGreevy:2003dn}.
\subsection{Matrix Model Dual}\label{11.3}
In this section, we discuss the matrix dual theory for the $\tilde{c} = 5$ two dimensional type II $\mathcal{N}=2$ super Liouville theory.\footnote{We have called the two dimensional noncritical string ``$c =1$" and ``$\hat{c}=1$" referring to the central charge of the matter section, but for the $\mathcal{N}=2$ super Liouville theory, the two dimension means no other matter sector. This leads to a somewhat different naming ``$\tilde{c} = 5$" theory referring to the central charge of the Liouville part.} The matrix model proposal for the discretized superstring was proposed in \cite{McGreevy:2003dn} and we will first review its properties in section \ref{11.3.1} and \ref{11.3.2}. Then we compare it with the continuum string theory in section \ref{11.3.3}.

\subsubsection{Marinari-Parisi model}\label{11.3.1}
We will review the basic facts about the Marinari-Parisi model \cite{Marinari:1990jc}. In this subsection, we follow the argument given by Dabholkar \cite{Dabholkar:1992te} (see also \cite{Rodrigues:1993by}), who truncated the model onto the diagonal gauge singlet sector as we will see. From the holographic point of view, this truncation is natural because it has been shown \cite{McGreevy:2003dn} that the gauged Marinari-Parisi model which is naturally obtained from the one dimensional theory on the D0-branes just corresponds to the Dabholkar's truncation.

Ungauged original Marinari-Parisi model has the following action
\begin{equation}
 S = -N\int dtd\bar{\theta}d\theta \mathrm{Tr} \left[\frac{1}{2}\bar{D}\Phi D\Phi + W_0 (\Phi)\right],
\end{equation}
where $\Phi$ is an Hermitian matrix valued superfield
\begin{equation}
\Phi = M + \bar{\theta} \Psi + \bar{\Psi} \theta + \theta\bar{\theta} F,
\end{equation}
and $D= \partial_{\bar{\theta}} + \theta \partial_t $, $\bar{D} = - \partial_{\theta}-\bar{\theta}\partial_t$ are super covariant derivatives. The Hamiltonian is given by
\begin{equation}
H = \frac{1}{2} \mathrm{Tr} \left(P^2 + \frac{\partial W_0(M)}{\partial M^*}\frac{\partial W_0(M)}{\partial M}\right) + \sum_{ijkl}[\Psi_{ji}^*,\Psi_{kl}]\frac{\partial^2W_0(M)}{\partial M^*_{ij}\partial M_{kl}},
\end{equation}
and the supercharges are given by
\begin{eqnarray}
Q &=& \sum_{ij} \Psi_{ij}^* \left(P^*_{ij}- i\frac{\partial W_0(M)}{\partial M^*_{ij}}\right), \cr
Q^\dagger &=& \sum_{ij} \Psi_{ij} \left(P_{ij} +i\frac{\partial W_0(M)}{\partial M_{ij}}\right).
\end{eqnarray}
The truncation done by Dabholkar is as follows. We diagonalize $\Phi$ by a unitary matrix. In general, this does \textit{not} diagonalize $\Psi$ and $F$ simultaneously (see \cite{Ferretti:1994fu} for a related discussion). Nevertheless, we treat them as if they were diagonalized,
\begin{equation}
(U\Phi U^\dagger)_{ii} = \lambda_i + \bar{\theta} \psi_i + \psi_i^\dagger \theta +\bar{\theta}\theta f_i.
\end{equation}
This truncated theory turns out to be a consistent subspace of the whole theory and makes a proper sense. The reduced Hilbert space is spanned by the following states:
\begin{equation}
f(\lambda) \prod_{k}\psi^\dagger_{m_k}|0\rangle.
\end{equation}
On these states, the supercharges act as
\begin{eqnarray}
Q &=& \sum_i \psi^\dagger_i \left(-i\frac{\partial}{\partial \lambda_i} - i\frac{\partial W_0}{\partial\lambda_i}\right) \cr
\bar{Q} &=& \sum_i \psi_i\left(-i\frac{\partial}{\partial \lambda_i} + i\frac{\partial W_0}{\partial\lambda_i}-i\sum_{l\neq i} \frac{1}{\lambda_i-\lambda_l}\right).
\end{eqnarray}
At first sight, this expression for the supercharges seems peculiar and looks inconsistent because $\bar{Q}$ is not Hermitian conjugate of $Q$ in a naive sense. However, this is because the innerproduct we should use is a nonstandard one due to the Jacobian from the change of variables. It is convenient to use a standard innerproduct by rescaling the wavefunction as $\phi \to J^{1/2} \phi$ and operators as $O \to J^{1/2} O J^{-1/2}$. With these rescaling we finally have
\begin{eqnarray}
Q &=& \sum_i \psi^\dagger_i \left(-i\frac{\partial}{\partial \lambda_i} - i\frac{\partial W}{\partial\lambda_i}\right) \cr
\bar{Q} &=& \sum_i \psi_i\left(-i\frac{\partial}{\partial \lambda_i} + i\frac{\partial W}{\partial\lambda_i}\right),
\end{eqnarray}
with an effective superpotential $W$ which is given by
\begin{equation}
W = W_0 -\sum_{i<j} \log (\lambda_i - \lambda_j). 
\end{equation}

\subsubsection{Double scaling limit and collective field theory}\label{11.3.2}
In the last subsection, we have introduced the Marinari-Parisi supermatrix model. Here we attempt to take the double scaling limit and obtain the (super)string theory if any. The alternative approach to obtain the string theory is to use the collective field theory method as we have discussed in section \ref{3.2.2}. The most difficult point here for the both methods is that the manifestly supersymmetric treatment demands that the Fermi level $\mu$ is not an independent parameter and the flight of time is finite so the Liouville like extra dimension seems to be compact. In contrast, as we will see in the next subsection, the continuum R-NS type construction of the two dimensional superstring demands that the Euclidean time is compact and the Liouville direction is noncompact. We will not solve this discrepancy here, but just review the facts so far obtained in the literature.

The double scaling limit of the Marinari-Parisi model in the singlet sector has been studied in \cite{Marinari:1990jc}, \cite{Dabholkar:1992te}. If we take the special choice of the cubic superpotential: $W_0 = \frac{1}{2}\mathrm{Tr}(g\Phi-\frac{1}{3}\Phi^3)$, the bosonic effective potential can be written as 
\begin{equation}
V(\lambda) = \frac{1}{2}\left(\lambda + \frac{1}{4}(\lambda^2 - g)^2\right).
\end{equation}
The double scaling limit can be obtained by the scaling ansatz:
\begin{align}
\lambda &= \lambda_c (1 +az) ,\  \lambda_c= 2^{-1/3}, \cr
g &= g_c ( 1+ a^2 z),	\ g_c = 3\cdot 2^{-2/3},
\end{align}
and we make $a\to 0$, $N\to \infty$ keeping $Na^{5/2} = \kappa$ fixed. Unlike in the bosonic case, we have to rescale the Euclidean time further\footnote{This rescaling also has been under debate \cite{Chaudhuri:1994yk}. In any case, the rescaling makes sense only when $t$ is noncompact. However, the naive holographic dual matrix model suggests that the Euclidean time variable should be compact in order to preserve the space-time SUSY.} as $t = -ia^{-1/2} \tau$. Then the effective action for the top eigenvalue can be written as (see \cite{Dabholkar:1992te} for the details)
\begin{equation}
S[z] = \frac{1}{\kappa} \int d\tau \left[\frac{1}{2}\left(\frac{\partial z}{\partial \tau}\right)^2 + \frac{1}{2}(v'(z))^2 + \kappa \psi^\dagger \frac{\partial \psi}{\partial \tau} + \kappa \psi^\dagger \psi v''(z)\right],
\end{equation}
where the superpotential $v(z)$ is given by
\begin{equation}
v(z) = \frac{1}{5}(3-z) (z+2)^{3/2}.
\end{equation}
The bosonic potential becomes $ V = \frac{1}{8}(z+2)(z-1)^2$ and looks like it is not positive definite, but we have to supply the infinite wall at $z=-2$. We will not study its properties any further, but we should remark that this effective action breaks supersymmetry by the nonperturbative quantum effect \cite{Dabholkar:1992te}. Indeed, this can be easily seen from the original superpotential. In the supersymmetric quantum mechanics, the supersymmetric condition uniquely fixes (if any) the wavefunction of the supersymmetric ground states \cite{Witten:1981nf,Witten:1982df} as 
\begin{equation}
\psi(z) = e^{-2 W(z)},
\end{equation}
but in our case, the highest power of $z$ in $W(z)$ is odd (apart from the $\log$ term which antisymmetrizes the wavefunction). Therefore this wavefunction is not normalizable and the ground state must be lifted via the tunneling effects. 

\begin{figure}[htbp]
	\begin{center}
	\includegraphics[width=0.5\linewidth,keepaspectratio,clip]{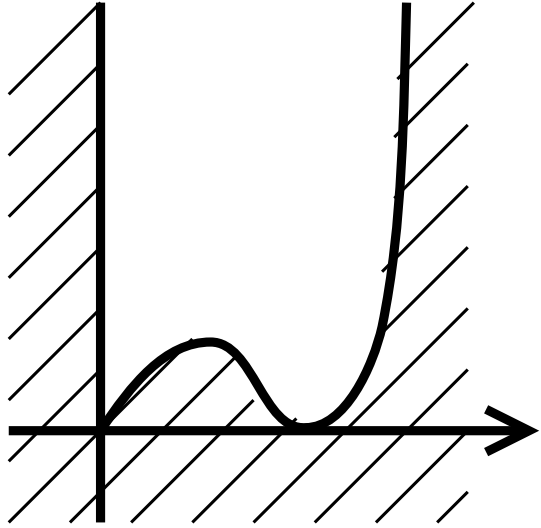}
	\end{center}
	\caption{The double scaled potential for the top eigenvalue of the Marinari-Parisi model. Because of the supersymmetry, the Fermi level $\mu$ is not an independent parameter.}
	\label{MPpot}
\end{figure}

Now let us go on to the collective field description of the Marinari-Parisi model \cite{Rodrigues:1993by}, \cite{Brustein:1994nc,Brustein:1994pu}. We expect that the collective fields yield the \textit{spacetime} fields of the double scaling limit theory as we have seen in the bosonic matrix model where the collective field describing the fluctuation of the Fermi surface becomes the spacetime tachyon field up to the nonlocal field redefinition. The ansatz here is
\begin{eqnarray}
\partial_x \varphi(x,t) &=& \sum_i \delta(x-\lambda_i) \cr
\psi (x,t) &=& -\sum_i \delta(x-\lambda_i) \psi_i(t) \cr
\bar{\psi}(x,t) &=& -\sum_i \delta(x-\lambda_i) \psi^\dagger_i(t) 
\end{eqnarray}

The collective Lagrangian is given by 
\begin{eqnarray}
L &=& \int dx\left[\frac{1}{2\phi} \dot{\varphi}^2 -\frac{1}{2} \phi (W')^2 + \frac{i}{2}\frac{\psi^\dagger\dot{\psi}-\dot{\psi}^\dagger\psi}{\phi} + \frac{i}{2}\frac{\dot{\varphi}}{\phi} \left[\partial_x\left(\frac{\psi^\dagger}{\phi}\right)\psi - \psi^\dagger\partial_x \left(\frac{\psi}{\phi}\right)\right] \right.\cr
& & \left.+ \frac{1}{2}\frac{1}{\phi} [\psi^\dagger,\psi]\partial_xW'\right] - \frac{1}{2}\int dx \int dy [\psi^\dagger(x),\psi(y)] W_{;xy},
\end{eqnarray}
where $\partial_x\varphi = \phi$, $W' = \frac{\delta{W}}{\delta \varphi(x)}$ and $W_{;xy} = \frac{\delta^2{W}}{\delta \varphi(x) \delta \varphi(y)}$.

To go further, we reexpand the action around the classical solution and change the variable $\tau' = \frac{1}{\sqrt{w^2x^2-\mu_f}}$, where $w$ is the quadratic coefficient of the potential and $\mu_f$ is the fermi energy, in order to avoid the tadpole contribution. Then we obtain an interacting field theory with a massless boson and a massless Majorana (left and right) fermion but do not have a supersymmetry (however in \cite{Brustein:1994nc} it has been shown that this theory can be rewritten as the spontaneously broken supersymmetric theory where one chiral super field has a nontrivial space dependent background). We will not discuss the properties of the collective field theory any further. In the next subsection, we consider the world sheet continuum description of the two dimensional superstring theory. Unfortunately, the current situation states that the correspondence between the continuum description and the collective field theory (or the double scaling limit of the supermatrix model) is not well established yet.

\subsubsection{Connection with the world sheet superstring theory}\label{11.3.3}
The Marinari-Parisi model has unusual properties as a two dimensional superstring theory. It would be interesting to see the connection with the continuum description of the two dimensional superstring. We first construct the two dimensional superstring from the R-NS formulation by using the $\mathcal{N}=2$ super Liouville theory technique which we have reviewed in the last section. 

Let us first consider the spectrum of the two dimensional superstring theory \cite{Kutasov:1990ua}, \cite{Kutasov:1991pv}. To do this it is convenient to use the free field description of the $\mathcal{N}=2$, $\tilde{c} = 5$ super Liouville theory and impose the chiral GSO projection. We first bosonize the fermionic partner of $\phi$: $\psi$ and the fermionic partner of $Y$: $\psi_Y$ as
\begin{equation}
\psi = \frac{1}{\sqrt{2}}(e^{ih} + e^{-ih}), \ \ \psi_Y = \frac{1}{\sqrt{2}}(ie^{ih} - ie^{-ih}),
\end{equation}
where $h$ is a canonically normalized bosonic field ($h(z)h(0) \sim -\log z$). Then the vertex operator for the tachyon (in the $(-1,-1)$ picture) is 
\begin{equation}
T_k =  \exp[-(\sigma + \bar{\sigma}) + ikY + \alpha \phi],
\end{equation}
where $\sigma$ is the usual bosonized superconformal ghost $\beta\gamma = \partial\sigma$, and $-\alpha +1 = |k|$. The (left part of the) $R$ vertex is given by
\begin{equation}
V_{-\frac{1}{2}} = \exp\left[-\frac{1}{2}\sigma + \frac{i}{2}\epsilon h + i kY + \alpha\phi\right],
\end{equation}
in the $-\frac{1}{2}$ picture. To perform the chiral GSO projection, we define the spacetime SUSY operator $S$:
\begin{equation}
S(z) = \exp\left[-\frac{1}{2}\sigma + \frac{i}{2}(h + 2Y)\right]. \label{eq:2susy}
\end{equation}
Note that the zero-mode charge $Q = \oint S(z)$ satisfies $Q^2 = 0$ in the two dimension, so the spacetime SUSY algebra is rather a BRST algebra \cite{Kutasov:1990ua}, \cite{Kutasov:1991pv}. Therefore, the connection between the collective field of the Marinari-Parisi model is not clear because the matrix model SUSY algebra has the usual relation $Q^2 = H$.
The GSO projection demands that every physical states should have a local OPE with $(\ref{eq:2susy})$. This projects onto 
\begin{itemize}
	\item NS sector: $k \in \mathbf{Z} + \frac{1}{2}$.
	\item R sector: $\epsilon = -1$, $k\in \mathbf{Z} > 0$, or $\epsilon = +1$, $0>k \in \mathbf{Z} + \frac{1}{2}$.
\end{itemize}

Combining the left and right part, we have for the type IIB theory one (massless) tachyon with half integer momentum, one left moving R-R scalar with integer momentum and one left moving complex fermion with half integer momentum.\footnote{In \cite{Murthy:2003es}, \cite{McGreevy:2003dn}, the same theory has been studied in the dual $SL(2,\mathbf{R})/U(1)$ coset construction which should be dual by the mirror symmetry discussed in section \ref{12.1}.} This is just the manifestation of the fact that the spacetime SUSY requires that $Y$ should be compactified with the radius $R=2$.

Since we have learned the contents of the unstable branes in section \ref{11.2}, we can conjecture the dual matrix theory by the holographic correspondence. The natural object here is the class 1 (ZZ type) B-brane, which can be interpreted as the D0-brane localized in the strong coupling region \cite{Eguchi:2003ik}, \cite{McGreevy:2003dn}. The spectrum is given by the (genuine) tachyon and a massless (complex) fermion. On the other hand, if we expand the Marinari-Parisi model with the superpotential $W_0 = \frac{1}{2}\Phi^2 -\frac{1}{3}\lambda^2 \Phi^3$ around the nonsupersymmetric critical point, we have
\begin{equation}
S = - N \int d\tau \mathrm{Tr} \left[\frac{1}{2}(D_\tau M)^2 + \bar{\Psi}D_\tau\Psi + \frac{1}{2}\lambda^2 M^2 -\frac{1}{16\lambda^2}\right].
\end{equation}
Comparing this with the spectrum obtained from the unstable class 1 B-brane, we can see this is the same. It is worthwhile mentioning, however, that the Euclidean time for the B-brane is compactified unlike the Marinari-Parisi model.\footnote{The author is not quite sure whether this fact is crucial or not. In \cite{McGreevy:2003dn}, the Wick rotation is performed and the odd winding modes are discarded. Also, if we decompactify $Y$, the space-time SUSY from the world sheet perspective becomes obscure.} 

Despite many subtleties, some other supporting arguments are given in \cite{McGreevy:2003dn}. They include,
\begin{itemize}
	\item Symmetry consideration: there are a fermion number $(-1)^F$ and a $\mathbf{Z}_2$ R symmetry both in the Marinari-Parisi model (for the odd superpotential, hence the broken supersymmetry) and the worldsheet theory.
	\item Ground states and instantons: The (supersymmetric) perturbative vacua for the Marinari-Parisi model are classified by the fermion number. However, the instanton effect ruins the fermion number conservation and the actual vacuum is something like the $\theta$ vacuum which is the summation over the different fermionic states and diagonalizes the Hamiltonian. On the other hand, since we have a BPS instanton in this theory (the class 1 BPS A-brane), we expect the similar things happen in the two dimensional space-time theory.
\end{itemize}

Perhaps the conjecture here has a much weaker support than any other matrix model dualities in this review, and we should study further both the Marinari-Parisi model itself and the continuum theory. Before closing this subsection, we have a few remarks:
\begin{itemize}
	\item The biggest mystery is what the actual target space is. This is a difficult problem since the string theory admits a T-duality if we compactify the theory, and we do have a compactified ``time" direction $Y$ in this case. Also, the matrix model collective field is related to the string field via the non-local field redefinition, which may cause the difficulty in understanding the space-time SUSY in the dual theory. Concerning this point, we should note that the Hamiltonian in the matrix model is a conserved quantity, but the ``Hamiltonian" in the space-time theory from the $\mathcal{N}=2$ super Liouville cannot be conserved because of the nontrivial background in the $Y$ direction.
	\item The space-time SUSY algebra is not the same in the matrix model and in the world sheet continuum theory. This has already been recognized in \cite{Dabholkar:1992te}, and they have stated that the dual world sheet theory is rather a Green-Schwarz type superstring than the R-NS formulation considered so far. However the Green-Schwarz superstring in the two dimension also has a subtlety concerning the realization of the kappa symmetry. The hybrid formalism of the two dimensional superstring is given in  \cite{Berkovits:2001tg}, and it would be interesting to see the connection. Also \cite{Gukov:2003yp} has proposed a matrix model dual of the type II superstring regularized by the  R-R background in contrast to the $\mathcal{N} =2$ super Liouville background. Since the world sheet description of the R-R background is beyond the scope of this review, we do not discuss this interesting subject here.
	
	\item Finally, the existence of the nontrivial double scaling limit as a naive continuum limit (superspace $\tau,\theta,\bar{\theta}$ sigma model coupled to the Liouville theory) has been questioned in \cite{Chaudhuri:1994yk}. The world sheet  theory is interacting and in the IR, all the coordinates become massive. The subtle point here is the rescaling of the time $t$ in the double scaling limit taken in \cite{Dabholkar:1992te}. It would be fair to say what the continuum theory (R-NS noncritical superstring or a Green-Schwarz type superstring or something else) corresponds to the double scaling limit of the Marinari-Parisi model is an open problem yet.

\end{itemize}

\subsection{Literature Guide for Section 11}\label{11.4}
The bulk $\mathcal{N} =2 $ super Liouville theory has been studied in \cite{Ivanov:1983wp}, \cite{Otto:1985dp}, \cite{Distler:1990nt}, \cite{Antoniadis:1990mx}, \cite{Ketov:1996es}, \cite{Kutasov:1990ua}, \cite{Kutasov:1991pv}, \cite{Murthy:2003es}, \cite{Mussardo:1989av}, \cite{Ahn:2002sx}. Also we can find many references on the $\mathcal{N} =2 $ super Liouville theory in connection with the holographic dual of the little string theory or singular CY space as we will see in the next section.

The boundary states for the $\mathcal{N} =2 $ super Liouville theory has been studied in \cite{Eguchi:2000cj}, \cite{McGreevy:2003dn} and further developed in \cite{Eguchi:2003ik}, \cite{Ahn:2003tt}.

The earlier studies on the Marinari-Parisi \cite{Marinari:1990jc} model have been done in \cite{Greensite:1984yc}, \cite{Mikovic:1990qf}, \cite{Dabholkar:1992te}, \cite{Cohn:1992zj}, \cite{vanTonder:1992vc}, \cite{Rodrigues:1993by}, \cite{Karliner:1990cd}, \cite{Gonzalez:1991yu}, \cite{Bellucci:1990gk}, \cite{Bellucci:1991ee}, \cite{Bellucci:1992ki}, \cite{Ferretti:1994fu}, \cite{Brustein:1994nc}, \cite{Chaudhuri:1994yk}, \cite{Brustein:1994pu}. Based on the unstable D0-brane spectrum, the duality between the Marinari-Parisi model and the $\mathcal{N}=2$ super Liouville theory has been proposed in \cite{McGreevy:2003dn}. The superstring theory from the $\mathcal{N}=2$ super Liouville theory has been reviewed in \cite{Kutasov:1991pv}, \cite{Murthy:2003es}.

Theoretically speaking, the $\mathcal{N}=4$ extended super Liouville theory is possible \cite{Ivanov:1984ht}, \cite{Ivanov:1985fe}, \cite{Ketov:1996es} (see also \cite{Ivanov:1988mz,Ivanov:1988rt} for related discussions). As far as the author knows, however, the application to the physical string theory is still in its infancy and study on the subject such as structure constants or brane contents etc is still limited.

\sectiono{Applications}\label{sec:12}
In this section, we discuss applications of the $\mathcal{N}=2$ super Liouville theory. The organization of this section is as follows.

In section \ref{12.1}, we discuss the fermionic string on the 2D black hole.  In subsection \ref{12.1.1} we review the basic properties of the string spectrum of the theory, and in subsection \ref{12.1.2}, we discuss the duality of the 2D fermionic black hole and the $\mathcal{N}=2$ super Liouville theory. In subsection \ref{12.1.3}, the matrix model dual proposal for the 2D fermionic black hole is reviewed.

In section \ref{12.2}, we review the application of the $\mathcal{N}=2$ super Liouville theory to the (non)critical string propagating in the singular Calabi-Yau space. In subsection \ref{12.2.1}, we discuss the bulk theory and in subsection \ref{12.2.2}, we discuss the branes present in the singular Calabi-Yau space from the branes in the $\mathcal{N}=2$ super Liouville theory we have discussed in section \ref{sec:11}.

In section \ref{12.3}, we provide an explanation of the duality in the $\mathcal{N}=2$ super Liouville theory by using other dualities discussed in this section.

\subsection{Fermionic String on 2D Black Hole}\label{12.1}
In this section, we discuss the relation between $\mathcal{N}=2$ super Liouville theory and the type 0 string theory propagating on the 2D black hole. 
\subsubsection{$SL(2,\mathbf{R})/U(1)$ supercoset model}\label{12.1.1}
Since the bosonic string propagating on the 2D (Euclidean) black hole is described by the string theory on the $SL(2,\mathbf{R})/U(1)$ coset theory quantum mechanically (see section \ref{6.5}), we naturally expect that the fermionic string propagating on the 2D black hole is described by the $SL(2,\mathbf{R})/U(1)$ Kazama-Suzuki supercoset theory \cite{Kazama:1989uz}. In this section, we review its construction following \cite{Giveon:1999px,Giveon:2003wn}.

Consider first the superconformal parent $SL(2,\mathbf{R})$ theory at level $k$. This theory has the left $SL(2,\mathbf{R})$ super current
\begin{equation}
\psi^a + \theta \sqrt{\frac{2}{k}} J^a,
\end{equation}
with $a = 1,2,3$, where 
\begin{equation}
J^a = j^a - \frac{i}{2} \epsilon^a_{\ bc} \psi^b\psi^c.
\end{equation}
The free field OPE of $\psi^a$ is given by
\begin{equation}
\psi^a(z) \psi^b (0) \sim \frac{\eta^{ab}}{z}
\end{equation}
where the metric is given by $\eta^{ab} = \mathrm{diag} (+,+,-)$. The current $j^a$ satisfies the usual affine $SL(2,\mathbf{R})$ algebra at level $k_B = k +2$. Then the total current satisfies
\begin{equation}
J^a(z) J^b(0) \sim \frac{\frac{k}{2}\eta^{ab}}{z^2} + \frac{i\epsilon^{ab}_{\ \ c} J^c}{z}.
\end{equation}

The central charge of the theory is given by
\begin{equation}
c(SL(2,\mathbf{R})) = \frac{9}{2} + \frac{6}{k}.
\end{equation}
It is convenient to define $\psi^{\pm} = \frac{1}{\sqrt{2}}(\psi^1 \pm i\psi^2)$ and bosonize them via
\begin{equation}
\partial H = \psi^2\psi^1 = i\psi^- \psi^+,
\end{equation}
with a canonically normalized bosonic field $H$; $H(z)H(0) \sim -\log(z)$. Then $J^3 = j^3 + i\partial H$.

In order to make a quotient, we introduce two canonically normalized independent scalars $X_3$ and $X_R$. Our gauging condition is 
\begin{eqnarray}
J^3 = -\sqrt{\frac{k}{2}} \partial X_3.
\end{eqnarray}
$X_R$ is defined as
\begin{equation}
iH = \sqrt{\frac{2}{k}}X_3 + i\sqrt{\frac{c}{3}}X_R.
\end{equation}
where $c = c(SL(2,\mathbf{R})/U(1)) = 3 + \frac{6}{k}$. Let $\Phi_{jm}$ be the holomorphic part of the primary field in the bosonic $SL(2,\mathbf{R})$ theory. Its property is 
\begin{equation}
j^3(z) \Phi_{jm}(0) \sim \frac{m \Phi(0)_{jm}}{z}.
\end{equation}
Therefore 
\begin{equation}
J^3(z) \Phi_{jm}(0) \sim \frac{m \Phi(0)_{jm}}{z}.
\end{equation}
Using this operator, we can construct a primary of the bosonic quotient CFT on $SL(2,\mathbf{R})/U(1)$: $V_{jm}$ and that of the superconformal quotient: $U_{jm}$ as
\begin{eqnarray}
V_{jm}  e^{i\frac{2m}{k+2} \sqrt{\frac{c}{3}} X_R} e^{m\sqrt{\frac{2}{k}} X_3} &=& \Phi_{jm} \cr
 U_{jm} e^{m\sqrt{\frac{2}{k}}X_3}  &=& \Phi_{jm}.
\end{eqnarray}
Furthermore we have the decomposition
\begin{equation}
e^{inH} = e^{n(\sqrt{\frac{2}{k}}X_3 + i\sqrt{\frac{c}{3}}X_R)}.
\end{equation}
Therefore, from a parent operator $\Phi_{jm} e^{inH}$, we can construct a coset primary operator
\begin{equation}
V^{n}_{jm} = \Phi_{jm} e^{inH} e^{-\sqrt{\frac{2}{k}} (m+n) X_3}.
\end{equation}
Alternatively, we can eliminate the explicit $X_3$ dependence and then we have
\begin{equation}
V^n_{jm} = V_{jm} e^{i(\frac{2m}{k+2} + n)\sqrt{\frac{c}{3}}X_R}.
\end{equation}
Their scaling dimensions are given by
\begin{equation}
\Delta(V^n_{jm}) = \frac{-j(j+1)+(m+n)^2}{k} + \frac{n^2}{2}.
\end{equation}
In the physical setup, we have $k=1/2$ to obtain $c = 15$. The BRST invariant operator contents with the GSO projection and their reflection property have been studied in \cite{Giveon:2003wn}.

\subsubsection{2D fermionic black hole and $\mathcal{N}=2$ super Liouville duality}\label{12.1.2}
We have discussed the conjectural duality between the 2D black hole and the sine-Liouville theory in section \ref{6.5}. Actually, the supersymmetric extension of this duality exists and its duality was proved by using the mirror symmetry \cite{Hori:2001ax}. We would like to review the proof here. The gist is that we can find the gauged linear sigma model which flows to the $SL(2,\mathbf{R})/U(1)$ supercoset model (which is the strict definition of what we call by the 2D fermionic black hole) in the IR, and we dualize this action so that we obtain the dual action which flows to the $\mathcal{N}=2$ super Liouville theory in the IR. In this way we can prove the duality between the 2D fermionic black hole and the $\mathcal{N}=2$ super Liouville theory (see also \cite{Giveon:1999px}, \cite{Tong:2003ik}).

The relevant gauged linear sigma model action is
\begin{equation}
S = \frac{4}{2\pi}\int d^2z d^4\theta \left[\bar{\Phi}e^V\Phi + \frac{k}{4}(P+\bar{P}+V)^2 -\frac{1}{2e^2}|\Sigma|^2 \right]\label{GNLS},
\end{equation}
where $\Phi$ and $P$ are chiral superfields, $V$ is a $U(1)$ vector superfield and $\Sigma$ is its field strength twisted chiral superfield.\footnote{See appendix \ref{a-2} for our conventions of the superfield. $4$ in front of the action is needed to connect our conventions with those in \cite{Hori:2001ax}.} It has a gauge symmetry which transforms as follows:
\begin{eqnarray}
\Phi &\to& \Phi e^{i\Lambda} \cr
P &\to& P +i\Lambda \cr
V &\to& V -i\Lambda +i\bar{\Lambda}\label{gauge}.
\end{eqnarray}
The imaginary part of $P$ is compactified, so $P\sim P +2i\pi$. In the following, we present the rough sketch of the proof in the three steps. In the step-1 we show the IR equivalence of the gauged linear sigma model and the $SL(2,\mathbf{R})/U(1)$ supercoset model. In the step-2 we dualize this gauged linear sigma model. In the step-3 we show the IR equivalence of the dualized model and the $\mathcal{N}=2$ super Liouville theory.

{\bfseries Step-1:}
To show the IR equivalence of the gauged linear sigma model and the $SL(2,\mathbf{R})/U(1)$ supercoset model, we first integrate out the massive field classically. 
The potential of the action (\ref{GNLS}) is given by
\begin{equation}
V = -D|\phi|^2 -|F|^2 +|\sigma|^2|\phi|^2 - \frac{k}{2}\left(|F_p|^2 - |\sigma|^2+D(p+\bar{p}) \right) -\frac{1}{2e^2} D^2.
\end{equation}
We integrate out $\sigma$ and set $D=F=F_p =0$. Then the $D$ term condition is
\begin{equation}
D = |\phi|^2 + k\mathrm{Re} p = 0.
\end{equation}
Setting $\mathrm{Im} p = 0 $ by the gauge choice, we have a non-linear sigma model of $\phi$ as the effective action. The target space metric is
\begin{equation}
ds^2 = \left(1+\frac{r^2}{k}\right)dr^2 + \frac{r^2}{1+\frac{r^2}{k}}d\theta^2, 
\end{equation}
where $\phi = \frac{r}{\sqrt{2}}e^{i\theta}$. However, this metric is not a 2D black hole metric. To see this, we set $r = \sqrt{k}\sinh \rho $ then the metric becomes
\begin{equation}
ds^2 = k(\cosh^4\rho d\rho^2 + \tanh^2\rho d\theta^2)\label{blm}.
\end{equation}
Note this metric is not Ricci flat, so there must be non-trivial flow to obtain a conformal fix point in the IR. Indeed there is such a correction and the one loop correction was studied in \cite{Hori:2001ax}. We will not reproduce the calculation here, but the result is that the perturbative correction changes the metric (\ref{blm}) into that of the 2D black hole 
\begin{equation}
ds^2 = k(d\rho^2 + \tanh^2\rho d\theta^2)
\end{equation}
and it also generates the dilaton background so that the one-loop beta function vanishes.

Actually we can show this IR flow even quantum mechanically. To show this, we use the similar trick which is very common in the four dimensional superconformal gauge theory. First, we find the anomaly free $R$ current in the UV theory which is assumed to be the superconformal $R$ current in the IR. Then we use the 't Hooft anomaly matching argument \cite{'tHooft:1980xb} to calculate the central charge $c$ of the IR theory. The anomaly matching states that this can be done in the UV theory which is free. 

In this case, the naive $R$ current is anomalous, but we can improve it by adding the current of $p$ which transforms anomalously because of the unusual gauge transformation (\ref{gauge}). The improved current is (see the original paper \cite{Hori:2001ax} for a thorough derivation)
\begin{equation}
j^+ = \psi^+\bar{\psi}^+ + \frac{k}{2}\chi^+\bar{\chi}^+ +\frac{i}{e^2}\sigma \partial\bar{\sigma} + i(D_zp-D_z\bar{p}).
\end{equation}
The free OPE can be used to calculate the OPE of this current in the IR
\begin{equation}
j^+(z) j^+(0) = - \frac{1+\frac{2}{k}}{z^2} +\cdots.
\end{equation}
As a result we obtain the central charge of the IR theory:
\begin{equation}
c = 3\left(1+\frac{2}{k}\right),
\end{equation}
which coincides with that of the level $k$ $SL(2,\mathbf{R})/U(1)$ supercoset model.

However, there is a subtly in the above reasoning. As is often the case, the anomaly free $R$ current is not unique. There is a freedom to add $ia D_z\bar{p}$ to the above $R$ current which makes the central charge $c' = 3+6(1-a)/k$. The question is what value of $a$ corresponds to the actual superconformal current in the IR. In \cite{Hori:2001ax}, it was shown that the mild assumption of the reality of the current uniquely chooses $a = 0$.\footnote{In the four dimensional case, it has been recently proposed that there is a procedure to find the correct conformal $R$ current in the UV theory \cite{Intriligator:2003jj}, \cite{Kutasov:2003iy}, \cite{Kutasov:2003ux}. This is done by maximizing $a$ which is supposed to be the four dimensional analog of the monotony decreasing Zamolodchikov $c$ function \cite{Zamolodchikov:1986gt}. However, in this case this does not seem to apply (maximizing $c$ does not make sense). Note that even the Zamolodchikov $c$ theorem does not hold because $c$ of the UV theory is 9, but that of the IR theory $3\left(1+\frac{2}{k}\right)$ can be arbitrarily large. This is because $c$ does not count the actual degrees of freedom of the theory with the linear dilaton coupling. }

To complete the proof, we have to show this IR theory which is superconformal and has a central charge $3\left(1+\frac{2}{k}\right)$ and other parity symmetry is unique. In other words the question is whether there is a marginal deformation which preserves all the symmetries of the $SL(2,\mathbf{R})/U(1)$ super coset theory. After a thorough investigation, it can be shown this is the case unlike the bosonic $SL(2,\mathbf{R})/U(1)$ coset.

 {\bfseries Step-2:}
In this step, we dualize the action of the gauged linear sigma model. Here we treat the action classically and in the third step we consider the quantum correction and the IR flow of the dualized action.

We would like to dualize the phase of $\Phi$ and the imaginary part of $P$. Concentrating on $\Phi$ for the time being, let us consider the following action
\begin{equation}
S' = \frac{4}{2\pi}\int d^2zd^4\theta \left(e^{V+B} - \frac{1}{2}(Y+\bar{Y})B\right), 
\end{equation}
where $B$ is a real superfield and $Y$ is a twisted chiral superfield. If we integrate out $Y$ first, this enforces $B$ to be decomposed as $B = \Psi + \bar{\Psi}$, where $\Psi$ is a chiral superfield. Substituting back this into the action, we obtain 
\begin{equation}
S = \frac{4}{2\pi}\int d^2zd^4\theta e^{V+\Psi+\bar{\Psi}} = \frac{4}{2\pi}\int d^4\theta \bar{\Phi} e^V\Phi,
\end{equation}
where we have introduced another chiral superfield $\Phi = e^\Psi$. This is just the first part of the action (\ref{GNLS}). Alternatively we can integrate out $B$ first, which gives
\begin{equation}
B = - V+\log\frac{Y+\bar{Y}}{2}.
\end{equation}
Substituting back this into the action, we obtain the dual action 
\begin{eqnarray}
\tilde{S} &=& \frac{4}{2\pi}\int d^2zd^4\theta\left(\frac{V}{2}(Y+\bar{Y})-\frac{1}{2}(Y+\bar{Y})\log(Y+\bar{Y}) \right)\cr
&=& \int \frac{4}{2\pi}d^2z d^4\theta\left[ -\frac{1}{2}(Y+\bar{Y})\log(Y+\bar{Y}) \right]+\frac{1}{2}\left(\frac{4}{2\pi}\int d^2z d\theta^+d\bar{\theta}^- \Sigma Y + h.c.\right).
\end{eqnarray}
We can do the same dualization for $P$. We start with the following action
\begin{equation}
S' = \frac{4}{2\pi}\int d^2z d^4\theta \left(\frac{k}{4}(C+V)^2 - \frac{1}{2}(Y_p+\bar{Y}_p)C \right).
\end{equation}
If we integrate out $Y_p$ then $ C = P_Y +\bar{P}_Y$ as above so that we obtain the second part of the original action 
\begin{equation}
S = \frac{4}{2\pi}\int d^2z d^4\theta \frac{k}{4}(P_Y + \bar{P}_Y + V)^2.
\end{equation}
To dualize this action, we integrate out $C$ instead, which yields $P_Y+\bar{P}_Y + V = \frac{1}{k}(Y_p+\bar{Y}_p)$. Substituting back this into the action,
\begin{eqnarray}
\tilde{S} &=& \frac{4}{2\pi}\int d^2z d^4\theta \frac{-1}{4k}(Y_p+\bar{Y}_p)^2 + \frac{1}{2}\left(\frac{4}{2\pi}\int d^2zd\theta^+d\bar{\theta}^- \Sigma Y_p +h.c.\right)\cr
&=& \frac{4}{2\pi}\int d^2z d^4\theta \frac{-1}{2k}Y_p\bar{Y}_p + \frac{1}{2}\left(\frac{4}{2\pi}\int d^2zd\theta^+d\bar{\theta}^- \Sigma Y_p + h.c.\right)
\end{eqnarray}
This is the dual action which we wanted. In the second line, we have used the fact that the $F$ term of the chiral superfield is a total derivative.

Finally, collecting all the terms, we obtain the dual action
\begin{eqnarray}
\tilde{S} = \frac{4}{2\pi} \int d^2z\left\{\int d^4\theta\left[-\frac{1}{2e^2}|\Sigma|^2 - \frac{1}{2}(Y+\bar{Y})\log(Y+\bar{Y})-\frac{1}{2k}|Y_p|^2\right] \right.\cr
\left.+ \frac{1}{2}\left(\int d\theta^+d\bar{\theta}^- \Sigma(Y+Y_p) +h.c.\right)\right\}
\end{eqnarray}
where $Y$ and $Y_p$ have $2\pi i$ periodicity.
Remember this action is derived classically since we have neglected all the Jacobians of the transformation for the path integral measure and other quantum corrections. The quantum corrections are discussed in the step-3.

{\bfseries Step-3:} We consider here the quantum corrections of the dualization procedure and the IR flow of the effective action. We first note that the nonrenormalization theorem for the $(2,2)$ supersymmetry suggests that the twisted superpotential does not receive a perturbative correction but it may have a non-perturbative correction and the K\"ahler potential does receive perturbative and non-perturbative corrections in general.

The origin of the nonperturbative correction for the twisted superpotential is instantons (or vortices in the two dimensions) as was discussed in \cite{Polyakov:1977fu} for the bosonic case. The usual supersymmetric holomorphy and symmetry argument shows this is the only correction. Therefore the twisted superpotential from the instanton correction is
\begin{equation}
\tilde{W} = \mu e^{-Y},
\end{equation}
where $\mu$ is a dynamically generated scale. Note $Y_p$ does not have this kind of superpotential because it is not charged under $U(1)$.

For a K\"ahler potential, one loop renormalization makes the metric 
\begin{equation}
ds^2 = \frac{|dy|^2}{2\log(\Lambda)+2\mathrm{Re}y} + \frac{1}{k}|dy_p|^2 .
\end{equation}
Integrating out $\Sigma$ gives a constraint $Y+Y_p =0$. Then we obtain the K\"ahler potential of $Y$ 
\begin{equation}
K(Y,\bar{Y}) = -\frac{1}{2k} |Y|^2 + \cdots,
\end{equation}
where omitted terms are possible quantum corrections. The twisted superpotential of $Y$ is now (because of the nonrenormalization theorem this does not suffer a quantum correction during the IR flow.)
\begin{equation}
\tilde{W}(Y) = \mu e^{-Y}.
\end{equation}
This is just the Liouville potential of $Y$! If the K\"ahler potential is flat, then we have a (2,2) superconformal field theory with a central charge
\begin{equation}
c = 3\left(1+\frac{2}{k}\right).
\end{equation}

Actually even the flatness of the K\"ahler potential can be proven by deforming the UV theory without destroying any symmetry and using the uniqueness of the IR theory as discussed in the step-1. An interested reader should consult the original paper \cite{Hori:2001ax}.

To summarize the result, we obtain the dual action of the $SL(2,\mathbf{R})/U(1)$ supercoset model:
\begin{equation}
\tilde{S} = \frac{4}{2\pi} \int d^2z \left[\int d^4\theta \frac{-1}{2k}|Y|^2 + \frac{1}{2}\left(\int d^2\theta \mu e^{-Y} + h.c.\right)\right],
\end{equation}
where $Y$ has a period $2\pi i$. This is the  $\mathcal{N}=2$ super Liouville theory compactified on the $r = \sqrt{2/k}$ circle. Its central charge is $ c= 3\left(1+\frac{2}{k}\right)$. Precisely speaking, we should redefine $Y \to -S$. In addition, $Y$ is a twisted chiral superfield and not a chiral superfield. However, this does not make any difference when we deal with only a twisted chiral field, for we can simply redefine the supercoordinate $\theta^-$ to $\bar{\theta}^-$.

\subsubsection{Matrix model for 2D fermionic black hole}\label{12.1.3}
The matrix model dual for the fermionic string propagating on the 2D black hole has been proposed in \cite{Giveon:2003wn}. In the bosonic case, it is the $r=3/2$ KKK matrix model which we have studied in section \ref{6.5.2}. We recall that this conjecture has depended on the further duality between the bosonic 2D black hole theory and the sine-Liouville theory. In the fermionic case, we expect the duality (which we have provided the proof in the last subsection) between the 2D fermionic black hole and the $\mathcal{N}=2$ super Liouville theory plays an important role.

In the critical case, the dual super Liouville theory has the following super potential:
\begin{equation}
L = \mu \int d^2 \theta e^{\frac{1}{2}(\phi + iY)} + c.c = \mu \int d^2 \theta e^{\frac{Q}{4}(\phi + iY)} + c.c. ,
\end{equation}
where $Q=2$, and the radius of $Y$ is 2 (in the $\alpha' = 2$ unit). The conjecture made in \cite{Giveon:2003wn} is that the dual matrix theory for the type 0B theory is the KKK matrix model at the selfdual radius $R=1$ (in the bosonic $\alpha'=1$ unit) with a symmetric potential. The selfdual radius is selected because at this point, the sine-Liouville potential behaves (see \ref{eq:sLe}) like $e^{(\phi + iX)}$, which reminds us of the \textit{superpotential} in the super Liouville theory. Therefore, we expect that the bosonic string theory on the $R=1$ sine-Liouville theory has similar correlators to those in the type 0B string on the $\mathcal{N} = 2$ Liouville theory (up to leg factors and rescaling of momenta). The symmetric property  of the potential is from the stability of the theory and the analogy to the duality between the $\mathcal{N} = 1$ type 0B super Liouville theory and the Hermitian matrix model with the symmetric potential.

Recalling that the $\mathcal{N }=2$ super Liouville theory is related to the fermionic black hole via duality, we have come to the following conjecture: the type 0A string theory on the $SL(2,\mathbf{R})_{\frac{1}{2}}/U(1)$ black hole is dual to the $R=1$ KKK matrix model with a symmetric potential and eigenvalues filling both sides of the potential.

It would be interesting to study this conjecture further from many points of view. The scattering amplitudes or partition function may be calculable.\footnote{The calculation of the partition function from the matrix point of view seems a little bit difficult. This is because at $R=1$, the method we have used in section \ref{6.5.2} is not applicable.} Also the study of the brane in this fermionic string theory may yield the foundation of the conjecture as has been done in the 0A/B $\mathcal{N} = 1$ Liouville - matrix model duality. For the bosonic black hole geometry, we have briefly reviewed the possible branes in the theory in section \ref{6.5.3}.

\subsection{Strings and Branes in Singular Calabi-Yau Space}\label{12.2}
In this section, we discuss the application of the $\mathcal{N}=2$ super Liouville theory to the superstrings propagating in the singular Calabi-Yau space (times the flat Minkowski space). Unlike most of the part in this review, we consider the $D>2$ higher dimensional theory. This is possible because the $\mathcal{N}=2$ super Liouville theory admits a lower central charge owing to the nonrenormalization of the background charge. 
\subsubsection{Strings propagating in singular CY and $\mathcal{N}=2$ Liouville theory}\label{12.2.1}
We start this subsection by reviewing the world sheet description of the near horizon limit of the NS5-brane -- Calan-Harvey-Strominger (CHS) background \cite{Callan:1991at,Callan:1991dj}\cite{Rey:1989xj} in the superstring theory. Remarkably, this geometry can be described by an exact solvable CFT.

First, we recall that the NS5-brane solution (for coincident branes) is given by
\begin{eqnarray}
ds^2 &=& e^{2\phi} dx^2 + dy^2  \cr	
e^{2\phi(x)} &=& C + \frac{N}{x^2} \cr
H &=& - N \epsilon, \label{eq:NS5sol}
\end{eqnarray}
where $\epsilon$ is the volume form of the unit-3sphere whose normalization is  $\int \epsilon = 2\pi^2$ and integrally quantized $N$ is the number of NS5-branes, and $C$ is the integration constant which determines the asymptotic string coupling.  $y$ is the tangential six dimensional coordinate and $x$ is the transverse four dimensional coordinate. 
In the near horizon limit $ x \to 0$, setting $t = \sqrt{N} \log \sqrt{N/x^2}$, we have
\begin{eqnarray}
ds^2 &=& dt^2 +  N d\Omega^2_3 \cr 
\phi &=& t/\sqrt{N} \cr
 H &=& - N \epsilon,  
\end{eqnarray}

We can rephrase this geometry in the exact CFT language as follows. The internal CFT is given by the super linear dilaton theory with the background charge $Q = \sqrt{\frac{2}{N}}$  whose central charge is $ c_{LD} = \frac{3}{2} + 3Q^2$. On the other hand, the $\mathrm{S}^3$ part with $H$ flux can be described by the (supersymmetric) $SU(2)$ WZW model with the level $k=N-2$ whose central charge is $ c_{WZW} = \frac{3k}{k+2} + \frac{3}{2} = \frac{3(N-2)}{N} + \frac{3}{2}$. The total central charge becomes $c= c_{WZW} + c_{LD} = 6$, irrespective of the number of NS5-brane as it should be (we can also see that the tangential dimension is six). This internal theory has (at least) an $\mathcal{N} =2 $ superconformal algebra, so the six dimensional theory has a space-time SUSY.

However, the string theory propagating in this CHS limit is actually singular. This is because the linear dilaton background forces us to have a strong coupling region as $t \to \infty$ and nothing prevents strings from propagating in the strong coupling region. In other words, the Liouville theory in the $\mu \to 0$ limit is singular as is easily seen from the identification $g_s \sim \mu^{-1}$.

The easiest way to regularize the situation is to turn on a Liouville potential. However, we should be careful to preserve the space-time SUSY. For this purpose, the world sheet $\mathcal{N}=2$ superconformal symmetry is necessary, and as a result the deformation by the $\mathcal{N} =2$ super Liouville potential becomes the best candidate. To do this, we have to supply compactified $Y$ direction from somewhere, which is from the $\mathrm{S}^3$ part as we will explain in the following.

Before doing this, it is interesting to see the geometrical nature of turning on the $\mathcal{N}=2$ Liouville potential in this theory \cite{Giveon:1999px}. First of all, the $N$ coincident NS5-branes are T-dual \cite{Ooguri:1996wj}, \cite{Kutasov:1996te} (see appendix \ref{b-7} for a brief review) to the string theory propagating on the singular $\mathrm{K}3$ whose complex structure is given by
\begin{equation}
z_1^N + z_2^2 + z_3^2 = 0.
\end{equation}
It has been proposed that turning on the Liouville potential corresponds to the deformation of the singularity by the most relevant perturbation as 
\begin{equation}
z_1^N + z_2^2 + z_3^2 = \mu. \label{eq:aale}
\end{equation}
To go back to the NS5-brane picture, we rewrite (\ref{eq:aale}) as 
\begin{equation}
\prod_{n=1}^N (z_1 - r_0 e^{\frac{2\pi in}{N}}) + z_2^2 + z_3^2 = 0,
\end{equation}
where $\mu = (-r_0)^N$. If we perform the T-duality, we have a system of $N$ NS5-branes which are distributed uniformly on a circle of radius $r_0$.

\begin{figure}[htbp]
	\begin{center}
	\includegraphics[width=0.8\linewidth,keepaspectratio,clip]{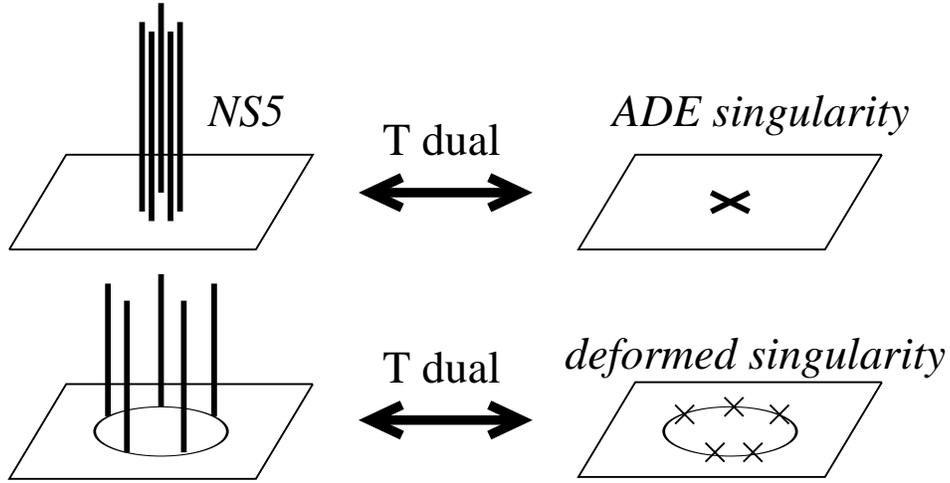}
	\end{center}
	\caption{The T-dual of NS5-branes is given by the ADE singularity. Turning on the Liouville potential corresponds to deforming the ADE singularity.}
	\label{Tdual}
\end{figure}

The general proposal \cite{Aharony:1998ub}, \cite{Giveon:1999px} for the near horizon limit of this system (for $\mu = 0$) is the string theory propagating on
\begin{equation}
\mathbf{R}^{d-1,1}\times \mathbf{R}_\phi \times \mathrm{S}^1 \times LG(W=F), \label{eq:lstdual}
\end{equation}
where $F=0$ is the defining equation of the (ADE type singular) CY manifold such as (\ref{eq:aale}) and technically speaking, the GSO projection on the integer $\mathcal{N}=2$ $U(1)$ charge sector is needed in the super Liouville ($\mathbf{R}_\phi$ and $\mathrm{S}^1$ part whose dilaton dependence is given by $\frac{Q}{2}\phi$) times the Landau-Ginzburg sector in order to ensure the space-time SUSY. This should be the dual theory of the decoupling limit of the nongravitational string theory which propagates on the singular Calabi-Yau space --- what is called the little string theory (LST).

At first sight, this geometry is not related to the CHS background considered above, but as we will see this is actually the same theory. The Landau-Ginzburg model with the super potential $W= z_1^N + z_2^2 + z_3^2$ corresponds to the $\mathcal{N}=2$ minimal model, which can be also constructed as the coset SCFT $\frac{SU(2)_N}{U(1)}$. The GSO projection acts as a $\mathbf{Z}_N$ orbifold on $\frac{SU(2)_N}{U(1)} \times \mathrm{S}^1$, turning it into $SU(2)_N$. As a result, the target space (\ref{eq:lstdual}) becomes $\mathbf{R}^{5,1} \times \mathbf{R}_\phi \times SU(2)$, which is just the CHS background.

Now that we have a dual description (\ref{eq:lstdual}) for the singular CY theory, it is easy to deform this theory while preserving the $\mathcal{N}=2$ superconformal properties. Turning on the $\mathcal{N}=2$ super Liouville deformation: 
\begin{equation}
\delta L_c = \mu \int d^2\theta e^{\frac{1}{Q}(\phi + iY)} + c.c.
\end{equation}
plays such a role. As we have stated earlier the nonchiral deformation:
\begin{equation}
\delta L_{nc} = \tilde{\mu} \int d^4\theta e^{Q\phi}
\end{equation}
may be related to the K\"ahler deformation.

We will not discuss an interesting physics of this bulk theory any further, but we have three comments in order.

\begin{itemize}
	\item This construction is closely related to the mirror duality of the $\mathcal{N}=2 $ super Liouville theory and the $SL(2,\mathbf{R})/U(1)$ super coset model discussed in the last subsection. This is because the sigma-model whose target space is the non-compact deformed CY manifold $F=\mu$ is formerly described by the Landau-Ginzburg model with the superpotential
\begin{equation}
W = -\mu z_0^{-k} + F,
\end{equation}
where $k = \frac{2}{Q^2}$. For the noninteger $k$, it has been proposed to interpret this theory as an $SL(2,\mathbf{R})/U(1)$ supercoset theory at level $k$. The proof of the duality between this theory and the $\mathcal{N}=2$ super Liouville theory is the strong support of the dual description of the (double scaling limit\footnote{The double scaling limit here means $\mu \to 0$ and $g_s \to 0$ with $x \equiv \mu^{\frac{1}{N}}g_s$ fixed, which is just the KPZ scaling. In the singular limit, the physical amplitudes depend only on $x$ as in the usual Liouville theory. The original definition of LST corresponds to $\mu \to 0$ and then $g_s \to 0$, hence $x =0$.} of the) LST.
	\item The original ($\mu = 0$) LST is the decoupling limit of the non-gravitational theory on the coincident NS5-branes. They have 16 space-time SUSY and do not have a dimensionless expansion parameter (for a review, see \cite{Aharony:1999ks}, \cite{Kutasov:2001uf}). The type IIA theory is a (2,0) supersymmetric conformal theory which has a tensor multiplet in the low energy limit which seems very mysterious. On the other hand the type IIB theory is a (1,1) supersymmetric theory and has a vector multiplet in the low energy limit and more familiar. The deformation considered above corresponds to higgsing the adjoint complex scalar as 
	\begin{equation}
	\langle \Phi \rangle = C \mathrm{diag} (e^{\frac{2\pi i}{N}}, e^{\frac{4\pi i}{N}},\cdots,e^{\frac{2\pi i N}{N}}).	
	\end{equation}
The double scaling limit is the decoupling limit with fixed W-boson mass in this perspective \cite{Giveon:1999px}.
	\item If we take the A-type singularity, we can replace the Landau-Ginzburg part with the $\mathcal{N}=2$ minimal model of level $k$ ($\tilde{c} =k/(k+2)$). Then the criticality condition \cite{Eguchi:2003ik}
	\begin{equation}
	\frac{d}{2} + \frac{k}{k+2} + (1+\frac{2K}{N}) = 5 
	\end{equation}
	 leads to the condition
	 \begin{equation}
	 d = 6 : N = k+2, \ K=1,
	 \end{equation}
	 which is just the CHS background,
	 \begin{eqnarray}
	 d =4&:& k = even, \ N = k+2, \ K = (k+4)/2 \cr
	 	& & k = odd , \  N = 2(k+2),\ K = k+4 
	 \end{eqnarray}
	 which is related to the NS5-brane wrapped around the Riemann surface, and 
	 \begin{equation}
	 d = 2: N = k+2, \ K =k+3.
	 \end{equation}
	 which is related to the NS5-brane wrapped around the K3.
\end{itemize}

\subsubsection{Branes wrapped around vanishing SUSY cycles}\label{12.2.2}
We have briefly reviewed the connection between the string theory propagating in the singular Calabi-Yau space in the double scaling limit and the $\mathcal{N} =2$ super Liouville theory in the last subsection. Now we have learned the bulk physics correspondence, it is good time to discuss the role of branes in the $\mathcal{N}=2$ super Liouville theory in the singular Calabi-Yau spaces \cite{Lerche:2000uy}, \cite{Lerche:2000jb}, \cite{Sugiyama:2000id}, \cite{Eguchi:2000cj}, \cite{Eguchi:2003ik}. In the following we focus on the A-type boundary condition which corresponds to the D$n$-brane wrapped around the middle-dimensional SUSY cycles in $\mathrm{CY}_n$ (special Lagrangian submanifold).

In our simple singular CY space, the special Lagrangian submanifold is easily obtained \cite{Gukov:1999ya}, \cite{Klemm:1996bj}, \cite{Eguchi:2000cj}. Let us parameterize the pair of roots of $P(X) = X^{k+2} + \mu$ in the complex $X$ plane as
\begin{equation}
X_a = (-\mu)^{1/(k+2)} e^{i\pi(M+L)/(k+2)}, X_b = (-\mu)^{1/(k+2)} e^{i\pi(M-L-2)/(k+2)}, \label{eq:vapara}
\end{equation}
where $L = 0,1,\cdots, \left[\frac{k}{2}\right]$, $M \in \mathbf{Z}_{2(k+2)}$, and $L+M \in 2\mathbf{Z}$. The vanishing Lagrangian submanifold corresponds to the curve $C_{ab}$ connecting these pair of points, which is parameterized by $(L,M)$. The BPS condition is given by
\begin{equation}
\int_{C_{a,b}} |P(X)^{\frac{6-d}{4}}| dX = |\int_{C_{a,b}} P(X)^{\frac{6-d}{4}} dX|.
\end{equation}
which states that the phase of $P(X)^{(6-d)/2}dX$ is constant along the path $C_{ab}$. It can be shown that the solution for the path $C_{ab} = \gamma_{ab}$ is unique in our simple setting $P(X) = X^{k+2} + \mu$ (for example, when $d=6$ the solution $\gamma_{ab}$ is just the straight line). The Lagrangian submanifold is given by the fibration of sphere $S^{n-1}$ over $\gamma_{ab}$ whose radius vanishes at the end points $X_a$, $X_b$.

The brane wrapped around the Lagrangian submanifold is supersymmetric. Therefore, if we find the (A-type) supersymmetric boundary states in the $\mathcal{N}=2$ super Liouville times super minimal model as we have seen in the last subsection, they should correspond to the vanishing cycles considered above one to one. Since the description of the branes in the super Liouville theory has been reviewed in section \ref{sec:11}, the remaining ingredient needed here is the branes for the minimal model. Without going into the detail of the derivations, we just collect the relevant facts (see e.g. \cite{Recknagel:1998sb}).

The Cardy states for the minimal model are labeled by two integers $L$ and $M$. For the NS sector, we should have $L + M \in 2\mathbf{Z}$ and for the R sector, we should have $L+M \in 2\mathbf{Z} +1$. They are made from the Ishibashi states by the Cardy's formula.

The Liouville part brane has been discussed in the last section, but the relevant brane here is the class 1 (ZZ type) brane which is localized in the $\phi \to \infty$ strong coupling region. This is natural since the vanishing cycle is localized in the strong coupling region near the singularity of the CY space.

Then the total boundary states is given by the closed sector GSO projected tensor product:
\begin{equation}
\sqrt{N} P_{GSO} \left(|L,M\rangle \otimes |B;r\rangle\right), \label{eq:ban}
\end{equation}
where $P_{GSO}$ projects onto the integral $U(1)$ charge sector in order to ensure the space-time SUSY. We should note, however, that the $P_{GSO}$ does not necessary guarantee the open GSO projection needed for the supersymmetric cycle. Our remaining task is to impose this condition on the open spectrum and determine supersymmetric cycles.

Before doing this, we restrict our range of discrete parameters. First, since we find that parameters $M$ and $r$ appear only through the combination $M+2r$, we can simply set $r=0$ ($1/2$) for NS (R) sector. Also if we consider only branes (no anti-branes) we can restrict ourselves to the range $L = 0,1,\cdots,\left[\frac{k}{2}\right],$ because the minimal model branes have a property that the R part changes just the sign when we perform the transformation $L \to k-L$ and $M \to M+k+2$ (for the NS part this change does nothing, therefore this transformation exchange D-branes with anti D-branes).

To evaluate the overlap integral (cylinder amplitude) with the boundary states (\ref{eq:ban}), we can simply borrow the results in the last section and insert the GSO projection 
\begin{equation}
\delta^{(N)}\left(\frac{N}{k+2}m + \beta\right) = \frac{1}{N} \sum_{r\in \mathbf{Z}_N} e^{-2\pi ir(\frac{1}{k+2}m + \frac{1}{N}\beta)}, 
\end{equation}
where $\beta = j$ for the massive representations and $\beta = s + 2Kr$ for the massless representations. The result is given by the summation over the character of the minimal model multiplied by the graviton representations of the Liouville character:
\begin{eqnarray}
e^{\pi\frac{\tilde{c}z^2}{\tau_c}} \langle L_1,M_1| e^{-\pi \tau_c H^c} e^{i\pi z(J_0 +\bar{J}_0)} | L_2,M_2\rangle \cr
 = \sum_{L=0}^{k} \sum_{r\in \mathbf{Z}_N} N^{L}_{L_1,L_2}\mathrm{ch}_{L,M_2-M_1-2r}(it_o,z') \chi_G(r;it_o,z'),
\end{eqnarray}
where $N^{L}_{L_1,L_2}$ is the fusion coefficients of $SU(2)_k$ and $\mathrm{ch}_{L,M_2-M_1-2r}(it_o,z')$ is the character for the minimal model whose precise form can be found in \cite{Eguchi:2003ik} and we do not use it here. The important point is that in order to ensure the open SUSY, we should have $M_1 \equiv M_2 \ mod \ 2(k+2)$. This parameter restriction is one to one corresponding to the vanishing cycle argument given in the beginning of this subsection. Therefore we expect that the minimal model parameter $L,M$ is just the same parameter which specifies the vanishing cycle in (\ref{eq:vapara}).

As a further consistency check, the open Witten indices, which can be obtained by the spectral flow of the above results, have been calculated in \cite{Eguchi:2003ik}. These show desirable properties as intersection numbers of the Lagrangian submanifold. For example, if we take $L=0$ and $d=2,6$, we have
\begin{equation}
I(0,M_1|0,M_2) = 2\delta^{(2(k+2))}(M_1-M_2) -\delta^{(2(k+2))}(M_1-M_2-2)-\delta^{(2(k+2))}(M_1-M_2+2),
\end{equation}
which is the $A_N$ type extended Cartan matrix.

\subsection{On the Duality of the $\mathcal{N}=2$ Super Liouville Theory}\label{12.3}

In section \ref{11.1.2}, we have discussed the conjectured duality \cite{Ahn:2002sx} of the $\mathcal{N} =2 $ super Liouville theory and its consequent result. This duality, if it holds, provides us a method to calculate various structure constants which we cannot determine from Teschner's trick otherwise. The claim of the duality is that the $\mathcal{N} =2$ super Liouville theory with the perturbation:
\begin{equation}
S_+ + S_- = \int d^2z \mu b^2 \psi^+ \psi^- e^{bS} + \mu b^2 \bar{\psi}^+\bar{\psi}^- e^{b S^*} + \pi \mu^2 b^2 :e^{bS}::e^{bS^*}:
\end{equation}
with a real $\mu$ is dual to the theory with the perturbation:
\begin{equation}
S_{nc} = \tilde{\mu}\int d^2 z (\partial \phi -i\partial Y -\frac{1}{2b}\psi^+\bar{\psi^+})(\bar{\partial}\phi + i\bar{\partial}Y -\frac{1}{2b}\psi^-\bar{\psi}^-) e^{\frac{1}{b}\phi},
\end{equation}
with the same $b$, hence the same central charge. 

In section \ref{11.1.2}, we have remarked that this duality is related to the hyper K\"ahler geometry in the T-dualized NS5-brane perspective, where the complex moduli deformation and the K\"ahler moduli deformation is related. Here we would like to prove the duality by combining other dualities which we have learned so far.\footnote{Technically speaking, we prove the duality only when $Y$ is compactified on the smallest circle, which is precisely related to the K3 geometry. It would be interesting to prove the duality with more general radii, which would provide a more nontrivial duality in the NS5-brane or singular CY perspective.} Note that this duality states that a certain K\"ahler potential deformation can be described by a superpotential deformation, which is interesting from the field theoretical point of view (because we believe that the physics depending on the superpotential is much easier to handle than those depending on the K\"ahler potential). See also \cite{Giveon:2001up}, \cite{Aharony:2003vk} for related arguments.

Let us start from the $\mathcal{N}=2$ super Liouville theory with the smallest compactification radius for $Y$. From the mirror symmetry we have discussed in section \ref{12.1.3}, it is equivalent to the $SL(2,\mathbf{R})/U(1)$ supercoset model (fermionic two dimensional black hole). Now we represent the  $SL(2,\mathbf{R})/U(1)$ supercoset model by the Wakimoto free field representation as we have done for the bosonic $SL(2,\mathbf{R})/U(1)$ in section \ref{6.5.1}. 

The supersymmetric Wakimoto representation is almost the same as that of the bosonic case if we recall the construction in section \ref{12.1} (see \cite{Eguchi:2000tc} for a complete description). The bosonic part of the current is the same as that of the bosonic coset model except the level $k$ becomes $k+2$. We add three free fermions for the $SL(2,\mathbf{R})$ fermionic partner and one more fermion for the $U(1)$ current which is the super partner of $X$. Finally, we need a bosonic ghost for the super partner of the $U(1)$ gauge fixed ghost. However, the fermionic part of the gauge fixing is almost trivial and the net result is we have only two fermionic freedom. Since the $U(1)$ BRST operator has become the same form as in the bosonic black hole, we have
\begin{equation}
Q^{U(1)} = \oint C(J^{3} + i\sqrt{\frac{k}{2}}\partial Y) + \mathrm{fermionic} \ \mathrm{term}.
\end{equation}

Then we can eliminate the Wakimoto coordinates $\beta$,$\gamma$ and ghost $B,C$ (and their super partners) as has been done in the bosonic case (see \ref{eq:sqf}). Since the bosonic part of the $SL(2,\mathbf{R})$ current and the $U(1)$ BRST charge are the same, the actual result can be easily obtained by simply supersymmetrizing the finial result of the bosonic theory. However, after eliminating the Wakimoto coordinates, the screening operator becomes nothing but the nonchiral deformation \cite{Eguchi:2000cj}:
\begin{equation}
S_{nc} = \tilde{\mu}\int d^2 z (\partial \phi -i\partial Y -\frac{1}{2b}\psi^+\bar{\psi}^+)(\bar{\partial}\phi + i\bar{\partial}Y -\frac{1}{2b}\psi^-\bar{\psi}^-) e^{\frac{1}{b}\phi}.
\end{equation}

This completes our proof of the conjectured duality of the $\mathcal{N}=2$ super Liouville theory. It would be interesting to extend the result to the boundary theory and the more general radius case. Also it is a challenging problem to obtain the precise relation between the cosmological constant $\mu$ and the dual cosmological constant (black hole mass parameter) $\tilde{\mu}$, which can be obtained from Teschner's trick argument as has been done in \cite{Ahn:2002sx}, along the same line of reasoning above. However, we find one difficulty; the mirror duality used here is essentially a nonperturbative effect and we have not been able to obtain the precise value for the dynamically generated scale so far.

\subsection{Literature Guide for Section 12}\label{12.4}
Materials we have discussed in this section are closely related to each other via various dualities. The fermionic black hole has been discussed in \cite{Giveon:1999px,Giveon:2003wn}, but it is dual to the $\mathcal{N} =2 $ super Liouville theory \cite{Hori:2001ax}, \cite{Giveon:1999px}, \cite{Giveon:2001up}, \cite{Tong:2003ik}. Originally, this duality was conjectured from the dual little string theory which has been studied in  \cite{Aharony:1998ub}, \cite{Giveon:1999px} in this context. This chain of duality and geometrical perspective leads us to the $\mathcal{N} = 2$ duality conjectured in \cite{Ahn:2002sx} as we have seen in section \ref{12.3}. As far as the author knows, the proof or explanation of the $\mathcal{N} = 2$ duality is first presented in this review, though the individual dualities have been known for years. See also \cite{Aharony:2003vk} for related discussions.

On the other hand, the exact description of the branes in the $\mathcal{N} = 2$ super Liouville theory \cite{Eguchi:2003ik}, \cite{Ahn:2003tt} enables us to extend the earlier works \cite{Lerche:2000uy}, \cite{Lerche:2000jb}, \cite{Sugiyama:2000id}, \cite{Eguchi:2000cj} and study the detailed physics of the branes in the singular CY space. Since the space-time is higher dimensional, we do not expect that the matrix model-like description is possible, but even the world sheet analysis of the Liouville theory will reveal important physics of the NS5-brane and the singular CY space.

\part{Unoriented Liouville Theory}

\sectiono{Liouville Theory on Unoriented Surfaces}\label{sec:13}
The bosonic Liouville theory on unoriented surfaces is discussed in this section. One of the motivations of this setup is to understand the properties of the orientifold plane under the nontrivial background. We use the $\alpha'=1$ notation in this section. The organization of this section is as follows.

In section \ref{13.1}, we derive the exact crosscap state in the Liouville theory  from the modular bootstrap method and the conformal bootstrap method, and we discuss its properties.

In section \ref{13.2}, the tadpole cancellation in the $c=1$ two dimensional unoriented string theory is reviewed. In subsection \ref{13.2.1}, the free field calculation is presented and in subsection \ref{13.2.2}, the boundary-crosscap state calculation is presented. From both approaches we conclude that two D1-branes are necessary to cancel the tadpole divergence. In subsection \ref{13.2.3}, on the other hand, alternative Fischler-Susskind mechanism is reviewed to deal with the crosscap tadpole.

In section \ref{13.3}, the matrix model dual of the unoriented Liouville theory is discussed. In subsection \ref{13.3.1}, we discuss the unoriented $c=0$ matrix model and compare some amplitudes with the continuum Liouville calculation. In subsection \ref{13.3.2}, the unoriented $c=1$ matrix quantum mechanics is discussed and particularly the thermodynamic aspects of the unoriented $c=1$ matrix model are investigated.

\subsection{Crosscap State}\label{13.1}
The most fundamental object for the unoriented Liouville theory is the one-point function on the projective plane. As in the one-point function on the disk, we can calculate it either by the direct calculation or by the crosscap state from the modular bootstrap. In this section, following \cite{Hikida:2002bt}, we obtain the projective plane one-point function first as a crosscap state from the modular bootstrap method and then check its consistency by the direct conformal bootstrap method.

Let us first review the building block of the crosscap state --- the Ishibashi crosscap states \cite{Ishibashi:1989kg}. They satisfy the crosscap condition
\begin{equation}
(L_n - (-1)^n \bar{L}_{-n}) |P,C\rangle = 0.
\end{equation}
This equation is solved formally by
\begin{equation}
|P,C\rangle = \left(1-\frac{L_{-1}\bar{L}_{-1}}{2 \Delta_P} + \cdots\right)|P\rangle
\end{equation}
with any bulk primary state $|P\rangle $. In the bosonic Liouville case considered here, we use  $e^{(Q+2iP)\phi}$ as the (normalizable) primary states. The defining properties of the Ishibashi states are
\begin{align}
\langle P, C| e^{-\pi\tau_c(L_c + \bar{L}_c -\frac{c}{12})}|P', C\rangle &= \delta(P-P')\frac{q^{P^2}}{\eta(i\tau_c)}\cr
\langle P, B| e^{-\pi\tau_c(L_c + \bar{L}_c -\frac{c}{12})} |P',C\rangle &= \delta(P-P')\frac{q^{P^2}}{\eta(i\tau_c+\frac{1}{2})}
\end{align}
where $q=e^{-2\pi \tau_c}$ is the closed modular parameter.

With these Ishibashi states, we expand the Cardy crosscap state as 
\begin{equation}
\langle C| = \int_{-\infty}^{\infty} dP\Psi_c(P) \langle P,C|
\end{equation}
Our task is to determine the wavefunction $\Psi_c(P)$ which is proportional to the projective plane one-point function. 

To do this, let us consider the M\"obius strip amplitude bounded by the $(1,1)$ ZZ brane and the crosscap state. This amplitude is highly constrained
from the sewing constraint \cite{Lewellen:1991tb},\cite{Fioravanti:1993hf}, which is the crosscap analogue of the Cardy condition. In the exchange channel, it becomes
\begin{equation}
Z_{M,(1,1)}(\tau_c)= \int_{-\infty}^{\infty} dP \Psi_{(1,1)}(-P) \Psi_c(P) \frac{q^{P^2}}{\eta(i\tau_c+\frac{1}{2})},
\end{equation}
where we can find the $(1,1)$ boundary wavefunction $\Psi_{(1,1)}(-P)$ in (\ref{eq:11wf}). The open/closed duality enables us to calculate the same amplitude in the open loop channel. Since the $(1,1)$ boundary state contains only the identity operator, we expect that the same amplitude becomes in the open channel,
\begin{equation}
Z_{M,(1,1)}(\tau_o) = \mathrm{Tr}_{o,(1,1)} [\Omega e^{-2\pi \tau_o(L_o -\frac{c}{24})}] = \frac{e^{-\frac{\pi \tau_o}{2} (b +b^{-1})^2} + e^{-\frac{\pi\tau_o}{2}(b-b^{-1})^2}}{\eta(i\tau_o + \frac{1}{2})}. \label{eq:m11t}
\end{equation}
In the more general case where the open spectrum is constructed by the $(n,m)$ degenerate state, we have \cite{Hikida:2002bt}
\begin{equation}
\mathrm{Tr}_{o,(n,m)} [\Omega e^{-2\pi \tau_o(L_o -\frac{c}{24})}] = \frac{e^{-\frac{\pi \tau_o}{2} (nb +mb^{-1})^2} -(-1)^{nm} e^{-\frac{\pi\tau_o}{2}(nb-mb^{-1})^2}}{\eta(i\tau_o + \frac{1}{2})}.
\end{equation}
Note that the factor $(-1)^{nm}$ comes from the fact that the level $nm$ null state has $\Omega = (-1)^{nm}$. In the M\"obius strip case, the modular transformation is given by $\frac{1}{4\tau_c} = \tau_o$. Using the modular transformation formula in appendix (\ref{eq:modun}), we can translate it into the closed exchange channel as
\begin{equation}
\mathrm{Tr}_{o,(1,1)} [\Omega e^{-2\pi \tau_o(L_o -\frac{c}{24})}] = \int_{-\infty}^{\infty} dP \frac{e^{-2\pi \tau_cP^2}}{\eta(i\tau_c + \frac{1}{2})} \left(\cosh(\pi(b+b^{-1})P) + \cosh (\pi (b-b^{-1})P)\right).
\end{equation}
Comparing this with (\ref{eq:m11t}) and substituting the (1,1) boundary state wavefunction \eqref{eq:11wf}, we have
\begin{equation}
\Psi_c(P) = (\pi\mu\gamma(b^2))^{-iPb^{-1}}\frac{2^{1/4}\Gamma(1+2iPb)\Gamma(1+2iPb^{-1})}{2\pi i P} \cosh(\pi Pb) \cosh(\pi Pb^{-1}). \label{eq:pp1p}
\end{equation}

To check its consistency, we use Teschner's trick to derive the constraint on the crosscap one-point function and see whether this is satisfied or not. As we will see, the solution of the functional relation is not unique, but in any case we can check that the above wavefunction derived from the modular bootstrap is a consistent one.

First, let us review several basic facts about the correlation functions on the projective plane. The projective plane can be seen as a complex plane with the identification: 
\begin{equation}
 z \sim -\frac{1}{\bar{z}}.
\end{equation}
Therefore, the primary fields on one-point and on its image point are related as
\begin{equation}
\phi(z,\bar{z}) \sim z^{-2\Delta} \bar{z}^{-2\bar{\Delta}} \phi (-1/\bar{z},-1/z).
\end{equation}
This constrains (with $SL(2,\mathbf{R})$ invariance) the form of the correlation functions as 
\begin{equation}
\langle \phi(z,\bar{z})\rangle_{\mathrm{RP}_2} = \frac{c}{(1+z\bar{z})^{2\Delta}}
\end{equation}
and 
\begin{equation}
\langle \phi(z_1,\bar{z}_1) \phi(z_2,\bar{z}_2) \rangle_{\mathrm{RP}_2} = \frac{(1+z_2\bar{z}_2)^{2\Delta_1-2\Delta_2}}{|1+z_1\bar{z_2}|^{4\Delta_1}} F(\eta),
\end{equation}
where $\eta$ is the crossratio:
\begin{equation}
\eta \equiv \frac{|z_1 - z_2|^2}{(1+z_1\bar{z}_1)(1+z_2\bar{z}_2)}
\end{equation}

As in the disk case, we consider the auxiliary two-point function:
\begin{equation}
G(\eta) \sim \langle V_\alpha(z_1,\bar{z}_1) V_{-b/2}(z_2,\bar{z}_2)\rangle_{\mathrm{RP}_2},
\end{equation}
where the trivial conformal factor is omitted from the left-hand side.
With the method of image, this amplitude is effectively given by
\begin{equation}
\langle V_\alpha(z_1) V_{-b/2}(z_2) V_{\alpha}(-1/\bar{z}_1)\bar{z}_1^{-2\Delta_\alpha} V_{-b/2}(-1/\bar{z}_2) \bar{z}_2^{-2\Delta_{2,1}}\rangle.
\end{equation}
Thus, evaluating this amplitude either by approaching $z_1 \to z_2$ first or by approaching $z_1 \to -1/\bar{z}_2$ first, and comparing their results, we obtain the functional relation for the one-point function:
\begin{equation}
\langle V_{\alpha}(z,\bar{z})\rangle_{\mathrm{RP}_2} = \frac{U(\alpha)}{|1+z\bar{z}|^{2\Delta_\alpha}}.
\end{equation}

\begin{figure}[htbp]
	\begin{center}
	\includegraphics[width=0.8\linewidth,keepaspectratio,clip]{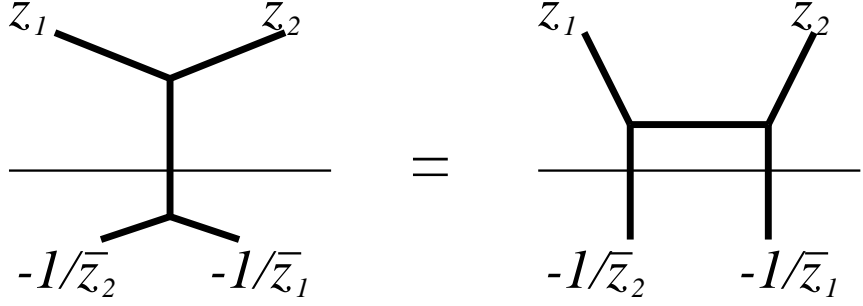}
	\end{center}
	\caption{We can calculate the two-point function on the $\mathrm{RP}_2$ by the method of image. Changing the order of taking OPE enables us to obtain the two different expression for the two-point function.}
	\label{crosssym}
\end{figure}

The direct OPE as $\eta \to 0$ yields
\begin{equation}
G(\eta) = C_+ U(\alpha-b/2)F_+(\eta) + C_- U(\alpha+b/2) F_-(\eta), 
\end{equation}
where we have already calculated the OPE coefficient $C_\pm$ in section \ref{5.1.1} (\ref{eq:OPE1})(\ref{eq:OPE2}). The conformal block $F_\pm$ is determined by the differential equation which the degenerate operator $V_{-b/2}$ should satisfy. These are given by\footnote{Note that these functions are different from (\ref{eq:bhg}). This is because from our definition of the projective plane crossratio $\eta$, the operator identification with the four-point bulk operator is different from that has been done in the disk case. In terms of the differential equation (\ref{eq:4hg}), we identify $\phi_4 = V_{\alpha}$, $\phi_3 = V_{-b/2}$ and $\phi_1 = V_{\alpha}$ here, but in the disk case, we have had  $\phi_3 = V_{\alpha}$, $\phi_4 = V_{-b/2}$ and $\phi_1 = V_{\alpha}$.}
\begin{eqnarray}
F_+(\eta) &=& \eta^{b\alpha} (1-\eta)^{b\alpha} F(2b\alpha,2b\alpha-2b^2-1,2b\alpha-b^2,\eta) \cr
F_-(\eta) &=& \eta^{1+b^2-b\alpha} (1-\eta)^{b\alpha} F(-b^2,1+b^2,2+b^2-2b\alpha,\eta) \label{eq:hggg}
\end{eqnarray}

On the other hand, in the cross channel, we have
\begin{equation}
G(\eta) = C_+ U(\alpha-b/2)F_+(1-\eta) + C_- U(\alpha+b/2) F_-(1-\eta), 
\end{equation}
where $F_{\pm}(1-\eta)$ is the same function as (\ref{eq:hggg}) with a different argument. From the hypergeometric function identity (\ref{eq:hyperg}) and the actual form of $C_\pm$ we can write the functional relation for the one-point function $U(\alpha)$ as
\begin{eqnarray}
U(\alpha + b/2) &=& \frac{\Gamma(2-2b\alpha+b^2)\Gamma(-1+2b\alpha - b^2)}{\Gamma(1+b^2)\Gamma(-b^2)}U(\alpha+b/2) \cr
& & -\frac{\gamma(-b^2)}{\pi\mu} \frac{\Gamma(2-2b\alpha+b^2)\Gamma(2b\alpha-b^2)}{\Gamma(1-2b\alpha)\Gamma(-1+2b\alpha-2b^2)} U(\alpha-b/2), \cr 
U(\alpha - b/2) &=& \frac{\Gamma(2b\alpha-b^2)\Gamma(1-2b\alpha + b^2)}{\Gamma(1+b^2)\Gamma(-b^2)}U(\alpha-b/2) \cr
& & -\frac{\pi\mu}{\gamma(-b^2)} \frac{\Gamma(1-2b\alpha+b^2)\Gamma(-1+2b\alpha-b^2)}{\Gamma(2b\alpha)\Gamma(2-2b\alpha + 2b^2)} U(\alpha+b/2). \label{eq:cfunc}
\end{eqnarray}
Actually, the first equation is not independent of the second one as we can see they give the same constraint from a simple algebra. We can solve these constraints with the duality assumption $b\to b^{-1}$ as,
\begin{equation}
U(\alpha) = \frac{2}{b}(\pi\mu\gamma(b^2))^{(Q-2\alpha)/2b} \Gamma(2b\alpha -b^2)\Gamma\left(\frac{2\alpha}{b}-\frac{1}{b^2}-1\right) f(\alpha),
\end{equation}
where $f(\alpha)$ is given by the linear combination of 
\begin{equation}
\cos((b+b^{-1})\pi(\alpha-Q/2)) , \ \ \ \ \cos((b-b^{-1})\pi(\alpha-Q/2)).
\end{equation}
Of course, we have already known that the actual combination is given by (\ref{eq:pp1p}), but we will use this constraint in the following to address some peculiar issues.

Using this crosscap state, we can calculate other M\"obius strip amplitudes and Klein bottle amplitude. In the next section, we discuss their divergences and the possibility of the tadpole cancellation. For the time being, we ignore the divergences and consider their general properties.

First, we can calculate the $(m,n)$ M\"obius strip amplitude which is given by
\begin{eqnarray}
Z_{m,n}^M &=& \int_{-\infty}^{\infty} dP \Psi_{m,n}(P)\Psi_C(-P) \frac{q^{P^2}}{\eta(i\tau_c+\frac{1}{2})} \cr
&=& 2\int_{-\infty}^{\infty} dP \frac{q^{P^2}}{\eta(i\tau_c+\frac{1}{2})} \frac{\sinh(2\pi mPb^{-1})\sinh(2\pi nPb)\cosh(\pi Pb^{-1})\cosh(\pi bP)}{\sinh(2\pi Pb^{-1})\sinh(2\pi bP)}.
\end{eqnarray}
With the trigonometric function formula:
\begin{equation}
\frac{\sinh(2\pi nbP) \cosh(\pi bP)}{\sinh(2\pi bP)} = \sum_{l=0}^{n-1} \cosh (\pi b P(2l+1)),
\end{equation}
we have \cite{Hikida:2002bt}
\begin{equation}
Z_{m,n}^M (\tau_c) = \sum_{k=0}^{m-1}\sum_{l=0}^{n-1} \mathrm{Tr}_{o(2k+1,2l+1)}[\Omega e^{-2\pi  \tau_o(L_o -\frac{c}{24})}].
\end{equation}
This is expected because this M\"obius strip amplitude should yield the $\Omega$ projection of the cylinder amplitude (\ref{eq:mnmn}) which is bounded both by the $(m,n)$ boundary state.

Things are all consistent so far. However, when we consider the M\"obius strip amplitude with the FZZT brane or the Klein bottle amplitude, situation becomes subtler. For example, we find that the M\"obius strip partition function with the FZZT brane whose boundary parameter is $s$ becomes
\begin{eqnarray}
Z_s^M &=& \int_{-\infty}^{\infty} dP \frac{q^{P^2}}{\eta(i\tau_c+\frac{1}{2})} \Psi_s(P) \Psi_{C}(-P) \cr
    &=& \int_{-\infty}^{\infty} dP'\frac{q^{(P')^2}}{\eta(i\tau_o+\frac{1}{2})}\rho_s(P'), 	
\end{eqnarray}
where the density of states $\rho_s(P')$ is given by
\begin{equation}
\rho_s(P') = \int_{-\infty}^{\infty} dP e^{2\pi i PP'} \Psi_s(P) \Psi_C(-P).
\end{equation}
On the other hand, the density of states from the cylinder partition function which is bounded by the same two FZZT branes with the boundary parameter $s$ is given by
\begin{equation}
\rho'_s(P') = \int_{-\infty}^{\infty} dP e^{4\pi i PP'} \Psi_s(P) \Psi_s(-P).
\end{equation}
For the $\Omega$ projection to make sense in the open spectrum, we expect
\begin{equation}
\rho'_s(P') = \rho_s(P').
\end{equation}
Otherwise, the ``unphysical states" propagate in the open loop channel. However, this identity \textit{does not} hold with the actual wavefunctions which have
 been obtained so far (even no matter how we adjust the boundary parameter $s$). This is very curious.\footnote{We have encountered the same problem in the GSO projection for the FZZT brane in section \ref{sec:10}.} Note that the actual evaluation of the density of states suffers an infrared divergence as $P \to 0$. In the free limit where we just focus on the constant part of the density of states which is proportional to the Liouville volume $-\frac{1}{2b}\log\mu$, this mismatch does not make any difference. This is because the $-\frac{1}{2b}\log\mu$ term comes from the infrared cutoff procedure and the strength of the divergence is the same. In the $c=1$ language, the mismatch appears in the non-bulk scattering but not in the bulk scattering. We will further study its implication in the next section \ref{13.2}.

With the puzzling issue raised above, one may wonder whether we can obtain the correct M\"obius amplitude for the FZZT brane. In other words, can we start from the FZZT brane to obtain the crosscap wavefunction? If this is the case, the crosscap should depend on the boundary parameter $s$, which seems odd and the proper projection on the closed sector is implausible. In any case, we can easily see that  the crosscap wavefunction naively derived from the FZZT brane projection by the modular bootstrap \textit{does not} satisfy the functional relation (\ref{eq:cfunc}).

Also we can calculate the Klein bottle amplitude. The density of states from the Klein bottle is given by
\begin{equation}
\rho^c_s(P') = \int_{-\infty}^{\infty} dP e^{4\pi i PP'} \Psi_C(P) \Psi_C(-P),
\end{equation}
which can be compared with the density of states for the closed Liouville theory  whose nonsingular part can be calculated by the reflection amplitude (see section \ref{4.2}). Since the above expression suffers an infrared divergence, we need a good regularization scheme in order to discuss whether or not this density of states is the same as in the closed one and the $\Omega$ projection works properly.

\subsection{Tadpole Cancellation}\label{13.2}
In this section, we discuss the tadpole cancellation in the $c=1$ unoriented Liouville theory \cite{Nakayama:2003ep}, \cite{Bergman:2003yp}. In the following we take two different approaches --- the free field method and the boundary-crosscap state method to calculate the one-loop divergence (hence, the tadpole divergence). The both methods yield the same tadpole cancellation condition, two D1-branes with the symplectic gauge group. However, there is a finite difference between these methods, which is somewhat related to the  $\Omega$ projection problem on the FZZT spectrum mentioned at the end of the last section. We will also discuss the recently advocated possibility that the Fischler-Susskind mechanism cancels the tadpole divergence in the unoriented Liouville theory \cite{Gomis:2003vi}.

\subsubsection{Free field calculation}\label{13.2.1}
First let us review the free field path integral method to calculate the Klein bottle partition function. As we have seen in section \ref{sec:2}, the Liouville  amplitude obey the exact WT identity

\begin{equation}
\langle e^{2\alpha_1 \phi} e^{2\alpha_2\phi} \cdots e^{2\alpha_N\phi} \rangle \propto \mu^{\frac{(1-g)Q-\sum_i \alpha_i}{b}}. \label{eq:ewt2}
\end{equation}
Therefore, the perturbation in $\mu$ is only possible when the power of $\mu$ is an integer. Furthermore, we have assumed that the perturbative calculation yields the correct result when this is the case. The simplest example is the torus partition function which we have studied in section \ref{2.2}.

Then it is natural to suppose that the Klein bottle partition function can be obtained in the same manner because it has the same $\mu$ dependence and the world sheet curvature is also zero. The free field path integration results in
\begin{equation}
Z_{K_2} = V_\phi V_X \int_0^\infty \frac{dt}{4t} \frac{1}{4\pi^2 t},
\end{equation}
which diverges when $t \to 0$. If we do the (formal) modular transformation $ s = \pi/2t$, it becomes more clear
\begin{equation}
Z_{K_2} = V_\phi V_X \int_0^\infty \frac{ds}{8\pi^3}.
\end{equation}
This is nothing but the massless tadpole amplitude.

As in the ordinary string theory, the massless tadpole should be canceled by other D-branes \cite{Green:1984sg}, \cite{Ohta:1987nq}, \cite{Polchinski:1988tu}. For this purpose, we introduce the cylinder and the M\"obius strip partition function as follows ($n$ is a number of D1-branes and $+,-$ sign corresponds to $SO(2n)$ and $Sp(2n)$ respectively):
\begin{equation}
Z_{C_2} = n^2V_\phi V_X  \int_0^\infty \frac{dt}{4t}\frac{1}{8\pi^2t} = n^2 V_\phi V_X  \int_0^\infty \frac{ds}{8\pi^3 4},
\end{equation}
\begin{equation}
Z_{M_2} = \pm nV_\phi V_X  \int_0^\infty \frac{dt}{4t}\frac{1}{8\pi^2t} = \pm n V_\phi V_X  \int_0^\infty \frac{ds}{8\pi^3 }.
\end{equation}
In the cylinder case, the modular transformation\footnote{Because these partition functions are divergent, the modular transformation is rather formal. We do a kind of ``dimensional regularization" to determine their conventional modular transformation (see \cite{Polchinski:1998rq}) which is needed to cancel the tadpole.} is $s = \pi/t$, and in the M\"obius strip case, it is $s = \pi/4t$. Combining these amplitudes, in order to cancel the tadpole divergence, we should take $n=2$ and $Sp(2)$ gauge group.

This is just the same tadpole cancellation in the free 2D string (without a ``tachyon background"). There might be some questions. What kind of D1-branes did we use?
 As we have seen in section \ref{sec:5}, D1-branes in the Liouville theory have a continuous parameter $s$ which is related to the boundary cosmological constant $\mu_B$ as 
 \begin{equation}
 \cosh^2\pi bs = \frac{\mu_B^2}{\mu} \sin\pi b^2.
 \end{equation}
We treat D1-branes such that on which strings have the Neumann boundary condition. Therefore, we might guess $\mu_B = 0$ on these D1-branes. Is this interpretation consistent with the disk one-point function? What happens if we turn on the boundary cosmological constant? We answer these questions after investigating the boundary-crosscap state formalism in the next subsection.

\subsubsection{Boundary-Crosscap state calculation}\label{13.2.2}
First, we calculate the Klein bottle partition function as follows,
\begin{equation}
Z_{K_2} =V_X \int_{0}^\infty  d\tau \left(\int_{-\infty}^{\infty} dP \Psi_C(P)\Psi_C(-P) \frac{q^{P^2}}{\eta(q)}\right)\frac{2}{\sqrt{2}\eta(q)}\eta(q)^2 ,
\end{equation}
where $q = e^{-2\pi\tau}$ is the closed channel modular parameter, and because of the explicit momentum integration we do not have the Liouville volume factor here. Substituting (\ref{eq:pp1p}) and setting $b=1$, we obtain
\begin{equation}
Z_{K_2} = V_X \int_{-\infty}^\infty  \frac{dP}{4\pi} \frac{[\cosh(2\pi P)+1]^2}{[\sinh(2\pi P)]^2P^2}.
\end{equation}
This partition function is ultraviolet finite but infrared divergent. This divergence shows the existence of the massless tadpole.

To cancel the tadpole, we should calculate other Euler number zero partition functions. The cylinder partition function \cite{Martinec:2003ka} is (with $n$ D1-branes whose boundary cosmological constants are the same for simplicity)
\begin{eqnarray}
Z_{C_2} &=& n^2V_X \int_{0}^\infty  d\tau \left(\int_{-\infty}^{\infty} dP \Psi_s(P)\Psi_s(-P) \frac{q^{P^2}}{\eta(q)}\right)\frac{1}{\sqrt{2}\eta(q)}\eta(q)^2 \cr
&=& n^2 V_X \int_{-\infty}^\infty \frac{dP}{4\pi} \frac{[\cos(2\pi sP)]^2}{[\sinh(2\pi P)]^2P^2}, \label{eq:anu}
\end{eqnarray}
which diverges as $P \to 0$. This is a closed channel infrared divergence, i.e. a tadpole divergence. 

The M\"obius strip partition function is\footnote{The correspondence between the $\pm$ here and the gauge group choice (orientifold operation on the Chan-Paton indices) is not so obvious. To fix this, consider the open sector calculation $\pm n \mathrm{Tr}_{O} [\Omega e^{-2\pi tH}]$. If we take the limit $t \to \infty$, this contribution is positive for the $SO(n)$ and negative for $Sp(n)$. Then we modular transform this and compare it with the boundary-crosscap state calculation in the $s = \pi/4t \to 0$ limit. The relevant sign is determined by the $\lim_{P \to 0} \Psi_c(P)\Psi_s(-P)$ which is positive in our normalization of $\Psi_c(P)$ and $\Psi_s(P)$. Thus $+$,$-$ indeed corresponds to $SO(n)$ and $Sp(n)$ respectively.}
\begin{equation}
Z_{M_2} = \pm n V_X \int_{-\infty}^\infty \frac{dP}{2\pi} \frac{[\cos(2\pi sP)][\cosh(2\pi P)+1]}{[\sinh(2\pi P)]^2P^2},
\end{equation}
which also diverges as $P \to 0$. Combining all these partition functions we obtain,
\begin{equation}
Z_{1loop} = V_X \int_{-\infty}^\infty \frac{dP}{4\pi} \frac{[\cosh(2\pi P)+1  \pm n \cos(2\pi s P)]^2}{[\sinh(2\pi P)]^2P^2}.
\end{equation}
The infrared tadpole divergence can be canceled if we choose the $Sp(2)$ gauge group for two D1-branes, irrespective of the value of $s$.

Using the boundary-crosscap state formalism, we have obtained the same tadpole cancellation condition i.e. two D1-branes with the gauge group $Sp(2)$. However, the finite part left is different in each method. In the free field calculation, the partition function completely vanishes. On the other hand, in the boundary-crosscap calculation, there is a finite part left even if we choose $s$ to be $i\frac{\pi}{2}$, which corresponds to $\mu_B=0$. It is interesting to see whether this finite part can be seen from the matrix quantum mechanical point of view (but it is unlikely as we will see). For the unoriented Liouville theory, the corresponding dual matrix quantum mechanics should be $SO(2N)$ or $Sp(2N)$ quantum mechanics ``living on the D0-branes"\footnote{Actually the tadpole canceled matrix quantum mechanics should choose $Sp(2N)$ gauge group. Once we choose the normalization of the crosscap state to be minus that of the D1-brane so as to cancel the tadpole, the D0-brane gauge field should also be $Sp(2N)$ \cite{Bergman:2003yp}. Intuitive argument in the matrix model point of view is that, for $Sp$ theory, a twisted loop has an extra minus sign and this is necessary to cancel the fundamental loop.}, which is also considered as a discretized version of the 2D unoriented quantum gravity with a boson on it. It is important to note that this matrix quantum mechanics is finite. Therefore, the matrix quantum mechanics predicts that the Klein bottle partition function should be finite. In contrast, the matrix model loop amplitude reproduces that of the divergent boundary-boundary annular diagram (\ref{eq:anu}) as was studied in \cite{Martinec:2003ka}. Consequently, the need for the tadpole cancellation from the matrix model point of view is rather puzzling and further study is needed. We will see in the next subsection that the pure $SO/Sp$ matrix model without D1-branes (vector) calculation might be related to the tadpole canceled theory by the Fischler-Susskind mechanism.

Some applications (e.g. the correction to the rolling tachyon amplitudes) of the above results are studied in \cite{Nakayama:2003ep}. Instead of repeating the arguments therein, here we discuss the uncanceled finite part left in the one-loop diagram further. The direct interpretation of the uncanceled part is that the open (massless) tachyon is not properly projected. The free field guess is that the open tachyon on the D1-brane has the quaternionic selfdual (real) representation of the $Sp(2)$ D1-brane Chan-Paton factor. However, in the intermediate loop diagram we find that there is a quaternionic antiselfdual representation which should be projected out. If we cut the diagram by the unitarity argument, we are forced to have an ``unphysical" degree of freedom in the spectrum.

However, since we are dealing with a theory in a nontrivial background (Liouville potential), the notion of unitarity should be used with a grain of salt. We should also note that in the $P \to 0$ limit which corresponds to the bulk physics, we do not have such a problem.\footnote{In the matrix model, the finite part is expected to appear as a nonuniversal term, so we might not find any difference in the matrix model. The author would like to thank J.~Gomis and A.~Kapustin for a valuable discussion on this point.} Therefore there is a possibility that the interaction which becomes larger in the deep $\phi$ region just affects the one-loop diagram as a ``boundary" contribution. At the same time, however, it is important to note that the $(m,n)$ ZZ brane which lives in the deep $\phi$ region, where the coupling is strong, has a proper $\Omega$ projection on the contrary.

One of the biggest motivations for considering the Liouville theory on unoriented surfaces is to understand the physics of the orientifold plane in a nontrivial background. This finite part and projection mismatch may yield an important clue to understand the general behavior of the orientifold plane under a nontrivial background. We hope that the further study on the subject and the solution to this puzzle will reveal an interesting physics of the orientifold plane dynamics.

\subsubsection{Tadpole cancellation by Fischler-Susskind mechanism}\label{13.2.3}
Unlike the R-R tadpole, the NS-NS tadpole (or the bosonic tachyon tadpole) can be canceled by the deformation of the background. This is what is called the Fischler-Susskind mechanism \cite{Fischler:1986ci,Fischler:1986tb}. The Fischler-Susskind mechanism cancels the higher Euler number diagram divergence by adding counterterms (vertices) to the lower Euler number diagram. In the Liouville case, the zero Liouville momentum divergence can be canceled by varying the tree-level cosmological constant \cite{Gomis:2003vi}. 

To see this, let us examine the tadpole part of the crosscap one-point function (which emits the zero energy/momentum on-shell tachyon).
\begin{equation}
\lim_{P\to 0} \langle V_{P} \rangle_{\mathrm{RP}_2} = V_x \lim_{P\to0}(\mu_R)^{-iP}\frac{2^{1/4}\Gamma(1+2iP)\Gamma(1+2iP)}{2\pi i P} \cosh(\pi P) \cosh(\pi P), \label{eq:0lim}
\end{equation}
where $\mu_R = \lim_{b\to 1} \pi \mu \gamma(b^2)$ is the renormalized\footnote{This is the tree level renormalization which we should do even in the sphere level. In the following, we renormalize $\mu_R$ further in order to absorb the divergence from the projective plane tadpole.} cosmological constant. Since the zero momentum limit in (\ref{eq:0lim}) is divergent, we need a regularization to obtain a sensible result. The natural way to do this is to put the system in a box of length $V_{\phi} = -\frac{1}{2}\log\mu$ and interpret the zero momentum mode as the lowest momentum mode $p_{min} = \frac{\pi}{V_\phi}$. Then we have
\begin{equation}
\lim_{P\to 0} \langle V_{P} \rangle_{\mathrm{RP}_2} = \frac{V_\phi}{2\pi} \frac{V_x}{2\pi} C,
\end{equation}
where $C$ is a numerical constant, which ultimately depends on our regularization procedure. Note that our choice of the size of box is natural from the argument in section \ref{4.3}. Also, by integrating by $\mu$ once, we will obtain the $\mathrm{RP}_2$ partition function which should be proportional to $\mu\log \mu$ from the general KPZ scaling argument.

Because of the existence of this tadpole, the typical scattering amplitudes on the projective plane or the Klein bottle amplitude diverge. In the last subsection we have introduced D1-branes to cancel such divergences. However, as we stated earlier in this subsection, the Fischler-Susskind mechanism may as well cancel them. For instance, let us consider the free-field calculation\footnote{With a proper regularization, we expect that the Klein bottle calculation from the exact wavefunction yields the same results (up to a subtle cut off dependent part which is not proportional to $\log\mu$ which may cause the unprojection problem on the spectrum).} of the Klein bottle amplitude. We assume $X$ is compactified on a circle whose radius is $R$. Taking into consideration that only momentum modes contribute to the amplitude, we obtain
\begin{eqnarray}
Z_{KB} &=& \frac{V_\phi}{2\pi} \int_0^\infty \frac{d\tau}{2\tau^{3/2}} \sum_{n=-\infty}^{\infty} \exp\left(-\pi \tau\frac{1}{R^2}n^2\right) \cr
	&=& \frac{V_{\phi} R}{4\pi}\int_0^\infty ds \frac{2}{\pi}\left(1+2\sum_{n=1}^{\infty} \exp\left(- 2R^2 n^2 s\right)\right) = \frac{V_\phi R}{4\pi}\left(\int_0^\infty \frac{2ds}{\pi} + \frac{\pi}{3R^2}\right),
\end{eqnarray}
where we have modular transformed to the exchange channel in the second line $(\tau=\pi/2s)$.

Combining this with the torus amplitude (we should recall further $1/2$ is needed for \textit{both} torus and Klein bottle amplitudes with the above convention), we write the whole one-loop amplitude as
\begin{equation}
Z = V_{\phi} \frac{1}{24}\left(R + \frac{2}{R}\right) + \frac{V_\phi R}{4\pi} \int_0^\infty \frac{ds}{\pi} .
\end{equation}

\begin{figure}[htbp]
	\begin{center}
	\includegraphics[width=0.7\linewidth,keepaspectratio,clip]{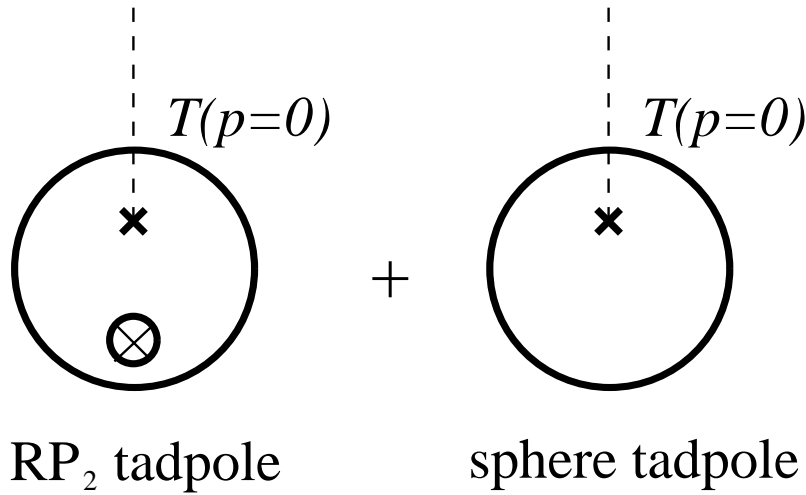}
	\end{center}
	\caption{The $\mathrm{RP}_2$ tadpole divergence can be canceled by introducing the tree level counter term (Fischler-Susskind mechanism).}
	\label{FS}
\end{figure}

To cancel the second part (or the $\mathrm{RP}_2$ tadpole), we add the counterterm:
\begin{equation}
\delta S = (\log\Lambda)  C g_s\int d^2z e^{2\phi}, 
\end{equation}
where $g_s$ can be set to $1$ if we have used the usual Liouville identification $\mu \sim 1/g_s$.
This term cancels the zero momentum emission of massless tachyons.
Using this counterterm perturbatively, we can obtain a finite string theory at least to the first order in the string perturbation theory.

\subsection{Matrix Model Dual}\label{13.3}
\subsubsection{$c=0$ matrix model}\label{13.3.1}
The matrix model dual for the $c=0$ unoriented Liouville theory is given by the double scaling limit of either $SO(2N)$ or $Sp(2N)$ matrix model. Just as in the  bosonic oriented Liouville theory, we can see this duality in two ways: the discretized surface construction of the two dimensional gravity or the holographic dual by the ZZ branes.

\begin{figure}[htbp]
	\begin{center}
	\includegraphics[width=0.8\linewidth,keepaspectratio,clip]{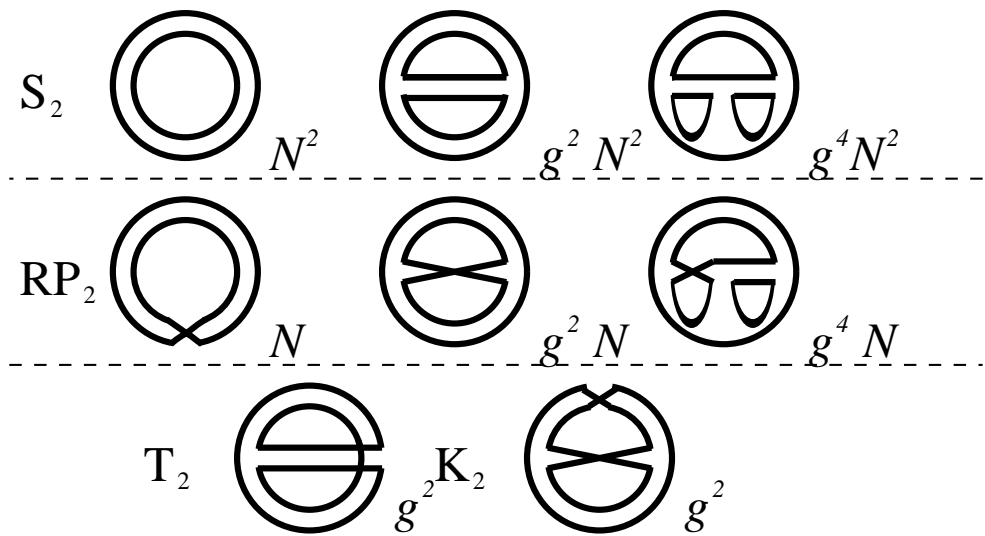}
	\end{center}
	\caption{Some oriented and unoriented diagrams for the $SO/Sp$ matrix model. The sign of the $\mathrm{RP}_2$ diagram is $-$ for the $Sp$ model and $+$ for the $SO$ model.}
	\label{unM}
\end{figure}

From the former viewpoint, unoriented surface comes from crossing of the 't Hooft double lines (see figure \ref{unM}). It is important to note that the major difference between $SO(2N)$ theory and $Sp(2N)$ theory lies in the fact that the $Sp(2N)$ theory has an extra minus sign whenever the 't Hooft lines cross. 

An intuitive argument is as follows. Consider the Wick contraction of the symmetric matrix $\langle S_{ab} S_{cd} \rangle = \delta_{ac}\delta_{bd} + \delta_{ad}\delta_{bc}$. In this case, crossing of the 't Hooft lines has the same sign as the noncrossing case. On the other hand, the Wick contraction of the antisymmetric matrix has an extra minus sign if we cross the 't Hooft lines. In the $Sp(2N)$ matrix model, the Wick contraction is given by one symmetric matrix and three antisymmetric matrices as we can see in appendix \ref{a-5}. The net result is $-2$ when we cross the 't Hooft lines. The complete argument can be found in   \cite{Mulase:2002cr}. Some examples of the (unoriented) diagrams are shown in figure \ref{unM}. 

Therefore, the amplitude with odd crosscaps in the $Sp(2N)$ matrix model yields simply the negative contribution of the corresponding $SO(2N)$ amplitude. If we interpret the unoriented theory as the oriented theory with a space-filling orientifold introduced, this just corresponds to introducing $O^{+}$ plane or $O^{-}$ plane which projects the Chan-Paton indices in a different way ($SO$ or $Sp$ respectively).

From the holographic dual point of view, this is more natural. Under the $O^{+}$ plane or $O^{-}$ plane, the ZZ brane has a tachyon which transforms as a real symmetric matrix or a quaternionic real selfdual matrix respectively. This dynamics provides the dual matrix model.

Let us first consider the $SO(2N)$ case \cite{Brezin:1990xr}, \cite{Harris:1990kc}. The partition function is given by
\begin{equation}
e^{Z_s} = \int d\Phi \exp[-\beta\mathrm{Tr} V(\Phi)], \label{eq:paom}
\end{equation}
where $\Phi$ is a real symmetric $N\times N$ matrix. The simplest potential is given by
\begin{equation}
V(x) = \frac{1}{2} x^2 + \frac{b}{4N} x^4.
\end{equation}
We first try to diagonalize the real symmetric matrix. Then the partition function (\ref{eq:paom}) can be rewritten as
\begin{equation}
e^{Z_s} = \int dx_i |\Delta(x)| \exp\left(-\sum_{i=1}^N \beta V(x_i)\right),
\end{equation}
where $\Delta(x) = \prod_{i<j} (x_i-x_j)$ is the usual Vandermonde determinant. However, note that the Jacobian is not the square of the Vandermonde determinant as in the Hermitian case, but its absolute value. See appendix \ref{b-10} for the derivation of the Jacobian. Because of this factor, the naive orthogonal polynomial method does not seem to work. 

To cure the situation we rewrite the (matrix) partition function as 
\begin{equation}
Z_{2l} = (2l)! \int_{-\infty<x_1 \cdots<x_{2l} < \infty} dx_i \prod_{i<j} (x_i-x_j)\exp\left(-\sum_{i=1}^{2l} \beta V(x_i)\right),
\end{equation}
with the same potential $V(x)$ with fixed $N \ge 2l$. At the cost of the complicated integration range, we now have the usual Vandermonde factor. Then we introduce orthogonal polynomials as
\begin{equation}
\int_{-\infty}^\infty dx P_i(x) P_j(x) e^{-2\beta V} = h_i \delta_{ij}, \label{eq:ortht}
\end{equation}
which begins like $P_n (x) = x^n + \cdots$. We further introduce $\phi_i(x) = P_i(x) e^{-\beta V(x)}$ as the corresponding wavefunctions. With these variables, one can show that the matrix partition function can be rewritten as \cite{Harris:1990kc}
\begin{equation}
Z_{2l} = (2l)! |\mathrm{det}(g_{ij})|,
\end{equation}
where 
\begin{equation}
g_{ij} = 2 \int_{-\infty}^\infty dy \int_{-\infty}^y dx \phi_{2i-2}(y) \phi_{2j-1}(x).
\end{equation}

As in the Hermitian matrix model (see section \ref{3.1}), we can derive the following recursion relation for the orthogonal polynomials $P_i(x)$:
\begin{eqnarray}
xP_j(x) &=& P_{j+1}(x) + R_j P_{j-1}(x) \cr
P'_j(x) &=& jP_{j-1}(x) + 2\beta(b/N) R_j R_{j-1}R_{j-2} P_{j-3}(x) \cr
\frac{jN}{2\beta b} &=& R_j (N/b + R_{j+1} + R_j + R_{j-1}),\label{eq:strq}
\end{eqnarray}
where $R_j = h_j/h_{j-1}$.
The above three equations allow us to obtain the recursion relation for the wavefunction $\phi$ as 
\begin{eqnarray}
\phi_{j+3} + \frac{N(j+1)}{2\beta b R_{j+1}} \phi_{j+1} - \frac{N}{2\beta b} \phi_{j-1} -R_jR_{j-1}R_{j-2} \phi_{j-3} = -\frac{N}{\beta b} \phi'_j.
\end{eqnarray}
For the later convenience, we define $W_{2l}$ as
\begin{equation}
W_{2l} = \frac{Z_{2l+4}|\beta b|}{Z_{2l+2} 2 h_{2l} (2l+4)(2l+3)}.
\end{equation}
Evaluating the matrix $g_{ij}$ with a considerable effort \cite{Harris:1990kc}, we obtain the following recursion relation for $W_{2l}$
\begin{equation}
R_{2l+1}R_{2l}R_{2l-1} + \frac{N(2l+1)W_{2l-2}}{2\beta b} -\frac{N(2l+2)W_{2l}W_{2l-2}}{2\beta b R_{2l+2}} - W_{2l+2}W_{2l}W_{2l-2} = 0. \label{eq:strq2}
\end{equation}
In terms of these variables the partition function is given by
\begin{equation}
e^{Z_s} = N!\left(\frac{2}{\beta b}\right)^{N/2-1} g_{11} \prod_{k=0}^{N/2-2} h_{2k} W_{2k}
\end{equation}

In order to obtain the unoriented noncritical string theory, we take the double scaling limit of above quantities. The relevant scaling ansatz are
\begin{eqnarray}
N &=& \frac{1}{2} a^{5/2} \cr
z &=& \frac{2}{a^2}\left(1-\frac{2j}{\beta N}\right) \cr
r &=& 2\frac{1}{a}\left(1-\frac{R_{2j}}{N}\right)
\end{eqnarray}
with $b = -\frac{1}{6}$. Substituting these into the string equations (\ref{eq:strq}) we have the Painlev\'e I equation
\begin{equation}
z = 2r^2 -\frac{2}{3}r'',
\end{equation}
as in the Hermitian matrix model. The scaling ansatz for $W_{2l}$ is given by
\begin{equation}
W_{2l} = N \exp\left[\left(\frac{1}{2}\beta N\right)^{-1/5} w(z)\right].
\end{equation}
Then from (\ref{eq:strq}), we have 
\begin{equation}
6r' - 6rw + w^3 - 6ww' + 4w'' = 0.
\end{equation}
Finally, the second derivative of the partition function is given by
\begin{equation}
-Z''_{s} =  \frac{1}{2} (r(z) + w'(z)).
\end{equation}
The last expression has a nice interpretation. The first term comes from the Painlev\'e I equation which contributes to the oriented diagram as in the Hermitian case. The second term represents purely the unoriented diagram contribution.

Finally, let us study the perturbative expansion of the partition function. To obtain a rational coefficients, it is convenient to rescale variables
\begin{equation}
y \equiv 6^{2/5} z/2, \ \ \ \ \tilde{r} \equiv 6^{1/5} r, \ \ \ \ \tilde{w} \equiv 6^{-2/5} w.
\end{equation}
Then the second derivative of the partition function becomes
\begin{equation}
-Z''_{s} \sim \frac{1}{3} (\tilde{r} + 3 \partial_y \tilde{w}).
\end{equation}

The perturbative solution is given by
\begin{eqnarray}
\tilde{r} &=& y^{1/2}\left(1-\frac{1}{16}y^{-5/2} -\frac{49}{512} y^{-5} \cdots\right) \cr
\tilde{w} &=& y^{1/4}\left(\pm1 +\frac{1}{4} y^{-5/4}  \mp\frac{5}{16} y^{-5/2}+\frac{25}{32} y^{15/4} \cdots\right).
\end{eqnarray}
As a result, the second derivative of the partition function is given by
\begin{equation}
-3Z''_{s} = y^{1/2} \pm \frac{3}{4} y^{-3/4} - \frac{13}{16}y^{-2} \pm\frac{135}{64}y^{-13/4} - \frac{4249}{512} y^{-9/2} + \cdots. \label{eq:unma}
\end{equation}
where the plus sign is for the $Sp$ model and minus sign is for the $SO$ model. At first glance, it may seem strange that we have also $Sp$ type solution from the $SO$ matrix model. However, as we will see, the recursion relation for $Sp$ matrix model is actually the same. Only the initial (perturbative) condition makes a difference between them.

Similarly we can discuss the $Sp(2N)$ case \cite{Myers:1990yd}. From the Jacobian which we have derived in appendix \ref{b-10}, we can write the string partition function as
\begin{equation}
e^{Z_s}= Z_N = \int dx_i \prod_{i<j} (x_i-x_j)^4 \exp\left[-2\beta\sum_{k=1}^N V(x_k)\right].
\end{equation}
By introducing the orthogonal polynomials as in (\ref{eq:ortht}), we can rewrite the Jacobian part as
\begin{equation}
\prod_{i<j} (x_i-x_j)^4 = \mathrm{det} \begin{bmatrix}
P_0 (x_1) & P_1 (x_1) & P_2(x_1)& \cdots & P_{2N-1}(x_1) \cr
P'_0 (x_1) & P'_1 (x_1) & P'_2(x_1)& \cdots & P'_{2N-1}(x_1) \cr
P_0 (x_2) & P_1 (x_2) & P_2(x_2)& \cdots & P_{2N-1}(x_2) \cr
	&	&	\cdots& 	&	 \cr
P'_0 (x_N) & P'_1 (x_N) & P'_2(x_N)& \cdots & P'_{2N-1}(x_N) \cr
\end{bmatrix}.
\end{equation}

With the simplest potential $V = \frac{ax^2}{2} + \frac{bx^4}{4}$, we have the recursion relation (\ref{eq:strq}). Then simple matrix manipulations\footnote{First, we try to eliminate the right side columns. The orthogonality states that there are only two possibilities. Combining $P'_{2N+1}$ with $P_{2N}$ yields the first contribution. On the other hand, using the recursion relation, we can combine $P'_{2N+1}$ and $P_{2N-2}$. In this case, further two options exist: combining $P'_{2N}$ with $P_{2N-1}$ or combining $P'_{2N}$ with $P_{2N-3}$, which yield the second and third term contribution.}imply
\begin{align}
Z_{N+1} &= (N+1) (2N+1) h_{2N} Z_N + 4N^2(N+1) \beta b h_{2N-1}h_{2N+1} Z_{N-1} \cr
&- 8N(N+1)(N-1) \beta^3 b^3 h_{2N-1}h_{2N} h_{2N+1} Z_{2N-2}. \label{eq:Wrec}
\end{align}
Now we would like to take the double scaling limit.
The string equation for $R$ yields the usual Painlev\'e I equation:
\begin{equation}
z = 2r^2 - \frac{2}{3} r''.
\end{equation}
On the other hand, if we define
\begin{equation}
W_N \equiv -\frac{Z_N}{Z_{N+1}} \frac{1}{2\beta N b h_{2N-1}}
\end{equation}
and assume the following scaling ansatz
\begin{equation}
W_N = \exp (\beta^{-1/5} w(z)), 
\end{equation}
then from (\ref{eq:Wrec}) we obtain 
\begin{equation}
6r' - 6rw + w^3 - 6ww' + 4w'' = 0.
\end{equation}
The second derivative of the partition function is given by
\begin{equation}
Z''_{s} = \frac{1}{2} (r+ w'),
\end{equation}
which is the \textit{same} expression as we have derived from the orthogonal ensemble. Therefore, only the physical perturbative input decides whether we are dealing with the orthogonal type or symplectic type in the double scaling limit. It is interesting to note that there is another branch of the solutions of above equations which does not have an unoriented string interpretation \cite{Myers:1990yd}, \cite{Harris:1990kc}. As we have seen in the unitary matrix model, if there is such a smooth interpolation between these branches, we have a geometric transition type duality which interpolates an unoriented string (orientifold) theory and an oriented string theory. It deserves further studying.

As we have seen in section \ref{6.6}, the matrix model calculation provides not only the perturbative string partition function but also (one of the candidates of) its nonperturbative completion. However, since it is defined by the differential equation, the ambiguity necessarily exists. Let us consider this point and the possible explanation of these instanton corrections \cite{Myers:1990yd}, \cite{Harris:1990kc}. 

The first nonperturbative correction comes from the Painlev\'e equation, which is  the same as in the oriented case. Its behavior is 
\begin{equation}
\delta\tilde{r} \sim y^{-1/8} \exp\left(-\frac{8}{5}y^{5/4}\right),
\end{equation}
which can be explained by the ZZ brane instanton as in the oriented case (see section \ref{6.6}). This correction induces $w$ instanton solution:
\begin{equation}
\delta \tilde{w} \sim y^{-3/8} \exp\left(-\frac{8}{5}y^{5/4}\right),
\end{equation}
which should be explained by the ZZ brane instanton as well. 

In addition, in the $SO$ case, there is a truly $w$ origin nonperturbative correction:
\begin{eqnarray}
\delta \tilde{w} &\sim& y^{1/4}\exp\left(-\frac{4}{5}y^{5/4}\right) \cr
\delta \tilde{w} &\sim& y^{1/4}\exp\left(-\frac{8}{5}y^{5/4}\right),
\end{eqnarray}
which is independent of $r$. From the counting of the KPZ scaling, this should be attributed to the ZZ brane instanton type correction, but its nature is not clear so far. At the same time, it is interesting to note that in the $Sp$ case, such corrections should not exist because the would-be instanton corrections behave as
\begin{eqnarray}
\delta \tilde{w} &\sim& y^{1/4}\exp\left(+\frac{4}{5}y^{5/4}\right) \cr
\delta \tilde{w} &\sim& y^{1/4}\exp\left(+\frac{8}{5}y^{5/4}\right),
\end{eqnarray}
which does not make any sense in the weak coupling limit. Therefore, although the difference of the perturbative expansion is just the alternating sign, the nonperturbative completion can be drastically different between $SO$ type and $Sp$ type. It would be interesting to understand this point from the continuum Liouville perspective. The crosscap state and various boundary states should play important roles.

To conclude this subsection, let us compare the matrix model results with the continuum Liouville results. As is the case with the nonperturbative effects which we have discussed in section \ref{6.6}, the normalization of $\mu$ independent quantity which we can unambiguously compare is 
\begin{equation}
 r = \frac{\langle e^{2b\phi} \rangle_{\mathrm{RP}_2}^2}{-\langle e^{2b\phi} e^{2b\phi} \rangle_{\mathrm{S}_2}} =  \frac{ (Z'_{\mathrm{RP}_2})^2}{-Z''_{\mathrm{S}_2}} = 3.
\end{equation}
The last equality is the matrix model prediction from (\ref{eq:unma}).

From the Liouville calculation we have (see section \ref{6.6})
\begin{equation}
\langle e^{2b\phi} e^{2b\phi} \rangle_{\mathrm{S}_2} = \frac{b^{-2}-1}{\pi b} [\pi \mu \gamma(b^2)]^{b^{-2}-1}\gamma(b^2)\gamma(1-b^{-2}).
\end{equation}
and from the result in the last subsection we have
\begin{equation}
\langle e^{2b\phi} \rangle_{\mathrm{RP}_2} = \frac{C}{\sqrt{2\pi}} (\pi\mu\gamma(b^2))^{\frac{b^{-2}-1}{2}}\frac{2^{1/4}\Gamma(b^2)\Gamma(2-b^{-2})}{\pi(b-b^{-1})} \sin(\pi b^2/2) \sin(\pi b^{-2}/2),
\end{equation}
with the \textit{same} normalization convention in section \ref{6.6}. We can calculate $r$ with $b= \sqrt{\frac{2}{3}}$ which corresponds to the $c=0$ pure gravity, and we find that $r = 3$ precisely when $C= 2$ as in the disk nonperturbative calculation in section \ref{6.6}. For general $b$, we have
\begin{equation}
r = \frac{C^2 b \tan\left(\frac{\pi}{2b^2}\right) \tan\left(\frac{\pi b^2}{2}\right)}{4\sqrt{2}(-1+b^2)}.
\end{equation}
 The factor $C$ should be the same in order to preserve the tadpole cancellation condition. Therefore this calculation is the nontrivial check of the unoriented matrix model-Liouville theory duality. It would be an intriguing problem to extend this calculation to the more general $c<1$ unoriented Liouville theories coupled to the (unitary) minimal models. However, the multimatrix model for $Sp/SO$ matrix which we believe to yield the dual description is a more complicated problem and as far as the author knows it is not explicitly solved yet.\footnote{The good starting point of this problem is \cite{Neuberger:1991vv}, where the general prescription to solve the multi unoriented matrix model which corresponds to the $c<1$ gravity coupled to $(p,q)$ minimal models is presented.}

Moreover we can calculate the torus partition function and the Klein-bottle partition function from the continuum formalism (free field path integral). The torus partition function is given by
\begin{equation}
Z_{T_2} = \frac{-\log\mu}{2b}\int_{\mathcal{F}} \frac{d^2\tau}{2\tau_2} \frac{1}{(4\pi^2 \tau_2)^{\frac{1}{2}}} |\eta(\tau)|^2 = -\frac{1}{48}\log\mu,
\end{equation}
where $b =\sqrt{\frac{2}{3}}$ for $c=0$ and we need a further division by two to obtain an unoriented torus partition function. The integration over the moduli has been reviewed in appendix \ref{b-8}. Similarity, the free path integral calculation for the Klein bottle is given by

\begin{equation}
Z_{K_2}=\frac{-\log\mu}{2b}\int_0^\infty \frac{dt}{2t} \frac{1}{\sqrt{4\pi^2 t}} \eta(2it) = -\frac{1}{4}\log\mu ,
\end{equation}
where we need a further division by two.

On the other hand, the matrix model prediction is that the ratio between the torus amplitude and the Klein bottle amplitude is given by $12$, which is precisely the same. In order to obtain the same result from the exact crosscap state as $Z_{K_2} = \langle C | q^H|C\rangle $, we need a good regularization to draw a $\log\mu$ part.

\subsubsection{$c=1$ matrix model}\label{13.3.2}
In this subsection, we consider the matrix dual for the $c=1$ unoriented string. From the general holographic argument (or in the bosonic case here, the discretized surface argument), we have either real symmetric matrix quantum mechanics for the $\Omega^+$ projection or real selfdual quaternionic matrix quantum mechanics for the $\Omega^-$ projection. Following \cite{Gomis:2003vi} we show that the system is tantamount to the Calogero-Moser system and reduced to the thermodynamics of the general statistics. The surprising and curious result is that the system does not yield the unoriented diagram contribution to the string partition function in the decompactifying limit, so it corresponds to the tadpole canceled version of the unoriented theory. This is probably done by the Fischler-Susskind mechanism we have reviewed in section \ref{13.2.3}. It is interesting to see what happens when we try to cancel the tadpole by introducing D1-branes, in which case, the dual matrix model becomes a matrix vector model.

The matrix model action is given by
\begin{equation}
S = \int dt \mathrm{Tr} \left(\frac{1}{2} (D_t M)^2 + U(M) \right),
\end{equation}
where the covariant derivative $D_t$ simply truncates the spectrum onto the gauge invariant sector.

First, we diagonalize the matrix as we have done in the Hermitian matrix quantum mechanics in section \ref{sec:3}. The Jacobians are given by
\begin{eqnarray}
U(N): J&=& \Delta(\lambda)^2 \cr
SO(N): J&=& |\Delta(\lambda)| \cr
Sp(N): J&=& \Delta(\lambda)^4 ,
\end{eqnarray}
where $\Delta(\lambda) = \prod_{i<j} (\lambda_i-\lambda_j)$ is the usual Vandermonde determinant. 

If we concentrate exclusively on the gauge invariant sector, the Hamiltonian can be written as 
\begin{equation}
H = -\frac{1}{2J}\sum_i \frac{d}{d\lambda_i} J \frac{d}{d\lambda_i} + \sum_iU(\lambda_i),
\end{equation}
where the Jacobian factor comes from the Pauli's prescription (or the general covariance). It is convenient to introduce a new antisymmetric wavefunction $f(\lambda) = \Psi(\lambda) \mathrm{sign}(\Delta)|\Delta|^{-\alpha/2}$ where $\Psi(\lambda)$ is the original symmetric wavefunction and $\alpha = 2,1,4$ corresponds to $U(N),SO(N)$ and $Sp(N)$ respectively. Then the new Hamiltonian for $f(\lambda)$ is given by
\begin{equation}
H = -\frac{1}{2}\sum_i \frac{d^2}{d\lambda^2_i} + \sum_i (U(\lambda_i)) +\frac{\alpha}{2}\left(\frac{\alpha}{2}-1\right) \sum_{i<j} \frac{1}{(\lambda_i-\lambda_j)^2}. \label{eq:CalMo}
\end{equation}
If we take $U(\lambda) = -\frac{\lambda^2}{2}$ as the double scaling limit potential, this becomes what is called the Calogero-Moser system.

Now let us calculate the string partition function or the free energy of the system. The relation between them is $\frac{F}{T} =  -Z$ and $T = \frac{1}{2\pi R}$.
 Since the Calogero-Moser system (\ref{eq:CalMo}) is not free but interacting theory, the calculation seems difficult. Fortunately, it can be shown by using the asymptotic Bethe ansatz method that the thermodynamics of the Calogero-Moser system is equivalent to that of the free inverted harmonic oscillator potential system obeying the more general statistics \cite{Polychronakos:1999sx}, \cite{Gomis:2003vi}. Here we just admit this statement and discuss the consequent results.

The thermal density of states for the general $g$ statistics and its properties are reviewed in appendix \ref{b-11} (see also \cite{Wu:1994it}, \cite{Haldane:1991xg}, \cite{Polychronakos:1999sx}). In this particular case, we have $g= \frac{1}{2}$ for the $SO(N)$ theory and $g = 2$ for the $Sp(N)$ theory (and of course $g=1$ for $U(N)$ theory). The free energy is given by
\begin{equation}
F = - T\int_{-\infty}^{\infty} \rho_0(\epsilon) \log \left( 1 +\frac{n(\epsilon)}{1-gn(\epsilon)}\right),
\end{equation}
where $\rho_0(\epsilon)$ is the zero temperature distribution for the free inverted harmonic oscillator potential which we have discussed in the $U(N)$ case in section \ref{sec:3} (\ref{eq:fden}).

For $g=1/2$, the probability density becomes
\begin{equation}
n(\epsilon) = \frac{1}{\sqrt{\frac{1}{4}+ \exp\left(\frac{2(\epsilon-\mu)}{T}\right)}},
\end{equation}
so the free energy is given by
\begin{equation}
F = -T \int_{-\infty}^{\infty} d\epsilon \rho_0(\epsilon) \log\left(\frac{2+n(\epsilon)}{2-n(\epsilon)}\right).
\end{equation}

Similarly for $g=2$,  the probability density becomes
\begin{equation}
n(\epsilon) = \frac{1}{2}\left(1-\frac{1}{\sqrt{1+4\exp(\frac{\mu-\epsilon}{T})}}\right),
\end{equation}
so the free energy is given by
\begin{equation}
F = -T \int_{-\infty}^{\infty} d\epsilon \rho_0(\epsilon) \log\left(\frac{1-n(\epsilon)}{1-2n(\epsilon)}\right).
\end{equation}

These integral expressions for the free energy suffer divergences, but their derivatives with respect to $T$ are finite. The easiest way to evaluate them is to differentiate with respect to $T$ and $\mu$, and then use the Sommerfeld expansion formula reviewed in appendix \ref{b-11}. This enables us to expand the free energy in terms of $1/\mu$ as 
\begin{equation}
F(T,\mu) - F(0,\mu) = \frac{\pi T^2}{12} \log\mu + \frac{\zeta(3)T^3}{4\pi \mu} + \mathcal{O}(\mu^{-2})
\end{equation}
for $g =1/2$ and
\begin{equation}
F(T,\mu) - F(0,\mu) = \frac{\pi T^2}{12} \log\mu - \frac{\zeta(3)T^3}{2\pi \mu} + \mathcal{O}(\mu^{-2})
\end{equation}
for $g=2$.

The zero temperature free energy is divergent, but the regularized form can be obtained from the thermodynamics formula
\begin{equation}
\frac{\partial E}{\partial \mu} = \mu \rho(\mu) = \frac{1}{g}\mu\rho_0(\mu).
\end{equation}
Thus we obtain
\begin{equation}
E = F(0,\mu) = \frac{1}{2\pi g}\left(-\frac{\mu^2}{2}\log\mu + \frac{1}{24}\log\mu - \sum_{m=1}^\infty \frac{(2^{2m+1}-1)|B_{2m+2}|}{8m(m+1)(2\mu)^{2m}}\right).
\end{equation}
Note that this is the direct consequence of the fact that the generalized $g$ statistics forms a Fermi surface as $T \to 0$ (unless $g = 0$). Therefore the zero temperature free energy should be the same as in the $U(N)$ case up to a multiplication of the filling factor.\footnote{To compare it with the result in section 3, the renormalization of $\mu$ is needed.} 

Some comments are in order.
\begin{itemize}
	\item The contribution to the projective plane diagram is zero irrespective of the temperature (compactification radius). The thermodynamical explanation from the generalized statistics is that this is due to the third law of the thermodynamics. Actually, the only possible projective plane contribution is from the linear $T$ dependent part in the free energy. However, since the entropy $S$ is the derivative of free energy with respect to $T$: $S =\frac{\partial F}{\partial T}$, we cannot have a linear term in $T$ in the free energy without violating the third law of the thermodynamics: ``Entropy becomes zero when the temperature goes to zero". The physical space-time interpretation of this result is that the tadpole is completely canceled (possibly up to a nonsingular term which corresponds to the finite part in the continuum theory). This is probably done by the Fischler-Susskind mechanism.
	
	\item As we have stated in the $c=0$ case in the last subsection, the difference between $Sp/SO$ model is just the alternating sign which appears in the unoriented surface with odd crosscaps. This can be also explained from the particle hole duality in the Calogero-Moser system \cite{Gomis:2003vi}. Actually we can show
	\begin{equation}
	F\left(T,\mu ,g = \frac{1}{g_0}\right) = g_0^2 F\left(\frac{T}{g_0},-\mu, g = g_0\right) + \mathrm{Nonsingular} \ \mathrm{terms} \ \mathrm{in} \ \mu.  
	\end{equation}
	For $g_0 = 2$, we obtain
	\begin{equation}
	F(T,\mu,g=1/2) = 4 F(T/2,-\mu, g=2) + \mathrm{Nonsingular} \ \mathrm{terms} \ \mathrm{in}\  \mu.
	\end{equation}
	Since the quaternionic selfdual matrix has a pair-wise eigenvalue, the Hamiltonian for the quaternionic-Hermitian matrix model is twice as large as that of the Calogero-Moser system with $g=2$. Therefore the temperature differs by a factor of two. Hence, the matrix quantum mechanics partition function is related as
	\begin{equation}
	\log Z^{SO}_{MQM}(T,\mu) = 2\log Z^{Sp}_{MQM}(T,-\mu).
	\end{equation}
	Therefore if we define the string partition function as $Z^{SO}_{s}(T,\mu) = \log Z^{SO}_{MQM}(T,\mu)$ and $Z^{Sp}_{s}(T,\mu) = 2\log Z^{Sp}_{MQM}(T,\mu)$, we have a relation
	\begin{equation}
	Z^{SO}_{s}(T,\mu) = Z^{Sp}_{s} (T,-\mu), 
	\end{equation}
	which is expected from the string theory perspective.
	
	\item From the above calculation, it is obvious that the partition function does not contain any unoriented surface contribution in the decompactifying limit (or the zero temperature limit). Though the reason behind may be the tadpole cancellation via the Fischler-Susskind mechanism, this looks rather peculiar. For example, the Feynman diagrammatic calculation has an apparent unoriented diagram contribution. In the double scaling limit, the contribution is suddenly gone (at least in the $c=0$ matrix model, they remains to exist). Recall that in the $c=0$ model the integration over the moduli is properly reproduced by the matrix model in contrast. It would be interesting to study what is actually happening and the fate of the unoriented diagrams. In the quantum mechanics setup here, it will lead to the physical origin of the Fischler-Susskind counterterm. Also, it would be interesting to study unoriented matrix-vector models which should cancel the tadpole by vector loops and compare its result with the double scaling limit considered here.
	
\end{itemize}

\subsection{Literature Guide for Section 13}\label{13.4}
The earlier studies on the unoriented matrix model can be found in \cite{Andric:1983jk}, \cite{Brezin:1990xr}, \cite{Harris:1990kc}, \cite{Myers:1990yd}, \cite{Brezin:1991dk}, \cite{Neuberger:1991vv}, \cite{Martin-Delgado:1993bt}, \cite{Anderson:1992js}, \cite{Nakazawa:1995jq}.

From the Liouville point of view, \cite{Hikida:2002bt} derived the crosscap one-point function. For $c=1$, the tadpole cancellation condition and the relation to the matrix model have been further studied in \cite{Nakayama:2003ep}, \cite{Gomis:2003vi}, \cite{Bergman:2003yp}. The comparison between the exact crosscap state and the matrix model calculation for the $c=0$ theory is first presented by the author in this review.

\sectiono{Super Liouville Theory on Unoriented Surfaces}\label{sec:14}
In this section, we extend the results obtained in the last section to the unoriented super Liouville theory. We use the $\alpha'=2 $ convention in this section. The organization of this section is as follows.

In section \ref{14.1}, the crosscap states for the unoriented $\mathcal{N}=1$ super Liouville theory are presented, where we derive them from the modular bootstrap method. In section \ref{14.2}, the tadpole cancellation is studied and possible tadpole free $\hat{c}=1$ unoriented type 0 string theories are classified. In section \ref{14.3}, the proposals of the matrix model dual of these models are reviewed, where we determine them from the holographic dual gauge theories on the unstable D0-branes.

\subsection{Crosscap State}\label{14.1}
In this section, we will derive the crosscap states for the supersymmetric Liouville theory \cite{Bergman:2003yp}. Our basic strategy here is again the modular bootstrap method. As we have done in section \ref{13.1}, we assume the $\Omega$ projected open spectrum (M\"obius strip amplitude) for the basic $(1,1)$ boundary state and modular-transform it into the closed exchange channel to obtain the crosscap states. By this method we will obtain the $\eta = \pm$ NS crosscap states, but not R crosscap states. This is because the R crosscap states are derived from the open fermion loop channel (see appendix \ref{a-4} for the relation between the open channel and closed channel spectrum), which cannot be obtained from the \textit{same} $(1,1)$ ZZ brane.\footnote{We should recall that $(1,1)$ brane does not have an $\eta = -$ part.} 

In principle, we can obtain the R crosscap states from Teschner's trick and the conformal bootstrap method as we have done in the bosonic case. Nevertheless we will not consider it for two reasons. The first one is the practical one: we will not use the R crosscap states to make any sensible type 0 $\hat{c} = 1$ unoriented fermionic string theory (this is because the orientifold projection involving R crosscap states does not keep the super Liouville deformation invariant as we will see in the next section, so this projection is actually irrelevant). The second reason is that even when we have obtained the functional relation from Teschner's trick, we do not expect that the solution is unique as we have seen in the bosonic crosscap state. 

Keeping these in mind, let us begin with the crosscap states for the type 0B case. To begin with, we introduce the following GSO projected character for the $\Omega$ projection as
\begin{eqnarray}
\chi^{GSO\pm}_1(\Delta) &=& \mathrm{Tr}_1 (P_{GSO}^\pm \Omega q^H) \cr
 &=& -\frac{1}{2}i\left[(1-iq^{1/2}+\cdots) \pm (-1-iq^{1/2}+\cdots)\right]q^\Delta \cr
\chi^{GSO\pm}_{\Sigma_1}(\Delta) &=& \mathrm{Tr}_{\Sigma_1} (P_{GSO}^\pm \Omega q^H) \cr&=& \frac{1}{2}\left[(1-iq^{1/2}+\cdots) \pm (-1-iq^{1/2}+\cdots)\right]q^{\Delta}, \label{eq:subti}
\end{eqnarray}
where the meaning of the subscript $1$ and $\Sigma_1$ will be clarified in the next section \cite{Bergman:2003yp}. For the time being the difference is simply the overall factor.

The orientifolded quantity for the basic $(1,1)$ state can be obtained by subtracting the null module:
\begin{eqnarray}
\chi^{GSO\pm}_1(\Delta+1/2) &=& -\frac{1}{2}iq^{1/2}\left[(1-iq^{1/2}+\cdots) \pm (-1-iq^{1/2}+\cdots)\right]q^{\Delta} \cr
\chi^{GSO\pm}_{\Sigma_1}(\Delta+1/2) &=& \frac{1}{2}q^{1/2}\left[(1-iq^{1/2}+\cdots) \pm (-1-iq^{1/2}+\cdots)\right]q^{\Delta}, \label{eq:subt}
\end{eqnarray}
The subtracted module is given by
\begin{equation}
\chi^{GSO\pm}_{1,\Sigma_1}(\Delta) + i\chi^{GSO\mp}_{1,\Sigma_1} (\Delta +1/2),
\end{equation}
which can be obtained by carefully comparing (\ref{eq:subti}) with (\ref{eq:subt}) \cite{Bergman:2003yp}.

Then we define the basic $(1,1)$ GSO and orientifold projected M\"obius strip partition function as 
\begin{eqnarray}
Z_{(1,1)} (t) &=& \mathrm{Tr}_{1;(1,1)} P_{GSO}^+ \Omega e^{-2\pi t (L_o- \frac{c}{24})} \cr
& & \pm \mathrm{Tr}_{\Sigma_1;(1,1)} P_{GSO}^- \Omega e^{-2\pi t (L_o- \frac{c}{24})}
\end{eqnarray}
where $\pm$ corresponds to the projection by the $\Omega$ and $\hat{\Omega}$ respectively, anticipating the results in the next section, and the trace is taken over the $(1,1)$ degenerate state. Note that this combination makes the subtracted and the unsubtracted characters real. $Z_{(1,1)}$ can be calculated explicitly by using the following relation:
\begin{equation}
L_o - \frac{c}{24} = -\frac{1}{8}(b+b^{-1})^2 -\frac{1}{16} + \mathrm{oscillator},
\end{equation}
which yields
\begin{equation}
Z_{(1,1)} (t) = -i\left(e^{\frac{\pi t}{4}(b+b^{-1})^2} \pm e^{\frac{\pi t}{4}(b-b^{-1})^2}\right) Z_{GSO+} \mp\left(e^{\frac{\pi t}{4}(b+b^{-1})^2} \mp e^{\frac{\pi t}{4}(b-b^{-1})^2}\right) Z_{GSO-}, \label{eq:untree}
\end{equation}
where 
\begin{equation}
Z_{GSO\pm} = \frac{\theta_3(it+1/2)^{1/2} \mp \theta_4(it+1/2)^{1/2}}{2 e^{-i\pi/16} \eta(it+1/2)^{3/2}}.
\end{equation}

Now we have learned the basic properties of the orientifold projected $(1,1)$ brane, let us obtain the wavefunction for the crosscap states. To do this, we introduce the following Ishibashi states\footnote{Note that the right-hand side here is not modular-transformed into the loop channel.}:
\begin{eqnarray}
\langle B;P' \pm|e^{-\pi t(L_0 + \bar{L}_0 -c/12)}|B;P \pm\rangle_{NSNS} &=& \delta(P-P') \frac{q^{-\frac{P^2}{2}}\theta_3(it)^{1/2}}{\eta(it)^{3/2}} \cr
\langle B;P' \pm|e^{-\pi t(L_0 + \bar{L}_0 -c/12)}|B;P \mp\rangle_{NSNS} &=& \delta(P-P') \frac{q^{-\frac{P^2}{2}}\theta_4(it)^{1/2}}{\eta(it)^{3/2}} \cr
\langle B;P' \pm|e^{-\pi t(L_0 + \bar{L}_0 -c/12)}|C;P \pm\rangle_{NSNS} &=& \delta(P-P') \frac{q^{-\frac{P^2}{2}}\theta_3(it+1/2)^{1/2}}{e^{-i\pi/16}\eta(it+1/2)^{3/2}} \cr
\langle B;P' \pm|e^{-\pi t(L_0 + \bar{L}_0 -c/12)}|C;P \mp\rangle_{NSNS} &=& \delta(P-P') \frac{q^{-\frac{P^2}{2}}\theta_4(it+1/2)^{1/2}}{e^{-i\pi/16}\eta(it+1/2)^{3/2}} \cr
\langle C;P' \pm|e^{-\pi t(L_0 + \bar{L}_0 -c/12)}|C;P \pm\rangle_{NSNS} &=& \delta(P-P') \frac{q^{-\frac{P^2}{2}}\theta_3(it)^{1/2}}{\eta(it)^{3/2}} \cr
\langle C;P' \pm|e^{-\pi t(L_0 + \bar{L}_0 -c/12)}|C;P \mp\rangle_{NSNS} &=& \delta(P-P') \frac{q^{-\frac{P^2}{2}}\theta_4(it)^{1/2}}{\eta(it)^{3/2}}. 
\end{eqnarray}
As we have explained in the beginning of this section, we will not treat the R Ishibashi states here.

We modular-transform (\ref{eq:untree}) into the tree exchange channel by using the modular transformation formula (\ref{eq:modun}) and we find
\begin{equation}
Z_{1,1} = \sqrt{\frac{1}{s}} \left[-i\left(e^{\frac{\pi}{16s}(b+b^{-1})^2} \pm e^{\frac{\pi}{16s}(b-b^{-1})^2}\right)Z^{BC-} \mp \left (e^{\frac{\pi}{16s}(b+b^{-1})^2} \mp e^{\frac{\pi}{16s}(b-b^{-1})^2}\right)Z^{BC+}\right]
\end{equation}
where we have introduced
\begin{equation}
Z^{BC\pm} = \frac{\theta_4(is+1/2)^{1/2} \pm e^{-\pi i/4} \theta_3(is+1/2)^{1/2}}{2 e^{-3i\pi/16}\eta(is + 1/2)^{3/2}},
\end{equation}
which can be obtained from the modular transformation.

It is convenient to transform it further into
\begin{eqnarray}
Z_{1,1} &=& \frac{\sqrt{2}}{2} \int_{-\infty}^{\infty} e^{-s\pi P^2}\left[-i(\cosh(\pi(b+b^{-1})P/2) \pm \cosh(\pi(b-b^{-1})P/2) )Z^{BC-} \right.\cr
& &\left. \mp(\cosh(\pi(b+b^{-1})P/2) \mp \cosh(\pi(b-b^{-1})P/2)) Z^{BC+}\right]. \label{eq:unzz}
\end{eqnarray}

Now we obtain the desired crosscap states \cite{Bergman:2003yp} by the superposition of the Ishibashi states as 
\begin{eqnarray}
_{NS}\langle C| = \frac{1}{2} \int_{-\infty}^{\infty} dP \left[ \Psi_{C+}(P) \langle C;P,+| + \Psi_{C-}(P)\langle C;P,-|\right],
\end{eqnarray}
where 
\begin{eqnarray}
\Psi_{C-}(P) &=& -\sqrt{2}(\pi\mu\gamma(bQ/2))^{-iP/b}\frac{\Gamma(1+iPb)\Gamma(1+iPb^{-1})}{-2i\pi P} \cr
& &(i\cosh(\pi Pb/2)\cosh(\pi Pb^{-1}/2) +\sinh(\pi Pb/2)\sinh(\pi Pb^{-1}/2)) \cr
\Psi_{C+}(P) &=& -\sqrt{2}(\pi\mu\gamma(bQ/2))^{-iP/b}\frac{\Gamma(1+iPb)\Gamma(1+iPb^{-1})}{-2i\pi P} \cr
& &(-i\cosh(\pi Pb/2)\cosh(\pi Pb^{-1}/2) +\sinh(\pi Pb/2)\sinh(\pi Pb^{-1}/2)) \label{eq:omegab}
\end{eqnarray}
for the $\Omega$ projection and 
\begin{eqnarray}
\Psi_{C-}(P) &=& \sqrt{2}(\pi\mu\gamma(bQ/2))^{-iP/b}\frac{\Gamma(1+iPb)\Gamma(1+iPb^{-1})}{-2i\pi P} \cr
& &(\cosh(\pi Pb/2)\cosh(\pi Pb^{-1}/2) -i\sinh(\pi Pb/2)\sinh(\pi Pb^{-1}/2)) \cr
\Psi_{C+}(P) &=& \sqrt{2}(\pi\mu\gamma(bQ/2))^{-iP/b}\frac{\Gamma(1+iPb)\Gamma(1+iPb^{-1})}{-2i\pi P} \cr
& &(\cosh(\pi Pb/2)\cosh(\pi Pb^{-1}/2) +i\sinh(\pi Pb/2)\sinh(\pi Pb^{-1}/2))\label{eq:omegahb}
\end{eqnarray}
for the $\hat{\Omega}$ projection. This can be obtained \textit{up to phase factors} by taking the overlap with the $(1,1)$ ZZ boundary states with $\eta = +$ (\ref{eq:szzw}) and comparing it with (\ref{eq:unzz}).\footnote{Technically, we also impose the condition that the Klein bottle amplitude can be properly interpreted in the loop channel.} The annoying phase here can be canceled when we consider the two dimensional case (because all oscillator cancels out with the ghost contribution). However, in the more general case, these phases cannot be canceled and to make the amplitude real, we should supply its complex conjugation to the amplitude, which ruins the sensible channel duality. In the following we will not consider the $b \neq 1$ case.

This wavefunction satisfies the desirable overlap property with the general $(m,n)$ branes with $m-n = even$ as we have seen in the bosonic case. On the other hand, with $m-n$ odd branes, it seems that the degeneration subtraction reverses and becomes an addition, which does not make sense. This is because the $m-n$ odd branes have a wavefunction in the $-$ sector instead of the $+$ sector. Also, we remark here that the overlap with the FZZT brane is subtler, for the FZZT branes do not have a proper open GSO projection from the beginning.

Similarly we can do the same calculation for the type 0A case again. The relevant $(1,1)$ projected partition function (again, the explanation is left to the next section) is given by 
\begin{equation}
Z_{1,1} = -ie^{\frac{\pi t}{4}(b+b^{-1})^2}Z_{GSO+} + e^{\frac{\pi t}{4}(b-b^{-1})^2} Z_{GSO -}
\end{equation}
for the $\hat{\Omega}$ projection and the wavefunction becomes (up to subtle phases)
\begin{eqnarray}
\Psi_{C+}(P) &=& - \frac{\sqrt{2}}{2}(\pi\mu\gamma(bQ/2))^{-iP/b}\frac{\Gamma(1+iPb)\Gamma(1+iPb^{-1})}{-2i\pi P} \cr
	&& (-i\cosh(\pi(b+b^{-1})P/2) + \cosh(\pi(b-b^{-1})P/2)) \cr
\Psi_{C-}(P) &=& - \frac{\sqrt{2}}{2}(\pi\mu\gamma(bQ/2))^{-iP/b}\frac{\Gamma(1+iPb)\Gamma(1+iPb^{-1})}{-2i\pi P} \cr
	&& (i\cosh(\pi(b+b^{-1})P/2) + \cosh(\pi(b-b^{-1})P/2)).
\end{eqnarray}
For the $\Omega$ projection, we have in the loop channel,
\begin{equation}
Z_{1,1} = \pm 2 \left(e^{\frac{\pi t}{4}(b+b^{-1})^2}Z_{GSO-} +ie^{\frac{\pi t}{4}(b-b^{-1})^2} Z_{GSO +}\right)
\end{equation}
where the sign corresponds to the two possible orientifold projection on the Chan Paton indices as we will explain in the next section. The straightforward calculation yields
\begin{eqnarray}
\Psi_{C+}(P) &=& \mp \sqrt{2}(\pi\mu\gamma(bQ/2))^{-iP/b}\frac{\Gamma(1+iPb)\Gamma(1+iPb^{-1})}{-2i\pi P} \cr
	&& (\cosh(\pi(b+b^{-1})P/2) + i\cosh(\pi(b-b^{-1})P/2)) \cr
\Psi_{C-}(P) &=& \mp \sqrt{2}(\pi\mu\gamma(bQ/2))^{-iP/b}\frac{\Gamma(1+iPb)\Gamma(1+iPb^{-1})}{-2i\pi P} \cr
	&& (\cosh(\pi(b+b^{-1})P/2) - i\cosh(\pi(b-b^{-1})P/2)),\label{eq:omegaa}
\end{eqnarray}
up to phases.

\subsection{Tadpole Cancellation}\label{14.2}
In this section, we discuss the tadpole cancellation condition for the two dimensional fermionic string. First we classify all the possible (space filling) orientifold planes for the type 0 string theory, following \cite{Gomis:2003vi}, \cite{Bergman:2003yp} (for the ten dimensional case, see e.g. \cite{Bergman:1997rf}, \cite{Klebanov:1998yy}, \cite{Bergman:1999km}, \cite{Bianchi:1990yu}, \cite{Gimon:1996rq}). Then we introduce D1-branes to cancel the tadpole divergence \cite{Bergman:2003yp}. It is important to note that for the super Liouville theory considered here, all the orientifold tadpoles come from the NS-NS sector and there is no R-R tadpole. Therefore in principle we can use the Fischler-Susskind mechanism to cancel the NS-NS tadpole as we have seen in the bosonic case in section \ref{13.2.3}. We will briefly discuss its possibility in the next section \cite{Gomis:2003vi}. In this section, we concentrate on the tadpole cancellation by introducing the D-brane which is the source of the NS-NS tadpole. 

Before the actual analysis, we have one remark here. In this section, we deal with most of the loop amplitudes by the free field method. As we have seen in part II of this review, the allowable D1-branes in the theory seem different from the free field consideration and the super Liouville consideration, which might affect the results in the following. We will not delve into this subtle point but it is important to keep this possibility in mind,

The type 0 theory admits three different kinds of orientifold projections. First let us consider the type 0B theory.

\textbf{0B/$\Omega$}

The R-R scalar $C$ is odd under $\Omega$ and the tachyon $T$ is even, so only the propagating degree of freedom is the tachyon \cite{Gomis:2003vi}, \cite{Bergman:2003yp}. In the free language, the crosscap state is given by 
\begin{equation}
|C \rangle = \frac{1}{\sqrt{2}}(|C,+ \rangle -|C,- \rangle).
\end{equation}
We can easily see that the Klein bottle amplitudes yield the projection onto the $\Omega$ invariant states considered above. The corresponding super Liouville crosscap state should be (\ref{eq:omegab}) which shows the $\Omega$ projection considered here in the $p \to 0$ limit.\footnote{For a finite $p$ there seem to be unprojected states, but we have encountered the same subtlety in the bosonic case and we will ignore it here for the moment.} Thus we can conclude that in the 0B/$\Omega$ theory, we do not have any (NS-NS or R-R) tadpole. The partition function comes only from the torus graph (with an extra $\frac{1}{2}$) \cite{Gomis:2003vi}
\begin{equation}
Z = V_{\phi} \frac{1}{24} \left(\frac{R}{\sqrt{\alpha'}} + 2 \frac{\sqrt{\alpha'}}{R}\right). \label{eq:kk0b}
\end{equation}

\textbf{0B/$\hat{\Omega}$}

$\hat{\Omega}$ is defined as $\hat{\Omega} \equiv \Omega\cdot (-1)^{F_L}$, where $F_L$ is the spacetime fermion number (or R-NS parity). The closed spectrum is give by $C$ and $T$ because $C$ has an extra minus sign for R-R fields. In the free field language, the crosscap state is given by
\begin{equation}
|C \rangle = \frac{1}{\sqrt{2}}(|C,+ \rangle +|C,- \rangle).
\end{equation}
Therefore it has a tachyon tadpole. The corresponding super Liouville crosscap state is given in (\ref{eq:omegahb}). To cancel this tachyon tadpole, we introduce the $\hat{\Omega}$ invariant D1-$\bar{\mathrm{D}}1$ pair which we call it $\widehat{\mathrm{D}1}$. With $n_1$ such pairs, the Chan-Paton labels are described by the $2n_1 \times 2n_1$ matrices $\Lambda = \{1,\Sigma_1,\Sigma_2,\Sigma_3\}$, where $\Sigma_i = 1 \otimes \sigma_i$. The cylinder amplitude is given by \cite{Bergman:2003yp}
\begin{equation}
A_C = \mathrm{Tr}_{NS} (P^+_{GSO} + P^-_{GSO} +P^{-}_{GSO} + P^+_{GSO}),
\end{equation}
and the M\"obius strip amplitude is given by
\begin{equation}
A_M =  \mathrm{Tr}_{NS} (P^+_{GSO} \pm P^-_{GSO} \pm P^{-}_{GSO} - P^+_{GSO})\Omega,
\end{equation}
where the sign corresponds to the two possible $\hat{\Omega}$ projection on the Chan-Paton factor as 
\begin{eqnarray}
\hat{\Omega}_+ \Lambda &=& \Sigma_1\Lambda^T\Sigma_1 \cr
\hat{\Omega}_- \Lambda &=& \Sigma_2\Lambda^T\Sigma_2. 
\end{eqnarray}
Then the tadpole from a $\widehat{\mathrm{D}1}$ brane is given by
\begin{equation}
\langle T \rangle_{\tilde{O1}} = \pm 2 \langle T \rangle_{\widehat{\mathrm{D}1}}.\end{equation}
which can also be obtained from the exact boundary-crosscap states analysis done in the last section (with an additional $\sqrt{2}$ contribution from the free $X,\psi$ sector). The tadpole cancellation can be done in the $\hat{\Omega}_-$ projection with two $\widehat{\mathrm{D}1}$ branes. There are further two possibilities: we can choose either the same type (electric or magnetic) branes or different type branes. In the former case, we have a $U(2)$ antisymmetric (hence the singlet) open (massless) tachyon, and in the latter case, we have $U(1) \times U(1)$ two charged massless fermions with their complex conjugates.

\textbf{0B/$\Omega'$}

In the ten dimensional 0B theory, yet another orientifold projection is possible \cite{Bianchi:1990yu}, \cite{Bergman:1999km}; $\Omega' = \Omega (-1)^{f_L} $, where $f_L$ is the left-hand world sheet fermion number. This has an R-R crosscap state, but the super Liouville interaction is not invariant under $\Omega'$ which projects out the tachyon. Thus we do not consider this projection any further.

\textbf{0A/$\Omega$}

$\Omega$ exchanges $(R-,R+)$ and $(R+,R-)$ with each other, so the R-R one-form $C^-_1$ is even and $C^+_1$ is odd and projected out. The massless tachyon is even, so it is invariant. In the free field language, the crosscap state is given by (for example)
\begin{equation}
|C \rangle = |C+\rangle,
\end{equation}
and in the super Liouville exact treatment, the crosscap state is given by (\ref{eq:omegaa}),\footnote{There may seem to be some difference here, but if we had started from the free crosscap state: $|C\rangle = (1+i)|C+\rangle + (1-i) |C-\rangle$, it would have given the same Klein bottle amplitudes.} which has an NS-NS tachyon tadpole. To cancel the tadpole, there are two possible ways to do that. The first one is the $SO(2)$ with an antisymmetric (hence the singlet) tachyon. The second one is a no Chan-Paton group (or $SO(1) \times SO(1)$) and a massless fermion.

Note that in this case only $1$ and $\Sigma_1$ sector\footnote{This is just the statement that in the type 0A theory the R-R uncharged brane is not the D1-$\bar{\mathrm{D}}$1 pair but the genuine ``unstable" brane. Although in terms of the boundary states they look very similar, the spectrum is different. When we say $|B +\rangle$, actually there are three possibilities which we should distinguish. 1, if Dp-brane (with R-R sector) is both invariant under the GSO and the orientifold projection, we name it D$\bar{\mathrm{D}}$ pair. 2, if Dp-brane is invariant under the GSO but not invariant under the orientifold projection, we should make an invariant combination of D-brane and $\bar{\mathrm{D}}$-brane and we name it $\widehat{\mathrm{D}p}$. 3, if R-R charged Dp-brane is not GSO invariant, we should have a ``wrong dimensional" brane, which we name it $\tilde{\mathrm{D}}$ brane. It is useful to remember that the gauge theory on it with an orientifold projection is $SO/Sp \times SO/Sp$, $U$, $SO/Sp$ respectively.} appear in the $\tilde{\mathrm{D}}1$ spectrum and the $\Omega$ projection is given in the standard manner:
\begin{equation}
\Omega\Lambda = \Gamma\Lambda^T \Gamma^{-1},
\end{equation}
where $\Gamma = 1$ or $\Sigma_2$ which corresponds to $SO/Sp$ model respectively. $\Omega$ projection on the ``vector" states are odd, so the symmetry group is given by $SO$ with $\Gamma =1$ and $Sp$ with $\Gamma = \Sigma_2$. On the other hand we find that the tachyon state is also odd. Therefore the massless tachyon is in the (anti) symmetric representation of $Sp(SO)$ model. In order to cancel the tadpole, we choose the $SO(2)$, and the tachyon is antisymmetric (hence the singlet). Instead of this, it is also possible to introduce different types ($\eta = \pm$) of branes simultaneously, where we have no symmetry group (actually $SO(1) \times SO(1)$) with a single massless fermion propagating between $\eta = + $ and $\eta =-$ branes.

\textbf{0A/$\hat{\Omega}$}

$\hat{\Omega} = \Omega (-1)^{F_L}$ is another possible orientifold projection. The superficial difference from $\Omega$ projection is that now we have $C^-_1$ is odd and $C^+_1$ is even. To cancel the tachyon tadpole, we introduce two $\tilde{\mathrm{D}}$1-branes as we have done in the $\Omega$ case. The difference is the projection on the tachyon ground states which becomes even. The net result is we introduce the $Sp(2)$ Chan-Paton group brane with symmetric (hence the singlet) tachyon on it.

\textbf{0A/$\Omega'$}

As in the 0B/$\Omega'$ case, this is impossible in the super Liouville theory, so we will not consider it here. Actually, this is not a $Z_2$ symmetry in the free theory and the orbifold by it yields a type 0B theory.

\subsection{Matrix Model Dual}\label{14.3}
As we have seen in the bosonic and fermionic (un)oriented two dimensional string theories, we consider the dual matrix model from the matrix quantum mechanics of tachyons living on unstable D0-branes in the double scaling limit. In this section, we list all the proposed matrix model dual of the (tadpole canceled) unoriented fermionic two dimensional string theory according to \cite{Gomis:2003vi}, \cite{Bergman:2003yp}. Since we have introduced the D1-branes to cancel the crosscap tadpole, we have also vectors in the theory on the D0-branes (see section \ref{6.1}). The candidate D0-brane can be made by tensoring the Neumann free $X,\psi$ brane with the super ZZ brane discussed in section \ref{7.2}. As we have seen in the oriented case, we expect only $(1,1)$ type brane will contribute to the dual matrix model. 

\textbf{0B/$\Omega$}

This model is orientifold tadpole free, so there is no vector in the dual matrix model. We introduce unstable $n_0$ $\tilde{\mathrm{D}}0$-branes to describe the system. With the choice of the Chan-Paton rotation $\Gamma$, we have either $SO(n_0)$ for $\Gamma = 1$ or $Sp(n_0)$ for $\Gamma = \Sigma_2$. By examining the $\Omega$ projection property on the tachyonic sector, which states $\Omega|(1,1)\rangle = - |(1,1)\rangle$, we have a symmetric representation for $Sp(n_0)$ and antisymmetric representation for $SO(n_0)$ whose net result is given by (\ref{eq:untree}). This model is related to the \textit{free} inverted harmonic oscillator as follows \cite{Gomis:2003vi}. Take the $SO$ theory for example and consider the diagonalization of the matrix. Then the Jacobian is given by
\begin{equation}
J = \prod_{i<j} (\lambda_i-\lambda_j)^2(\lambda_i + \lambda_j)^2.
\end{equation}
By redefining the wavefunction as $\Psi(\lambda) = \mathrm{sign}(J)|J|^{-1/2} f(\lambda) $ the Hamiltonian can be rewritten as 
\begin{equation}
\tilde{H} = -\frac{1}{2}\sum_i\frac{d^2}{d\lambda_i^2} + \sum_i U(\lambda_i)
\end{equation}
for the singlet sector. The constraint on the wavefunction (in addition to the antisymmetry) is 
\begin{equation}
f(-\lambda_i) = f(\lambda_i).
\end{equation}
Therefore, the fluctuation of this system is the same as that of the tachyon fluctuation of the oriented $\hat{c} = 1$ theory, which is expected from the space-time theory. Note that the partition function is simply the half of the oriented $\hat{c} =1$ theory, which is indeed the case in the direct calculation (\ref{eq:kk0b}).

\textbf{0B/$\hat{\Omega}$}

Since we have $n_0$ unstable $\tilde{\mathrm{D}}$0-branes, the gauge group is either $SO(n_0)$ or $Sp(n_0)$. By comparing the various M\"obius strip amplitudes,\footnote{Another heuristic way to see their relation is to construct a mass term of 0-1 NS string which can be obtained from the world sheet theory. For example, if we have vectors $v^{a}_{i}$, where $a$ is the gauge indices on $Sp$ 0-brane and $i$ is the global symmetry indices on $Sp$ 1-brane, the mass term $mvv$ is possible. However, if the 0-brane gauge group is $SO$, we cannot make a mass term because of the antisymmetry. This works for the bosonic string, 0B/$\Omega$, and 0A/$\hat{\Omega}$ model.} we can find a relation between the orientifold projection for the D1-branes which we have introduced to cancel the tadpole and that for the D0-branes. For the tadpole free theory, we can find (see \cite{Gomis:2003vi} for a detailed discussion) $Sp(n_0)$ gauge group with antisymmetric (or real selfdual) tachyons. The vector sectors differ according to the type of D1-branes introduced to cancel the tadpole. We should recall here that the D0-brane must be $\eta = +$. Therefore, for two $\widehat{\mathrm{D}1}$-branes with $\eta =+$, we have four fundamental bosons, and for two $\widehat{\mathrm{D}1}$-branes with $\eta=-$, we have four fundamental fermions, and finally for one $\widehat{\mathrm{D}1}$-brane with $\eta=+$ and one $\widehat{\mathrm{D}1}$-brane with $\eta=-$, we have two fundamental bosons and fermions.

We have two comments here. First is the cancellation of the tadpole from the matrix model point of view. The combination of the $Sp$ matrix model with fundamental vectors is the same as in the bosonic case and we expect that the fundamental loop cancels the unoriented cross loop of the $Sp$ matrix model. However, the number of the fundamental loop is doubled here and we should study the details of the interaction to verify whether the loop actually cancels or not. Furthermore, this is the second comment, the fermionic loop is possible from the holographic point of view, but it is doubtful that the fermionic loop (which has an extra $-$ sign) cancels the $Sp$ loop. We should study the details further to see what is actually happening here.\footnote{Indeed, the interaction cannot be same for the boson and fermion. For example, if $BTB$ is possible, $FTF$ vanishes because of the antisymmetry.}

\textbf{0A/$\Omega$}

Since the R-R charged D0-brane (with $\eta = +$) is mapped to the $\bar{\mathrm{D}}$0-brane under $\Omega$, we have to introduce $n_0$ invariant $\widehat{\mathrm{D}0}$-branes. With the tadpole canceled orientifold projection and two unstable $\tilde{\mathrm{D}}1$-branes, we have a $U(n_0)$ gauge group with antisymmetric tachyons. The vector sector is different according to the type of $\tilde{\mathrm{D}}1$-branes introduced. For the two $\eta = +$ case, we have two fundamental bosons and two antifundamental bosons. For the two $\eta = -$ case, we have two fundamental fermions and two antifundamental fermions. For the one $\eta = +$ and one $\eta = -$ case, we have one fundamental boson, one fundamental fermion, one antifundamental boson, and one antifundamental fermion. 

\textbf{0A/$\hat{\Omega}$}

Since R-R charged D0-brane (with $\eta = +$) is invariant under $\hat{\Omega}$, we introduce $n_0$ D0-$\bar{\mathrm{D}}0$ pairs. The gauge group and the representation of the tachyon in the tadpole canceled case can be obtained by examining various M\"obius strip amplitudes. The result is that the matrix model is $Sp(n_0)\times Sp(n_0)$ with a bifundamental (stretching between branes and antibranes) tachyon. The vector from the 0-1 string is given by two fundamentals ($\square$,1) and two fundamentals (1,$\square$) either bosonic or fermionic according to the choice of the D1-brane. 

We collect these results in table 1, which we have taken from \cite{Bergman:2003yp}.

\begin{table}
\begin{center}
\begin{tabular}[h]{|c|c|c|c|}
\hline
theory & closed & D1-brane & D0-brane matrix(-vector) model\\
       & strings & open strings & \\
\hline
 &&& \\[-10pt]
bos/$\Omega_-$ & $T$ & 
$Sp(2) ,\,  {\bf 1}$ & $Sp({n_0}) ,\,  \asymm + 2\,\funda$ \\[10pt]
\hline
&&&\\[-10pt]
0B/$\Omega_\pm$ & $T$ & -- &
$
\begin{array}{l}
SO(n_0),\, \asymm_{\,b}\\[5pt]
Sp({n_0}) ,\, \symm_{\,b}
\end{array}
$
\\[20pt]
\hline
&&&\\[-10pt]
0B/$\widehat{\Omega}_-$ & $T, C_0$ & 
$
\begin{array}{l}
 U(2) ,\, {\bf 1}_b^0 \\[5pt]
 U(1)^2 ,\, 2(\pm,\mp)_f 
\end{array}
$
&
$
Sp({n_0}) ,\,  \asymm_{\,b} +
\left\{
\begin{array}{l}
4\,\funda_{\,b\,or\, f} \\[5pt]
2\,\funda_{\,b} + 2\,\funda_{\,f}
\end{array}\right.
$\\[20pt]
\hline
&&&\\[-10pt]
0A/$\Omega_-$ & $T, C_1^+$ &
$
\begin{array}{l}
SO(2) ,\, {\bf 1}_b\\[5pt]
- ,\, {\bf 1}_f
\end{array}
$ &
$U(n_0) ,\, \asymm_{\,b}  + 
\left\{
\begin{array}{l}
2[\funda 
+ \,\overline{\funda}]_{\,b\, or\, f}\\[5pt]
[\funda + \,\overline{\funda}]_{\,b} 
+ [\funda + \,\overline{\funda}]_{\,f}
\end{array}\right.$
\\[10pt]
\hline
&&&\\[-10pt]
0A/$\widehat{\Omega}_-$ & $T, C_1^-$ &
$Sp(2) ,\, {\bf 1}_b$ & 

$ Sp({n_0})\times Sp({\bar{n}_0}) , (\funda,\funda)_{\,b} 
+ 2[(\funda,{\bf 1}) 
+ ({\bf 1},\funda)]_{b\, or \, f}$
\\[20pt]
\hline
\end{tabular}
\end{center}
\caption{Tadpole-free unoriented open and closed string theories in two 
dimensions and their dual matrix(-vector) models \cite{Bergman:2003yp}.}
\label{tadpole_free_models}
\end{table}

\subsection{Literature Guide for Section 14}\label{14.4}
Unoriented $\mathcal{N}=1$ super Liouville theory ($\hat{c} = 1$) has been extensively studied in \cite{Bergman:2003yp}, \cite{Gomis:2003vi}, where the matrix dual model and the tadpole cancellation condition have been proposed. It would be interesting to study further these matrix models to understand the dynamics of the orientifold plane in a non-trivial background. Also it is worth extending this section's results to the $\mathcal{N}=2$ super Liouville theory (or $SL(2,\mathbf{R})/U(1)$ supercoset model), where we expect that the subtlety concerning the R wavefunction and the GSO projection is none because of the spectral flow. Physically, it will correspond to the orientifold plane in the singular CY space.

\sectiono{Conclusion and Future Outlook}\label{sec:15}
In this thesis, we have discussed both the old formulation and the recent developments of the Liouville field theory (coupled to the minimal model or $c=1$ matter) and the corresponding matrix models. To conclude the whole thesis, we would like to ask once again why and how these theories are dual to each other. This question has been one of the main themes of the Liouville-matrix theory (and of course of this review). 

The matrix model exact solution states that the Liouville theory coupled to the minimal model or $c=1$ matter is solvable. We here mean ``solvable" as not only being solvable as a CFT, but also mean that the integration over the moduli space can be explicitly performed. This statement seems peculiar in two ways. First, the Liouville theory itself is not easy to solve as a CFT as we have seen in this review. The consistent three-point function is proposed \textit{after} the matrix revolution. The second peculiarity is the problem with the integration over the moduli space. For example, the free 26 bosons with ghosts ($=$ critical string) are trivial as a CFT, yet the integration over the moduli space is another story --- we have not been able to perform it to all orders of perturbation.\footnote{Of course, in the bosonic critical string, the integration over the moduli space diverges because of the tachyon contribution.} In the matrix revolution era, we have not solved the world sheet theory on the one hand, but we have had a full genus answer on the other hand.

After the accumulation of the knowledge of the nonperturbative string theory and the developments in the Liouville field theory, we have come close to the proof of this duality. On the Liouville side, the central objects (it happens only when we couple it to the minimal model or $c=1$ matter) are the ground ring structure and the higher equations of motion. The ground ring structure helps us to understand the integrable structure of the correlation function, and the higher equations motion will become the first step to understand why the integration over the moduli space is reduced to the cohomological integration (such as the intersection form of the Mumford-Morita-Miller classes).

On the matrix side, we have seen in this review that the matrix, which has been originally considered as the discretization of the Riemann surface (or in the topological model, the moduli of the Riemann surface), has now a holographic dual interpretation of the bulk theory. The double scaling matrix model is the dual description by the localized D0-branes (of the ZZ branes) as we have seen in this review, and very recently the Kontsevich matrix model is proposed to be the dual description by the extending D1-brane (or the FZZT brane) in \cite{Gaiotto:2003yb}.  Actually, the physical double scaling matrix model and the topological Kontsevich matrix model yield the same description of the noncritical string theory, which has been yet another mystery. The recent paper \cite{Aganagic:2003qj} has proposed that the topological B-model on the CY manifold gives the unifying perspective, where the compact B-branes correspond to the physical matrix and the noncompact B-branes correspond to the topological matrix. It would be very interesting to see their connections with the ZZ brane and the FZZT brane.

With these rapid developments based on the knowledge over this decade, we expect that this longstanding question will be solved from the first principle in the near future. In fact, some relevant papers cited above have appeared in the last month of our preparation of this review thesis\footnote{December 2003.}. Some subjects reviewed in this thesis may become obsolete within several years, but in a sense the author hopes so.

\begin{figure}[htbp]
	\begin{center}
	\includegraphics[width=0.6\linewidth,keepaspectratio,clip]{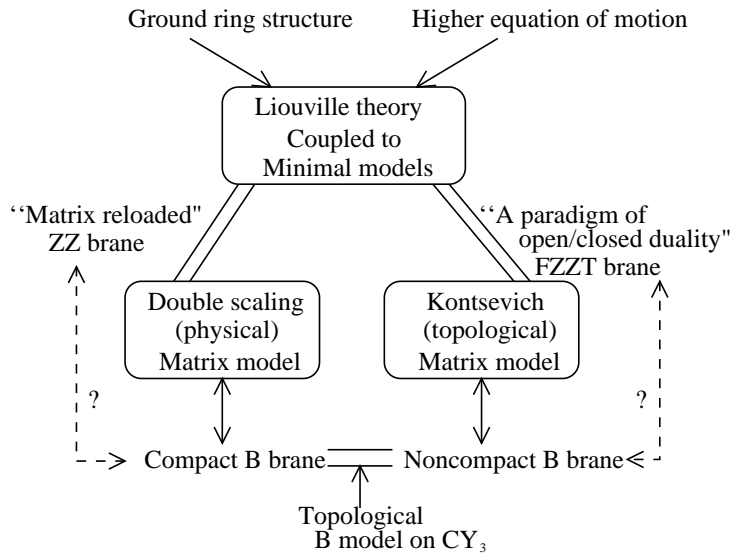}
	\end{center}
	\caption{Why is the Liouville field theory coupled to the minimal model integrable? The available map in the year 2003.}
	\label{integ}
\end{figure}

Once we accept the Liouville-matrix duality, we encounter many surprising consequences as we have seen in this review (particularly in the application sections). Among these, we have found various exact and explicit realizations of the nonperturbative string physics which we have listed in the introduction section. Actually, the matrix formulation of the Liouville theory not only reproduces the world sheet description (integration of the CFT over the moduli space) but also provides a nonperturbative formulation of the theory. This includes, for instance, the instanton effects, the R-R background physics and the string field theory. The matrix formulation extends our world sheet description so far available (i.e. the Liouville CFT coupled to the minimal model) to the nonperturbative level, so it would be an interesting remaining problem to understand their nature from the world sheet (or another first principle of the string theory) because we can apply them to the higher dimensional theory.

Historically speaking, we cannot say that the matrix revolution in a decade ago  triggered the second revolution of the string theory though it taught us the importance of the nonperturbative effects, as far as the author knows. On the contrary, developments after the second revolution have enabled us to understand the full nonperturbative physics of the matrix-Liouville theory. We hope that the word ``the matrix theory has delivered much less than it seemed promise" \cite{Johnson:2003hy} will no longer be true. In order to ensure this, we should find new physics in the noncritical string theory which can be applied to the more realistic (higher dimensional) string theory \textit{before} they will be found in their higher dimensional setup. We have seen some of them in this review, but we believe that the matrix-Liouville model has not revealed its full potential yet.

Concerning this point, we have also discussed the slight extension of the Liouville theory --- the $\mathcal{N}=1,2$ super Liouville theory and the Liouville theory on unoriented surfaces. These subjects drew less attention than the original bosonic matrix model in a decade ago, but they have now caught more attention in order to understand variable nonperturbative aspects of the theory which the bosonic model does not have. One may say that this is the first string revolution in the Liouville-matrix model, namely the discovery of the ``superstring". 

To conclude the whole thesis, we would like to emphasize once again that we hope the matrix reloaded this year (2003) does not lead to nowhere (or the prison) as we have seen in ``The Game of Everything" by Dijkgraaf \cite{Dijkgraaf:2003cy}.\footnote{He has placed the square ``2D black hole" before the big depression, but we have already seen some applications to the 2D black hole in this thesis!} We will not be caught into the limit cycle in the hep-th space because we have a ``c-function" on it.

\section*{Acknowledgements}
The author would like to thank his advisor T.~Eguchi for his hospitable guidance and advice. The author is grateful to Y.~Sugawara for reading the first draft and commenting on the contents. A part of this thesis is based on the note of the informal seminar held at the university of Tokyo high energy physics theory group. The author would like to thank all the participants in the seminar: Y.~Imamura, I.~Kishimoto, F.~Koyama, Y.~Matsuo, R.~Nobuyama, K.~Ohmori, K.~Sakai, H.~Takayanagi, E.~Watanabe and S.~Yamaguchi. The author also acknowledges valuable discussions with  J.~Gomis, Y.~Hikida, S.~Hirano, K.~Hosomichi, A.~Kapustin, B.~Ponsot, Y.~Tachikawa and T.~Takayanagi.

\appendix\sectiono{Appendices I: Conventions and Useful Formulae}\label{sec:A}
\subsection{Conventions}\label{a-1}
In the most of part I of this review (for the bosonic theory) we use $\alpha'=1$ convention unless otherwise stated. In part II (for the supersymmetric theory) we use $\alpha'=2$ convention mostly. 

When we regard the world sheet sphere as the complex plane, we use the complex coordinate $z=x_1+ix_2$, $\bar{z} = x_1-ix_2$. In this coordinate, the differentials become $\partial = \partial_{z}$, $\bar{\partial} = \partial_{\bar{z}}$. Then the Laplacian becomes $\Delta = 4\partial\bar{\partial}$. The volume element is $d^2 z = dx_1 dx_2$, and the delta function is defined as $\delta(z) = \delta(x_1)\delta(x_2)$\footnote{Note this is different from Polchinski's notation \cite{Polchinski:1998rq}.}

Let us present the action and OPE for the free field in our convention. The action of the Coulomb gas free boson is
\begin{equation}
\int d^2 z \sqrt{g}\left( \frac{1}{4\pi\alpha'}g^{ab}\partial_a\phi \partial_b \phi+ \frac{QR}{4\pi}\phi\right)
\end{equation}
and the central charge is given by $c = 1+ 6 \alpha'Q^2 $. The relevant OPEs are 
\begin{equation}
\phi(z) \phi (0) = -\frac{\alpha'}{2}\log|z|^2 +\cdots,
\end{equation}
and 
\begin{equation}
e^{ik_1 \phi(z)}e^{ik_2\phi(0)} = |z|^{\alpha'k_1k_2}e^{i(k_1+k_2)\phi(0)} + \cdots.
\end{equation}
The energy-momentum tensor is given by
\begin{equation}
T = -\frac{1}{\alpha'}(\partial \phi)^2 + Q\partial^2 \phi.
\end{equation}

Next, we consider the $bc$ ghost system. The action is
\begin{equation}
\frac{1}{\pi} \int d^2 z b\bar{\partial} c.
\end{equation}
The OPE is given by
\begin{equation}
b(z) c(0) = \frac{1}{z} + \cdots.
\end{equation}
Let us set the weight of $b$ to be $\lambda$ and that of $c$ to be $1-\lambda$. Then the energy-momentum tensor is given by
\begin{equation}
T = (\partial b)c - \lambda\partial(bc),
\end{equation}
and the central charge is written as $c = -3(2\lambda-1)^2 + 1 $.

Similarly the $\beta\gamma$ ghost is described by the action
\begin{equation}
\frac{1}{\pi} \int d^2 z \beta\bar{\partial}\gamma.
\end{equation}
The OPE is given by 
\begin{equation}
\beta(z)\gamma(0) = -\frac{1}{z} +\cdots.
\end{equation}
If we set the weight of $\beta$ to be $\lambda-\frac{1}{2}$ and that of $\gamma$ to be $\frac{3}{2}-\lambda$, the energy-momentum tensor is given by
\begin{equation}
T = (\partial \beta)\gamma- \frac{1}{2}(2\lambda-1)\partial(\beta\gamma).
\end{equation}
The central charge becomes $c= 3(2\lambda-2)^2-1$.

When we do the bosonization, we take the convention $\alpha'=2$ so that $v(z) v(0) \sim -\log(z)$. For example, the bosonization of the $(1,0)$ $\beta\gamma$ system, which was discussed in the main text, is given by $\beta = i\partial v e^{iv-u}$ and $\gamma = e^{u-iv}$, where $u$ and $v$ are normalized bosons in this $\alpha'=2$ convention.

Finally, we show the action and currents of the supersymmetric linear dilaton system in the component formalism (for the superspace formalism, see the next subsection). The action is
\begin{equation}
S = \frac{1}{2\pi} \int d^2z \left(\partial X\bar{\partial}X + \psi\bar{\partial}\psi + \bar{\psi}\partial\bar{\psi}+ \frac{1}{4}QRX \right)
\end{equation}
The energy-momentum tensor and the superconformal current are given by
\begin{eqnarray}
T(z) &=& -\frac{1}{2}\partial X\partial X + \frac{Q}{2}\partial^2 X - \frac{1}{2}\psi\partial \psi \cr
G(z) &=& i\psi\partial X - iQ\partial \psi.
\end{eqnarray}
The central charge of this system is $c = \frac{3}{2} \hat{c} = \frac{3}{2} + 3 Q^2$.

To evaluate the free path integral on the sphere, the following formula is useful
\begin{equation}
\int d^2z |z|^{2\alpha-2} |1-z|^{2\beta-2} = \pi\gamma(\alpha)\gamma(\beta) \gamma(\delta),\label{eq:kbnl}
\end{equation}
where $\alpha + \beta + \delta = 1$.

For more general correlation functions, we have the Dotsenko-Fateev formula \cite{Dotsenko:1984nm,Dotsenko:1985ad}:
\begin{eqnarray}
\frac{1}{m!} \int dz_i^2 \prod_{i=1}^m |z_i|^{2\alpha} |1-z_i|^{2\beta} \prod_{i<j}^m|z_i-z_j|^{4\rho} = \pi^m(\gamma(1-\rho))^m \cr
\prod_{i=1}^{m} \gamma(i\rho) \prod_{i=0}^{m-1} \gamma(1+\alpha+i\rho)\gamma(1+\beta+i\rho)\gamma(-1-\alpha-\beta-(m-1+i)\rho). \label{eq:DFF}
\end{eqnarray}

\subsection{Superspace Convention}\label{a-2}
In this appendix, we present our conventions for $\mathcal{N}=1,2$ supersymmetries in the two dimensional Euclidean space.
\subsubsection{$\mathcal{N}=1$ Supersymmetry}
For $(1,1)$ supersymmetry, we introduce two fermionic coordinates $\theta, \bar{\theta}$. The integration measure is $\int d^2z d\theta d\bar{\theta}$. The supercharges are defined as
\begin{equation}
Q = \frac{\partial}{\partial \theta} - \theta \partial, \ \ \ \ \bar{Q} = \frac{\partial}{\partial\bar{\theta}} - \bar{\theta}\bar{\partial},
\end{equation}
and the covariant derivatives are given by
\begin{equation}
D = \frac{\partial}{\partial \theta} + \theta \partial, \ \ \ \ \bar{D} = \frac{\partial}{\partial\bar{\theta}} + \bar{\theta}\bar{\partial}.
\end{equation}
General superfields  are expanded as 
\begin{equation}
\Phi = \phi + i\theta\psi + i\bar{\theta}\bar{\psi} + i\theta\bar{\theta}F,
\end{equation}
where $\phi$ is a real scalar and $\psi$ is a Majorana Weyl spinor (if we were in the Minkowski space) and $F$ is an auxiliary field.

For example we consider the free action 
\begin{eqnarray}
S &=& \frac{1}{2\pi}\int d^2z d\bar{\theta}d\theta D\Phi \bar{D} \Phi \cr
 &=& \frac{1}{2\pi} \int d^2z \left(\partial \phi\bar{\partial}\phi + \psi\bar{\partial}\psi + \bar{\psi}\partial\bar{\psi} -F^2 \right).
\end{eqnarray}
We can eliminate the auxiliary field $F$ simply by setting it zero.

\subsubsection{$\mathcal{N}=2$ Supersymmetry}
For $(2,2)$ supersymmetry, we introduce four fermionic coordinates $\theta^+, \bar{\theta}^+,\theta^-,\bar{\theta}^-$. Note that as opposed to the $\mathcal{N}=1$ case, the $+,-$ represents the chirality of the spinor, and the bar indicates the formal ``(Euclidean) complex conjugation" which does \textit{not} change the chirality. The covariant derivatives are 
\begin{equation}
D_+ = \frac{\partial}{\partial\theta^+} + \frac{1}{2}\bar{\theta}^+\partial,\ \ \ \ \bar{D}_+ = \frac{\partial}{\partial\bar{\theta}^+} + \frac{1}{2}\theta^+\partial,
\end{equation}
\begin{equation}
D_- = \frac{\partial}{\partial\theta^-} + \frac{1}{2}\bar{\theta}^-\bar{\partial},\ \ \ \ \bar{D}_- = \frac{\partial}{\partial\bar{\theta}^-} + \frac{1}{2}\theta^-\bar{\partial}.
\end{equation}

A chiral superfield is defined as
\begin{equation}
\bar{D}_\pm \Phi= 0, 
\end{equation}
and expanded as 
\begin{equation}
\Phi = \phi + i\theta^+\psi^- - i\theta^-\psi^+ + i\theta^+\theta^- F+ \cdots
\end{equation}
where the omitted part contains a derivative of the fields. The anti-chiral superfield is defined as 
\begin{equation}
D_\pm \bar{\Phi}= 0, 
\end{equation}
and expanded as 
\begin{equation}
\bar{\Phi} = \phi^* + i\bar{\theta}^+\bar{\psi}^- - i\bar{\theta}^-\bar{\psi}^+ + i\bar{\theta}^+\bar{\theta}^- F^*+ \cdots,
\end{equation}
which is the complex conjugation of $\Phi$ (in the Minkowski sense). These superfields can be seen as dimensional reductions of the four dimensional chiral and anti-chiral superfields respectively.

We can also introduce a twisted chiral superfield as 
\begin{equation}
\bar{D}_+ Y = D_- Y = 0
\end{equation}
which can be expanded as 
\begin{equation}
Y = y + i\theta^+\bar{\chi}^- - i\bar{\theta}^-\chi^+ +i\theta^+\bar{\theta}^- G+\cdots .
\end{equation}
Its ``complex conjugation" is a twisted anti-chiral superfield defined as
\begin{equation}
D_+ \bar{Y} = \bar{D}_- \bar{Y} = 0
\end{equation}
which can be expanded as 
\begin{equation}
\bar{Y} = y^* + i\bar{\theta}^+{\chi}^- - i\theta^-\bar{\chi}^+ +i\bar{\theta}^+\theta^- G^*+ \cdots .
\end{equation}

As in the four dimension, we introduce a vector superfield. In the Wess-Zumino gauge, it can be expanded as 
\begin{eqnarray}
V &=& \theta^-\bar{\theta}^- \bar{v} + \theta^+\bar{\theta}^+v - \theta^-\bar{\theta}^+\sigma - \theta^+\bar{\theta}^-\bar{\sigma} \cr
&+& i \theta^-\theta^+(-\bar{\theta}^-\bar{\lambda}^++\bar{\theta}^+\bar{\lambda}^-) + i \bar{\theta}^+\bar{\theta}^-(-\theta^-\lambda^++\theta^+\lambda^-) +\theta^-\theta^+\bar{\theta}^+\bar{\theta}^- D,
\end{eqnarray}
where $v = v_1-iv_2$, $\bar{v} = v_1+iv_2$ and $\sigma$ is a complex scalar which corresponds to $v_3$ and $v_4$ in the four dimensional vector multiplet. The field strength superfield is a twisted chiral superfield, and in the Abelian case, it is defined as 
\begin{eqnarray}
\Sigma &=& -\bar{D}_+D_- V \cr
&=& \sigma + i\theta^+\bar\lambda^- - i\bar{\theta}^-\lambda^+ + \theta^+\bar{\theta}^-(D-iv_{12}) +\cdots
\end{eqnarray}

\subsection{Special Functions}\label{a-3}
In this appendix, we explain special functions which are needed in the main text. We are not careful about the mathematical rigor such as a convergence and an exchange of the order of the differentiations etc. The useful reference of the special functions is \cite{WW} for example.

{\bfseries Basic definitions/formulae}
\begin{equation}
\zeta(s,a) \equiv  \sum_{n=0}^\infty \frac{1}{(a+n)^s}
\end{equation}
\begin{equation}
\zeta(0,a)=\frac{1}{2}-a \label{eq:zeta1}
\end{equation}
\begin{equation}
\zeta'(0,a)=\log\Gamma(a)-\frac{1}{2}\log(2\pi) \label{eq:zeta2}
\end{equation}
\begin{equation}
\Gamma(z+1) = z\Gamma(z)
\end{equation}
\begin{equation}
\Gamma(z)\Gamma(1-z) = \frac{\pi}{\sin\pi z}
\end{equation}
\begin{equation}
\Gamma(\frac{1}{2}+z)\Gamma(\frac{1}{2}-z) = \frac{\pi}{\cos\pi z}
\end{equation}
\begin{equation}
\Gamma(2x) =(2\pi)^{-1/2}2^{2x-1/2}\Gamma(x)\Gamma(x+1/2) \label{eq:ledgam}
\end{equation}
\begin{equation}
\zeta(1-2m) = \frac{B_{2m}}{2m} \label{eq:berz}
\end{equation}
\begin{equation}
\frac{x}{e^x-1}\equiv \sum_{n=0}^{\infty}\frac{B_nx^n}{n!}
\end{equation}
\begin{equation}
\lim_{s\to1}\left[\zeta(s,a)-\frac{1}{s-1}\right] = -\psi(a)
\end{equation}
\begin{equation}
\psi(z) \equiv \frac{\Gamma'(z)}{\Gamma(z)} = \mathrm{Re}\left(\int_0^\infty dt\frac{e^{-t}}{t} -\frac{e^{-zt}}{1-e^{-t}}\right)
\end{equation}
\begin{equation}
\gamma(x) \equiv \frac{\Gamma(x)}{\Gamma(1-x)}
\end{equation}
\begin{equation}
\gamma(1+x) = -x^2\gamma(x)
\end{equation}
\begin{equation}
\gamma(x)\gamma(-x) = -\frac{1}{x^2}
\end{equation}
\begin{equation}
\gamma(x)\gamma(1-x) = 1
\end{equation}
\begin{eqnarray}
F(\alpha,\beta,\gamma,z) = \frac{\Gamma(\gamma)\Gamma(\gamma-\alpha-\beta)}{\Gamma(\gamma-\alpha)\Gamma(\gamma-\beta)}F(\beta,\alpha,\alpha+\beta+1-\gamma,1-z) \cr
+ \frac{\Gamma(\gamma)\Gamma(\alpha+\beta-\gamma)}{\Gamma(\alpha)\Gamma(\beta)}(1-z)^{\gamma-\alpha-\beta}F(\gamma-\alpha,\gamma-\beta,\gamma+1-\alpha-\beta,1-z) 
\label{eq:hyperg} 
\end{eqnarray}
\begin{eqnarray}
F(\alpha,\beta,\gamma,z) &=& \frac{\Gamma(\gamma)\Gamma(\beta-\alpha)}{\Gamma(\beta)\Gamma(\gamma-\alpha)}(-z)^{-\alpha}F(\alpha,\alpha+1-\gamma,\alpha+1-\beta,1/z) \cr
&+& \frac{\Gamma(\gamma)\Gamma(\alpha-\beta)}{\Gamma(\alpha)\Gamma(\gamma-\beta)}(-z)^{-\beta}F(\beta,\beta+1-\gamma,\beta+1-\alpha,1/z) \label{eq:inv} 
\end{eqnarray}
\begin{equation}
I_\nu(z) \equiv \left(\frac{z}{2}\right)^\nu \sum_{n=0}^{\infty}\frac{(z/2)^{2n}}{n!\Gamma(\nu+n+1)}
\end{equation}
\begin{equation}
I_{-\nu}(z) - I_{\nu}(z) = \frac{2\sin \nu\pi}{\pi}\int_0^\infty e^{-z\cosh t} \cosh(\nu t) dt
\end{equation}
\begin{equation}
K_\nu(z) \equiv \frac{\pi}{2}\frac{I_{-\nu}(z)-I_{\nu}(z)}{\sin \nu \pi}
\end{equation}
\begin{equation}
K_\nu(z) = \int_0^\infty e^{-z \cosh t}\cosh(\nu t)dt
\end{equation}
{\bfseries Double $\zeta$ Function}\footnote{Those functions presented below have been introduced and studied in \cite{Barnes}, \cite{Shin}, \cite{Kuro}. A good review on them can be found in appendix of \cite{Jimbo:1996ss}.}

Definition:
\begin{equation} \zeta_2(s,t|w_1,w_2) \equiv \sum_{n_1,n_2 \ge 0} (s+n_1w_1+n_2w_2)^{-t},
\end{equation}
which can be analytically continued to $t=0$. We use this fact to define the double gamma function as follows.

{\bfseries Double $\Gamma$ Function}

Definition:
\begin{equation} \log\left[\Gamma_2(s|w_1,w_2)\right] \equiv \frac{\partial}{\partial t}\zeta_2(s,t|w_1,w_2)|_{t=0}.
\end{equation}
This function satisfies the following important ``difference formula"
\begin{equation} \frac{\Gamma_2(s+w_1|w_1,w_2)}{\Gamma_2(s|w_1,w_2)} = \frac{\sqrt{2\pi}}{w_2^{\frac{s}{w_2}-\frac{1}{2}}\Gamma\left(\frac{s}{w_2}\right)} \label{eq:diff},
\end{equation}
which is the root of the name ``double gamma function". While the ordinary gamma function is multiplied by $x$ when we add one to the argument, the double gamma function is multiplied essentially by $\Gamma(x)$. Similarly we can define the multiple gamma functions.

The proof of (\ref{eq:diff}):
\begin{eqnarray*}
(LHS)& = &\exp[\log\Gamma_2(s+w_1)-\log\Gamma_2(s)] \cr
&=&\exp\left(\frac{\partial}{\partial t}\left[\zeta_2(s+w_1,t)-\zeta_2(s,t)\right]_{t=0} \right)\cr
&=&\exp\left(\frac{\partial}{\partial t}\left[-\sum_{n \ge 0}(s+nw_2)^{-t}\right]_{t=0}\right) \cr
&=&\exp\left(-\frac{\partial}{\partial t}\left[w_2^{-t}\zeta\left(t,\frac{s}{w_2}\right)\right]_{t=0}\right) \cr
&=&\exp\left[\log(w_2) \zeta\left(0,\frac{s}{w_2}\right)-\zeta'\left(0,\frac{s}{w_2}\right) \right] \cr
&=&\exp\left[\left(\frac{1}{2}-\frac{s}{w_2}\right)\log(w_2) -\log\Gamma\left(\frac{s}{w_2}\right)+\frac{1}{2}\log (2\pi) \right] = (RHS)\cr 
\end{eqnarray*}
where we have used the formulae of the $\zeta$ function (\ref{eq:zeta1},\ref{eq:zeta2}). From this double gamma function, we can define $\Gamma_b$ function as follows.

Definition:
\begin{equation}
\Gamma_b(x) \equiv \frac{\Gamma_2(x|b,b^{-1})}{\Gamma_2(Q/2|b,b^{-1})}, 
\end{equation}
where $Q=b+b^{-1}$. The difference formula (\ref{eq:diff}) states immediately
\begin{align}
\Gamma_b(x+b) &= \frac{\sqrt{2\pi}b^{bx-1/2}}{\Gamma(bx)}\Gamma_b(x) \cr
\Gamma_b(x+b^{-1}) &= \frac{\sqrt{2\pi}b^{-x/b+1/2}}{\Gamma(x/b)}\Gamma_b(x).
\end{align}

$\Gamma_b$ function has an integral representation:
\begin{equation}
\log \Gamma_b(x) = \int_0^\infty\frac{dt}{t}\left[\frac{e^{-xt}-e^{-\frac{Qt}{2}}}{(1-e^{-bt})(1-e^{-t/b})}-\frac{\left(\frac{Q}{2}-x\right)^2}{2}e^{-t}-\frac{\frac{Q}{2}-x}{t}\right].
\end{equation}
The proof can be given by using the same technique we will explain in the $S_b(x)$.

{\bfseries $\Upsilon$ Function}

Definition:
\begin{equation} 
\Upsilon_b(x) \equiv \frac{1}{\Gamma_b(x)\Gamma_b(Q-x)}
\end{equation}

Difference formula:
\begin{align}
\Upsilon_b(x+b) &= \frac{\Gamma(bx)}{\Gamma(1-bx)}b^{1-2bx}\Upsilon_b(x) \cr
\Upsilon_b(x+b^{-1}) &= \frac{\Gamma(x/b)}{\Gamma(1-x/b)}b^{\frac{2x}{b}-1}\Upsilon_b(x), \label{eq:updif}
\end{align}
which can be proven by using (\ref{eq:diff}) such that
\begin{eqnarray}
\frac{\Upsilon_b(x+b)}{\Upsilon_b(x)} &=& \frac{\Gamma_b(x)\Gamma_b(b+b^{-1}-x)}{\Gamma_b(x+b)\Gamma_b(b^{-1}-x)} \cr
&=& \frac{\Gamma(bx)}{b^{bx-\frac{1}{2}}}\frac{b^{b(b^{-1}-x)-\frac{1}{2}}}{\Gamma(1-bx)} \cr
&=& \frac{\Gamma(bx)}{\Gamma(1-bx)}b^{1-2bx}.
\end{eqnarray}

The zeros of the $\Upsilon_b(x)$ are important for the applications and located where $\Gamma_b(x)$ or $\Gamma_b(Q-x)$ has poles by its definition. 
When we consider the fact that $\Gamma_2(s)$ is an analytic continuation of
\begin{equation}
\prod_{n_1,n_2} \frac{1}{s+n_1w_1+n_2w_2} \exp[(s+n_1w_1+n_2w_2)^{-t}]|_{t\to 0}
\end{equation}
in terms of $t$, it is easy to see that $\Gamma_b(x)$ has poles at $x=-nb-\frac{m}{b}$, where $n$ and $m$ are positive integers. Then $\Upsilon_b(x)$ has simple zeros when $x= -nb-mb^{-1},Q+nb+mb^{-1}$.

In fact, there is an analytic continued formula:
\begin{equation}
\Gamma_2^{-1}(s|w_1,w_2) = e^{\frac{s^2}{2\gamma_{21}}+s\gamma_{22}}s\prod_{m,n \ge 0}\left(1+\frac{s}{\Omega}\right)e^{-\frac{s}{\Omega}+\frac{s^2}{\Omega^2}}, \label{eq:infpg}
\end{equation}
where $\Omega = mw_1+nw_2$. $\gamma_{12}$ and $\gamma_{22}$ are very complicated functions of $w_1$ and $w_2$ \cite{Barnes}. 

$\Upsilon_b(x)$ has also an integral representation:
\begin{equation}
\log\Upsilon_b(x) = \int_0^\infty \frac{dt}{t}\left[\left(\frac{Q}{2}-x\right)^2e^{-t}-\frac{\sinh^2\left(\frac{Q}{2}-x\right)\frac{t}{2}}{\sinh\frac{bt}{2}\sinh\frac{t}{2b}}\right].
\end{equation}

The following formulae are the direct consequence of the definition
\begin{align}
\Upsilon_b(x) &= \Upsilon_{b^{-1}}(x), \cr
\Upsilon_b(x) &= \Upsilon_b(Q-x). 
\end{align}

{\bfseries Double Sine Function $S_b(x)$ }

Definition:
\begin{equation}
S_b(x) \equiv \frac{\Gamma_b(x)}{\Gamma_b(Q-x)}
\end{equation}

Difference formula:
\begin{align}
S_b(x+b) &= 2\sin(\pi bx)S_b(x) \cr
S_b(x+b^{-1}) &= 2\sin(\pi x b^{-1}) S_b(x), \cr
\end{align}
which can be proven as follows; by using (\ref{eq:diff})
\begin{eqnarray}
\frac{S_b(x+b)}{S_b(x)} &=& 2\pi \frac{\Gamma_b(x+b)}{\Gamma_b(b^{-1}-x)}\frac{\Gamma_b(b+b^{-1}-x)}{\Gamma_b(x)} \cr
&=& \frac{2\pi}{\Gamma(bx)\Gamma(1-bx)} \cr
&=& 2\sin(\pi bx).
\end{eqnarray}

\begin{figure}[htbp]
	\begin{center}
	\includegraphics[width=0.5\linewidth,keepaspectratio,clip]{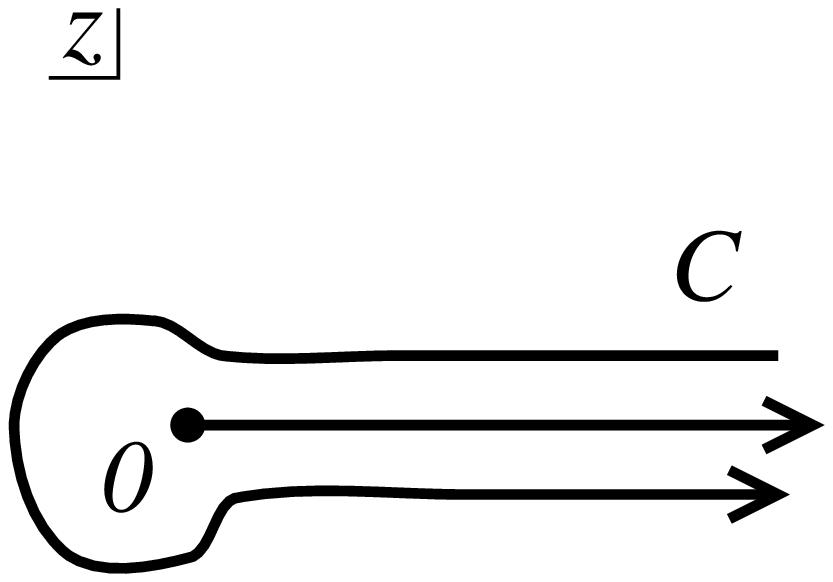}
	\end{center}
	\caption{An integration contour for the double sine (gamma) function.}
	\label{figb-1}
\end{figure}

Now, let us introduce the integral representation of $S_b(x)$. First, we consider an analytic continuation of the double $\zeta$ function as follows:
\begin{eqnarray}
 \zeta_2(s,t|w_1,w_2) &\equiv& \sum_{n_1,n_2 \ge 0} (s+n_1w_1+n_2w_2)^{-t} \cr
 &=& \frac{1}{\Gamma(t)}\sum_{\vec{n}\ge 0}\int_0^\infty dy y^{t-1} e^{-(s+\vec{w}\cdot\vec{n})y} \cr
 &=& \frac{1}{\Gamma(t)}\int_0^\infty dy \frac{y^{t-1} e^{-sy}}{(1-e^{-w_1y})(1-e^{-w_2y})} \cr
 &=& -\Gamma(1-t)\int_C \frac{e^{-sz}(-z)^{t-1}}{(1-e^{-w_1z})(1-e^{-w_2z})} \frac{dz}{2\pi i}
\end{eqnarray}
In the last line we have used the typical method of the analytic continuation of the $\zeta$ function. Then we differentiate this by $t$, and obtain the integral formula of the double gamma function:
\begin{equation}
\log \Gamma_2(s|w_1,w_2) = -\gamma \int_C \frac{e^{-sz}}{(1-e^{-w_1z})(1-e^{-w_2z})}\frac{dz}{2\pi iz} + \int_C \frac{e^{-sz}\log(-z)}{(1-e^{-w_1z})(1-e^{-w_2z})} \frac{dz}{2\pi i z},
\end{equation}
where $\gamma$ is Euler's constant. When we make the double sine function following its definition by using this formula, we can see that the first term of this integral formula cancels out. Thus we have
\begin{equation}
\log S_b(x) = \int_C \frac{\sinh(x- Q/2)z}{2\sinh(bz/2)\sinh(z/2b)} \log(-z)\frac{dz}{2\pi iz}.
\end{equation}
This contour integral can be evaluated by splitting an integral around zero and an integral along the real line. Though the integral around zero has a divergence such as $\mathcal{O}(1/\epsilon)$, this can be rewritten as an integral along the real line.  Combining these integrals we obtain 
\begin{equation}
\log S_b(x) = \frac{1}{2} \int_0^\infty \frac{dt}{t}\left[ \frac{\sinh(Q-2x)t}{\sinh(bt)\sinh(t/b)}+\frac{2x-Q}{t}\right], \label{eq:intdsin}
\end{equation}
which is finite.
\subsection{Modular Functions}\label{a-4}
In this section, we review our conventions and properties of the modular functions in the Liouville theory. We use the notation $q \equiv e^{2\pi i \tau}$. For the bosonic theory, we have
\begin{equation}
\mathrm{Tr} q^{L_0 - \frac{c}{24}} = \chi_P (\tau) = q^{P^2 - \frac{1}{24}} \prod_{n=1}^\infty (1-q^n)^{-1} =\frac{q^{P^2}}{\eta(\tau)} 
\end{equation}

On the other hand, for the supersymmetric theory, we have
\begin{eqnarray}
\mathrm{Tr}_{NS} q^{L_0 - \frac{c}{24}} &=& \chi^+_{P(NS)} (\tau) = q^{\frac{P^2}{2} - \frac{1}{16}} \prod_{n=1}^\infty (1-q^n)^{-1}(1+q^{n-\frac{1}{2}}) = \sqrt{\frac{\theta_3(\tau)}{\eta(\tau)}}\frac{q^{\frac{P^2}{2}}}{\eta(\tau)} \cr
\mathrm{Tr}_{NS} (-1)^Fq^{L_0 - \frac{c}{24}} &=& \chi^-_{P(NS)} (\tau) = q^{\frac{P^2}{2} - \frac{1}{16}} \prod_{n=1}^\infty (1-q^n)^{-1}(1-q^{n-\frac{1}{2}}) = \sqrt{\frac{\theta_4(\tau)}{\eta(\tau)}}\frac{q^{\frac{P^2}{2}}}{\eta(\tau)} \cr
\mathrm{Tr}_{R} q^{L_0 - \frac{c}{24}} &=& \chi^+_{P(R)} (\tau) = 2q^{\frac{P^2}{2}} \prod_{n=1}^\infty (1-q^n)^{-1}(1+q^n) = \sqrt{\frac{2\theta_2(\tau)}{\eta(\tau)}}\frac{q^{\frac{P^2}{2}}}{\eta(\tau)} \cr
\mathrm{Tr}_{R} (-1)^Fq^{L_0 - \frac{c}{24}}& =& \chi^-_{P(R)} (\tau) = 0,
\end{eqnarray}
where we have introduced the following $\theta$ functions
\begin{eqnarray}
\theta_2 &=& 2 q^{\frac{1}{8}}\prod_{n=1}^\infty (1-q^n)(1+q^n)^2 \cr
\theta_3 &=& \prod_{n=1}^\infty (1-q^n)(1+q^{n-\frac{1}{2}})^2 \cr 
\theta_4 &=& \prod_{n=1}^\infty (1-q^n)(1-q^{n-\frac{1}{2}})^2 \cr
\eta &=& q^{\frac{1}{24}} \prod_{n=1}^\infty (1-q^n).
\end{eqnarray}
The modular transformation is given by
\begin{eqnarray}
\theta_3(\tau + 1) = \theta_4(\tau) \cr
\theta_4(\tau + 1) = \theta_3(\tau) \cr
\theta_2(\tau + 1) = e^{\pi i/4} \theta_2(\tau) \cr
\eta(\tau + 1) = e^{\pi i/12} \eta(\tau) \label{eq:mot}
\end{eqnarray}
for the T-transformation, and 
\begin{eqnarray}
\theta_3(-1/\tau) &=& \sqrt{-i\tau} \theta_3(\tau) \cr
\theta_4(-1/\tau) &=& \sqrt{-i\tau} \theta_2(\tau) \cr
\theta_2(-1/\tau) &=& \sqrt{-i\tau} \theta_4(\tau) \cr
\eta(-1/\tau) &=& \sqrt{-i\tau} \eta(\tau), \label{eq:mos}
\end{eqnarray}
for the S-transformation.
Therefore the modular transformation for the super character is given by
\begin{eqnarray}
\chi_{P' (NS)}^+ (\tau) &=& \int_{-\infty}^\infty dP e^{2\pi i PP'} \chi^+_{P(NS)} (-1/\tau) \cr
\sqrt{2}\chi_{P'(NS)}^- (\tau) &=& \int_{-\infty}^\infty dP e^{2\pi i PP'} \chi^+_{P(R)} (-1/\tau) \cr
\frac{1}{\sqrt{2}}\chi_{P' (R)}^+ (\tau) &=& \int_{-\infty}^\infty dP e^{2\pi i PP'} \chi^-_{P(NS)} (-1/\tau) . \label{eq:modult}
\end{eqnarray}

For the application to the unoriented Liouville theory, the following modular transformation is needed,
\begin{eqnarray}
\eta\left(\frac{i}{4t} + \frac{1}{2}\right) &=& \sqrt{2t}\eta\left(it + \frac{1}{2}\right) \cr
\theta_3\left(\frac{i}{4t} + \frac{1}{2}\right) &=& \sqrt{2t}e^{\frac{\pi i}{4}}\theta_4\left(it + \frac{1}{2}\right) \cr
\theta_4\left(\frac{i}{4t} + \frac{1}{2}\right) &=& \sqrt{2t}e^{\frac{-\pi i}{4}}\theta_3\left(it + \frac{1}{2}\right). \label{eq:modun}
\end{eqnarray}
To obtain these, we have to perform the succeeding modular transformations $TSTTS$. We can show this explicitly by setting $t_4 = \frac{i}{4t} + \frac{1}{2}$; then
\begin{eqnarray}
t_3 &=& -\frac{1}{t_4} = -\frac{2it}{it-\frac{1}{2}} \cr
t_2 &=& t_3 + 2 = -\frac{1}{it-\frac{1}{2}} \cr
t_1 &=& -\frac{1}{t_2} = it - \frac{1}{2} \cr
t_0 &=& t_1 + 1 = it + \frac{1}{2}.
\end{eqnarray}
Therefore, the modular transformation becomes (we take $\eta$ for example)
\begin{eqnarray}
\eta(t_4) &=& \eta (-1/t_3) \cr
	  &=& \sqrt{-it_3}\eta(t_3) \cr
	&=& \sqrt{-i(t_2-2)}\eta(t_2-2) \cr
	&=& \sqrt{-i(t_2-2)}e^{-\frac{i\pi}{6}}\eta(t_2) \cr
	&=& \sqrt{-i(-\frac{1}{t_1} - 2)}e^{-\frac{i\pi}{6}} \sqrt{-it_1}\eta(t_1) \cr
	&=& \sqrt{1+2t_1} e^{-\frac{i\pi}{6}}\eta(t_1)\cr
	&=& \sqrt{2t_0 -1} e^{-\frac{i\pi}{4}}\eta(t_0) \cr
	&=& \sqrt{2t} \eta(it + \frac{1}{2}).
\end{eqnarray}
Similarly we can obtain the transformation for $\theta$.

For the $\mathcal{N} = 2$ theory, we use the following theta functions with characteristics,
\begin{eqnarray}
\theta_1(\nu,\tau) &=& i\sum_{n=-\infty}^\infty (-1)^n q^{(n-1/2)^2/2} z^{n-1/2} = 2\exp(\pi i\tau/4)\sin\pi\nu \prod_{m=1}^\infty(1-q^m)(1-zq^m)(1-z^{-1}q^m) \cr\theta_2(\nu,\tau) &=& \sum_{n=-\infty}^\infty q^{(n-1/2)^2/2} z^{n-1/2} = 2\exp(\pi i\tau/4)\cos\pi\nu \prod_{m=1}^\infty(1-q^m)(1+zq^m)(1+z^{-1}q^m) \cr
\theta_3(\nu,\tau) &=& \sum_{n=-\infty}^\infty q^{n^2/2}z^n = \prod_{m=1}^\infty(1-q^m)(1+zq^{m-1/2})(1+z^{-1}q^{m-1/2}) \cr
\theta_4(\nu,\tau) &=& \sum_{n=-\infty}^\infty (-1)^n q^{n^2/2} z^{n} =\prod_{m=1}^\infty(1-q^m)(1-zq^{m-1/2})(1-z^{-1}q^{m-1/2}),
\end{eqnarray}
where $z = \exp(2\pi i\nu)$. The modular transformation of these functions are the same as we have in the case with $\nu = 0$ \eqref{eq:mot},\eqref{eq:mos} except the factor $\exp(\pi i\nu^2/\tau)$ in the S-transformation.

Let us finally list the general channel duality property of the Ishibashi states we use in the main text. The easiest way to derive the following results is to calculate directly those quantities by using the free field Ishibashi states. The left-hand side is the tree exchange channel and the right-hand side is the loop channel which can be obtained by the modular transformation: $s =1/t$ for cylinder $s=1/2t$ for the Klein bottle and $s=1/4t$ for the M\"obius strip.

\begin{eqnarray}
\langle B \pm| e^{-\pi Ht} |B \pm\rangle_{RR} &\to& NS; \ (-1)^f \cr
\langle B \pm| e^{-\pi Ht} |B \mp\rangle_{RR} &\to& R; \ (-1)^f \cr
\langle B \pm| e^{-\pi Ht} |B \pm\rangle_{NSNS} &\to& NS; \ 1 \cr
\langle B \pm| e^{-\pi Ht} |B \mp\rangle_{NSNS} &\to& R; \ 1 
\end{eqnarray}
for the cylinder.
\begin{eqnarray}
\langle C \pm| e^{-\pi Ht} |C \pm\rangle_{RR} &\to& NS-NS; \ ((-1)^f+(-1)^{\tilde{f}})\cdot\Omega \cr
\langle C \pm| e^{-\pi Ht} |C \mp\rangle_{RR} &\to& R-R; \ ((-1)^f+(-1)^{\tilde{f}})\cdot\Omega \cr
\langle C \pm| e^{-\pi Ht} |C \pm\rangle_{NSNS} &\to& NS-NS; \ (1+(-1)^{f+\tilde{f}})\cdot\Omega \cr
\langle C \pm| e^{-\pi Ht} |C \mp\rangle_{NSNS} &\to& R-R; \ (1+(-1)^{f+\tilde{f}})\cdot\Omega 
\end{eqnarray}
for the Klein bottle. 
\begin{eqnarray}
\langle B \pm| e^{-\pi Ht} |C \pm\rangle_{RR} &\to& R; \ (-1)^f\cdot\Omega \cr
\langle B \pm| e^{-\pi Ht} |C \mp\rangle_{RR} &\to& R; \ 1 \cdot\Omega \cr
\langle B \pm| e^{-\pi Ht} |C \pm\rangle_{NSNS} &\to& NS; \ (-1)^f \cdot\Omega \cr
\langle B \pm| e^{-\pi Ht} |C \mp\rangle_{NSNS} &\to& NS; \ 1 \cdot \Omega
\end{eqnarray}
for the M\"obius strip (up to a phase factor).
\subsection{Conventions for Quaternionic Matrix}\label{a-5}
In this appendix we review our conventions of quaternionic matrices. We follow the conventions used in Mehta's book \cite{Mehta}.
First we begin with the convention for one quaternion. 

We define a quaternion $q$ as a $2\times 2$ matrix which can be written as 
\begin{equation}
q = q^{(0)} + q^{(i)}e_i,
\end{equation}
where $i=1,2,3$ and coefficients $q$ are complex numbers in general. The elements of quaternion $e_i$ are realized by the Pauli matrices as $e_i = i\sigma_i$. We call a quaternion is real if all $q^{0},q^{1},q^{2},q^{3}$ are real numbers. In the main text, we exclusively use real quaternions, so we sometimes omit the adjective ``real". It is important to distinguish the following three possible conjugations of quaternions.
\begin{eqnarray}
\bar{q} &\equiv& q^{0} - q^{(i)}e_i \cr
q^* &\equiv& q^{0*} + q^{(i)*}e_i \cr
q^\dagger &\equiv& q^{0*} - q^{(i)*}e_i,
\end{eqnarray}
which are called ``quaternionic conjugation", ``complex conjugation" and ``Hermitian conjugation" respectively. We say that a quaternion is Hermitian when $q^\dagger = q$ holds. 

Let us now go on to a quaternionic matrix, which is defined as an $N \times N$ matrix with quaternions as each element (hence $2N \times 2N$ matrix). In terms of the quaternion language, some of the conjugation properties become
\begin{equation}
(Q^T)_{kj} = -e_2 \bar{q}_{jk} e_2,
\end{equation}
for the transposed matrix, where $k,j$ runs from $1$ to $N$ and represents the quaternionic indices. The Hermitian conjugation is
\begin{equation}
(Q^\dagger)_{kj} = q^\dagger_{jk}.
\end{equation}
Note that $\dagger$ is acted only on the single quaternion element $q_{ij}$, and does not exchange the indices $ij$.
Finally the ``time reversal"\footnote{The notion of time reversal comes from a quantum mechanical transformation of operators which act on the time reversed spinors. See any good text book on the quantum mechanics (e.g. \cite{LL}).} is given by
\begin{equation}
(Q^R)_{kj} = \bar{q}_{jk} = e_2(Q^T)_{kj} e_2^{-1}.
\end{equation}
We say that a quaternionic matrix is selfdual when it satisfies $Q^R = Q$.

In the actual application to the unoriented string theory, we encounter the following matrix condition (see appendix \ref{b-9}):
\begin{eqnarray}
e_2 Q^T e_2 &=& - Q \cr
Q^\dagger &=& Q,
\end{eqnarray}
It is easy to see that $Q$ is a quaternionic matrix, and from the definition we have
\begin{equation}
[Q^R]_{kj} = \bar{q}_{jk} = q_{kj} = q^\dagger_{jk},
\end{equation}
which is a real selfdual quaternionic matrix. We rephrase this as ``an Hermitian matrix $Q$ which is skew antisymmetric, namely $JQ = -(JQ)^T$, is a real selfdual quaternionic matrix". It is also easy to see that a real selfdual quaternionic matrix can be decomposed as 
\begin{equation}
Q_{jk} = S_{jk} + A_{jk}^{(i)} e_i,
\end{equation}
where $S$ is real symmetric and $A$ is real antisymmetric. This decomposition is necessary to understand the Jacobian of the diagonalization and the Feynman rule for the $Sp$ matrix model.

\sectiono{Appendices II: Miscellaneous Topics}\label{sec:B}
\subsection{Super Liouville Theory from 2D Supergravity}\label{b-1}
We introduce the super Liouville theory by quantizing the 2D supergravity \cite{Distler:1990nt}.\footnote{We follow the notation of D'Hoker-Phong \cite{D'Hoker:1988ta} in this section.}
The notation for the $(1,1)$ supergravity is as follows. We use the (super)coordinate $\mathbf{z}^M = (z,\bar{z};\theta,\bar{\theta})$ and the (super)differential $\partial_M = (\partial_z,\partial_{\bar{z}};\partial_\theta,\partial_{\bar{\theta}})$. The tangent space indices are written as $A= (\xi,\bar{\xi};+,-)$. As is usual in the supergravity, the local Lorentz transformation is restricted to the $SO(1,1)$ subgroup of the more general tangent bundle (super)rotation.\footnote{That is to say, $\theta$ and $z$ do not mix each other.} Thus every tensor transforms with the $U(1)$ charge under the local Lorentz transformation. The covariant derivative for the Lorentz tensor is given by
\begin{eqnarray}
\mathcal{D}_M V &=& \partial_M V + in\Omega_M V \cr
\mathcal{D}_M V_A &=& \partial_M V_A + \Omega_M E_A^{\ B} V_B \cr
\mathcal{D}_M V^A &=& \partial_M V^A -(-1)^{mb}V^BE_B^{\ A} \Omega_M,
\end{eqnarray}
where we have set the $U(1)$ charge of the scalar to be $n$ and introduced the generator of the Lorentz transformation as $ E_a^{\ b} = \epsilon_{a}^{\ b}$, $E_a^{\ \beta} = E_\alpha^{\ b} = 0$, $E_{\alpha}^{\ \beta} = \frac{1}{2}(\gamma_5)_\alpha^{\ \beta}$.

When we fix the genus of the world sheet, the partition function of the string theory with the local supersymmetry on the world sheet can be schematically written as 
\begin{equation}
Z_{\chi} = \int[\mathcal{D}E_M^{\ A}][\mathcal{D}X^I] e^{-S[X,E]} \label{eq:SGA},
\end{equation}
where the matter part is arbitrary, but for definiteness, we take the supersymmetric free boson-fermion system
\begin{eqnarray}
S &=& \frac{1}{8\pi}\int d^2{\mathbf{z}} E \mathcal{D}^\alpha X^I\mathcal{D}_\alpha X^I \cr
&=& \frac{1}{4\pi}\int d^2{\mathbf{z}} E \mathcal{D}_{-}X^I\mathcal{D}_+ X^I. 
\end{eqnarray}

Before performing the path integral, we review the correspondence with the component representation, though it is not necessary in the following calculation.
The matter field is decomposed as 
\begin{equation}
X^I = x^I +\theta^\alpha \psi^I_\alpha + i\theta\bar{\theta}F^{I}.
\end{equation}
$\psi^I$ is ``Majorana".\footnote{Since we are in the Euclidean signature, the Weyl spinor cannot be simultaneously Majorana. However, if we Wick rotate to the Minkowski space, $\theta^\alpha\psi_\alpha = \theta^+\psi_++\theta^{-}\psi_{-}$. Though the complex conjugation of $\theta^+$ is $\theta^{-}$ in the Euclidean signature, we can think $\theta^+$ is real in the Minkowski signature.} We next consider the component representation of the supervielbein. To do so, it is natural to do in the Wess-Zumino gauge. In general, the two dimensional supervielbein can be removed (locally) except one superfield freedom by the gauge transformation as we will see later. However, it is convenient, because of the analogy to the four dimensional supergravity, to use the Wess-Zumino gauge where only the algebraically removable components are eliminated. Though we leave the details to the reference \cite{D'Hoker:1988ta}, the idea is that we fix the gauge by using the freedom of the superdiffeomorphism and the super Lorentz transformation whose transformation parameter is proportional to $\theta$. Consequently the residual gauge freedoms are the conventional (bosonic) diffeomorphism and the local supersymmetry from the superdiffeomorphism, and the local Lorentz transformation from the super Lorentz transformation.\footnote{As we have stated above, super Lorentz transformation does not have a transformation which mixes boson and fermion from the beginning.}.

This gauge fixing yields the following component expansion:
\begin{eqnarray}
E_m^{\ a} &=& e_m^{\ a}+ \theta\gamma^a\chi_m -\frac{i}{2}\theta\bar{\theta}e_m^{\ a}A \cr
E_m^{\ \alpha} &=& -\frac{1}{2}\chi_m^{\ \alpha}-\frac{i}{2}\theta^{\beta}(\gamma_m)_\beta^{\ \alpha}A - \frac{1}{2} \theta^\beta(\gamma_5)^{\ \alpha}_\beta\omega_m+i\theta\bar\theta\left[\frac{1}{2}(\gamma_m)^{\alpha\beta}\Lambda_\beta-\frac{3}{8}\chi_m^{\ \alpha} A\right] \cr
E_\mu^{\ a} &=& (\gamma^a)_\mu^\beta\theta_\beta \cr
E_\mu^{\ \alpha} &=& \delta_\mu^\alpha(1+i\theta\bar{\theta}A/4) \cr
\mathrm{Sdet}E_{M}^{\ A} &=&  e\left(1+\frac{1}{2}\theta\gamma^n\chi_n-\frac{i}{2}\theta\bar{\theta} A+\frac{1}{8}\theta\bar{\theta}\epsilon^{mn}\chi_m\gamma_5\chi_n\right) \cr
\omega_m &=& -e_m^{\ a} \epsilon^{pq} \partial_pe_q^{\ a}-\frac{1}{2}\chi_m \gamma_5 \gamma^p \chi_p \cr
\Lambda &=& -i\gamma_5\epsilon^{mn}D_m\chi_n-\frac{1}{2}\gamma^m\chi_m A, 
\end{eqnarray}
In addition, the super Euler term becomes
\begin{equation}
\chi(M) = \frac{i}{2\pi} \int d^2\mathbf{z} E R_{+-} = \frac{1}{2}\int d^2z \sqrt{g} R ,
\end{equation}
in this gauge. The integration measure is given by
\begin{equation}
 d^2\mathbf{z} E= d^2z d\theta d\bar{\theta}\mathrm{Sdet} E_M^{\ A}. 
\end{equation}
Note that in the Wess-Zumino gauge, we cannot distinguish whether the $\theta$ component is Einstein or Lorentz, owing to the structure of the above supervielbein (usually we set $A=0$).

For a reference, we present the action for the matter field in this gauge after we eliminate the auxiliary field
\begin{equation}
S[X] = \frac{1}{4\pi} \int d^2z \sqrt{g}\left[\frac{1}{2}g^{mn}\partial_mx^I\partial_nx^I +\psi^I\gamma^m\partial_m\psi^I-\psi^I\gamma^a\gamma^m\chi_a\partial_mx^I-\frac{1}{4}\psi^I\gamma^a\gamma^b\chi_a(\chi_b\psi^I)\right] 
\end{equation}
We should notice that the spin connection does not appear in the kinetic term of the 2D spinor.

Let us turn back to the quantum supergravity. Since the matter field in (\ref{eq:SGA}) is Gaussian, the path integration can be performed explicitly. Taking the number of scalar to be $d$, we obtain
\begin{equation}
e^{-S_m[E]} = V\left[\frac{8\pi^2\mathrm{Sdet}'(\square)_E}{\int d^2\mathbf{z} E}\right]^{-\frac{d}{2}},
\end{equation}
where $\square = \mathcal{D}^\alpha\mathcal{D}_\alpha$. This effective action is not invariant under the following super Weyl transformation:
\begin{eqnarray}
E_M^a &=& e^\Phi \hat{E}_M^a \cr
E_M^\alpha &=& e^{\Phi/2}[\hat{E}^\alpha_M +\hat{E}^a_M(\gamma_a)^{\alpha\beta}\hat{\mathcal{D}}_\beta\Phi].
\end{eqnarray}
We abbreviate this transformation as $E_M^A = e^\Phi \hat{E}_M^A $.\footnote{Let us briefly explain the nature of the additional second term. The supervielbein should have its torsion tensor satisfy the definite constraint conditions. However, the arbitrary transformation of the field ruins this condition, so the second term is necessary to hold this constraint. By the way, in the usual Riemannian geometry, the constraint on the torsion tensor just yields the formula from which we can make the connection by the vielbein. In contrast, since the supergravity restricts its connection in the $U(1)$ subgroup as we have stated above, the constraint on the torsion does restrict the supervielbein. This is one of the most important points of the superfield formalism of the supergravity.}

The claim here is that the variation of the effective action under this super Weyl transformation is given by
\begin{equation}
\log\left(\frac{\mathrm{Sdet}'\square}{\mathrm{Sdet}'\hat{\square}}\right) = c - S_{SL}(\Phi), 
\end{equation}
where $c$ is related to the zero-mode which we will not delve into here. $S_{SL}(\Phi)$ is the unrenormalized super Liouville action:
\begin{equation}
S_{SL} = \frac{1}{4\pi}\int d^2 \mathbf{z} \hat{E} (\hat{\mathcal{D}}_+\Phi\hat{\mathcal{D}}_-\Phi - i\hat{R}_{+-}\Phi).
\end{equation}

To calculate this variation, we use the super heat-kernel method. To regularize the expression, we first define\footnote{The reason why $\square$ is squared is to ensure the positivity of the norm in the superspace.}
\begin{equation}
\log \delta(s) \equiv \log \mathrm{Sdet}[(\square)^2+s] = -\int_\epsilon^\infty\frac{dt}{t}e^{-ts}\mathrm{STr} e^{-t(\square)^2}.
\end{equation}
The Weyl variation of this expression is given by
\begin{equation}
\delta\log\delta(s) = 2\int_\epsilon^\infty dt e^{-ts} \mathrm{STr}\left[(\delta\square)\square e^{-t(\square)^2}\right].
\end{equation}
Using the relation $\delta \square = -\delta\Phi \square $, we can rewrite this as
\begin{equation}
\mathrm{STr}\left[(\delta\square)\square e^{-t(\square)^2}\right] = -\left[(\delta\Phi)\square^2 e^{-t(\square)^2}\right].
\end{equation}
We further transform this as follows (the second line is the partial integration),
\begin{eqnarray}
\delta\log\delta(s) &=&2\int_\epsilon^\infty e^{-ts} \frac{\partial}{\partial t} \mathrm{STr} \delta\Phi e^{-t\square^2} \cr
&=& 2e^{-ts}\mathrm{STr}[\delta\Phi e^{-t\square^2}]|^\infty_\epsilon + 2s\int_\epsilon^\infty dt e^{-ts}\mathrm{STr} [\delta\Phi e^{-t\square^2}]. 
\end{eqnarray}
If we take the $s\to 0$ limit, the second term almost vanishes.\footnote{Only the zero-modes of $\square^2$, which we do not consider here, contribute. In the final expression, these become the supermoduli integrations.} The first term leaves just the $\epsilon$ part. Thus we have
\begin{equation}
\delta\log\mathrm{Sdet}'\square = - \mathrm{STr}[ \delta\Phi e^{-\epsilon \square^2}] + C.
\end{equation}
Evaluating the supertrace by the heat-kernel method, we can show
\begin{equation}
\mathrm{STr} \delta\Phi e^{-\epsilon \square^2} = -i\frac{1}{4\pi} \int d^2\mathbf{z} E R_{+-} \delta\Phi. 
\end{equation}
Note that as opposed to the bosonic case, we do not have a $\epsilon^{-1}$ divergence because of the supersymmetry. Namely, there is no renormalization of the ``cosmological constant" at this step. Integrating this, we obtain
\begin{equation}
\log\left(\frac{\mathrm{Sdet}'\square}{\mathrm{Sdet}'\hat{\square}}\right) = c - S_{SL}(\Phi).
\end{equation}
As a result, we can show $S_m[E] \to S_m[e^\Phi E] = S_m[E] - \frac{2}{3}c_m S_{SL} [\Phi ,E]$, where $c_m = \frac{3d}{2}$.

Going on to the quantization of the 2D supergravity, we fix the gauge of $E_M^{\ A}$. In the superconformal gauge, we fix it by demanding
\begin{equation}
f^*(E) = e^{\Phi}\hat{E} (\Upsilon)
\end{equation}
symbolically. In this gauge, the variation of the supervielbein $\mathcal{D}E_M^{\ A}$ is decomposed into (after the torsion constraint) the superdiffeomorphism $V$, the super local Lorentz transformation $L$, the super Weyl transformation $\Phi$ and the moduli  $\Upsilon$. The Jacobian of this transformation is given by the Fadeev-Popov superdeterminant. After dividing by the volume of the gauge group, this is given by
\begin{equation}
\int[\mathcal{D} E_M^{\ A}] = \int\{ d\Upsilon\} [\mathcal{D}_E \Phi] \mathcal{D}_E B \mathcal{D}_E C e^{-S_{gh}[B,C,E]}, 
\end{equation}
where, using $B(\mathbf{z}) = \beta(z) +\theta b(z)$, $C(\mathbf{z}) = c(z) +\theta \gamma(z)$, we can write the action of the superghost system as
\begin{equation}
S_{gh} = \frac{1}{2\pi} \int d^2\mathbf{z} E (B\mathcal{D}_-C+\bar{B}\mathcal{D}_+\bar{C}).
\end{equation}
We have set here the Lorentz $U(1)$ charge of $B$ to be $3/2$ and that of $C$ to be $-1$. The path integral of the superghost is also Gaussian, so it can be carried out, which yields the further Weyl anomaly. The calculation is almost the same \cite{D'Hoker:1988ta}:
\begin{equation}
S_{gh}[E] \to S_{gh}[e^\Phi E] = S_{gh}[E] - \frac{2}{3}c_{gh} S_{SL} [\Phi ,E]
,
\end{equation}
where $c_{gh} = -15$.

Combining all these contributions, the final path integral of the 2D supergravity in this superconformal gauge becomes
\begin{equation}
Z = \int{d\Upsilon} [\mathcal{D}_E \Phi] e^{-S_m[e^\Phi \hat{E}] -S_{gh}[e^\Phi \hat{E}]}.
\end{equation}
If we have $d=10$, the $\Phi$ integration simply yields a constant, so we can safely discard it (critical superstring). For the noncritical dimension, we have to evaluate this integral.

However, note the functional measure of the bare super Liouville field. Originally, the measure has been given by
\begin{equation}
||\delta\Phi ||^2_E = \int d^2 \mathbf{z} E (\delta\Phi)^2 = \int d^2 \mathbf{z} \hat{E} e^{\Phi} (\delta\Phi)^2,
\end{equation}
in order to preserve the superdiffeomorphism of the functional measure. Since this measure is neither Gaussian nor invariant under the translation in the functional space, it is very inconvenient. Thus we would like to transform it into the conventional form
\begin{equation}
|| \delta\Phi||^2_{\hat{E}} = \int d^2 \mathbf{z} \hat{E} (\delta\Phi)^2.
\end{equation}
We should obtain the Jacobian of this transformation and include it in the action. Since it is difficult to obtain this Jacobian from the first principle, we will find the consistent action by imposing ``locality", ``diffeomorphism invariance", ``renormalizability" and ``superconformal invariance". We assume its form as the renormalized super Liouville theory.\footnote{See \cite{D'Hoker:1990md,D'Hoker:1991ac} for a justification of this assumption from the functional integration.} We can see that this assumption is valid as long as $d \le 1$, just as in the bosonic string theory. The action is
\begin{equation}
S[\Phi,\hat{E}] = \frac{1}{4\pi} \int d^2\mathbf{z} \hat{E} \left(\frac{1}{2}\hat{\mathcal{D}}^\alpha \Phi \hat{\mathcal{D}}_\alpha + Q \hat{R}_{+-} \Phi+2i\mu e^{b \Phi}\right),
\end{equation}
and its component form is
\begin{equation}
S = \frac{1}{2\pi} \int d^2 z\left( \partial\phi\bar{\partial}\phi + \frac{1}{4} Q \hat{R} \phi + \bar{\psi}\partial\bar{\psi}+\psi\bar{\partial}\psi -F^2-2\mu bFe^{b\phi} +2i\mu b^2\psi\bar{\psi}e^{b\phi}\right).
\end{equation}
We would like to determine $Q$ and $b$. First, we recall that the decomposition between $E$ and $\hat{E}$ is arbitrary. Therefore, the theory should be invariant under $\hat{E} \to e^\sigma \hat{E}$ and $\Phi \to \Phi - \frac{\sigma}{b}$. The latter transformation is trivially invariant from the new definition of the measure. For the former to be the symmetry of the theory, the central charge of the theory should be zero.\footnote{In other words, the super Liouville theory should be invariant under the super Weyl transformation. In the quantization procedure, we should preserve the superconformal symmetry, which removes the arbitrariness of the quantization.}
\begin{equation}
c_{tot} = c_m +c_{gh} + c_{\Phi} = 0,
\end{equation}
which results in $c_\Phi = \frac{3}{2}(10-d)$. We can show that the central charge of $\Phi$ is independent of $\mu$ from the canonical quantization or from the discussion we will see below, so it is given by the super Coulomb gas result: $c_\Phi = \frac{3}{2} + 3Q^2$. Therefore, we have
\begin{equation}
 Q = \sqrt{\frac{9-d}{2}}.
\end{equation}

For this theory to be consistent as the superconformal theory, it is necessary that the ``interaction term" $e^{b\Phi}$ is the $(1/2,1/2)$ tensor. From the super Coulomb gas method, the weight is calculated as 
\begin{equation}
\Delta (e^{b \Phi}) = -\frac{1}{2}b(b-Q) = \frac{1}{2}.
\end{equation}
Hence, we obtain the famous formula $Q = b+b^{-1}$. Furthermore, in order to keep $b$ real (or this is equivalent to maintain the metric real), it is necessary to have $c_m \le \frac{3}{2}$.

\subsection{Coulomb Gas System}\label{b-2}
In this appendix we discuss the calculation method for the correlation function of the Coulomb gas (linear dilaton) system on the sphere. The Coulomb gas theory \cite{Feigin:1988se} is a close cousin of the Liouville theory and this calculation plays an important role in the evaluation of the Liouville correlation function.
\footnote{This appendix is just the problem 6.2 of Polchinski's text book \cite{Polchinski:1998rq}.} We consider the generating function of the correlation function
\begin{equation}
Z[J] = \langle \exp(i\int d^2\sigma J(\sigma)X(\sigma))\rangle_{Q,free}.
\end{equation}
The action is given by
\begin{equation}
S = \frac{1}{4\pi} \int d^2\sigma \sqrt{g} (g^{ab}\partial_aX\partial_bX+QRX).
\end{equation}
We expand $X$ in the eigenfunction of the Laplacian as follows
\begin{eqnarray}
X(\sigma) &=& \sum_I x_I X_I(\sigma) \cr
\nabla^2 X_I &=& - \omega_I^2 X_I \cr
\int d^2 \sigma \sqrt{g} X_I X_{I'} &=& \delta_{II'}.
\end{eqnarray}

For the world sheet metric, we take the conformal gauge which is given by
\begin{eqnarray}
 g_{ab} &=& e^{2\omega(\sigma)} \delta_{ab} \cr
 \sqrt{g} R &=& -2\nabla^2 \omega(\sigma). 
\end{eqnarray}
Then the generating function can be written as
\begin{equation}
Z(J) = \prod_I \int dx_I \exp\left(-\frac{\omega_Ix_Ix_I}{4\pi} + i x_IJ_I + \frac{QR_Ix_I}{4\pi}\right),
\end{equation}
where $J_I$ and $R_I$ are defined as
\begin{eqnarray}
J_I &=& \int d^2\sigma J(\sigma)X_I(\sigma), \cr
R_I &=& \int d^2\sigma \sqrt{g} R(\sigma) X_I(\sigma).
\end{eqnarray}
First, we evaluate the zero-mode integration. On the sphere there is only one zero-mode which is a constant: $X_0=(\int d^2\sigma \sqrt{g})^{1/2} $, so the integration over the zero-mode only yields the delta-function $2\pi\delta(J_0 - i2Q)$, where the Gauss-Bonnet theorem is used. Since the non-zero modes are Gaussian integral, it can be easily done,
\begin{eqnarray}
Z[J] &=& 2\pi\delta(J_0-i2Q) \left(\det\frac{-\nabla^2}{4\pi^2}\right)^{-1/2} \cr
&\times& \exp\left[-\frac{1}{2}\int d^2\sigma d^2\sigma'\left(J(\sigma)-\frac{QiR(\sigma)}{4\pi}\right)G'(\sigma,\sigma')\left(J(\sigma')-\frac{QiR(\sigma')}{4\pi}\right)\right],\label{eq:exp}
\end{eqnarray}
where we have introduced the Green function on the sphere which is given by
\begin{equation}
G'(\sigma_1,\sigma_2) = -\frac{1}{2}\log|z_{12}|^2 + f(z_1,\bar{z}_1) + f(z_2,\bar{z}_2),
\end{equation}
\begin{equation}
f(z,\bar{z}) = \frac{X_0^2}{2} \int d^2 z' \exp(2\omega(z'))\log|z-z'|^2 + k, 
\end{equation}
where $k$ is a certain constant which we will not need in the following.
The normalization of the Green function is 
\begin{equation}
\frac{1}{2\pi}\nabla^2 G'(\sigma,\sigma') = \frac{1}{\sqrt{g}}\delta(\sigma-\sigma') -X_0^2.
\end{equation}

Now let us calculate the tachyon amplitude 
\begin{equation}
\langle e^{ik_1X_1} \cdots e^{ik_N X_N} \rangle_{Q,free},
\end{equation}
by substituting  $J(\sigma) = \sum_{i=1}^N k_i \delta^2(\sigma-\sigma_i)$.
The argument of $\exp$ in (\ref{eq:exp}) becomes
\begin{eqnarray}
& &\exp\left[- \sum_{i<j} k_ik_j G'(\sigma_i,\sigma_j) - \frac{1}{2}\sum_{i=1}^N k_i^2 G'_r(\sigma_i,\sigma_i)\right] \cr
&\times &\exp\left[\int d^2\sigma \sum_j \frac{QiR(\sigma)}{4\pi} G'(\sigma,\sigma_j)k_j\right] \cr
&\times &\exp\left[\frac{1}{2}\int d^2\sigma d^2\sigma' Q^2\frac{R(\sigma)R(\sigma')}{16\pi^2} G'(\sigma,\sigma')\right].
\end{eqnarray}
We note that the renormalization scheme used in Polchinski's textbook \cite{Polchinski:1998rq} states
\begin{equation}
G'_r(\sigma,\sigma) = 2f(z,\bar{z}) +\omega(z,\bar{z}). 
\end{equation}
Furthermore, substituting $R=-2\nabla^2\omega$, we find some of the Green functions cancel, and we obtain
\begin{eqnarray}
& & \exp\left(-\frac{1}{2}\sum_i k_i^2 \omega(\sigma_i)\right)\prod_{i<j} |z_{ij}|^{k_ik_j}\cr
&\times & \exp\left(Q i \sum_j \omega(\sigma_j)k_j\right) \cr
&\times & \exp\left[-\frac{1}{4\pi} \int d^2\sigma Q^2 \omega(\sigma) \nabla^2 \omega(\sigma)\right].
\end{eqnarray}

Collecting all of these, we finally obtain
\begin{eqnarray}
\langle e^{ik_1X_1} \cdots e^{ik_N X_N} \rangle_{Q,free} &=& C_{\mathrm{S}^2}^Z 2\pi \delta\left(\sum_{j}k_j - i2Q\right) \exp\left(-\frac{1}{2}\sum_j k_j(k_j-i2Q) \omega(\sigma_j)\right) \cr
&\times& \prod_{i<j}|z_{ij}|^{k_ik_j} \exp\left[-\frac{1}{4}\int d^2\sigma Q^2\omega(\sigma)\nabla^2\omega(\sigma)\right].
\end{eqnarray}
The term proportional to $\exp(\omega)$ shows the conformal anomaly of the inserted operators (which means we have calculated the weight of the $\exp$ vertex operator). When these are BRST invariant, they should vanish by the integration or other contributions. The last term is the Liouville action from the difference of the conformal anomaly between the free field and the Coulomb gas. The conformal anomaly of the free field is contained in $C_{\mathrm{S}^2}^X$. We can read the central charge of the Coulomb gas system from this. In Polchinski's notation \cite{Polchinski:1998rq}, the central charge is related to the Weyl anomaly as follows
$$ \exp\left(-\frac{c}{24\pi}\int d^2\sigma \omega\nabla^2\omega\right).$$
Therefore we have derived the formula
\begin{equation}
c_{Q} = 1 + 6Q^2.
\end{equation}

\subsection{Asymptotic Expansion of the Liouville Partition Function}\label{b-3}
The density of states of the $c=1$ Liouville theory is given by
\begin{equation}
\frac{\partial \rho}{\partial (\beta \mu)} = \frac{1}{\pi} \mathrm{Im} \int_0^\infty dT e^{-i\beta\mu T} \frac{T/2}{\sinh(T/2)}.
\end{equation}
In this appendix, we derive the asymptotic expansion of this quantity in terms of $(\beta\mu)^{-1}$ \cite{Gross:1990ay}. First of all, we integrate this quantity with respect to $\mu$ and we find
\begin{equation}
\rho = \frac{1}{\pi}\left[\log\left(\frac{1}{2}\beta\right)-\mathrm{Re}\psi\left(\frac{1}{2}(1+2i\beta\mu)\right)\right], \label{eq:fden}
\end{equation}
where the integration constant has been determined from the genus 0 expression. This can be verified by differentiating it and using the integral formulae which are collected in appendix \ref{a-3}. This can be further rewritten as the $\zeta$ function,
\begin{eqnarray}
\rho &=& \frac{1}{\pi} \mathrm{Re}[\zeta(1,(1+2i\beta\mu)/2)-\infty] \cr
 &=& -\frac{1}{\pi}\log\mu + \frac{1}{\pi\beta\mu} \mathrm{Im} \sum_{n=0}^\infty \frac{1}{1-i(2n+1)/2\beta\mu}.
\end{eqnarray}
This formula can be intuitively understood as follows. The density of states for the harmonic oscillator is given by $\rho(\mu) = \frac{1}{\pi} \mathrm{Im}\sum_n\frac{1}{(n+1/2)\omega-\beta\mu-i\epsilon}$, so we substitute $\omega = -i$ into this expression. Then we heuristically obtain the above formula (up to $\log$ term).

Now, we expand this formula as follows, though it is not valid in a rigorous sense (but we should recall we attempt to derive the asymptotic formula),
\begin{eqnarray}
\rho &=& -\frac{1}{\pi}\log\mu +\frac{1}{\pi\beta\mu} \mathrm{Im} \sum_{n=0}^\infty \sum_{k=0}^\infty i^k \left(\frac{2n+1}{2\beta\mu}\right)^k \cr
&=& \frac{1}{\pi} \left[-\log{\mu} + 2\sum_{k=1}^\infty (-1)^k (2\beta\mu)^{-2k}(2^{2k-1}-1)\zeta(1-2k)\right].
\end{eqnarray}
In the second line, we have changed the order of the infinite summation and substituted $\sum_n(2n+1)^{2k-1} = (2^{2k-1}-1)\zeta(1-2k)$ formally.
If we substitute the formula \eqref{eq:berz}, which relates the Bernoulli number to $\zeta(1-2k)$, into the above expression, we finally obtain the desired asymptotic expansion
\begin{equation}
\rho(\mu) = \frac{1}{\pi}\left[-\log\mu + \sum_{m=1}^\infty (2^{2m-1}-1)\frac{|B_{2m}|}{m}(2\beta\mu)^{-2m}\right].
\end{equation}

\subsection{Exact Marginality of Boundary Sine-Gordon Theory}\label{b-5}
In this appendix we discuss the exactly marginal perturbation in the boundary conformal field theory by taking the boundary Sine-Gordon theory \cite{Callan:1994mw} for example. The necessary and sufficient condition for the exact marginality of the perturbation is that the dimension of the operator is one and it is self-local \cite{Recknagel:1998ih}, \cite{Alekseev:2000fd}.

Here, we briefly review the procedure to construct the BCFT for the self-local perturbation. Since it is given by the perturbation theory, we cannot apply this method directly to the general correlation functions of the boundary Liouville theory. However, the residue at the pole in the Liouville correlation functions can be calculated perturbatively as usual.
We take the upper half plane as the world sheet and write the perturbation as 
$\lambda\int_{-\infty}^\infty \psi \frac{dx}{2\pi} $. Naively, the perturbative expansion of the bulk correlation function is given by
\begin{equation}
\langle \phi_1(z_1,\bar{z}_1) \cdots \phi_N(z_N,\bar{z}_N) \rangle = \sum_n \frac{\lambda^n}{n!} \int_{-\infty}^\infty\frac{dx_1}{2\pi}\cdots \frac{dx_n}{2\pi} \langle \psi(x_1)\cdots\psi(x_n) \phi_1\cdots\phi_N\rangle_0.
\end{equation}
Since this is divergent from where $x$ approaches each other, we need to regularize the expression. When all the inserted operators are in the bulk, we can redefine the expression by shifting the contour of the $x$ integration as follows
\begin{equation}
\langle \phi_1(z_1,\bar{z}_1) \cdots \phi_N(z_N,\bar{z}_N) \rangle_r \equiv \lim_{\epsilon \to 0} \sum_n\frac{\lambda^n}{n!} \int_{\gamma_1}\cdots\int_{\gamma_n} \frac{dx_1}{2\pi}\cdots \frac{dx_n}{2\pi} \langle \psi(x_1)\cdots\psi(x_n) \phi_1\cdots\phi_N\rangle_0,
\end{equation}
where the $p$th contour $\gamma_p$ is given by the parallel line with respect to the real line, and its imaginary part is given by $\mathrm{Im} \gamma_p = i\epsilon/p$. This is manifestly finite and does not depend on $\epsilon$ (as long as $\epsilon$ is small enough). Note that it is necessary that $\psi(x)$ is self-local, namely
\begin{equation}
\psi(x_1) \psi(x_2) = \psi(x_2)\psi(x_1)
\end{equation}
or in the language of OPE,
\begin{equation}
\psi(x_1)\psi(x_2) = \frac{K}{(x_1-x_2)^2} + reg
\end{equation}
should be hold in order for this regularization to make sense (because of the existence of cuts etc). We have assumed the self-locality here.

In addition, the perturbative series for the correlation functions which include boundary operators can be defined similarly though the situation becomes slightly complicated. If the boundary operators are local with respect to the perturbation $\psi(x)$, the similar prescription works.
\begin{equation}
\langle \psi_1(u_1) \cdots \psi_N(u_N)\rangle \equiv \sum_n \frac{\lambda^n}{n!} \int_{\gamma_1}\cdots\int_{\gamma_n} \frac{dx_1}{2\pi} \cdots \frac{dx_n}{2\pi} \langle \psi(x_1)\cdots \psi(x_n) \tilde{\psi}_1\cdots\tilde{\psi}_N\rangle_0,
\end{equation}
where, with $C_i$ being the small circle around $u_i$, the renormalized operators have been defined as
\begin{equation}
\tilde{\psi}_i \equiv \sum_{n=0}^\infty \frac{\lambda^n}{2^nn!} \oint_{C_1} \frac{dx_1}{2\pi} \cdots \oint_{C_n} \frac{dx_n}{2\pi} \psi_i(u_i)\psi(x_n)\cdots\psi(x_1).
\end{equation}
Note $\psi_i$ and $\tilde{\psi}_i$ have the same dimension. The gist is that we have just taken the pole of the OPE between $\psi_i$ and $\psi$. As we have emphasized, this definition does not make sense unless $\psi_i$ is local with $\psi$.
This ``definition" of the perturbative BCFT by the above prescription enables us to prove that the perturbation is exactly marginal if the perturbing operator is self-local and its dimension is one. To do so, we substitute $\psi$ itself into the definition of the boundary correlation function. Then, $\tilde{\psi} = \psi$ from the self-locality. Furthermore, the contour integral vanishes because there are no poles on the upper-half plane. Therefore, there is no perturbative correction to the two-point function of $\psi$, which means the exact marginality of the perturbation in this renormalization prescription (of course, other correlation functions have perturbative corrections).

Finally if we apply this method to the boundary sine-Gordon theory, we can understand that the boundary sine-Gordon perturbation is an exactly marginal perturbation with respect to the free theory.
\subsection{Zero-mode Algebra for $\hat{c} = 1$ String Theory and Allowed Branes}\label{b-6}
The allowed brane in the $\hat{c} = 1$ theory has a little subtlety. This appendix is intended to show the author's current understanding of the subject.
Let us first review the allowed branes\footnote{``Allowed" here means the stable brane with R-R part. Of course ``unstable" branes always exist no matter how R zero-mode behaves.} in the free two dimensional theory. All the relevant consideration is zero-modes, so we concentrate on them. The zero-mode algebra\footnote{For simplicity we take the Euclidean signature.} is given by
\begin{equation}
\{\psi^i,\psi^j\} = \delta^{ij}, \ \ \ \{\bar{\psi}^i,\bar{\psi}^j\} = \delta^{ij}, 
\end{equation}
where $i,j = 0,1$. The left (right) realization of this algebra is done by the usual Clifford algebra
\begin{equation}
\psi^0 = \sigma^1, \ \ \ \psi^1 = \sigma^2, \ \ \  (-1)^{F_L} = \sigma^3.
\end{equation}
Combining left and right (with the proper cocycle) we have
\begin{eqnarray}
\psi^0 = \sigma^1_L \otimes 1, \ \ \ \bar{\psi}^0 = \sigma^3_L\otimes \sigma^1_R \cr
\psi^1 = \sigma^2_L \otimes 1, \ \ \ \bar{\psi}^1 = \sigma^3_L\otimes \sigma^2_R \cr
(-1)^{F_L+F_R} = \sigma^3_L\otimes \sigma^3_R.
\end{eqnarray}
This algebra is represented by the four dimensional tensor product of left and right two dimensional Clifford algebras.

The allowed states are as follows; in the type 0B theory we have $|++\rangle$ and $|--\rangle$, while in type 0A theory we have  $|+-\rangle$ and $|-+\rangle$

Then let us check the allowed branes in the theory. We always take $X^0$ to be Neumann\footnote{In the Dirichlet case, we have $(\psi^0 + i\eta \psi^0)|k\rangle$.}, so
\begin{equation}
(\psi^0 - i\eta \bar{\psi}^0)|k=0\rangle,
\end{equation}
where $\eta$ is related to the remaining superconformal current.
In case of the D1-brane, we have 
\begin{equation}
(\psi^1 - i\eta \bar{\psi}^1)|k=0\rangle.
\end{equation}
With the explicit realization of the algebra we have introduced, we should have, for the 0B theory,
\begin{eqnarray}
&&|++\rangle + i|--\rangle \ \ \ \ \mathrm{for} \ \ (\eta = 1) \cr
&&|++\rangle - i|--\rangle \ \ \ \ \mathrm{for} \ \ (\eta = -1).
\end{eqnarray}
On the other hand for the 0A theory, there is no solution.

In case of the D0-brane, the situation is reversed. For the 0A, we have two kinds of (electric and magnetic) $\eta =\pm$ brane, but none for the 0B.
This is just the well known fact about the allowed D-branes in the 0A/B theory.

Now let us go on to the super Liouville $\otimes$ $\hat{c} = 1$ theory. In this case, we have four zero-modes: matter part $\psi^0,\bar{\psi}^0$ and Liouville part $G_0$,$\bar{G_0}$. The Liouville part $G_0$ satisfies the following algebra (unless the energy is zero)
\begin{equation}
\{ G_0, G_0 \} = c.
\end{equation}
We rescale $G_0$ so that $c = 1$.\footnote{We here consider the nonzero energy Ishibashi states.} To classify the allowed Ishibashi states, we should realize these zero-mode algebras. At first sight, the algebra is just the same as in the free case considered above. However, mimicking the procedure above does not seem to work. This is perhaps because $(-1)^{F_L}$ and $(-1)^{F_R}$ are not separately conserved under the Liouville potential, so the $G_0$, $\psi_0$, $(-1)^{F_L}$ Clifford algebra actually does not make sense. As a consequence we have to combing left and right separately for each sector.\footnote{The same procedure is done in \cite{Douglas:2003up}.}

Thus, the realization of $G_0$ and $\bar{G}_0$ becomes
\begin{equation}
G_0 = \sigma^1_{Liouville} \ \ \ \ \ \bar{G_0} = \sigma^2_{Liouville} \ \ \ \ \ (-1)^{F_{Liouville}} = \sigma^3_{Liouville}
\end{equation}
and the $\psi^0$ algebra is given by 
\begin{equation}
\psi^0 = \sigma^1_M \ \ \ \ \ \bar{\psi}^0 = \sigma^2_M \ \ \ \ \ (-1)^{F_M} = \sigma^3_M
\end{equation}
Tensor products (with cocycles) make the whole zero-mode algebra of the theory.

Now we can consider the allowed branes (or Ishibashi states) in the two dimensional super Liouville background. We take the Neumann condition on $\psi^0$ as usual.
\begin{equation}
(\psi^0 - i\eta\bar{\psi}^0 )|k=0 \rangle = 0.
\end{equation}
On the other hand, for the super Liouville part, the only sensible condition is 
\begin{equation}
(G_0 + i\eta\bar{G}_0)|\mathrm{Ishibashi;\eta}\rangle = 0 .\label{1}
\end{equation}
Note that once we choose $\eta$, there is no freedom to change the sign in front of $i\bar{G}_0$ as opposed to the free case where we can change the sign by changing Neumann to Dirichlet.

This is the very puzzling point. With the explicit realization of algebra, we can conclude that in the type 0A theory we allow two kinds of Ishibashi states $(\eta = \pm)$ for the super Liouville part and none in 0B theory. We should recall that the difference between the super FZZT brane and the ZZ brane is just the wavefunction (or the coefficient of the Ishibashi states) and not related to the allowed Ishibashi states.\footnote{Of course ZZ (1,1) brane should have $\eta = +1$.} As a result we have the following (perhaps wrong?) conclusion.

In the type 0A theory, we have both Neumann $X$ times FZZT R-R charged brane (two kinds) and R-R charged ZZ brane (one kind). In the type 0B theory, there is no R-R charged brane at all. What a contradictory result compared with the free field analysis! Note, however, that in the $\mathcal{N}=2$ super Liouville theory, which naturally incorporates the time-like boson, has the same brane contents as we have seen in this appendix \cite{Eguchi:2003ik}.


\subsection{T-duality of NS5-brane}\label{b-7}
In this appendix, we review the T-duality property of the NS5-brane solution (and its near horizon geometry). The basic geometrical knowledge can be found in e.g. \cite{Eguchi:1980jx}, \cite{Vafa:1997pm}. Anticipating the result, the T-dual geometry of the NS5-brane is described by the (multi-center) Taub-NUT (Newman-Unti-Tamburino) geometry. Thus we begin with the following Taub-NUT metric:
\begin{eqnarray}
ds^2_{TN} &=& V^{-1}(d\psi + \vec{A}\cdot d\vec{x})^2 + Vd\vec{x}^2 \cr
 V &=& 1+ \frac{1}{|\vec{x}-\vec{y}|} \cr
\nabla V &=& \nabla \times \vec{A}, \label{eq:TaubNUT}
\end{eqnarray}
where the range of $\psi$ is $\psi \in [0,4\pi]$, and vector $\vec{x}$ consists of three dimensional coordinate $(x_7,x_8,x_9)$. It can be easily extended to the multi-center Taub-NUT solution by replacing $V$ with $V = 1+ \sum_i\frac{1}{|\vec{x}-\vec{y_i}|}$. It is worthwhile mentioning that this is the solution of the non-linear Einstein equation (actually the hyper K\"ahler metric) though we take the linear combination of $V$, which is related to the BPS nature of the solution. We sometimes call this geometry as the KK monopole because the $\mathrm{S}^1$ fibered over an $\mathrm{S}^2$ in the base $\mathrm{R}^3$ surrounding a center, is a non-trivial $\mathrm{S}^1$ bundle over $\mathrm{S}^2$ with the first Chern class equal to 1 (for the unicenter Taub-NUT space). If we collide the $k$ centers of the Taub-NUT, the first Chern class becomes $k$ and a singularity appears. This also explains why the Taub-NUT and NS5-brane is related via the T-duality; the former is charged magnetically under  $g_{6,i}$ while the latter is charged magnetically under $B_{6,i}$, and the T-duality, roughly speaking, exchanges $g_{6,i}$ and $B_{6,i}$. 

Note that $\psi$ is a $U(1)$ isometry of this metric, so we perform the T-duality along the $U(1)$ isometry direction. The T-duality rule (for a review see e.g. \cite{Giveon:1994fu}) is summarized as 
\begin{eqnarray}
\tilde{G}_{\psi\psi} &=& G_{\psi\psi}^{-1} \cr
(\tilde{G}+\tilde{B})_{\psi i} &=& -G_{\psi\psi}^{-1} (G+B)_{\psi i} \cr
(\tilde{G}-\tilde{B})_{\psi i} &=& G_{\psi\psi}^{-1} (G-B)_{\psi i} \cr
(\tilde{G}+\tilde{B})_{ij} &=& (G+B)_{ij} - G_{\psi\psi}^{-1} (G-B)_{\psi i} (G+B)_{\psi j} \cr
\tilde{\phi}  + \frac{1}{2}\log(\det\tilde{G}) &=& \phi + \frac{1}{2}\log(\det G),
\end{eqnarray}
where the quantities with a tilde are T-dualized ones. Performing this T-duality to the Taub-NUT metric (\ref{eq:TaubNUT}), we obtain
\begin{eqnarray}
ds^2 &=& V(x) (dx^6 dx^6 + d\vec{x}^2) \cr
e^{2\phi} &=& V(x) =1+ \sum_i \frac{1}{|\vec{y}-\vec{x}_i|} \cr
H_{mns} &=& \epsilon_{mns}^{\ \ \ \ \ r}\partial_r \phi. \label{eq:Ttaub}
\end{eqnarray}
where $x^6$ is the dual coordinate with respect to $\psi$. The Kalb-Ramond field is given by $B_{6i} = A_{i}$. \eqref{eq:Ttaub} is almost the NS5-brane solution, but not quite. This solution is in fact the smeared NS5-brane solution in the $x_6$ direction (because we do not have any structure in the $x_6$ direction). By localizing the $x_6$ direction, we obtain the precise NS5-brane metric (\ref{eq:NS5sol}).\footnote{This smeared brane from the T-duality always occurs when we make D$p$ brane from $\mathrm{D}(p-1)$ brane by the T-duality.}

In order to take the CHS limit (near horizon limit), we neglect the $1$ in the harmonic function $V$. In the T-dualized Taub-NUT geometry, this corresponds to focusing on the singularity of the geometry, which is described by the ALE space. The metric for the ALE space is described by the Gibbons-Hawking metric which is just (\ref{eq:TaubNUT}) with 1 omitted from $V$. Similarly the metric for the multi-ALE space is obtained by colliding the center of the Taub-NUT metric.

To see the singularity, it is convenient to introduce a complex structure into this geometry. The multi-center ALE space is described by the submanifold
\begin{equation}
u^n + \tilde{v}^2 + \tilde{w}^2 = 0, \label{eq:anALE}
\end{equation}
in $\mathbf{C}^3$. We can regard this defining equation as the orbifold singularity as follows.\footnote{One may wonder this geometry has something to do with the free description of the orbifold CFT, but it is important to realize that this is \textit{not} the case. The free orbifold CFT implicitly assumes the existence of a certain amount of the background Kalb-Ramond (twisted) field, which is none in the geometry we are considering here.} Let us take the complex coordinate $z_1$ and $z_2$ which describes $\mathbf{R}^4=\mathbf{C}^1 \otimes \mathbf{C}^1$. Introducing $\omega^n =1$, we identify
\begin{eqnarray}
z_1 &=& \omega z_1 \cr
z_2 &=& \omega^{-1} z_2. \label{eq:orbi}
\end{eqnarray}
It is natural to choose coordinates that are invariant under this action. For this purpose, we define
\begin{eqnarray}
u &=& z_1 z_2 \cr
v &=& z_1^n  \cr
w &=& z_2^n,
\end{eqnarray}
which are clearly invariant under the $\mathbf{Z}_n$ action (\ref{eq:orbi}). However $(u,v,w)$ has the following relation:
\begin{equation}
 u^n = vw.
\end{equation}
If we further define $vw = -\tilde{v}^2 - \tilde{w}^2$, we obtain (\ref{eq:anALE}), which shows the $A_{n-1}$ type singularity. Let us first take the $A_1$ for example and try to deform the singularity. The defining equation is 
\begin{equation}
f = u^2 + \tilde{v}^2 + \tilde{w}^2 = 0.
\end{equation}
We take the real part of this expression; then we have an equation of the 2-sphere with a vanishing radius. This is just the 2-sphere blown up in the Eguchi-Hanson metric which yields the deformation of the singular ALE metric (this is done by changing the K\"ahler metric on the manifold, so it is known as a K\"ahler deformation). Alternatively, we can remove this singularity by deforming the complex structure by
\begin{equation}
f = u^2 + \tilde{v}^2 + \tilde{w}^2 = \mu,
\end{equation}
which eliminates the singular point $u = \tilde{v} = \tilde{w} = 0$ and the manifold becomes smooth. However, if $\mu$ is real, this complex structure deformation  gives the vanishing sphere a finite volume. Thus, the resolution of the singularity by the complex structure deformation and the K\"ahler deformation is not unrelated in this special case (this is related to the fact that the ALE space is actually a hyper K\"ahler manifold and is not true in the more general situation such as the Calabi-Yau threefold).

The resolution of the singularity in the general $A_{n-1}$ singularity is given by
\begin{equation}
\prod_{i=1}^n (u-a_i) +\tilde{v}^2 + \tilde{w}^2 = 0. \label{eq:reso2}
\end{equation}
If we have two equal $a_i$, we have a singularity of the type $A_1$ which shows a vanishing 2-sphere. In general we can associate a vanishing 2-sphere to any pair of $a_i$ (see section \ref{12.2.2} and the arguments therein). However, homologically they are not independent. It can be shown that $n-1$ cycles are homologically independent cycles and can be represented as the $A_{n-1}$ Dynkin diagram, where every independent generator $S_j$ is assigned a node and every intersection between adjacent vanishing spheres is represented as a line connecting them. 

From the resolution (\ref{eq:reso2}) we can see that the multi-ALE singularity occurs when centers of the single ALE space collide. This corresponds to the multi-ALE metric which becomes singular when the sources of the harmonic function collide. In the T-dual language, the singularity comes when $n$ NS5 branes coincide with each other. The resolution center $a_i$ just yields the transverse coordinate of the NS5 brane position.

\subsection{Some Moduli Integrations}\label{b-8}
In this appendix, we derive some formulae of the moduli integration on the torus (see e.g. \cite{Danielsson:1992bx}). The most basic formula, concerning the 2D partition function, is given by
\begin{equation}
\int_{\mathcal{F}} \frac{d\tau^2}{\tau_2^2} = \int_{-\frac{1}{2}}^{\frac{1}{2}} dx \int_{\sqrt{1-x^2}}^\infty \frac{dy}{y^2} = \frac{\pi}{3}.
\end{equation}
For the 1D partition function, we can show the following formula holds:
\begin{equation}
\int_{\mathcal{F}} \frac{d\tau^2}{\tau_2^{3/2}}|\eta(q)|^2 = \frac{\pi}{3\sqrt{6}}.
\end{equation}
The proof of this formula is rather technical.

First, we derive the Euler pentagonal formula:
\begin{equation}
\eta(q) = \sum_{n=-\infty}^{\infty} (-1)^n q^{\frac{3}{2}(n-\frac{1}{6})^2}.
\end{equation}
This can be derived by the Jacobi triple product formula:
\begin{equation}
\prod_{n=1}^{\infty} (1+q^{n-1/2}w)(1+q^{n-1/2}w^{-1}) = \sum_{N=-\infty}^{\infty} \frac{q^{\frac{1}{2}N^2}}{\prod_{n=1}^{\infty} (1-q^{n})}w^N,
\end{equation}
whose physical interpretation (proof) is as follows \cite{Ginsparg:1988ui}. Consider the fermionic system whose energy level is $E = n-\frac{1}{2}, n \in \mathbf{Z}$ and whose number difference between particles and antiparticles is $N=N_+-N_-$. Setting $q=e^{-1/T}$ and $w=e^{\mu/T}$, we have the grand partition function:
\begin{equation}
Z(w,q) = \sum e^{-E/T + \mu N/T} = \prod_{n=1}^{\infty} (1+q^{n-1/2}w)(1+q^{n-1/2}w^{-1}),
\end{equation}
which is nothing but the left-hand side of the Jacobi triple product formula. On the other hand, the grand partition function can be decomposed into the fixed number canonical partition function as
\begin{equation}
\sum_{N=-\infty}^{\infty} w^N Z_N(q),
\end{equation}
which is the right-hand side. Hence, all we want is $Z_N(q)$. Let us first consider $Z_0$. The lowest energy state is given by when the states under the Fermi surface ($E=0$) is all occupied. The number of states with energy $E$ can be enumerated as follows. Take the series of positive integers which satisfy $k_1 \ge k_2 \ge\cdots \ge k_l >0$ and $\sum_i k_i = E $. With these numbers, we construct a state where we excite the first particle under the Fermi surface with $k_1$ units, $\cdots$ and excite the $i$th particle under the Fermi surface with $k_i$ units and so on. This state has the energy $E$ and there is no double counting nor missing. The partition of the positive integer is just the number of ways we divide $E$ into positive integers, whose generating function is just $Z_0$. We find that it is given by
\begin{equation}
Z_0 = \frac{1}{\prod_{n=1}^{\infty} (1- q^n)}.
\end{equation}
With the fermion number $N$, we can similarly obtain the partition function by starting with the new Fermi surface whose height is rising in $N$ units. The corresponding factor is just $q^{N^2/2}$. Thus, we finally obtain
\begin{equation}
Z_N = \frac{q^{N^2/2}}{\prod_{n=1}^{\infty} (1- q^n) }.
\end{equation}
This completes the proof of the Jacobi triple product formula. Substituting $q\to q^3$ and $w\to-q^{-\frac{1}{2}}$, we obtain
\begin{equation}
\prod_{n=1}^\infty(1-q^{3n})(1-q^{3n-2})(1-q^{3n-1}) = \sum_{n=-\infty}^{\infty}(-1)^n q^{3n^2/2-n/2}.
\end{equation}
Rewriting this yields the Euler pentagonal formula:
\begin{equation}
\eta(q) = \sum_{n=-\infty}^{\infty} (-1)^n q^{\frac{3}{2}(n-\frac{1}{6})^2}.
\end{equation}

By making use of this formula, we would like to derive the following equality:
\begin{equation}
|\eta(q)|^2 = \frac{1}{2}\sum_{s,t} \left[q^{\frac{3}{2}(\frac{s}{6}+t)^2}\bar{q}^{\frac{3}{2}(\frac{s}{6}-t)^2} - q^{\frac{3}{2}(\frac{s}{2}+\frac{t}{3})^2}\bar{q}^{\frac{3}{2}(\frac{s}{2}-\frac{t}{3})^2}\right]. \label{eq:st}
\end{equation}
From the Euler  pentagonal formula, we have
\begin{equation}
|\eta(q)|^2 = \sum_{n,m} (-1)^{n+m} q^{\frac{3}{2}(n-1/6)^2}\bar{q}^{\frac{3}{2}(m-1/6)^2}. \label{eq:nm}
\end{equation}
Comparing (\ref{eq:st}) with (\ref{eq:nm}) term by term, we can complete the proof. First, we can solve 
\begin{equation}
\left\{
\begin{array}{l}
(\frac{s}{6} +t )^2 = (n-1/6)^2  \\
(\frac{s}{6} -t )^2 = (m-1/6)^2 
\end{array}
\right.
\end{equation}
to obtain either
\begin{equation}
\left\{
\begin{array}{l}
t = \frac{n-m}{2} \\
s = 3(n+m)-1 
\end{array}
\right.
\end{equation}
or
\begin{equation}
\left\{
\begin{array}{l}
t = \frac{m-n}{2} \\
s = -3(n+m) +1
\end{array}
,\right.
\end{equation}
where $n\pm m$ is even. Thus when $n\pm m$ is even, this contributes to (\ref{eq:nm}). Similarly,
\begin{equation}
\left\{
\begin{array}{l}
t = \frac{3(m+n)-1}{2} \\
s = n-m
\end{array}
\right.
\end{equation}
\begin{equation}
\left\{
\begin{array}{l}
t = \frac{-3(m+n)-1}{2} \\
s = m-n
\end{array}
\right.
\end{equation}
is the solution when $n\pm m$ is odd. These terms explain all terms in (\ref{eq:nm}), but the corresponding terms in (\ref{eq:st}) are the first terms of $s$ odd and not divisible by 3 and the second terms of $s$ odd and $t$ not divisible by 3. Therefore, all the terms left should cancel with each other. The canceling condition can be written as
\begin{equation}
\left\{
\begin{array}{l}
(\frac{s_1}{6}+t_1)^2 = (\frac{s_2}{2}+\frac{t_2}{3})^2 \\
(\frac{s_1}{6}-t_1)^2 = (\frac{s_2}{2}-\frac{t_2}{3})^2.
\end{array}
\right.
\end{equation}
Its solution is given by either $s_1 = s_2$ and $t_2 = 3t_1$ which corresponds to the remaining second term or $s_1 = 2t_2$ and $s_2 = 2t_1$ which corresponds to the remaining first term.

Finally we use the Poisson resummation formula:
\begin{equation}
\sum_{n=-\infty}^{\infty} e^{-\pi n^2 a + 2\pi n ab} = \frac{1}{\sqrt{a}}e^{\pi a b^2} \sum_{m=-\infty}^{\infty} e^{-\frac{\pi m^2}{a} - 2\pi im b},
\end{equation}
whose proof is given as follows. By substituting $1 = \int dr \delta(n-r) $ into the left-hand side and using $\sum_me^{2\pi imr} = \sum_n\delta(n-r)$, we can perform the Gaussian integration to obtain the right-hand side. In this way, $|\eta(q)|^2$ becomes
\begin{equation}
|\eta(q)|^2 = \frac{1}{2\sqrt{\tau_2}}\left(\sqrt{6}\sum_{n,m}e^{-\frac{6\pi}{\tau_2}|n+m\tau|^2} - \sqrt{\frac{3}{2}}\sum_{n,m}e^{-\frac{3\pi}{2\tau_2}|n+m\tau|^2}\right).
\end{equation}
Then we would like to calculate
\begin{equation}
\sum_{n,m} \int_{\mathcal{F}}\frac{d^2\tau}{\tau_2^2} e^{-\frac{\pi x}{2\tau_2}|n+m\tau|^2}.
\end{equation}
The trick here is to exchange the summation over $m$ with the extension of the integration region by the modular transformation \cite{Polchinski:1986zf}. This is equivalent to extending the fundamental region $\mathcal{F}$ to the rectangular $-1/2<\tau_1<1/2$ with positive $\tau_2$. Therefore, we obtain,
\begin{equation}
\sum_{n,m} \int_{\mathcal{F}}\frac{d^2\tau}{\tau_2^2} e^{-\frac{\pi x}{2\tau_2}|n+m\tau|^2} = \int_{\mathcal{F}} \frac{d^2\tau}{\tau_2^2} + 2\int_0^\infty \frac{d\tau_2}{\tau_2^2} \sum_{n>0} e^{-\frac{\pi x}{2\tau_2}n^2} = \frac{\pi}{3}+ \frac{2\pi}{3x},
\end{equation}
where we have used $\sum_{n>0} n^{-2} = \pi^2/6$. Combining all these, we have obtained the desired formula
\begin{equation}
\int_{\mathcal{F}} \frac{d\tau^2}{\tau_2^{3/2}}|\eta(q)|^2 = \frac{\pi}{3\sqrt{6}}.
\end{equation}
\subsection{Projection on Chan-Paton Factor}\label{b-9}
In this appendix, we review the $\Omega$ projection on the Chan-Paton indices. We follow the argument given in section 6.5 of Polchinski's book \cite{Polchinski:1998rq}.

We denote the open string states with a Chan-Paton factor by $|n;ij\rangle$, where $n$ represents excitation level and $i,j$ are Chan-Paton indices. $\Omega$ operation reverses the orientation of string, but it can also rotate the Chan-Paton indices as 
\begin{equation}
\Omega |n;ij\rangle = (-1)^n \gamma_{jj'}|n;j'i'\rangle \gamma^{-1}_{i'i}.
\end{equation}
Acting twice with $\Omega$, we have 
\begin{equation}
\Omega^2 |n;ij\rangle = [(\gamma^T)^{-1}\gamma]_{ii'} |n;i'j'\rangle (\gamma^{-1}\gamma^T)_{j'j}.
\end{equation}
If we insist\footnote{This should be true only on the physical sector. Thus we must be careful when we consider the fermionic string \cite{Gimon:1996rq}.} $\Omega^2 =1$ then we have
\begin{equation}
\gamma^T = \pm \gamma,
\end{equation}
which is the direct consequence of the Schur's lemma. A general change of the Chan-Paton basis:
\begin{equation}
|n;i,j\rangle = U^{-1}_{ii'} |n;i'j'\rangle U_{j'j} 
\end{equation}
transforms $\gamma$ to
\begin{equation}
\gamma' = U^T \gamma U.
\end{equation}
In the symmetric case, it is always possible to find a basis such that $\gamma =1$. Then by using the Hermitian matrix realization of the Chan-Paton indices: $|n;a\rangle = \sum_{i,j} \lambda^a_{ij} |n;ij\rangle$, we have $\lambda^T = \lambda$ for the level $n$ even, and $\lambda^T = -\lambda $ for $n$ odd. This means that the odd level including ``tachyon" transforms as a real symmetric representation of the $SO(N)$ group and the even level including ``Lorentz vector" transforms as an adjoint representation of the $SO(N)$ group.

On the other hand, in the antisymmetric case\footnote{We assume $N$ is an even integer.}, we can choose a basis in which
\begin{equation}
\gamma = iJ \equiv i\begin{bmatrix}
0 & I \cr
-I & 0 \cr
\end{bmatrix}.
\end{equation}
Therefore for even level including ``Lorentz vector" we have $J\lambda^T J = +\lambda$ which transforms as an adjoint representation of the $Sp(N)$ group. For odd level including ``tachyon" we have $J\lambda^T J = -\lambda $ which transforms as a quaternionic real selfdual representation of the $Sp(N)$ group.

Let us finally remark on the connection between the sign in front of the crosscap states and the Chan-Paton projection considered so far. For the $SO(N)$ case, the symmetric part $N(N+1)/2$ of $U(N)$ original Chan-Paton degrees of freedom acquires $+1$ under the $\Omega$ operation while the antisymmetric part $N(N-1)/2$ of $U(N)$ original Chan-Paton degrees of freedom acquires $-1$ under the $\Omega$ operation. Then overall M\"obius amplitude $\mathrm{Tr}_o [\Omega q^{H}]$ obtains a factor of $+N$. Since $N$ comes from the boundary states normalization of $N$ coincident branes, we have $+$ sign for the $SO$ type crosscap state.

For the $Sp(N)$ case, these degeneracies are just reversed\footnote{The easiest way to see this is to recognize that if $J\Lambda$ is symmetric, $\Omega$ is $-$, and if antisymmetric, $\Omega$ is $+$.}, giving $-N$ total factor in the M\"obius amplitude. Since $N$ comes from the boundary states normalization of $N$ branes, we have $-$ sign for the $Sp$ type crosscap state.

\subsection{Jacobian for Diagonalization of $SO/Sp$ Matrix}\label{b-10}
In this appendix, we derive the Jacobian for the change of variables from $SO/Sp$ matrix integral to their eigenvalues. Most of the following argument is directly borrowed from Mehta's book \cite{Mehta}.
\subsubsection{Orthogonal ensemble}
First, it is well-known that any real symmetric matrix $H$ can be decomposed as \begin{equation}
H = U \Theta U^\dagger, \label{eq:odec}
\end{equation}
where $\Theta$ is a real diagonal matrix and $U$ is a real orthogonal matrix. This decomposition cannot be unique, but we parameterize $U$ by $N(N-1)/2$ parameters $p_\mu$ which uniquely determines $H$ with $N$ eigenvalues $\theta_i$. The purpose of this subsection is to obtain the $p$ independent\footnote{This is because in the actual calculation we exclusively consider the gauge invariant sector. Note that this enables us to parameterize the angular variables $p_\mu$ in any desired manner.} part of the Jacobian:
\begin{equation}
J(\theta,p) = \left|\frac{\partial (H_{11},H_{12} \cdots H_{NN})}{\partial (\theta_1\cdots \theta_N,p_1\cdots p_l)}\right|.
\end{equation}
The orthogonality condition of $U$ is given by
\begin{equation}
UU^T = U^T U = 1. \label{eq:oth}
\end{equation}
Differentiating (\ref{eq:oth}) by $p_\mu$ we have
\begin{equation}
\frac{\partial U^T}{\partial p_\mu} U + U^T \frac{\partial U}{\partial p_\mu} = 0.
\end{equation}
Thus we can define antisymmetric matrices $S^{(\mu)}$ as
\begin{equation}
S^{(\mu)} = U^T \frac{\partial U}{\partial p_\mu} = -\frac{\partial U^T}{\partial p_\mu} U. \label{eq:defs}
\end{equation}
On the other hand, differentiating (\ref{eq:odec}) by $p_\mu$, we have
\begin{equation}
\sum_{i,j} \frac{\partial H_{jk}}{\partial p_\mu} U_{j\alpha}U_{k\beta} = S^{(\mu)}_{\alpha\beta} (\theta_\beta -\theta_\alpha).\label{eq:o11}
\end{equation}
Similarly, differentiating (\ref{eq:odec}) by $\theta_\gamma$, we have
\begin{equation}
\sum_{i,j} \frac{\partial H_{jk}}{\partial \theta_\gamma} U_{j\alpha}U_{k\beta} = \delta_{\alpha\beta}\delta_{\alpha\gamma}. \label{eq:o12}
\end{equation}
The Jacobian matrix can be written in the following form:
\begin{equation}
[J(\theta,p)] = 
\begin{bmatrix}
\frac{\partial H_{jj}}{\partial \theta_\gamma} & \frac{\partial H_{jk}}{\partial \theta_{\gamma}} \cr
\frac{\partial H_{jj}}{\partial p_\mu} & \frac{\partial H_{jk}}{\partial p_\mu}\label{eq:jacoo}
\end{bmatrix},
\end{equation}
The two columns in (\ref{eq:jacoo}) correspond to $N$ and $N(N-1)/2$ actual columns: $1 \le j<k \le N$. The two rows in (\ref{eq:jacoo}) correspond again to $N$ and $N(N-1)/2$ actual rows: $\gamma = 1,2, \cdots,N$; $\mu = 1,2, \cdots, N(N-1)/2$. If we multiply the matrix $[J]$ on the right by the $N(N+1)/2 \times N(N+1)/2$ matrix written in partitioned form as
\begin{equation}
[V] = \begin{bmatrix}
(U_{j\alpha} U_{j\beta}) \cr
(2U_{j\alpha} U_{k\beta})
\end{bmatrix},
\end{equation}
where the two rows correspond to $N$ and $N(N-1)/2$ actual rows: $1\le j <k \le N$, and the column corresponds to $N(N+1)/2$ actual columns: $1 \le \alpha \le \beta \le N$, we obtain by using (\ref{eq:o11}) and (\ref{eq:o12})
\begin{equation}
[J][V] = \begin{bmatrix}
\delta_{\alpha\beta}\delta_{\alpha\gamma} \cr
S^{(\mu)}_{\alpha\beta}(\theta_\beta-\theta_\alpha)
\end{bmatrix}.
\end{equation}
Taking the determinant of the both sides, we have
\begin{equation}
J(\theta,p) = \prod_{\alpha<\beta} |\theta_{\beta}-\theta_{\alpha}| f(p).
\end{equation}
Therefore the angular variables $p$ independent part is given by 
\begin{equation}
J(\theta) = \prod_{\alpha<\beta} |\theta_{\beta}-\theta_{\alpha}|.
\end{equation}
\subsubsection{Symplectic ensemble}
Next, we consider the symplectic matrix model. The integral variable is the matrix which satisfies
\begin{equation}
JH^\dagger J = -H,
\end{equation}
where $J$ is the $Sp(2N)$ invariant tensor. In this case, $H$ can be decomposed as 
\begin{equation}
H = U\Theta U^\dagger,
\end{equation}
where $U$ is a symplectic matrix and $\Theta$ is a diagonal matrix whose form is 
\begin{equation}
\Theta = \begin{bmatrix}
\theta_1 & 0       &     &      &      & \cr
0 	&  \theta_1 &	&	&	& \cr
	&	& \theta_2&	0&	& \cr
	&	&	0&	\theta_2&	&\cr
 	&	&	&	& 	\cdots&
\end{bmatrix}.
\end{equation}
We should note that every eigenvalue appears twice. As in the orthogonal case, we parameterize angular variables by $p_\mu$. Then, the Jacobian is given by
\begin{equation}
J(\theta,p) =  \left|\frac{\partial (H^{(0)}_{11},\cdots H^{(0)}_{NN},H^{(0)}_{12},\cdots,H_{12}^{(3)},\cdots, H_{N-1,N}^{(0)},\cdots,H^{(3)}_{N-1,N})}{\partial (\theta_1\cdots \theta_N,p_1\cdots p_{2N(N-1)})}\right|,
\end{equation}
where we have used the quaternionic decomposition,
\begin{equation}
H_{jk} = H^{(0)}_{jk} + H^{(1)}_{jk} e_1 + H^{(2)}_{jk} e_2 + H_{jk}^{(3)} e_3, 
\end{equation}
with $e_i = i\sigma_i$ and 
\begin{equation}
S^{\mu}_{\alpha\beta} = S^{(0\mu)}_{\alpha\beta} + S^{(1\mu)}_{\alpha\beta} e_1 + S^{(2\mu)}_{\alpha\beta} e_2 + S_{\alpha\beta}^{(3\mu)} e_3 ,
\end{equation}
where $S$ is defined in the same way as in (\ref{eq:defs}).

As in the orthogonal case, we can multiply a matrix $V$ which only depends on $U$ so that the $J V$ becomes
\begin{equation}
JV = \begin{bmatrix} \rho_{\gamma,\alpha} & 0 & \cdots & 0 \cr
\epsilon_{\alpha}^\mu & S_{\alpha\beta}^{(0\mu)} (\theta_\beta-\theta_\alpha) &
\cdots &S_{\alpha\beta}^{(3\mu)} (\theta_\beta-\theta_\alpha)
\end{bmatrix},
\end{equation}
where $\rho$ does not depend on $\theta$. Taking the determinant of both sides, we finally obtain
\begin{equation}
J(p,\theta) = \prod_{\alpha<\beta} (\theta_{\alpha} - \theta_{\beta})^4f(p),
\end{equation}
whose $p$ independent part is given by
\begin{equation}
J(\theta) = \prod_{\alpha<\beta} (\theta_{\alpha} - \theta_{\beta})^4.
\end{equation}

\subsection{Sommerfeld Expansion for Various Statistics}\label{b-11}
In this appendix, we derive the Sommerfeld expansion formula, first for the Fermi-Dirac statistics, and then for the more general statistics. The Fermi-Dirac distribution is defined as
\begin{equation}
f(\epsilon) = \frac{1}{1+\exp\left(\frac{\epsilon -\mu}{T}\right)}.
\end{equation}
We would like to derive the low temperature expansion of the following quantity
\begin{equation}
G(T,\mu) = \int_{-\infty}^{\infty} d\epsilon g(\epsilon) f(\epsilon).
\end{equation}
First, we decompose this expression as follows
\begin{equation}
G(T) = \int_{-\infty} ^\mu d\epsilon g(\epsilon) + T\int_0^\infty dy g(\mu-Ty)[f(\mu-Ty)-1] + T\int_0^\infty dy g(\mu+Ty)[f(\mu+Ty)],\label{eq:zmf}
\end{equation}
where we have set $\epsilon = \mu + Tx$ so that $f(\mu+Tx) = \frac{1}{1+\exp(x)}$. Owing to the identity $1-\frac{1}{1+\exp(-y)} = \frac{1}{1+\exp(y)}$, we can rewrite (\ref{eq:zmf}) as
\begin{equation}
G(T) = \int_{-\infty} ^\mu d\epsilon g(\epsilon) + T\int_0^\infty dy \frac{1}{1+\exp(y)}[g(\mu+Ty)-g(\mu-Ty)].
\end{equation}
Expanding $g(\mu+Ty)$ by $T$, we obtain the Sommerfeld expansion formula:
\begin{equation}
G(T) = \int_{-\infty} ^\mu d\epsilon g(\epsilon) +2T\int_0^\infty dy\sum_{n=0}^\infty \frac{y^{2n+1}}{1+e^y}\frac{T^{2n+1}}{(2n+1)!}\left(\frac{\partial}{\partial\mu}\right)^{2n+1}[g(\mu)].
\end{equation}
In fact, the $y$ integration can be performed explicitly,
\begin{equation}
\int_0^\infty dy \frac{y^{2n+1}}{1+\exp(y)} = (2n+1)!\frac{2^{2n+1}-1}{2^{2n+1}}\zeta(2n+2).
\end{equation}

Next let us consider the more general statistics \cite{Wu:1994it}, \cite{Haldane:1991xg}. Probability density with the $g$ statistics is given by
\begin{equation}
n = \frac{1}{\omega(\xi) +g}
\end{equation}
where $\omega(\xi)$ satisfies
\begin{equation}
\omega(\xi)^g (1+\omega(\xi))^{1-g} = \xi \equiv e^{(\epsilon-\mu)/T}.
\end{equation}
For $g = 0$, we obtain the usual Bose-Einstein distribution, while for $g=1$, we obtain the usual Fermi-Dirac distribution. Let us evaluate the general integral\begin{eqnarray}
G(T,\mu) = \int_{-\infty}^\infty d\epsilon g(\epsilon) n(\epsilon)
\end{eqnarray}
by the Sommerfeld low temperature expansion. Setting $\epsilon = \mu +Ty$ as before, we can rewrite this as
\begin{equation}
G(T,\mu) = \int_{-\infty}^\mu d\epsilon g(\epsilon) + T\int_0^\infty dy\left[g(\mu-Ty)[n(\mu-Ty)-1] + g(\mu+Ty)n(\mu+Ty)\right]
\end{equation}
Then, we expand the integrand $g$ by $T$. Noticing that $n(\mu+Ty) = n(y)$ does not depend on $T$, we obtain 
\begin{eqnarray}
G(T,\mu) = \int_{-\infty}^\mu d\epsilon g(\epsilon) &+& T g(\mu) \int_0^\infty dy \left[n(-y) + n(y)-1\right] \cr
&+& T^2 g'(\mu) \int_0^\infty dy y\left[n(-y) - n(y)-1\right] \cr
&+& \frac{T^3}{2} g''(\mu) \int_0^\infty dy y^2\left[n(-y) + n(y)-1\right] + \cdots
\end{eqnarray}
The $y$ integration can be performed. The first order $T$ correction is actually zero (this corresponds to the third law of the thermodynamics). The second order integral becomes $\frac{\pi^2}{6}$, which does not depend on the statistics. The third order integral becomes $2\zeta(3)(1-g)$. The numerical or analytical evaluation of these higher integrals can be found in the literature \cite{TG}, \cite{SS}. This is the Sommerfeld expansion for the general statistics.

\bibliographystyle{utcaps}
\bibliography{liouville}

\end{document}